\documentclass[SIG,biber]{nowfnt} 

\usepackage[utf8]{inputenc}

\usepackage{amsmath}
\usepackage{rotating}
\usepackage{amsthm}
\usepackage{amssymb}
\usepackage{longtable}
\usepackage{subcaption}
\usepackage{bookmark}
\usepackage{overpic}
\definecolor{blue}{RGB}{0,0,255}

\usepackage{graphicx}
\usepackage{multirow}
\usepackage{psfrag}
\usepackage{url}
\usepackage[framemethod=TikZ]{mdframed}
\usepackage{lipsum}
\usepackage[chapter]{algorithm}
\usepackage{algpseudocode}

\mdfdefinestyle{GrayFrame}{%
    outerlinewidth=0pt,
    roundcorner=10pt,
    innertopmargin=\baselineskip,
    innerbottommargin=\baselineskip,
    innerrightmargin=20pt,
    innerleftmargin=4pt,
    backgroundcolor=yellow!25!white}

\newcommand{\mathacr}[1]{\mathsf{#1}}

\newcommand{\vect}[1]{\mathbf{#1}}

\newcommand{\condSum}[3]{\overset{#3}{\underset{\underset{#2}{#1}}{\sum}}}

\def\Psiv{\vect{\Psi}}
\def\diag{\mathrm{diag}}

\def\tr{\mathrm{tr}}

\def\Htran{\mbox{\tiny $\mathrm{H}$}}
\def\Ttran{\mbox{\tiny $\mathrm{T}$}}
\def\CN{\mathcal{N}_{\mathbb{C}}} 
\def\taupu{\tau_{p}}

\def\sigmaDL{\sigma^2_{\mathrm{dl}}} 
\def\sigmaUL{\sigma^2_{\mathrm{ul}}}

\def\imagunit{\mathsf{j}}

\def\cent{\mathsf{c}}
\def\centuatf{\mathsf{c-UatF}}
\def\dist{\mathsf{d}}
\def\cell{\textrm{cellular}}
\def\pmax{p_{\textrm{max}}}
\def\rhomax{\rho_{\textrm{max}}}

\usepackage[acronym, nonumberlist, nopostdot]{glossaries}
\glsdisablehyper 
\usepackage{glossary-tree}
\makenoidxglossaries
\setacronymstyle{long-short}

\renewcommand{\glossarysection}[2][]{}
\setglossarystyle{long}

\newacronym[description=3rd Generation Partnership Project]{3GPP}{3GPP}{3rd Generation Partnership Project}

\newacronym[first={analog-to-digital converter (ADC)}, description={Analog-to-Digital Converter},plural= {ADCs},firstplural={analog-to-digital converters (ADCs)}]{ADC}{ADC}{analog-to-digital converter}

\newacronym[first={digital-to-analog converter (DAC)}, description={Digital-to-Analog Converter},plural= {DACs},firstplural={digital-to-analog converters (DACs)}]{DAC}{DAC}{digital-to-analog converter}

\newacronym[first={angle of arrival (AoA)}, description={Angle of Arrival},plural= {AoAs},firstplural={angles of arrival (AoAs)}]{AoA}{AoA}{angle of arrival}

\newacronym[first={angle of departure (AoD)}, description={Angle of Departure},plural={AoDs},firstplural={angles of departure (AoDs)}]{AoD}{AoD}{angle of departure}

\newacronym[description=Adjacent-Channel Leakage Ratio]{ACLR}{ACLR}{adjacent-channel leakage ratio}

\newacronym[description=Angular Standard Deviation]{ASD}{ASD}{angular standard deviation}

\newacronym[first={access point (AP)},description={Access Point},plural={APs}, firstplural={access points (APs)}]{AP}{AP}{access point}

\newacronym[description=Additive White Gaussian Noise]{AWGN}{AWGN}{additive white Gaussian noise}

\newacronym[description=Binary Phase-Shift Keying]{BPSK}{BPSK}{binary phase-shift keying}

\newacronym[description=Base Station]{BS}{BS}{base station}

\newacronym[description=Bit Error Rate]{BER}{BER}{bit error rate}

\newacronym[description=Code-Division Multiple Access]{CDMA}{CDMA}{code-division multiple access}

\newacronym[description=Circuit Power]{CP}{CP}{circuit power}

\newacronym[description=Cumulative Distribution Function]{CDF}{CDF}{cumulative distribution function}

\newacronym[description=Coordinated Multipoint]{CoMP}{CoMP}{coordinated multipoint}

\newacronym[first={central processing unit (AP)},description={Central Processing Unit},plural={CPUs}, firstplural={central processing units (CPUs)}]{CPU}{CPU}{central processing unit}
\newacronym[description=Cloud Radio Access Network]{C-RAN}{C-RAN}{cloud radio access network}
\newacronym[description=Channel State Information]{CSI}{CSI}{channel state information}

\newacronym[description=Digital Audio Broadcast]{DAB}{DAB}{digital audio broadcast}
\newacronym[description=Dynamic Cooperation Clustering]{DCC}{DCC}{dynamic cooperation clustering}
\newacronym[description=Downlink]{DL}{DL}{downlink}

\newacronym[description=Decibel-Milliwatt]{dBm}{dBm}{decibel-milliwatt}

\newacronym[description=Discrete Fourier Transform]{DFT}{DFT}{discrete Fourier transform}

\newacronym[description= Effective Transmit Power] {ETP} {ETP} {effective transmit power}

\newacronym[description=Energy Efficiency]{EE}{EE}{energy efficiency}

\newacronym[description=Electromagnetic]{EM}{EM}{electromagnetic}

\newacronym[description=Element-Wise MMSE]{EW-MMSE}{EW-MMSE}{element-wise MMSE}

\newacronym[description=Error Vector Magnitude]{EVM}{EVM}{error vector magnitude}

\newacronym[description=Filter Bank Multi-Carrier]{FBMC}{FBMC}{filter bank multi-carrier}

\newacronym[description=Finite Impulse Response]{FIR}{FIR}{finite impulse response}

\newacronym[description=Frequency-Division Duplex]{FDD}{FDD}{frequency-division duplex}

\newacronym[description=Global Positioning System]{GPS}{GPS}{Global Positioning System}

\newacronym[description=Global System for Mobile Communications]{GSM}{GSM}{Global System for Mobile Communications}

\newacronym[description=Hardware Efficiency]{HE}{HE}{hardware efficiency}

\newacronym[description=Hybrid Automatic Repeat Request]{HARQ}{HARQ}{hybrid automatic repeat request}

\newacronym[description=Information and communication technology]{ICT}{ICT}{information and communication technology}

\newacronym[description=Independent and Identically Distributed]{iid}{i.i.d.\@}{independent and identically distributed}

\newacronym[description=In-Phase/Quadrature]{I/Q}{I/Q}{in-phase/quadrature}

\newacronym[description=Institute of Electrical and Electronics Engineers]{IEEE}{IEEE}{Institute of Electrical and Electronics Engineers}

\newacronym[description=Joint Spatial-Division and Multiplexing]{JSDM}{JSDM}{joint spatial-division and multiplexing}

\newacronym[description=Joint Processing]{JP}{JP}{joint processing}

\newacronym[description=Karush-Kuhn-Tucker]{KKT}{KKT}{Karush-Kuhn-Tucker}

\newacronym[description=Large-Scale Fading Decoding]{LSFD}{LSFD}{large-scale fading decoding}

\newacronym[description=Large-Scale Fading Precoding]{LSFP}{LSFP}{large-scale fading precoding}

\newacronym[description=Linear MMSE]{LMMSE}{LMMSE}{linear MMSE}

\newacronym[description=Line-of-Sight]{LoS}{LoS}{line-of-sight}

\newacronym[description=Least-Squares]{LS}{LS}{least-squares}

\newacronym[description=Local MMSE]{L-MMSE}{L-MMSE}{Local MMSE}

\newacronym[description=Local P-MMSE]{LP-MMSE}{LP-MMSE}{local P-MMSE}

\newacronym[description=Local Oscillator]{LO}{LO}{local oscillator}

\newacronym[description=Long Term Evolution]{LTE}{LTE}{Long Term Evolution}

\newacronym[description=Millimeter Wavelength]{mmWave}{mmWave}{millimeter wavelength}

\newacronym[description=Multiple-Input Multiple-Output]{MIMO}{MIMO}{multiple-input multiple-output}

\newacronym[description=Multi-User Multiple-Input Multiple-Output]{MU-MIMO}{MU-MIMO}{multi-user multiple-input multiple-output}

\newacronym[description=Multiple-Input Single-Output]{MISO}{MISO}{multiple-input single-output}

\newacronym[description=Minimum Mean-Squared Error]{MMSE}{MMSE}{minimum mean-squared error}

\newacronym[description=Multicell Minimum Mean-Squared Error]{M-MMSE}{M-MMSE}{multicell minimum mean-squared error}

\newacronym[description=Mean-Squared Error]{MSE}{MSE}{mean-squared error}

\newacronym[description=Maximum Ratio]{MR}{MR}{maximum ratio}

\newacronym[description=Normalized MSE]{NMSE}{NMSE}{normalized MSE}

\newacronym[description=Non-Line-of-Sight]{NLoS}{NLoS}{non-line-of-sight}
\newacronym[description=Nearly Optimal]{n-opt}{n-opt}{nearly optimal}

\newacronym[description=Optimal]{opt}{opt}{optimal}

\newacronym[description=Max-Min Fairness]{MMF}{MMF}{}

\newacronym[description=Fractional Power Allocation]{FPA}{FPA}{}

\newacronym[description=Fractional Power Control]{FPC}{FPC}{}

\newacronym[description=Orthogonal Frequency-Division Multiplexing]{OFDM}{OFDM}{orthogonal frequency-division multiplexing}

\newacronym[description=Partial MMSE]{P-MMSE}{P-MMSE}{partial MMSE}

\newacronym[description=Partial Regularized Zero-Forcing]{P-RZF}{P-RZF}{partial regularized zero-forcing}

\newacronym[description=Power Amplifier]{PA}{PA}{power amplifier}

\newacronym[description=Probability Density Function]{PDF}{PDF}{probability density function}

\newacronym[description=Phase-Shift Keying]{PSK}{PSK}{phase-shift keying}

\newacronym[description=Poisson Point Process]{PPP}{PPP}{Poisson point process}

\newacronym[description=Power Consumption]{PC}{PC}{power consumption}

\newacronym[description=Quadrature Amplitude Modulation]{QAM}{QAM}{quadrature amplitude modulation}

\newacronym[description=Quality of Service]{QoS}{QoS}{quality of service}
\newacronym[description=Quality of Experience]{QoE}{QoE}{Quality of Experience}

\newacronym[description=Random Access]{RA}{RA}{random access}

\newacronym[description=Random Access Channel]{RACH}{RACH}{random access channel}

\newacronym[description= Small Cell]{SC}{SC}{small cell}

\newacronym[description= Small-Cell Base Station]{SBS}{SBS}{small-cell base station}

\newacronym[description=Space-Division Multiple Access]{SDMA}{SDMA}{space-division multiple access}

\newacronym[description=Radio Frequency]{RF}{RF}{radio frequency}

\newacronym[first={spectral efficiency (SE)}, description=Spectral Efficiency, plural={SEs},firstplural={spectral efficiencies (SEs)}]{SE}{SE}{spectral efficiency}

\newacronym[description=Symbol Error Rate]{SER}{SER}{symbol error rate}

\newacronym[description=Signal-to-Interference-plus-Noise Ratio]{SINR}{SINR}{signal-to-interference-plus-noise ratio}

\newacronym[description=Signal-to-Interference Ratio]{SIR}{SIR}{signal-to-interference ratio}

\newacronym[description=Single-Input Single-Output]{SISO}{SISO}{single-input single-output}

\newacronym[description=Single-Input Multiple-Output]{SIMO}{SIMO}{single-input multiple-output}

\newacronym[description=Single-Cell Minimum Mean-Squared Error]{S-MMSE}{S-MMSE}{single-cell minimum mean-squared error}

\newacronym[description=Signal-to-Noise Ratio]{SNR}{SNR}{signal-to-noise ratio}

\newacronym[description=Time-Division Duplex]{TDD}{TDD}{time-division duplex}

\newacronym[description=Time-division multiple access]{TDMA}{TDMA}{time-division multiple access}

\newacronym[description= Area Transmit Power] {ATP} {ATP} {area transmit power}

\newacronym[description=Macro User Equipment]{MUE}{MUE}{macro user equipment}

\newacronym[description=Small-Cell User Equipment]{SUE}{SUE}{small-cell user equipment}

\newacronym[description=User Equipment]{UE}{UE}{user equipment}

\newacronym[description=Use-and-then-Forget]{UatF}{UatF}{use-and-then-forget}

\newacronym[description=Uplink]{UL}{UL}{uplink}

\newacronym[description=Urban Microcell]{UMi}{UMi}{urban microcell}

\newacronym[description=Uniform Circular Array]{UCA}{UCA}{uniform circular array}
\newacronym[description=Uniform Linear Array]{ULA}{ULA}{uniform linear array}
\newacronym[description=Universal Mobile Telecommunications System]{UMTS}{UMTS}{Universal Mobile Telecommunications System}

\newacronym[description=Trademark used for \gls{WLAN}]{WiFi}{WiFi}{WiFi}

\newacronym[description=Reverse Time-Division Duplex]{RTDD}{RTDD}{reverse time-division duplex}

\newacronym[description=Regularized Zero-Forcing]{RZF}{RZF}{regularized zero-forcing}

\newacronym[description=Cross-Polar Isolation]{XPI}{XPI}{cross-polar isolation}

\newacronym[description=Cross-Polar Discrimination]{XPD}{XPD}{cross-polar discrimination}

\newacronym[description=Zero-Forcing]{ZF}{ZF}{zero-forcing}

\newacronym[description=Wireless Local Area Network]{WLAN}{WLAN}{wireless local area network}

\title{Foundations of User-Centric Cell-Free Massive MIMO}

\maintitleauthorlist{
\"{O}zlem Tu\u{g}fe Demir \\
KTH Royal Institute of Technology \\
and Link\"{o}ping University \\
ozlemtd@kth.se
\and
Emil Bj\"{o}rnson \\
KTH Royal Institute of Technology \\
and Link\"{o}ping University \\
emilbjo@kth.se
\and
Luca Sanguinetti \\
University of Pisa \\
luca.sanguinetti@unipi.it
}

\issuesetup
{%
 copyrightowner={\"{O}. T. Demir and E. Bj\"{o}rnson and L. Sanguinetti},
 volume        = 14,
 issue         = 3-4,
 pubyear       = 2021,
 isbn          = 978-1-68083-790-2,
 eisbn         = 978-1-68083-791-9,
 doi           = {10.1561/2000000109},
 firstpage     = 162, 
 lastpage      = 472
 }

\usepackage{mwe}

\author[1]{Demir, \"{O}zlem Tu\u{g}fe}
\author[2]{Bj\"{o}rnson, Emil}
\author[3]{Sanguinetti,Luca}

\affil[1]{KTH Royal Institute of Technology and Link\"{o}ping University; ozlemtd@kth.se}
\affil[2]{KTH Royal Institute of Technology and Link\"{o}ping University; emilbjo@kth.se}
\affil[3]{University of Pisa; luca.sanguinetti@unipi.it\vspace{-0.5cm}}

\articledatabox{\nowfntstandardcitation}

\begin{document}

\makeabstracttitle
\begin{abstract}Imagine a coverage area where each mobile device is communicating with a preferred set of wireless access points (among many) that are selected based on its needs and cooperate to jointly serve it, instead of creating autonomous cells. This effectively leads to a user-centric post-cellular network architecture, which can resolve many of the interference issues and service-quality variations that appear in cellular networks. This concept is called User-centric Cell-free Massive MIMO (multiple-input multiple-output) and has its roots in the intersection between three technology components: Massive MIMO, coordinated multipoint processing, and ultra-dense networks. The main challenge is to achieve the benefits of cell-free operation in a practically feasible way, with computational complexity and fronthaul requirements that are scalable to enable massively large networks with many mobile devices. 
This monograph covers the foundations of User-centric Cell-free Massive MIMO, starting from the motivation and mathematical definition. It continues by describing the state-of-the-art signal processing algorithms for channel estimation, uplink data reception, and downlink data transmission with either centralized or distributed implementation.
The achievable spectral efficiency is mathematically derived and evaluated numerically using a running example that exposes the impact of various system parameters and algorithmic choices.
The fundamental tradeoffs between communication performance, computational complexity, and fronthaul signaling requirements are thoroughly analyzed.
Finally, the basic algorithms for pilot assignment, dynamic cooperation cluster formation, and power optimization are provided, while open problems related to these and other resource allocation problems are reviewed. 
All the numerical examples can be reproduced using the accompanying Matlab code.
\end{abstract}


\chapter{Introduction and Motivation}
\label{c-introduction} 

The purpose of mobile networks is to provide devices with wireless access to a variety of data services anywhere in a wide geographical area. For many years, the main service of these networks was voice calls, but nowadays transmission of data packets is the dominant service \cite{EricssonMobility}. Hence, the service quality of contemporary networks is mainly determined by the data rate (measured in bit per second) that can be delivered at different locations in the coverage area. The range of  wireless transmission is determined by the propagation environment. Since the received signal power decays quadratically, or even faster, with the propagation distance, a traditional mobile network infrastructure consists of a set of geographically distributed transceivers that the connecting device can choose between. These are typically deployed at elevated locations (e.g., in masts and at rooftops) to provide unobstructed propagation to many places in the area. Each transceiver will be called an \emph{\gls{AP}} and each user device will be called a \emph{\gls{UE}} in this monograph.

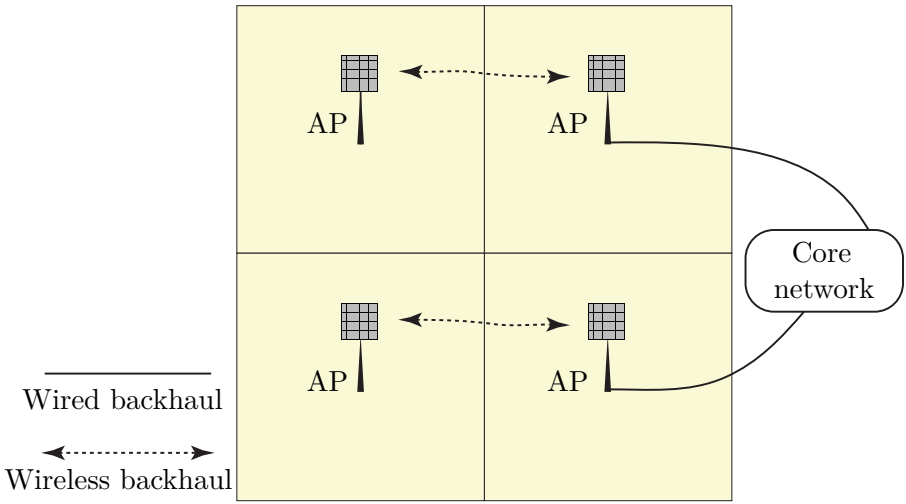
\begin{figure}[t!]
	\centering 
	\begin{overpic}[width=\columnwidth,tics=10]{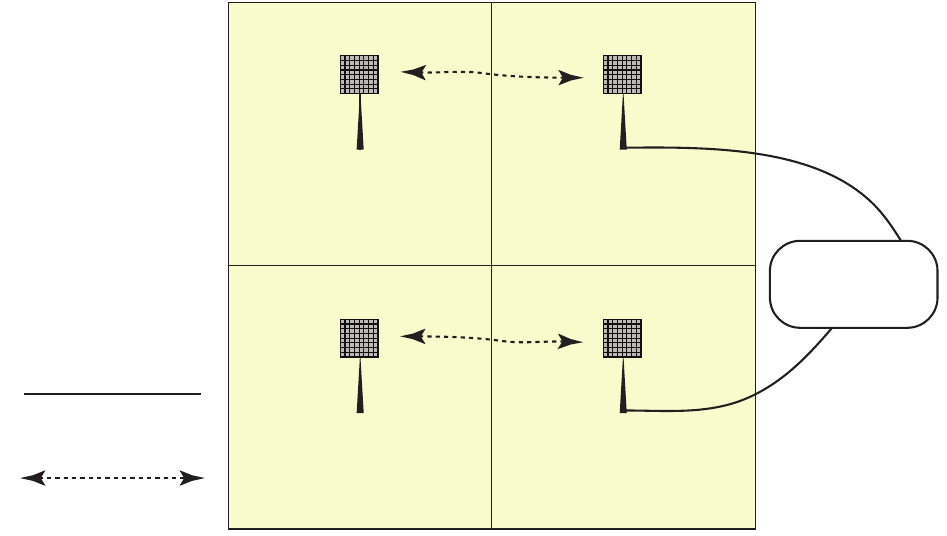}
		\put(59,13){AP}
		\put(59,42){AP}
		\put(32,13){AP}
		\put(32,42){AP}
		\put(0,11){Wired backhaul}
		\put(-2,2){Wireless backhaul}
		\put(86.5,27.5){Core}
		\put(84.5,23.5){network}
	\end{overpic} 
	\caption{Cellular network with four \glspl{AP}, which are all connected to the core network via wired backhaul. Some \glspl{AP} are also interconnected by wireless backhaul.} 
	\label{fig:cellular-basics} 
\end{figure}

Current mobile networks are built as \emph{cellular networks}, which means that each \gls{UE} connects to one \gls{AP}, namely the one that provides the strongest signal. The \gls{UE} locations for which a particular \gls{AP} is selected is called a \emph{cell}. Figure~\ref{fig:cellular-basics} shows the basic infrastructure of a cellular network with four \glspl{AP}, each equipped with a planar antenna array containing both the antenna elements and the associated radio units (also known as transceiver chains). The antenna elements emit and receive \gls{RF} waves, while the radios generate the analog RF signals to be emitted and process the received RF signals. The radios are connected to a baseband unit that processes the transmitted and received signals in the digital domain. This monograph is focused on the digital signal processing associated with the baseband, thus we will simply refer to each radio and its associated antenna element(s) as \emph{an antenna}. The exact hardware implementation is thereby abstracted away. It is the number of such antennas that determines the dimensionality of the signals that will be generated and processed in the baseband.

The square area around each \gls{AP} illustrates the cell that the AP provides service to. In reality, the cells will not have symmetric shapes (such as squares, triangles, or hexagons), but it is commonly illustrated like that when describing the fundamentals. The infrastructure of contemporary cellular networks can be divided into two parts: an edge and a core. The edge consists of \glspl{AP} and other hardware units that are directly involved in the physical-layer communication with the \glspl{UE}. The core network facilitates all the services requested by the UEs, including routing of data packages and connection to the Internet. The connections between the edge and core are called \emph{backhaul} links and can either be fully wired (e.g., using fiber cables) or partially wireless (e.g., using fixed microwave links). Figure~\ref{fig:cellular-basics} shows an example where the \glspl{AP} to the right are connected via wired backhaul links to the core network. The \glspl{AP} to the left are connected wirelessly to the  \glspl{AP} to the right, thus their backhaul traffic flows over both wireless and wired links.

\begin{figure}[t!]
	\centering 
        \begin{subfigure}[b]{\columnwidth} \centering
	\begin{overpic}[width=.8\columnwidth,tics=10]{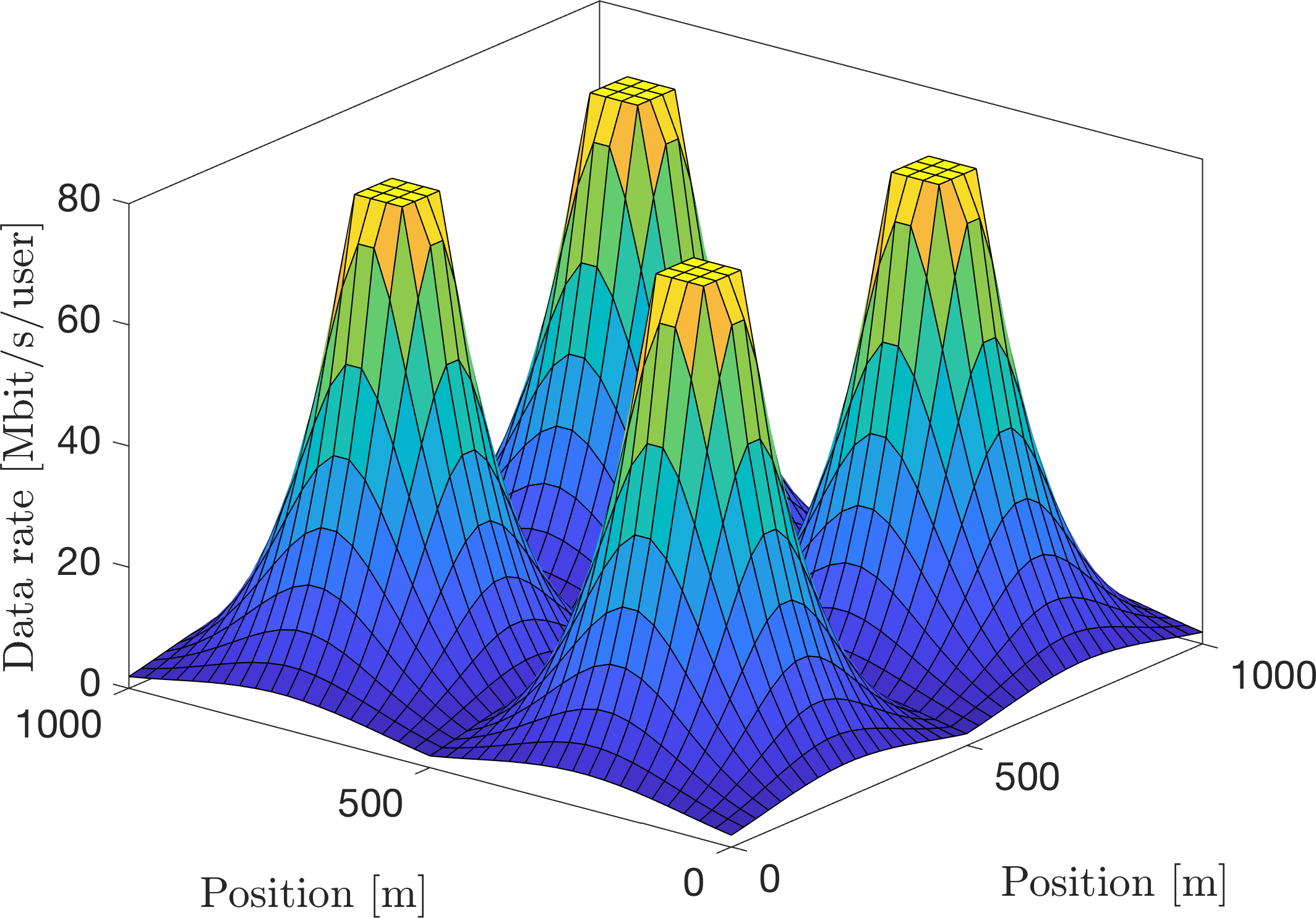}
\end{overpic} 
                \caption{Each \gls{AP} has a 9\,dBi fixed-gain antenna.}    \vspace{+2mm}
        \end{subfigure} 
        \begin{subfigure}[b]{\columnwidth} \centering 
	\begin{overpic}[width=.8\columnwidth,tics=10]{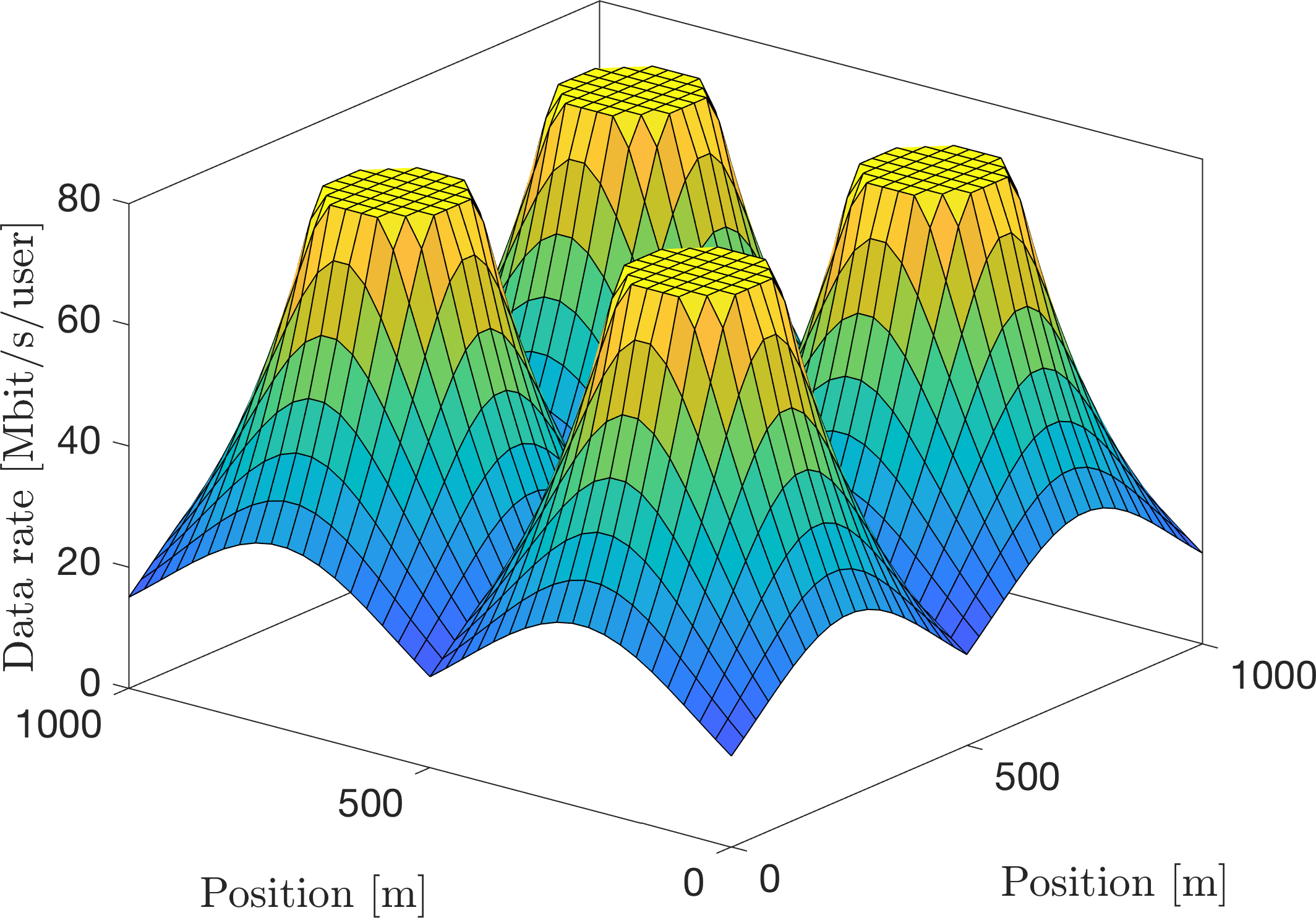}
\end{overpic} 
                \caption{Each \gls{AP} is equipped with 64 omni-directional antennas.} 
        \end{subfigure} 
	\caption{Example of the downlink data rate achieved by a \gls{UE} at different locations in the cellular network in Figure~\ref{fig:cellular-basics}, assuming each \gls{AP} transmits with full power.
	The cell-edge SNR is 0\,dB in (a) and the power is assumed to decay as the distance to the power of four. The bandwidth is 10\,MHz, and the maximum \gls{SE} is 8\,bit/s/Hz. The key observation is that the rates vary substantially in the network.} 
	\label{fig:cellular-rates} 
\end{figure}

\enlargethispage{\baselineskip} 

An important consequence of the fact that the received signal power rapidly decays with the propagation distance is that the \glspl{UE} that happen to be close to an \gls{AP} (i.e., in the cell center) will experience a higher \gls{SNR} than those that are close to the edge between two cells. A $10\,000$ times (40\,dB) difference is common between the cell center and cell edge. Moreover, \glspl{UE} at the cell edge are also affected by interference from neighboring \glspl{AP}, thus the \gls{SINR} can be substantially lower than the \gls{SNR} at these locations. The data rate is an increasing function of the \gls{SINR}, thus there are large rate variations in each cell. Figure~\ref{fig:cellular-rates}(a) exemplifies this behavior by showing the data rate achieved in the downlink by a \gls{UE} at different locations, when each \gls{AP} uses a traditional fixed-gain antenna and transmits with maximum power. When the \gls{UE} is close to one of the \glspl{AP}, it achieves the maximum rate that is supported by the system, which is 80\,Mbit/s in this example. In contrast, \glspl{UE} at the cell edges achieve rates below 1\,Mbit/s. This is insufficient for many data services but is nevertheless enough for making voice calls. Depending on the codec, a voice call requires as little as 10-100\,kbit/s and this is supported everywhere in this example.
Cellular networks were initially designed with this property in mind; we needed the \gls{SNR} to be above a threshold everywhere in the coverage area to prevent dropped calls, but there was no benefit from being far above that threshold. This basic property has  changed entirely when we started using cellular technology for data transmission.
Since the \glspl{UE} request the same data services everywhere in the coverage area, cell-center \glspl{UE} only need to be connected part of the time, while the cell-edge \glspl{UE} must be turned on for a much larger fraction of time (if the requested service can even be provisioned). Hence, at a given time instance, the majority of active \glspl{UE} are at the cell edges and their performance will determine how the customers perceive the service quality of the network as a whole.

The large data rate variations are inherent to the cellular network architecture and remain even if the \glspl{AP} are equipped with advanced hardware, such as \emph{Massive \gls{MIMO}} \cite{massivemimobook}, \cite{Marzetta2010a}, \cite{Marzetta2016a}. The \gls{MIMO} technology enables each \gls{AP} to use an array of antennas (with integrated radios) to serve multiple \glspl{UE} in its cell by directional transmission, which also increases the \gls{SNR} and reduces inter-cell interference.
More precisely, in the uplink, multiple \glspl{UE} transmit data to the \glspl{AP} in the same time-frequency resource. The \glspl{AP} exploit the massive number of channel observations (made on the receive antennas) to apply linear receive combining, which discriminates the desired signal from the interfering signals using the spatial domain. 
In the downlink, the \glspl{UE} are coherently served by all the antennas, in the same time-frequency resource, but separated in the spatial domain by receiving very directive signals. 
 Figure~\ref{fig:cellular-rates}(b) shows the downlink data rate achieved by a \gls{UE} at different locations when each \gls{AP} has an array of 64 antennas. The data rates are generally higher than in Figure~\ref{fig:cellular-rates}(a). The cell-center area where the maximum data rate is delivered grows and large improvements are also seen at the  cell-edge \glspl{UE}, since beamforming from the antenna array at the \gls{AP} can increase the \gls{SNR} without increasing the inter-cell interference. Despite these gains, there are still substantial rate variations in each cell. Each AP could, in principle, optimize its transmit power to even out the differences (e.g., by reducing the power when serving UEs in the cell center) but this is undesirable since it results in serving all the UEs using the relatively low rates that can be delivered at the cell edge.
 
 \enlargethispage{\baselineskip} 
 
Current cellular networks can achieve high peak data rates in the cell centers, but the large variations within each cell make the service quality unreliable. Even if the rates are sufficiently high at, say, 80\% of the locations in a cell, this is not sufficient when we are creating a society where wireless access is supposed to be ubiquitous. When payments, navigation, entertainment, and control of autonomous vehicles are all relying on wireless connectivity, we must raise the uniformity of the data service quality. 
In summary, the primary goal for future mobile networks should not be to increase the peak rates, but the rates that can be guaranteed to the very vast majority of the locations in the geographical coverage area. 
The cellular network architecture was not designed for high-rate data services but for low-rate voice services, thus it is time to look beyond the cellular paradigm and make a clean-slate network design that can reach the performance requirements of the future. 
This monograph considers the cell-free network architecture that is designed to reach the aforementioned goal of uniformly high data rates everywhere.

The cell-free concept for wireless communication networks is defined in Section~\ref{sec:cell-free-networks}, which briefly describes how to operate such networks. Section~\ref{sec:historical-background} puts the new technology into a historical perspective. Section~\ref{sec:benefits-cellular-networks} describes three basic benefits that cell-free networks have compared to   cellular networks. The key points are summarized in Section~\ref{c-introduction-definition-summary}.

\section{Cell-Free Networks}\label{sec:cell-free-networks}

We will now describe the basic architecture and terminology of a  cell-free network. The system and channel propagation models, including the mathematical notation, will be introduced in Section~\ref{c-cell-free-definition} on p.~\pageref{c-cell-free-definition}.

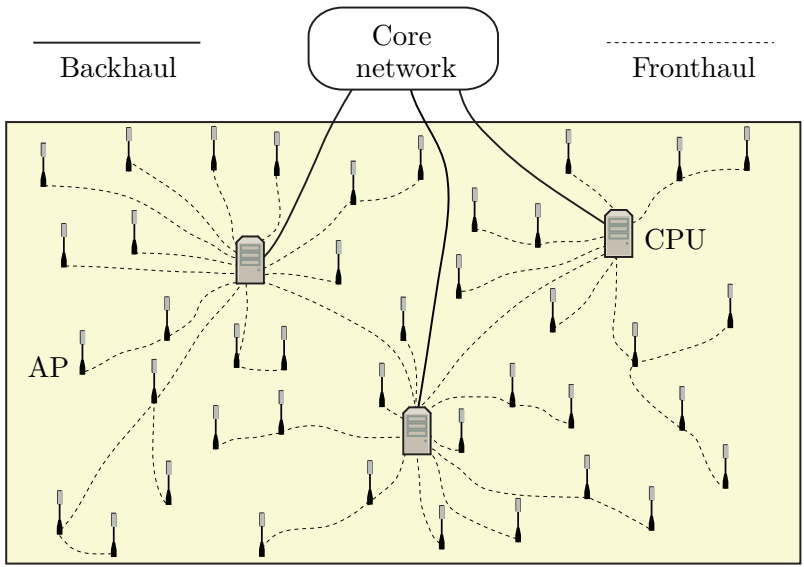
\begin{figure}[t!]
	\centering 
	\begin{overpic}[width=0.9\columnwidth,tics=10]{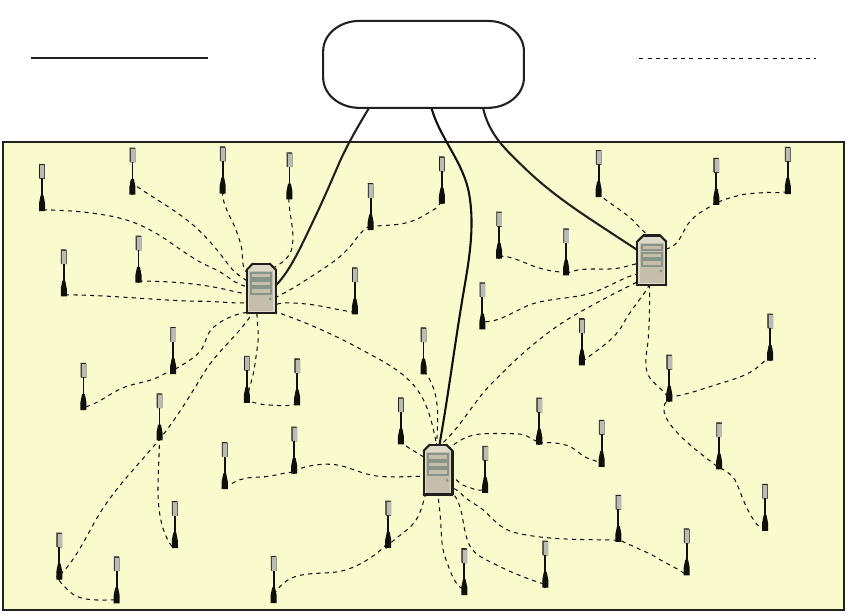}
		\put(80,40){CPU}
		\put(3,24){AP}
		\put(7,61.5){Backhaul}
		\put(78.5,61.5){Fronthaul}
		\put(46,65.5){Core}
		\put(44,61.5){network}
	\end{overpic} 
	\caption{Illustration of a cell-free network with many geographically distributed APs connected to CPUs via fronthaul links. The CPUs are connected to the core network via backhaul links. The \glspl{AP} are jointly serving all the \glspl{UE} in the coverage area.} 
	\label{fig:cell-free} 
\end{figure}

A cell-free network consists of $L$ geographically distributed \glspl{AP} that are jointly serving the \glspl{UE} that reside in the area. Each \gls{AP} is connected via a \emph{fronthaul} to a \gls{CPU}, which is responsible for the \gls{AP} cooperation. There can be multiple \glspl{CPU} all connected via fronthaul links, which can be wired or wireless. An illustration of a cell-free network with single-antenna \glspl{AP} is provided in Figure~\ref{fig:cell-free}. A cell-free network can be divided into an edge and a core, just as cellular networks. The \glspl{AP} and \glspl{CPU} are at the edge and the connections between them are called fronthaul links, while the connections between the edge and core are still called backhaul links. Hence, the \glspl{CPU} are connected to the core network via backhaul links, which are used to send/receive data from the Internet and other sources, to facilitate various data services. In contrast, the fronthaul links can be used for: 1) sharing physical-layer signals that will be transmitted in the downlink; 2) forwarding received uplink data signals that are yet to be decoded; and 3) sharing \gls{CSI} related to the physical channels. The fronthaul also facilitates phase-synchronization between geographically distributed \glspl{AP}, for example, by providing a common phase reference.

A particular fronthaul topology is illustrated in Figure~\ref{fig:cell-free}, where some \glspl{AP} are directly connected to a \gls{CPU} while other \glspl{AP} are connected via a neighboring \gls{AP}.
We stress that this is only for illustration purposes. No specific assumption on the topology will be made in this monograph, 
except that the fronthaul links exist, have infinite capacity, negligible latency, and introduce no errors. This allows us to quantify~the ultimate physical-layer performance of the cell-free network architecture. 
Practical constraints on the fronthaul infrastructure are briefly reviewed in Section~\ref{sec:implementation-constraints} on p.~\pageref{sec:implementation-constraints}.
We also note that a \gls{CPU} may not be a separate physical unit but may be viewed as a logical entity; for example, the \glspl{CPU} may represent a set of local processors that can be either located at a subset of the \glspl{AP} or at other physical locations, and which are connected via fronthaul links. 
Aligned with the ongoing cloudification of wireless networks \cite{6882182}, \cite{Peng2016a}, known as \emph{\gls{C-RAN}}, the \gls{CPU}-related processing tasks can be distributed between the local processors in different ways \cite{Bjornson2013d}.

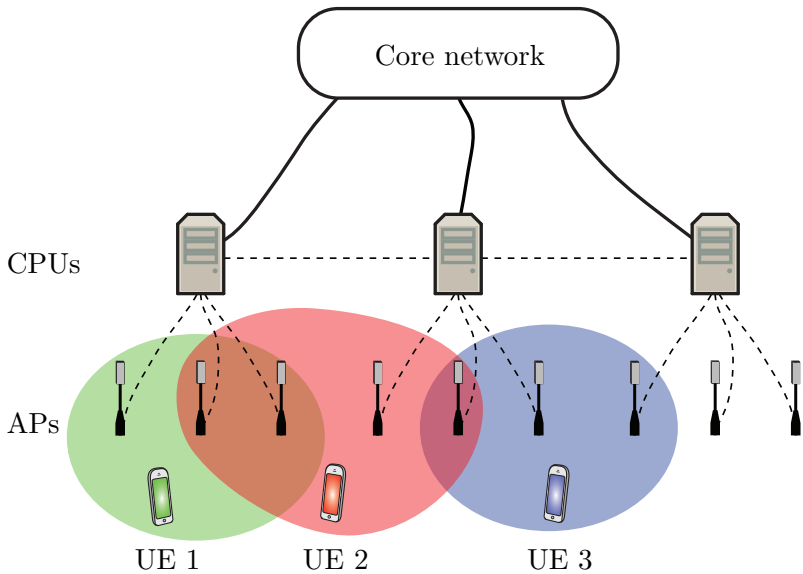
\begin{figure}[t!]
	\centering 
	\begin{overpic}[width=0.9\columnwidth,tics=10]{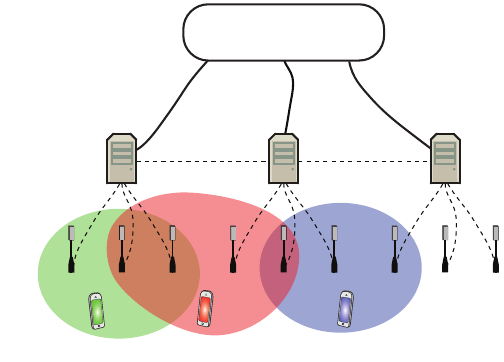}
		\put(46,64){Core network}
		\put(0,38){CPUs}
		\put(0,18){APs}
		\put(16,1){UE 1}
		\put(37,1){UE 2}
		\put(65,1){UE 3}
	\end{overpic} 
	\caption{Illustrations of the different layers in a cell-free network. Each \gls{UE} connects to a subset of the \glspl{AP}, which is illustrated by the shaded regions. Each \gls{AP} is connected to one \gls{CPU} via fronthaul. The \glspl{CPU} are interconnected either directly or via the core network.} 
	\label{fig:cell-free_distribution_processing} 
\end{figure}

Generally speaking, C-RAN is a network deployment architecture where a group of APs is connected to the same CPU, which carry out most of the APs' baseband processing. By sharing computational resources, the total computational capacity can be reduced since it is unlikely that all APs need the maximum capacity simultaneously. One can also make use of general-purpose hardware and open protocols. Recently, the C-RAN abbreviation has started to stand for \emph{centralized RAN}, since the word ``cloud'' gives the impression that the CPU is owned by another vendor than the wireless network and can be located anywhere in the world. However, to meet the latency constraints of baseband processing, the CPU is rather an edge-cloud processor located in the same geographical area as the APs.
Many different physical-layer technologies can be implemented using the C-RAN architecture. So far, it has mainly been used for cellular networks but it is also the foundation for cell-free networks.
Figure~\ref{fig:cell-free_distribution_processing} gives a schematic view of a cell-free network that uses the C-RAN architecture. It is divided into different layers: the core network, the \gls{CPU} layer, the \gls{AP} layer, and the \gls{UE} layer.
Each \gls{UE} is served by a subset of the \glspl{AP}, for example, all the neighboring ones. These subsets are illustrated by the shaded regions in Figure~\ref{fig:cell-free_distribution_processing}.
For each \gls{UE}, one of the selected \glspl{AP} is the so-called \emph{Master \gls{AP}} that is responsible for serving the \gls{UE} and appointing a \gls{CPU} where the uplink data decoding and downlink data encoding will be carried out. That \gls{CPU} delivers the downlink data to all \glspl{AP} that are transmitting to the \gls{UE} and combines/fuses the uplink received signals obtained at those \glspl{AP} in a final decoding step. 
A \gls{UE} can be served by \glspl{AP} connected to different \glspl{CPU}; there exists a fronthaul link between every pair of \glspl{AP} even if it might go via other entities.
The signal processing required for communication can be divided between the \glspl{AP} and \gls{CPU} in different ways, which will be explored in later sections of this monograph. 
As the \gls{UE} moves around, the Master \gls{AP} assignment, selection of \gls{CPU}, and selection of cooperating \glspl{AP} may change dynamically.

\enlargethispage{\baselineskip} 

The word ``cell-free'' signifies that no cell boundaries exist from a \gls{UE} perspective during uplink and downlink transmission since all \glspl{AP} that affect a \gls{UE} will take an active part in the communication. For example, when a \gls{UE} transmits an uplink data signal then all \glspl{AP} that receive it, with an \gls{SNR} that is above a threshold, will collaborate in decoding the signal. The partially overlapping shaded regions in Figure~\ref{fig:cell-free_distribution_processing} can be created in that way.
The network is jointly serving all the $K$ \glspl{UE} that are active in the coverage area of the network, even if not all \glspl{AP} might serve every single \gls{UE}. The differences between cellular and cell-free networks exist at the infrastructure and signal processing side, but can be transparent to the \glspl{UE}. It should be possible for the same \gls{UE} to connect to both types of networks without upgrading its software.

\begin{figure}[t!]
	\centering 
	\begin{subfigure}[b]{\columnwidth} \centering
		\begin{overpic}[width=.8\columnwidth,tics=10]{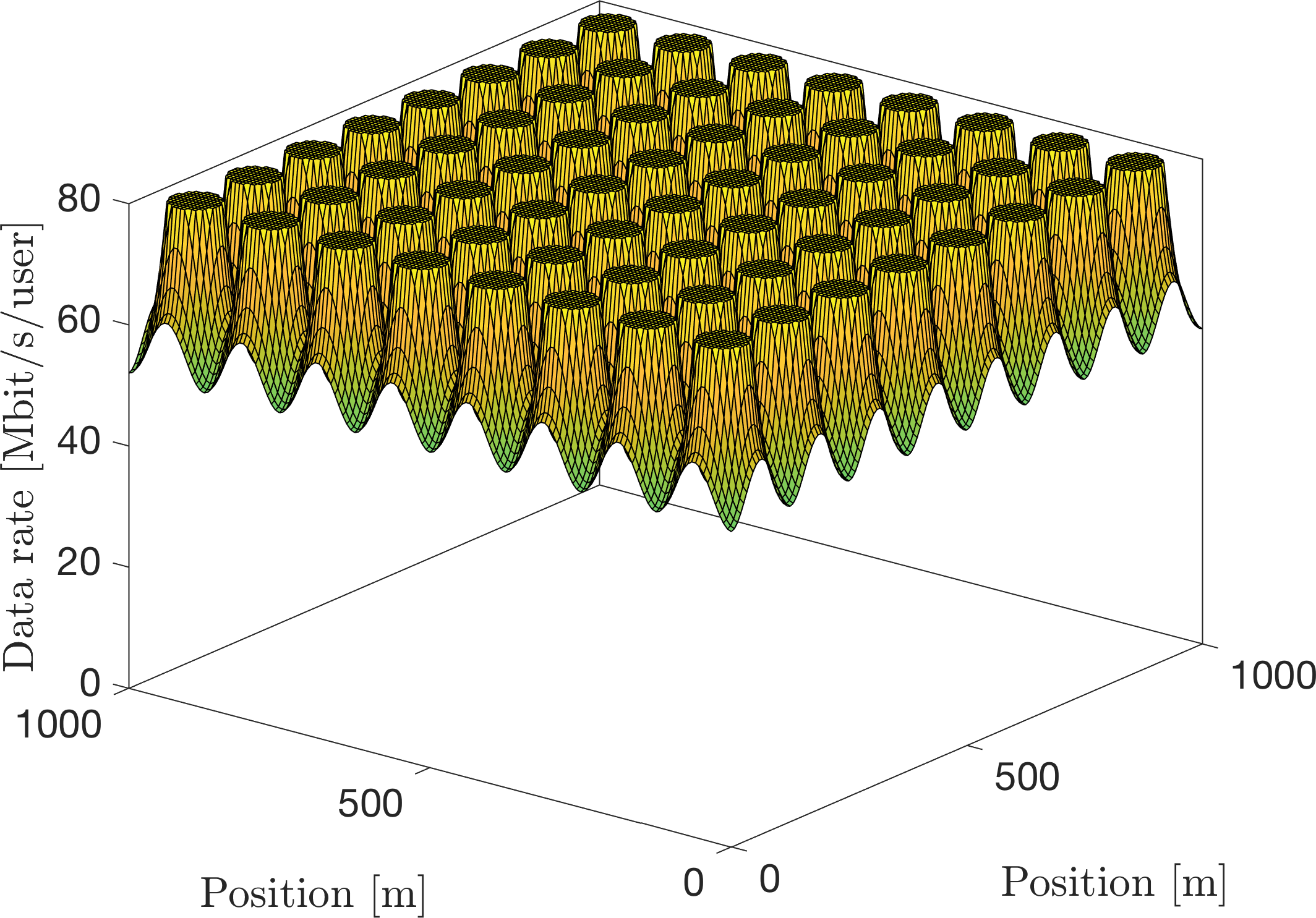}
		\end{overpic} 
		\caption{Cell-free network.}    \vspace{+1mm}
	\end{subfigure} 
	\begin{subfigure}[b]{\columnwidth} \centering 
		\begin{overpic}[width=.8\columnwidth,tics=10]{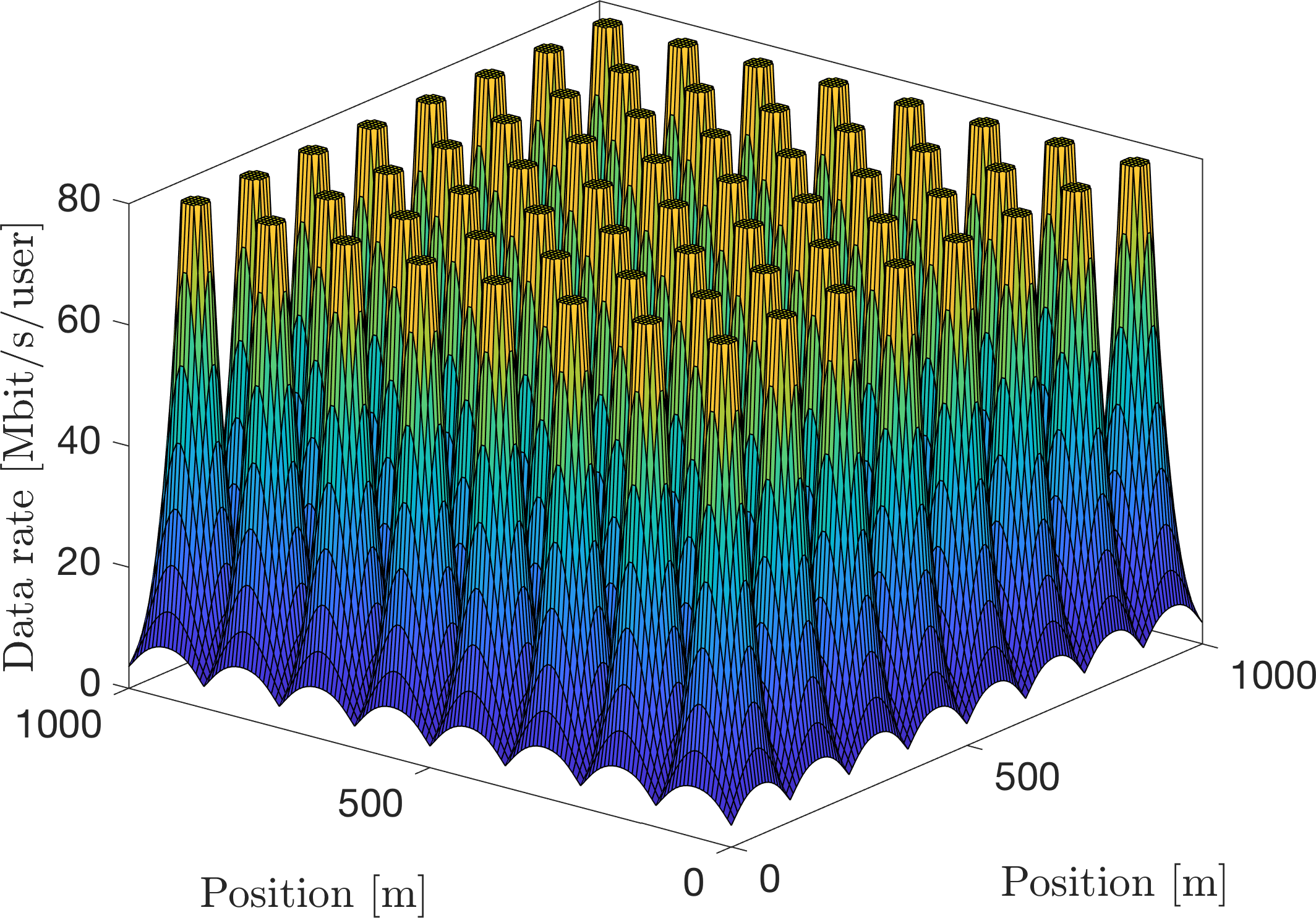}
		\end{overpic} 
		\caption{Cellular network with the same AP locations.} 
	\end{subfigure} 	
	\caption{Example of the downlink data rate achieved by a \gls{UE} at different locations in a network with 64 \glspl{AP} with omni-directional antennas deployed on a square grid and jointly transmitting to the UE. The propagation parameters are otherwise the same as in Figure~\ref{fig:cellular-rates}. A cell-free network operation is considered in (a), while a cellular network operation is considered in (b). The key observation is that only the cell-free operation can provide almost uniformly high data rates in the entire network.} 
	\label{fig:cellfree-rates} 
\end{figure}

To give a first impression of the goal of creating cell-free networks, Figure~\ref{fig:cellfree-rates}(a) shows the downlink data rate achieved by a \gls{UE} at different locations in a setup that resembles the cellular example in Figure~\ref{fig:cellular-rates}. For simplicity, an ideal deployment with 64 \glspl{AP} deployed on an $8 \times 8$ square grid is considered.
The figure shows that the rates vary between 52 and 80\,Mbit/s everywhere in the coverage area. One contributing factor is the denser deployment, which greatly reduces the average propagation distance between a \gls{UE} and the closest \gls{AP}.
 However, the main reason is that all the surrounding \glspl{AP} are jointly transmitting to the \gls{UE}, thereby alleviating the inter-cell interference issue that is one of the main causes of the large rate variations in cellular networks.
This is evident when comparing Figure~\ref{fig:cellfree-rates}(a) with Figure~\ref{fig:cellfree-rates}(b), where a cellular network with the same AP locations is considered. The inter-cell interference then gives rise to large rate variations.

\section{Historical Background}\label{sec:historical-background}

The cellular architecture has played a key role in enabling mobile communications, from the early concepts developed in the 1950s and 1960s \cite{Bullington1953a}, \cite{Frefkiel1970a}, \cite{Schulte1960a} to the first commercial deployment in 1979 \cite{Kinoshita2018a}.
The motivating factor of building a cellular network was to make efficient use of the limited frequency spectrum by enabling many concurrent transmissions in the geographical area covered by the network. To control the interference between the transmissions, the coverage area was divided into predefined geographical zones, known as cells, where a fixed \gls{AP} takes care of the service. In the beginning, a predefined frequency plan was utilized so that adjacent cells use different frequency resources, thereby limiting the inter-cell interference. Over the years, commercial cellular networks have been densified by deploying more \glspl{AP} per area unit \cite{Cooper2010a}. By using steerable multi-antenna panels at each \gls{AP}, instead of fixed-beam antennas, the interference between adjacent cells can be partially controlled so that the traditional frequency plans can be alleviated. 
Depending on the deployment scenario (e.g., indoor/outdoor, frequency band, coverage area, and distance from the AP to the closest UE location), different types of AP hardware are utilized \cite{Kamel2016a}. The resulting parts of the cellular networks are sometimes categorized as microcells, picocells, and femtocells. We will use the overarching term \emph{small cells} when referring to such networks \cite{Hoydis2011c}.
The use of smaller and smaller cells has been an efficient way to increase the network capacity, in terms of the number of bits per second that can be transferred in a given area. 
Ideally, the network capacity grows proportionally to the number of \glspl{AP} (with active \glspl{UE}), but this trend gradually tapers off due to the increasing inter-cell interference \cite{Andrews2017a}, \cite{Zhou2003a}. After a certain point, further network densification can actually reduce rather than increase the network capacity. 
This is particularly the case in the \emph{ultra-dense network} regime \cite{Hwang2013a}, \cite{Kamel2016a}, \cite{Stefanatos2014a}, where the number of \glspl{AP} is larger than the number of \emph{simultaneously active} \glspl{UE}. Even if each \gls{AP} would have a handful of antennas, this is not enough to suppress all the interference in such a dense scenario.
A cell-free network is an attempt to move beyond those limits \cite{Chen2016a}, \cite{Interdonato2018}, \cite{Ngo2017b}, \cite{Zhang2020a}, \cite{Zhang2019a}. Before explaining how that can be achieved, we will give a detailed historical background.

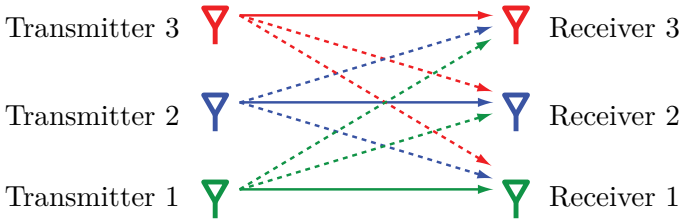
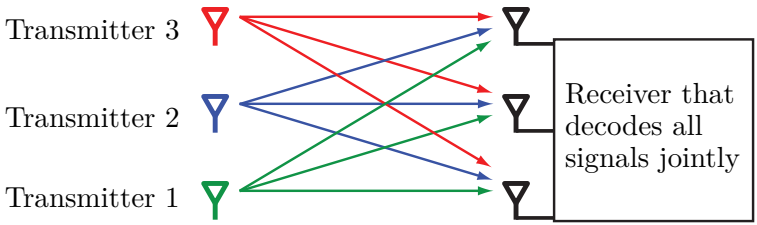
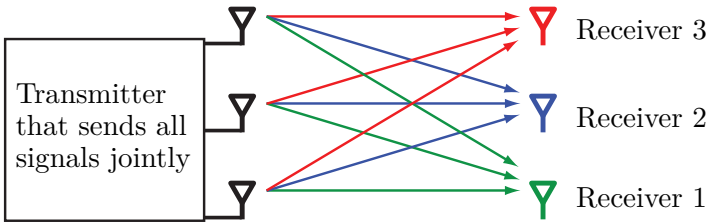
\begin{figure} 
        \centering \vspace{-4mm}
        \begin{subfigure}[b]{\columnwidth} \centering
	\begin{overpic}[width=.9\columnwidth,tics=10]{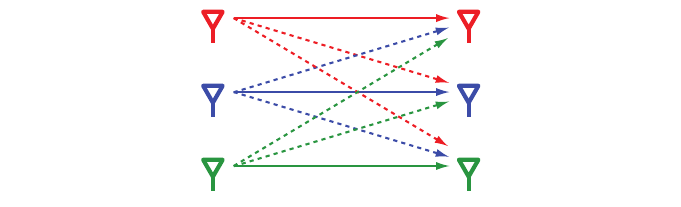}
 \put(5,4){Transmitter 1}
  \put(5,14){Transmitter 2}
   \put(5,25){Transmitter 3}
   \put(73,4){Receiver 1}
   \put(73,14){Receiver 2}
   \put(73,25){Receiver 3}
\end{overpic} 
                \caption{An \emph{interference channel} representing how cellular networks are conventionally operated. Each receiver wants to decode the data sent from its transmitter, subject to the dashed interfering signals from simultaneous transmissions.}    \vspace{+4mm}
        \end{subfigure} 
        \begin{subfigure}[b]{\columnwidth} \centering 
	\begin{overpic}[width=.9\columnwidth,tics=10]{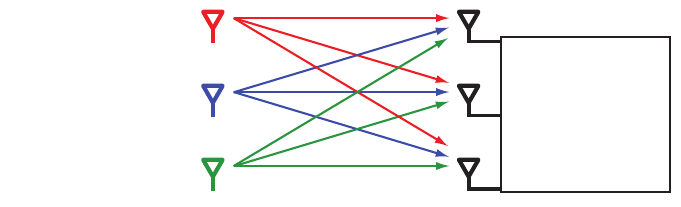}
 \put(5,4){Transmitter 1}
  \put(5,14){Transmitter 2}
   \put(5,25){Transmitter 3}
   \put(75,17){Receiver that}
   \put(75,13){decodes all}
   \put(75,9){signals jointly}
\end{overpic} 
                \caption{A \emph{multiaccess channel} representing how distributed receive antennas can cooperate to jointly decode the signals from all transmitters. The information contained in all received signals can be utilized. There are no unusable interfering signals. This describes the ideal uplink operation of a cell-free network.}  \vspace{+4mm}
        \end{subfigure} 
        \begin{subfigure}[b]{\columnwidth} \centering
	\begin{overpic}[width=.9\columnwidth,tics=10]{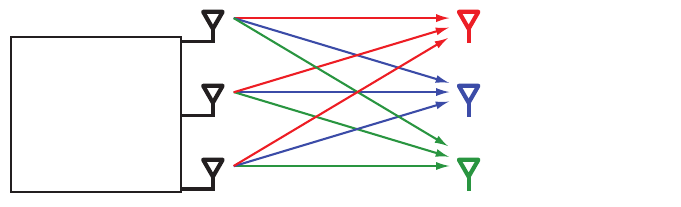}
   \put(3,17){Transmitter}
   \put(3,13){that sends all}
   \put(3,9){signals jointly}
   \put(73,4){Receiver 1}
   \put(73,14){Receiver 2}
   \put(73,25){Receiver 3}
\end{overpic} 
                \caption{A \emph{broadcast channel} representing how distributed transmit antennas can cooperate to jointly send the signals to all receivers. The signals sent from all antennas can be utilized at each receiver. There are no unusable interfering signals. This describes the ideal downlink operation of a cell-free network.}    \vspace{+2mm}
        \end{subfigure} 
        \caption{A cellular network is conventionally operated as an interference channel, which is shown in (a). To alleviate inter-cell interference, the uplink of a cell-free network is instead operated as the multiaccess channel shown in (b) and the downlink is operated as the broadcast channel shown in (c).} 
        \label{figure_interference_MAC_broadcast}  
\end{figure}

\enlargethispage{\baselineskip} 

As mentioned earlier, a key property of conventional cellular networks is that each \gls{UE} is assigned to one cell and only served by its \gls{AP}. This is known as an \emph{interference channel} in information theory and is illustrated in Figure~\ref{figure_interference_MAC_broadcast}(a) for the case of three single-antenna transmitters and three single-antenna receivers. Each receive antenna obtains  a signal containing the information sent from one desired transmitter (solid line) plus two interfering signals (dashed lines) sent from the undesired transmitters. 
Even in the absence of noise, identifying the desired signal is like solving an ill-conditioned linear system of equations with three unknowns but only one equation.
Hence, the inter-cell interference is unusable in this case; it only limits the performance.
When operating such a cellular network, the transmit powers might be adjusted to determine which of the cells will be most affected by the interference. There is no other cooperation between the \glspl{AP}; neither \gls{CSI} nor transmitted/received signals are shared between cells.
These assumptions were challenged by Wyner in \cite{Wyner1994a} from 1994, where the uplink was studied and the benefit of jointly decoding the data from all \glspl{UE} using the received signals in all cells was explored. In this way, the interference channel is turned into a \emph{multiaccess channel}, where all the receive antennas collaborate. Even if each antenna receives a superposition of multiple signals, there is no unusable interference but the task of the receiver is to extract the information contained in all the received signals.
This alternative way of operating the system is illustrated in Figure~\ref{figure_interference_MAC_broadcast}(b). In this example, the receiver has access to three observations that contain linear combinations of the three desired signals. 
In the absence of noise, signal detection can be viewed as solving a linear system of equations with three unknowns and three equations, which is a well-conditioned problem.
Importantly, the interference is not only canceled by this approach, but the observations made at multiple receive antennas are combined to increase the \gls{SNR} compared to the case where there was no interference between the transmissions \cite{Gesbert2010a}. \emph{Interference is turned from being bad to being good!}

Similarly, Shamai and Zaidel proposed a downlink co-processing framework in \cite{Shamai2001a} from 2001. Using information-theoretic terminology, the cellular downlink was transformed from an interference channel to a \emph{broadcast channel}, where all the geographically distributed transmit antennas collaborate. This case is illustrated in Figure~\ref{figure_interference_MAC_broadcast}(c). Each antenna  transmits a linear combination of the downlink signals intended for the UEs in all cells, where the linear combination is designed based on the channels to limit inter-cell interference. 
For example, in the setup shown in Figure~\ref{figure_interference_MAC_broadcast}(c) with three geographically distributed transmitters (\glspl{AP}) and three distributed receivers (\glspl{UE}), \gls{ZF} precoding can be utilized to completely avoid interference. This is not possible in the interference channel in Figure~\ref{figure_interference_MAC_broadcast}(a), where each signal is only sent from one transmitter and no precoding can be used.

\enlargethispage{\baselineskip} 

While the premise of  \cite{Shamai2001a}, \cite{Wyner1994a} was to add co-processing to an existing cellular network, the idea of building a cell-free network from the outset was pioneered by Zhou, Zhao, Xu, Wang, and Yao in \cite{Zhou2003a} from 2003. Their concept was called \emph{Distributed Wireless Communication System} and resembles the architecture described in Section~\ref{sec:cell-free-networks} with geographically distributed antennas and processing, and a \gls{CPU} that controls the system. The paper proposes that a \gls{UE} should not be served by all the antennas but only by the nearest set of distributed antennas, as illustrated by the shaded regions in Figure~\ref{fig:cell-free_distribution_processing}. This is an early step towards a user-centric assignment of network infrastructure, where each \gls{UE} is served by the user-preferred set of \glspl{AP} instead of by a predefined set. Similar ideas appeared for soft handoff in \gls{CDMA} systems \cite{Viterbi1994a}, where UEs at cell edges are jointly served by all the nearest \glspl{AP}.

Many other researchers contributed to this topic during the 2000s and a variety of terminologies have been used to refer to systems where the \glspl{AP} are jointly processing the transmitted and received signals. We will provide some key examples in this paragraph, without attempting to provide an exhaustive list. 
Non-linear co-processing schemes were developed by Jafar, Foschini, and Goldsmith  \cite{Jafar2004a} with the goal of enabling new \glspl{UE} to be added to a cellular network without affecting the rates of existing \glspl{UE}.
Cooperative downlink processing with multi-antenna \glspl{AP} was studied by Zhang and Dai in \cite{Zhang2004a}. 
The concept of \emph{Group Cell} was introduced by Zhang, Tao, Zhang, Wang, Li, and Wang in \cite{Zhang2005b} to serve mobile UEs by multiple cells to enable smooth handover during mobility. Multi-cell detection features were also discussed using the group cell name \cite{Tao2005a}.
Coherent coordinated transmission from the \glspl{AP} based on linear \gls{ZF} precoding and non-linear dirty paper coding was studied by Foschini, Karakayali, and Valenzuela in \cite{Foschini2006a}, \cite{Karakayali2006a}.
The term \emph{Network MIMO} was coined by Venkatesan, Lozano, and Valenzuela in \cite{Venkatesan2007a} to describe a cellular network where all the \glspl{AP} within the range of a \gls{UE} share their received signals over a backhaul network, to turn the cellular uplink from an interference channel to a multiaccess channel.
Soft handover between distributed antennas in \gls{OFDM} systems was studied by T\"olli, Codreanu, and Juntti in \cite{Tolli2008a}. While \gls{AP} cooperation with infinite-capacity backhaul links was assumed in the above-mentioned works, implementation of joint uplink detection with limited-capacity backhaul was considered by Sanderovich, Somekh, Poor, and Shamai in \cite{Sanderovich2009a}, while the downlink counterpart was studied by Simeone, Somekh, Poor, and Shamai in \cite{Simeone2009a}. 
Iterative data detection methods, where the \glspl{AP} exchange soft information to reduce the inter-cell interference, were considered by Khattak, Rave, and Fettweis in \cite{Khattak2008a}.
Finally, Bj\"ornson, Zakhour, Gesbert, and Ottersten showed in \cite{Bjornson2010c} that coherent joint transmission can be implemented in \gls{TDD} systems without sharing \gls{CSI} between the \glspl{AP}, at the cost of increased interference since the \gls{AP} cannot cancel each others' signals at undesired receivers.

\subsection{Towards Standardization in 4G}

The multi-cell cooperation concepts were considered in the 4G standardization of LTE-Advanced in the late 2000s \cite{Parkvall2008a}, under the umbrella term of \emph{\gls{CoMP}} transmission/reception. 
The co-processing of data at multiple \glspl{AP}, which is the focus of this monograph, is called \emph{\gls{JP}} in \gls{CoMP} \cite{Boldi2011a}. Other \gls{CoMP} options are coordinated scheduling/precoding where each cell only serves its own \glspl{UE}, which fall into the category of methods that can be also implemented in conventional cellular networks. Both centralized and decentralized architectures for facilitating \gls{JP} were explored in the context of \gls{CoMP}. In the centralized approach, the cooperating \glspl{AP} are connected to a \gls{CPU} (which might be co-located with an \gls{AP}) and send their information to it. Hence, the \glspl{AP} can be also viewed as relays that facilitate communication between \glspl{UE} and the \gls{CPU} \cite{Estella2019a}.
In the decentralized approach, the cooperating \glspl{AP} only acquire \gls{CSI} from the \glspl{UE} \cite{Papadogiannis2009a}, but data must still be shared between \glspl{AP}.

\begin{figure} 
        \centering \vspace{-4mm}
        \begin{subfigure}[b]{\columnwidth} \centering
	\begin{overpic}[width=.75\columnwidth,tics=10]{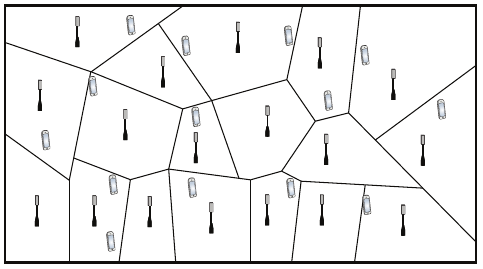}
\end{overpic} 
                \caption{A conventional cellular network, where each \gls{UE} is only served by one \gls{AP}.}    \vspace{+2mm}
        \end{subfigure} 
        \begin{subfigure}[b]{\columnwidth} \centering 
	\begin{overpic}[width=.75\columnwidth,tics=10]{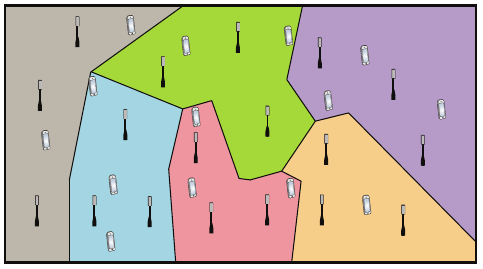}
\end{overpic} 
                \caption{A network-centric implementation of \gls{CoMP} in a cellular network, where the \glspl{AP} are divided into disjoint clusters. The \glspl{UE} in a cluster are jointly served by the \glspl{AP} in that cluster.}  \vspace{+2mm}
        \end{subfigure} 
        \begin{subfigure}[b]{\columnwidth} \centering
	\begin{overpic}[width=.75\columnwidth,tics=10]{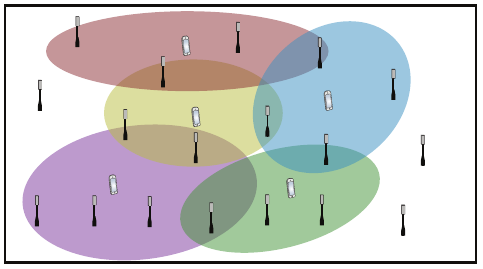}
\end{overpic} 
                \caption{A user-centric implementation of \gls{CoMP} in a cellular network, where each \gls{UE} selects a set of preferred \glspl{AP} that will serve it. This is the approach taken also in cell-free networks.}    \vspace{+1mm}
        \end{subfigure} 
        \caption{Comparison between a conventional cellular network and two ways of implementing multi-cell cooperation.} 
        \label{figure_cellular_clusters}  
\end{figure}

\subsubsection{Network-Centric Clustering}
\enlargethispage{\baselineskip} 
Since each \gls{UE} in a conventional cellular network would only be affected by interference from its own cell and a set of neighboring cells, it is only the corresponding cluster of \glspl{AP} that needs to cooperate to alleviate inter-cell interference for this \gls{UE}. 
Different ways to implement the \gls{AP} clustering was explored alongside the development of LTE-Advanced \cite{Boldi2011a}.
The starting point for the clustering is that a cellular network already exists and needs to be improved. We will use the example in Figure~\ref{figure_cellular_clusters}(a) to explain the clustering approaches.
The first option is \emph{network-centric clustering} where the \glspl{AP} are divided into disjoint clusters \cite{Huang2009b}, \cite{Marsch2008a}, \cite{Zhang2009b}, each serving a disjoint set of \glspl{UE}. For example, groups of three neighboring cells can be clustered into a joint region, as illustrated by the colored regions in  Figure~\ref{figure_cellular_clusters}(b). Compared to the conventional cellular network in Figure~\ref{figure_cellular_clusters}(a), the cell edges within each cluster are removed, but interference will still occur between clusters. Hence, \glspl{UE} that are close to a cluster edge might not benefit from the network-centric clustering. 
The clusters can be changed over time or frequency in an effort to make sure that most of the served \glspl{UE} are in the center of a cluster and not at the edges \cite{Cayirci2002a}, \cite{Jungnickel2014a}, \cite{Marsch2011a}, \cite{Papadogiannis2009a}.
The network-centric clustering is conceptually similar to having a conventional cellular network where each cell contains a set of distributed antennas that are controlled by a single \gls{AP} \cite{Choi2007a}. Each cell in such a setup corresponds to one cluster in the network-centric clustering.

\subsubsection{User-Centric Clustering}

Another option is \emph{user-centric clustering} where each \gls{UE} selects a set of preferred  \glspl{AP}  \cite{Bjornson2011a}, \cite{Bjornson2013d}, \cite{Chen2016a}, \cite{Garcia2010a}, \cite{Kaviani2012a}, \cite{Xu2013a}, \cite{Zhang2005b}. This is illustrated for five \glspl{UE} in  Figure~\ref{figure_cellular_clusters}(c), where each colored region corresponds~to the set of \glspl{AP} selected by the corresponding \gls{UE}. Note that the sets are partially overlapping between neighboring \glspl{UE}, thus disjoint \gls{AP} clusters cannot be created to achieve the same result. Irrespective of the \gls{UE}'s location, user-centric clustering will guarantee the control~of interference.
In this monograph, we will make use of the \emph{\gls{DCC}} framework for user-centric clustering,~which was introduced by Bj{\"{o}}rnson, Jald{\'e}n, Bengtsson, and Ottersten in \cite{Bjornson2011a}.

If the clusters are well designed, user-centric clustering outperforms network-centric clustering since the latter is essentially a special case of the former.
However, both approaches are complicated to add to an existing cellular network since the interfaces between the \glspl{AP} must be standardized to enable cooperation among \gls{AP} equipment from different vendors. When potential solutions were simulated in the 4G standardization body, the performance gains were often so small that the additional control signaling might remove the gains \cite{Boldi2011a}. 
An important reason was that the algorithms were jointly designed for \gls{FDD} and \gls{TDD} systems, thus they could not exploit the particular features that only exist in one of these duplexing modes. In particular, \gls{CSI} for downlink precoding had to be sent around between the \glspl{AP} over low-latency backhaul links to make the system work \cite{Fantini2016a}. It is only in a pure \gls{TDD} implementation exploiting uplink-downlink channel reciprocity that the \gls{CSI} necessary for downlink precoding can be obtained at each \gls{AP} without backhaul signaling \cite{Bjornson2010c}. We return to this later in this section.
\enlargethispage{\baselineskip} 

In Release 10 of LTE-Advanced, only a special case of network-centric clustering was supported \cite{Boldi2011a}: each cluster consists of \glspl{AP} that are deployed on the same physical site to cover different geographical sectors. Such clustering can only limit the interference between cell sectors, but not between \glspl{UE} at cell edges. Despite the lack of standardization, the major vendors of \gls{AP} hardware have made proprietary implementations of \gls{CoMP} with \gls{JP} that can only be applied among their own \glspl{AP}. 
These solutions are often implemented using the \gls{C-RAN} architecture, which was briefly introduced in Section~\ref{sec:cell-free-networks}. In this cellular context, a set of neighboring \glspl{AP} is connected via a low-latency fronthaul to an edge-cloud processor where the baseband processing is carried out. \gls{CoMP} algorithms can be conveniently implemented in such a setup.
It is not publicly known what \gls{CoMP} methods are used by different vendors and how well the implementations perform.
However, the pCell technology from Artemis \cite{Perlman2015a} is claimed to utilize user-centric clustering.

\subsection{Cellular Massive \gls{MIMO} in 5G}

Instead of focusing on \gls{CoMP}, the new feature in the 5G cellular networks is Massive \gls{MIMO}. This concept was introduced by Marzetta in \cite{Marzetta2010a} from 2010 and essentially means that each \gls{AP} \emph{operates individually} and is equipped with an array of a very large number of active low-gain antennas that can be individually controlled using separate radios (transceiver chains). This stands in contrast to the passive high-gain antennas traditionally used in cellular networks, which might have similar physical dimensions but only a single radio.
Massive \gls{MIMO} has its roots in \emph{space-division multiple access} \cite{Anderson1991a}, \cite{Richard1996a}, \cite{Swales1990a}, \cite{Winters1987a}, which was introduced in the 1980s and 1990s to enable multiple \glspl{UE} to be served by an \gls{AP} at the same time and frequency. The antenna arrays enable directional transmission to each \gls{UE} (and directional reception from them), thus \glspl{UE} located at different locations in the same cell can be served simultaneously with little interference.
This technology has later been known as multi-user \gls{MIMO}.

\enlargethispage{\baselineskip} 

\subsubsection{Benefits}

The characteristic feature of Massive \gls{MIMO}, compared to traditional multi-user MIMO, is that each \gls{AP} has many more antennas than there are active \glspl{UE}  in the cell. Two important propagation phenomena~appear in those cases \cite{Larsson2014a}, \cite{Rusek2013a}: \emph{channel hardening} and \emph{favorable propagation}. The former means that fading channels behave almost as deterministic channels if the antenna signals are processed properly to neutralize~the small-scale fading. In principle, the processing makes use of the massive spatial diversity offered by having many antennas. 
Favorable propagation means that the channels of spatially separated \glspl{UE} are nearly orthogonal in the spatial domain, since transmission and reception are very spatially directive. We will describe these phenomena in detail in Section~\ref{sec:hardening-favorable} on p.~\pageref{sec:hardening-favorable}.
Motivated by the second phenomenon, it was initially claimed that low-complexity interference-ignoring signal processing methods, such as \gls{MR} processing, are close-to-optimal when each \gls{AP} is equipped with a large number of antennas.
It has later been established that more advanced linear signal processing methods,  such as \emph{\gls{MMSE} processing}, are needed to make efficient use of Massive \gls{MIMO} \cite{BjornsonHS17}, \cite{Hoydis2013a}, \cite{Neumann2017a}, \cite{Sanguinetti2019a}. In essence, this means that interference must be actively suppressed (one cannot rely on it disappearing automatically when there are many antennas), but the loss in desired signal power is small and there is little need for non-linear methods such as successive interference cancellation \cite{massivemimobook}.

A rigorous framework for analyzing the achievable data rates under imperfect \gls{CSI} was developed in the Massive \gls{MIMO} literature and is summarized in recent textbooks, such as \cite{massivemimobook}, \cite{Marzetta2016a}. Many tools from this framework will be also utilized in later sections of this monograph.

\subsubsection{Limitations}

As illustrated in Figure~\ref{fig:cellular-rates} earlier in this chapter, Massive \gls{MIMO} can increase the data rates in a cellular network compared to conventional technology, but large rate variations and inter-cell interference will still remain. Moreover, the 64-antenna panels that have been deployed in 5G cellular networks are not \glspl{ULA}, as is normally explicitly or implicitly assumed in the Massive \gls{MIMO} literature \cite{massivemimobook}, \cite{Marzetta2016a}, but compact planar arrays that can be deployed in the same way as conventional antennas. 
Since the horizontal width of an array determines its ability to separate \glspl{UE} located in different azimuth angles with respect to the array (i.e., wider arrays mean better spatial resolution), the service quality provided by planar arrays is far from what is presented in the literature \cite{Aslam2019a}, \cite{Bjornson2019d}.
In summary, Massive \gls{MIMO} is a solution to some of the interference problems that are faced in conventional cellular networks. However, a cellular deployment of physically wide horizontal  \glspl{ULA} is practically questionable since it greatly deviates from the form factor of conventional cellular \glspl{AP}. Even if this practical barrier is overcome, the large variations in the distance to the served \glspl{UE} will still lead to large rate variations of the kind illustrated in Figure~\ref{fig:cellular-rates}. Hence, a different deployment architecture is required to deliver a more uniform service quality over the coverage~area.

\subsection{Cell-Free Networks Beyond 5G}
\label{subsec-history-cellfree}

The \emph{cell-free} terminology was coined by Yang and Marzetta in \cite{Yang2013b} from 2013, while the name \emph{Cell-free Massive \gls{MIMO}} first appeared in \cite{Ngo2015a} by Ngo, Ashikhmin, Yang, Larsson, and Marzetta from 2015. 
While most of the research described earlier adds multi-cell cooperation to an existing cellular network architecture, Cell-free Massive \gls{MIMO} instead follows in the footsteps of the Distributed Wireless Communication System concept from \cite{Zhou2003a}, where a network consisting of distributed cooperating antennas is designed from the outset. 
The word ``massive'' refers to an envisioned operating regime with many more \glspl{AP} than \glspl{UE} \cite{Ngo2015a}, and is as an analogy to the conventional Massive MIMO regime in cellular networks; that is, having many more antennas at the infrastructure side than UEs to be served. Interestingly, the envisioned operating regime coincides with that of ultra-dense networks \cite{Hwang2013a}, \cite{Kamel2016a}, \cite{Stefanatos2014a}, but with the core difference that the APs are cooperating to form a distributed antenna array. The original motivation of Cell-free Massive \gls{MIMO} was to provide an almost uniformly high service quality in a given geographical area \cite{Ngo2015a}, as illustrated in Figure~\ref{fig:cellfree-rates}. 

\enlargethispage{\baselineskip} 

\subsubsection{Background}

The cell-free architecture, shown in Figure~\ref{fig:cell-free} and Figure~\ref{fig:cell-free_distribution_processing}, was analyzed in the early works \cite{Nayebi2017a}, \cite{Ngo2017b} with the focus on a distributed operation where the \glspl{AP} perform all the signal processing tasks, except for those that critically require central coordination. The system operates in \gls{TDD} mode, which means that the uplink and downlink take place in the same frequency band but are separated in time. Hence, the downlink/uplink channels can be jointly estimated by sending known pilot signals from the \glspl{UE} to the \glspl{AP}. In this way, each \gls{AP} obtains local \gls{CSI} regarding the channels between itself and the different \glspl{UE}. In the downlink setup studied in \cite{Nayebi2017a}, \cite{Ngo2017b}, each data signal is encoded at a \gls{CPU} and sent over the fronthaul to the \glspl{AP}, which transmit the signals using \gls{MR} precoding based on the locally available \gls{CSI}. Similarly, in the uplink, each \gls{AP} applies \gls{MR} combining locally and sends its soft data estimates over the fronthaul to the \gls{CPU}, which makes the final decoding without having access to any \gls{CSI}.
This concept is well aligned with the cellular joint transmission framework from \cite{Bjornson2010c}, where the \glspl{AP} only make use of local \gls{CSI} obtained from uplink pilots in \gls{TDD} mode. 
Variations of this type of distributed processing can be found in \cite{Bashar2019a}, \cite{Bjornson2019c}, \cite{Buzzi2017a}, \cite{Fan2019a}, \cite{Ozdogan2018a}, \cite{Yang2019a}, \cite{Zhang2018a}. One key insight from the more recent works is that the performance can be greatly improved by using \gls{MMSE} processing instead of \gls{MR} \cite{Bjornson2019c}, which is in line with what has also been observed in the Cellular Massive \gls{MIMO} literature \cite{BjornsonHS17}, \cite{Sanguinetti2019a}.
Hence, even if favorable propagation effects can be observed also in cell-free networks with many distributed antennas \cite{Chen2018b}, it remains important to design the signal processing schemes to actively suppress interference.

The data rates can be also improved by semi-centralized implementations, potentially, at the cost of additional fronthaul signaling. One option is to provide the \gls{CPU} with statistical \gls{CSI} so that it can optimize how the uplink data estimates from the \glspl{AP} are combined by taking their relative accuracy into account \cite{Adhikary2017a}, \cite{Bjornson2019c}, \cite{Nayebi2016a}, \cite{Ngo2018b}. 
For example, an \gls{AP} that is close to the \gls{UE} should have more influence than an \gls{AP} that is further away or that is subject to strong interference.
Another option is to let the \gls{CPU} take care of all the processing while the \glspl{AP} only act as relays \cite{Bashar2018a}, \cite{Bjornson2019c}, \cite{Chen2018b}, \cite{Nayebi2016a}, \cite{Riera2018a}, \cite{Yang2019a}. These different options will be analyzed in detail in later sections of this monograph.

\subsubsection{The Roots of Cell-Free Massive MIMO}

The first papers on Cell-free Massive \gls{MIMO} assumed all \glspl{UE} are served by all \glspl{AP}, while the user-centric clustering from the \gls{CoMP} literature was first considered in the cell-free context in \cite{Buzzi2017a}, \cite{Buzzi2017b}. A new practical implementation of such clustering was proposed in \cite{Interdonato2019a}, by first dividing the \glspl{AP} into clusters in a network-centric fashion and then let each \gls{UE} select a preferred subset of the network-centric clusters. A framework for creating the user-centric clusters in a decentralized fashion was proposed in \cite{Bjornson2020a}, where the scalability of the different signal processing tasks was also analyzed. Similar user-centric clustering concepts exist in the literature on ultra-dense networks \cite{Chen2016a}.

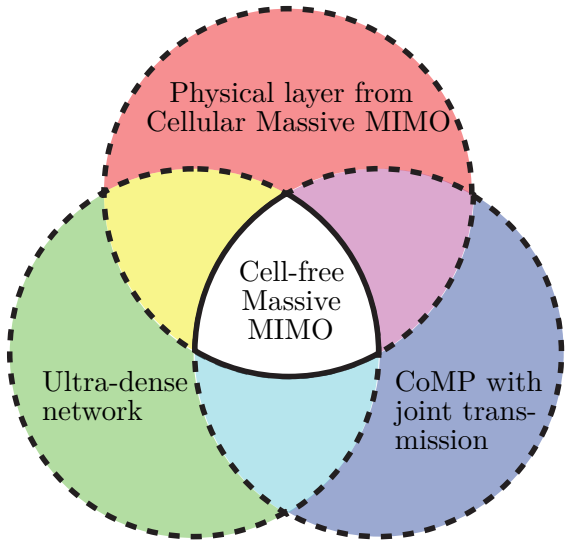
\begin{figure}[t!]
	\centering 
	\begin{overpic}[width=.65\columnwidth,tics=10]{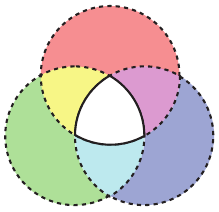}
		\put(42,47){Cell-free}
		\put(42.5,42){Massive}
		\put(43.5,37){MIMO}
		\put(8,28){Ultra-dense}
		\put(8,23){network}
		\put(69,28){CoMP with}
		\put(69,23){joint trans-}
		\put(69,18){mission}
		\put(30,78){Physical layer from}
		\put(26,73){Cellular Massive MIMO}
	\end{overpic} 
	\caption{Cell-free Massive MIMO can be defined as the intersection between three technology components: The physical layer from Cellular Massive MIMO, the  joint transmission concept for distributed APs in the CoMP literature, and the deployment regime of ultra-dense networks.} 
	\label{fig:venn-diagram} 
\end{figure}

Many of the concepts described in the Cell-free Massive \gls{MIMO} literature have previously (or simultaneously) appeared and been analyzed in the cellular literature; for example, in some of the papers mentioned earlier in this section. 
With this in mind, there are two approaches to defining Cell-free Massive \gls{MIMO}. 
The first approach is to specify its unique characteristics. As illustrated by the Venn diagram in Figure~\ref{fig:venn-diagram}, it can be viewed as the intersection between the physical layer from the Cellular Massive MIMO literature, the joint transmission concept for distributed APs in the CoMP literature, and the deployment regime of ultra-dense networks. This corresponds to the inner region in the diagram. In other words, we take the best aspects from three technologies, combine them into a single network, and then jointly optimize them to achieve an ultimate embodiment of a wireless network. The second approach is to view Cell-free Massive \gls{MIMO} as the union of the three circles; that is, an overarching concept focused on cell-free networks but which contains conventional Massive MIMO, conventional CoMP, and conventional ultra-dense networks as three special cases. The presentation of the technical content of this monograph will follow the first approach, thus it is that narrow definition that should be remembered when reading the term ``Cell-free Massive MIMO'' in later sections.
We will focus on describing the foundations of Cell-free Massive \gls{MIMO}, including the state-of-the-art signal processing and optimization methods. We will focus on how a user-centric viewpoint can be used to identify a scalable implementation, which are two dimensions that are not captured by the Venn diagram. We will compare the achievable performance with that of Cellular Massive MIMO and small cells, which we will extract as two special cases from our analytical formulas. The presentation is not based on a particular set of papers, but is an attempt to summarize the topic as a whole.

\enlargethispage{2\baselineskip} 

\section{Three Benefits over Cellular Networks} \label{sec:benefits-cellular-networks}

We will end this section by showcasing three major benefits that cell-free networks have compared to conventional cellular networks. More precisely, we compare the setups illustrated in Figure~\ref{fig:basic-comparison-cellular-cellfree}. The first one is a single-cell setup with a 64-antenna Massive \gls{MIMO} \gls{AP}, the second one consists of 64 small cells deployed on a square grid, and the last one is a cell-free network where the same 64 \gls{AP} locations are used. The comparison of these setups will be made by presenting basic mathematical expressions and simulation results, while a more in-depth analysis of cell-free networks will be provided in later sections.

\subsection{Benefit 1: Higher \gls{SNR} With Smaller Variations}\label{sec:benefits-distributed-antennas}

\begin{figure}[t!]
	\centering  \vspace{-4mm}
        \begin{subfigure}[b]{\columnwidth} \centering
	\begin{overpic}[width=.4\columnwidth,tics=10]{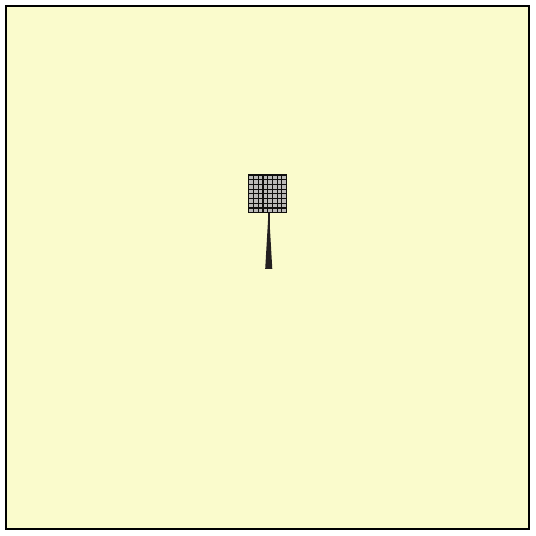}
\end{overpic} 
                \caption{One cell with a 64-antenna \gls{AP} in a cellular setup.}    \vspace{+2mm}
        \end{subfigure} 
        \begin{subfigure}[b]{\columnwidth} \centering 
	\begin{overpic}[width=.4\columnwidth,tics=10]{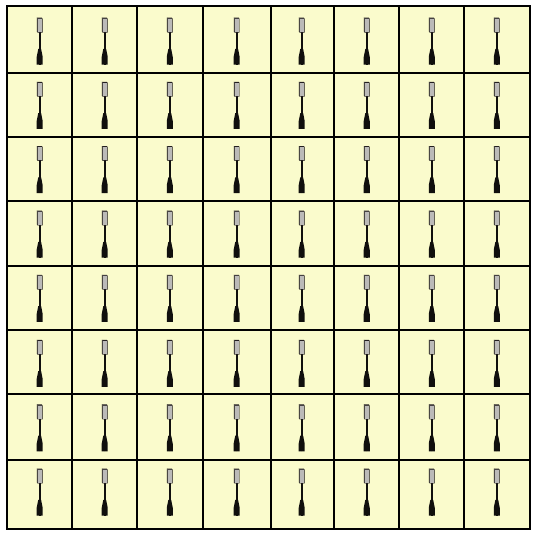}
\end{overpic} 
                \caption{Cellular setup with 64 single-antenna \glspl{AP}.} 
        \end{subfigure}
        \begin{subfigure}[b]{\columnwidth} \centering 
	\begin{overpic}[width=.4\columnwidth,tics=10]{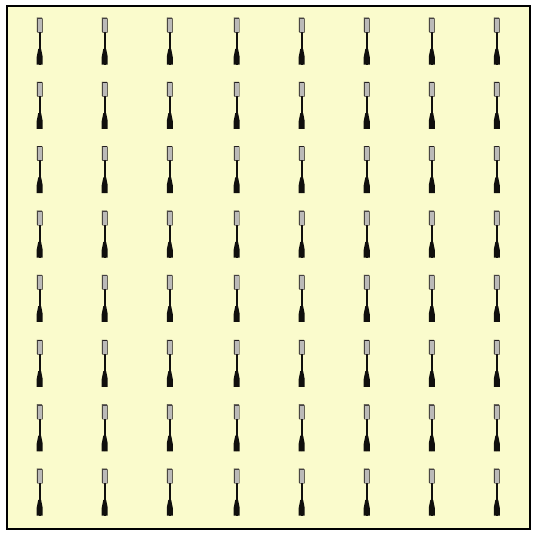}
\end{overpic} 
                \caption{Cell-free setup with the same \gls{AP} locations as in (b).} 
        \end{subfigure} 
	\caption{Three basic setups are compared in Section~\ref{sec:benefits-cellular-networks}: Two cellular networks and one cell-free network (connected to the CPU via fronthaul links, not shown for simplicity).} 
	\label{fig:basic-comparison-cellular-cellfree} 
\end{figure}

The first benefit of the cell-free architecture is that it achieves a higher and more uniform  \gls{SNR} within the coverage area than conventional cellular networks.
To explain this, we assume there is only one active \gls{UE} in the network and quantify the \gls{SNR} that the \gls{UE} achieves in the uplink, when the \gls{UE}'s transmit power is $p$ and the noise power is $\sigmaUL$.
In each of the three setups in Figure~\ref{fig:basic-comparison-cellular-cellfree}, there are $64$ antennas. 
The received power is substantially lower than the transmit power in wireless communications. For the sake of argument, we model the \emph{channel gain} (also known as pathloss or large-scale fading coefficient) for a propagation distance $d$ as (in decibels)
\begin{equation}\label{eq:large-scale-model-simple}
\beta(d) \,  [\textrm{dB}] = -30.5 - 36.7 \log_{10}\left( \frac{d}{1\,\textrm{m}} \right).\end{equation}
The first term says that 30.5\,dB of the power is lost at 1\,m distance while the second term says that another 36.7\,dB of power is lost for every ten-fold increase in the propagation distance. All channels are deterministic and thus known to the transmitters and receivers in this section. A more realistic channel model is provided in Section~\ref{sec:channel-modeling} on p.~\pageref{sec:channel-modeling} and is then used in the remainder of the monograph.

\subsubsection{Massive \gls{MIMO} Setup}

We first consider the single-cell Massive \gls{MIMO} setup in Figure~\ref{fig:basic-comparison-cellular-cellfree}(a), where the \gls{AP} is equipped with $M=64$ antennas. This represents one cell in a cellular network.
We denote by $\vect{g} = [g_1 \, \ldots \, g_M]^{\Ttran} \in \mathbb{C}^M$ the channel response between the \gls{UE} and the $M$ antennas.
The received uplink signal $\vect{y}^{\textrm{MIMO}} \in \mathbb{C}^M$ at the \gls{AP} is
\begin{equation} \label{eq:basic-uplink}
\vect{y}^{\textrm{MIMO}} = \vect{g} s + \vect{n}
\end{equation}
where $s \in \mathbb{C}$ is the information signal with transmit power $\mathbb{E}\{ | s|^2 \} = p$ and $\vect{n} \sim \CN ( \vect{0}_M, \sigmaUL \vect{I}_M )$ is the receiver noise.

The main task for the \gls{AP} is to estimate $s$ and this can be done by applying a receive combining vector $\vect{v} \in \mathbb{C}^M$ to \eqref{eq:basic-uplink}, which leads to
\begin{equation} \label{eq:s-estimate-MIMO}
\hat{s}^{\textrm{MIMO}}  = \vect{v}^{\Htran} \vect{y}^{\textrm{MIMO}}  =  \vect{v}^{\Htran} \vect{g} s + \vect{v}^{\Htran}\vect{n}.
\end{equation}
From this expression, it is clear that the \gls{SNR} is
\begin{equation} \label{eq:mMIMO-basic}
 \frac{\mathbb{E}\{ | \vect{v}^{\Htran} \vect{g} s|^2   \} }{ \mathbb{E} \{ | \vect{v}^{\Htran}\vect{n} |^2 \}} = \frac{p}{\sigmaUL} \frac{ | \vect{v}^{\Htran} \vect{g} |^2   }{ \| \vect{v} \|^2}.
\end{equation}
The \gls{AP} can select $\vect{v}$ based on the channel $\vect{g}$ to maximize the \gls{SNR}. It follows from the Cauchy-Schwartz inequality that \eqref{eq:mMIMO-basic} is maximized when $\vect{v}$ and $\vect{g}$ are parallel vectors. In particular, the unit-norm \gls{MR} combining vector $\vect{v} = \vect{g} / \| \vect{g} \|$ can be used to obtain the maximum \gls{SNR}
\begin{equation}
\mathacr{SNR}^{\textrm{MIMO}}  = \frac{p}{\sigmaUL} \| \vect{g}\|^2.
\end{equation}
Since all the antennas are co-located in a big array at the \gls{AP}, there is the same propagation distance $d$ from the \gls{UE} to all antennas. Hence, $|g_m|^2 = \beta(d) $ for $m=1,\ldots,M$ using the channel gain model in \eqref{eq:large-scale-model-simple}.
We then obtain
\begin{equation} \label{eq:SNR-simple-mMIMO}
\mathacr{SNR}^{\textrm{MIMO}}  = \frac{p}{\sigmaUL} M \beta(d)
\end{equation}
which shows that, in a single-cell Massive \gls{MIMO} system, the \gls{SNR} is proportional to the number of antennas, $M$.

\subsubsection{Cellular Setup With Small Cells}

In the cellular setup in Figure~\ref{fig:basic-comparison-cellular-cellfree}(b), there are $L=64$ geographically distributed \glspl{AP}. Each one has a single antenna and the \gls{UE} will only be served by one of them. We let $h_l\in \mathbb{C}$ denote the channel response between the \gls{UE} and AP $l$. In the uplink, the received signal $y_l^{\textrm{small-cell}} \in \mathbb{C}$ at AP $l$ is 
\begin{equation} \label{eq:basic-uplink-small-cell}
y_l^{\textrm{small-cell}} = h_l s + n_l
\end{equation}
where $s \in \mathbb{C}$ denotes the information signal that satisfies $\mathbb{E}\{ | s|^2 \} = p$ and $n_l \sim \CN ( 0, \sigmaUL )$ is the receiver noise.
The \gls{SNR} at \gls{AP} $l$ is
\begin{equation}
\mathacr{SNR}_l^{\textrm{small-cell}} =   \frac{\mathbb{E}\{ | h_l  s|^2   \} }{ \mathbb{E} \{ | n_l |^2 \}}  = \frac{p}{\sigmaUL} |h_l|^2.
\end{equation}
The \gls{UE} needs to choose only one of the \glspl{AP} since this is a conventional cellular network with no cooperation among \glspl{AP}. The \gls{UE} will naturally select the one providing the largest \gls{SNR}. Hence, the \gls{SNR} experienced by the UE becomes
\begin{align}  \notag
\mathacr{SNR}^{\textrm{small-cell}} &= \max_{l \in \{1,\ldots,L\} }
\mathacr{SNR}_l^{\textrm{small-cell}} \\ & =  \frac{p}{\sigmaUL} \max_{l \in \{1,\ldots,L\} } |h_l|^2. \label{eq:SNR-simple-small}
\end{align}
If we let $d_l$ denote the distance between the \gls{UE} and \gls{AP} $l$, then $|h_l |^2 = \beta(d_l)$ and the \gls{SNR} in \eqref{eq:SNR-simple-small} can be rewritten as
\begin{equation}
\mathacr{SNR}^{\textrm{small-cell}} =  \frac{p}{\sigmaUL} \max_{l \in \{1,\ldots,L\} } \beta(d_l).
\end{equation}

\subsubsection{Cell-Free Setup}

In the cell-free setup in Figure~\ref{fig:basic-comparison-cellular-cellfree}(c), we have the same $L$ \glspl{AP} as in the previous small-cell setup, but the \glspl{AP} are now cooperating to serve the \gls{UE}. We can write the received signals in \eqref{eq:basic-uplink-small-cell} jointly as
\begin{equation} \label{eq:basic-uplink-cell-free}
\vect{y}^{\textrm{cell-free}} = \vect{h} s + \vect{n}
\end{equation}
where $\vect{h} = [ h_1 \, \ldots \, h_L ]^{\Ttran}$ and $\vect{n} = [ n_1 \, \ldots \, n_L ]^{\Ttran}$.
Similar to the single-cell Massive \gls{MIMO} case above, a receive combining vector $\vect{v} \in \mathbb{C}^L$ can be applied to \eqref{eq:basic-uplink-cell-free} in an effort to estimate $s$. This leads to
\begin{equation}
\hat{s}^{\textrm{cell-free}}  = \vect{v}^{\Htran} \vect{y}^{\textrm{cell-free}}   =  \vect{v}^{\Htran} \vect{h} s + \vect{v}^{\Htran}\vect{n}.
\end{equation}
Since this equation has the same structure as \eqref{eq:s-estimate-MIMO}, it follows that \gls{MR} combining with $\vect{v} = \vect{h} / \| \vect{h} \|$ provides the maximum \gls{SNR}:
\begin{equation} \label{eq:SNR-simple-cell-free}
\mathacr{SNR}^{\textrm{cell-free}}  = \frac{p}{\sigmaUL} \| \vect{h}\|^2 =  \frac{p}{\sigmaUL}\sum_{l=1}^L  |h_l|^2.
\end{equation}
If we compare this expression with that for the small-cell network in \eqref{eq:SNR-simple-small}, we observe that the cell-free network obtains an \gls{SNR} proportional to $\sum_{l=1}^L  |h_l|^2$, while the small-cell setup only contains the largest term in that sum.
Hence, the cell-free network will always obtain a larger \gls{SNR}, but the difference will be small if there is one term that is much larger than the sum of the others.

\enlargethispage{\baselineskip} 

If we instead compare the cell-free setup with the single-cell Massive \gls{MIMO} setup, the main difference is due to the channels $\vect{h}$ and $\vect{g}$. The \glspl{SNR} are proportional to $\| \vect{h}\|^2$ and $\| \vect{g}\|^2$, respectively. We cannot conclude from the mathematical expressions which of these squared norms is the largest. It will depend on the \gls{UE} location. Therefore, we need to continue the comparison using simulations. Recall that $d_l$ denotes the distance between \gls{AP} $l$ and the \gls{UE}, thus we can also write \eqref{eq:SNR-simple-cell-free} as
 
\begin{equation} \label{eq:SNR-simple-cell-free2}
\mathacr{SNR}^{\textrm{cell-free}}  =\frac{p}{\sigmaUL}\sum_{l=1}^L  \beta(d_l).
\end{equation}

\subsubsection{Numerical Comparison}

\enlargethispage{\baselineskip} 

We will now compare the three setups in Figure~\ref{fig:basic-comparison-cellular-cellfree} by simulation when the total coverage area is $400$\,m $\times\,$ $400$\,m. We will drop one \gls{UE} uniformly at random in the area and compute the uplink \glspl{SNR} as described above, assuming the transmit power is $p=10$\,dBm and the noise power is $\sigmaUL = -96$\,dBm, which are reasonable values when the bandwidth is 10\,MHz.
When computing the propagation distances, we assume the \glspl{AP} are deployed 10\,m above the \glspl{UE}.

\begin{figure}[t!]
	\centering 
	\begin{overpic}[width=\columnwidth,tics=10]{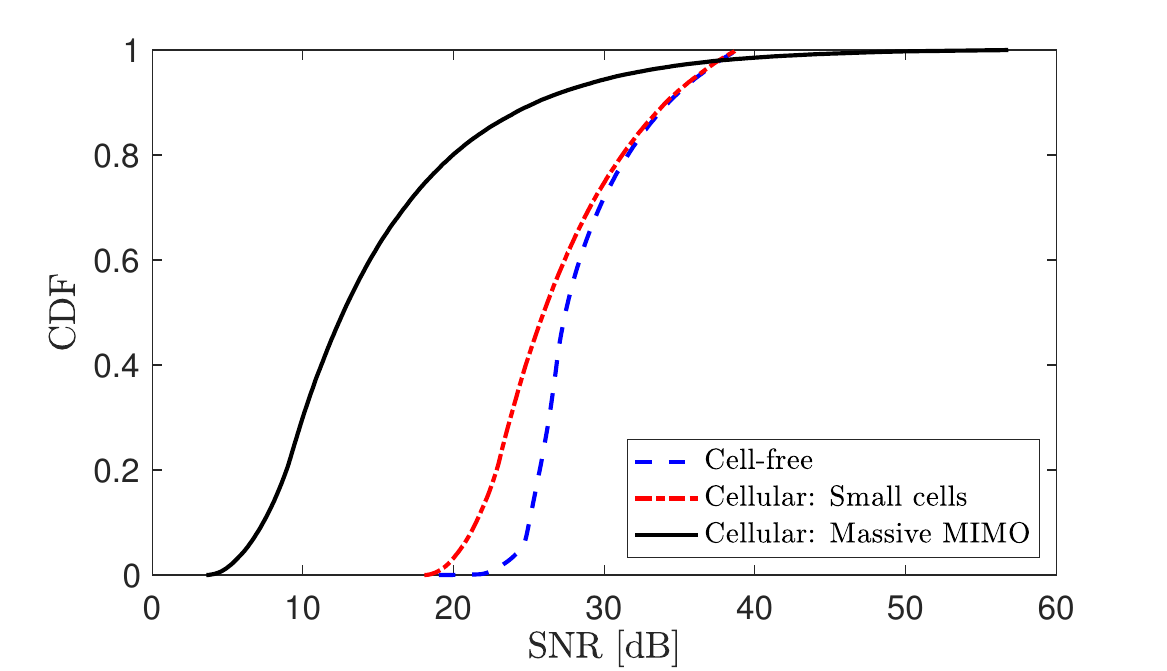}
		\put(24,12){95\% likely}
		\put(13,9.65){$- - - - - - - - - - - - $}
	\end{overpic}  
	\caption{The \gls{SNR} achieved by a \gls{UE} in each of the setups illustrated in Figure~\ref{fig:basic-comparison-cellular-cellfree}. The  \gls{UE} location is selected uniformly at random in the area, which gives rise to the \glspl{CDF}.} 
	\label{fig:section1_figure9} 
\end{figure}

Figure~\ref{fig:section1_figure9} shows the \gls{CDF} of the \gls{SNR} achieved by 
the \gls{UE} at different random locations. In the single-cell Massive \gls{MIMO} case, there are 50\,dB \gls{SNR} variations, where the largest values are achieved when the \gls{UE} is right underneath the \gls{AP} and the smallest values are achieved when the \gls{UE} is in the corner.
The \gls{SNR} variations are much smaller for the cell-free network, since the distances to the closest \gls{AP} is generally much shorter than in the Massive \gls{MIMO} case. Moreover, the \gls{SNR} is higher at the vast majority of \gls{UE} locations. If we look at the 95\% likely \gls{SNR}, indicated by the dashed line where the CDF value is 0.05, there is an 18\,dB difference. More precisely, the cell-free network guarantees an SNR of 24.5\,dB (or higher) at  95\% of all \gls{UE} locations, while Massive \gls{MIMO} only guarantees 6.5\,dB. It is only in the upper end of the \gls{CDF} curves (representing the most fortunate \gls{UE} locations) that Massive \gls{MIMO} is the preferred option. This represents the case when the \glspl{UE} are very close to the 64-antenna Massive \gls{MIMO} array, while a \gls{UE} can only be close to a few \gls{AP} antennas at a time in the cell-free network.

As expected from the analytical expressions, the cell-free network always achieves a higher \gls{SNR} than the corresponding small-cell setup. The difference is negligible in the upper end of the \gls{CDF} curves, when the \gls{UE} is very close to only one of the \glspl{AP} so there is a single dominant term in \eqref{eq:SNR-simple-cell-free2}, while there is a 4\,dB gap in the 95\% likely \gls{SNR}. Based on this example, we can conclude that distributed antennas are preferred over large co-located arrays, but the cell-free architecture only has a minor benefit compared to the cellular small-cell network having the same \gls{AP} locations. To observe a more convincing practical benefit of the cell-free approach, we need to consider a setup with multiple \glspl{UE} so that there is interference between the concurrent transmissions.

\subsection{Benefit 2: Better Ability to Manage Interference}
\label{subsec:benefit2}

We will now demonstrate that cell-free networks have the ability to manage interference, which is what small-cell networks are lacking. For the sake of argument, we once again consider the uplink of the three setups shown in Figure~\ref{fig:basic-comparison-cellular-cellfree} but now with $K=8$ \glspl{UE}.
We let $p$ denote the transmit power used by each \gls{UE}, while $\sigmaUL$ denotes the noise power. 

\subsubsection{Massive \gls{MIMO} Setup}

In the single-cell Massive \gls{MIMO} setup in Figure~\ref{fig:basic-comparison-cellular-cellfree}(a), we let $\vect{g}_k \in \mathbb{C}^M$ denote the channel from \gls{UE} $k$ to the \gls{AP}. Similar to \eqref{eq:basic-uplink}, the received uplink signal becomes
\begin{equation} \label{eq:basic-uplink-multiuser}
\vect{y}^{\textrm{MIMO}} = \sum_{i=1}^{K} \vect{g}_i s_i + \vect{n}
\end{equation}
where $s_i \in \mathbb{C}$ is the information signal transmitted by \gls{UE} $i$ (with $\mathbb{E}\{ | s_i|^2 \} = p$) and $\vect{n} \sim \CN ( \vect{0}_M, \sigmaUL \vect{I}_M )$ is the receiver noise. The \gls{AP} applies the receive combining vector $\vect{v}_k \in \mathbb{C}^M$ to the received signal in \eqref{eq:basic-uplink-multiuser} in an effort to obtain the estimate 
\begin{equation} \label{eq:s-estimate-MIMO-multiuser}
\hat{s}_k^{\textrm{MIMO}}  = \vect{v}_k^{\Htran} \vect{y}^{\textrm{MIMO}}  =   \sum_{i=1}^{K} \vect{v}_k^{\Htran} \vect{g}_i s_i + \vect{v}_k^{\Htran}\vect{n}
\end{equation}
of the signal $s_k$ from \gls{UE} $k$. The corresponding \gls{SINR} is
\begin{align} \notag
\mathacr{SINR}_k^{\textrm{MIMO}} &\!= \frac{\mathbb{E}\{ | \vect{v}_k^{\Htran} \vect{g}_k s_k  |^2 \}}{ \mathbb{E} \left\{ \left|
\condSum{i=1}{i \neq k}{K} \vect{v}_k^{\Htran} \vect{g}_i s_i \!+\! \vect{v}_k^{\Htran}\vect{n}
\right|^2 \right\} } \!=\! \frac{| \vect{v}_k^{\Htran} \vect{g}_k|^2 p }{\vect{v}_k^{\Htran} \left( p  \condSum{i=1}{i \neq k}{K} \vect{g}_i \vect{g}_i^{\Htran} \!+\! \sigmaUL \vect{I}_M  \right) \vect{v}_k} \\
&\leq p \vect{g}_k^{\Htran} \left( p  \condSum{i=1}{i \neq k}{K} \vect{g}_i \vect{g}_i^{\Htran} + \sigmaUL \vect{I}_M  \right)^{-1}\!\!\!\!\!\! \vect{g}_k \label{eq:SINRk-basic-uplink-multiuser}
\end{align}
where the upper bound is achieved by  \cite[Lemma B.10]{massivemimobook}
\begin{align}
\vect{v}_k = \left(  p  \condSum{i=1}{i \neq k}{K} \vect{g}_i \vect{g}_i^{\Htran} + \sigmaUL \vect{I}_M  \right)^{-1} \!\!\!\!\!\!\vect{g}_k.
\end{align}
We will provide a more detailed derivation later in this monograph. For now, the important thing is that the maximum uplink \gls{SINR} of \gls{UE} $k$ is given by \eqref{eq:SINRk-basic-uplink-multiuser}.

\enlargethispage{\baselineskip} 

\subsubsection{Cellular Setup With Small Cells}

In the small-cell setup in Figure~\ref{fig:basic-comparison-cellular-cellfree}(b), we let $h_{kl} \in \mathbb{C}$ denote the channel response between \gls{UE} $k$ and AP $l$. Similar to \eqref{eq:basic-uplink-small-cell}, the received uplink signal at AP $l$ becomes 
\begin{equation} \label{eq:basic-uplink-small-cell-multiuser}
y_l^{\textrm{small-cell}} = \sum_{i=1}^{K} h_{il} s_i + n_l
\end{equation}
where $s_i \in \mathbb{C}$ denotes the information signal from \gls{UE} $i$ (with $\mathbb{E}\{ | s_i|^2 \} = p$) and $n_l \sim \CN ( 0, \sigmaUL )$ is the receiver noise.
The \gls{SINR} at \gls{AP} $l$ with respect to the signal from \gls{UE} $k$ is
\begin{equation}
\mathacr{SINR}_{kl}^{\textrm{small-cell}} =  
\frac{\mathbb{E}\{ | h_{kl} s_k  |^2 \}}{ \mathbb{E} \left\{ \left|
\condSum{i=1}{i \neq k}{K}  h_{il} s_i + n_l
\right|^2 \right\} }  = \frac{p |h_{kl}|^2}{p\condSum{i=1}{i \neq k}{K} | h_{il}|^2 +\sigmaUL}. \vspace{-1mm}
\end{equation}
Each \gls{UE} selects to receive service from the AP  that provides the largest \gls{SINR}. Hence, the \gls{SINR} of \gls{UE} $k$ is
\begin{align}  \label{eq:basic-small-cell-SINR-multiuser}
\mathacr{SINR}_k^{\textrm{small-cell}} = \max_{l \in \{1,\ldots,L\} }
\mathacr{SINR}_{kl}^{\textrm{small-cell}}.
\end{align}
The preferred \gls{AP} might not be the one with the largest \gls{SNR} due to the interference. It can happen that one \gls{AP} serves multiple \glspl{UE}.

\vspace{-3mm}
\enlargethispage{\baselineskip} 
\subsubsection{Cell-Free Setup}

In the cell-free setup in Figure~\ref{fig:basic-comparison-cellular-cellfree}(c), the $L$ \glspl{AP} from the small-cell setup are cooperating in detecting the information sent from the $K$ \glspl{UE}. We can write the received signals in \eqref{eq:basic-uplink-small-cell-multiuser} jointly as
\begin{equation} \label{eq:basic-uplink-cell-free-multiuser}
\vect{y}^{\textrm{cell-free}} = \sum_{i=1}^{K} \vect{h}_i s_i + \vect{n}
\end{equation}
where $\vect{h}_i = [ h_{i1} \, \ldots \, h_{iL} ]^{\Ttran}$ and $\vect{n} = [ n_1 \, \ldots \, n_L ]^{\Ttran}$.
Similar to the single-cell Massive \gls{MIMO} case, a receive combining vector $\vect{v} _k \in \mathbb{C}^L$ is applied to \eqref{eq:basic-uplink-cell-free-multiuser} to detect the signal from \gls{UE} $k$. This leads to the estimate
\begin{equation} \label{eq:s-estimate-cellfree-multiuser}
\hat{s}_k^{\textrm{cell-free}}  = \vect{v}_k^{\Htran} \vect{y}^{\textrm{cell-free}}  =   \sum_{i=1}^{K} \vect{v}_k^{\Htran} \vect{h}_i s_i + \vect{v}_k^{\Htran}\vect{n}
\end{equation}
of $s_k$. The corresponding \gls{SINR} is
\begin{align} \notag
\mathacr{SINR}_k^{\textrm{cell-free}} &= \frac{\mathbb{E}\{ | \vect{v}_k^{\Htran} \vect{h}_k s_k  |^2 \}}{ \mathbb{E} \left\{ \left|
\condSum{i=1}{i \neq k}{K} \vect{v}_k^{\Htran} \vect{h}_i s_i + \vect{v}_k^{\Htran}\vect{n}
\right|^2 \right\} } = \frac{| \vect{v}_k^{\Htran} \vect{h}_k|^2 p }{\vect{v}_k^{\Htran} \left( p  \condSum{i=1}{i \neq k}{K} \vect{h}_i \vect{h}_i^{\Htran} + \sigmaUL \vect{I}_M  \right) \vect{v}_k} \\
&\leq p \vect{h}_k^{\Htran} \left( p  \condSum{i=1}{i \neq k}{K} \vect{h}_i \vect{h}_i^{\Htran} + \sigmaUL \vect{I}_M  \right)^{-1} \!\!\!\!\!\!\vect{h}_k \label{eq:SNR-simple-cell-free-multiuser}
\end{align}
where the upper bound is achieved by \cite[Lemma B.10]{massivemimobook}
\begin{align}
\vect{v}_k = \left( p  \condSum{i=1}{i \neq k}{K} \vect{h}_i \vect{h}_i^{\Htran} + \sigmaUL \vect{I}_M  \right)^{-1} \!\!\!\!\!\!\vect{h}_k.
\end{align} 
Compared to the single \gls{UE} case studied earlier, it is harder to utilize the \gls{SINR} expressions derived in this section to deduce which setup will provide the best performance. Intuitively, the cell-free setup will provide higher \gls{SINR} than the small-cell setup since we are using the optimal combining vector, while one suboptimal option is to let $\vect{v}_k$ contain $1$ at the position representing the \gls{AP} with the highest local \gls{SINR} and $0$ elsewhere. That suboptimal selection would lead to the same SINR as in the small-cell setup. To compare the cell-free setup with the single-cell Massive \gls{MIMO} setup, we need to run simulations since the \gls{SINR} expressions in \eqref{eq:SINRk-basic-uplink-multiuser} and \eqref{eq:SNR-simple-cell-free-multiuser} have a similar form, but contain channel vectors that are generated differently.

\subsubsection{Numerical Comparison}

We will now simulate the performance of this multi-user setup using the channel gain model in \eqref{eq:large-scale-model-simple} and the same parameter values as in Figure~\ref{fig:section1_figure9}. More precisely, if the propagation distance is $d$, then the channel is generated as $\sqrt{\beta(d)} e^{\imagunit \phi}$ where $\phi$ is an independent random variable uniformly distributed between $0$ and $2\pi$. This variable models the random phase shift between the transmitter and receiver. This phase was omitted in the previous simulation since the result was determined only by the norms of the channels. However, it is important to include the phases when considering multi-user interference, which is also determined by the directions of the channel vectors.

\begin{figure}[t!]
	\centering 
	\begin{overpic}[width=\columnwidth,tics=10]{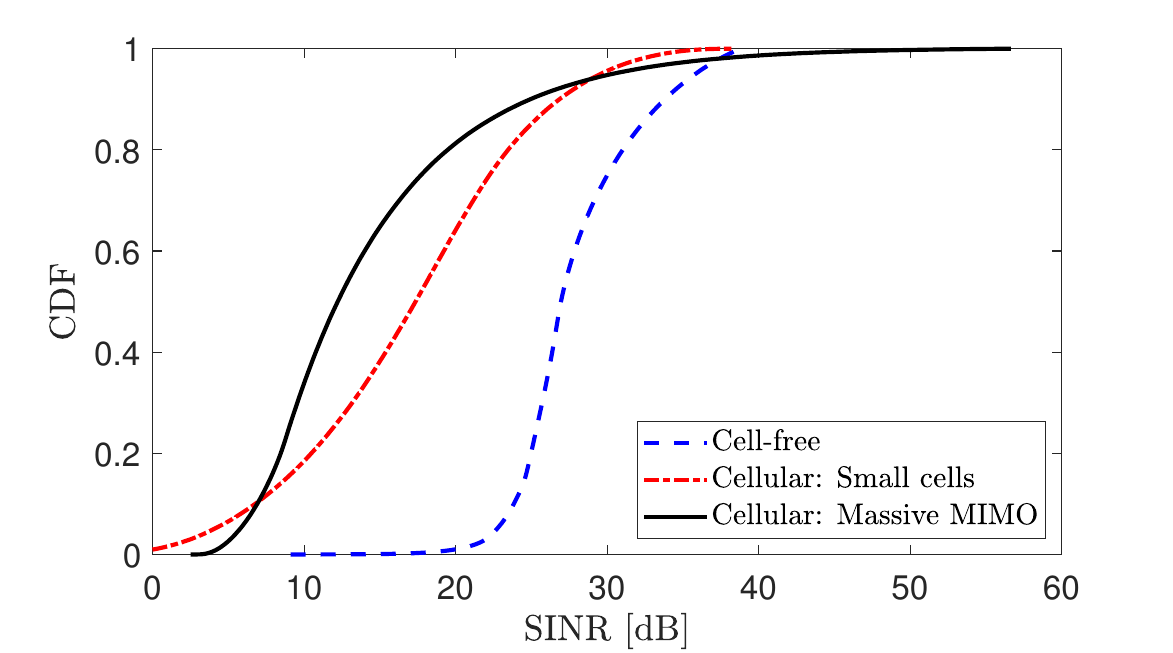}
	\put(25,12){95\% likely}
	\put(13,9.7){$- - - - - - - - - - - - $}
	\end{overpic} 
	\caption{The \gls{SINR} achieved by an arbitrary \gls{UE} in each of the setups illustrated in Figure~\ref{fig:basic-comparison-cellular-cellfree}. There are $K=8$ \glspl{UE} that are distributed uniformly at random in the area. This gives rise to the CDF.} 
	\label{fig:section1_figure10} 
\end{figure}

Figure~\ref{fig:section1_figure10} shows the \gls{CDF} of the \gls{SINR} achieved by a randomly selected \gls{UE} in a random realization of the $K=8$ uniformly distributed \gls{UE} locations. As compared to Figure~\ref{fig:section1_figure9}, all the curves are moved to the left in Figure~\ref{fig:section1_figure10} due to the interference among the \glspl{UE}. The Massive \gls{MIMO} case is barely affected by the interference, which demonstrates that this technology has the ability to separate the \glspl{UE}' channels spatially using the large array of co-located antennas. However, due to the large variations in distances to the \gls{AP}, there are 50\,dB variations in the \gls{SINR} between different \gls{UE} locations.
The cell-free network is also barely affected by the interference, but one can see a tiny lower tail that corresponds to the random event that two \glspl{UE} are randomly deployed at almost the same location.

The major difference from the single-\gls{UE} case is that the small-cell curve is moved far to the left and the 95\%-likely \gls{SINR} is even lower than with Massive \gls{MIMO}. The reason is that each \gls{AP} only has a single antenna and thus cannot suppress inter-cell interference. The cell-free setup is greatly outperforming the small-cell setup in this multi-user setup. This is what will occur in practice since mobile networks are deployed to serve multiple \glspl{UE} in the same geographical area.

\subsection{Benefit 3: Coherent Transmission Increases the \gls{SNR}}

The previous two benefits were exemplified in the uplink but there are also counterparts in the downlink, which lead to similar results but for partially different reasons. One important difference is that the received power in the uplink increases with the number of receive antennas (i.e., a larger fraction of the transmit power is collected), thus it is always beneficial to have more antennas. Consider now a downlink scenario where we can deploy any number of antennas, but constrain the total downlink transmit power to be constant (to not change the energy consumption). We then need to determine how the power should be divided between the \glspl{AP} to maximize the \gls{SNR}.
Suppose a \gls{UE} is in the vicinity of two \glspl{AP} but one has a substantially better channel. It might then seem logical that all the transmit power should be assigned to the \gls{AP} with the better channel, but we will show that this is not the optimal strategy.

\enlargethispage{\baselineskip} 
Suppose, for the sake of argument, that AP 1 has the channel response $h_1=\sqrt{\alpha}$ to the \gls{UE}, while AP 2 has the channel response $h_2=\sqrt{\alpha/2}$. If we compare the channel gains $|h_1|^2=\alpha$ and $|h_2|^2=\alpha/2$, it is clear that AP 1 has the best channel.
Let $\rho$ denote the total downlink transmit power and $\sigmaDL$ denote the receiver noise power. If only \gls{AP} 1 transmits to the \gls{UE}, then the \gls{SNR} at the receiver is
\begin{equation}
\frac{\rho \alpha}{\sigmaDL}.
\end{equation}
However, if \gls{AP} 1 instead transmits with power $2\rho/3$ and \gls{AP} 2 transmits with power $\rho/3$, which also corresponds to a total power of $\rho$, then the \gls{SNR} is
\begin{equation}\label{eq:downlink-SNR}
\frac{1}{\sigmaDL} \left( \sqrt{\frac{2\rho}{3}} h_1 + \sqrt{\frac{\rho}{3}} h_2 \right)^2 =1.5 \frac{\rho \alpha}{\sigmaDL}.
\end{equation}
Hence, the \gls{SNR} is higher when we transmit from both \glspl{AP}. This is a consequence of the coherent combination (i.e., constructive interference) of the signals from the two \glspl{AP}. The power gain is reminiscent of the beamforming gain from co-located arrays, where the transmit power is stronger in some angular directions than in other ones, but the physical interpretation is somewhat different. 
The coherent combination of signals that are transmitted from different geographical points does not give rise to beam patterns but rather local signal amplification in a region around the receiver.
Moreover, all the antennas in a co-located array will experience (roughly) the same channel gain so it is logical that they should be jointly utilized for coherent transmission. In contrast, distributed antennas can experience very different channel gains but are anyway useful for coherent transmission.

\begin{figure}[t!]
	\centering  \vspace{-4mm}
        \begin{subfigure}[b]{\columnwidth} \centering
	\begin{overpic}[width=.8\columnwidth,tics=10]{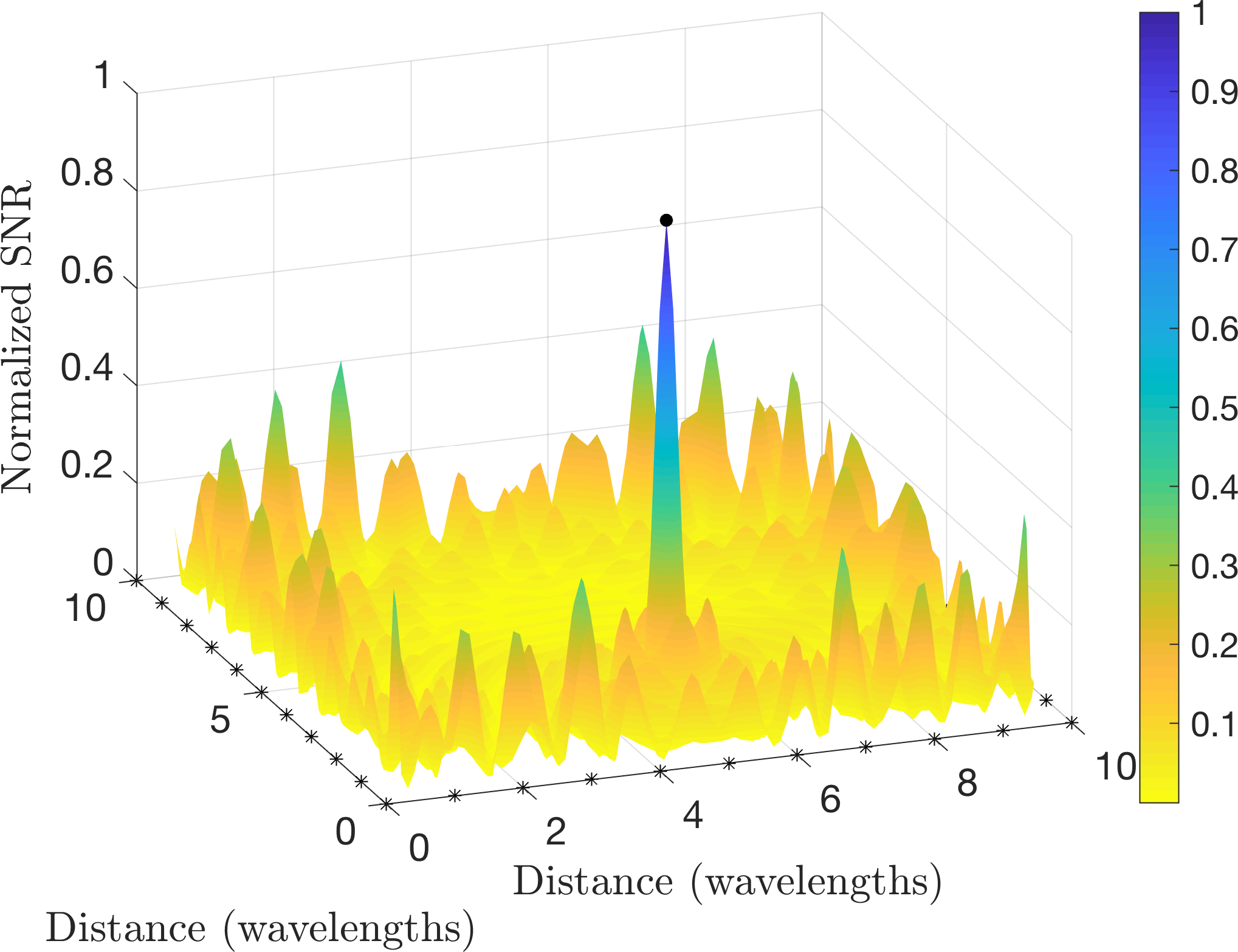}
\end{overpic} 
                \caption{View with normalized \gls{SNR} (between 0 and 1) on the vertical axis.}    \vspace{+2mm}
        \end{subfigure} 
        \begin{subfigure}[b]{\columnwidth} \centering 
	\begin{overpic}[width=.8\columnwidth,tics=10]{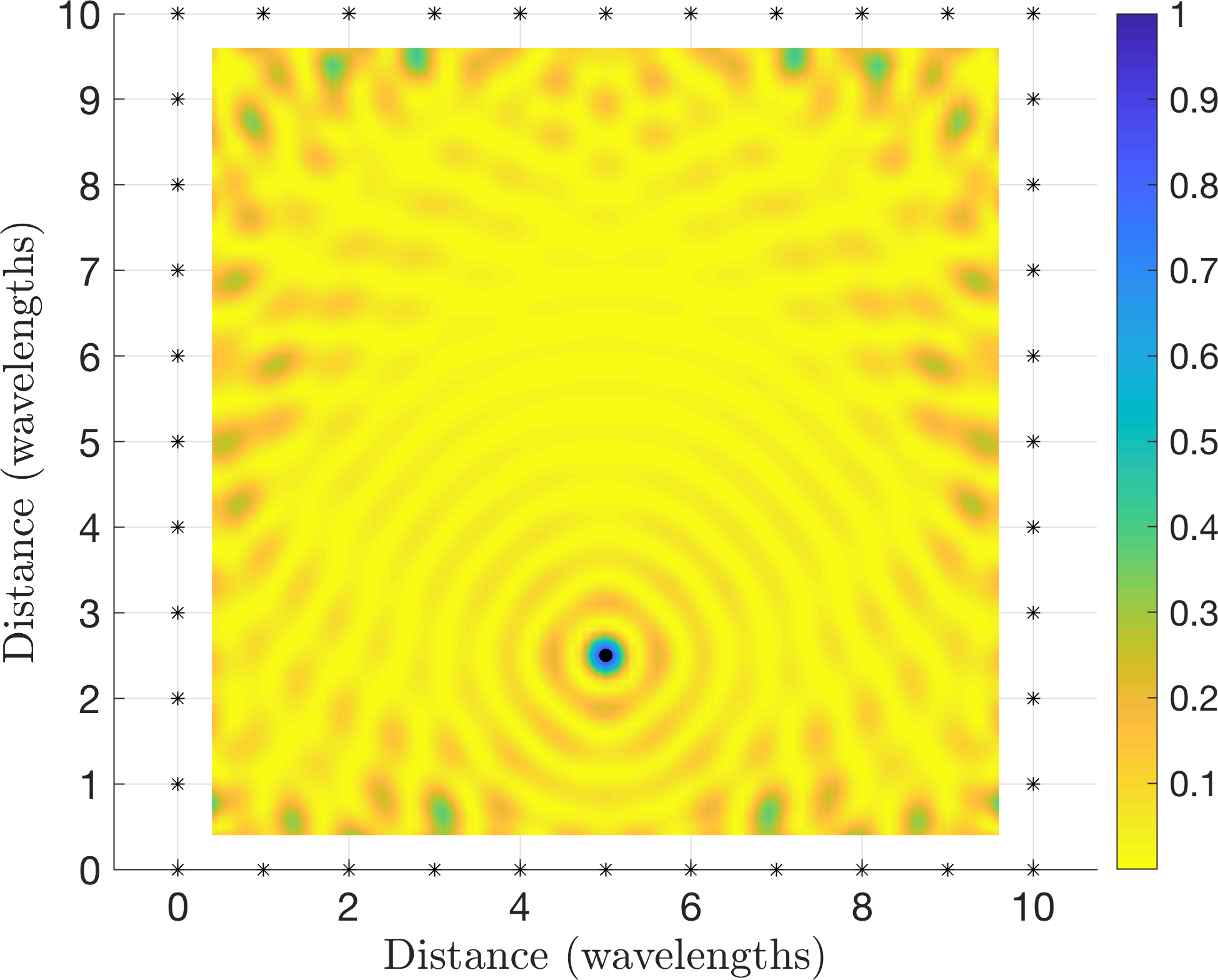}
\end{overpic} 
                \caption{Same as in (a) but viewed from above.} 
        \end{subfigure} 
	\caption{The received signal power at different locations when transmitting from one-wavelength-spaced antennas along the walls (each marked with a star). The antennas transmit with equal power and the signals are phase-shifted to achieve coherent combination at the point where the normalized \gls{SNR} is 1.} 
	\label{fig:distributed-focusing} \vspace{-4mm}
\end{figure}

We will now illustrate that the signal focusing obtained by a distributed array does not give rise to signal beams.
Figure~\ref{fig:distributed-focusing} shows the \gls{SNR} variations when transmitting from $M=40$ antennas that are equally spaced along the perimeter of a square. The adjacent antennas are one-wavelength-spaced and transmit with equal power. The received power decays as the square of the propagation distance (as would be the case in free space). If a narrowband signal is considered, the received signals will be phase-shifted (time-delayed) by the ratio between the propagation distance and the signal's wavelength. For the sake of argument, the distances in Figure~\ref{fig:distributed-focusing} are therefore measured as fractions of the wavelength (each side of the square is ten wavelengths). To achieve a coherent combination at the point of the receiver, each antenna must phase-shift (time-delay) its signal before transmission to make sure that all the $M$ signals are reaching the receiver perfectly synchronized.

In Figure~\ref{fig:distributed-focusing}(a), we show the \glspl{SNR} measured at different locations when focusing all the signals into a single point. The \gls{SNR} values are normalized so that they are equal to 1 at the point-of-interest (i.e., the location of the receiver) and smaller elsewhere. We observe that the \gls{SNR} is much larger at that point than on all the surrounding points, where the 40 signals are not coherently combined. There are some points near the edges of the simulation area where the \gls{SNR} is also strong but this is not due to a coherent combination of multiple signal components. Instead, it is because these points are close to some of the transmit antennas. Figure~\ref{fig:distributed-focusing}(b) shows the same results but from above. The figure reveals that the \gls{SNR} is strong in a circular region around the point-of-interest. The diameter of this region is roughly half-a-wavelength. In summary, the signal focusing from distributed arrays will not give rise to angular beams (as in Cellular Massive MIMO) but local signal focusing around the receiver in a region that is smaller than the wavelength. When considering a three-dimensional propagation environment, the \gls{SNR} will be large within a sphere around the point-of-interest with the diameter being half-a-wavelength. When transmitting multiple signals, we can focus each one at a different point and if these points are several wavelengths apart, the mutual interference will be small according to Figure~\ref{fig:distributed-focusing}.

\newpage

\section{Summary of the Key Points in Section~\ref{c-introduction}}
\label{c-introduction-definition-summary}

\begin{mdframed}[style=GrayFrame]

\begin{itemize}
\item Traditional wireless networks use the cellular architecture. The cellular approach was conceived for providing wide-area coverage to low-rate voice services. Each \gls{AP} is surrounded by \glspl{UE} at very different distances, having widely different \glspl{SNR}. This architecture is badly suited for providing ubiquitous access to high-rate data services.

\item The cell-free architecture turns the situation around: each \gls{UE} is surrounded by \glspl{AP}. Each \gls{AP} has relatively simple hardware and cooperates with surrounding APs to jointly serve the UEs in their area of influence. A cell-free network is user-centric if each UE is served by its nearest APs.

\item The name \emph{Cell-free Massive MIMO} signifies that it is the combination of three previously known components: The physical layer of Massive MIMO, the vision of creating ultra-dense networks with many more APs than UEs, and the coordinated multipoint methods for achieving a cell-free network. The main novelty lies in how to co-design these components to achieve a user-centric operation that is sufficiently scalable to enable large-scale deployments.

\item The first key benefit of the cell-free architecture is the smaller SNR variations compared to cellular networks with a sparse deployment of APs and Massive MIMO.

\item The second key benefit is the ability to manage interference by joint processing at multiple APs, which is not done in cellular networks with an equally dense AP deployment.

\item The third key benefit is that coherent transmission increases the SNR. It is better to involve APs with weaker channels in the transmission than only using the AP with the best channel.

\end{itemize}

\end{mdframed}


\chapter{User-Centric Cell-Free Massive MIMO Networks}
\label{c-cell-free-definition} 

This section introduces the basic system model and concepts related to a User-centric Cell-free Massive MIMO network, which will be used in the remainder of this monograph. Section~\ref{def:cell-free-Massive-MIMO} provides a formal definition of Cell-free Massive MIMO. The user-centric operation is introduced in Section~\ref{sec:dynamic-clustering} based on the \gls{DCC} framework. The modulation and communication protocol are defined in Section~\ref{sec:signal-model-UL-and_DL}, along with the uplink and downlink system models. Section~\ref{sec:scalability} defines the meaning of network scalability and reviews the system model from that perspective. A sufficient condition for a cell-free network being scalable is provided. Section~\ref{sec:channel-modeling} introduces the channel model that will be used for analysis and performance evaluation in later sections. Section~\ref{sec:hardening-favorable} defines the concepts of channel hardening and favorable propagation, and analyzes them in the context of cell-free networks.
The key points are summarized in Section~\ref{c-cell-free-definition-summary}.

\enlargethispage{\baselineskip} 

\section{Definition of Cell-Free Massive MIMO}\label{def:cell-free-Massive-MIMO}

In Section~\ref{subsec-history-cellfree} on p.~\pageref{subsec-history-cellfree}, we defined Cell-free Massive MIMO as an ultra-dense network (i.e., many more APs than UEs) where the APs are cooperating to serve the UEs by joint coherent transmission and reception, while making use of the physical layer concepts from the Cellular Massive MIMO area. We will now turn this into a technical framework and then progressively add more details to it in later sections.

\begin{figure}[t!]
	\centering 
	\begin{overpic}[width=\columnwidth,tics=10]{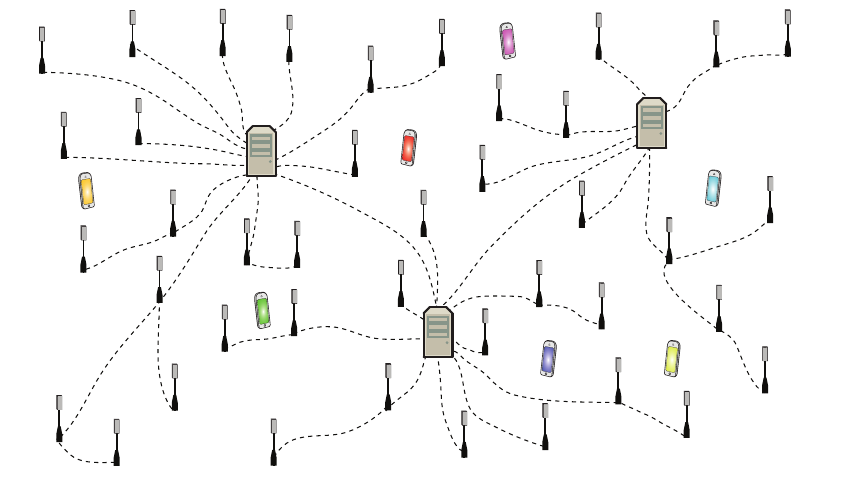}
		\put(79,40){\footnotesize{CPU}}
		\put(2,23){\footnotesize{AP $l$}}
		\put(45,42){\footnotesize{UE $k$}}
	\end{overpic} 
	\caption{A Cell-free Massive MIMO network consists of $L$  geographically distributed APs that jointly serve $K$ geographically distributed UEs. Each AP is equipped with $N$ antennas, while each UE has a single antenna.} 
	\label{fig:cell-free-repeated}  
\end{figure}

We consider a network with $L$ \glspl{AP}, each equipped with $N$ antennas, that are geographically distributed over the coverage area, as shown in Figure~\ref{fig:cell-free-repeated}. We let $M=NL$ denote the total number of AP antennas in the network. The \glspl{AP}  are jointly serving $K$ single-antenna \glspl{UE}. More precisely, each \gls{UE} is communicating with a subset of the \glspl{AP}, which is selected based on the \gls{UE}'s needs. The \glspl{AP} are connected via fronthaul links to CPUs, which facilitate the \gls{AP} coordination.
The cell-free architecture can also handle multi-antenna UEs, where the extra antennas are utilized to suppress interference or spatially multiplex several signals per UE. This case is not covered in this monograph because it substantially complicates the analysis and one can easily apply the presented theory to the multi-antenna case by treating each UE antenna as a separate UE. More intricate solutions can be found in \cite{Alonzo2019a}, \cite{Buzzi2017b}, \cite{Buzzi2020b}, \cite{Jin2019a}, \cite{Li2016b}, \cite{Mai2020a}.

\pagebreak
The intended operating regime of Cell-free Massive \gls{MIMO} is $L \gg K$  \cite{Nayebi2017a}, \cite{Ngo2017b}, which is the mathematical definition of an ultra-dense network. This also implies $M \gg K$; that is, the total number of \gls{AP} antennas is much larger than the total number of \glspl{UE}, as in conventional Cellular Massive \gls{MIMO} systems \cite{Marzetta2010a}.
The intention is that the cell-free network will, under these circumstances, have sufficiently many spatial degrees-of-freedom so that the \glspl{UE} can be separated in space by processing the transmitted and received signals. For example, whenever the \glspl{AP} focus the transmission towards a particular \gls{UE}, the focus area will be sufficiently small so that no other UE receives high interference. Moreover, the number of antennas $N$ per AP is intended to be small and each AP might serve more UEs than its antennas, thus AP cooperation is essential to suppress interference between the UEs.
However, we stress that there is no need for a formal assumption of $M \gg K$ or $L \gg K$ in the analysis of User-centric Cell-free Massive \gls{MIMO} networks. The concepts and signal processing methods presented in this monograph hold for any value of $L$, $K$, and $N$. In fact, we will later extract Cellular Massive MIMO and small-cell networks as two special cases of the framework presented in this monograph, which enables a fair comparison between the technologies.

\section{User-Centric Dynamic Cooperation Clustering}\label{sec:dynamic-clustering}

The first papers on Cell-free Massive MIMO assumed all UEs were served by all APs, as mentioned in Section~\ref{subsec-history-cellfree} on p.~\pageref{subsec-history-cellfree}. This is both impractical and unnecessary in a geographically large network, where each UE is only physically close to a subset of APs. 
By limiting the set of \glspl{AP} that is allowed to transmit to the \gls{UE}, the service quality will inevitably be reduced. However, if all APs that can reach the \gls{UE} with a signal power that is non-negligible compared to the thermal noise power take an active part in serving that \gls{UE}, then the performance loss should be negligible. This calls for a user-centric approach to the formation of the \gls{AP} clusters that serve each \gls{UE}. The practical benefits are that the fronthaul signaling is reduced when only a subset of the \glspl{AP} must receive the downlink data intended for the \gls{UE} and send their corresponding estimates of the uplink data to the \gls{CPU}. Moreover, the computational complexity is reduced when each \gls{AP} only needs to process signals related to a subset of the \glspl{UE}.

To model which \glspl{AP} are serving which \glspl{UE}, we will make use of the \gls{DCC} framework initially proposed in \cite{Bjornson2011a}, \cite{Bjornson2013d}.\footnote{The \gls{DCC} framework was introduced to enable ``\emph{unified analysis of anything from interference channels to Network MIMO}''. Therefore, the user-centric cooperation clusters considered in Cell-free Massive MIMO is only one of the many instances of this framework.}

\begin{definition}\label{definition:DCC}
Dynamic cooperation clustering (DCC) means that \gls{UE} $k$ is served only by the \glspl{AP} with indices in the set $\mathcal{M}_k \subset \{ 1, \ldots, L\}$.
\end{definition}

\begin{figure}[t!]
	\centering 
	\begin{overpic}[width=\columnwidth,tics=10]{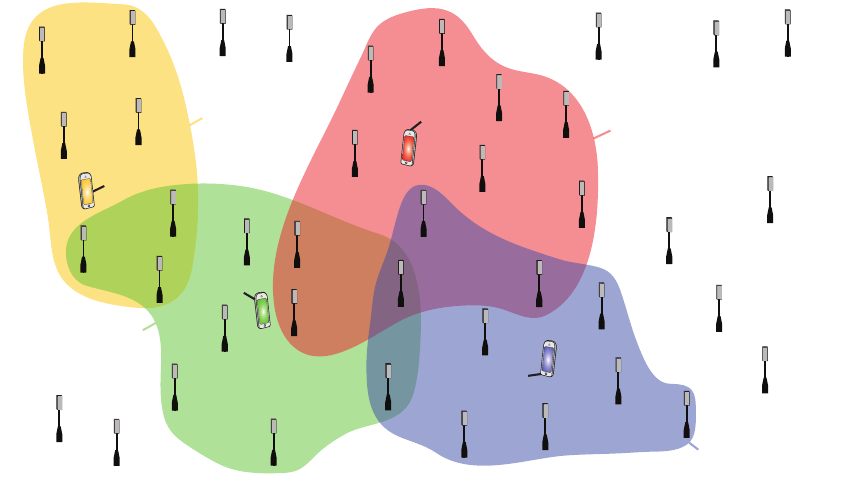}
		\put(13,35){\footnotesize{UE 1}}
		\put(24,43){\footnotesize{AP cluster}}
		\put(24,40){\footnotesize{for UE 1}}
		\put(22,22){\footnotesize{UE 2}}
		\put(3,17){\footnotesize{AP cluster}}
		\put(3,14){\footnotesize{for UE 2}}
		\put(50,42){\footnotesize{UE 3}}
		\put(73,41){\footnotesize{AP cluster}}
		\put(73,38){\footnotesize{for UE 3}}
		\put(55,11){\footnotesize{UE 4}}
		\put(83,3){\footnotesize{AP cluster}}
		\put(83,0){\footnotesize{for UE 4}}
	\end{overpic}  
	\caption{Example of dynamic cooperation clusters for four \glspl{UE} in a Cell-free Massive MIMO network with a large number of \glspl{AP}.}
	\label{fig:illustrateCooperation}  
\end{figure}
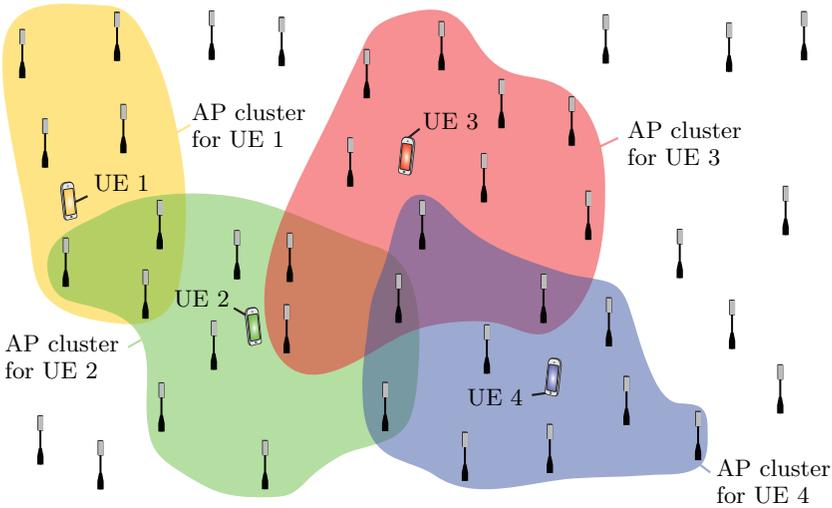

The word \emph{dynamic} refers to the fact that the sets $\mathcal{M}_k$ can be adapted to time-variant characteristics, such as UE locations, service requirements, interference situation, etc. Figure~\ref{fig:illustrateCooperation} illustrates one instance of the DCC framework with four \glspl{UE} and a large number of \glspl{AP}. The colored regions illustrate which clusters of \glspl{AP} are serving which \glspl{UE}. The \glspl{AP} within a colored region are those with indices in the set $\mathcal{M}_k$ and we select them to involve all the neighboring ones. The fact that the clusters are partially overlapping between the \glspl{UE} is showcasing that this is a user-centric cell-free network; we cannot divide the \glspl{AP} into disjoint sets that serve disjoint subsets of the UEs. More generally, every AP is co-serving UEs with a set of other APs, which in turn are co-serving other UEs with another set of APs, and the chain continues like this until all APs have been considered (at least when many UEs are distributed over the coverage area).

When analyzing the performance of the data transmission, we assume the sets $\mathcal{M}_k$, for $k=1,\ldots,K$, are fixed and known everywhere needed. The reason is that the dynamic variations occur over much larger time intervals than channel fading variations. More precisely, the data is transmitted in blocks that are sufficiently large to be subject to many fading realizations but only one realization of $\mathcal{M}_1,\ldots,\mathcal{M}_K$. Different ways to select these sets are later discussed in Section~\ref{sec:selecting-DCC} on p.~\pageref{sec:selecting-DCC}.

For notational convenience, we also define a set of diagonal matrices $\vect{D}_{kl} \in \mathbb{C}^{N \times N}$, for $k=1,\ldots,K$ and $l=1,\ldots,L$, determining which APs communicate with which UEs. More precisely, $\vect{D}_{kl}$ is the identity matrix ${\bf I}_N$ if AP $l$ is allowed to transmit to and decode signals from UE $k$ and ${\bf 0}_{N\times N}$ otherwise. Following the notation from Definition~\ref{definition:DCC}, we have that
\begin{equation}\label{eq:user-centric}
\vect{D}_{kl}= \begin{cases}
\vect{I}_N &  l \in \mathcal{M}_k \\ 
\vect{0}_{N \times N} &  l \not \in \mathcal{M}_k. \end{cases}
\end{equation}
Note that if we select $\mathcal{M}_k=\{ 1, \ldots, L\}$, for $k=1,\ldots,K$ and, hence, all matrices $\vect{D}_{kl}=\vect{I}_N$, then we obtain the original form of Cell-free Massive \gls{MIMO} from \cite{Nayebi2017a}, \cite{Ngo2017b}, where all \glspl{AP} jointly serve all \glspl{UE} in the network.
Another special case of the \gls{DCC} framework is a small-cell network, which is obtained by selecting $\mathcal{M}_k$ to have a single element for $k=1,\ldots,K$, so that each \gls{UE} is only served by one \gls{AP}.

\section{System Models for Uplink and Downlink}\label{sec:signal-model-UL-and_DL}

In electrical engineering, a \emph{system} is something that takes an input signal and filters it to produce an output signal. A wireless communication channel is an example of such a system: it filters the transmitted signal and outputs the received signal. In this subsection, we motivate and describe the system model for User-centric Cell-free Massive MIMO that is used in the remainder of this monograph. We begin by describing the block-fading model and introducing the coherence block concept.

\subsection{Block Fading and Coherence Blocks}\label{sec:block-fading}

The type of systems that are convenient to analyze is linear and time-invariant.
Wireless channels are linear due to the superposition principle prescribed by Maxwell's equations. However, the channels are generally time-varying since the transmitter, receiver, and objects in the wireless propagation environment can move. Even minor movements over a distance proportional to the wavelength (i.e., a few millimeters or centimeters) will substantially change the channel. For example, the UE at the focus point in Figure~\ref{fig:distributed-focusing} on p.~\pageref{fig:distributed-focusing} only needs to move a quarter-of-a-wavelength to leave the area where the SNR is high. 
However, if we consider a sufficiently short time interval, the channel can be approximated as constant and, therefore, the communication system is time-invariant. That interval is called \emph{channel coherence time}. This time is determined by the speed of movement and is typically a few milliseconds in mobile networks. Since it is important for the communication system to know the channel properties, it needs to estimate them once per coherence time interval.

Within the coherence time, the channel can be described by a \gls{FIR} filter where each term of the impulse response describes one distinguishable propagation path in a multipath environment with a distinct time delay and pathloss. Such a filter reacts differently to input signals with different frequencies, but the variations are rather smooth when varying the signal frequency. Hence, if we consider a sufficiently narrow frequency range, the channel can be considered constant. The \emph{channel coherence bandwidth} defines the frequency interval in which the frequency response is approximately constant.

\begin{figure}[t!]
	\centering 
	\begin{overpic}[width=\columnwidth,tics=10]{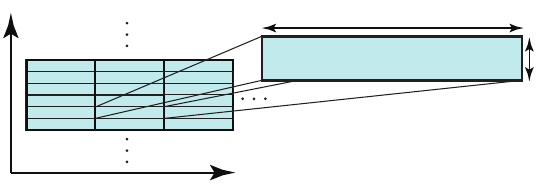}
		\put(0,32){Frequency}
		\put(44,1){Time}
		\put(97,22){$B_c$}
		\put(69,30){$T_c$}
		\put(56,22){One coherence block}
	\end{overpic}  
	\caption{In the block-fading model, the time-frequency resources are divided into coherence blocks in which the channel is time-invariant and frequency-flat. The channel is modeled as independent between different coherence blocks.}
	\label{basic_block_fading}  
\end{figure}
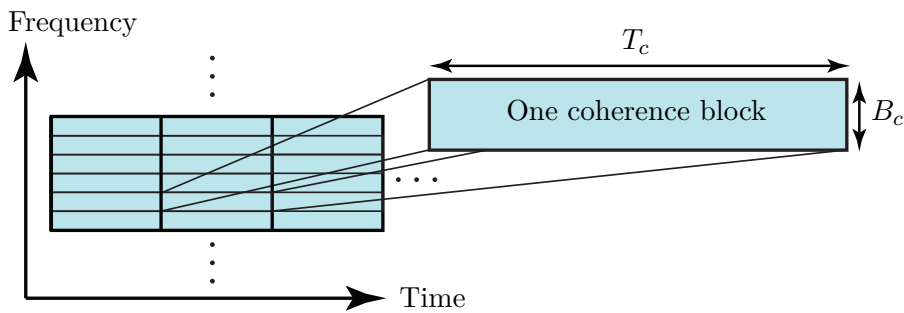

A \emph{channel coherence block} is a time-frequency block with a time duration that equals the coherence time and a frequency width that equals the coherence bandwidth. The channel between two antennas is constant and frequency-flat within a coherence block, which implies that it can be described by only one scalar coefficient. The time-frequency resources of a communication system can be divided into different coherence blocks as illustrated in Figure~\ref{basic_block_fading}. Note that the blocks are distributed over both time and frequency. When developing communication algorithms, it is convenient to break down the operation into blocks like this and then design, analyze, and optimize them separately. If one further assumes that the channel-describing scalar coefficient takes an independent random realization in each coherence block, then one can study one block at a time without loss of generality. This is called the \emph{block-fading model} since the channel fading takes one independent realization in each coherence block. The block-fading model is assumed throughout this monograph.

Let $T_c$ denote the coherence time (in seconds) and $B_c$ denote the coherence bandwidth (in Hertz). According to the Nyquist-Shannon sampling theorem, a signal that fits into this block is uniquely described by
$\tau_c = T_cB_c$ complex-valued samples. These are the parameters that can be used to convey information in a communication system. We will call them \emph{transmission symbols} or just symbols.

In practice, the coherence time and bandwidth depend on many factors such as the channel delay spread (i.e., the time difference between the shortest and longest resolvable paths), UE mobility, and carrier frequency. The exact values are user-dependent and can be measured experimentally, but we will provide some rules-of-thumb.
A common approximation of the coherence time is $T_c = \lambda/(4 \upsilon)$ \cite[Sec.~2.3]{Tse2005a}, where $\lambda$ is the wavelength and $\upsilon$ is the velocity of the UE. This is the time it takes to move a quarter-of-a-wavelength. Hence, if the carrier frequency is high, the channel changes more rapidly for a given velocity. The same happens if the UE velocity increases for a given carrier frequency. The coherence bandwidth can be approximated by $B_c=1/(2 T_d)$ where $T_d$ is the channel delay spread; that is, the time difference between the earliest and last propagation path.

We need a common block size when operating the system but every UE has different values of $T_c$ and $B_c$. A practical solution is to fix $\tau_c$ based on the worst-case scenario that the network should support \cite[Sec.~2.1]{massivemimobook}. To give quantitative numbers, suppose the carrier frequency is 2\,GHz, which gives the wavelength $\lambda = 15$\,cm. We consider two examples:

\begin{itemize}
\item Outdoor scenario: Suppose the delay spread is up to 2.5\,$\mu$s (i.e., 750\,m path differences) and we want to support mobility up to $\upsilon = 37.5 \, \textrm{m/s} = 135 \, \textrm{km/h}$. The approximations above then become  $T_c = 1$\,ms and $B_c = 200$\,kHz. The coherence block contains $\tau_c=200$ transmission symbols in this scenario that supports fairly high mobility and high channel dispersion.

\item Indoor scenario: Suppose the delay spread is up to 0.1\,$\mu$s (or 30\,m path differences) and we want to support mobility up to $\upsilon = 0.75 \, \textrm{m/s} = 2.7 \, \textrm{km/h}$. The approximations above then become $T_c = 50$\,ms and $B_c = 5$\,MHz. The coherence block contains $\tau_c=250\,000$ transmission symbols in this scenario with low mobility and low channel dispersion. 
\end{itemize}

In summary, the number of transmission symbols per coherence block ranges from hundreds (with high mobility and high channel dispersion) to hundreds of thousands of samples (with low mobility and low channel dispersion). Setups similar to the outdoor scenario with  $\tau_c=200$ are commonly studied in the literature on Cellular Massive MIMO \cite{massivemimobook}, \cite{Marzetta2016a}. If a cell-free network is deployed to serve the same type of \glspl{UE} as in conventional cellular networks, we can safely assume that $\tau_c \geq 200$. However, the coherence blocks can be much larger in cell-free networks if we primarily target low-mobility use cases and deploy the antennas closer to the \glspl{UE}, so that the delay spread shrinks.

\begin{remark}[Relation to practical systems]
The main reason for study\nolinebreak ing\linebreak block-fading channels, instead of a system with a practical multicarrier modulation scheme, is that we can neglect many of the technicalities that appear in practical systems. In this way, we can focus on the~main concepts and develop a theory that applies to many types of systems. In a practical system where \gls{OFDM} is utilized, the available band\nolinebreak width\linebreak is divided into many subcarriers and each one features a scalar chan\nolinebreak nel.\linebreak Several subcarriers will then fit into what we have defined as a coherence block \cite[Sec.~2.1.6]{Marzetta2016a}. These subcarriers will not have identical~channels, but there is a known transformation between them, thus if~one~learns\linebreak the channel coefficient at some of the subcarriers, the other ones can be obtained by interpolation \cite{Kashyap2016a}.
Moreover, the channel coefficients~will~not change abruptly every coherence block but gradually. In particular,~the channel realizations will be correlated between adjacent coherence blocks, which can actually be utilized to improve the performance compared to the methods described in this monograph. This is especially important for low-mobility UEs for which the channel might be constant for a much longer time than the worst-case value of $T_c$ that is used by the system.
\end{remark}

\enlargethispage{\baselineskip} 

\subsection{Time-Division Duplex Protocol}\label{sec:tdd-protocol}

The channel between AP $l$ and UE $k$ in an arbitrary coherence block is represented by the vector $\vect{h}_{kl} \in \mathbb{C}^N$. It is an $N$-dimensional vector where the $n$th element represents the complex-valued scalar channel coefficient between the UE and the $n$th antenna of the AP. We also define the collective channel $\vect{h}_k \in \mathbb{C}^{M}$ from all APs to UE $k$ as
\begin{equation}\label{eq:collective-channel}
\vect{h}_k = \begin{bmatrix} \vect{h}_{k1} \\ \vdots \\  \vect{h}_{kL}
\end{bmatrix}. 
\end{equation}
According to the block-fading model assumption, the channel vector $\vect{h}_{kl}$ takes one independent realization in each coherence block from a stationary random distribution. Due to the channel reciprocity of physical channels, the channel realization is the same in both uplink and downlink. We will define the channel distribution later in Section~\ref{sec:channel-modeling}.

The APs must know the channel realizations, at least partially, to perform coherent processing both between the antennas on each AP and across the cooperating APs. The most efficient way to estimate the channels is to consider a \gls{TDD} protocol where each coherence block is used for both uplink and downlink transmissions. It is then sufficient to transmit pilots only in the uplink. Based on the received uplink signals, each AP can then estimate the channels between itself and all the UEs. These channel estimates can be utilized for both uplink and downlink transmissions, thanks to the channel reciprocity. By applying coherent precoding in the downlink based on the channel estimates, the complex-valued $M$-dimensional channel vector to each UE is transformed into a scalar channel that is positive (except for minor perturbations due to estimation errors). This scalar channel can be deduced at the UE from the downlink data signals, without sending explicit pilots \cite{Ngo2017a}. We will elaborate more on this later in this monograph.
We refer to \cite[Sec.~1.3.5]{massivemimobook} for a more detailed discussion on this and a comparison with the alternative \gls{FDD} protocol, in terms of signaling overhead. While the uplink can operate identically in \gls{TDD} and \gls{FDD} systems, the downlink must be implemented differently in \gls{FDD}.
Applications of \gls{FDD} protocols to Cell-free Massive MIMO can be found in 
\cite{Abdallah2020a}, \cite{Kim2020a}, \cite{Kim2018a} and will not be covered in this monograph.

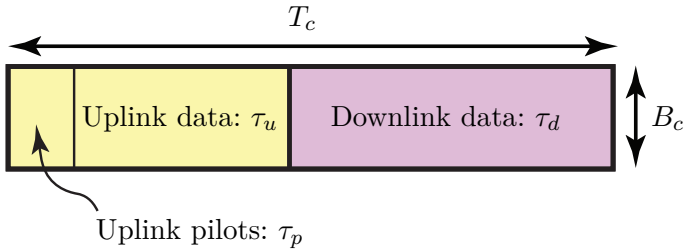
\begin{figure}[t!]
	\centering 
	\begin{overpic}[width=.8\columnwidth,tics=10]{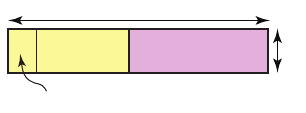}
		\put(13,19){Uplink data: $\tau_u$}
		\put(48,19){Downlink data: $\tau_d$}
		\put(93,19){$B_c$}
		\put(42,33){$T_c$}
		\put(15,3){Uplink pilots: $\tau_p$}
	\end{overpic}  
	\caption{A \gls{TDD} protocol is considered where each coherence block is used for both uplink and downlink transmissions.}
	\label{figure:basic_protocol}  
\end{figure}

We consider the standard Cellular Massive \gls{MIMO} \gls{TDD} protocol, where the $\tau_c$ transmission symbols of a coherence block are used for three purposes:
\begin{enumerate}
\item $\tau_p$ symbols for uplink pilots; 
\item $\tau_u$ symbols for uplink data; 
\item $\tau_d$ symbols for downlink data.
\end{enumerate}
This protocol is illustrated in Figure~\ref{figure:basic_protocol}. The three sets of transmission symbols can be divided between the three purposes in different ways, under the constraint that $\tau_c = \tau_p + \tau_u + \tau_d$. Note that the pilots and data are transmitted at different times and that the TDD protocol is synchronized across the APs. 
The uplink pilots must be transmitted before the downlink data so that the downlink transmission can be precoded based on the uplink estimates. However, it is plausible to transmit uplink data before the uplink pilots, or to change the order between the uplink and downlink data transmissions, as long as it all fits into a single coherence block. The drawback with the former variation is that the latency increases since the data detection cannot be initiated until after the pilots have been received. The drawback with the latter variation is that the switch between downlink and uplink requires a guard time interval so that all UEs receive the downlink data before they switch to uplink transmission mode.

\subsection{Uplink System Model}\label{sec:uplink-system-model}

During the uplink data transmission, all APs will receive a superposition of the signals sent from all UEs. The received signal ${\bf y}_{l}^{\rm{ul}} \in \mathbb {C}^{N}$ at AP $l$ is
\begin{align}\label{eq:DCC-uplink}
	{\bf y}_{l}^{\rm{ul}} = \sum\limits_{i=1}^{K} {\bf h}_{il}s_{i} + {\bf n}_{l}
\end{align}
where $s_{i} \in \mathbb{C}$ is the signal transmitted from UE $i$ with a power that we denote as $p_i = \mathbb{E} \{ |s_i|^2 \}$ during the uplink data transmission and as $\eta_i = |s_i|^2$ in the uplink pilot transmission.
The independent additive receiver noise is ${\bf n}_{l}  \sim \CN({\bf 0}_N,\sigmaUL{\bf I}_{N})$. The uplink signals can either contain data or pilots. While the channels are constant within a coherence block, the signals and noise take new realizations at every transmission symbol.

Based on the received signal in \eqref{eq:DCC-uplink}, AP $l$ can compute an estimate $\widehat{s}_{kl}$ of the signal $s_k$ transmitted by UE $k$. This is the first step towards decoding the information that the signal contains.
However, the AP should only do that if $l\in \mathcal {M}_k$ or, equivalently, $\vect{D}_{kl}=\vect{I}_N$. For notational convenience, we want to represent both cases jointly by setting $\widehat{s}_{kl}=0$ if  $l \not \in \mathcal {M}_k$.
This is achieved by defining a receive combining vector $\vect{v}_{kl} \in \mathbb {C}^{N}$ that AP $l$ could use if it serves UE $k$. We then define the \emph{effective receive combining vector}
\begin{equation}
\vect{D}_{kl} \vect{v}_{kl} = \begin{cases}
\vect{v}_{kl} & l\in \mathcal {M}_k \\
\vect{0}_N & l \not \in \mathcal {M}_k. \label{eq:effective-receive-combining}
\end{cases}
\end{equation}
Note that it is equal to the actual receive combining vector for APs that serve UE $k$ and zero otherwise. The estimate of $s_k$ at AP $l$ can then be computed by taking the inner product between $\vect{D}_{kl}  \vect{v}_{kl}$ and ${\bf y}_{l}^{\rm{ul}}$ (noting that $\vect{D}_{kl}= \vect{D}_{kl}^{\Htran}$):
\begin{align}
\widehat{s}_{kl} &= \vect{v}_{kl}^{\Htran} \vect{D}_{kl} {\bf y}_{l}^{\rm{ul}}\label{eq:uplink-data-estimate}.
\end{align}
By utilizing \eqref{eq:effective-receive-combining}, the expression in \eqref{eq:uplink-data-estimate} can be rewritten as
\begin{equation}
\widehat{s}_{kl} =\begin{cases}
\vect{v}_{kl}^{\Htran}  {\bf h}_{kl}s_{k} + \condSum{i=1}{i \neq k}{K} \vect{v}_{kl}^{\Htran} {\bf h}_{il}s_{i} + \vect{v}_{kl}^{\Htran} {\bf n}_{l} & l\in \mathcal {M}_k \\
0 & l \not \in \mathcal {M}_k.
\end{cases}
\label{eq:uplink-data-estimate-2}
\end{equation}
However, we will be utilizing the general expression in \eqref{eq:uplink-data-estimate} in the majority of this monograph since it allows us to cover both cases in a joint manner, instead of treating them separately.

The receive combining vectors ${\bf v}_{kl}$, for $l\in \mathcal {M}_k$, should be selected based on the \gls{CSI} available at the APs. Different designs based on various degrees of cooperation among the APs will be introduced and analyzed in detail in Section~\ref{c-uplink} on p.~\pageref{c-uplink}. Channel estimation is considered in Section~\ref{c-channel-estimation} on p.~\pageref{c-channel-estimation}.

\subsection{Downlink System Model} \label{sec:downlink-system-model-ch2}

In the downlink, we let $\vect{x}_l \in \mathbb{C}^N$ denote the signal transmitted by AP $l$.~The received signal at UE $k$ can then be written as
\begin{align} \label{eq:DCC-downlink-0}
y_k^{\rm{dl}} = \sum_{l=1}^{L}  \vect{h}_{kl}^{\Htran}\vect{x}_l + n_k
\end{align}
where $n_k \sim \CN(0,\sigmaDL)$ is the receiver noise.
Note that we have represented the downlink channel by $\vect{h}_{kl}^{\Htran}$, although there will only be a transpose and not any complex conjugate in practice.
However, the \linebreak additional conjugation is a standard way of simplifying the notation without changing the performance. We will therefore adopt this convention.\footnote{When applying the downlink results of this monograph in practical systems, we should transmit $\vect{x}_l^{\star}$ instead of $\vect{x}_l$, we should treat $(y_k^{\rm{dl}})^{\star}$ as the true received signal, and the true noise term will be $n_k^{\star}  \sim \CN(0,\sigmaDL)$. Hence, it is only a few conjugates that differ, while the communication performance is identical.}

We let $\varsigma_i \in \mathbb{C}$ denote the independent unit-power data signal intended for UE $i$, thus $\mathbb{E} \{{|\varsigma_i |^2}\} = 1$.
The signal transmitted by AP $l$ consists of a precoded superposition of the signals intended for the different UEs that this AP serves. This can be expressed as
\begin{align} 
\vect{x}_l = \sum_{i=1}^{K} \vect{D}_{il}  \vect{w}_{il} \varsigma_i \label{eq:DCC-transmit-downlink}
\end{align}
where  $\vect{w}_{il} \in \mathbb{C}^N$ denotes the transmit precoding vector that AP $l$ assigns to UE $i$. This vector has two important impacts on the transmission: the direction $\vect{w}_{il}/\|\vect{w}_{il}\|$ determines the spatial directionality and the average squared norm $\mathbb{E}\{ \| \vect{w}_{il} \|^2 \}$ determines the average transmit power.
We can define the \emph{effective transmit precoding vector} as
\begin{equation} \label{eq:effective-precoding}
\vect{D}_{il}  \vect{w}_{il} = \begin{cases} \vect{w}_{il} & l \in \mathcal{M}_i \\
\vect{0}_{N} & l \not \in \mathcal{M}_i
\end{cases}
\end{equation}
since $ \vect{D}_{il}=\vect{0}_{N\times N}$ implies $\vect{D}_{il}  \vect{w}_{il}=\vect{0}_N$. Hence, each AP only needs to select precoding vectors for the UEs that it serves.

By substituting the transmitted signal in \eqref{eq:DCC-transmit-downlink} into \eqref{eq:DCC-downlink-0}, the received downlink signal at UE $k$ becomes
\begin{align} \notag
y_k^{\rm{dl}} & = \sum_{l=1}^{L}  \vect{h}_{kl}^{\Htran} \left( \sum_{i=1}^{K} \vect{D}_{il}  \vect{w}_{il} \varsigma_i \right)+ n_k \\&= \sum_{l=1}^{L}  \vect{h}_{kl}^{\Htran} \vect{D}_{kl}  \vect{w}_{kl} \varsigma_k + \condSum{i=1}{i \neq k}{K}  \sum_{l=1}^{L}  \vect{h}_{kl}^{\Htran}\vect{D}_{il}  \vect{w}_{il} \varsigma_i + n_k.\label{eq:DCC-downlink}
\end{align}
The first term in \eqref{eq:DCC-downlink} is the desired signal, the second term is inter-user interference, and the third term is noise. The property of the effective transmit precoding vectors in \eqref{eq:effective-precoding} can be used to rewrite \eqref{eq:DCC-downlink} as
\begin{align} 
y_k^{\rm{dl}} = \sum_{l \in \mathcal{M}_k}  \vect{h}_{kl}^{\Htran}  \vect{w}_{kl} \varsigma_k + \condSum{i=1}{i \neq k}{K}  \sum_{l \in \mathcal{M}_i}  \vect{h}_{kl}^{\Htran} \vect{w}_{il} \varsigma_i + n_k
\end{align}
which will not simplify but rather complicate the notation later in this monograph. We will therefore utilize the general form in \eqref{eq:DCC-downlink} for performance analysis and keep in mind that one can always utilize \eqref{eq:effective-precoding} to get alternative expressions.

As in the uplink, the selection of the precoding vectors $ \vect{w}_{kl}$ depends on the available channel estimates and the degree of cooperation among the \glspl{AP}. This will be thoroughly discussed in Section~\ref{c-downlink} on p.~\pageref{c-downlink}.

\section{Network Scalability}\label{sec:scalability}

Since the geographical coverage area of the network can be huge, the technology must be designed to be scalable in the sense that one can add more APs and/or UEs to the network without having to increase the capabilities of the existing ones.
Conventional cellular networks achieve scalability through a divide-and-conquer approach. The coverage area is divided into cells and each cell operates autonomously. Within a cell, there is a maximum number of UEs that can be simultaneously served by spatial multiplexing (if there are more UEs, then time-frequency scheduling must be used). However, the cell operation is unaffected by the addition of new APs or UEs belonging to other cells. Therefore, cellular technology is inherently scalable and this is also confirmed by the fact that nation-wide cellular networks exist all over the world.

The situation is more complicated in cell-free networks since all APs are somehow involved in the service of all UEs. A given AP can either be directly involved in the service of a particular UE $k$, in the sense of sending/receiving data to/from the UE, or indirectly involved; for example, by serving another UE together with an AP that transmits to UE $k$. When we deploy an AP, it will have a maximum computational capability and a maximum fronthaul capacity. These resources must remain sufficient even as the network size increases, in terms of deploying new APs, and as the network load grows, in terms of adding more UEs.

To determine if a cell-free network is scalable or not, it is instrumental to let $K \to \infty$ and see which of the following tasks remain practically implementable at each AP:
\begin{enumerate}
\item \emph{Signal processing for channel estimation};
\item\emph{Signal processing for data reception and transmission};
\item \emph{Fronthaul signaling for data and \gls{CSI} sharing};
\item \emph{Power allocation optimization}.
\end{enumerate}

Based on this list, we make the following definition.

\begin{definition}[Scalability] \label{def:scalable}
A Cell-free Massive MIMO network is\linebreak \emph{scalable} if all the four above-listed tasks have finite computational complexity and resource requirements with respect to each AP as $K \to \infty$.
\end{definition}

\enlargethispage{0.5\baselineskip} 

The intention with this definition is \emph{not} that a cell-free network will serve an infinitely large number of UEs in practice, but to uncover fundamental scalability issues that become clear in the asymptotic regime. There are many algorithms that have manageable complexity for generating simulations in academic papers, but cannot be used in practice where the number of APs and UEs will be much larger than in a basic simulation. Definition~\ref{def:scalable} might give the impression that all the signal processing and optimization must be carried out locally at the AP, but this is not necessary. An implementation where each AP sends the data and CSI to a CPU, which carries out those tasks, can be scalable as long as the corresponding complexity and fronthaul signaling do not grow with $K$.

\enlargethispage{\baselineskip}

\subsection{Example of Unscalable Network Operation}

Before explaining how to achieve scalability in a cell-free network, we will exemplify the opposite: an unscalable network operation. Suppose AP $l$ is serving all the $K$ UEs in the entire cell-free network, which implies that $l \in \mathcal{M}_k$, for $k=1,\ldots,K$. The AP carries out the signal processing and optimization locally. Four potential scalability issues were mentioned in Definition~\ref{def:scalable} and we will go through them one after the other to explain why they are not satisfied.

Firstly, we have the complexity of the signal processing related to channel estimation. If AP $l$ should serve $K$ UEs, it will have to learn the channels of all these UEs.
The details on channel estimation are provided in Section~\ref{c-channel-estimation} on p.~\pageref{c-channel-estimation}, but since there are $K$ channel vectors ${\bf h}_{1l},\ldots,{\bf h}_{Kl}$ to estimate, the computational complexity will for sure grow at least linearly with the number of UEs. Hence, the complexity approaches infinity as $K \to \infty$. Moreover, the memory size needed to store these channel estimates will be proportional to $K$ and, thus, any finite-sized memory module will be insufficient as $K \to \infty$.

Secondly, AP $l$ needs to create the downlink signal $\sum_{i=1}^{K} \vect{w}_{il} \varsigma_i$ in \eqref{eq:DCC-transmit-downlink}, where the summation implies infinite complexity as $K \to \infty$. The complexity of computing the $K$ precoding vectors $\{\vect{w}_{kl}: k=1,\ldots,K\}$ depends on the precoding scheme, but will also grow at least linearly with $K$ since there are $NK$ coefficients to compute, and every vector also needs to be properly scaled. The same scalability issue appears in the uplink, where the AP needs to compute $\{\vect{v}_{kl}^{\Htran}{\bf y}_{l}^{\rm{ul}}: k=1,\ldots,K\}$ using $K$ different combining vectors $\{\vect{v}_{kl}: k=1,\ldots,K\}$.

Thirdly, AP $l$ needs to receive the $K$ downlink data signals $\{\varsigma_k: k=1,\ldots,K\}$ from a CPU and must forward its $K$ processed received signals $\{\vect{v}_{kl}^{\Htran}{\bf y}_{l}^{\rm{ul}}: k=1,\ldots,K\}$ over the fronthaul links. The number of scalars to be sent over the fronthaul grows unboundedly as $K \to \infty$, thus the AP's fronthaul capacity will be insufficient. Moreover, the memory capacity required to temporarily store the data symbols of the $K$ UEs at AP $l$ will also grow unboundedly with $K$.

Fourthly, AP $l$ needs to select how to allocate its transmit power between the $K$ UEs. Any power allocation algorithm that makes use of channel knowledge related to all UEs will have a complexity that grows at least linearly with $K$, which is not scalable as $K \to \infty$. 

This example demonstrates that the core of the scalability issue is that AP $l$ serves all the UEs. Based on this observation, we will provide a sufficient condition for achieving a scalable implementation of cell-free networks.

\subsection{A Sufficient Condition for Scalability}

Recall that $\vect{D}_{il}$ is non-zero only if AP $l$ serves UE $i$. Hence, the set of \glspl{UE} served by AP $l$ can be defined as
\begin{equation}\label{eq:definition-of-the-set-D_l}
\mathcal{D}_l = \bigg\{ i : \tr(\vect{D}_{il})\geq 1, i \in \{ 1, \ldots, K \} \bigg\}.
\end{equation}
The following lemma presents a condition on $\mathcal{D}_l$ that partially guarantees scalability in a cell-free network.

\begin{lemma} \label{lemma:D-requirement}
	If the cardinality $|\mathcal{D}_l |$ remains finite as $K \to \infty$ for $l=1,\ldots,L$, then the cell-free network satisfies the first three scalability tasks in Definition~\ref{def:scalable}.
\end{lemma}
\begin{proof}
	\gls{AP} $l$ only needs to compute the channel estimates and precoding/combining vectors for $|\mathcal{D}_l |$ \glspl{UE}. This has a finite complexity as $K \to \infty$ if $|\mathcal{D}_l |$ remains finite. Moreover, \gls{AP} $l$ only needs to send/receive data related to these $|\mathcal{D}_l |$ \glspl{UE} over the fronthaul links, which is a finite number as $K \to \infty$.
\end{proof}

The implication of Lemma~\ref{lemma:D-requirement} is that we need to limit the number of active UEs that each AP can serve. Ideally, we should assign APs to UEs so that every UE is served by all the surrounding APs; that is, the APs that will influence the UE's performance noticeably. At the same time, we need to make sure that $|\mathcal{D}_l |$ remains small and finite. How to determine sets $\mathcal{D}_l$ that guarantee both scalability and good service to the \glspl{UE} will be explored first in Section~\ref{sec:selecting-DCC} on p.~\pageref{sec:selecting-DCC} and then in further detail in Section~\ref{sec:selecting-DCC-advanced} on p.~\pageref{sec:selecting-DCC-advanced}. Until then, we  assume that $\mathcal{D}_1,\ldots,\mathcal{D}_L$ are given.

Note that only the fourth scalability condition (regarding the complexity of the power allocation) is not guaranteed by the finite cardinality of $\mathcal{D}_l$ in Lemma~\ref{lemma:D-requirement}. The next lemma provides the missing piece.

\begin{lemma} \label{lemma:scalability2}
Suppose every AP $l$ selects its downlink transmit power based only on information about the $|\mathcal{D}_l|$ UEs served by that AP. Furthermore, suppose every UE is assigned a transmit power by only one of the serving APs and this AP selects that power based only on information about the $|\mathcal{D}_l|$ UEs that it serves. If Lemma~\ref{lemma:D-requirement} is satisfied, then the cell-free network satisfies all the scalability conditions in Definition~\ref{def:scalable}.
\end{lemma}
\begin{proof}
\gls{AP} $l$ only needs to compute the downlink transmit power for $|\mathcal{D}_l|$ UEs using an information set that is proportional to $|\mathcal{D}_l|$, but independent of $K$. Furthermore, every AP will at most compute the uplink transmit powers for $|\mathcal{D}_l|$ UEs. If the condition in Lemma~\ref{lemma:D-requirement} is satisfied, then the fourth condition in Definition~\ref{def:scalable} will also be satisfied as $K \to \infty$ by following the power allocation requirements specified in this lemma.
\end{proof}

The implication of Lemma~\ref{lemma:scalability2} is that the transmit powers should be selected in a distributed manner to achieve scalability. There exist several network-wide power allocation algorithms that jointly select the uplink or downlink powers. 
These can, for instance, be the solutions to linear or convex optimization problems, whose computational complexity grows polynomially in the number of optimization variables and, thus, are unscalable.
Some key examples are provided in Section~\ref{sec:optimized-power-allocation} on p.~\pageref{sec:optimized-power-allocation} but these are meant as benchmarks rather than as practical solutions.

Scalable power allocation algorithms should preferably be implemented separately for every AP, with no or limited interaction between the APs. It is easy to design such algorithms; for example, every UE can transmit with full power in the uplink and every AP can divide its transmit power equally between the UEs that it serves.
However, it is much harder to create algorithms that operate close to the network-wide power allocation benchmarks. State-of-the-art algorithms for scalable power allocation will be reviewed in Section~\ref{sec:scalable-power-allocation} on p.~\pageref{sec:scalable-power-allocation}. Until then, we will assume the transmit powers are fixed and given.

\begin{remark}[Scalability in centralized operation]
The sufficient condition for scalability in Lemma~\ref{lemma:scalability2} states that APs make decisions related to the transmit powers. This does not mean that each decision has to be made in a local processor at an AP, but it can also be carried out at a neighboring CPU that is associated with the AP. In other words, a centralized operation where the CPUs carry out almost all the signal processing and optimization can also be scalable if designed to satisfy Definition~\ref{def:scalable}.
\end{remark}

 \vspace{-3mm}
 \section{Channel Modeling}
\label{sec:channel-modeling}  

A realistic performance assessment of any MIMO technology requires the use of a channel model that reflects its main characteristics. The wireless channel is deterministic in nature but is often so complicated, due to multipath propagation, that the channel variations appear as random. Hence, both deterministic and stochastic channel models can be used for analysis (cf.~\cite[Sec.~7.3]{massivemimobook}). Deterministic models can be based on ray-tracing or recorded channel measurements. The free-space \gls{LoS} propagation model is another example of a deterministic model \cite[Sec.~1.3.2]{massivemimobook}. The general drawback of deterministic models is that they are only valid for specific scenarios. In contrast, stochastic models are independent of a particular propagation environment and, consequently, allow for more far-reaching quantitative analysis. Nevertheless, the random distribution and its parameter values limit the scope of each stochastic model to a certain category of propagation conditions, such as urban, suburban, or rural scenarios.

\enlargethispage{-\baselineskip}

A classical stochastic model for \gls{NLoS} communications is \emph{uncorrelated Rayleigh fading} where the channel between \gls{AP} $l$ and \gls{UE} $k$ is generated as $\vect{h}_{kl} \sim \CN( \vect{0}_N, \beta_{kl}\vect{I}_N)$.
The complex Gaussian distribution accounts for the random small-scale fading caused by small-scale movements of the transmitter, receiver, or other objects in the propagation environment. More precisely, the received signal is the summation of signal copies arriving from a large number of propagation paths with seemingly random phase shifts. This gives rise to constructive and destructive interference between the paths that can be modeled by  a Gaussian distribution thanks to the central limit theorem. This fading model is called Rayleigh fading since the magnitude of each element of $\vect{h}_{kl}$ has a Rayleigh distribution.

When Rayleigh fading is used in a block-fading model, a new independent realization is drawn from the Gaussian distribution in each coherence block.
Moreover, the variance $\beta_{kl}$ describes the large-scale fading and determines the average channel quality as the UE moves around in a relatively small area.
In other words, $\beta_{kl}$ is determined by the geometric pathloss and shadow fading. The uncorrelated Rayleigh fading model has been (and still is) the basis of most theoretical research in multiple antenna communications since it is an analytically tractable model.
However, the physical consequence of having independently and identically distributed elements in $\vect{h}_{kl}$ is that the AP can transmit signals in any direction and the average received power at the UE will always be the same. 
Measurements campaigns have repeatedly shown that this model is inadequate in practical systems since the elements of the channel vector will be statistically correlated \cite{Sanguinetti2019a}. This is called \emph{spatial correlation} and originates from two phenomena: 1) Transmissions in some spatial directions are more likely to lead to the UE than other directions; 2) The geometry of the antenna array (i.e., shape and antenna spacing) makes it more suited to transmit/receive signals in some directions than in other directions. 
It is only under very strict technical conditions that perfectly uncorrelated fading can be achieved \cite{Pizzo2020}.\footnote{The classical example is a \gls{ULA} with half-wavelength-spaced omni-directional antennas that are placed in an isotropic scattering environment \cite[Sec.~7.2.1]{Marzetta2016a}, where the scattering objects are uniformly distributed over all angles in three dimensions. Note that both the omni-directionality and isotropic scattering are theoretical constructs that are not practically occurring.}

\enlargethispage{\baselineskip} 

 \subsection{Correlated Rayleigh Fading}\label{sec:correlated-rayleigh-fading}

In this monograph, we capture the spatial correlation characteristics by making use of the relatively tractable \emph{correlated Rayleigh fading model}. This model is suitable for NLoS channels with substantial multipath propagation so that the elements of the channel vector can be modeled as being complex Gaussian distributed. The channel between \gls{AP} $l$ and \gls{UE} $k$ is generated as
\begin{equation}
\vect{h}_{kl} \sim \CN( \vect{0}_N, \vect{R}_{kl})
\end{equation}
where $\vect{R}_{kl} \in \mathbb{C}^{N \times N}$ is the spatial correlation matrix between \gls{AP} $l$ and \gls{UE} $k$. Note that this is the correlation matrix of $\vect{h}_{kl}$ since $\mathbb{E} \{ \vect{h}_{kl} \vect{h}_{kl}^{\Htran} \} = \vect{R}_{kl}$.
The Gaussian distribution models the small-scale fading whereas the positive semi-definite correlation matrix $\vect{R}_{kl}$ describes the large-scale fading, including geometric pathloss, shadowing, antenna gains, and spatial channel correlation \cite[Sec.~2.2]{massivemimobook}. 

The diagonal elements of $\vect{R}_{kl}$ might be different but, in analogy with uncorrelated Rayleigh fading, we define the large-scale fading coefficient as the average value:
 \begin{equation} \label{eq:beta-definition-lk}
 \beta_{kl} = \frac{1}{N} \tr \left( \vect{R}_{kl} \right).
 \end{equation}
This is the average channel gain between an antenna at \gls{AP}~$l$ and \gls{UE}~$k$. Moreover, we assume the channel vectors of different APs are independently distributed, thus ${\mathbb{E}}\{\vect{h}_{kn} \vect{h}_{kl}^{\Htran}\} = {\bf 0}_{N\times N}$ for $l\ne n$. This is a reasonable assumption whenever the APs are separated by tens of wavelengths or more, which will be an underlying assumption in this monograph. 
The collective channel is thus distributed as follows:
\begin{equation}\label{eq:collective_channel}
\vect{h}_k \sim \CN(\vect{0}_M, \vect{R}_{k}) 
\end{equation}
where $\vect{R}_{k} = \diag(\vect{R}_{k1}, \ldots,  \vect{R}_{kL}) \in \mathbb{C}^{M \times M}$
is the block-diagonal collective spatial correlation matrix. 

The correlated Rayleigh fading model is used throughout this monograph.
Furthermore, we assume the spatial correlation matrices $\vect{R}_{kl}$ are available wherever needed; see \cite{BSD16A}, \cite{CaireC17a}, \cite{NeumannJU17}, \cite{Sanguinetti2019a}, \cite{UpadhyaJU17} for practical methods for estimating spatial correlation matrices.

 \subsection{Large-Scale Fading Model for Urban Deployments}
 \label{subsec:large-scale-fading-model}

The mathematical analysis in this monograph is applicable for arbitrary values of the large-scale fading coefficients but we will now define a model that will be used for simulations.
We assume that the APs are deployed in urban environments to serve a dense population of UEs. Moreover, the APs are deployed ten meters above the plane where the UEs are located. This matches well with the 3GPP Urban Microcell model that is defined in \cite[Table~B.1.2.1-1]{LTE2017a}. 
The model is designed for the traditional 2\,GHz band but is representative also for other sub-6 GHz frequency bands.
We follow that model in the simulations and compute the large-scale fading coefficient (channel gain) in dB as 
\begin{equation}\label{eq:pathloss-coefficient}
\beta_{kl} \,  [\textrm{dB}] = -30.5 - 36.7 \log_{10}\left( \frac{d_{kl}}{1\,\textrm{m}} \right)  + F_{kl}
\end{equation}
where $d_{kl}$\,[m] is the three-dimensional distance between \gls{AP} $l$ and \gls{UE} $k$, taking the height difference into account.
Moreover, $F_{kl} \sim \mathcal{N}(0,4^2)$ is the shadow fading. The shadowing terms from an AP to different UEs are correlated as \cite[Table B.1.2.2.1-4]{LTE2017a}
\begin{equation} \label{eq:shadowing-decorrelation}
\mathbb{E} \{ F_{kl} F_{ij} \} = 
\begin{cases}
4^2 2^{-\delta_{ki}/9\,\textrm{m}} & l = j \\
0 & l \neq j
\end{cases}
\end{equation}
where $\delta_{ki}$ is the distance between UE $k$ and UE $i$. The second row in \eqref{eq:shadowing-decorrelation} accounts for the correlation of the shadowing terms related to two different APs, which is assumed to be zero. The reason is that APs are typically distributed in the network at a distance larger than tens of meters and purposely deployed to view the deployment area from different directions than other APs. Note that it is only the simulations that rely on these specific assumptions, while all the analytical results in this monograph can be applied to any propagation environment that features Rayleigh fading.

 \subsection{Local Scattering Model for Spatial Correlation}
 \label{subsec:local-scattering-model}

The spatial correlation matrix depends on two main factors: the array geometry and the angular distribution of the multipath components. For small-sized APs, which are envisioned to be used in cell-free networks, it is common to utilize a \gls{ULA} where the $N$ antennas are equally spaced on a horizontal line.
We will consider a \gls{ULA} with half-wavelength-spacing in the simulations of this monograph. If all the multipaths arrive from the far-field of the array, the $(m,\ell)$th element of a generic spatial correlation matrix $\vect{R}$ can be computed as
\begin{equation} \label{eq:elements-of-R}
\left[\vect{R}\right]_{m\ell} = \beta \int \int e^{\imagunit \pi (m-\ell) \sin(\bar{\varphi}) \cos (\bar{\theta})} f(\bar{\varphi},\bar{\theta})  d\bar{\varphi} d\bar{\theta}
\end{equation}
where $\beta$ is the common large-scale fading coefficient while $\bar{\varphi}$ denotes the azimuth angle and $\bar{\theta}$ denotes the elevation angle of a multipath component, both computed with respect to the broadside of the array \cite[Sec.~7.3.2]{massivemimobook}. Moreover, $f(\bar{\varphi},\bar{\theta})$ is the joint \gls{PDF} of $\bar{\varphi}$ and $\bar{\theta}$. Hence, the double-integral in \eqref{eq:elements-of-R} computes $\mathbb{E}\{ e^{\imagunit \pi (m-\ell) \sin(\bar{\varphi}) \cos (\bar{\theta})}  \}$ with respect to randomly located multipath components that are distributed according to $f(\bar{\varphi},\bar{\theta})$.

 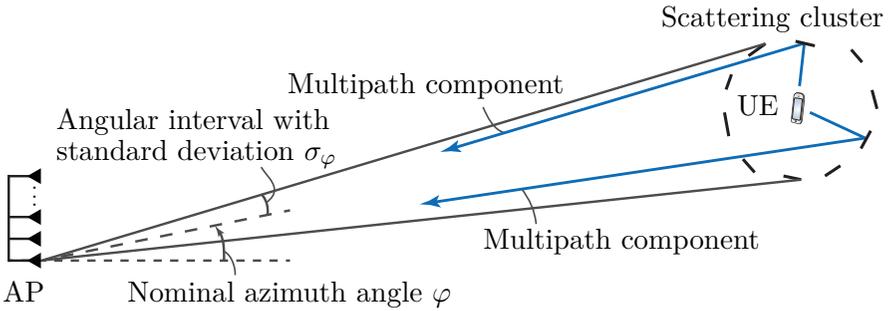
\begin{figure}[t!]
  	\begin{center}
 	\begin{overpic}[width=\columnwidth,tics=10]{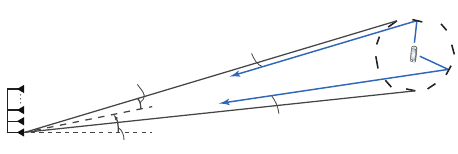}
		\put(33,23.5){Multipath component}
		\put(55,6){Multipath component}
		\put(1,0){AP}
		\put(83.5,21){UE}
		\put(15,0){Nominal azimuth angle $\varphi$}
		\put(75,31){Scattering cluster}
		\put(7,19.5){Angular interval with}
		\put(7,16){standard deviation $\sigma_{\varphi}$}
	\end{overpic} 
 	\end{center} \vskip-5mm
 	\caption{Illustration of \gls{NLoS} propagation under the local scattering model, where the scattering is localized around the \gls{UE}. The figure only shows the azimuth plane and two of the many multipath components are indicated. The nominal angle $\varphi$ and the {ASD} $\sigma_{\varphi}$ of the multipath components are key parameters to model the spatial correlation matrix.} \label{figure:local_scattering} 
 \end{figure} 
 
 \enlargethispage{\baselineskip} 

The integrals in \eqref{eq:elements-of-R} can be computed numerically for any \gls{PDF}. In the simulations, we will adopt the local scattering model, which is a common way to select the \gls{PDF} by assuming the multipath components are symmetrically distributed around a straight line drawn between the AP and UE. The intention with this model is that there is a scattering cluster centered around the UE and the multipath components are arriving from nearby angles.
 The model has been utilized in numerous studies along with Gaussian \cite{Adachi1986a},  \cite{massivemimobook}, \cite{Trump1996a}, \cite{Yin2013a}, \cite{Zetterberg1995a}, Laplace \cite{jiang2015achievable}, \cite{Molisch2007}, \cite{Pedersen1997a}, and uniform \cite{Adhikary2013a}, \cite{Salz1994a}, \cite{Shiu2000a}, \cite{Yin2013a} angular distributions.
We will consider the jointly Gaussian distribution
\begin{equation} \label{eq:Gaussian-ASD}
f(\bar{\varphi},\bar{\theta}) = \frac{1}{2\pi \sigma_{\varphi} \sigma_{\theta}} e^{-\frac{(\bar{\varphi} - \varphi)^2}{2 \sigma_{\varphi}^2 }} e^{-\frac{(\bar{\theta} - \theta)^2}{2 \sigma_{\theta}^2 }}
\end{equation}
where $\varphi$ and $\theta$ are the nominal azimuth and elevation angles, computed by drawing a straight line between the AP and UE. 
Note that the random variations in the azimuth and elevation angles are assumed to be independent.
The standard deviations $\sigma_{\varphi}\ge 0$ and $\sigma_{\theta}\ge 0$ are called the \emph{\gls{ASD}}. This model is illustrated from the above in Figure~\ref{figure:local_scattering}, where the multipath variations in the azimuth angle are visible.

\begin{remark}[Beyond correlated Rayleigh fading]
The theoretical framework of this monograph applies to any scenario with correlated Rayleigh fading channels and is not limited to the local scattering model. However, not all practical channels are well modeled by Rayleigh fading since the Gaussian distribution can only be motivated when there are many propagation paths and these have pathlosses drawn from a common distribution.
These conditions are not satisfied when there are few propagation paths or in LoS scenarios where the direct path is substantially stronger than all other paths.
Under those circumstances, one can either consider channel models with a finite number of paths \cite{Femenias2019a}, \cite{Jin2019a} or Rician fading where there is one strong path plus Rayleigh fading that describes all the other paths
\cite{Alonzo2019a}, \cite{Demir2020b}, \cite{Jin2019a}, \cite{Ngo2018b}, \cite{Ozdogan2019a}, \cite{Zhang2020b}. We will not cover any of these models in this monograph to keep the presentation simple, but we note that correlated Rayleigh fading can be viewed as a worst-case approximation of the aforementioned models if one uses the same spatial correlation matrices but replaces the more structured fading distribution with the Gaussian distribution \cite[Sec.~2.2]{massivemimobook}.
\end{remark}

\begin{figure}[t!]
	\begin{center}
		\includegraphics[width=\columnwidth]{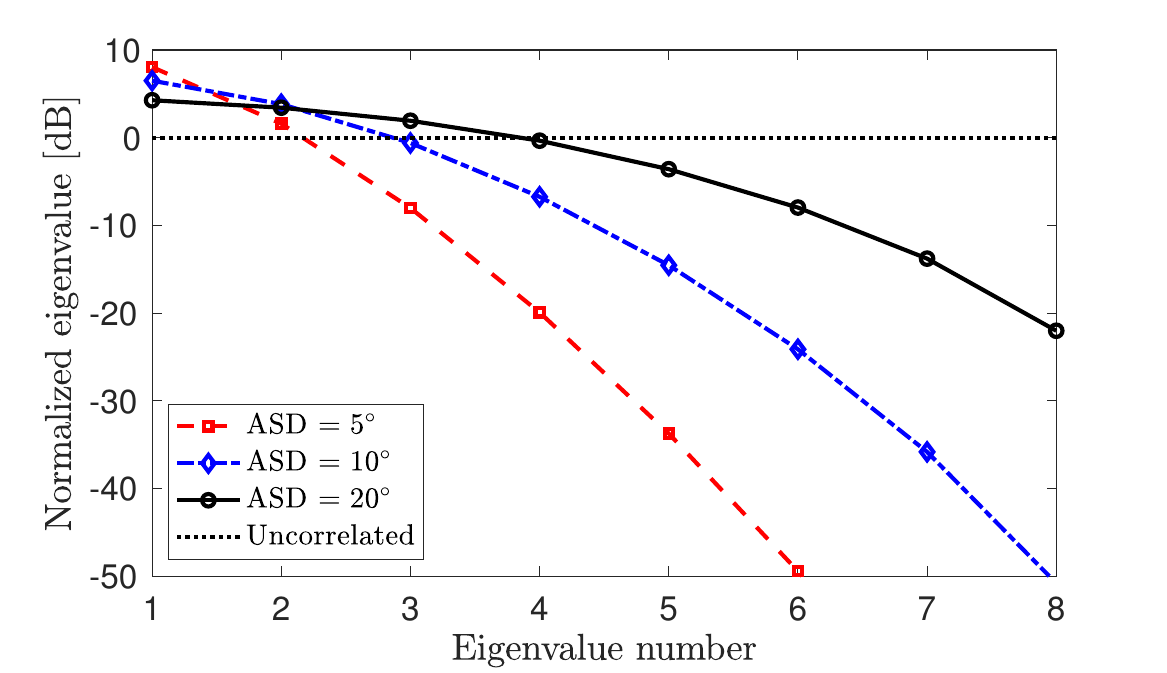}
	\end{center} \vskip-5mm
	\caption{Eigenvalues of the spatial correlation matrix $\vect{R}$ when using the local scattering model with $N=8$, the nominal azimuth angle ${\varphi} = 30^\circ$, and the nominal elevation angle ${\theta} = -15^\circ$. The correlation matrix is computed based on \eqref{eq:elements-of-R} using the Gaussian local scattering model in \eqref{eq:Gaussian-ASD}. Three different ASDs are considered: $\sigma_{\varphi}=\sigma_{\theta} \in \{ 5^\circ, 10^\circ, 20^\circ\}$. Uncorrelated Rayleigh fading is shown as a reference case.} \label{figureSec1_eigenvalue_variations} 
\end{figure}

One way to quantify the effect of spatial channel correlation is~by studying the eigenvalues of $\vect{R}$. Figure~\ref{figureSec1_eigenvalue_variations} shows the eigenvalues in decreasing order for an \gls{AP} equipped with $N=8$ antennas. 
 The nominal azimuth and elevation angles are set as ${\varphi} = 30^\circ$ and ${\theta} = -15^\circ$, respectively. We consider the Gaussian local scattering model described above with three different sets of \gls{ASD} values: $\sigma_{\varphi}=\sigma_{\theta}=5^\circ$, $\sigma_{\varphi}=\sigma_{\theta}=10^\circ$, and $\sigma_{\varphi}=\sigma_{\theta}=20^\circ$. The correlation matrices are normalized such that $\tr(\vect{R})/N=1$. The figure also shows the reference case of uncorrelated Rayleigh fading with $\vect{R}= \vect{I}_N$ for which all eigenvalues are equal to one.

The main observation from Figure~\ref{figureSec1_eigenvalue_variations} is that the eigenvalues are widely different when the \gls{ASD} is small, but the eigenvalue spread decreases with an increasing \gls{ASD}. In the case of $\sigma_{\varphi}=\sigma_{\theta} = 5^\circ$, the first four eigenvalues contribute to $99.99\%$ of the sum of the eigenvalues. Hence, the correlation matrix is nearly rank-deficient and all channel realizations will essentially be linear combinations of the corresponding four eigenvectors. In fact, the single largest (dominant) eigenvalue contributes to 80\% of the sum. Therefore, most channel realizations will point in similar directions as the dominant eigenvector. The eigenvalue variations are substantially smaller for $\sigma_{\varphi}=\sigma_{\theta} = 20^\circ$, but there is still a 425 times difference between the largest and smallest eigenvalues. This can be compared to the extreme case of uncorrelated fading, represented by the dotted curve, where all eigenvalues are identical.

Spatial correlation is of fundamental importance in Cellular Massive MIMO \cite{Sanguinetti2019a}, where the number of antennas is so large that the eigenvalue variations play a key role in mitigating interference between UEs \cite{BjornsonHS17}, \cite{Huh2012a},  \cite{Yin2016a}, \cite{Yin2013a}.
We refer to \cite[Sec.~2]{massivemimobook} for more details on the basic impact that spatial correlation can have on a multi-user system. Since the number of antennas per AP is envisaged to be fairly small in Cell-free Massive MIMO, spatial correlation cannot be utilized as efficiently but we will anyway consider it throughout this monograph.
In particular, we will analyze the effect of the eigenvalue spread on the channel estimation performance in Section~\ref{c-channel-estimation} on p.~\pageref{c-channel-estimation}.

\section{Channel Hardening and Favorable Propagation}
\label{sec:hardening-favorable}

In Cellular Massive \gls{MIMO} systems, two important properties appear when employing a large number of antennas at an AP: \emph{channel hardening} and \emph{favorable propagation}. These properties provide an insightful explanation for the performance gain of Cellular Massive MIMO and are elaborated for channels featuring correlated fading in \cite[Sec.~2.5]{massivemimobook} and for some other types of channels in \cite[Ch.~7]{Marzetta2016a}. Whether these properties appear or not depends strongly on the number of antennas and the spatial channel correlation properties.

In this section, we define and  analyze channel hardening and favorable propagation in Cell-free Massive \gls{MIMO}, which are rather different from the cellular case since multiple APs serve each UE. There are several different definitions in the cellular literature and we have selected those that extend most naturally to the cell-free case. We will observe that the two properties are now affected not only by the spatial correlation and the total  number of antennas in the network, $M=LN$, but also by the variations in the large-scale fading coefficients among the APs and the power allocation.
Recall from Section~\ref{subsec:large-scale-fading-model} that the large-scale fading is determined by the distances between the APs and UEs, as well as other macroscopic propagation effects (e.g., shadow fading). 
We stress that, although we dedicate a section to describe when channel hardening and favorable propagation might appear in practice, these properties are not formally required to make use of any of the results presented in this monograph. We refer to \cite{Chen2018b} for a more detailed analysis of the two properties in Cell-free Massive MIMO.

\subsection{Channel Hardening}
\label{subsec-channel-hardening}

Each element of a channel vector takes a new independent realization in every coherence block, but these small-scale fading variations will not necessarily affect the communication performance. As can be seen in \eqref{eq:DCC-downlink}, the received downlink signal at  \gls{UE} $k$ contains the desired signal component $\sum_{l=1}^{L}  \vect{h}_{kl}^{\Htran} \vect{D}_{kl}  \vect{w}_{kl} \varsigma_k$, where the \emph{effective channel} is
\begin{equation} \label{eq:effective-DL-channel}
\sum_{l=1}^{L}  \vect{h}_{kl}^{\Htran} \vect{D}_{kl}  \vect{w}_{kl} = \sum_{l \in \mathcal{M}_k}  \vect{h}_{kl}^{\Htran}  \vect{w}_{kl}.
\end{equation}
This is the summation of the inner product between the channel vector $\vect{h}_{kl}$ and the precoding vector $\vect{D}_{kl} \vect{w}_{kl}$ from all the APs.
The value of this effective channel can potentially remain almost constant even if the individual elements of the $L$ channel vectors $ \vect{h}_{k1},\ldots, \vect{h}_{kL}$ are changing.
More precisely, when the random realizations of the effective channel in \eqref{eq:effective-DL-channel} are close to the mean value (i.e., the variance is small), then approximately the same effective scalar channel will appear in every coherence block and we can, thus, operate the system as if we were communicating over a deterministic channel.
When this happens, we say that we have achieved \emph{channel hardening}, where the word ``harden'' refers to the fact that the effective channel is becoming more deterministic.

\subsubsection{Definition and Sufficient Condition for Channel Hardening}

To give a formal definition of the channel hardening property, we assume all serving APs apply  \gls{MR} precoding, which we define as
\begin{equation} \label{eq:MR-precoding-hardening}
\vect{w}_{kl} = \sqrt{\frac{\rho_{kl}}{\mathbb{E} \{\| \vect{h}_{kl} \|^2 \}} } \vect{h}_{kl}
\end{equation}
with $\rho_{kl}\geq 0$ determining how much power AP $l$ is assigning for transmission to UE $k$. Note that MR is the precoding method that maximizes the received signal power at the UE. If we insert this precoding into \eqref{eq:effective-DL-channel}, the effective channel becomes
\begin{equation} \label{eq:effective-DL-channel2}
\sum_{l \in \mathcal{M}_k} \sqrt{\frac{\rho_{kl}}{\mathbb{E} \{\| \vect{h}_{kl} \|^2 \}} } \| \vect{h}_{kl} \|^2.
\end{equation}
Note that the following channel hardening definition considers the combined channel from all the serving APs to a particular UE $k$, which is the natural extension of the definition in Cellular Massive \gls{MIMO} where only a single AP serves each UE \cite[Sec.~2.5.1]{massivemimobook}.

\begin{definition}[Channel hardening] \label{def:channel-hardening-new}
For a given set $\mathcal{M}_k$ of serving APs and downlink power allocation coefficients $\rho_{k1},\ldots,\rho_{kL}$, the effective channel to \gls{UE} $k$ is said to provide
asymptotic channel hardening if
\begin{equation} \label{eq:channel-hardening-definition}
\frac{ \sum\limits_{l \in \mathcal{M}_k} \sqrt{\frac{\rho_{kl}}{\mathbb{E} \{\| \vect{h}_{kl} \|^2 \}} } \| \vect{h}_{kl} \|^2 }{ \mathbb{E} \left\{ \sum\limits_{l \in \mathcal{M}_k} \sqrt{\frac{\rho_{kl}}{\mathbb{E} \{\| \vect{h}_{kl} \|^2 \}} } \| \vect{h}_{kl} \|^2  \right\} } \rightarrow 1
\end{equation}
in the mean-squared sense as $N \to \infty$.\footnote{Convergence in the mean-squared sense also implies convergence in probability and in distribution. The exact type of convergence is of limited importance since we are not interested in the asymptotic regime but to get a structured way to evaluate when we can obtain approximate channel hardening for a finite number of antennas.}
\end{definition}

The expectation in~\eqref{eq:channel-hardening-definition} is computed with respect to the independent channel realizations in different coherence blocks, while $\mathcal{M}_k$ and the power allocation coefficients are assumed to be constant.

This definition says that the effective channel in \eqref{eq:effective-DL-channel2} is close to its mean value when the number of AP antennas grows large.
Since the $N$-length channel vectors become larger as more antennas are added, the underlying assumption is that the antenna arrays are deployed according to some distribution that determines the evolution of the channel statistics (e.g., spatial correlation matrices). This is why the convergence is specified in the mean-squared sense. When considering correlated Rayleigh fading, a sufficient condition for channel hardening is as follows.

\begin{lemma}
For correlated Rayleigh fading channels, a sufficient condition for the effective channel of \gls{UE} $k$ to provide asymptotic channel hardening is that
\begin{equation} \label{eq:sufficient-hardening}
\frac{ \sum\limits_{l \in \mathcal{M}_k} \rho_{kl} \frac{\tr ( \vect{R}_{kl}^2 ) }{N \beta_{kl} }
}{N \bigg( \sum\limits_{l \in \mathcal{M}_k} \sqrt{\rho_{kl} \beta_{kl}} 
\bigg)^2 } \to 0 \quad  \textrm{ as } N \to \infty.
\end{equation}
\end{lemma}
\begin{proof}
The ratio in the left-hand side of \eqref{eq:channel-hardening-definition} has a mean value of one, thus we want to prove convergence to the mean value. In this case, convergence in the mean-squared sense requires that the variance of the ratio goes asymptotically to zero. The variance can be computed as
\begin{equation}
\mathbb{V}\left\{ 
\frac{ \sum\limits_{l \in \mathcal{M}_k} \sqrt{\frac{\rho_{kl}}{\mathbb{E} \{\| \vect{h}_{kl} \|^2 \}} } \| \vect{h}_{kl} \|^2 }{ \mathbb{E} \left\{ \sum\limits_{l \in \mathcal{M}_k} \sqrt{\frac{\rho_{kl}}{\mathbb{E} \{\| \vect{h}_{kl} \|^2 \}} } \| \vect{h}_{kl} \|^2  \right\} } 
\right\} = \frac{ \sum\limits_{l \in \mathcal{M}_k} \rho_{kl} \frac{\tr ( \vect{R}_{kl}^2 ) }{\tr ( \vect{R}_{kl} ) }
}{ \bigg( \sum\limits_{l \in \mathcal{M}_k} \sqrt{\rho_{kl} \tr ( \vect{R}_{kl} )} 
\bigg)^2 } 
\end{equation}
by utilizing the statistical independence of the channels to the different \glspl{AP} and by applying \cite[Lemma B.14]{massivemimobook} to compute the expectations. We then obtain the sufficient condition in \eqref{eq:sufficient-hardening} by utilizing the definition of $\beta_{kl}$ in \eqref{eq:beta-definition-lk}, which implies that $ \tr ( \vect{R}_{kl} ) = N \beta_{kl}$.
\end{proof}

Although the definition of channel hardening relies on asymptotic arguments, we can use it to evaluate the \emph{degree of channel hardening} for setups with a finite number of antennas. If the expression in \eqref{eq:sufficient-hardening} is close to zero, this means that the left-hand side of \eqref{eq:channel-hardening-definition} is close to one and thus channel hardening is approximately achieved.
 
It has been shown in the Cellular Massive MIMO literature that spatial correlation reduces the convergence rate to channel hardening \cite[Sec.~2.5.1]{massivemimobook}; that is, more antennas are needed to achieve a certain degree of channel hardening.
This can be also observed in \eqref{eq:sufficient-hardening} where $\tr ( \vect{R}_{kl}^2 )$ can be computed as the sum of the squared eigenvalues of the spatial correlation matrix $\vect{R}_{kl}$. Recall from Figure~\ref{figureSec1_eigenvalue_variations} that spatial correlation is represented by large eigenvalue variations.
Since the sum of the eigenvalues equals $\tr(\vect{R}_{kl}) = N \beta_{kl}$ by definition, the squared sum of the eigenvalues can take values between $N \beta_{kl}^2$ (when all eigenvalues equal to $ \beta_{kl}$) and $N^2\beta_{kl}^2$ (when there is only one non-zero eigenvalue that equals $N \beta_{kl}$).
The latter represents an extreme case of high spatial correlation.
To achieve a high degree of channel hardening, we want the expression in \eqref{eq:sufficient-hardening} to be small and the case when all eigenvalues are the same (i.e., no spatial correlation) achieves the smallest value.

\subsubsection{Impact of the Number and Geographical Distribution of APs}

In the context of Cell-free Massive MIMO, where each AP is supposed to have a relatively small number of antennas but there will be many APs instead, it is interesting to evaluate if one can compensate for having a small $N$ by having many APs.
To analyze this further, we assume that $\vect{R}_{kl} = \beta_{kl} \vect{I}_N$ so that there is no spatial correlation. The expression  in \eqref{eq:sufficient-hardening} then simplifies to
\begin{equation} \label{eq:sufficient-hardening-iid}
\frac{ \sum\limits_{l \in \mathcal{M}_k} \rho_{kl}  \beta_{kl} 
}{ N \bigg( \sum\limits_{l \in \mathcal{M}_k} \sqrt{\rho_{kl}  \beta_{kl}} 
\bigg)^2 } .
\end{equation}
If $|\mathcal{M}_k |=1$, or one AP has a much larger value on $\rho_{kl}  \beta_{kl} $ than the other APs that serve UE $k$, then \eqref{eq:sufficient-hardening-iid} becomes approximately $1/N$. If we instead consider the case when all the serving APs have the same value of $\rho_{kl}  \beta_{kl} $, then \eqref{eq:sufficient-hardening-iid} becomes $1/( N | \mathcal{M}_k|)$. Hence, it is plausible to achieve the same degree of channel hardening by having many APs with few antennas as with one AP with many antennas in some cases. If $\rho_{kl}  \beta_{kl} $ is the same for all the serving APs, it is only the total number of AP antennas that determines the degree of channel hardening.
However, when the APs have different values of  $\rho_{kl}  \beta_{kl} $, then the total number of antennas needed to achieve a certain degree of channel hardening will likely be larger than in Cellular Massive MIMO.

The geographical distribution of the APs, with respect to the UE, determines the values of the large-scale fading coefficients $\{ \beta_{kl} \}$. There can naturally be tens of dB differences between the serving APs, even if they have rather similar distances to the UE.
One can compensate for this by applying a power allocation algorithm where the powers $\{\rho_{kl} \}$ are selected to even out the differences. More precisely, we can reduce the power from nearby APs (with larger $\beta_{kl}$) and increase the power from APs that are further away  (with smaller $\beta_{kl}$) to achieve roughly the same value of $\rho_{kl}  \beta_{kl} $ for all the serving APs. This will increase the degree of channel hardening but will not necessarily be desirable from an end-to-end performance perspective since we can achieve a higher total received power at the UE by allocating the same power in the opposite way (more power from the nearby APs).

 \begin{figure}[t!]
	\begin{center}
		\includegraphics[width=\columnwidth]{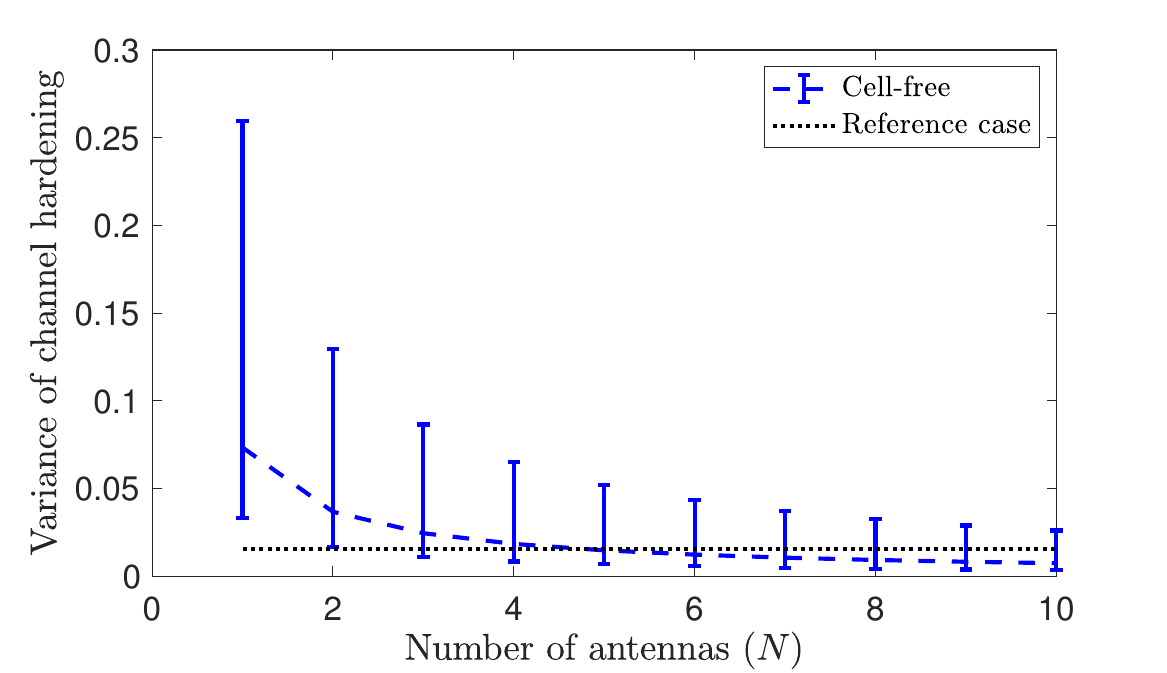}
	\end{center} \vskip-5mm
	\caption{The value of \eqref{eq:sufficient-hardening-iid}, which represents the variance of the ratio between the effective channel and its mean value, is plotted for different $N$. When \eqref{eq:sufficient-hardening-iid} is small, there is a high degree of channel hardening. The reference case corresponds to a 64-antenna Cellular Massive MIMO setup. The cell-free setup consists of 64 randomly distributed APs and a varying number of antennas per AP. The line shows the median value while the bars show the interval containing 90\% of the random realizations.} \label{figureSec2_hardening} 
\end{figure}

Figure~\ref{figureSec2_hardening} exemplifies how the degree of channel hardening depends on the geographical AP distribution and the number of antennas per AP. We consider the channel to a UE in the center of $400\,$m $\times$ $400$\,m area. $L=64$ APs are uniformly distributed in the area and transmit with equal power to the UE. We use the large-scale fading model in \eqref{eq:pathloss-coefficient} and assume \gls{iid} Rayleigh fading. The figure shows the value of the expression in \eqref{eq:sufficient-hardening-iid} for different values of $N$. Since the value depends on the AP distribution, the dashed line shows the median value while the vertical bars indicate the interval in which 90\% of all realizations occur. As a reference, the dotted horizontal line shows the value of \eqref{eq:sufficient-hardening-iid} for a single AP with 64 antennas, which represents a Cellular Massive MIMO setup that would achieve a high degree of channel hardening.
If single-antenna APs are used, then the degree of channel hardening is far from that of the reference case, even if the total number of antennas is the same. The median reaches the reference case for $N=5$ but half of the realizations give substantially larger values. It is first when the number of antennas per AP reaches 8 or 10 that  the degree of channel hardening is comparable to that of Cellular Massive MIMO. Note that the total number of antennas is then an order of magnitude larger than in the reference case.

 \subsubsection{Takeaways}
In summary, channel hardening is a desirable property but not mandatory for the operation of Cell-free Massive MIMO. In fact, we can expect a substantially lower degree of channel hardening in cell-free networks than in cellular networks when the number of serving antennas is the same. Channel hardening makes some capacity lower bounds (that we will present later in this monograph) closer to the true capacity and also serves as a motivation for not trying to optimize the power allocation in every coherence block but rather on average. We will return to these things later in this monograph.

Note that we defined channel hardening from a downlink precoding perspective in this section. The same argumentation can be made also in the uplink, in which case the precoding vectors are replaced by combining vectors, and the power allocation coefficients are replaced by weights that the CPU uses when fusing the signals $\hat{s}_{kl}$ from different APs. The details on this will be provided in Section~\ref{c-uplink} on p.~\pageref{c-uplink}.

\subsection{Favorable Propagation}
\label{subsec-favorable-propagation}

When multiple UEs are spatially multiplexed in the system, there will generally be interference between their transmissions, and the precoding/combining can be selected to strike a balance between achieving strong desired signals and causing little interference. 
However, if the UEs' channels are spatially orthogonal, then the inter-user interference is automatically mitigated by MR processing so one can basically ignore it. When this happens, we say that \emph{favorable propagation} is experienced.

If we consider the received downlink signal of UE $k$ in \eqref{eq:DCC-downlink}, then the interference caused by the concurrent transmission to UE $i$ is $ \sum_{l=1}^{L}  \vect{h}_{kl}^{\Htran}\vect{D}_{il}  \vect{w}_{il} \varsigma_i $. If the MR precoding defined in \eqref{eq:MR-precoding-hardening} is utilized, then the \emph{effective interfering channel} is
\begin{equation}\label{effective-interfering-channel}
 \sum_{l=1}^{L}   \sqrt{\frac{\rho_{il}}{\mathbb{E} \{\| \vect{h}_{il} \|^2 \}} }   \vect{h}_{kl}^{\Htran}\vect{D}_{il}\vect{h}_{il} =  \sum_{l \in \mathcal{M}_i}   \sqrt{\frac{\rho_{il}}{\mathbb{E} \{\| \vect{h}_{il} \|^2 \}} } \vect{h}_{kl}^{\Htran}  \vect{h}_{il}.
\end{equation}
It is the magnitude of this term compared to the magnitude of the effective channel of the desired signal in \eqref{eq:effective-DL-channel2} that determines how strong the interference is in relative terms. If the ratio is close to zero, then the interference will be negligible and we benefit from favorable propagation.

 \subsubsection{Definition and Sufficient Condition for Favorable Propagation}

There are several different formal definitions of favorable propagation in the Cellular Massive MIMO literature but none of them is a perfect fit for the case when multiple APs are transmitting coherently. Hence, we provide the following new definition that is tailored to the cell-free scenario.

\begin{definition}[Favorable propagation]\label{def:favorable-propagation-new} \index{favorable propagation}
A given UE $k$ experiences asymptotically favorable propagation with respect to UE $i$ (with $i \neq k$) if
	\begin{equation} \label{eq:asympt-favorable-new}
	\frac{ \sum\limits_{l \in \mathcal{M}_i}   \sqrt{\frac{\rho_{il}}{\mathbb{E} \{\| \vect{h}_{il} \|^2 \}} } \vect{h}_{kl}^{\Htran}  \vect{h}_{il} }{ \mathbb{E} \left\{ \sum\limits_{l \in \mathcal{M}_k} \sqrt{\frac{\rho_{kl}}{\mathbb{E} \{\| \vect{h}_{kl} \|^2 \}} } \| \vect{h}_{kl} \|^2  \right\} } \rightarrow 0
	\end{equation}
	in the mean-squared sense as  $N  \rightarrow \infty$.
\end{definition}

This definition says that the effective interfering channel in~\eqref{effective-interfering-channel} (i.e., a weighted sum of the inner products between the channel vectors of the two UEs) should go asymptotically to zero, when it is normalized by the average value of the desired effective channel in~\eqref{eq:effective-DL-channel2}. The weights are the downlink transmit powers.
Note that the numerator depends on the set $\mathcal{M}_i$ of APs that serve the interfering UE $i$, while the denominator depends on the set $\mathcal{M}_k$ of APs serving the considered UE $k$. For correlated Rayleigh fading channels, a sufficient condition for favorable propagation is obtained as follows.

\begin{lemma}
For correlated Rayleigh fading channels, a sufficient condition for UE $k$ to experience 
asymptotically favorable propagation with respect to UE $i$ is that 
\begin{equation} \label{eq:sufficient-favorable}
\frac{ \sum\limits_{l \in \mathcal{M}_i} \rho_{il} \frac{\tr ( \vect{R}_{il} \vect{R}_{kl} ) }{N \beta_{il} }
}{ N \bigg( \sum\limits_{l \in \mathcal{M}_k} \sqrt{\rho_{kl}  \beta_{kl}} 
\bigg)^2 } \to 0 \quad  \textrm{ as } N \to \infty.
\end{equation}
\end{lemma}
\begin{proof}
The ratio in the left hand side of \eqref{eq:asympt-favorable-new} has a mean value of zero, thus we want to prove convergence to the mean value. In this case, convergence in mean-squared sense requires that the variance of the ratio goes asymptotically to zero. The variance can be computed as
\begin{equation}
\mathbb{V}\left\{ 
\frac{ \sum\limits_{l \in \mathcal{M}_i}   \sqrt{\frac{\rho_{il}}{\mathbb{E} \{\| \vect{h}_{il} \|^2 \}} } \vect{h}_{kl}^{\Htran}  \vect{h}_{il} }{ \mathbb{E} \left\{ \sum\limits_{l \in \mathcal{M}_k} \sqrt{\frac{\rho_{kl}}{\mathbb{E} \{\| \vect{h}_{kl} \|^2 \}} } \| \vect{h}_{kl} \|^2  \right\} }  
\right\} = \frac{ \sum\limits_{l \in \mathcal{M}_i} \rho_{il} \frac{\tr ( \vect{R}_{il} \vect{R}_{kl} ) }{\tr ( \vect{R}_{il} ) }
}{ \bigg( \sum\limits_{l \in \mathcal{M}_k} \sqrt{\rho_{kl} \tr ( \vect{R}_{kl} )} 
\bigg)^2 } 
\end{equation}
by utilizing the statistical independence of the channels to the different \glspl{AP} and by applying \cite[Lemma B.14]{massivemimobook} to compute the expectations. We then obtain the sufficient condition in \eqref{eq:sufficient-favorable} by utilizing the definition of $\beta_{kl}$ in \eqref{eq:beta-definition-lk}, which implies that $ \tr ( \vect{R}_{kl} ) = N \beta_{kl}$.
\end{proof}

Although the definition of favorable propagation relies on asymptotic arguments, we can use it to evaluate the degree of favorable propagation for setups with a finite number of antennas. If the expression in \eqref{eq:sufficient-favorable} is close to zero, this means that the left-hand side of \eqref{eq:asympt-favorable-new} is close to zero and thus favorable propagation is approximately achieved.

The value of the expression in \eqref{eq:sufficient-favorable} depends on the spatial correlation properties in a nontrivial manner. The term $\tr ( \vect{R}_{il} \vect{R}_{kl} )$ in the numerator takes its largest value when the spatial correlation matrices $\vect{R}_{il}$ and $\vect{R}_{kl}$ are identical up to a scaling factor, which represents the case when the two UEs have almost the same spatial directivity of their channels (statistically speaking).
In contrast, the term takes its smallest value when  the spatial correlation matrices have identical eigenvectors but the eigenvalues are matched together in the opposite order (i.e., large eigenvalues from one matrix are multiplied with small eigenvalues from the other matrix). This represents the case when the two UEs have very different spatial directivity. Clearly, it is the latter case that leads to the highest degree of favorable propagation.

\subsubsection{Impact of the Geographical Distribution of APs}

To instead focus on the different large-scale fading variations that the UEs experience, we can assume that $\vect{R}_{kl} = \beta_{kl} \vect{I}_N$ so that there is no spatial correlation. The expression in \eqref{eq:sufficient-favorable} then simplifies to
\begin{equation} \label{eq:sufficient-favorable-2}
\frac{ \sum\limits_{l \in \mathcal{M}_i} \rho_{il} \beta_{kl} 
}{ N \bigg( \sum\limits_{l \in \mathcal{M}_k} \sqrt{\rho_{kl}  \beta_{kl}} 
\bigg)^2 } .
\end{equation}
This expression decays as $1/N$ so more antennas lead to a higher degree of favorable propagation. Moreover, the set $\mathcal{M}_i$ of APs that serve UE $i$ plays an important role in determining the value of \eqref{eq:sufficient-favorable-2}. If those APs are far from UE $k$, all the $\beta_{kl} $ terms in the numerator will likely be very small and then we can achieve favorable propagation even with $N=1$. However, if $\mathcal{M}_i = \mathcal{M}_k$, the numerator and denominator might be of comparable size and then we might need many antennas per AP to achieve favorable propagation.
The power allocation also plays a key role; to make the term $ \rho_{il} \beta_{kl} $ in the numerator small, we can either have a small value of $\beta_{kl} $ and/or use a low transmit power $ \rho_{il}$ on that AP.

 \begin{figure}[t!]
	\begin{center}
		\includegraphics[width=\columnwidth]{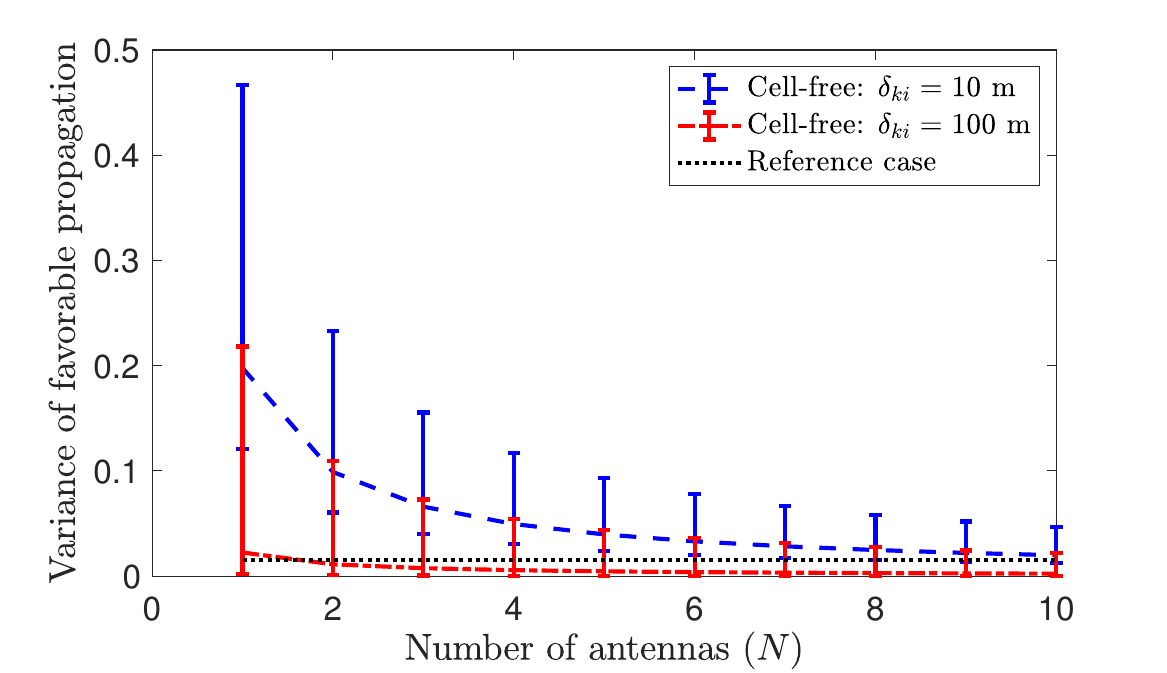}
	\end{center} \vskip-5mm
	\caption{The value of \eqref{eq:sufficient-favorable-2}, which represents the variance of the ratio between the effective interfering channel and the mean of the desired effective channel, is plotted for different $N$. When \eqref{eq:sufficient-favorable-2} is small for a selected UE pair, there is a high degree of favorable propagation between them. The reference case corresponds to a 64-antenna Cellular Massive MIMO setup. The cell-free setup consists of 64 randomly distributed APs and a varying number of antennas per AP. The UEs are either 10 or 100\,m apart, and are each served by the 8 APs providing the best channels. The line shows the median value while the bars show the interval containing 90\% of the random realizations.} \label{figureSec2_favorable} 
\end{figure}

Figure~\ref{figureSec2_favorable} exemplifies how the degree of favorable propagation depends on the geographical AP distribution, the distance between the UEs, and the number of antennas per AP. We consider the same setup as in Figure~\ref{figureSec2_hardening} but add an interfering UE that is either 10\,m or 100\,m to the east of the desired UE. We assume each UE is served by the 8 APs that provide the eight largest large-scale fading coefficients and these APs transmit with equal power.
The figure shows the value of the expression in \eqref{eq:sufficient-favorable-2} for different $N$. Since the value depends on the AP distribution, the dashed line shows the median value while the vertical bars indicate the interval in which 90\% of all realizations occur. As a reference, the dotted horizontal line shows the value of \eqref{eq:sufficient-favorable-2} when a single AP with 64 antennas serves both UEs, which represents a Cellular Massive MIMO setup. 
We notice that the distance between the UEs strongly determines the degree of favorable propagation. When the UEs are close together, we need around $N=10$ antennas per AP before favorable propagation is approximately achieved. When the UEs are 100\,m apart, most realizations of the AP locations lead to favorable propagation even for $N=1$, but there are still large variations. We can expect that, in practice, some UEs that are far apart might still not achieve favorable propagation with respect to each other.

\subsubsection{Takeaways}

In summary, favorable propagation is a desirable property that makes inter-user interference disappear when using MR precoding (which does not actively suppress any interference). However, favorable propagation is not a necessary property. We have observed that UEs that are served by partially the same APs might not experience it with respect to each other. In those cases, the interference can instead be suppressed by using precoding methods that are actively suppressing interference. This is covered in detail in Section~\ref{c-downlink} on p.~\pageref{c-downlink}.

Note that we defined favorable propagation from a downlink precoding perspective in this section. The same argumentation can be made also in the uplink, as previously discussed in relation to channel hardening. When there is a lack of favorable propagation in the uplink,  we can use receive combining to deal with the interference instead; see Section~\ref{c-uplink} on p.~\pageref{c-uplink}. Although there are clear mathematical similarities between channel hardening and favorable propagation, the properties are different. In particular, a UE might exhibit channel hardening but not favorable propagation or vice versa. It is also likely that some pairs of UEs will experience favorable propagation with respect to each other, while other pairs will not.

\newpage
\section{Summary of the Key Points in Section~\ref{c-cell-free-definition}}
\label{c-cell-free-definition-summary}

\begin{mdframed}[style=GrayFrame]

\begin{itemize}
\item A Cell-free Massive \gls{MIMO} network consists of $L$ \glspl{AP}, each equipped with $N$ antennas, that are arbitrarily distributed over the coverage area. The \glspl{AP}  may jointly serve $K$ single-antenna \glspl{UE} in each channel coherence block. More precisely, each \gls{UE} is communicating with a subset of the \glspl{AP}, which is selected based on the \gls{UE}'s needs.

\item  By having a total number of $M=LN$ AP antennas much larger than $K$, the cell-free network typically has sufficiently many spatial degrees-of-freedom to separate  \glspl{UE} in space by linearly processing the transmitted and received signals. 

\item Since a cell-free network may have a large geographical coverage area, it must be designed to be scalable in the sense that the computational capability and fronthaul capacity of existing APs must remain sufficient as new APs are deployed and as more UEs are being served.
A cell-free network in which each AP serves all the UEs is not scalable.

\item Although the number of co-located antennas per AP is envisaged to be fairly small, the spatial correlation must be anyway considered in the analysis since the channel fading is spatially correlated in practice.

\item When employing a large number of co-located antennas at an AP, two important properties appear: channel hardening and favorable propagation. In cell-free networks, the antennas at multiple APs contribute to these properties. However, a lower degree of channel hardening is expected than in cellular networks because of the fairly small number of antennas per AP. Cell-free networks are expected to provide a high degree of favorable propagation between UEs that are relatively far apart, but not among those that are closely spaced.
\end{itemize}

\end{mdframed}


\chapter{Theoretical Foundations}
\label{c-theoretical-foundations} 

\vspace{-5mm}

This section describes some basic results from estimation, information, and optimization theory which will be utilized later in this monograph. Section~\ref{sec:estimation-theory} describes the foundations of estimating Gaussian distributed random variables, which might be the channels in a cell-free network. Section~\ref{sec:capacity-bounds} defines the channel capacity and describes standard methods for computing achievable lower bounds, which are also known as achievable spectral efficiencies (SEs). 
In Section~\ref{sec:Rayleigh-quotient}, we consider the maximization of a ratio known as a generalized Rayleigh quotient.
In Section~\ref{sec:optimization-algorithms}, two different utility maximization algorithms are provided, which are of particular interest for the optimization of UE performance in a cell-free network. The key points are summarized in Section~\ref{c-theoretical-foundations-summary}.


\enlargethispage{\baselineskip} 

\section{Estimation Theory for Gaussian Variables}
\label{sec:estimation-theory}

The estimation of channel coefficients is an important aspect of any coherent communication system. In this section, we provide the theoretical foundation for estimating the realization of a Gaussian random variable from an observation that is corrupted by independent additive Gaussian noise. We only present the main concepts and results that will be used in later chapters and refer to textbooks such as \cite{massivemimobook}, \cite{Kay1993a} for further details and explanations.

\begin{definition}[Minimum mean-squared error estimator] 
Consider a\linebreak random variable $\vect{x} \in \mathbb{C}^N$ with support in $\mathcal{X}$ and let $\hat{\vect{x}}(\vect{y})$ denote an arbitrary estimator of $\vect{x}$ based on the observation $\vect{y} \in \mathbb{C}^{M}$. The particular choice of  $\hat{\vect{x}}(\vect{y}) : \mathbb{C}^{M} \to \mathbb{C}^{N}$  that minimizes the \gls{MSE}
\begin{equation}
\mathbb{E} \left\{ \| \vect{x} - \hat{\vect{x}}(\vect{y}) \|^2 \right\}
\end{equation}
is called the \gls{MMSE} estimator of $\vect{x}$.
It can be computed as
\begin{equation}
\hat{\vect{x}}_{\mathrm{MMSE}}(\vect{y}) = \mathbb{E} \{ \vect{x} | \vect{y} \} = \int_{\mathcal{X}}  \vect{x} f(\vect{x} | \vect{y} ) d \vect{x}
\end{equation}
where $f(\vect{x} | \vect{y} ) $ is the conditional \gls{PDF} of $\vect{x}$ given the observation $\vect{y}$.
\end{definition}

The \gls{MMSE} estimator of a complex Gaussian random variable from an observation that is corrupted by independent additive complex Gaussian noise (and interference) can be computed in closed form.

\begin{lemma} \label{lemma:MMSE-estimator}
Consider estimation of the $N$-dimensional vector $\vect{x} \sim \CN(\vect{0}_N,\vect{R})$, with a positive semi-definite correlation matrix $\vect{R}$, from the observation $\vect{y} = \vect{x} q + \vect{n} \in \mathbb{C}^{N}$.
The pilot signal $q \in \mathbb{C}$ is known and $\vect{n} \sim \CN(\vect{0}_N,\vect{S})$ is an independent noise/interference vector with a positive definite correlation matrix. 

The \gls{MMSE} estimator of $\vect{x}$ is
\begin{equation} \label{eq_MMSE_channel_matrix-rankdef}
\begin{split}
\hat{\vect{x}}_{\mathrm{MMSE}}(\vect{y}) = q^\star  \vect{R}
\left( |q|^2 \vect{R} + \vect{S} \right)^{-1} \vect{y}.
\end{split}
\end{equation}
The estimation error correlation matrix is
\begin{equation} \label{eq_error_covariance_channel_matrix-rankdef}
 \vect{C}_{\mathrm{MMSE}} 
= \vect{R} -  |q|^2 \vect{R} 
 \left(  |q|^2 \vect{R} + \vect{S} \right)^{-1} \vect{R}
\end{equation}
and the \gls{MSE} is
\begin{equation} \label{eq_MSE_channel_matrix-rankdef}
\mathacr{MSE} = \tr \left( \vect{R} -  |q|^2 \vect{R} 
 \left(  |q|^2 \vect{R} + \vect{S} \right)^{-1} \vect{R} \right).
\end{equation}
\end{lemma}
\begin{proof}
The detailed proof can be found in \cite[App.~B.4]{massivemimobook}.
\end{proof}

For brevity, we will denote the MMSE estimate as $\hat{\vect{x}}_{\mathrm{MMSE}}$, without explicitly specifying what observation it was based on.
A consequence of Lemma~\ref{lemma:MMSE-estimator} is that the MMSE estimate is distributed as
\begin{equation}
\hat{\vect{x}}_{\mathrm{MMSE}} \sim \CN ( \vect{0}_N , \vect{R} - \vect{C}_{\mathrm{MMSE}}).
\end{equation}
Moreover, the estimation error $\tilde{\vect{x}} = \vect{x} - \hat{\vect{x}}$ is distributed as 
\begin{equation}
\tilde{\vect{x}}_{\mathrm{MMSE}} \sim \CN ( \vect{0}_N ,  \vect{C}_{\mathrm{MMSE}}).
\end{equation}
These random variables are independent, which is a useful property that will be utilized in the analysis in future sections.


\section{Capacity Bounds and Spectral Efficiency}
\label{sec:capacity-bounds}

The SE is the performance metric that will be used throughout this monograph.
This is a measure of the average amount of information that can be correctly transferred per complex-valued sample, when a very large block of data is transmitted. This is the natural metric in broadband applications, where the demand for data is large.

\begin{definition}[Spectral efficiency]\label{definition_spectral_efficiency} 
The achievable SE of an encoding/decoding scheme is the average number of bits of information, per complex-valued sample, that it can transmit reliably over the channel under consideration.
\end{definition}

When communicating over a bandwidth of $B$\,Hz, the Nyquist-Shannon sampling theorem specifies that the communication signal is fully determined by $B$ complex-valued samples per second \cite{Shannon1949b} (or $2B$ real-valued samples per second). These are the samples that the data is encoded into in communications and the SE describes the amount of data that can be transferred per such complex sample. Since there are $B$ samples per second, the unit of the SE is either bit per complex sample or bit per second per Hertz, which is abbreviated as bit/s/Hz.
We will use the latter unit in the remainder of this monograph.
A related metric is the \emph{information rate} [bit/s], which is defined as the product of the SE and the bandwidth $B$.

The channel between a given transmitter and receiver supports many different SEs (depending on the chosen encoding/decoding scheme), but the largest achievable SE is of key importance when designing communication systems. The maximum SE is determined by the so-called \emph{channel capacity}.
We refer to the original paper \cite{Shannon1948a} by Shannon and textbooks on information theory, such as \cite{Cover1991a}, for the general and formal definition. 
In wireless communications, we are particularly interested in channels where the received signal is a superposition of a scaled version of the desired signal and additive Gaussian noise. These channels are commonly referred to as discrete memoryless \gls{AWGN} channels, since each discrete input signal leads to a discrete output signal that is independent of previous and future inputs. We will provide exact capacity expressions and capacity lower bounds for two distinctly different cases: deterministic and random channels. Although the channels are modeled as random in this monograph, both cases will be utilized.

\subsection{Capacity Bounds for Deterministic Channels}

\begin{figure}[t!]
	\begin{center}
		\includegraphics{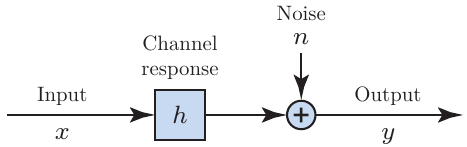}
	\end{center} \vskip-5mm
	\caption{A discrete memoryless channel with input $x$ and output $y = hx + n$, where $h$ is the channel response and $n$ is independent Gaussian noise.} \label{figure:AWGN-channel} 
\end{figure}

We begin with the canonical case when the channel is deterministic and the communication is only disturbed by Gaussian distributed noise. A block diagram of this setup is provided in Figure~\ref{figure:AWGN-channel}.

\begin{lemma} \label{lemma:capacity-memoryless-deterministic} 
Consider a discrete memoryless \gls{AWGN} channel with input $x \in \mathbb{C}$ and output $y \in \mathbb{C}$ given by
\begin{equation}
y = h x + n
\end{equation}
where $n \sim \CN(0,\sigma^2)$ is independent noise. The input distribution is power-limited as $\mathbb{E}\{ |x|^2 \} \leq p$ and the channel response $h \in \mathbb{C}$ is deterministic and known at the output.

In this case, any SE smaller or equal to the channel capacity
\begin{equation} \label{eq:capacity-fixed-channel}
C = \log_2 \left(1+\frac{p |h|^2}{\sigma^2} \right)
\end{equation}
is achievable. The capacity is achieved by selecting $x \sim \CN(0,p)$.
\end{lemma}
\begin{proof}
The detailed proof can be found in \cite{Shannon1949b}.
\end{proof}

The practical meaning of the channel capacity is that for any SE smaller or equal to the capacity, there exists an encoding/decoding scheme such that an arbitrarily low error probability can be achieved. More precisely, if we consider an information sequence with $N$ scalar inputs to the \gls{AWGN} channel, then the probability of error goes to zero as $N\to \infty$. This means that an infinite decoding delay is required to achieve the capacity. However, in practice, one can operate very close to the capacity whenever blocks of thousands of bits are transmitted \cite{Bjornson2016b}, which is typically the case in broadband applications.

The channel considered in Lemma~\ref{lemma:capacity-memoryless-deterministic} is called a \gls{SISO} channel because one input signal is sent and results in one output signal. However, we will demonstrate in later sections how the same result can be utilized in situations with multiple antennas. 
The capacity expression in \eqref{eq:capacity-fixed-channel} has a form that is typical for communications: the base-two logarithm of one plus the  \gls{SNR} defined as 
\begin{equation}
\mathacr{SNR} =  \frac{\overbrace{p |h|^2  }^{\textrm{Received signal power}} }{\underbrace{  \vphantom{p |h|^2} \sigma^2}_{\textrm{Noise power}}}.
\end{equation}

In the setups considered in this monograph, the communication is also disturbed by interference from other concurrent transmissions. In this case, the exact capacity is generally unknown, but convenient lower bounds can be obtained \cite{Biglieri1998a}. The following lemma provides a lower bound on the capacity that will be used repeatedly in this monograph.

\begin{lemma} \label{lemma:capacity-memoryless-deterministic-interference} 
Consider a discrete memoryless interference channel  with input $x \in \mathbb{C}$ and output $y \in \mathbb{C}$ given by
\begin{equation} 
y = h x + \upsilon + n
\end{equation}
where $n \sim \CN(0,\sigma^2)$ is independent noise and $\upsilon\in \mathbb{C}$ is random interference with an arbitrary distribution that has zero mean, variance $p_{\upsilon}$, and is uncorrelated with the input (i.e., $\mathbb{E}\{ x^{\star} \upsilon \} = 0$).

The input distribution is power-limited as $\mathbb{E}\{ |x|^2 \} \leq p$ and the channel response $h \in \mathbb{C}$ is deterministic and known at the output.
The channel capacity $C$ is then lower bounded as
\begin{equation} \label{eq:capacity-fixed-channel-interference}
C \geq \log_2 \left(1+\frac{p |h|^2}{p_{\upsilon} + \sigma^2} \right)
\end{equation}
where the bound is achieved using the input distribution $x \sim \CN(0,p)$.
\end{lemma}
\begin{proof}
The detailed proof can be found in \cite[App.~C.1.2]{massivemimobook}.
\end{proof}

The lower bound on the capacity in \eqref{eq:capacity-fixed-channel-interference} is achieved by encoding/decoding the signal as if the interference $\upsilon$ is independent Gaussian noise, because this is the worst case from a communication perspective \cite{Hassibi2003a}. There are no approximations involved but the achievable SE is generally smaller than the capacity, particularly when there are very strong interfering signals. However, treating non-Gaussian interference as independent Gaussian noise in the encoding/decoding is practically convenient and provably optimal in the low-interference regime \cite{Annapureddy2009a}, \cite{Annapureddy2011a}, \cite{Motahari2009a},  \cite{Shang2011b}, \cite{Shang2009b}. When we use the SE expression from Lemma~\ref{lemma:capacity-memoryless-deterministic-interference} in later sections, the random interference term $\upsilon$ will not only include interference from other concurrent transmissions but also unknown random variations in the desired channel.

The SE expression in \eqref{eq:capacity-fixed-channel-interference} has a similar form as the capacity expression in Lemma~\ref{lemma:capacity-memoryless-deterministic}. It is the base-two logarithm of one plus the expression  
\begin{equation} \label{eq:effective-SINR}
\mathacr{SINR} = \frac{\overbrace{p |h|^2  }^{\textrm{Received signal power}}}{\underbrace{  \vphantom{p |h|^2} p_{\upsilon}}_{\textrm{Interference power}} + \underbrace{ \vphantom{p |h|^2} \sigma^2}_{\textrm{Noise power}}}
\end{equation}
which is an \gls{SINR}. We will refer to it as an \emph{effective \gls{SINR}}, which means that the lower bound in \eqref{eq:capacity-fixed-channel-interference} is effectively the same as the capacity of an \gls{AWGN} channel with an \gls{SNR} equal to $\mathacr{SINR}$ in \eqref{eq:effective-SINR}. This implies that the SE in \eqref{eq:capacity-fixed-channel-interference} can be practically achieved using channel codes designed for \gls{AWGN} channels.

\subsection{Capacity Bounds for Random Channels}

We consider randomly fading channels in this monograph.
If the channel is a random variable that takes a new independent realization after a finite block of complex samples (e.g., a coherence block), then another capacity concept can be defined: the \emph{ergodic capacity}. The transmission then spans asymptotically many realizations of the random variable that describes the channel and the word ``ergodic'' identifies that all the statistical properties of the channel are deducible from a single sequence of channel realizations. We begin with the canonical case when the communication is only disturbed by Gaussian distributed noise. 

\begin{lemma} \label{lemma:capacity-memoryless-random} 
Consider a discrete memoryless channel with input $x \in \mathbb{C}$ and output $y \in \mathbb{C}$ given by
\begin{equation}
y = h x + n
\end{equation}
where $n \sim \CN(0,\sigma^2)$ is independent noise and the input distribution is power-limited as $\mathbb{E}\{ |x|^2 \} \leq p$.
The channel $h$ is a realization of a random variable $\mathbb{H}$ that is independent of the signal and noise, and the realization $\mathbb{H}=h$ is known at the receiver.

The ergodic channel capacity is
\begin{equation} \label{eq:capacity-fading-channel}
C = \mathbb{E} \left\{ \log_2 \left(1+\frac{p |h|^2}{\sigma^2} \right) \right\}
\end{equation}
where the expectation is with respect to $h$. The capacity is  achieved by selecting the input distribution $x \sim \CN(0,p)$.
\end{lemma}
\begin{proof}
The detailed proof can be found in \cite[App.~C.1.1]{massivemimobook}.
\end{proof}

The ergodic capacity of the block-fading channel is similar to the capacity in \eqref{eq:capacity-fixed-channel}, but the key difference is that there is an expectation in front of the logarithm in \eqref{eq:capacity-fading-channel}. In this case, $\mathacr{SNR} = \frac{p |h|^2}{\sigma^2}$ is called the \emph{instantaneous \gls{SNR}}.
We can also define the average \gls{SNR} as
\begin{equation} \label{eq:average-SNR-basic-def}
 \mathacr{SNR} = \frac{p \mathbb{E} \{ |h|^2 \}  }{ \sigma^2}
\end{equation}
where the expectation is computed with respect to the channel realizations. 

As mentioned above, the communication will be disturbed by interference from other concurrent transmissions in this monograph.
 In this case, the exact ergodic capacity is generally unknown, but a convenient lower bound can be obtained by once again treating the interference as noise \cite{Biglieri1998a}, \cite{Medard2000a}. The following lemma provides a lower bound on the ergodic capacity that will be used repeatedly in this monograph.

\begin{lemma} \label{lemma:capacity-memoryless-random-interference} 
Consider a discrete memoryless interference channel  with input $x \in \mathbb{C}$ and output $y \in \mathbb{C}$ given by
\begin{equation} \label{eq:interference-capacity}
y = h x + \upsilon + n
\end{equation}
where $n \sim \CN(0,\sigma^2)$ is independent noise, the channel response $h\in \mathbb{C}$ is known at the output, and $\upsilon\in \mathbb{C}$ is random interference with an arbitrary distribution satisfying the properties listed below. The input is power-limited as $\mathbb{E}\{ |x|^2 \} \leq p$.

Suppose $h\in \mathbb{C}$ is a realization of the random variable $\mathbb{H}$ and that $\mathbb{U}$ is a random variable with realization $u$ that affects the interference variance. The realizations of these random variables are known at the output. 
 If the noise $n$ is conditionally independent of $\upsilon$ given $h$ and $u$, the interference $\upsilon$ has conditional zero mean (i.e., $\mathbb{E}\{ \upsilon | h, u \}  = 0 $) and conditional variance denoted by $p_{\upsilon}(h,u) = \mathbb{E}\{ |\upsilon|^2 | h, u \}$, and the interference is conditionally uncorrelated with the input (i.e., $\mathbb{E}\{ x^{\star} \upsilon | h, u \} = 0$), then the ergodic channel capacity $C$ is lower bounded as
\begin{equation} \label{eq:capacity-fading-channel-interference}
C \geq \mathbb{E} \left\{ \log_2 \left(1+\frac{p |h|^2}{p_{\upsilon}(h,u) + \sigma^2} \right) \right\}
\end{equation}
where the expectation is taken with respect to $h$ and $u$, and the bound is achieved using the input distribution $x \sim \CN(0,p)$.
\end{lemma}
\begin{proof}
The detailed proof can be found in \cite[App.~C.1.2]{massivemimobook}.
\end{proof}

Note that in Lemma~\ref{lemma:capacity-memoryless-random-interference}, we used the shorthand notation $\mathbb{E}\{ \upsilon | h, u \} $ for the conditional expectation $\mathbb{E}\{ \upsilon | \mathbb{H}= h, \mathbb{U} = u \} $. For notational convenience, we omit the random variables in similar expressions in the remainder of this monograph and only write out the realizations.

The lower bound on the ergodic capacity in \eqref{eq:capacity-fading-channel-interference} is an achievable SE and will often be called that in this monograph. Except for the list of technical conditions that must be satisfied, the SE in \eqref{eq:capacity-fading-channel-interference} is essentially the same as for the case of deterministic channels in \eqref{eq:capacity-fixed-channel-interference}, but with an expectation in front of the logarithm.
The expression $\frac{p |h|^2}{p_{\upsilon}(h,u) + \sigma^2}$ in \eqref{eq:capacity-fading-channel-interference} will be called the \emph{effective instantaneous SINR}, because it takes the same role as the instantaneous SNR in the ergodic capacity expression in \eqref{eq:capacity-fading-channel}. In other words, the capacity lower bound is equivalent to the ergodic capacity of a fading \gls{AWGN} channel where the instantaneous SNR is $\frac{p |h|^2}{p_{\upsilon}(h,u) + \sigma^2}$. This implies that the SE in \eqref{eq:capacity-fading-channel-interference} can be practically achieved using channel codes designed for fading \gls{AWGN} channels.

When we use the SE expression result in Lemma~\ref{lemma:capacity-memoryless-random-interference} in later sections, the random interference term $\upsilon$ will not only include interference from other concurrent transmissions but also channel uncertainty due to estimation errors regarding the desired channel. In those cases, $h$ represents an estimate of the desired channel and $u$ represents the estimates of the interfering channels.

\section{Maximization of Rayleigh Quotients}
\label{sec:Rayleigh-quotient}

A commonly appearing problem formulation in multi-antenna communications is that of maximizing a power ratio between the desired signal and interference plus noise. The optimization variable is a vector determining how the signal observations made at the different receive antennas are weighted together. For example, consider the received signal $\vect{y} \in \mathbb{C}^N$ at $N$ antennas. It is modeled as
\begin{equation} \label{eq:Rayleigh-problem1}
\vect{y} = \vect{h} x + \vect{n}
\end{equation}
where $\vect{h}  \in \mathbb{C}^N$ is a deterministic channel vector, $x$ is the random information signal with power $\mathbb{E}\{ |x|^2\} = 1$, and $\vect{n}\sim \CN(\vect{0}_N,\vect{I}_N)$ is independent noise.
If we apply a receive combining vector $\vect{v}$ to \eqref{eq:Rayleigh-problem1}, we get
\begin{equation}
\vect{v}^{\Htran} \vect{y} =  \vect{v}^{\Htran} \vect{h} x +\vect{v}^{\Htran}  \vect{n}
\end{equation}
and the \gls{SNR} becomes
\begin{equation} \label{eq:Rayleigh-problem1b}
\frac{ \mathbb{E}\{ | \vect{v}^{\Htran} \vect{h} x  |^2 \} }{ \mathbb{E}\{ | \vect{v}^{\Htran}  \vect{n}  |^2 \} } = \frac{ | \vect{v}^{\Htran} \vect{h}  |^2  }{\vect{v}^{\Htran} \vect{v} }.
\end{equation}
The type of expression in the right-hand side of  \eqref{eq:Rayleigh-problem1b} is known as a \emph{Rayleigh quotient}\footnote{The expression in \eqref{eq:Rayleigh-problem1b} is also known as the Rayleigh--Ritz ratio. Note that the only connection between Rayleigh fading and Rayleigh quotient is that they have been named after the same researcher.} and can be maximized with respect to $\vect{v}$ as follows.

\begin{lemma} \label{lemma:max-Rayleigh-quotient}
For a given channel vector $\vect{h} \in \mathbb{C}^{N}$, it holds that 
\begin{equation}
\max_{\vect{v} \in \mathbb{C}^N} \quad \frac{  |  \vect{v}^{\Htran} \vect{h} |^2  }{ \vect{v}^{\Htran}  \vect{v}} = \| \vect{h} \|^2
\end{equation}
where the maximum is attained by $\vect{v} = \vect{h}$.
\end{lemma}
\begin{proof}
The Cauchy-Schwarz inequality for two vectors $\vect{v}$ and $\vect{h}$ says~that \begin{equation}
| \vect{v}^{\Htran} \vect{h} |^2 \leq \| \vect{v} \|^2 \, \| \vect{h} \|^2
\end{equation}
with equality if and only if $\vect{v}$ and $\vect{h}$ are linearly dependent. It follows that $\frac{| \vect{v}^{\Htran} \vect{h} |^2}{\| \vect{v} \|^2 } \leq  \| \vect{h} \|^2$ where the maximum is attained for $\vect{v} = \vect{h}$ (among other solutions where the vectors are linearly dependent).
\end{proof}

\enlargethispage{0.5\baselineskip} 

The result in Lemma~\ref{lemma:max-Rayleigh-quotient} is commonly used in signal processing and the optimal vector $\vect{v}$ is known as \gls{MR} combining or matched filtering. This method maximizes the SNR, however, it is a generalization that will be particularly useful in this monograph. Let us generalize \eqref{eq:Rayleigh-problem1} by replacing the \gls{iid} noise term $\vect{n}$ with the colored noise term $\vect{n}' \sim \CN(\vect{0}_N,\vect{B})$ having a positive definite covariance matrix $\vect{B} \in \mathbb{C}^{N \times N}$. The new received signal is
\begin{equation} \label{eq:Rayleigh-problem2}
\vect{y} = \vect{h} x + \vect{n}'
\end{equation}
and by applying a receive combining vector $\vect{v}$ to \eqref{eq:Rayleigh-problem2}, the SNR becomes
\begin{equation}
\frac{ \mathbb{E}\{ | \vect{v}^{\Htran} \vect{h} x  |^2 \} }{ \mathbb{E}\{ | \vect{v}^{\Htran}  \vect{n}'  |^2 \} } = \frac{ | \vect{v}^{\Htran} \vect{h}  |^2  }{\vect{v}^{\Htran} \vect{B} \vect{v} }.
\end{equation}
This type of expression is known as a \emph{generalized Rayleigh quotient} and can be maximized with respect to $\vect{v}$ as follows.

\begin{lemma} \label{lemma:gen-rayleigh-quotient}
For a given channel vector $\vect{h} \in \mathbb{C}^{N}$ and Hermitian positive definite matrix $\vect{B} \in \mathbb{C}^{N \times N}$, it holds that
\begin{equation} \label{eq:gen-rayleigh-quotient}
\max_{\vect{v} \in \mathbb{C}^N} \quad \frac{  |  \vect{v}^{\Htran} \vect{h} |^2  }{ \vect{v}^{\Htran} \vect{B} \vect{v}} = \vect{h}^{\Htran} \vect{B}^{-1} \vect{h}
\end{equation}
where the maximum is attained by $\vect{v} = \vect{B}^{-1} \vect{h}$ or $\vect{v} = (\vect{h} \vect{h}^{\Htran}+\vect{B})^{-1} \vect{h}$.
\end{lemma}
\begin{proof}
The matrix square root $\vect{B}^{\frac12}$ of $\vect{B}$ exists since $\vect{B}$ is positive definite, thus we can define the new variable $\bar{\vect{v} } = \vect{B}^{\frac12} \vect{v}$ and establish that
\begin{equation}
\frac{  |  \vect{v}^{\Htran} \vect{h} |^2  }{ \vect{v}^{\Htran} \vect{B} \vect{v}}  = \frac{  |  \bar{\vect{v}}^{\Htran} \vect{B}^{-\frac12} \vect{h} |^2  }{  \bar{\vect{v}}^{\Htran} \bar{\vect{v}}  }.
\end{equation}
The expression at the right-hand side is a Rayleigh quotient of the kind in Lemma~\ref{lemma:max-Rayleigh-quotient} and, thus, maximized by $\bar{\vect{v} } = \vect{B}^{-\frac12} \vect{h}$. Hence, the solution to \eqref{eq:gen-rayleigh-quotient} is $\vect{v} = \vect{B}^{-1} \vect{h}$ and the maximum value follows accordingly.
Since any vector that is parallel to $ \vect{B}^{-1} \vect{h}$ also achieves the maximum value, we can also use
$\vect{v} = (\vect{h} \vect{h}^{\Htran}+\vect{B})^{-1} \vect{h}$ since
\begin{equation}
(\vect{h} \vect{h}^{\Htran}+\vect{B})^{-1} \vect{h} = \frac{1}{1+\vect{h}^{\Htran} \vect{B}^{-1} \vect{h}} \vect{B}^{-1} \vect{h}
\end{equation}
according to the matrix-inversion lemma (Lemma~\ref{lemma:matrix-inversion} on p.~\pageref{lemma:matrix-inversion}).
\end{proof}

This lemma shows how to maximize a generalized Rayleigh quotient and what the maximum value is. The solution to \eqref{eq:gen-rayleigh-quotient} is not unique but  $\vect{v} = c \vect{B}^{-1} \vect{h}$ for any $c \neq 0$ will also maximize the expression.
This lemma will be utilized several times in Section~\ref{c-uplink} on p.~\pageref{c-uplink} to optimize different kinds of receiver processing in the uplink.

\section{Optimization Algorithms for Utility Maximization}
\label{sec:optimization-algorithms}

A utility function is a metric of system performance and can be formulated in different ways.
In this section, we provide state-of-the-art optimization algorithms for solving two main types of utility maximization problems. These algorithms will be utilized in Section~\ref{c-resource-allocation} on p.~\pageref{c-resource-allocation} to solve several resource allocation problems.

The SE is used in this monograph to measure the communication performance of each UE. When designing the network operation, there are $K$ different SEs to take into consideration and these are mutually conflicting due to interference. For example, we can improve the SE of one UE by reducing the transmit powers assigned to other UEs, but with the side-effect that the SEs achieved by the other UEs are then reduced. To identify a good tradeoff between the UEs' individual performance, a scalar-valued utility function can be defined to take all the SEs into consideration. This is called scalarization in the field of multi-objective optimization \cite{Bjornson2014c}. We will consider two utilities: max-min fairness and sum SE.

Since we have not provided any explicit SE expressions for User-centric Cell-free Massive MIMO systems yet, the presentation in this section is based on a generic \gls{SISO} scenario. Suppose the received signal used for decoding the signal of UE $k$ can be expressed as 
\begin{align}
y_k=\mathrm{DS}_k\left(\vect{p}\right)s_k+\sum_{i=1}^K\mathrm{I}_{ki}\left(\vect{p}\right)s_i+n_k \label{eq:SISO-received}
\end{align}
where $s_k \in \mathbb{C}$ denotes the normalized, independent random data signal of UE $k$ with $\mathbb{E}\{|s_k|^2\}=1$ and $\vect{p}=[p_1 \, \ldots \, p_K]^{\Ttran}$ is the set of transmit power coefficients for all the $K$ UEs.\footnote{As we see in Section~\ref{c-downlink} on p.~\pageref{c-downlink}, there can be more than one transmit power coefficient per UE in the downlink. Although here we assume the length of the vector $\vect{p}$ is $K$, we can simply change it for the mentioned downlink power allocation optimization.} The transmit powers are non-negative: $p_k\geq0$, for $k=1,\ldots,K$. We write this as $\vect{p} \geq \vect{0}_K$.

The received signal in \eqref{eq:SISO-received} contains three parts. The term $\mathrm{DS}_k(\vect{p})s_k$ is the desired part, containing the data signal $s_k$ multiplied with a deterministic amplitude $\mathrm{DS}_k(\vect{p})$ that is a function of $\vect{p}$ (it usually only depends on $p_k$). The interference term $\sum_{i=1}^K\mathrm{I}_{ki}(\vect{p})s_i$ contains the interfering data signal $s_i$ of UE $i$, for $i=1,\ldots,K$, and the term $\mathrm{I}_{ki}(\vect{p})$ is random and determines the strength of the interference caused to UE $k$ by UE $i$. This term is also a function of $\vect{p}$. Note that we have included a self-interference term $\mathrm{I}_{kk} (\vect{p})s_k$ with a random amplitude $\mathrm{I}_{kk}(\vect{p})$, which we further assume to have zero mean. This term can model the uncertainty that the receiver has regarding the channel that the desired signal propagated over, which will be utilized and further explained in later sections. Finally, $n_k \sim \mathcal{N}_{\mathbb{C}}\left(0,\sigma^2\right)$ is the independent additive noise.

We obtain the following result by using Lemma~\ref{lemma:capacity-memoryless-deterministic-interference}. 

\begin{lemma} \label{lemma:generic-SINR-expression}
An achievable SE of the channel for UE $k$ in \eqref{eq:SISO-received} is 
\begin{align} \label{eq:SISO-received-SE}
\mathrm{SE}_k\left(\vect{p}\right)=\log_2\left(1+\mathrm{SINR}_k\left(\vect{p}\right)\right)
\end{align}
where the effective SINR is
\begin{align}\label{eq:SISO-received-SINR}
\mathrm{SINR}_k\left(\vect{p}\right)=\frac{\left|\mathrm{DS}_k\left(\vect{p}\right)\right|^2}{\sum\limits_{i=1}^K\mathbb{E}\left\{\left|\mathrm{I}_{ki}\left(\vect{p}\right)\right|^2\right\}+\sigma^2}.
\end{align}
\end{lemma}
\begin{proof}
We obtain this result from Lemma~\ref{lemma:capacity-memoryless-deterministic-interference} by letting $y=y_k$ be the received signal in the lemma, $x= s_k$ is the desired signal, $h=\mathrm{DS}_k(\vect{p})$ is the channel response, and $n=n_k$ is the noise. The interference term $\upsilon = \sum_{i=1}^K\mathrm{I}_{ki}(\vect{p})s_i$ is random with zero mean and variance  $p_{\upsilon} = \sum_{i=1}^K\mathbb{E}\{|\mathrm{I}_{ki}(\vect{p})|^2\}$ since the data signals have zero mean, unit variance, and are independent. The condition $\mathbb{E}\{ x^{\star} \upsilon \} = 0$ is satisfied since $\mathrm{I}_{kk}(\vect{p})$ was assumed to have zero mean.
\end{proof}

\begin{figure}[t!]
	\centering 
	\begin{overpic}[width=\columnwidth,tics=10]{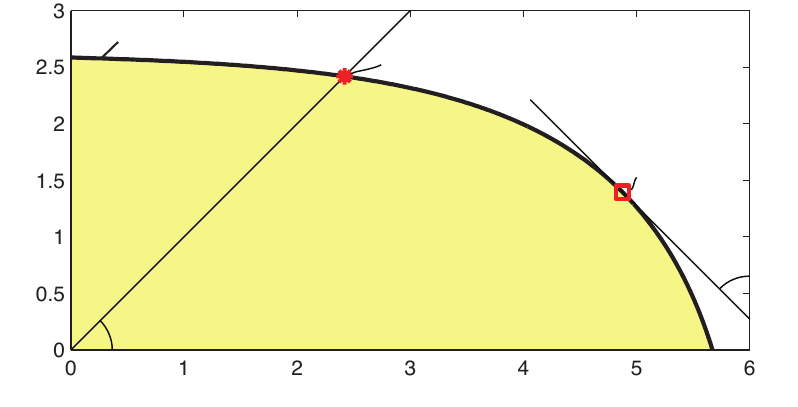}
		\put(15,46){\small{Outer boundary}}
		\put(80.5,29){\small{Maximum}}
		\put(80.5,26){\small{sum SE}}
		\put(89,17){\small{$45^\circ$}}
		\put(14,9){\small{$45^\circ$}}
		\put(48,42){\small{Max-min fairness}}
		\put(48,0){\small{$\mathrm{SE}_1(\vect{p})$}}
		\put(0,23){\rotatebox{90}{\small{$\mathrm{SE}_2(\vect{p})$}}}
	\end{overpic}  
	\caption{Example of the SE region (shaded) containing all the points $(\mathrm{SE}_1(\vect{p}),\mathrm{SE}_2(\vect{p}))$ that can be obtained by different feasible selections of the power coefficient vector $\vect{p}$. The points at the outer boundary that give max-min fairness and maximum sum SE are indicated.}
	\label{figure:basic_rate_region}  
\end{figure}
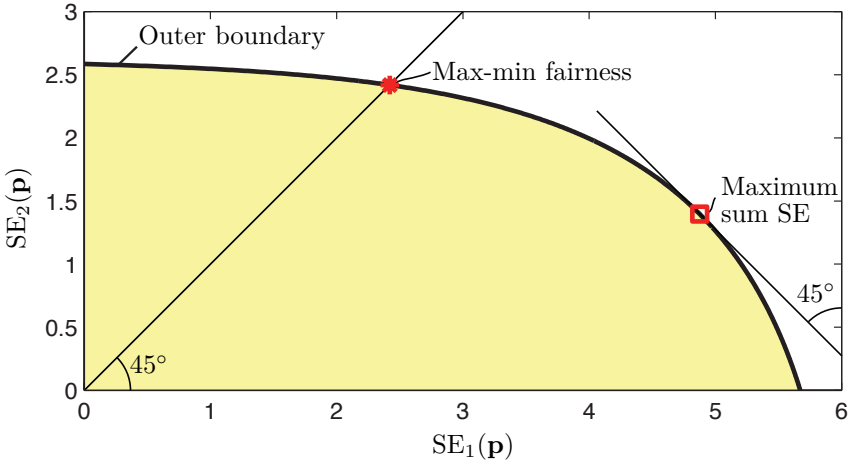

With the notation provided by Lemma~\ref{lemma:generic-SINR-expression}, the individual performance of the UEs are represented by their $K$ SEs $\{\mathrm{SE}_k(\vect{p}):k=1,\ldots,K\}$, which are $K$ different performance metrics. 
For each choice of the transmit power vector $\vect{p}$, the $K$ UEs will achieve a certain combination of SEs that can be gathered in a $K$-dimensional vector $[ \mathrm{SE}_1(\vect{p}) \, \ldots \, \mathrm{SE}_K(\vect{p})]^{\Ttran}$. By considering the set of all such vectors that can be obtained with feasible transmit power vectors (i.e., those satisfying a set of power constraints that exist in the system), we can generate a $K$-dimensional region that represents all the possible ways of operating the system. This region is called an SE region and is exemplified in Figure~\ref{figure:basic_rate_region} for $K=2$. The shaded region contains all the feasible operating points. We need to select one of these points and there is no truly optimal way to do it. All the points in the interior of the region are strictly suboptimal since there are points at the outer boundary where all the UEs achieve higher SEs. On the other hand, each point on the outer boundary represents one possible tradeoff between the performance achieved by the different UEs. When comparing two points on the outer boundary, one UE will prefer the first point and another UE will prefer the second point. This conflict cannot be resolved in a fully objective manner but the network operator must inject a subjective opinion of what kind of operating point is preferred by the network as a whole.

As mentioned at the beginning of this section, we will adopt the scalarization approach to identify a tradeoff between the $K$ metrics \cite{Bjornson2014c}. This means that we combine the  $K$ SE metrics into a single scalar utility function representing the network-wide performance. We consider the two most common utility functions from the literature: max-min fairness and sum SE. These utilities represent two extremes in how to balance between transmitting many bits and obtaining user fairness.
The max-min fairness utility requires every UE to get the same SE, which means that the optimal operating point in Figure~\ref{figure:basic_rate_region} lies on a line from the origin with $45^\circ$ slope. As indicated in the figure, the max-min fairness solution is the point where this line intersects the outer boundary.
The maximum sum SE is generally obtained at another operating point, as illustrated in Figure~\ref{figure:basic_rate_region}. It lies on a line with equation $\mathrm{SE}_1(\vect{p})+\mathrm{SE}_2(\vect{p})=c$, where $c$ is the maximum achievable sum SE. While this point maximizes the number of bits that are transmitted in total, the downside is that the bits are unequally distributed among the UEs. In this example, UE 2 gets a smaller SE than at the max-min fairness point, while UE 1 gets a larger SE. In the remainder of this section, we will describe algorithms for finding these operating points in networks with arbitrarily many UEs.

\subsection{Max-Min SE Fairness}

The aim of max-min SE fairness is to maximize the lowest SE among all the UEs in the network, which is known as max-min fairness. Since the SE of UE $k$ in \eqref{eq:SISO-received-SE} is an increasing function of the effective SINR, $\mathrm{SINR}_k(\vect{p})$, maximizing the minimum SE is the same as maximizing the minimum effective SINR among all the UEs.
We assume there are $R$ linear constraints on the transmit power vector:
\begin{equation}
\vect{a}_r^{\Ttran}\vect{p}\leq p_{\textrm{max}}, \quad r=1,\ldots,R,
\end{equation}
where the fixed vector $\vect{a}_r\in \mathbb{R}_{\geq 0}^{K}$ specifies the weighting coefficient for each UE's power coefficient and $p_{\textrm{max}}$ is the maximum allowed power,\footnote{This model can be utilized for both downlink and uplink power allocations. In the downlink, per-AP power constraints are considered, which leads to $L$ power constraints (one per AP). In this case, the non-zero elements of $\vect{a}_{l}$ are multiplied with the power coefficients for the UEs that are served by AP $l$. On the other hand, in the uplink, there are $K$ individual power constraints (one per UE). Hence, only the $k$th element of the vector $\vect{a}_k$ is non-zero while all other elements are zero.} the max-min fairness problem can be formulated as
\begin{align} \label{eq:maxmin-fairness-problem}
&\underset{\vect{p} \geq \vect{0}_K}{\textrm{maximize}} \quad \min\limits_{k\in \{1,\ldots,K\}} \quad \mathrm{SINR}_k\left(\vect{p}\right) \\
&\textrm{subject to} \quad \vect{a}_r^{\Ttran}\vect{p}\leq p_{\textrm{max}}, \quad r=1,\ldots,R.  \nonumber
\end{align}
To solve this problem, we first introduce the auxiliary variable $t$ that represents the lowest SINR among all the UEs. We can then obtain the following optimization problem that is equivalent to \eqref{eq:maxmin-fairness-problem} in terms of having the same maximum utility value and optimal value of $\vect{p}$:
\begin{align} \label{eq:maxmin-fairness-problem-t}
&\underset{\vect{p}\geq \vect{0}_K, t \geq 0}{\textrm{maximize}} \quad t \\
&\textrm{subject to} \quad  \mathrm{SINR}_k\left(\vect{p}\right)\geq t, \quad k=1,\ldots,K \nonumber \\
&\hspace{21mm} \vect{a}_r^{\Ttran}\vect{p}\leq p_{\textrm{max}}, \quad r=1,\ldots,R.\nonumber
\end{align}
The equivalence of the two problems can be verified by noting that $t$ is at most equal to the lowest SINR among the UEs and it should be equal to that value to maximize the objective function. The reformulation in \eqref{eq:maxmin-fairness-problem-t} is known as the epigraph form of \eqref{eq:maxmin-fairness-problem} \cite[Sec.~3.1.7]{Boyd2004a}.

An optimal solution to the problem in \eqref{eq:maxmin-fairness-problem-t} can be obtained by a bisection search over $t$. To see this, let $t^{\rm opt}$ denote the optimal objective value of the problem in \eqref{eq:maxmin-fairness-problem-t}. If we select $t>t^{\rm opt}$ in \eqref{eq:maxmin-fairness-problem-t}, the problem will obviously be infeasible.
On the other hand, if we select any $t<t^{\rm opt}$, the problem in \eqref{eq:maxmin-fairness-problem-t} is feasible. Hence, we can apply a bisection search over $t$ to find $t^{\rm opt}$ \cite[Sec.~4.2.5]{Boyd2004a}.
At each iteration, we consider a value $t=t^{\rm candidate}$ and need to solve the following feasibility problem:
\begin{align} \label{eq:maxmin-fairness-feasibility}
&\textrm{find} \quad \vect{p} \\
&\textrm{subject to} \quad  \mathrm{SINR}_k\left(\vect{p}\right)\geq t^{\rm candidate}, \quad k=1,\ldots,K \nonumber \\
&\hspace{21mm} \vect{a}_r^{\Ttran}\vect{p}\leq p_{\textrm{max}}, \quad r=1,\ldots,R \nonumber \\
&\hspace{21mm} \vect{p} \geq \vect{0}_K \nonumber.
\end{align}
The goal of solving the feasibility problem in \eqref{eq:maxmin-fairness-feasibility} is to find any solution $\vect{p}$ that satisfies the constraints. The numerical precision that is needed to find the value of $t^{\rm candidate}$ that corresponds to the global optimum can be extremely high, particularly if some variables in $\vect{p}$ have a much smaller impact on the SINR of the UE with the worst channel conditions than the other variables.
One can prove by contradiction that all UEs should get identical SINRs at the global optimum but this is only one of the many feasible solutions to the feasibility problem in \eqref{eq:maxmin-fairness-feasibility} for any $t^{\rm candidate}<t^{\rm opt}$. We have noticed experimentally that this is a real issue when optimizing large cell-free networks, where some of the serving APs have much weaker channels than other serving APs.
To improve the convergence rate, we will therefore replace the feasibility problem with a problem having the same constraints but where the total power $\sum_{k=1}^{K}p_k $ is minimized as well. This will encourage numerical solvers to identify the feasible point where all UEs get the same SINR, while retaining optimality. By utilizing this property, we obtain Algorithm~\ref{alg:bisection} where the revised subproblem is given in \eqref{eq:maxmin-fairness-feasibility2}. 
\begin{algorithm}
	\caption{Bisection search algorithm for solving the max-min fairness problem in \eqref{eq:maxmin-fairness-problem-t}.} \label{alg:bisection}
	\begin{algorithmic}[1]
\State {\bf Initialization:} Set the solution accuracy $\epsilon>0$
\State Set the initial lower and upper bounds for the max-min SINR:
\State $t^{\rm lower}\gets0$
\State $t^{\rm upper}\gets \underset{k\in \{1,\ldots,K\}}{\min}\quad  \underset{\vect{p} \geq \vect{0}_K}{\max} \quad  \textrm{SINR}_k\left(\vect{p}\right)$
\State Initialize solution variables: $\vect{p}^{\rm opt}=\vect{0}_K$, $t^{\rm opt}=0$
\While{$t^{\rm upper}-t^{\rm lower}>\epsilon$}
\State $t^{\rm candidate}\gets \frac{t^{\rm lower}+t^{\rm upper}}{2}$
\State Solve the following problem for fixed $t=t^{\rm candidate}$:
\begin{align} \label{eq:maxmin-fairness-feasibility2}
&\underset{\vect{p} \geq \vect{0}_K}{\textrm{minimize}} \quad \, \sum_{k=1}^Kp_k \\
&\textrm{subject to} \quad  \mathrm{SINR}_k\left(\vect{p}\right)\geq t^{\rm candidate}, \quad k=1,\ldots,K \nonumber \\
&\hspace{21mm} \vect{a}_r^{\Ttran}\vect{p}\leq p_{\textrm{max}}, \quad r=1,\ldots,R \nonumber \\
&\hspace{21mm} \vect{p} \geq \vect{0}_K \nonumber
\end{align}
\If{\eqref{eq:maxmin-fairness-feasibility2} is feasible}
\State $t^{\rm lower}\gets t^{\rm candidate}$
\State $\vect{p}^{\rm opt} \gets \vect{p}$, which is the solution to \eqref{eq:maxmin-fairness-feasibility2}
\Else
\State  $t^{\rm upper}\gets t^{\rm candidate}$
\EndIf
\EndWhile
		\State {\bf Output:} $\vect{p}^{\rm opt}$, $t^{\rm opt}=\underset{k\in \{1,\ldots,K\}}{\min} \textrm{SINR}_k\left(\vect{p}^{\rm opt}\right)$
	\end{algorithmic}
	\end{algorithm}

Instead of solving a sequence of optimization problems, as in Algorithm~\ref{alg:bisection}, a fixed-point algorithm can be implemented when the SINR expression satisfies certain additional conditions stated in the next lemma from \cite[Lemma~1, Theorem~1]{Hong2014}. 

\newpage

\begin{lemma}\label{lemma:fixed-point}
Suppose $\left\{\textrm{SINR}_k\left(\vect{p}\right):k=1,\ldots,K\right\}$ satisfy the following:
\begin{enumerate}	
   \item $\textrm{SINR}_k\left(\vect{p}\right)>0$ if $\vect{p}>\vect{0}_K$ and $\textrm{SINR}_k\left(\vect{p}\right)=0$ if and only if $p_k=0$, $\forall k$;
   \item $\textrm{SINR}_k\left(\vect{p}\right)$ is strictly increasing with respect to $p_k$ and is strictly decreasing with respect to $p_i$, for $i\neq k$, when $p_k>0$, $\forall k$;
  \item For $\lambda > 1$ and $p_k > 0$, $\textrm{SINR}_k\left(\lambda\vect{p}\right)>\textrm{SINR}_k\left(\vect{p}\right)$, $\forall k$. 
  \end{enumerate}
The optimal solution to the optimization problem in \eqref{eq:maxmin-fairness-problem-t} then satisfies $t^{\rm opt}>0$ and $\vect{p}^{\rm opt}>\vect{0}_K$. Moreover, $\textrm{SINR}_k\left(\vect{p}^{\rm opt}\right)=t^{\rm opt}$, for all $k$ and $\vect{a}_r^{\Ttran}\vect{p}^{\rm opt}= p_{\textrm{max}}$, for at least one $r$. 

Define $\vect{T}\left(\vect{p}\right)=\left[ p_1\big/\textrm{SINR}_1\left(\vect{p}\right) \ \ldots \  p_K\big/\textrm{SINR}_K\left(\vect{p}\right) \right]^{\Ttran}$ and let $\mathcal{U}$ denote the set of feasible $\vect{p}$ that satisfies this last property:
\begin{align}
\mathcal{U}=\big\{\vect{p}\geq \vect{0}_K: \, & \vect{a}_r^{\Ttran}\vect{p}\leq p_{\textrm{max}}, \quad r=1,\ldots,R, \nonumber \\ 
&\text{ with at least one equality for some }r \big\}.
\end{align} 
The power vector $\vect{p}$ computed by Algorithm~\ref{alg:fixed-point} converges to the optimal solution to \eqref{eq:maxmin-fairness-problem-t} if the following additional conditions are satisfied:
\begin{itemize}
\item  There exist numbers $a > 0$, $b > 0$, and a vector $\vect{e} > \vect{0}_K$ such that $a\vect{e} \leq \vect{T}\left(\vect{p}\right) \leq b\vect{e}$, for all $\vect{p}\in \mathcal{U}$;
\item For any $\vect{p}_1, \vect{p}_2 \in \mathcal{U}$ and $0\leq \lambda \leq 1$: $\lambda\vect{p}_1\leq \vect{p}_2 \implies \lambda\vect{T}\left(\vect{p}_1\right)\leq \vect{T}\left(\vect{p}_2\right)$;
\item For any $\vect{p}_1, \vect{p}_2 \in \mathcal{U}$ and $0\leq \lambda < 1$:  $\lambda\vect{p}_1\leq \vect{p}_2$ and $\lambda\vect{p}_1\neq \vect{p}_2$  $\implies \lambda\vect{T}\left(\vect{p}_1\right)< \vect{T}\left(\vect{p}_2\right)$.
\end{itemize}
\end{lemma}

Whenever the conditions stated in Lemma~\ref{lemma:fixed-point} are satisfied, the fixed-point algorithm in Algorithm~\ref{alg:fixed-point} should be utilized to solve the max-min fairness problem since it fully utilizes the SINR structure.

\begin{algorithm}
	\caption{Fixed-point algorithm for solving the max-min fairness problem in \eqref{eq:maxmin-fairness-problem-t}.} \label{alg:fixed-point}
	\begin{algorithmic}[1]
		\State {\bf Initialization:} Set arbitrary $\vect{p}>\vect{0}_K$ and the solution accuracy $\epsilon>0$
		\While{$\underset{k\in\{1,\ldots,K\}}{\max} \textrm{SINR}_k\left(\vect{p}\right)-\underset{k\in\{1,\ldots,K\}}{\min} \textrm{SINR}_k\left(\vect{p}\right) > \epsilon$} 
		\State $p_k \gets \frac{p_k}{\textrm{SINR}_k\left(\vect{p}\right)}$, $k=1,\ldots,K$
		\State $\vect{p} \gets \frac{p_{\rm max}}{\underset{r\in\{1,\ldots,R\}}{\max}\vect{a}_r^{\Ttran}\vect{p}}\vect{p}$
		\EndWhile
		\State {\bf Output:} $\vect{p}$, $t=\underset{k\in\{1,\ldots,K\}}{\min} \textrm{SINR}_k\left(\vect{p}\right)$
	\end{algorithmic}
\end{algorithm}

\subsection{Sum SE Maximization}

A potential drawback of the max-min SE fairness problem is that all the emphasis is put on the UEs with the worst channel conditions. When having a large network where every UE only causes interference to a small subset of neighboring UEs, it is likely that many UEs can achieve substantially larger SEs while barely affecting the UEs having the worst conditions. 
In these situations, it might be better to maximize the sum SE, which represents the total number of bits that are transmitted without considering how the bits are assigned between the UEs.
The sum SE maximization problem can be expressed as
 \begin{align}
 &\underset{\vect{p} \geq \vect{0}_K}{\textrm{maximize}} \quad \sum\limits_{k=1}^K  \log_2\left(1+\mathrm{SINR}_k\left(\vect{p}\right)\right) \nonumber\\
 &\textrm{subject to} \quad \vect{a}_r^{\Ttran}\vect{p}\leq p_{\textrm{max}}, \quad r=1,\ldots,R. \label{eq:sum-SE-maximization}
 \end{align}
Note that the above problem is usually not convex and, hence, it is hard to obtain the optimal solution \cite{Luo2008a}. There exist global optimization methods \cite{Bjornson2013d}, \cite{Weeraddana2012a} that find the optimum but their computational complexity is unsuitable for real-time applications. A pragmatic solution is to instead settle for a local optimum. A common approach for finding a local optimum to sum SE maximization problems is the \emph{weighted MMSE} method \cite{Shi2011}, which results in iterative algorithms.

To obtain the weighted MMSE reformulation of the sum SE maximization problem at hand, we recall the \gls{SISO} channel for UE $k$ in \eqref{eq:SISO-received} where the received signal is $y_k$. The receiver can compute an estimate $\widehat{s}_{k}=u_{k}^{\star}y_{k}$ of the desired signal $s_{k}$ using a scalar combining coefficient $u_{k} \in\mathbb{C}$ that can amplify it and rotate the phase. The corresponding \gls{MSE} is given by
\begin{align} \nonumber
&e_k\left(\vect{p}, u_k\right)=\mathbb{E}\left\{\left|\widehat{s}_{k}-s_{k}\right|^2\right\} \\
&=\left|u_{k}\right|^2\left(\left|\textrm{DS}_{k}\left(\vect{p}\right)\right|^2+\sum\limits_{i=1}^K\mathbb{E}\left\{\left|\textrm{I}_{ki}\left(\vect{p}\right)\right|^2\right\}+\sigma^2\right) -2\Re\left(u_{k}^{\star}\textrm{DS}_{k}\left(\vect{p}\right)\right)+1  \label{eq:ek-MSE}
\end{align}
and it is a convex function of $u_k$. One can show that the value of $u_k$ that minimizes the MSE for UE $k$ in \eqref{eq:ek-MSE} for a given $\vect{p}$ is 
\begin{align} \label{eq:uk}
u_{k}\left(\vect{p}\right)=\frac{\textrm{DS}_{k}\left(\vect{p}\right)}{\left|\textrm{DS}_{k}\left(\vect{p}\right)\right|^2+\sum\limits_{i=1}^K\mathbb{E}\left\{\left|\textrm{I}_{ki}\left(\vect{p}\right)\right|^2\right\}+\sigma^2}.
\end{align} 
After plugging $u_{k}\left(\vect{p}\right)$ into \eqref{eq:ek-MSE}, it follows that the corresponding $e_k$ is equal to $1/\left(1+\textrm{SINR}_{k}\left(\vect{p}\right)\right)$. 

The main idea of the weighted MMSE method is to introduce the auxiliary weight $d_{k}\geq 0$ for the MSE $e_{k}$ and attempt to solve the following optimization problem:
\begin{align} \label{eq:sum-SE-maximization-weighted-MMSE} 
&\underset{\vect{p} \geq \vect{0}_K,  \left\{u_{k}, \ d_{k}\geq 0 : \, k=1,\ldots,K\right\}}{\textrm{minimize}} \quad \sum_{k=1}^K\Big(d_{k}e_{k}\left(\vect{p},u_k\right)-\ln\left( d_{k}\right)\Big)  \\
&\hspace{1.1cm}\textrm{subject to} \quad \vect{a}_r^{\Ttran}\vect{p}\leq p_{\textrm{max}}, \quad r=1,\ldots,R. \nonumber
\end{align}
The problem is equivalent to the original sum SE maximization problem in \eqref{eq:sum-SE-maximization} in the sense that they have the same global optimal solution.
The equivalence of the two problems follows from the fact that the optimal $d_{k}$ in \eqref{eq:sum-SE-maximization-weighted-MMSE} is $1/e_{k}$, which is equal to $1+\textrm{SINR}_{k}$. Hence, we obtain the original sum SE maximization problem  in \eqref{eq:sum-SE-maximization} with the same constraints. The benefit of the reformulation in \eqref{eq:sum-SE-maximization-weighted-MMSE} is that we have the following result that is adapted from \cite[Theorem 3]{Shi2011}.

\begin{lemma}  \label{theorem:weighted-MMSE-convergence} The block coordinate descent algorithm given in Algorithm~\ref{alg:sum-SE} converges to a local optimum (stationary point) of the problem in \eqref{eq:sum-SE-maximization-weighted-MMSE} by alternating between optimizing three blocks of variables: $\left\{u_{k}:k=1,\ldots,K\right\}$, $\left\{d_{k}:k=1,\ldots,K\right\}$, and $\vect{p}$. The obtained $\vect{p}$ is also a local optimum to the problem in \eqref{eq:sum-SE-maximization}.
	\end{lemma}

\begin{algorithm}
	\caption{Block coordinate descent algorithm for solving the sum SE maximization problem in \eqref{eq:sum-SE-maximization-weighted-MMSE}.} \label{alg:sum-SE}
	\begin{algorithmic}[1]
		\State {\bf Initialization:} Set the solution accuracy $\epsilon>0$
		\State Set an arbitrary feasible $\vect{p}$
		\While{ the objective function is either improved more than $\epsilon$ or not improved at all}
		\State $u_{k} \gets \frac{\textrm{DS}_{k}\left(\vect{p}\right)}{\left|\textrm{DS}_{k}\left(\vect{p}\right)\right|^2+\sum\limits_{i=1}^K\mathbb{E}\left\{\left|\textrm{I}_{ki}\left(\vect{p}\right)\right|^2\right\}+\sigma^2}, \quad k=1,\ldots,K$
		\State $d_k \gets 1/e_k\left(\vect{p},u_k\right), \quad k=1,\ldots,K$
		\State Solve the following problem for the current values of  $u_k$ and $d_k$:
		\begin{align}
		&\underset{\vect{p} \geq \vect{0}_K}{\textrm{minimize}} \quad \sum_{k=1}^Kd_{k}e_{k}\left(\vect{p},u_k\right) \label{eq:sum-SE-maximization-weighted-MMSE-subproblem} \\
		&\hspace{0cm}\textrm{subject to} \quad \vect{a}_r^{\Ttran}\vect{p}\leq p_{\textrm{max}}, \quad r=1,\ldots,R  \nonumber
		\end{align}
		\State Update $\vect{p}$ by the obtained solution to \eqref{eq:sum-SE-maximization-weighted-MMSE-subproblem}
		\EndWhile
		\State {\bf Output:} $\vect{p}$
	\end{algorithmic}
\end{algorithm}

When implementing Algorithm~\ref{alg:sum-SE}, the subproblem in \eqref{eq:sum-SE-maximization-weighted-MMSE-subproblem} is required to be solved optimally with respect to $\vect{p}$ (when the other variables are kept constant), otherwise, the block coordinate descent algorithm will not converge to a local optimum. Fortunately, the problem in \eqref{eq:sum-SE-maximization-weighted-MMSE-subproblem} is convex for the power allocation problems we consider in Section~\ref{c-resource-allocation} on p.~\pageref{c-resource-allocation} and the optimal solution can, thus, be obtained by any of the standard algorithms from convex optimization theory \cite{Boyd2004a}.
We elaborate further on this in Remark~\ref{remark:convex-solvers} on p.~\pageref{remark:convex-solvers}.

\newpage

\section{Summary of the Key Points in Section~\ref{c-theoretical-foundations}}
\label{c-theoretical-foundations-summary}

\begin{mdframed}[style=GrayFrame]

\begin{itemize}
\item The realization of a complex Gaussian distributed random variable can be estimated using the \gls{MMSE} estimator presented in Lemma~\ref{lemma:MMSE-estimator}. It will be used in later sections for channel estimation.

\item The channel capacity is a suitable performance metric in broadband applications and will be utilized throughout this monograph. The exact capacity is hard to compute in situations with interference or limited channel knowledge, but there are lower bounds that can be utilized and achieved in practice. Lemma~\ref{lemma:capacity-memoryless-deterministic-interference} presents a lower bound for deterministic channels and Lemma~\ref{lemma:capacity-memoryless-random-interference} presents a lower bound for fading channels. Both will be utilized in later sections.

\item A lower bound on the capacity is called an achievable SE if there is a known way to achieve it. The capacity lower bounds presented in this section will be called SEs in later sections and are achievable using channel codes designed for AWGN channels.

\item A commonly appearing problem in multi-antenna communications is that of maximizing a power ratio between the desired signal and interference plus noise. This problem can be formalized as a generalized Rayleigh quotient. Lemma~\ref{lemma:gen-rayleigh-quotient} shows how to maximize it and what its maximum value is.

\item The $K$ SEs of the UEs can be combined into a single scalar utility function that can be maximized, for example, with respect to the transmit power coefficients. Two examples of utility functions are max-min fairness and sum SE. The corresponding utility maximization problems are solved by Algorithm~\ref{alg:bisection} and Algorithm~\ref{alg:sum-SE}, respectively.

\end{itemize}

\end{mdframed}


\chapter{Channel Estimation}
\label{c-channel-estimation}

This section describes how channel estimation is carried out in User-centric Cell-free Massive \gls{MIMO} systems based on the transmission of uplink pilots. The system model during pilot transmission is defined in Section~\ref{sec:uplink-pilot}. The \gls{MMSE} channel estimation scheme is presented in Section~\ref{sec:MMSE-channel-estimation}.  The impact of the cell-free architecture, interference from pilot-sharing UEs, and spatial correlation in the estimation process is evaluated in Section~\ref{sec:impact-on-estimation-accuracy}. 
A basic algorithm for pilot assignment and dynamic cooperation cluster formation is described in Section~\ref{sec:pilot-assignment}.
The key points are summarized in Section~\ref{c-channel-estimation-summary}.

\enlargethispage{\baselineskip} 

\section{Uplink Pilot Transmission}
\label{sec:uplink-pilot}

We consider the cell-free network defined in Section~\ref{c-cell-free-definition} on p.~\pageref{c-cell-free-definition} where $L$ \glspl{AP}, each equipped with $N$ antennas, are arbitrarily distributed over the coverage area. The \glspl{AP} are connected to CPUs via fronthaul links. These can be used for sharing the \gls{CSI} and channel statistics  needed to decode the uplink data signals and precode the downlink data.

\gls{UE} $k$ is served only by the APs with indices in the set $\mathcal{M}_k \subset \{1,\ldots,L\}$, which is assumed fixed and known wherever needed. 
To perform coherent transmission processing (both between the antennas at each AP and across the cooperating APs), knowledge of the channel responses from the UEs to the serving APs is required. It is particularly important for AP $l$ to have estimates of the channel vector ${\bf h}_{kl}$ from UE $k$ if $l \in \mathcal{M}_k$. Since the channels are assumed to be constant throughout one coherence block and change independently from one block to another, following the block fading model described in Section~\ref{sec:block-fading} on p.~\pageref{sec:block-fading}, they need to be estimated once per each coherence block. According to the TDD protocol defined in Section~\ref{sec:tdd-protocol} on p.~\pageref{sec:tdd-protocol}, $\tau_p$ samples are reserved for uplink pilot signaling in each coherence block. Therefore, each UE can transmit a pilot sequence that spans these $\tau_p$ samples and each AP can use the received signals to estimate the channel.

Ideally, we would like every UE to use a pilot sequence that is orthogonal to the pilots of all other UEs so there is no interference between the transmissions.
However, since the pilots are $\taupu$-dimensional vectors, we can only find a set of at most $\taupu$ mutually orthogonal sequences (i.e., a set of vectors that forms an orthogonal basis that spans $\mathbb{C}^{\tau_p}$). 
The finite length of the coherence blocks imposes the constraint $\taupu\leq \tau_c$ that makes it impossible to assign mutually orthogonal pilots to all \glspl{UE} in practical networks with $K \gg \tau_c$. 
Hence, we instead assume the network utilizes a set of $\tau_p$ mutually orthogonal pilot sequences $\boldsymbol{\phi}_{1},\ldots,\boldsymbol{\phi}_{\tau_p}\in\mathbb{C}^{\tau_p}$ that are designed to satisfy $\| \boldsymbol{\phi}_{t} \|^2 = \tau_p$, for $t=1,\ldots,\tau_p$. One option is to use the columns of $\sqrt{\taupu} \vect{I}_{\taupu}$ as pilot sequences.
We refer to \cite[Sec.~3.1.1]{massivemimobook} for further examples of how to select pilot sequences. The only important aspect in the context of this monograph is that
\begin{equation}
\boldsymbol{\phi}_{t_1} ^{\Htran} \boldsymbol{\phi}_{t_2} = \begin{cases} \tau_p & t_1 = t_2\\
0 & t_1 \neq t_2. \end{cases}
\end{equation}
These $\taupu$ pilots are assigned to the UEs in a deterministic way, for example, when they get access to the network. We return to the pilot assignment problem in Section~\ref{sec:pilot-assignment}. 

More than one UE might be assigned to each pilot sequence. We denote  the index of the pilot assigned to UE $k$ as $t_k \in \{ 1, \ldots, \tau_p\}$ and define
\begin{equation}
\mathcal{P}_k = \left\{ i : t_i = t_k, i=1,\ldots,K  \right\} \subset \{ 1, \ldots, K\} 
\end{equation}
as the set of UEs that use the same pilot as UE $k$, including itself. The elements of $\boldsymbol{\phi}_{t_k}$ are scaled by the square-root of the uplink transmit power of UE $k$, which we denote as $\eta_k \geq 0 $ in the pilot transmission phase.
The elements of $\sqrt{\eta_k}\boldsymbol{\phi}_{t_k}$ are transmitted as the signal $s_k$ in \eqref{eq:DCC-uplink} over $\tau_p$ consecutive samples.
The received signal at \gls{AP} $l$ during the entire pilot transmission is denoted as
$\vect{Y}_{l}^{{\rm pilot}} \in \mathbb{C}^{N \times \tau_p}$ and
 is given by
\begin{equation} \label{eq:received-pilot-matrix}
\vect{Y}_{l}^{{\rm pilot}} = \sum_{i=1}^{K} \sqrt{\eta_i } \vect{h}_{il} \boldsymbol{\phi}_{t_i}^{\Ttran}+ \vect{N}_{l}
\end{equation}
where $\vect{N}_{l}  \in \mathbb{C}^{N \times \tau_p}$ is the receiver noise with \gls{iid} elements distributed as $ \CN (0, \sigmaUL )$. The received uplink signal $\vect{Y}_{l}^{{\rm pilot}}$ is the observation that AP $l$ can use to estimate the channels $\{\vect{h}_{kl}:   l\in \mathcal{M}_k\}$ to all the UEs that it serves.
The estimation can either be carried out directly at \gls{AP} $l$ or  be delegated to the \gls{CPU}. In the latter case, \gls{AP} $l$ acts as a relay and sends the received pilot signal $\vect{Y}_{l}^{{\rm pilot}}$ to the \gls{CPU} via the fronthaul link. The CPU can in principle compute all the channel estimates $\{\widehat{\vect{h}}_{kl} : k=1,\ldots,K, l\in \mathcal{M}_k\}$ using the received pilot signals $\{\vect{Y}_{l}^{{\rm pilot}}:  l\in \mathcal{M}_k\}$. However, since the channel vectors are independent, there is no loss of optimality if the channel estimates are computed separately at each AP $l$ in the set $\mathcal{M}_k$. 
The estimation results presented in this section apply to both cases.

\section{MMSE Channel Estimation}\label{sec:MMSE-channel-estimation}
Suppose we want to estimate $\vect{h}_{kl}$ (either at AP $l$ or at the CPU), based on the received pilot signal $\vect{Y}_{l}^{{\rm pilot}}$ in \eqref{eq:received-pilot-matrix}. The first step is to remove the interference from UEs using orthogonal pilots by multiplying the received signal $\vect{Y}_{l}^{{\rm pilot}}$ with the normalized conjugate of the associated pilot $\boldsymbol{\phi}_{t_k}$. This yields $\vect{y}_{t_kl}^{{\rm pilot}} =\vect{Y}_{l}^{{\rm pilot}}  \boldsymbol{\phi}_{t_k}^{\star}/{\sqrt{\tau_p}} \in \mathbb{C}^N$, given by
\begin{align} \notag
\vect{y}_{t_kl}^{{\rm pilot}} & = 
\sum_{i=1}^{K}  \frac{\sqrt{\eta_i }}{\sqrt{\tau_p}}\vect{h}_{il} \boldsymbol{\phi}_{t_i}^{\Ttran}  \boldsymbol{\phi}_{t_k}^{\star}+ \frac{1}{\sqrt{\tau_p}} \vect{N}_{l}  \boldsymbol{\phi}_{t_k}^{\star}  \\
& = \underbrace{ \vphantom{\sum_{i \in \mathcal{P}_k\setminus \{k \}}} \sqrt{\eta_k \tau_p } \vect{h}_{kl}}_{\text{Desired part}} + \underbrace{\sum_{i \in \mathcal{P}_k\setminus \{k \}} \sqrt{\eta_i \tau_p } \vect{h}_{il}}_{\text{Interference}} + \underbrace{\vphantom{\sum_{i \in \mathcal{P}_k\setminus \{k \}}} \vect{n}_{t_k l}}_{\text{Noise}}
\label{eq:received-pilot}
\end{align}
where the first term contains the desired channel $ \vect{h}_{kl}$ scaled by $\sqrt{\eta_k \tau_p }$, which is the square root of the total pilot power $\| \sqrt{\eta_k}\boldsymbol{\phi}_{t_k} \|^2 = \eta_k \tau_p$. The second term is the interference generated by the pilot-sharing UEs and the third term is the noise
$\vect{n}_{t_k l} =  \frac{1}{\sqrt{\tau_p}} \vect{N}_{l}  \boldsymbol{\phi}_{t_k}^{\star} \sim \CN (\vect{0}_N, \sigmaUL \vect{I}_N)$. The noise term has independent elements with variance $\sigmaUL$ since $\vect{N}_l$ has \gls{iid} $ \CN (0, \sigmaUL )$-elements  and $\boldsymbol{\phi}_{t_k}^{\star}/ \sqrt{\tau_p}$ is a vector with unit norm. 

Note that $\vect{y}_{t_kl}^{{\rm pilot}}$ is sufficient statistics for the estimation of $\vect{h}_{kl}$ since no information  about $\vect{h}_{kl}$  or the pilot-sharing UEs' channels is lost  by multiplying $\vect{Y}_{l}^{{\rm pilot}}$ with $\boldsymbol{\phi}_{t_k}^{\star}/{\sqrt{\tau_p}}$. This is a positive consequence of using mutually orthogonal pilots; if a larger set of partially overlapping pilots would be used instead, then many more UEs are \emph{partially} sharing the same pilot dimensions and this would make the computation of the MMSE estimator more computationally involved.
In our case, the resulting signal $\vect{y}_{t_kl}^{{\rm pilot}}$ in \eqref{eq:received-pilot} matches with the MMSE estimation structure in Lemma~\ref{lemma:MMSE-estimator} on p.~\pageref{lemma:MMSE-estimator}, leading to the following result.

\enlargethispage{\baselineskip} 

\begin{corollary} \label{cor:MMSE-estimation}
The \gls{MMSE} estimate of $\vect{h}_{kl}$ based on $\vect{y}_{t_kl}^{{\rm pilot}}$ is
\begin{equation} \label{eq:estimates}
\widehat{\vect{h}}_{kl} = \sqrt{\eta_k \tau_p} \vect{R}_{kl} \vect{\Psi}_{t_kl}^{-1} \vect{y}_{t_kl}^{{\rm pilot}}  
\end{equation}
where
\begin{equation} \label{eq:Psitl}
\vect{\Psi}_{t_kl} = \mathbb{E} \left\{ \vect{y}_{t_kl}^{{\rm pilot}} {(\vect{y}_{t_kl}^{{\rm pilot}})}^{\Htran} \right\} = \sum_{i \in \mathcal{P}_k} \eta_i\tau_p \vect{R}_{il} + \sigmaUL \vect{I}_{N}
\end{equation}
is the correlation matrix of the received signal in \eqref{eq:received-pilot}.
The estimate $\widehat{\vect{h}}_{kl}$ and estimation error $\widetilde{\vect{h}}_{kl} = \vect{h}_{kl} - \widehat{\vect{h}}_{kl}$ are independent random variables distributed as
\begin{align} \label{eq:estimate-distribution}
\widehat{\vect{h}}_{kl} &\sim \CN \left( \vect{0}_N, \eta_k \tau_p \vect{R}_{kl} \vect{\Psi}_{t_kl}^{-1} \vect{R}_{kl} \right) \\
\widetilde{\vect{h}}_{kl} &\sim \CN(\vect{0}_N,\vect{C}_{kl})
\end{align}
with the error correlation matrix
\begin{equation} \label{eq:error-correlation}
\vect{C}_{kl} = \mathbb{E} \{ \widetilde{\vect{h}}_{kl} \widetilde{\vect{h}}_{kl}^{\Htran} \}= \vect{R}_{kl} - \eta_k \tau_p \vect{R}_{kl} \vect{\Psi}_{t_kl}^{-1} \vect{R}_{kl}.
\end{equation} 
\end{corollary}
\begin{proof}
We consider the received signal in \eqref{eq:received-pilot} and define $q = \sqrt{\eta_k \tau_p} $, ${\bf n} = \sum_{i \in \mathcal{P}_k\setminus \{k \}} \sqrt{\eta_i \tau_p } \vect{h}_{il} + \vect{n}_{t_k l}$, ${\bf R} = {\bf R}_{kl}$ and ${\bf S} = \sum_{i \in \mathcal{P}_k\setminus \{k \}}\eta_i\tau_p \vect{R}_{il} + \sigmaUL \vect{I}_{N}$. The MMSE estimate and its properties then follow directly from Lemma~\ref{lemma:MMSE-estimator} on p.~\pageref{lemma:MMSE-estimator}.
\end{proof}

The  \gls{MMSE} estimator has this name because it minimizes the MSE $\mathbb{E}\{ \| \vect{h}_{kl} - \widehat{\vect{h}}_{kl} \|^2 \}=\mathbb{E}\{ \| \widetilde{\vect{h}}_{kl} \|^2 \}$ among all possible estimators.
The \gls{MMSE} estimator in \eqref{eq:estimates} is linear in the sense that $\widehat{\vect{h}}_{kl}$ is computed by multiplying the processed received signal $\vect{y}_{t_kl}^{{\rm pilot}}$ with matrices. It is therefore sometimes
called the linear MMSE estimator. However, we prefer to use the pure MMSE notion to make it clear that one cannot further reduce the MSE by using a non-linear estimator. Note that
\begin{equation}
\eta_k \tau_p \vect{R}_{kl} \vect{\Psi}_{t_kl}^{-1} \vect{R}_{kl}  = \vect{R}_{kl} - \vect{C}_{kl}
\end{equation}
so the correlation matrix of the channel estimate $\widehat{\vect{h}}_{kl} $ can also be expressed as $\vect{R}_{kl} - \vect{C}_{kl}$. The distribution of $\widehat{\vect{h}}_{kl}$ in \eqref{eq:estimate-distribution} allows us to generate random realizations of the channel estimate without having to compute the received signal $\vect{y}_{t_kl}^{{\rm pilot}}$ as an intermediate step. It is convenient for both mathematical analysis and numerical simulations.

In practice, the computation of $\widehat{\vect{h}}_{kl}$ in \eqref{eq:estimates} requires knowledge of two matrices: 
\begin{enumerate}
\item The spatial correlation matrix $\vect{R}_{kl}=\mathbb{E}\{  \vect{h}_{kl} \vect{h}_{kl}^{\Htran} \}$ of $\vect{h}_{kl}$; 
\item
The sum $\vect{\Psi}_{t_kl} = \sum_{i \in \mathcal{P}_k} \eta_i\tau_p \vect{R}_{il} + \sigmaUL \vect{I}_{N}$ of the correlation matrices of pilot-sharing UEs and noise.
\end{enumerate}
Both matrices depend on the statistics of the channels and thus are constant.
Therefore, we can assume that the matrix $\sqrt{\eta_k \tau_p} \vect{R}_{kl} \vect{\Psi}_{t_kl}^{-1}$ can be precomputed. This matrix can be described by $N^{2}$ complex scalars, which can be exchanged via the fronthaul links to make $\sqrt{\eta_k \tau_p} \vect{R}_{kl} \vect{\Psi}_{t_kl}^{-1}$ available wherever needed (e.g., at AP $l$ or the CPU). The required fronthaul signaling is negligible since the channel statistics are deterministic and thus constant throughout the entire data transmission.\footnote{In practice, the channel statistics will change due to UE mobility or new scheduling decisions, but measurements suggest roughly two orders of magnitude slower variations compared to the channel vectors.}
If the matrix is precomputed, computing the MMSE estimate in a given coherence block entails first computing $\vect{y}_{t_kl}^{\rm{pilot}}$ and then multiplying it with $\sqrt{\eta_k \tau_p} \vect{R}_{kl} \vect{\Psi}_{t_kl}^{-1}$ of each UE served by AP $l$.
The first operation requires $N\tau_p$ complex multiplications per pilot sequence while the second needs $N^{2}$ complex multiplications per UE \cite[Sec.~3.4]{massivemimobook}. 
In summary, the computational complexity for channel estimation at AP $l$ is
\begin{equation} \label{eq:complexity-estimation}
|\mathcal{D}_l| (N\tau_p + N^{2}) \quad \textrm{complex multiplications}
\end{equation}
per coherence block, if every UE that it serves uses a different pilot sequence (which is preferable). This value is scalable as $K \to \infty$ if $|\mathcal{D}_l| $ remains finite, which is in line with the sufficient condition for scalability presented in Lemma~\ref{lemma:D-requirement} on p.~\pageref{lemma:D-requirement}.

\begin{remark}[Alternative estimation methods]
The MMSE estimator is the optimal choice when having full statistical knowledge, as assumed in this monograph (see Section~\ref{sec:correlated-rayleigh-fading} on p.~\pageref{sec:correlated-rayleigh-fading}). Note that each spatial correlation matrix $\vect{R}_{kl}$ contains $N^2$ elements and the computational complexity in \eqref{eq:complexity-estimation} is proportional to $N^2$. 
There are alternative channel estimators in the literature that provide larger MSEs, since the MMSE estimator is optimal, but can be useful to limit the computational complexity or deal with incomplete statistical knowledge.
This can be of interest in Cellular Massive MIMO systems, where $N$ is large, since the complexity or overhead required to learn all the matrix elements can be substantial. However, none of these issues are pressing in Cell-free Massive MIMO, where $N$ is small, and will thus not be covered in this monograph.
We refer to \cite[Sec.~3.4]{massivemimobook} for details on alternative estimators and to \cite{BSD16A}, \cite{CaireC17a}, \cite{NeumannJU17}, \cite{Sanguinetti2019a}, \cite{UpadhyaJU17} for methods to estimate spatial correlation matrices in practice.
\end{remark}

\subsection{Normalized MSE}

The value of the MSE depends on the average channel gain $ \beta_{kl} $ (among other factors) and a strong channel might have larger errors in absolute terms than a weaker channel. However, it is the relative size of the error that matters, not the absolute size.
Hence, the estimation accuracy is quantified by considering the \gls{NMSE}. When it comes to the channel between AP $l$ and UE $k$, the \gls{NMSE} is defined as
\begin{align}
\mathacr{NMSE}_{kl}=\frac{\mathbb{E}\{ \| \vect{h}_{kl} - \widehat{\vect{h}}_{kl} \|^2 \}}{ \mathbb{E}\{ \| \vect{h}_{kl}\|^2 \}} &= \frac{\tr( \vect{C}_{kl}  )}{ \tr( \vect{R}_{kl}  )} =1 - \frac{\eta_k \tau_p  \tr(\vect{R}_{kl} \vect{\Psi}_{t_kl}^{-1} \vect{R}_{kl}) }{ \tr( \vect{R}_{kl}  )} \label{eq:nmse-correlated-AP-l}
\end{align}
and measures the relative estimation error variance per
antenna.  The \gls{NMSE} is 0 when perfect estimation is achieved while it is 1 if $\eta_k=0$ so that we are only receiving noise and interference.
In general, any reasonable estimator will provide an \gls{NMSE} between 0 and 1, where smaller values are preferable.

In the case of single-antenna APs (i.e., $N=1$), the spatial correlation matrix $\vect{R}_{kl}$ reduces to the average channel gain $\beta_{kl}$ and \eqref{eq:nmse-correlated-AP-l} simplifies to 
\begin{align}
\mathacr{NMSE}_{kl}=1 - \frac{\eta_k \tau_p \beta_{kl}}{ \eta_k \tau_p \beta_{kl} + \hspace{-1cm}\underbrace{\sum_{i \in \mathcal{P}_k\setminus \{k \} }\eta_i \tau_p \beta_{il}}_{\text{Interference from pilot-sharing UEs}} \hspace{-1cm} + \,\sigmaUL} \label{eq:nmse-uncorrelated-AP-l}
\end{align}
which can be rewritten as
\begin{align}
\mathacr{NMSE}_{kl}=1 - \frac{\mathacr{SNR}_{kl}^{\text{pilot}}}{ \mathacr{SNR}_{kl}^{\text{pilot}} +\sum\limits_{i \in \mathcal{P}_k\setminus \{ k \}}\mathacr{SNR}_{il}^{\text{pilot}}  + 1} \label{eq:nmse-uncorrelated-AP-l-1}
\end{align}
where 
\begin{align}
\mathacr{SNR}_{kl}^{\text{pilot}}= \frac{\eta_k\tau_p \beta_{kl}}{\sigmaUL} \label{eq:effective-pilot-SNR}
\end{align}
is the \emph{effective} SNR of the pilot transmission from UE $k$ to AP $l$. The word ``effective'' implies that the pilot processing gain $\tau_p$ is included in the SNR expression, which is achieved since the receiver processing in \eqref{eq:received-pilot} coherently captures all the transmitted pilot energy without increasing the noise. If the pilot sequences are $\tau_p = 10$ samples long, then the effective SNR is $10$\,dB larger than the nominal SNR at a single sample. This gain is highly desirable for achieving good estimation quality also for UEs with limited transmit power and/or weak channel conditions.

Since it is the total pilot energy $\eta_k\tau_p$ that determines $\mathacr{SNR}_{kl}^{\text{pilot}}$, in practice, one can choose between spreading it over the $\tau_p$ samples of the pilot sequence or concentrating it into only one of them. The former solution keeps the peak-to-average-power ratio low but is sensitive to hardware imperfections; for example, a UE might only be able to approximately generate its pilot sequence, leading to only approximate orthogonality between the sequences that are meant to be orthogonal. This problem does not occur in the latter case since UEs with orthogonal pilots are transmitting at different samples of the coherence block. On the other hand, the concentration of pilot energy into specific samples corresponds to boosting their power, which is not possible in all implementations since it can increase the peak-to-average-power ratio. Generally speaking, the system can improve the effective SNR of the pilot transmission by either increasing the pilot length or boosting the pilot power on specific samples.

\subsection{Pilot Contamination}

From \eqref{eq:nmse-uncorrelated-AP-l}, it follows that the interference generated by the pilot-sharing UEs increases the NMSE, thereby reducing the channel estimation quality. In the case of multiple-antenna APs (i.e., $N > 1$), the NMSE in \eqref{eq:nmse-correlated-AP-l} is exactly the same in \eqref{eq:nmse-uncorrelated-AP-l} if uncorrelated Rayleigh fading is assumed (i.e., $\vect{R}_{kl}=\beta_{kl}\vect{I}_{N}$). This means that there is neither an advantage nor a disadvantage in using multiple antennas at the APs in the absence of spatial correlation. With spatially correlated channels, the NMSE in \eqref{eq:nmse-correlated-AP-l} has a more complicated structure. It is clearly increased by the interference from the pilot-sharing UEs, which enters into $\vect{\Psi}_{t_kl}$ in \eqref{eq:Psitl}. Unlike \eqref{eq:nmse-uncorrelated-AP-l}, \eqref{eq:nmse-correlated-AP-l} depends not only on the average channel gains but on the full spatial correlation matrices $\vect{R}_{il}$, for $i\in \mathcal{P}_k$. Intuitively, the NMSE should be better when the interfering UEs' channels have very different spatial correlation properties than the desired UE. For example, if we know a priori that the channels of two UEs contain multipath components distributed over widely different sets of angular directions, then their corresponding spatial correlation matrices should have different dominant eigenspaces and this should somehow be beneficial during channel estimation.
This intuition will be confirmed later on in Section~\ref{sec:impact-of-spatial-correlation}. 

The ``pilot interference'' generated by the pilot-sharing UEs is known as \emph{pilot contamination} in the Cellular Massive MIMO literature and behaves differently from the receiver noise; it does not only reduce the estimation quality, but also makes the channel estimates of pilot-sharing UEs correlated. To see this, let us compare the MMSE estimate $\widehat{\vect{h}}_{kl}$ in \eqref{eq:estimates} of UE $k$ with the estimate 
\begin{equation} \label{eq:estimates_il-first}
\widehat{\vect{h}}_{il} =  \sqrt{\eta_i \tau_p} \vect{R}_{il} \vect{\Psi}_{t_i l}^{-1} \vect{y}_{t_i l}^{{\rm pilot}}
\end{equation}
of another UE $i \in \mathcal{P}_{k}\setminus \{ k \}$ using the same pilot. In this case, we have that $\Psiv_{t_kl} = \Psiv_{t_il}$ and 
$\vect{y}_{t_kl}^{{\rm pilot}} = \vect{y}_{t_il}^{{\rm pilot}} $ such that \eqref{eq:estimates_il-first} can be rewritten as
\begin{align}\label{eq:estimates_il}
\widehat{\vect{h}}_{il} = \sqrt{\eta_i \tau_p} \vect{R}_{il} \vect{\Psi}_{t_kl}^{-1} \vect{y}_{t_kl}^{{\rm pilot}}.
\end{align}
By comparing \eqref{eq:estimates_il} with \eqref{eq:estimates}, we notice that only the scalar and the first matrix differ. If $\vect{R}_{kl}$ is invertible, we can write the relation as
\begin{equation} \label{eq:relation-i-k-estimates}
\widehat{\vect{h}}_{il}   =\sqrt{ \frac{\eta_i}{\eta_k} }  \vect{R}_{il}\vect{R}_{kl}^{-1}   \widehat{\vect{h}}_{kl}, \quad i\in\mathcal{P}_k.
\end{equation}
This implies that the two estimates are correlated, with the cross-correlation matrix given by
\begin{equation} \label{eq:cross-correlation-channel-estimates}
\mathbb{E}\left\{\widehat{\vect{h}}_{kl}\widehat{\vect{h}}_{il}^{\Htran}\right\}=\sqrt{\eta_k\eta_i}\tau_p\vect{R}_{kl}\vect{\Psi}_{t_kl}^{-1}\vect{R}_{il}, \quad i\in\mathcal{P}_k.
\end{equation}  
This happens although the true channels were assumed statistically independent (i.e., $
\mathbb{E}\left\{{\vect{h}}_{kl}{\vect{h}}_{il}^{\Htran}\right\} = {\bf 0}_{N\times N}$, for $i\neq k$). Observe that if there is no spatial correlation so that $\vect{R}_{kl}=\beta_{kl}\vect{I}_{N}$ and $\vect{R}_{il}=\beta_{il}\vect{I}_{N}$, then the channel estimate $\widehat{\vect{h}}_{il}$ in \eqref{eq:estimates_il} for any $i\in \mathcal{P}_k \setminus \{ k \}$ will be the same as the channel estimate $\widehat{\vect{h}}_{kl}$ except for a scaling factor. In this case, the channel estimates are fully correlated. 

In conclusion, UEs that transmit the same pilot sequence contaminate each others' channel estimates, in the same way as in Cellular Massive MIMO systems \cite{Marzetta2010a}, \cite{Sanguinetti2019a} and other types of cellular systems \cite{Niemela2004a}, \cite{Sapienza2001a}. The contamination not only reduces the estimation quality (i.e., increases the NMSE) but also makes the channel estimates statistically dependent~\cite{Sanguinetti2019a}. This has an important impact beyond channel estimation, since the contamination makes it harder to mitigate interference between UEs that use the same pilot in uplink and downlink. This will be investigated in Sections~\ref{c-uplink} and~\ref{c-downlink} on p.~\pageref{c-uplink} and p.~\pageref{c-downlink}, respectively.

\subsection{MMSE Estimation of the Collective Channel}\label{sec:mmse-of-collective-channel}

So far, we have considered the estimation of a channel vector $\vect{h}_{kl}$ between a generic AP $l$ and UE $k$. To perform coherent processing of the signals at multiple APs, it is necessary to have knowledge of the collective channel $\vect{h}_k \sim \CN(\vect{0}_M, \vect{R}_{k}) $ defined in \eqref{eq:collective_channel}, where $\vect{R}_{k} = \diag(\vect{R}_{k1}, \ldots,  \vect{R}_{kL}) \in \mathbb{C}^{M \times M}$
is the block-diagonal collective spatial correlation matrix. However, the estimate $\widehat{\vect{h}}_{k}  = [\widehat{\vect{h}}_{k1}^{\Ttran} \, \ldots \, \widehat{\vect{h}}_{kL}^{\Ttran}]^{\Ttran}$ of $\vect{h}_k$ from all APs to UE $k$ can be only partially computed since only the APs in $\mathcal{M}_k \subset\{1,\ldots,L\}$ are computing estimates of the channels and/or send their pilot signals to the CPU. This means that only the following partial channel estimate is known in a scalable cell-free network:
\begin{equation} \label{eq:estimates-central}
\vect{D}_{k}\widehat{\vect{h}}_{k} \triangleq \begin{bmatrix} \vect{D}_{k1}\widehat{\vect{h}}_{k1} \\ \vdots \\ \vect{D}_{kL}\widehat{\vect{h}}_{kL} 
\end{bmatrix} \sim \CN \left( \vect{0}_M, \eta_k \tau_p \vect{D}_{k}\vect{R}_{k} \vect{\Psi}_{t_k}^{-1} \vect{R}_{k}\vect{D}_{k} \right)
\end{equation}
where $\vect{D}_{k}  = \diag( \vect{D}_{k1} , \ldots, \vect{D}_{kL} )$ is a block-diagonal matrix with $\vect{D}_{kl}$ being defined in~\eqref{eq:user-centric} as $\vect{I}_N$ for APs that serve UE $k$ and zero otherwise.
Moreover,  $\vect{\Psi}_{t_k}^{-1} = \diag( \vect{\Psi}_{t_k1}^{-1}, \ldots, \vect{\Psi}_{t_kL}^{-1})$ contains the inverses of the received pilot signal correlation matrices defined in \eqref{eq:Psitl}.
 The estimation error is $\vect{D}_{k}\widetilde{\vect{h}}_k = \vect{D}_{k}\vect{h}_k - \vect{D}_{k}\widehat{\vect{h}}_{k} \sim \CN(\vect{0}_M,\vect{D}_{k}\vect{C}_k)$ with
$\vect{C}_k = \diag(\vect{C}_{k1}, \ldots, \vect{C}_{kL})$. Note that $\vect{D}_{k}\vect{C}_k = \vect{D}_{k}\vect{C}_k \vect{D}_k$ so we use the former expression for brevity.

The \gls{NMSE} of $\vect{D}_{k}\widehat{\vect{h}}_{k}$ can be computed as \begin{align}\notag
\mathacr{NMSE}_{k} &=\frac{\mathbb{E}\{ \| \vect{D}_{k}\vect{h}_{k} - \vect{D}_{k}\widehat{\vect{h}}_{k} \|^2 \}}{ \mathbb{E}\{ \| \vect{D}_{k}\vect{h}_{k}\|^2 \}} = \frac{\tr( \vect{D}_{k}\vect{C}_{k}  )}{ \tr( \vect{D}_{k}\vect{R}_{k}  )}\mathop{=}^{(a)}\frac{\sum\limits_{l=1}^L\tr( \vect{D}_{kl}\vect{C}_{kl}  )}{ \sum\limits_{l=1}^L\tr( \vect{D}_{kl}\vect{R}_{kl}  )}\\&=1 - \frac{\eta_k \tau_p \sum\limits_{l \in \mathcal{M}_k} \tr(\vect{R}_{kl} \vect{\Psi}_{t_kl}^{-1} \vect{R}_{kl}) }{ \sum\limits_{l \in \mathcal{M}_k}\tr( \vect{R}_{kl}  )} \label{eq:nmse-correlated-collective}
\end{align}
where ${(a)}$ follows from the block-diagonal structure of both $\vect{C}_{k}$ and $\vect{R}_{k}$. The NMSE of the collective channel in \eqref{eq:nmse-correlated-collective} has a similar form as the NMSE of an individual AP-to-UE channel in \eqref{eq:nmse-correlated-AP-l}. Note that \eqref{eq:nmse-correlated-collective} is not the sum or average of the individual NMSEs, but contains a summation of MSEs in the numerator normalized by a summation of channel gains.
\pagebreak

In the case of single-antenna APs or uncorrelated Rayleigh fading, \eqref{eq:nmse-correlated-collective} can be simplified as
\begin{align}\notag
\mathacr{NMSE}_{k}&=1 - \frac{\eta_k \tau_p \sum\limits_{l \in \mathcal{M}_k} \frac{\beta_{kl}^2}{{ \eta_k\tau_p  \beta_{kl} +\sum\limits_{i \in \mathcal{P}_k\setminus \{ k \}}\eta_i\tau_p  \beta_{il} + \sigmaUL}} }{ \sum\limits_{l \in \mathcal{M}_k}\beta_{kl} }\\&=
1 - \frac{\sum\limits_{l \in \mathcal{M}_k}\frac{(\mathacr{SNR}_{kl}^{\text{pilot}})^2}{ \mathacr{SNR}_{kl}^{\text{pilot}} + \sum\limits_{i \in \mathcal{P}_k\setminus \{k\}} \mathacr{SNR}_{il}^{\text{pilot}} +1}}{ \sum\limits_{l \in \mathcal{M}_k} \mathacr{SNR}_{kl}^{\text{pilot}}}. \label{eq:nmse-uncorrelated-collective}
\end{align}
In any case, the presence of pilot-sharing UEs will increase the NMSE and create a correlation between their respective channel estimates.

\section{Impact of Architecture, Contamination, \& Spatial Correlation}\label{sec:impact-on-estimation-accuracy}

We will exemplify how the estimation accuracy of the MMSE estimator is affected by the cell-free architecture, pilot contamination, and spatial correlation. The numerical examples will uncover some of the basic properties that determine the NMSE and its impact on communication performance. 

\enlargethispage{\baselineskip}

\subsection{Impact of the Cell-Free Architecture}

To gain insights into the basic impact of channel estimation errors in cell-free networks, we begin by considering  single-antenna APs and the estimation of the collective channel of an arbitrary UE $k$ that has a unique pilot sequence: $\mathcal{P}_k = \{k\}$. In this case, the NMSE in \eqref{eq:nmse-uncorrelated-collective} reduces to
\begin{align}
\mathacr{NMSE}_{k}=1 - \frac{\sum\limits_{l \in \mathcal{M}_k} \frac{ \mathacr{SNR}_{kl}^{\text{pilot}}}{1+\frac{1}{\mathacr{SNR}_{kl}^{\text{pilot}}}}}{\sum\limits_{l \in \mathcal{M}_k}  \mathacr{SNR}_{kl}^{\text{pilot}}}.\label{eq:nmse-uncorrelated-unique-pilot}
\end{align}
If one AP has a much larger value of its effective SNR than the other APs in $\mathcal{M}_k$, then \eqref{eq:nmse-uncorrelated-unique-pilot} can be approximated as
\begin{align}
\mathacr{NMSE}_{k}\approx 1 - \frac{1}{1+\frac{1}{\mathacr{SNR}_{kl_{\max} }^{\text{pilot}}}}
 = \frac{1}{\mathacr{SNR}_{kl_{\max} }^{\text{pilot}} +1}
\label{eq:nmse-uncorrelated-unique-pilot-unique-AP}
\end{align}
with $l_{\max} = \arg \max_{l \in \mathcal{M}_k} \mathacr{SNR}_{kl}^{\text{pilot}}$. This approximation is exact if
$|\mathcal{M}_k|=1$ or when all the serving APs have the same value of $\mathacr{SNR}_{kl}^{\text{pilot}}$. However, when the APs have different values of $\mathacr{SNR}_{kl}^{\text{pilot}}$, then the exact NMSE in \eqref{eq:nmse-uncorrelated-unique-pilot} will be larger than the approximation in \eqref{eq:nmse-uncorrelated-unique-pilot-unique-AP}.

\begin{figure}[t!]
	\begin{overpic}[width=\columnwidth,tics=10]{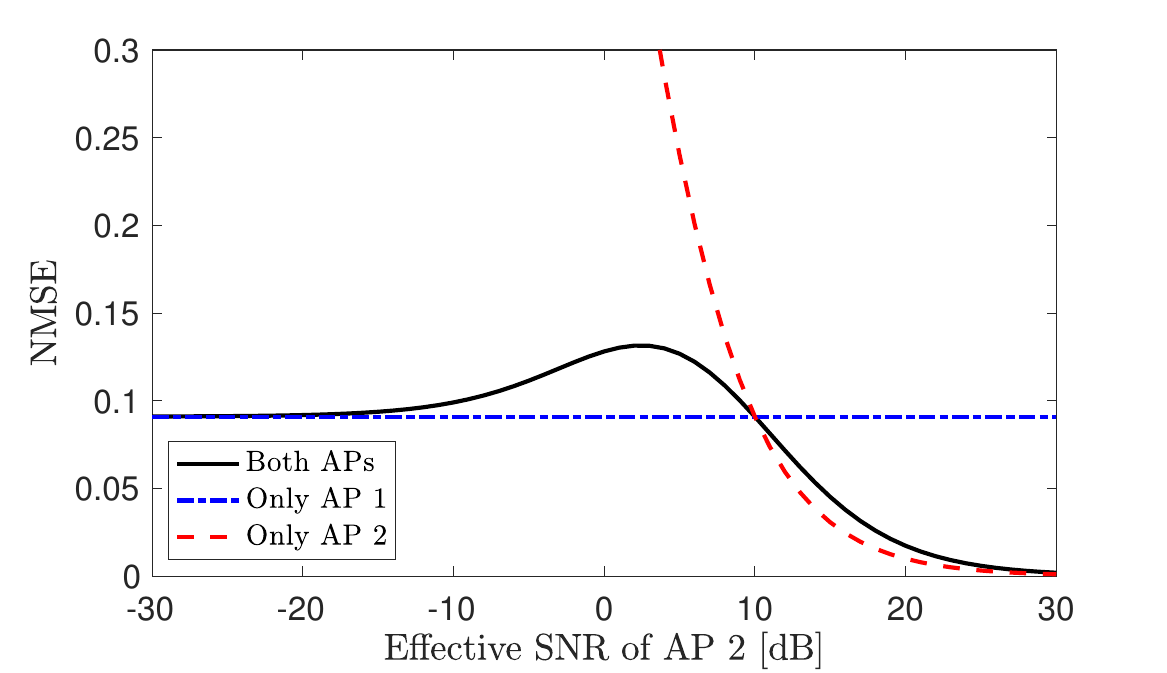}
\end{overpic} 
                \caption{NMSE when two APs may serve UE $k$. We assume that $\mathacr{SNR}_{k1}^{\text{pilot}}$ is fixed to $10$\,dB while $\mathacr{SNR}_{k2}^{\text{pilot}}$ varies from $-30$\,dB to $30$\,dB. The cases in which only AP $1$ or $2$ is serving UE $k$ is reported for comparison.}    \vspace{+2mm}
	\label{fig:basic-NMSE-analysis-twoAPs} 
\end{figure}

This fact is exemplified in Figure~\ref{fig:basic-NMSE-analysis-twoAPs} where \eqref{eq:nmse-uncorrelated-unique-pilot} is plotted for the case in which UE $k$ is served by $|\mathcal{M}_k|=2$ APs, or just one of them. We assume that $\mathacr{SNR}_{k1}^{\text{pilot}}$ is fixed to $10$\,dB while we let $\mathacr{SNR}_{k2}^{\text{pilot}}$ vary from $-30$\,dB to $30$\,dB.  There is one NMSE curve for the case when both APs serve the UE and two curves for the case when only one of the APs serves the UE.
We observe that the NMSE with both APs is higher than the NMSE with only AP $1$ serving the UE if $\mathacr{SNR}_{k2}^{\text{pilot}} < \mathacr{SNR}_{k1}^{\text{pilot}}$. In contrast, a lower NMSE is obtained for $\mathacr{SNR}_{k2}^{\text{pilot}} > \mathacr{SNR}_{k1}^{\text{pilot}}$. This shows that having more than one serving AP is not necessarily improving the estimation accuracy of the collective channel.

To gain further insights into this, we revisit the simulation scenario from Figure~\ref{fig:section1_figure9} on p.~\pageref{fig:section1_figure9}, where a cell-free network was compared~with a corresponding small-cell setup and a single-cell Massive MIMO setup. In the cell-free case, we consider $L=64$ single-antenna APs that are deployed on a square grid in a coverage area of $400$\,m$\,\times\, 400$\,m as shown in Figure~\ref{fig:basic-comparison-cellular-cellfree}(c) on p.~\pageref{fig:basic-comparison-cellular-cellfree}. The bandwidth is 10\,MHz and the noise power is $\sigmaUL = -96$\,dBm. The channels are Rayleigh fading, the channel gains are modelled as in \eqref{eq:large-scale-model-simple}, and the propagation distances are computed assuming that the \glspl{AP} are deployed 10\,m above the \gls{UE}. Comparisons are made with a single-cell setup with a 64-antenna Massive MIMO AP (with uncorrelated Rayleigh fading) and with a cellular network consisting of $64$ small cells deployed at the same $64$ AP locations.

\begin{figure}[t!]
	\begin{overpic}[width=\columnwidth,tics=10]{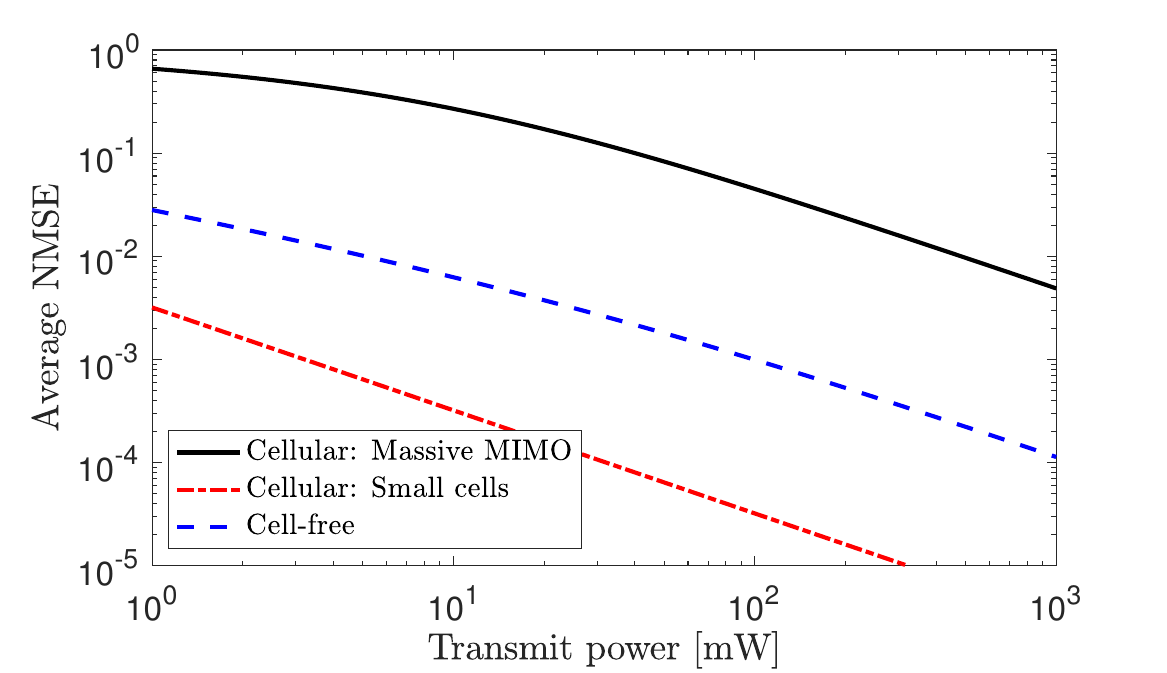}
\end{overpic} 
                \caption{Average NMSE for a single UE in the three different setups illustrated in Figure~\ref{fig:basic-comparison-cellular-cellfree} on p.~\pageref{fig:basic-comparison-cellular-cellfree}.}    \vspace{+2mm}
	\label{fig:basic-NMSE-comparison-cellular-cellfree} 
\end{figure}

Figure~\ref{fig:basic-NMSE-comparison-cellular-cellfree} shows the average of the NMSE in~\eqref{eq:nmse-uncorrelated-unique-pilot} for different uniformly distributed UE positions. The pilot transmit power $\eta_k$ varies from $1$\,mW to $1$\,W and the pilot length is $\tau_p = 10$. When comparing the three setups, we notice that it is preferable to have many single-antenna APs rather than a single AP with a large co-located array (as in conventional Cellular Massive MIMO). However, the small-cell network achieves better average estimation accuracy than the cell-free architecture since the AP with the lowest NMSE is always chosen in the small-cell network. This is in agreement with the results of Figure~\ref{fig:basic-NMSE-analysis-twoAPs} where the lowest NMSE is achieved when only using the AP with the best channel. However, this does not mean that the cell-free architecture will achieve a lower communication performance, but only that the varying estimation quality must be taken into account when the cell-free network is combining the received signals from multiple APs.

\enlargethispage{\baselineskip}

The SE gain of using a cell-free architecture will be quantified in later sections of this monograph, but we will give an indication of the performance gain by considering the SNR that is achieved in the data detection.
Suppose only UE $k$ is active in the network, then the local estimate in \eqref{eq:uplink-data-estimate} of the data signal $s_{k}$  can be expressed as
\begin{align} \notag
\widehat{s}_{kl} &= \underbrace{\vect{v}_{kl}^{\Htran} \widehat{\vect{h}}_{kl}s_{k}}_{\textrm{Signal over estimated channel}} + \underbrace{\vect{v}_{kl}^{\Htran} \widetilde{\vect{h}}_{kl}s_{k}}_{\textrm{Signal over unknown channel}} +  \underbrace{\vect{v}_{kl}^{\Htran}{\bf n}_{l}}_{\textrm{Noise}}
\label{eq:uplink-data-estimate-single-UE}.
\end{align} 
This is the data estimate at AP $l$.
All the local estimates of the serving APs are sent to a CPU where the final estimate of $s_k$ is obtained as
\begin{align} \notag
\widehat s_{k} &= \sum_{l\in \mathcal{M}_k} \widehat s_{kl} \\
&= \sum_{l\in \mathcal{M}_k} \vect{v}_{kl}^{\Htran} \widehat{\vect{h}}_{kl}s_{k} +  \sum_{l\in \mathcal{M}_k}\vect{v}_{kl}^{\Htran} \widetilde{\vect{h}}_{kl}s_{k} + \sum_{l\in \mathcal{M}_k} \vect{v}_{kl}^{\Htran}{\bf n}_{l}.
\end{align}
For simplicity, we assume the APs apply \gls{MR} combining with $\vect{v}_{kl} = \widehat{{\bf h}}_{kl}$ and compute the resulting SNR as
\begin{align} \notag
\mathacr{SNR}_{k}^{\text{data}} &= 
\frac{\mathbb{E} \left\{ \left| \sum\limits_{l\in \mathcal{M}_k} \! \vect{v}_{kl}^{\Htran} \widehat{\vect{h}}_{kl}s_{k} \right|^2 \right\}}{ \mathbb{E} \left\{ \left| \sum\limits_{l\in \mathcal{M}_k} \! \vect{v}_{kl}^{\Htran}{\bf n}_{l} \right|^2 \right\}}  = 
\frac{p_k \mathbb{E} \left\{ \left| \sum\limits_{l\in \mathcal{M}_k} \! \| \widehat{\vect{h}}_{kl} \|^2 \right|^2 \right\}}{ \sigmaUL \sum\limits_{l\in \mathcal{M}_k}  \mathbb{E} \left\{ \| \widehat{{\bf h}}_{kl} \|^2 \right\}} 
\\ &=
\frac{\sum\limits_{l\in \mathcal{M}_k} \! p_k  \tr \left( (\vect{R}_{kl}-\vect{C}_{kl})^2 \right)  }{ \sum\limits_{l\in \mathcal{M}_k} \! \sigmaUL  \tr(\vect{R}_{kl}-\vect{C}_{kl}) } + \sum\limits_{l\in \mathcal{M}_k} \! \! \frac{p_k \tr(\vect{R}_{kl}-\vect{C}_{kl}) }{\sigmaUL} \label{eq:SNR-simple} 
\end{align}
where the expectations with respect to the channel estimates are computed using Lemma~\ref{lemma:fourth-order-moment-Gaussian-vector} on p.~\pageref{lemma:fourth-order-moment-Gaussian-vector}. We stress that this is only one of the many ways to compute SNRs in the presence of estimation errors.

\begin{figure}[t!]
	\begin{overpic}[width=\columnwidth,tics=10]{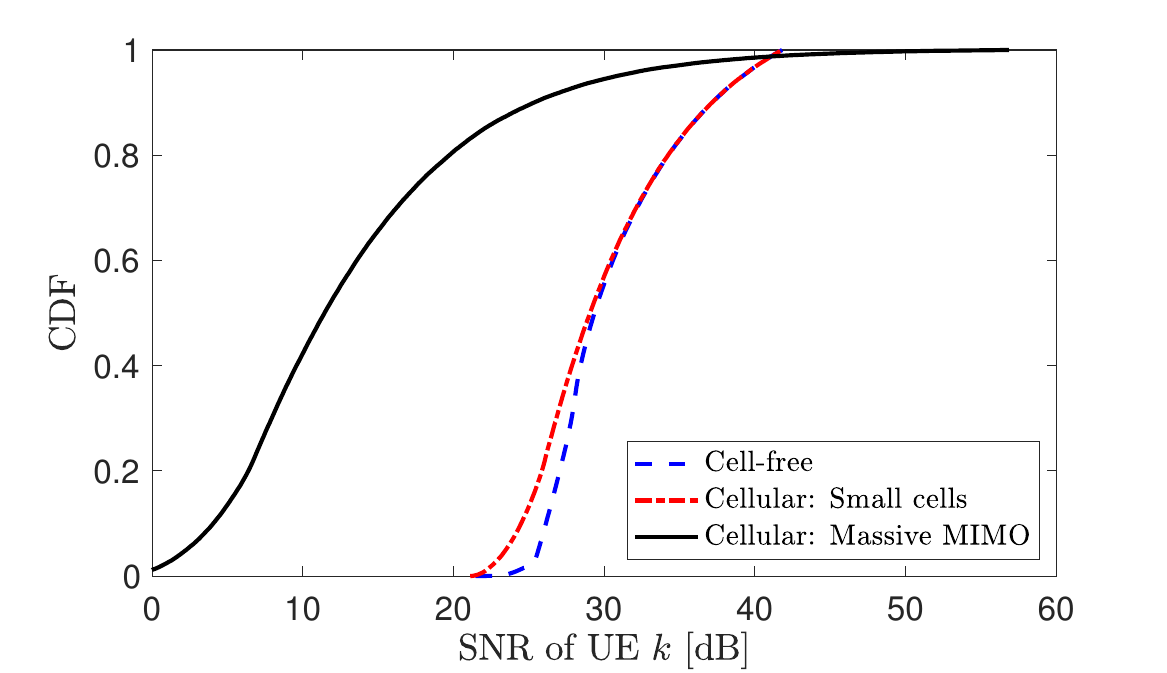}
\end{overpic} 
                \caption{CDF of the SNR~\eqref{eq:SNR-simple} achieved in the uplink by the UE with MR combining and MMSE estimation in the same three setups as in Figure~\ref{fig:basic-NMSE-comparison-cellular-cellfree}.}    \vspace{+2mm}
	\label{fig:basic-signal-gain-comparison-cellular-cellfree} 
\end{figure}

Figure~\ref{fig:basic-signal-gain-comparison-cellular-cellfree} shows the distribution of the SNR value in~\eqref{eq:SNR-simple} achieved at different random UE locations. We consider the same setups as in Figure~\ref{fig:basic-NMSE-comparison-cellular-cellfree} and let $p_k = \eta_k = 10$\,mW.
The results show that the cell-free network achieves a higher \gls{SNR} in the data detection than the corresponding small-cell setup (and Cellular Massive MIMO). This happens despite the lower average estimation quality of the channel estimates, as reported in Figure~\ref{fig:basic-NMSE-comparison-cellular-cellfree}. The reason is that the cell-free network achieves a beamforming gain by coherently processing the signals at multiple APs. 
An even larger benefit will be achieved in a  setup with multiple UEs, where the cell-free system excels at mitigating interference, as first illustrated in Section~\ref{subsec:benefit2} on p.~\pageref{subsec:benefit2}. This will be thoroughly analyzed in later sections.

\enlargethispage{\baselineskip}

\subsection{Impact of Pilot Contamination}

\begin{figure}[t!]
	\begin{overpic}[width=\columnwidth,tics=10]{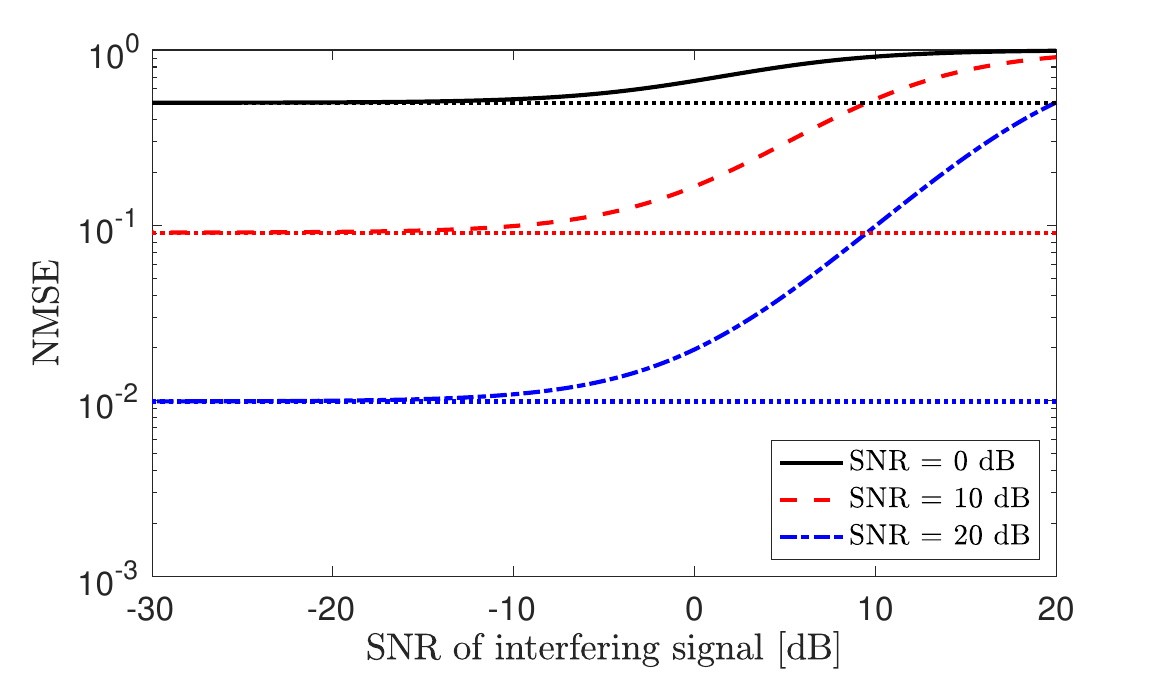}
\end{overpic} 
                \caption{NMSE as a function of the SNR ${\mathacr{SNR}_{2l}^{\text{pilot}}}$ of the interfering pilot signal. The effective SNR of the desired UE is $\mathacr{SNR}_{1l}^{\text{pilot}} =0, 10$, or $20$\,dB. The dotted lines correspond to the NMSE in the absence of pilot contamination.}    \vspace{+2mm}
	\label{fig:basic-NMSE-pilot-contamination-effect} 
\end{figure}

We now shift focus to the impact of pilot contamination. We concentrate on an arbitrary single-antenna AP $l$ and assume that it wants to \linebreak estimate the channel of UE $1$, while UE $2$ uses the same pilot. From \eqref{eq:nmse-uncorrelated-AP-l-1}, we obtain
\begin{align}
\mathacr{NMSE}_{1l}=1 - \frac{ \mathacr{SNR}_{1l}^{\text{pilot}} }{ \mathacr{SNR}_{1l}^{\text{pilot}} +\mathacr{SNR}_{2l}^{\text{pilot}} + 1 } .\label{eq:nmse-uncorrelated}
\end{align}
The NMSE increment caused by pilot contamination depends on how much the interference is affecting the term $\mathacr{SNR}_{2l}^{\text{pilot}} + 1$ in the denominator of \eqref{eq:nmse-uncorrelated}, which accounts for interference and noise. The contamination is small when the value is close to one.
Figure~\ref{fig:basic-NMSE-pilot-contamination-effect} shows the NMSE of the desired channel estimate for ${\mathacr{SNR}_{1l}^{\text{pilot}}}  \in \{0,10,20\}$\,dB  and varying values of ${\mathacr{SNR}_{2l}^{\text{pilot}}}$ from $-30$ to $20$\,dB.
The dotted lines correspond to the case without pilot contamination (i.e., ${\mathacr{SNR}_{2l}^{\text{pilot}}} =0$) and are reported as references. The NMSE increases when the interfering signal becomes stronger, particularly when the desired UE has a strong channel to the AP. However, the pilot contamination effect is negligible whenever the interfering signal is $10$\,dB weaker than the noise. We can thus conclude that each UE is only sensitive to pilot contamination from UEs that are relatively close to its serving APs. 
Another implication is that the UEs that an AP serves should preferably have different pilot sequences to limit pilot contamination \cite{Bjornson2020a}, \cite{Sabbagh2018a}. This is aligned with the Cellular Massive MIMO literature where orthogonal pilots~are normally utilized within every cell but reused in different cells \cite{massivemimobook},~\cite{Marzetta2016a}.

\begin{figure}[t!]
	\begin{overpic}[width=\columnwidth,tics=10]{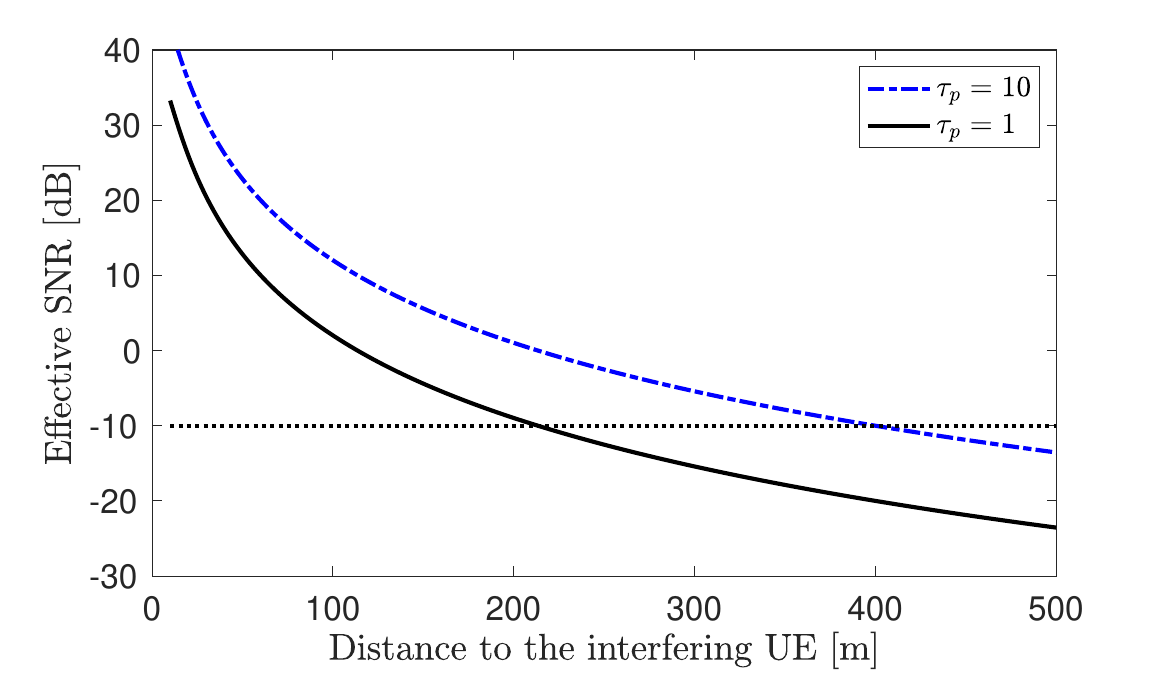}
		\put(15,18){Reference: $-10$\,dB}
\end{overpic} 
                \caption{The effective SNR~\eqref{eq:effective-pilot-SNR} of an interfering pilot signal depends on the distance between the AP and the interfering UE. In this simulation, the same propagation model as in Figure~\ref{fig:basic-NMSE-comparison-cellular-cellfree} is used. Pilot contamination has a negligible impact on the NMSE when the SNR is below the reference curve of $-10$\,dB.}    \vspace{+1mm}
	\label{fig:basic-NMSE-distance-effect} 
\end{figure}

The distance between the AP and the interfering UE is one of the main factors determining the SNR of the interfering signal. Figure~\ref{fig:basic-NMSE-distance-effect} shows the effective SNR~\eqref{eq:effective-pilot-SNR} as a function of the propagation distance using the same parameter values as in Figure~\ref{fig:basic-signal-gain-comparison-cellular-cellfree}.
The SNR reduces with the distance but can be above the $-10$\,dB reference curve for several hundred meters. Hence, pilot contamination is a non-negligible issue in practice. However, it is a gradual effect so the largest impact occurs when the interfering UE is rather close to the AP.
The effective SNR increases by 10\,dB when increasing the pilot length from $\tau_p=1$ to  $\tau_p=10$, which affects both the desired signal and contaminating signal. Even if the signal-to-contamination ratio remains the same, the propagation distances for which pilot contamination is a problem will grow since it is determined by comparing the contamination with the noise. On the other hand, each pilot can be reused less frequently among the UEs when  $\tau_p$ is increased. Hence, we can expect the use of longer pilots to be beneficial when it comes to limiting the pilot contamination, as long as we assign the pilots properly to the UEs. We return to this in Section~\ref{sec:pilot-assignment}.
Importantly, the curves in Figure~\ref{fig:basic-NMSE-distance-effect} will move up or down depending on the transmit power, bandwidth, and propagation model so it is not the exact numbers that matter but the general behavior.

\enlargethispage{\baselineskip}

\subsection{Impact of Spatial Correlation}
\label{sec:impact-of-spatial-correlation}
We now quantify the impact of spatial correlation among AP antennas on the channel estimation accuracy and pilot contamination. 

\subsubsection{Impact of Spatial Correlation on Estimation Accuracy}
Consider the spatially correlated channel $\vect{h} \sim \CN ( \vect{0}_N, \vect{R}  )$ between an arbitrary AP and an arbitrary UE, where the \gls{UE} and \gls{AP} indices are dropped for simplicity. We assume the UE uses a unique pilot sequence, thus the \gls{NMSE} in \eqref{eq:nmse-correlated-AP-l} can be simplified as
\begin{equation} \label{eq:nmse2}
\mathacr{NMSE}=1-\frac{\eta \tau_p \tr\left(\vect{R} \left(\eta\tau_p\vect{R}+\sigmaUL\vect{I}_N\right)^{-1}   \vect{R}\right) }{ \tr( \vect{R} )}.
\end{equation}
Let $\vect{R} = \vect{U} \vect{\Lambda} \vect{U}^{\Htran}$ denote the eigenvalue decomposition of the spatial correlation matrix, where the columns of the unitary matrix $\vect{U} \in \mathbb{C}^{N \times N}$ contain the eigenvectors (also called eigendirections) and the diagonal matrix $\vect{\Lambda} = \diag(\lambda_1,\ldots,\lambda_N)$ contains the corresponding non-negative eigenvalues where $\sum_{n=1}^N\lambda_n=\tr( \vect{R} )=N\beta$.
By using the eigenvalue decomposition, the NMSE in \eqref{eq:nmse2} can be rewritten as
\begin{align} \label{eq:nmse-without-pilot-contamination}
\mathacr{NMSE}&=1-\frac{\eta \tau_p \tr\left(\vect{U}\vect{\Lambda}\vect{U}^{\Htran} \left(\eta\tau_p\vect{U}\vect{\Lambda}\vect{U}^{\Htran}+\sigmaUL\vect{I}_N\right)^{-1}  \vect{U}\vect{\Lambda}\vect{U}^{\Htran}\right) }{ N\beta} \nonumber \\
&=1-\frac{\eta\tau_p}{N\beta}\tr\left(\vect{\Lambda} \left( \eta \tau_p \vect{\Lambda}+\sigmaUL\vect{I}_N \right)^{-1} \vect{\Lambda}\right) \nonumber \\
&=1-\frac{\eta\tau_p}{N\beta}\sum_{n=1}^N \frac{\lambda_n^2}{\eta\tau_p\lambda_n+\sigmaUL}\nonumber\\
&=1-\frac{1}{N\beta}\sum_{n=1}^N \frac{\mathacr{SNR}^{\text{pilot}}\lambda_n^2}{\mathacr{SNR}^{\text{pilot}}\lambda_n+\beta}
\end{align}
where the second equality follows from moving the matrices $\vect{U}$ and $\vect{U}^{\Htran}$ into the inverse, applying the cyclic shift property of the trace (see the second identity in Lemma~\ref{lemma-matrix-identities} on p.~\pageref{lemma-matrix-identities}), and utilizing that $\vect{U}\vect{U}^{\Htran}= \vect{U}^{\Htran}\vect{U}=\vect{I}_N$. An important observation from \eqref{eq:nmse-without-pilot-contamination} is that the NMSE depends on the eigenvalues but not on the eigenvectors. This is a direct consequence of the fact that the estimation error correlation matrix $\vect{C} = \vect{R} - \eta \tau_p \vect{R} \left(\eta\tau_p\vect{R}+\sigmaUL\vect{I}_N\right)^{-1}   \vect{R} $ has the same eigenvectors as the spatial correlation matrix $\vect{R}$ \cite[Sec.~3.3]{massivemimobook}.

We cannot change the spatial correlation matrices of a practical channel, but it is anyway important to understand what impact the correlation has. To this end, we now search for the eigenvalue distributions that maximize and minimize the NMSE under the constraint that $\sum_{n=1}^N\lambda_n=N\beta$. We will make use of the following result.

\begin{lemma}\label{lemma:nmse-concave} The function $\mathacr{NMSE} : \mathbb{R}_{\geq 0}^{N} \to \mathbb{R}_{\geq 0}$ of the eigenvalue vector $\boldsymbol{\lambda}=[ \lambda_1 \, \ldots \, \lambda_N ]^{\Ttran}$ given in \eqref{eq:nmse-without-pilot-contamination} is strictly concave on its domain.
\end{lemma}
\begin{proof}
The proof is available in Appendix~\ref{proof-of-lemma:nmse-concave} on p.~\pageref{proof-of-lemma:nmse-concave}.
\end{proof}

A consequence of the strict concavity stated in Lemma~\ref{lemma:nmse-concave} is that the maximum value of the NMSE in \eqref{eq:nmse-without-pilot-contamination} under the average channel gain constraint $\sum_{n=1}^N\lambda_n=N\beta$ is unique and achieved when all eigenvalues are equal: $\lambda_n=\beta$, for $n=1,\ldots,N$. This shows that, among all the spatial correlation matrices $\vect{R}$ that satisfy $\tr( \vect{R} )=N\beta$, the uncorrelated fading model with $\vect{R}=\beta\vect{I}_N$ provides the largest NMSE.

To identify the spatial correlation that instead minimizes the NMSE, we need the following lemma.

\begin{lemma}\label{lemma:nmse-eigenvalue-distribution} Suppose the eigenvalues of $\vect{R}$ are sorted in decaying order:
\begin{equation}
\lambda_1\geq\lambda_2\geq\ldots\lambda_r>\lambda_{r+1}=\ldots=\lambda_N=0
\end{equation}
where $r\leq N$ is the rank. Let $\boldsymbol{\lambda}=[\lambda_1 \, \ldots \, \lambda_{r-2} \ \lambda_{r-1} \ \lambda_{r} \ 0  \, \ldots \, 0]^{\Ttran}$ denote the corresponding eigenvalue vector. Then, $\mathacr{NMSE}\left(\boldsymbol{\lambda}\right)>\mathacr{NMSE}\left(\boldsymbol{{\lambda}^\prime}\right)$ where $\boldsymbol{{\lambda}^\prime}=[\lambda_1 \, \ldots \, \lambda_{r-2} \ \lambda_{r-1}+\lambda_{r} \ 0 \, \ldots \, 0 ]^{\Ttran}$.
\end{lemma}
\begin{proof}
The proof is available in Appendix~\ref{proof-of-lemma:nmse-eigenvalue-distribution} on p.~\pageref{proof-of-lemma:nmse-eigenvalue-distribution}.
\end{proof}

\begin{figure}[t!]
	\begin{overpic}[width=\columnwidth,tics=10]{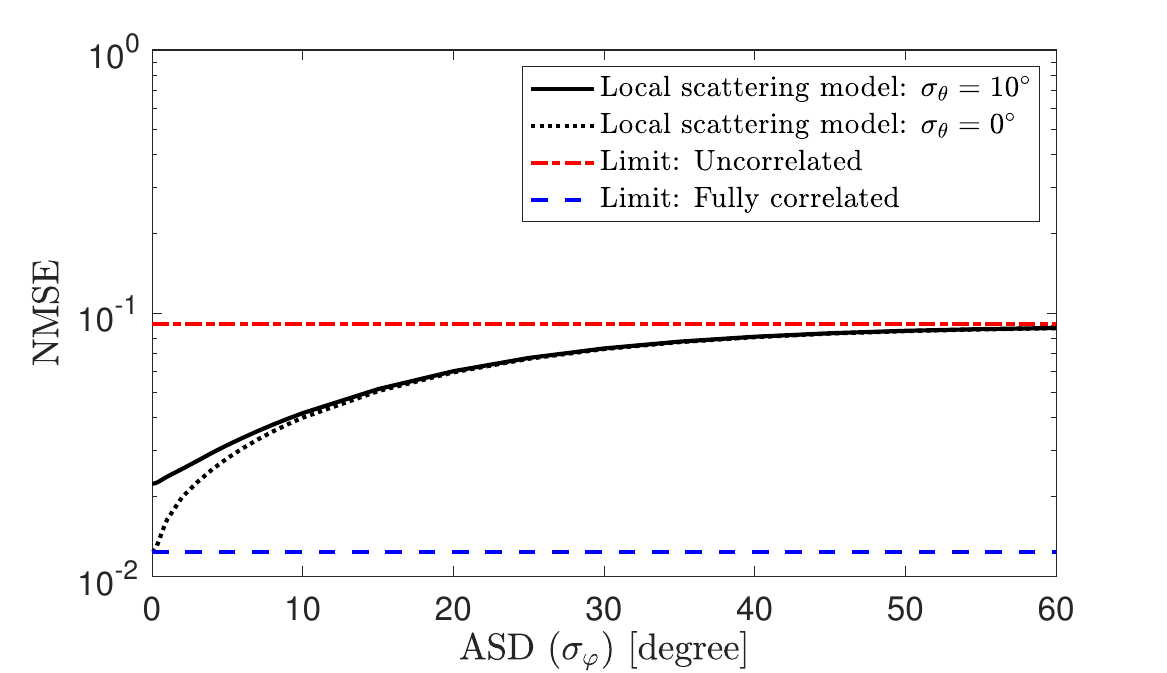}
\end{overpic} 
                \caption{NMSE in the estimation of an arbitrary UE's spatially correlated channel, as a function of the azimuth ASD $\sigma_{\varphi}$. The local scattering model~\eqref{eq:elements-of-R} with Gaussian angular distribution is used. The results are averaged over different nominal azimuth angles and the elevation angle is $\theta = -15^\circ$. The effective SNR~\eqref{eq:effective-pilot-SNR} is $10$\,dB and the AP is equipped with $N=8$ antennas.}    \vspace{+2mm}
	\label{fig:basic-NMSE-ASD-effect} 
\end{figure}

This lemma provides a procedure for reducing the NMSE by modifying the two smallest non-zero eigenvalues of the correlation matrix. More precisely, they are replaced with one eigenvalue being the sum of two smallest non-zero eigenvalues and one eigenvalue being identically zero.
If we repeat this procedure multiple times, we will progressively reduce the NMSE and eventually reach the case when $\lambda_1=N\beta$, and $\lambda_n=0$, for $n\geq 2$, which provides the smallest possible NMSE among all spatial correlation matrices $\vect{R}$ that satisfy $\sum_{n=1}^N\lambda_n=\tr( \vect{R} )=N\beta$.
This corresponds to the case where $\vect{R}=\lambda_1\vect{u}_1\vect{u}_1^{\Htran}$ is a rank-one matrix and thus represents the strongest type of spatial correlation.
Any rank-one correlation matrix with the same  $\lambda_1=N\beta$ achieves the same estimation accuracy. 
The conclusion is that the higher the spatial correlation is, the easier it is to estimate the channel since there is more structure to be utilized by the estimator.

\begin{figure}[t!]
	\begin{overpic}[width=\columnwidth,tics=10]{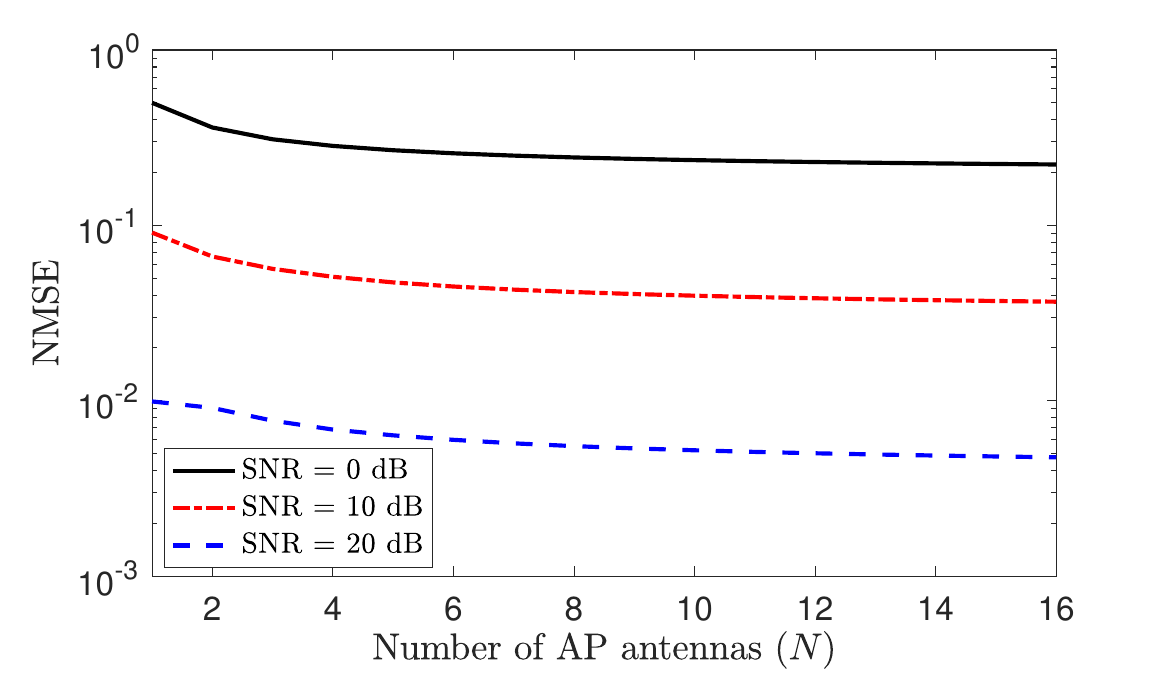}
\end{overpic} 
                \caption{NMSE in the estimation of an arbitrary spatially correlated channel, as a function of the number of antennas $N$ at the AP.
                The effective SNR~\eqref{eq:effective-pilot-SNR} is $0,10$, or $20$\,dB and the spatial correlation is modeled as in Figure~\ref{fig:basic-NMSE-ASD-effect} with $\sigma_{\varphi}=\sigma_{\theta}=10^\circ$.}    \vspace{+2mm}
	\label{fig:basic-NMSE-numOfAntennas-effect} 
\end{figure}

The above properties are illustrated numerically in Figure~\ref{fig:basic-NMSE-ASD-effect}, where the \gls{NMSE} is shown for the local scattering model in~\eqref{eq:elements-of-R}, as a function of the azimuth \gls{ASD} $\sigma_{\varphi}$. The results are averaged over different nominal azimuth angles, while the elevation angle is fixed at $\theta = -15^\circ$ and the elevation ASD is either $\sigma_{\theta} = 0^\circ$ or $\sigma_{\theta} = 10^\circ$.
The effective \gls{SNR}~\eqref{eq:effective-pilot-SNR} is 10\,dB and there are $N=8$ antennas at the AP. Figure~\ref{fig:basic-NMSE-ASD-effect} shows that the NMSE is smaller when the \gls{ASD} reduces (i.e., with higher spatial correlation). The \gls{NMSE} for uncorrelated channels is shown in Figure~\ref{fig:basic-NMSE-ASD-effect} as a reference.  For strongly spatially correlated channels, the NMSE can be one order of magnitude smaller than in the uncorrelated case, but this benefit is basically lost when $\sigma_{\varphi}\ge 40^\circ$.
The ASD in the elevation dimension only affects the NMSE when the ASD in the azimuth dimension is small. The lower bound for fully correlated channels is attained in the extreme case of $\sigma_{\varphi}=\sigma_{\theta} = 0^\circ$.

\enlargethispage{\baselineskip}

The impact of the number of AP antennas is studied in Figure~\ref{fig:basic-NMSE-numOfAntennas-effect} using the same correlation model as in Figure~\ref{fig:basic-NMSE-ASD-effect} but with the ASDs $\sigma_{\varphi}=\sigma_{\theta}=10^\circ$. The NMSE is shown as a function of the number of antennas for effective SNRs of $0$, $10$, and $20$\,dB. The NMSE reduces as the number of AP antennas increases since the channel realizations observed at adjacent antennas are strongly correlated, which is utilized by the MMSE estimator to improve the estimation quality.
However, only marginal gains are observed for $N\ge 8$, irrespective of the effective SNR. 
This happens when the ULA has become so wide that the outmost antennas experience almost uncorrelated channel realizations.

\subsubsection{Impact of Spatial Correlation on Pilot Contamination}

Next, we will analyze the impact of spatial correlation on pilot contamination. Assume UE $1$ and UE $2$ use the same pilot sequence and consider an arbitrary AP whose index is omitted.
The \gls{NMSE} of UE 1 in \eqref{eq:nmse-correlated-AP-l} takes the form
\begin{align} \label{eq:nmse-lower-bound}
\mathacr{NMSE}_1 =1-\frac{\eta_1 \tau_p \tr\left(\vect{R}_1 \left(\eta_1 \tau_p\vect{R}_1+ \eta_2\tau_p \vect{R}_2+\sigmaUL\vect{I}_N\right)^{-1}   \vect{R}_1\right) }{ \tr( \vect{R}_1 )}.
\end{align}
Unlike the NMSE in \eqref{eq:nmse-uncorrelated} for single-antenna APs, \eqref{eq:nmse-lower-bound} depends not only on the effective SNRs but also on the full spatial correlation matrices $\vect{R}_1$ and $\vect{R}_2$ of the UEs. We now observe that \eqref{eq:nmse-lower-bound} can be lower bounded by Lemma~\ref{lemma:scaling-mse} on p.~\pageref{lemma:scaling-mse} as
\begin{align} \label{eq:nmse-lower-bound-no-interference}
\mathacr{NMSE}_1 \geq 1-\frac{\eta_1 \tau_p \tr\left(\vect{R}_1 \left(\eta_1 \tau_p\vect{R}_1+\sigmaUL\vect{I}_N\right)^{-1}   \vect{R}_1\right) }{ \tr( \vect{R}_1 )}
\end{align}
where the equality occurs if and only if the correlation matrices are spatially orthogonal $\vect{R}_1\vect{R}_2 = \vect{0}_{N \times N}$. This  condition means that the eigenspaces are non-overlapping. If $\vect{R}_1 \neq \vect{0}_{N \times N}$ or $\vect{R}_2 \neq \vect{0}_{N \times N}$, the condition can only be satisfied if both $\vect{R}_1$ and $\vect{R}_2$ are rank-deficient.

 In the case of spatially orthogonal correlation matrices (i.e., $\vect{R}_1\vect{R}_2 = \vect{0}_{N \times N}$), the NMSE is completely unaffected by the interfering UE. Therefore, in theory, it is possible to completely avoid pilot contamination between the UEs. While the orthogonality condition is unlikely to hold in practice, a good rule-of-thumb is to assign pilots to the UEs such that
$\tr\left(\vect{R}_1\vect{R}_2\right)$ is small.

\begin{figure}[t!]
	\begin{overpic}[width=\columnwidth,tics=10]{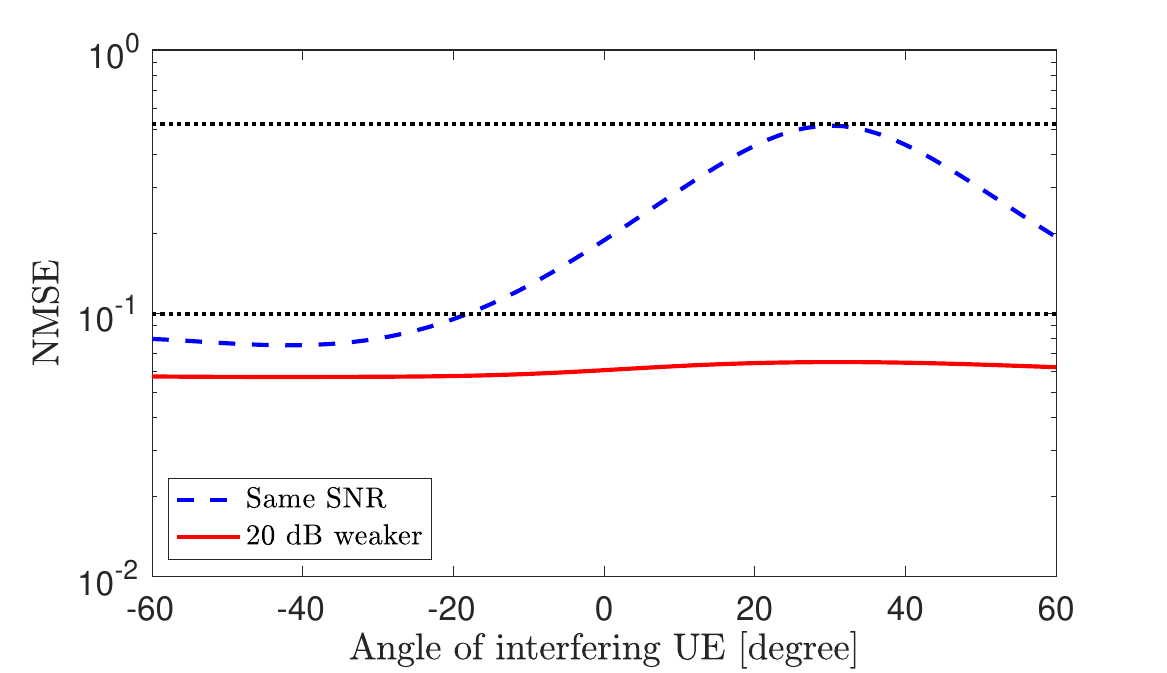}
	\put(15,48){\footnotesize{Uncorrelated fading: same SNR}}
\put(48,32){\footnotesize{Uncorrelated fading: 20\,dB weaker}}
\end{overpic} 
                \caption{NMSE in the estimation of the desired UE's channel when there is an interfering UE, which uses the same pilot. There are $N = 4$ antennas. The local scattering model is used with $\theta_1 = \theta_2= -15^\circ$, $\sigma_{\varphi}=10^\circ$, and $\sigma_{\theta}=0^\circ$.
The desired UE has a nominal azimuth angle of $30^\circ$, while the azimuth angle of the interfering UE is varied between $-60^\circ$ and $60^\circ$. The NMSE with uncorrelated fading is shown as references.}  
	\label{fig:basic-NMSE-spatial-correlation-effect-pilot-contamination} 
\end{figure}

\enlargethispage{\baselineskip} 

Figure~\ref{fig:basic-NMSE-spatial-correlation-effect-pilot-contamination} shows the \gls{NMSE} of the estimate of the desired channel with $N=4$ antennas. The spatial correlation is modeled by the local scattering model with elevation angles $\theta_1 = \theta_2= -15^\circ$ and ASDs $\sigma_{\varphi}=10^\circ$,  $\sigma_{\theta}=0^\circ$.
The azimuth angle of the desired UE is $\varphi_1 = 30^\circ$, while the azimuth angle $\varphi_2$ of the interfering UE is varied between $-60^\circ$ and $60^\circ$.
The effective \gls{SNR} of the desired \gls{UE} is 10\,dB and the interfering signal either has the same SNR or is 20\,dB weaker. When the \gls{UE} angles are well-separated, the \gls{NMSE} is below $0.1$, irrespective of how strong the interfering pilot signal is. This shows that pilot contamination has a minor impact on the estimation quality when the \glspl{UE} have well-separated eigenspaces. The \gls{NMSE} increases when the \glspl{UE} have similar angles and reaches a peak at $\varphi_1=\varphi_2$. The variations are large when the interfering \gls{UE} has a strong channel to the \gls{AP}, while the variations are small when the interfering signal is 20\,dB weaker. 
The figure also includes reference curves for the case of uncorrelated fading, where the NMSE is angle-independent.
We notice that the NMSE with correlated fading is always lower (or equal) to the NMSE with uncorrelated fading, even in the worst-case scenario where the desired and interfering UEs have identical angles.
Hence, spatial channel correlation is helpful in practice to improve the estimation quality under pilot contamination.

\section{Pilot Assignment and Dynamic Cooperation Cluster Formation}
\label{sec:pilot-assignment}
\label{sec:selecting-DCC}

The $\tau_p$ mutually orthogonal pilots must be reused among the UEs and can be assigned to them in an effort to limit the pilot contamination effect. The examples given earlier in this section have demonstrated that this entails keeping a large physical distance between the pilot-sharing UEs, so that each AP that serves a given UE will be subject to pilot contamination that drowns in the receiver noise. This is a good way of picturing what we want to achieve when performing a pilot assignment but it is a simplification since the observations were made using a propagation model in which the SNR is a strictly decreasing function of the distance. In reality, the SNR and propagation distance are correlated, but there is also a fair amount of random variations that can be modeled as shadow fading.
Hence, when designing pilot assignment algorithms, we should not take geometric parameters, such as the locations on a map, as input but instead consider the spatial correlation matrices $\vect{R}_{kl}$. Those matrices are describing both the average channel gains $\beta_{kl}$, including shadow fading, and the spatial channel characteristics.

The pilot assignment is a combinatorial problem. There are $(\tau_p)^K$ possible assignments in a setup with $K$ UEs and $\tau_p$ pilots, thus the complexity of evaluating all of them grows exponentially with the number of UEs.
The implication is that only suboptimal methods are feasible in practice. We will present one such algorithm in this section, and it will be utilized for performance evaluation in later sections. The algorithm will iteratively assign pilots to the UEs by always selecting the one that leads to the least pilot contamination. This is a so-called greedy algorithm and comes with no performance guarantees, but it will effectively avoid worst-case situations where closely spaced UEs (i.e., UEs with similar SNRs) are assigned to the same pilot. It is also practically attractive since the pilot assignment decision for a UE is made locally at a neighboring AP, instead of involving all APs in the network.
We will provide a literature review of alternative pilot assignment methods in Section~\ref{sec:pilot-assignment-advanced} on p.~\pageref{sec:pilot-assignment-advanced}.

\enlargethispage{\baselineskip} 

The formation of the \gls{DCC} is tightly connected to the pilot assignment. We noted earlier in this section that a single-antenna AP can only reasonably serve one UE per pilot, namely the one having the strongest channel gain. The reason is that this UE would cause strong pilot contamination to all the pilot-sharing UEs if the AP tried to serve some of those UEs as well. Consequently, a basic algorithm for cluster formation is to first assign pilots to the UEs and then let each AP serve exactly $\tau_p$ UEs; for every pilot, the AP serves the UE with the strongest channel gain among the subset of UEs that have been assigned that pilot. 
This is most likely a suboptimal algorithm, particularly for multiple-antenna APs which could use spatial correlation to distinguish between UEs that send the same pilot and thereby serve multiple UEs per pilot if they have nearly non-overlapping eigenspaces (recall Figure~\ref{fig:basic-NMSE-spatial-correlation-effect-pilot-contamination}). However, it makes good practical sense since every AP can make its clustering decisions locally.
This is the clustering algorithm that will be utilized for performance evaluation in later sections. We will provide a literature review of alternative methods in Section~\ref{sec:selecting-DCC-advanced} on p.~\pageref{sec:selecting-DCC-advanced}.

\subsubsection{An Algorithm for Joint Pilot Assignment and Cluster Formation}

The basic pilot assignment algorithm consists of two steps. First, the $\tau_p$ UEs with indices from $1$ to $\tau_p$ are assigned mutually orthogonal pilots: UE $k$ uses pilot $k$ for $k=1,\ldots,\tau_p$.
The remaining UEs, with indices from $\tau_p+1$ to $K$, are then assigned pilots one after the other. UE $k$ begins by determining which AP it has the strongest channel to. The index of this AP is computed as
\begin{equation} \label{eq:Master-AP}
\ell = \underset{l\in\{1,\ldots,L\}}{\mathrm{arg \, max}} \,\,\,  \beta_{kl}.
\end{equation}
Since AP $\ell$ is expected to contribute strongly to the service of UE $k$, it is preferable to assign UE $k$ to the pilot for which AP $\ell$ experiences the least pilot contamination. Hence, for each pilot $t$, the AP can compute the sum of the average channel gains $\beta_{i \ell}$ of the UEs that have already been assigned to that pilot. The AP then identifies the index of the pilot where the pilot interference is minimized:
\begin{equation} \label{eq:preferred-pilot}
\tau = \underset{t\in\{1,\ldots,\tau_p\}}{\mathrm{arg \, min}} \,\,\, \condSum{i=1}{t_i = t}{k-1} \beta_{i \ell}.
\end{equation}
This pilot is assigned to the UE and the algorithm then continues with the next UE. When all the UEs have been assigned to pilots, the clusters can be created. Each AP goes through each pilot and identifies which of the UEs have the largest channel gain among those using that pilot. The AP will serve that UE.
The procedure is summarized in Algorithm~\ref{alg:basic-pilot-assignment}.

\begin{algorithm}
	\caption{Basic pilot assignment and cooperation clustering} \label{alg:basic-pilot-assignment}
	\begin{algorithmic}[1]
	\State {\bf Initialization:} Set $\mathcal{M}_1 = \ldots = \mathcal{M}_K = \emptyset$
	\For{$k=1,\ldots,\tau_p$} 
		\State $t_k \gets k$ \Comment{Assign orthogonal pilots to first $\tau_p$ UEs}
	\EndFor
	\For{$k=\tau_p+1,\ldots,K$}
		\State $\ell \gets \underset{l\in\{1,\ldots,L\}}{\mathrm{arg \, max}} \,\,\, \beta_{kl}$ \Comment{Find the best AP for UE $k$}
		\State $\tau \gets \underset{t\in\{1,\ldots,\tau_p\}}{\mathrm{arg \, min}} \,\,\, \condSum{i=1}{t_i = t}{k-1} \beta_{i \ell}$ \Comment{Find pilot with least interference at AP $\ell$}
		\State $t_k \gets \tau$ \Comment{Assign pilot $\tau$ to UE $k$}
	\EndFor
	\For{$l=1,\ldots,L$}
		\For{$t=1,\ldots,\tau_p$}
			\State $i \gets \underset{k\in\{1,\ldots,K\} : t_k = t}{\mathrm{arg \, max}}  \,\,\, \beta_{k l}$ \Comment{Find UE that AP $l$ serves on pilot $t$}
			\State $\mathcal{M}_i \gets \mathcal{M}_i  \cup \{ l \}$
		\EndFor
	\EndFor
		\State {\bf Output:} Pilot assignment $t_1,\ldots,t_K$ and DCCs $\mathcal{M}_1,\ldots,\mathcal{M}_K $
	\end{algorithmic}
\end{algorithm}

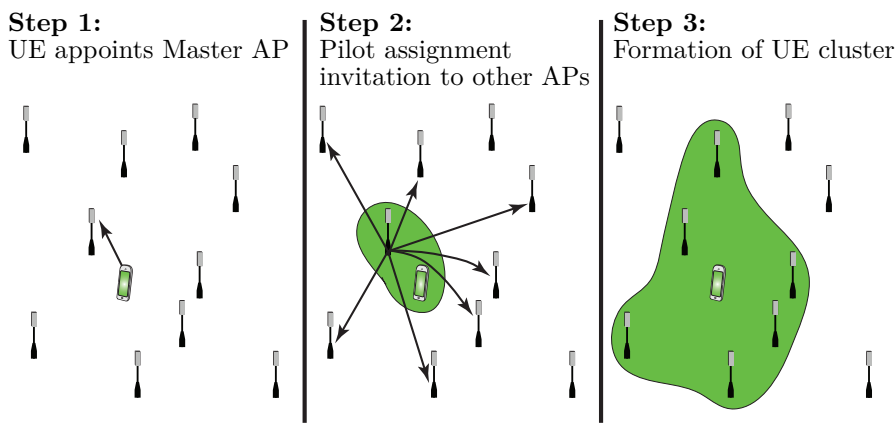
\begin{figure}[t!]
	\centering 
	\begin{overpic}[width=\columnwidth,tics=10]{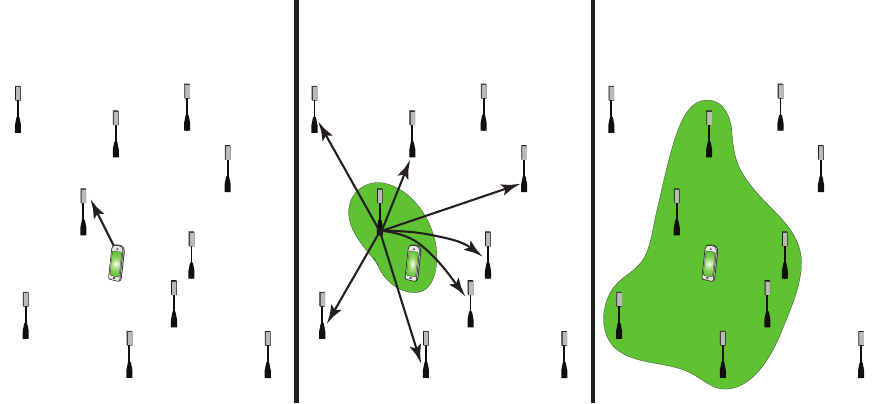}
	\put(0,44){\small{\textbf{Step 1:}}}
	\put(0,41){\small{UE appoints Master AP}}
	\put(35,44){\small{\textbf{Step 2:}}}
	\put(35,41){\small{Pilot assignment}}
	\put(35,38){\small{invitation to other APs}}
	\put(68,44){\small{\textbf{Step 3:}}}
	\put(68,41){\small{Formation of UE cluster}}
	\end{overpic} \vspace{-5mm}
	\caption{The procedure from  \cite{Bjornson2020a} for joint initial access, pilot assignment, and cluster formation. This is one way of implementing the steps in Algorithm~\ref{alg:basic-pilot-assignment} in a dynamic fashion.}
	\label{fig:cellfree_formation}  
\end{figure}

This algorithm resembles the ones proposed in \cite{Bjornson2020a}, \cite{Chen2016a}. The key idea was that whenever a new UE is admitted to the network, it begins by identifying the AP that it has the strongest channel gain to. In practical systems, the APs are periodically broadcasting synchronization signals to convey basic access information. The channel gains can be measured based on these signals. The accessing UE computes $\ell$ according to \eqref{eq:Master-AP} and appoints AP $\ell$ as its \emph{Master AP}. Note that this selection is made in a user-centric manner.
The UE also uses the broadcasted signal to synchronize to the AP. The UE can contact its Master AP via a standard random access procedure \cite{sanguinetti13bis}, \cite{Sesia09}. 
The Master AP computes the preferred pilot $\tau$ using \eqref{eq:preferred-pilot} and informs the surrounding APs of the existence of the new UE. The surrounding APs can then determine if they should change which UEs to serve on pilot $\tau$. This is a dynamic variation of Algorithm~\ref{alg:basic-pilot-assignment} that can be applied when new UEs are entering the system. It can also be applied when a UE has moved to such a large extent that its channel statistics have changed and, therefore, it can be treated as a new UE \cite{Bjornson2020a}, \cite{Chen2016a}. The three main steps of this initial access algorithm are illustrated in Figure~\ref{fig:cellfree_formation}.

\enlargethispage{\baselineskip} 

The algorithm described above is distributed in the sense that each UE is managed separately and each AP makes its pilot assignment decisions and cluster formation locally. A key challenge for any distributed cluster formation algorithm, including Algorithm~\ref{alg:basic-pilot-assignment}, is the risk that some UEs will not be served by any AP. This can happen when a particular AP has the strongest channel to more than $\tau_p$ UEs, while it is only allowed to serve at most $\tau_p$ of them.
A solution to this problem is to require that each UE $k$ is served by at least one AP, namely AP $\ell$ that is selected according to \eqref{eq:Master-AP} \cite{Bjornson2020a}. This will lead to pilot contamination but at least guarantees that each UE is being served.
Since the intended operating regime of Cell-free Massive MIMO is the ultra-dense regime of $L \gg K$, each AP will on average only have the strongest channel to $K/L < 1$ UEs. Hence, if $\tau_p=10$, there is a very low risk that one AP will have the strongest channel to more than ten UEs, but it can happen.

\begin{remark}[Pilot decontamination] \label{remark:decontamination}
Specific signal processing algorithms have been developed for situations where pilot contamination is the main performance-limiting factor. The general aim is to expand the dimension of the pilot sequences and we will briefly describe three main categories of such \emph{pilot decontamination} approaches. The first category utilizes channel statistics, particularly spatial channel correlation, to assign pilots to UEs to proactively limit the pilot contamination effect. Some key references from the Cellular Massive MIMO literature are \cite{BjornsonHS17}, \cite{Huh2012a}, \cite{Yin2013a}, \cite{Xu2015a}, but these are not directly applicable to cell-free networks where each AP has a small number of antennas. Approaches developed for cell-free networks will be described in Section~\ref{sec:pilot-assignment-advanced} on p.~\pageref{sec:pilot-assignment-advanced}.
The second approach is to make use of the uplink data as additional pilots, which is known as semi-blind or data-aided estimation \cite{Carvalho1997a}. The main complication is that the data is unknown, but since long sequences of random data samples are fairly orthogonal, the transmitted signals will fall into different subspaces that can be identified by the receiving AP.
Some key references from the Cellular Massive MIMO literature are \cite{Ma2014a}, \cite{Mueller2014b}, \cite{Neumann2014a}, \cite{Ngo2012a}, \cite{Vinogradova2016a}, \cite{Yin2016a}. The existing algorithms rely \linebreak on asymptotic properties that occur when $N$ and $\tau_c$ are jointly growing large, thus they are not applicable in cell-free networks where~$N$~is~small. The third approach is to superimpose low-power pilot sequences onto the uplink data, thereby making the pilot length equal to the number of uplink samples per coherence block. The benefit is that many more orthogonal pilot sequences can be generated, while the drawback is that the pilots and data will now mutually interfere. There are situations where it brings benefits, as exemplified in \cite{Upadhya2014a}, \cite{Upadhya2017b}, \cite{Verenzuela2017a}, \cite{Verenzuela2020a} for Cellular Massive MIMO. However, it cannot increase the capacity scaling at high SNR, which is achievable by reserving certain dimensions for pilots \cite{Zheng2002a}. We will not cover any of these methods in detail in this monograph, because it will later become apparent that pilot contamination is not a main limiting factor in Cell-free Massive MIMO.
\end{remark}

\newpage
\section{Summary of the Key Points in Section~\ref{c-channel-estimation}}
\label{c-channel-estimation-summary}

\begin{mdframed}[style=GrayFrame]

\begin{itemize}
\item Channel estimation is fundamental to perform coherent transmission processing both between AP antennas and across cooperating APs. Since the channels are constant throughout one coherence block and change from one block to another, they need to be estimated once per each
coherence block.

\item Channel estimation is carried out by sending uplink pilots. The estimates can be either computed at the CPU or at the APs. Both options give the same estimation quality.

\item The MMSE estimator exploits channel statistics to obtain good estimates. The interference generated by pilot-sharing UEs not only reduces the estimation quality but also makes the channel estimates statistically dependent. This phenomenon is called pilot contamination.

\item Spatial correlation is helpful to improve the estimation quality. With pilot contamination, the correlation between channel estimates is low when the correlation matrices of the pilot-sharing UEs are sufficiently different.
 
\item The estimation accuracy achieved with many single-antenna APs is higher than in the case of a single AP with a large co-located array. The highest estimation accuracy is achieved when the UE is served only by the APs that are close to it. However, coherent transmission processing among APs makes the network robust to lower-quality channel estimates.

\item As a rule-of-thumb, the pilots should be assigned to the UEs so that each AP is close to (i.e., has a high effective SNR to) at most one UE per pilot. Optimal assignment is impractical but a sequential algorithm was provided in Algorithm~\ref{alg:basic-pilot-assignment}.

\end{itemize}

\end{mdframed}


\chapter{Uplink Operation}
\label{c-uplink}

This section describes two different uplink implementations of User-centric Cell-free Massive \gls{MIMO}, which are characterized by different degrees of cooperation among the APs. Section~\ref{sec:level4-uplink} considers a centralized operation in which the pilot and data signals received at all APs are gathered (through the fronthaul links) at the CPU, which performs channel estimation and data detection. The achievable SE is derived based on the MMSE channel estimates and used to obtain the optimal, but unscalable, centralized receive combining. Alternative scalable combining schemes are then proposed. In Section~\ref{sec:level3-uplink}, the SE analysis and design of scalable receive combining schemes are extended to a distributed operation where the local MMSE channel estimates are used at the APs to obtain local estimates of the UE data, which are then gathered and linearly combined at the CPU for final detection. To exemplify the performance and properties of cell-free networks under somewhat realistic conditions, a simulation setup is defined in Section~\ref{sec:running-example} that will be used as a running example in the remainder of this monograph. Performance analysis and comparisons are carried out in Section~\ref{sec:uplink-performance-comparison}, while a summary of the key points is provided in Section~\ref{c-uplink-summary}.

\section{Centralized Uplink Operation}
\label{sec:level4-uplink}
  \vspace{-2mm}

\begin{figure}[t!]
	\centering 
	\begin{overpic}[width=\columnwidth,tics=10]{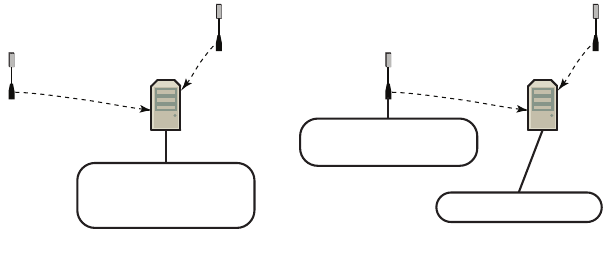}
		\put(3,28.5){\small{Received signals}}
		\put(66,28.5){\small{Data estimates}}
		\put(0,36){\small{AP}}
		\put(32,24){\small{CPU}}
		\put(15,13){\small{Channel estimation}}
		\put(15,10){\small{Receive combining}}
		\put(15,7){\small{Data detection}}
		\put(51.5,20){\small{Channel estimation}}
		\put(51.5,17){\small{Receive combining}}
		\put(62,36){\small{AP}}
		\put(94,24){\small{CPU}}
		\put(75.5,7.5){\small{Data detection}}
		\put(10,0){\small{(a) Centralized operation}}
		\put(60,0){\small{(b) Distributed operation}}
	\end{overpic}  
	\caption{The uplink signal processing tasks can be divided between the APs and CPU in different ways. In the centralized operation, the channel estimation, receive combining, and data detection are done at the CPU. In the distributed operation, everything, except the data detection, is done at the APs.}
	\label{figure:distributed_centralized}  \vspace{-3mm}
\end{figure}
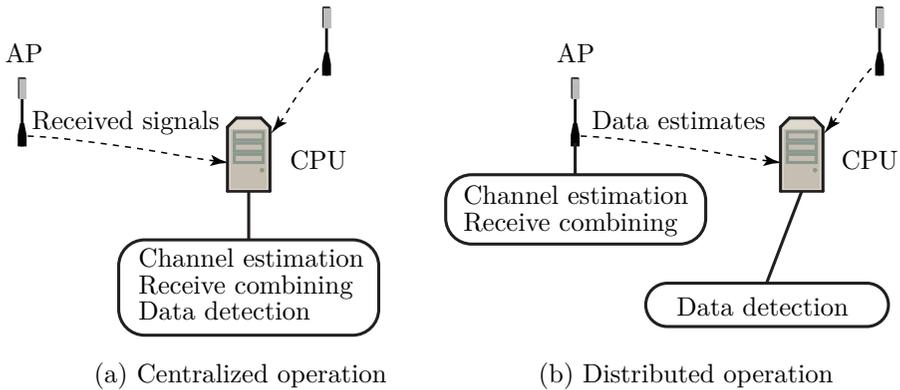

The most advanced uplink implementation of Cell-free Massive MIMO is a fully centralized 
operation, where the APs only act as remote-radio heads or relays that forward their received baseband signals to the CPU for processing.
More precisely, each AP $l$ sends the $\tau_p$ received pilot signals $\{ \vect{y}_{t l}^{\rm pilot} : t = 1,\ldots,\tau_p \}$ in \eqref{eq:received-pilot} and the received uplink data signal $\vect{y}_{l}^{\rm{ul}}$ in \eqref{eq:DCC-uplink} to the CPU, which performs channel estimation, receive combining and data detection. The centralized operation is illustrated in Figure~\ref{figure:distributed_centralized}(a).  Although we call it \emph{centralized operation}, this should not be interpreted as building a network that gathers the information related to all UEs at a single point at the center of the network. As described in Section~\ref{sec:cell-free-networks} on p.~\pageref{sec:cell-free-networks}, the CPU is a logical entity and its tasks can be divided between many geographically distributed edge-cloud processors \cite{Interdonato2019a}, as illustrated in Figure~\ref{fig:cell-free} on p.~\pageref{fig:cell-free}. Different UEs might use different processors. Hence, in practice, centralized operation means that each UE is associated with one nearby processor which we refer to as ``the CPU'' and all its serving APs send their respective received uplink signals to it. We will now explain what processing is done at the CPU and quantify the achievable uplink SE.

For each UE $k$, the pilot signals are individually used at the CPU  to compute the partial MMSE estimates $\{\vect{D}_{k}\widehat{\vect{h}}_{i} : i= 1, \ldots,K\}$ of the collective channels $\{{\vect{h}}_{i} : i=1,\ldots,K\}$ from all UEs to the APs that serve UE $k$ (i.e., the APs with indices $l\in \mathcal{M}_k$). 
The estimation procedure was described in Section~\ref{sec:MMSE-channel-estimation} on p.~\pageref{sec:MMSE-channel-estimation}.
Recall that $\vect{D}_{k}  = \diag( \vect{D}_{k1} , \ldots, \vect{D}_{kL} )$ is a block-diagonal matrix with $\vect{D}_{kl}$ being defined in~\eqref{eq:user-centric} as $\vect{I}_N$ for APs that serve UE $k$ and zero otherwise. 

The received uplink data signals at the serving APs are jointly used at the CPU to compute an estimate $\widehat{s}_{k}$ of the signal $s_k$ transmitted by UE $k$. From the signal model in Section~\ref{sec:uplink-system-model} on p.~\pageref{sec:uplink-system-model}, this is achieved by summing up the inner products between the effective receive combining vectors $\vect{D}_{kl}\vect{v}_{kl}$ and $\vect{y}_{l}^{\rm{ul}}$ for $l=1,\ldots,L$. This yields the estimate 
\begin{align} \notag
\widehat{s}_{k} &=  \sum\limits_{l=1}^L\widehat{s}_{kl} \\ &= \sum\limits_{l=1}^L\vect{v}_{kl}^{\Htran} \vect{D}_{kl} {\bf y}_{l}^{\rm{ul}} = \vect{v}_{k}^{\Htran} \vect{D}_{k} {\bf y}^{\rm{ul}}\label{eq:uplink-CPU-data-estimate}
\end{align}
where $\vect{v}_{k} = [\vect{v}_{k1}^{\Ttran} \, \ldots \, \vect{v}_{kL}^{\Ttran}]^{\Ttran} \in \mathbb{C}^{LN}$ is the centralized combining vector and $\vect{y}^{\rm{ul}} \in \mathbb{C}^{LN}$ is the collective uplink data signal given by
\begin{equation} \label{eq:received-data-central2}
\vect{y}^{\rm{ul}} = \begin{bmatrix} \vect{y}_{1}^{\rm{ul}} \\ \vdots \\ \vect{y}_{L}^{\rm{ul}} 
	\end{bmatrix} = \sum_{i=1}^{K} \vect{h}_{i} s_i + \vect{n}
\end{equation}
with $\vect{n} = [\vect{n}_{1}^{\Ttran} \, \ldots \, \vect{n}_{L}^{\Ttran}]^{\Ttran} \in \mathbb{C}^{LN}$ being the collective noise vector. Notice that~\eqref{eq:received-data-central2} depends on all the channel vectors $\{\vect{h}_{i}: i=1,\ldots,K\}$ since all the APs receive the signal from all UEs. Since $ \vect{D}_{kl}=\vect{0}_{N\times N}$ implies $\vect{v}_{kl}^{\Htran} \vect{D}_{kl} =\vect{0}_N^{\Ttran}$ in \eqref{eq:uplink-CPU-data-estimate}, the CPU will however apply receive combining 
using only the data signals of the APs with $ \vect{D}_{kl} \neq \vect{0}_{N \times N}$. In other words, \eqref{eq:uplink-CPU-data-estimate} can be equivalently written as  $\widehat{s}_{k}  =  \sum_{l \in \mathcal{M}_k} \vect{v}_{kl}^{\Htran} {\bf y}_{l}^{\rm{ul}}$ by including only the subset of APs that serves UE $k$ in the summation.
While this alternative expression is more compact at this point, it would require much more notation if used for performance analysis. That is why we utilize \eqref{eq:uplink-CPU-data-estimate} in this section.

The expression in \eqref{eq:received-data-central2} is mathematically equivalent to the signal model of an uplink single-cell Massive MIMO system with correlated fading \cite[Sec.~2.3.1]{massivemimobook}, if one treats the CPU as a receiver equipped with $LN$ antennas. However, there are some key differences:
\begin{enumerate}
\item Multiple UEs that are managed by the CPU are using the same pilot, which is normally avoided in single-cell systems.

\item The antennas are distributed over different geographical locations. Hence, the collective channel is distributed as $\vect{h}_k \sim \CN(\vect{0}_{LN}, \vect{R}_{k})$ where the spatial correlation matrix $\vect{R}_{k} = \diag(\vect{R}_{k1}, \ldots,  \vect{R}_{kL}) \in \mathbb{C}^{LN \times LN}$ has a block-diagonal structure, which is normally not the case in single-cell systems.

\item Not all antennas are being used for signal detection, but only those belonging to the APs that serve the particular UE.
\end{enumerate}

Despite these distinct differences, we can compute an achievable uplink SE, based on the knowledge of the partial MMSE estimates $\{\vect{D}_{k}\widehat{\vect{h}}_{i} : i= 1, \ldots,K\}$, using the methodology from the Cellular Massive MIMO literature \cite[Sec.~4.1]{massivemimobook}. From this expression, we can further obtain the optimal centralized receive combining vector $\vect{v}_{k}$. All the analysis in this section relies on the assumption that the spatial correlation matrices $\{\vect{R}_{kl}: k=1,\ldots,K, \, l=1,\ldots,L\}$ are perfectly known at the CPU to perform channel estimation, receive combining and data detection. We will return to this assumption in Section~\ref{sec:level4-fronthaul}.

\subsection{Spectral Efficiency With Centralized Operation}
\label{sec:level4-SE}
While the ergodic capacity of fully cooperative networks with perfect \gls{CSI} is known in some cases \cite{Estella2019a}, it is generally unknown in the considered case with imperfect CSI. However, we can rigorously analyze the performance by using the standard capacity lower bounds described in 
Section~\ref{sec:capacity-bounds} on p.~\pageref{sec:capacity-bounds}.
 As a first step, we use~\eqref{eq:received-data-central2} to rewrite~\eqref{eq:uplink-CPU-data-estimate} as 
\begin{align} \label{eq:uplink-data-estimate2}
\widehat{s}_k& = \hspace{-1cm}\underbrace{\vect{v}_k^{\Htran}\vect{D}_k{\widehat{\vect{h}}}_k s_k}_{\textrm{Desired signal over estimated channel}}+\underbrace{\vect{v}_k^{\Htran}\vect{D}_k{\widetilde{\vect{h}}}_k s_k}_{\textrm{Desired signal over unknown channel}}\nonumber \\ &+\underbrace{\condSum{i=1}{i \neq k}{K}\vect{v}_k^{\Htran}\vect{D}_k\vect{h}_i s_i}_{\textrm{Interference}}\;+\;\underbrace{\vect{v}_k^{\Htran}\vect{D}_k\vect{n}}_{\textrm{Noise}}
\end{align}
where the desired signal term has been divided into two parts: one
that is received over the known partially estimated channel $\vect{D}_k{\widehat{\vect{h}}}_k$ from UE $k$ and one that is received over the unknown part of the channel, represented by the estimation error $\vect{D}_k{\widetilde{\vect{h}}}_k$. The former part can be utilized straight away for signal detection, while the latter part is less useful since only the distribution of the estimation error is known, as reported in Section~\ref{sec:mmse-of-collective-channel} on p.~\pageref{sec:mmse-of-collective-channel}. An achievable SE can be computed by treating the latter part as additional interference in the signal detection, by utilizing Lemma~\ref{lemma:capacity-memoryless-random-interference} on p.~\pageref{lemma:capacity-memoryless-random-interference}. If the first term in \eqref{eq:uplink-data-estimate2} is treated as the desired part while the other three terms are treated as noise in the receiver, the following SE is achievable.

\begin{theorem} \label{proposition:uplink-capacity-general}
	An achievable SE of UE $k$ in the centralized operation is
	\begin{equation} \label{eq:uplink-rate-expression-general}
	\mathacr{SE}_{k}^{({\rm ul},\cent)} = \frac{\tau_u}{\tau_c} \mathbb{E} \left\{ \log_2  \left( 1 + \mathacr{SINR}_{k}^{({\rm ul},\cent)}  \right) \right\} \quad \textrm{bit/s/Hz}
	\end{equation}
	where the instantaneous effective \gls{SINR} is given by
	\begin{equation} \label{eq:uplink-instant-SINR-level4}
	\mathacr{SINR}_{k}^{({\rm ul},\cent)} = \frac{ p_{k} \left |  \vect{v}_{k}^{\Htran} \vect{D}_{k}\widehat{\vect{h}}_{k} \right|^2  }{ 
		\condSum{i=1}{i \neq k}{K} p_{i} \left | \vect{v}_{k}^{\Htran} \vect{D}_{k}\widehat{\vect{h}}_{i} \right|^2
		+ \vect{v}_{k}^{\Htran}  {\bf{Z}}_k \vect{v}_{k}  + \sigmaUL \| \vect{D}_k \vect{v}_{k} \|^2
	}
	\end{equation}
	with
	\begin{align}
	{\bf{Z}}_k = \sum\limits_{i=1}^{K} p_{i}  \vect{D}_k  \vect{C}_{i} \vect{D}_k
	\end{align}
	and the expectation is with respect to the channel estimates. The matrix $\vect{C}_{i}$ is the error correlation matrix of the collective channel ${\vect{h}}_{i}$.
\end{theorem}
\begin{proof}
	The proof is available in Appendix~\ref{proof-of-proposition:uplink-capacity-general} on p.~\pageref{proof-of-proposition:uplink-capacity-general}.
\end{proof}

The pre-log factor ${\tau_u}/{\tau_c}$ in \eqref{eq:uplink-rate-expression-general} is the fraction of each coherence block that is used for uplink data transmission. The term $\mathacr{SINR}_{k}^{({\rm ul},\cent)}$ takes the form of an ``effective instantaneous SINR'' \cite{massivemimobook}, with the desired signal power received over the estimated channel $\vect{D}_k{\widehat{\vect{h}}}_k$ in the numerator and the interference plus noise in the denominator. More precisely, the numerator contains $p_{k}|  \vect{v}_{k}^{\Htran} \vect{D}_{k}\widehat{\vect{h}}_{k} |^2$, while the first term in the denominator is the interference that is received over the estimated parts of the interfering channels, ${\bf{Z}}_k $ is the combined spatial correlation matrix of the signals received over the unknown parts of the channels, and the last term in the denominator is the noise power after receive combining. Notice that the matrix $\vect{C}_{i}$ is the error correlation matrix of the collective channel ${\vect{h}}_{i}$; see Section~\ref{sec:mmse-of-collective-channel} on p.~\pageref{sec:mmse-of-collective-channel}.
We call \eqref{eq:uplink-instant-SINR-level4}  an ``effective'' instantaneous SINR since it cannot be measured in the system at any particular point in time, because the averaging over the estimation errors does not occur instantaneously.
However, the SE is effectively the same as that of a fading single-antenna single-user channel where \eqref{eq:uplink-instant-SINR-level4} is the instantaneously measurable SNR and the receiver has perfect CSI. In particular, this means that the desired data signal can be encoded and the received signal can be decoded as if we would communicate over such a fading \gls{AWGN} channel.

The SE expression in Theorem~\ref{proposition:uplink-capacity-general} is not in closed form since there is an expectation in front of the logarithm in \eqref{eq:uplink-rate-expression-general}. However, the SE can be easily computed numerically using Monte Carlo methods, which means that we approximate the expectation with an average over a large number of random realizations. More precisely, we can generate realizations of the channel estimates in a large set of coherence blocks and compute the value of the logarithm for each one of them, followed by taking the sample mean of these values.

\enlargethispage{\baselineskip} 

Theorem~\ref{proposition:uplink-capacity-general} is rather general in the sense that the SE is derived using the DCC framework and considering multi-antenna APs with arbitrary correlated fading channels and arbitrary DCC. The case of $N=1$ was considered in \cite{Attarifar2020a}, \cite{Nayebi2016a}, \cite{Riera2018a}, \cite{Yang2019a}, while \cite{Bashar2018a}, \cite{Chen2018b} considered $N \geq 1$ with uncorrelated fading and \cite{Bjornson2019c}, \cite{Bjornson2020a} considered $N \geq 1$ with correlated fading.
In the extreme case when all APs are serving UE $k$ (i.e., $\vect{D}_k = \vect{I}_{LN}$), the SINR in \eqref{eq:uplink-instant-SINR-level4} can be simplified to 
\begin{equation} \label{eq:uplink-instant-SINR-level4-all}
	\mathacr{SINR}_{k}^{({\rm ul},\cent)} =  \frac{ p_{k} \left |  \vect{v}_{k}^{\Htran} \widehat{\vect{h}}_{k} \right|^2  }{ 
	\condSum{i=1}{i \neq k}{K} p_{i} \left | \vect{v}_{k}^{\Htran} \widehat{\vect{h}}_{i} \right|^2
		+ \vect{v}_{k}^{\Htran}  \left( \sum\limits_{i=1}^{K} p_{i}  \vect{C}_{i}  + \sigmaUL  \vect{I}_{LN}\right) \vect{v}_{k}  
	}.
\end{equation}

\subsection{An Alternative Spectral Efficiency Expression}
\label{subsec:alternative-bound}

The SE expression in Theorem~\ref{proposition:uplink-capacity-general} is a lower bound on the capacity that is derived by utilizing the property that the channel estimate $\widehat{\vect{h}}_{k}$ and the estimation error $\widetilde{\vect{h}}_{k} = {\vect{h}}_{k} -\widehat{\vect{h}}_{k}$ are independent random vectors. This condition is satisfied when using the MMSE estimator and Rayleigh fading channels as assumed in this monograph.
An alternative bound that can be applied along with any channel estimator and fading model is the so-called \emph{\gls{UatF}} that is widely used in Cellular Massive MIMO \cite[Theorem~4.4]{massivemimobook}, and also in Cell-free Massive MIMO \cite{Bashar2019a}, \cite{Bjornson2019c}, \cite{Nayebi2016a} with $\vect{D}_{kl}=\vect{I}_N$ for all $k,l$ and specific combining vectors. The name comes from the fact that the channel estimates are used for computing the receive combining vectors and then effectively ``forgotten'' before the signal detection takes place. In the considered network setting with multiple-antenna APs, arbitrary correlated fading channels, and arbitrary DCC, the following achievable SE is found by applying the \gls{UatF} bound.

\begin{theorem} \label{proposition:uplink-capacity-general-UatF}
	An achievable SE of UE $k$ in the centralized operation is
	\begin{equation} \label{eq:uplink-rate-expression-general-UatF}
	\mathacr{SE}_{k}^{({\rm ul},\centuatf)} = \frac{\tau_u}{\tau_c} \log_2  \left( 1 + \mathacr{SINR}_{k}^{({\rm ul},\centuatf)}  \right) \quad \textrm{bit/s/Hz}
	\end{equation}
	where the effective \gls{SINR} is given by
	\begin{align} \nonumber
	&\mathacr{SINR}_{k}^{({\rm ul},\centuatf)} 
	\\ &= \frac{ p_{k} \left |\mathbb{E} \left\{  \vect{v}_{k}^{\Htran} \vect{D}_{k}{\vect{h}}_{k} \right\}\right|^2  }{ 
		\sum\limits_{i=1}^K p_{i}  \mathbb{E} \left\{| \vect{v}_{k}^{\Htran} \vect{D}_{k}{\vect{h}}_{i} |^2\right\}
		- p_{k} \left |\mathbb{E} \left\{  \vect{v}_{k}^{\Htran} \vect{D}_{k}{\vect{h}}_{k} \right\}\right|^2  + \sigmaUL \mathbb{E} \left\{\| \vect{D}_k \vect{v}_{k} \|^2\right\}
	}  \label{eq:uplink-instant-SINR-level4-UatF}
	\end{align}
	and the expectation is with respect to the channel realizations.
\end{theorem}
\begin{proof}
	The proof is available in Appendix~\ref{proof-of-proposition:uplink-capacity-general-UatF} on p.~\pageref{proof-of-proposition:uplink-capacity-general-UatF}.	
\end{proof}

Intuitively, $\mathacr{SE}_{k}^{({\rm ul},\centuatf)}$ in \eqref{eq:uplink-rate-expression-general-UatF} should be smaller than $\mathacr{SE}_{k}^{({\rm ul},\cent)}$ in \eqref{eq:uplink-rate-expression-general} since it relies on a simplified implementation in which the channel estimates are not used at the CPU for signal detection. This conjecture is hard to prove analytically but has been verified by numerous numerical experiments. Except for Section~\ref{sec:uplink-performance-comparison} where will compare  $\mathacr{SE}_{k}^{({\rm ul},\centuatf)}$ with $\mathacr{SE}_{k}^{({\rm ul},\cent)}$, we will not use $\mathacr{SE}_{k}^{({\rm ul},\centuatf)}$ in the remainder of this section, to avoid underestimating the achievable performance. However, we stress that one benefit with the simplified bound in Theorem~\ref{proposition:uplink-capacity-general-UatF} is that there is no expectation in front of the logarithm, because the bounding technique is treating the channel as deterministic. We will return to the expression in Section~\ref{sec:optimized-power-allocation-uplink} on p.~\pageref{sec:optimized-power-allocation-uplink} where the uplink powers $\{p_k : k =1,\ldots,K\}$ will be optimized by making use of the fact that there are only expectations at the inside of the logarithm.

\subsection{Optimal Receive Combining}
\label{subsec:optimal-receive-combining}
The SE expression in \eqref{eq:uplink-rate-expression-general} holds for any receive combining vector $\vect{v}_{k}$ but we would like to use the one that maximizes its value.
We first note that we can multiply $\vect{D}_k \vect{v}_{k}$ with any non-zero scalar without affecting the SINR value, since all the terms in the ratio are scaled equivalently. Hence, the goal of designing the receive combining  is not to find the right length of the vector $\vect{D}_k \vect{v}_{k}$ but the right direction in the vector space.
We notice that \eqref{eq:uplink-instant-SINR-level4} has the form of a generalized Rayleigh quotient:
	\begin{equation} \label{eq:uplink-instant-SINR-level4-Rayleigh}
	\mathacr{SINR}_{k}^{({\rm ul},\cent)} = \frac{ p_{k} \left |  \vect{v}_{k}^{\Htran} \vect{D}_{k}\widehat{\vect{h}}_{k} \right|^2  }{\vect{v}_{k}^{\Htran} \left(   \condSum{i=1}{i \neq k}{K} p_{i}  \vect{D}_{k}\widehat{\vect{h}}_{i} \widehat{\vect{h}}_{i}^{\Htran}\vect{D}_{k} + \vect{Z}_{k}  + \sigmaUL \vect{D}_k \right) \vect{v}_{k}  
	}
	\end{equation}
where we have replaced $\vect{D}_k^2$ by $\vect{D}_k$ in the last term of the denominator by using the relation $\vect{D}_k^2=\vect{D}_k$. Hence, the SINR can be maximized as described in Lemma~\ref{lemma:gen-rayleigh-quotient} on p.~\pageref{lemma:gen-rayleigh-quotient}. This leads to the following optimal combining vector.

\begin{corollary} \label{cor:MMSE-combining}
The instantaneous SINR in \eqref{eq:uplink-instant-SINR-level4} for UE~$k$ is maximized by the MMSE combining vector
\begin{equation} \label{eq:MMSE-combining}
\vect{v}_{k}^{{\rm MMSE}} =    p_{k} \left( \sum\limits_{i=1}^{K} p_{i}  \vect{D}_{k} ( \widehat{\vect{h}}_{i} \widehat{\vect{h}}_{i}^{\Htran} + \vect{C}_{i})\vect{D}_{k} + \sigmaUL \vect{I}_{LN} \right)^{\!-1} \vect{D}_{k}\widehat{\vect{h}}_{k}
\end{equation}
which leads to the maximum value
\begin{align} \label{eq:uplink-maximized-SINR}
\!\!\!\!\!\!\mathacr{SINR}_{k}^{({\rm ul},\cent)} \! \!= p_{k}  \widehat{\vect{h}}_{k}^{\Htran}\vect{D}_{k} \left( \condSum{i=1}{i \neq k}{K} p_{i}  \vect{D}_{k}\widehat{\vect{h}}_{i} \widehat{\vect{h}}_{i}^{\Htran}\vect{D}_{k} + \vect{Z}_{k} + \sigmaUL \vect{I}_{LN}  \right)^{-1}  \vect{D}_{k}\widehat{\vect{h}}_{k}.\!\!
\end{align}
\end{corollary}
\begin{proof}
We can maximize \eqref{eq:uplink-instant-SINR-level4} by using Lemma~\ref{lemma:gen-rayleigh-quotient} on p.~\pageref{lemma:gen-rayleigh-quotient} with $\vect{v} = \vect{v}_{k}$, $\vect{h} = \sqrt{p_{k}} \vect{D}_{k}\widehat{\vect{h}}_{k}$, and $\vect{B} = \sum_{i=1, i \neq k}^{K} p_{i}  \vect{D}_{k}\widehat{\vect{h}}_{i} \widehat{\vect{h}}_{i}^{\Htran}\vect{D}_{k} + \vect{Z}_{k} + \sigmaUL \vect{D}_k$. 
 To handle that $\vect{B}$ might not be positive definite (but only positive semi-definite), we can regularize it by replacing $ \sigmaUL \vect{D}_k$ with  $\sigmaUL \vect{I}_{LN}$. This makes $\vect{B}$ invertible and will not affect the final result since the inverse is then multiplied with $\vect{D}_k$, which removes the extra noise that was added to the unused dimensions. 
\end{proof}

\begin{figure}[t!]
	\centering 
	\begin{overpic}[width=\columnwidth,tics=10]{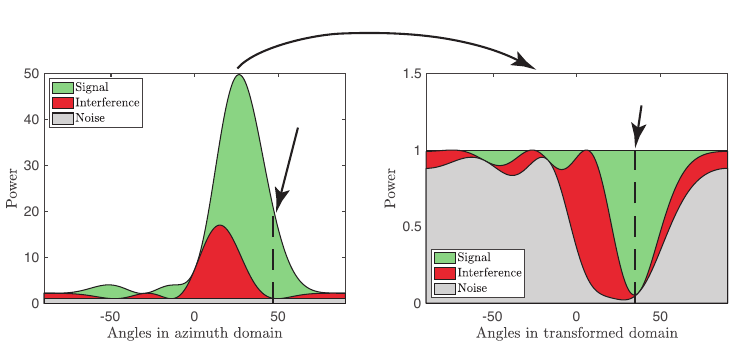}
		\put(34,34){\small{Maximum}}
		\put(37,30.5){\small{SINR}}
		\put(77,33.5){\small{Maximum SINR}}
		\put(40,43.5){\small{Spatial whitening}}
	\end{overpic}  
	\caption{This figure exemplifies how MMSE combining in  \eqref{eq:MMSE-combining} is maximizing the effective SINR when using an $N=4$ antenna AP. The left graph shows how the received signal power, interference, and noise are distributed over different azimuth angles. The curves are placed on top of each other, thus it is the colored height between them that measures the power.
	The SINR-maximizing direction of the combining vector finds a nontrivial tradeoff between having a strong signal and having low interference. 
This direction can be easily identified by transforming the received vector using a spatial whitening filter based on the conditional correlation matrix $\mathbb{E}\{\vect{D}_k \vect{y}^{\rm{ul}} (\vect{y}^{\rm{ul}})^{\Htran}\vect{D}_k | \{ \widehat{\vect{h}}_{i} \}  \}$. This results in the graph to the right where the sum of the signal, interference, and noise is the same in all directions. The maximum SINR is then achieved when the signal power is maximized. By transforming this signal back to the original domain, the MMSE combining is obtained.}
	\label{figure:whitening_example}  
\end{figure}
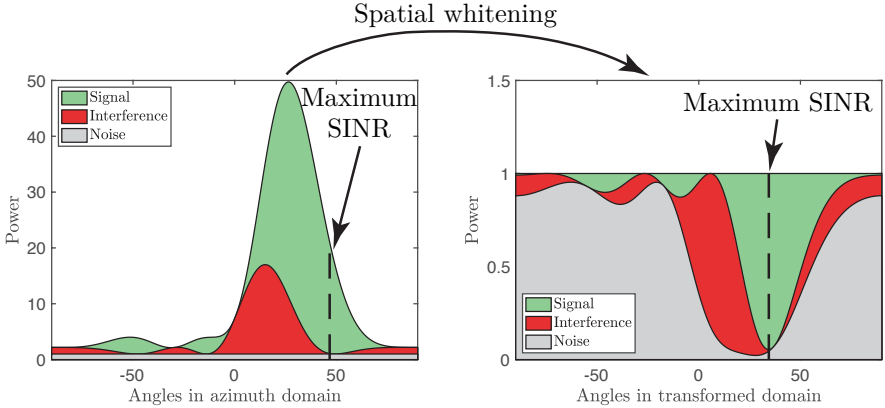

The optimal combining vector in \eqref{eq:MMSE-combining} consists of two parts: 1) the partial estimate $\vect{D}_{k}\widehat{\vect{h}}_{k}$ of the channel to UE $k$ and 2) the inverse of a matrix that equals $\mathbb{E}\{ \vect{D}_k \vect{y}^{\rm{ul}} (\vect{y}^{\rm{ul}})^{\Htran}  \vect{D}_k | \{ \widehat{\vect{h}}_{i} \}  \}$ (plus the regularization term that is added to the unused dimensions), which is the conditional correlation matrix of the received signal, given the channel estimates. The first part identifies the combining vector that would maximize the desired signal power, while the second part rotates that vector to strike a balance between achieving a strong signal and suppressing interference.
Note that the matrix inverse acts as a spatial whitening filter, as exemplified in Figure~\ref{figure:whitening_example}.
The left graph in this figure shows how the received power of the desired signal and interference are arriving from different azimuth angles, while the noise is equally distributed over all angles.
The goal of the receive combining is to identify the dashed direction where the SINR is maximized, which is clearly different from the $30^\circ$ direction where the signal is the strongest. By applying a spatial whitening filter, we obtain the graph at the right where the sum of the signal, interference, and noise powers is constant in all directions. We can then easily maximize the SINR by selecting the ``direction'' in the transformed domain where the signal is stronger, which automatically implies that the sum of the interference plus noise is at its lowest value.

It can be shown that the combining vector that maximizes the instantaneous SINR also minimizes the \gls{MSE} in the data detection, which is defined as 
\begin{align}{\rm{MSE}}_k = \mathbb{E} \left\{ \left| s_{k} - \widehat s_{k} \right |^2  \big| \left\{ \widehat{\vect{h}}_{i} : i=1,\ldots,K \right\}  \right\}.
\end{align}
This is the conditional MSE between the true data signal $s_k$ and the estimate  $\widehat s_{k} $ of it from \eqref{eq:uplink-data-estimate2}; see \cite[Sec.~4.1]{massivemimobook} for a deeper discussion. This is the reason why $\vect{v}_{k}^{{\rm MMSE}}$ is generally called \emph{MMSE combining}. This type of receive combining maximizes the mutual information of many types of channels with multiple receive antennas \cite{Park2013a}. However, the particular expression in \eqref{eq:MMSE-combining} is unique for Cell-free Massive MIMO with the centralized operation and the \gls{DCC} framework.

\enlargethispage{\baselineskip} 

 Note that $ \vect{{D}}_{k}\widehat{\vect{h}}_{i}$ in~\eqref{eq:MMSE-combining} is always non-zero except  when $ \vect{D}_{k}=\vect{0}_{LN \times LN}$, which is an uninteresting special case since the CPU only needs to apply receive combining when at least one AP serves the UE.
This implies that the CPU needs to compute all the $K$ MMSE channel estimates $\{\widehat{\vect{h}}_{il}: i =1,\ldots,K\}$ corresponding to any AP $l$ that is serving UE $k$ (i.e., APs with index $l\in \mathcal{M}_k$). The total number of complex multiplications required for channel estimation is reported in Table~\ref{table:complexity-with-level4} and is computed as discussed above~\eqref{eq:complexity-estimation}. In addition, we need to account for the complexity of computing the combining vector $\vect{v}_{k}^{{\rm MMSE}}$ in~\eqref{eq:MMSE-combining} for $k=1,\ldots,K$ once per coherence block. This complexity can be computed using the framework described in \cite[App.~B.1.1]{massivemimobook} and exploiting the fact that only a subset of the APs takes part in estimating the transmitted signal $s_k$; see Remark~\ref{remark:computational-complexity-opt-MMSE} below for further details. Table~\ref{table:complexity-with-level4} summarizes the total number of complex multiplications required for the computation of the MMSE combining vector of a generic UE $k$. Since this number grows linearly with $K$, the optimal MMSE combining makes the network unscalable according to Definition~\ref{def:scalable} on p.~\pageref{def:scalable}. We will, therefore, look for alternatives.

\begin{figure}[t!]
	\centering 
	\begin{overpic}[width=.7\columnwidth,tics=10]{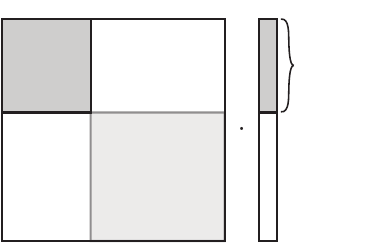}
		\put(5,63){$N|{\mathcal{M}}_k|$}
		\put(-7,40){\rotatebox{90}{$N|{\mathcal{M}}_k|$}}
		\put(4,49){Serving}
		\put(8,43){APs}
		\put(29,21){Only noise:}
		\put(27,15){$\sigmaUL \vect{I}_{LN - |{\mathcal{M}}_k|N}$}
		\put(87,46){Serving APs}
		\put(77,55){\rotatebox{-90}{$N|{\mathcal{M}}_k|$}}
		\put(77,33){\rotatebox{-90}{$LN - |{\mathcal{M}}_k|N$}}
		\put(87,16){Non-serving APs}
		\put(58,60){$-1$}
	\end{overpic}  
	\caption{Only the serving APs need to be considered when computing the MMSE combining vector $\vect{v}_{k}^{{\rm MMSE}}$  of UE $k$ in \eqref{eq:MMSE-combining}. By ordering the APs so that the $| \mathcal{M}_k |$ serving ones are listed first followed by the $L-| \mathcal{M}_k |$ non-serving APs, a block-diagonal matrix needs to be inverted. Only the shaded upper block of dimension $N | \mathcal{M}_k | \times N | \mathcal{M}_k |$ needs to be inverted and multiplied with a vector of length $N | \mathcal{M}_k |$. The computational complexity is identical to the case of having only $| \mathcal{M}_k |$ APs.}
	\label{figure:block-matrices}  
\end{figure}
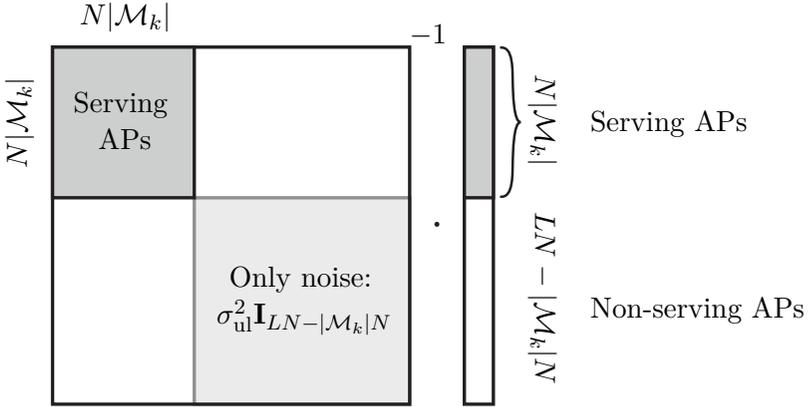

\begin{remark} [Computation of MMSE combining]\label{remark:computational-complexity-opt-MMSE}
The complexity of computing the MMSE combining vector of UE $k$ in \eqref{eq:MMSE-combining} depends on the number of serving APs $|\mathcal{M}_k|$, not the total number of APs. To observe this, recall that $\vect{D}_{k}  = \diag( \vect{D}_{k1} , \ldots, \vect{D}_{kL} )$ is a block-diagonal matrix with $\vect{D}_{kl}$ being $\vect{I}_N$ if AP $l$ serves UE $k$ (i.e., $l\in \mathcal{M}_k$) and zero otherwise. Therefore, the elements of $\{\widehat{\vect{h}}_i:i=1,\ldots,K\}$ and $\vect{Z}_k$ whose indices correspond to the zeros of $\vect{D}_k$ do not affect any term in \eqref{eq:MMSE-combining}. 
In particular, suppose the APs with index $1,\ldots,|\mathcal{M}_k|$ are the serving ones. The computations will then be as shown in Figure~\ref{figure:block-matrices}, where only the shaded parts are non-zero. Only the upper block of size 
$N|\mathcal{M}_k|\times N|\mathcal{M}_k|$ needs to be inverted and then multiplied with a vector of length $N|\mathcal{M}_k|$. The total number of complex multiplications required for these two operations can be obtained by using the framework described in \cite[App.~B.1.1]{massivemimobook} and is summarized in Table~\ref{table:complexity-with-level4}. 
If the $|\mathcal{M}_k|$ serving APs have arbitrary indices, the computational complexity remains the same since the implementation can simply reorder the APs so that the serving ones are listed first.
\end{remark}

\begin{sidewaystable} 
\centering
    \begin{tabular}{|c|c|c|} \hline
    Scheme & {Channel estimation} & Combining vector computation \\ \hline \hline
        \gls{MMSE} & $(N\tau_p + N^2)K |\mathcal{M}_k|$ & $\frac{\left(N|\mathcal{M}_k|\right)^2 + N|\mathcal{M}_k|}{2} K + \left(N|\mathcal{M}_k|\right)^2+\frac{\left(N|\mathcal{M}_k|\right)^3-N|\mathcal{M}_k|}{3} $ \\ \hline
            P-MMSE & $(N\tau_p + N^2)|\mathcal{S}_k| |\mathcal{M}_k|$ & $\frac{\left(N|\mathcal{M}_k|\right)^2 + N|\mathcal{M}_k|}{2} |\mathcal{S}_k| + \left(N|\mathcal{M}_k|\right)^2+\frac{\left(N|\mathcal{M}_k|\right)^3-N|\mathcal{M}_k|}{3} $  \\ \hline
                P-RZF & $(N\tau_p + N^2)|\mathcal{S}_k| |\mathcal{M}_k|$ & $\frac{|\mathcal{S}_k|^2+|\mathcal{S}_k|}{2}N|\mathcal{M}_k| +|\mathcal{S}_k|^2+ |\mathcal{S}_k|N|\mathcal{M}_k|+\frac{|\mathcal{S}_k|^3-|\mathcal{S}_k|}{3} $  \\ \hline
                    \gls{MR} & $(N\tau_p + N^2) |\mathcal{M}_k|$ & $-$ \\ \hline
    \end{tabular}
    \caption{The number of complex multiplications required per coherence block to compute the combining vector of UE $k$, including the computation of the channel estimates.
    Different combining schemes for the centralized operation are considered.} 
     \label{table:complexity-with-level4}
\end{sidewaystable}

\subsection{Scalable Combining Schemes for Centralized Operation}
\label{subsec:scalable-combining}

We will now develop centralized receive combining schemes that are scalable, in the sense that the computational complexity per UE is independent of $K$.
The simplest solution is to use MR combining with
\begin{align}\label{MR_combining_D_k}
\vect{v}_k^{{\rm MR}} = \vect{D}_{k}\widehat{\vect{h}}_k 
\end{align} 
 which maximizes the numerator $ |  \vect{v}_{k}^{\Htran} \vect{D}_{k}\widehat{\vect{h}}_{k} |^2$ of the effective SINR in \eqref{eq:uplink-instant-SINR-level4}.
 In other words, the power of the desired signal is maximized while the existence of interference and estimation errors is ignored. Since the MMSE channel estimate is directly used as the MR combining vector, the computational complexity is equal to that of obtaining the partial MMSE channel estimate $\vect{D}_{k}\widehat{\vect{h}}_k$. This requires a total of $(N\tau_p + N^2) |\mathcal{M}_k|$ complex multiplications, which is summarized in Table~\ref{table:complexity-with-level4}.
Since no additional computations are needed, MR combining is a scalable solution with low complexity.
However, we will observe later that the performance can be poor in the presence of interference, since a high degree of favorable propagation cannot be guaranteed between all UEs (see Section~\ref{subsec-favorable-propagation} on p.~\pageref{subsec-favorable-propagation}).
Two scalable combining schemes that are superior to MR, in terms of dealing with interference, are therefore derived next by taking inspiration from the optimal MMSE combining in~\eqref{eq:MMSE-combining}.

\subsubsection{Partial MMSE Combining}

In a large cell-free network with distributed UEs, it is reasonable to believe that the interference that affects UE $k$ is mainly generated by a small subset of the other UEs, which are located in the neighborhood of UE $k$. We want to utilize this observation to reduce the computational complexity of the optimal MMSE combining and thereby achieve a scalable alternative.
To this end, we assume that  only the UEs that are served by partially the same APs as UE $k$ should be included in the inverse matrix in \eqref{eq:MMSE-combining}. These UEs have indices in the set
\begin{equation} \label{eq:S_k}
\mathcal{S}_k = \left\{ i: \vect{D}_{k} \vect{D}_{i} \neq \vect{0}_{LN \times LN} \right\}.
\end{equation}
By utilizing $\mathcal{S}_k$, an alternative \emph{\gls{P-MMSE}} scheme can be defined as follows \cite{Attarifar2020a}, \cite{Bjornson2020a}, \cite{Nayebi2016a}:
\begin{equation} \label{eq:P-MMSE-combining}
\vect{v}_k^{{\rm P-MMSE}} =    p_k\left(\sum\limits_{i \in \mathcal {S}_k} p_{i}  \vect{D}_{k}\widehat{\vect{h}}_{i} \widehat{\vect{h}}_{i}^{\Htran}\vect{D}_{k} + {\bf{Z}}_{{\mathcal S}_k} + \sigmaUL  \vect{I}_{LN} \right)^{\!-1} \vect{{D}}_{k}\widehat{{\bf h}}_k
\end{equation}
with 
\begin{align}\label{eq:Z_k-prime}
{\bf{Z}}_{{\mathcal S}_k} 
= \sum\limits_{i \in \mathcal{S}_k} p_{i} \vect{{D}}_{k}
\vect{C}_{i} \vect{{D}}_{k}.
\end{align}
This scheme is not designed to be optimal from an SE perspective but as a scalable approximation of the optimal MMSE combining. We note that P-MMSE combining coincides with MMSE combining only when UE $k$ is served by all APs; that is, $\vect{D}_k=\vect{I}_{LN}$ and $\mathcal{S}_k=\{1,\ldots,K\}$.
Another way to view this is that P-MMSE combining would be optimal if only the UEs with indices in $\mathcal{S}_k$ were active. However, MMSE combining and P-MMSE combining are generally different schemes.

In Section~\ref{sec:selecting-DCC} on p.~\pageref{sec:selecting-DCC}, a DCC selection method was described where each AP serves exactly one UE per pilot sequence, which guarantees $|\mathcal{D}_l| = \tau_p$ for all $l$. Using this scheme, $|\mathcal{S}_k|=\tau_p$ if all the APs that serve UE $k$ co-serve the same set of UEs. This is the smallest value that $|\mathcal{S}_k|$ can take.
 Moreover, it holds that $|\mathcal{S}_k| \leq (\tau_p-1) |\mathcal{M}_k|+1$, where the upper bound is achieved in the unlikely case that all the APs in $\mathcal{M}_k$ serve UE $k$ but otherwise serve entirely non-overlapping sets of UEs. Importantly, even this unlikely upper bound is independent of $K$.
The number of complex multiplications required by P-MMSE for channel estimation and combining vector computation is reported in Table~\ref{table:complexity-with-level4}. In computing those numbers, we have taken into account the discussion from Remark~\ref{remark:computational-complexity-opt-MMSE} that only the number of serving APs affects the size of the matrix inverse and matrix-vector multiplication. The exact expression is rather lengthy but, importantly, it is independent of $K$, which makes P-MMSE a scalable combining scheme that supports an arbitrarily large number of UEs. 

In practice, some fine-tuning can be done to improve the performance of P-MMSE by including additional UEs in $\mathcal{S}_k$ to deal with strongly interfering UEs that are only served by other APs. A proper selection of these extra UEs will increase network performance without violating the scalability. However, in this monograph, we stick to the definition of $\mathcal{S}_k$ in \eqref{eq:S_k} for brevity.

\subsubsection{Partial Regularized Zero-Forcing Combining}

The complexity of computing the P-MMSE combining vector scales with $(N|\mathcal{M}_k|)^3$. Each AP is supposed to have a relatively small number of antennas $N$ in cell-free networks, but $|\mathcal{M}_k|$ might be large when there are many APs in the vicinity of UE $k$. Since the combining vector is computed at the CPU, which  is assumed to have high computational capability, the complexity might not be an issue. However, it is nevertheless interesting to explore if a further complexity reduction can be achieved without affecting the performance to any great extent.
We will therefore present an alternative receive combining scheme, which is obtained as a simplification of P-MMSE. To this end, we note that if the channel conditions of the interfering UEs in $\mathcal{S}_k$ are good, all the corresponding estimation error correlation matrices $\{{\bf C}_i : i \in {\mathcal S}_k\}$ in $\vect{Z}_{{\mathcal S}_k}$ will be small. If we neglect $\vect{Z}_{{\mathcal S}_k}$ in \eqref{eq:P-MMSE-combining}, we obtain
\begin{equation} \label{eq:P-RZF-combining_0}
\vect{v}_k^{{\rm P-RZF}} =   
p_k\left(\sum\limits_{i \in \mathcal {S}_k} p_{i}  \vect{D}_{k}\widehat{\vect{h}}_{i} \widehat{\vect{h}}_{i}^{\Htran}\vect{D}_{k} + \sigma^2_{\rm ul}\vect{I}_{LN} \right)^{\! -1} \vect{{D}}_{k}\widehat{{\bf h}}_k.
\end{equation}
This change has a negligible impact on the complexity but it enables further reformulation.
Let us construct $\vect{\widehat{H}}_{\mathcal{S}_k} \in \mathbb{C}^{LN\times |\mathcal{S}_k|}$ by stacking all the vectors $\vect{\widehat{h}}_i$ with indices $i\in \mathcal{S}_k$ with the first column being $\vect{\widehat{h}}_k$.
We further define $\vect{{P}}_{\mathcal{S}_k} \in \mathbb{R}^{|\mathcal{S}_k| \times |\mathcal{S}_k|}$ as a diagonal matrix containing the transmit powers $p_i$ for $i\in \mathcal{S}_k$, listed in the same order as the columns of $\vect{\widehat{H}}_{\mathcal{S}_k}$.
By letting $[\cdot]_{:,1}$ denote the operation of only keeping the first column of its matrix argument,
we can reformulate \eqref{eq:P-RZF-combining_0} as 
\begin{align} \nonumber
\vect{v}_{k}^{{\rm P-RZF}} &= \left [ \left( \vect{D}_{k} \vect{\widehat{H}}_{\mathcal{S}_k}  \vect{{P}}_{\mathcal{S}_k} \vect{\widehat{H}}_{\mathcal{S}_k}^{\Htran}\vect{D}_{k} + \sigma^2_{\rm ul}\vect{I}_{LN} \right)^{\!-1} \vect{{D}}_{k}\vect{\widehat{H}}_{\mathcal{S}_k} \vect{{P}}_{\mathcal{S}_k} \right]_{:,1} \\ \nonumber
&=  \left [\vect{{D}}_{k}\vect{\widehat{H}}_{\mathcal{S}_k} \vect{{P}}_{\mathcal{S}_k} \left( \vect{\widehat{H}}^{\Htran}_{\mathcal{S}_k}\vect{{D}}_{k}\vect{{D}}_{k}\vect{\widehat{H}}_{\mathcal{S}_k} \vect{{P}}_{\mathcal{S}_k} +\sigmaUL\vect{{I}}_{|\mathcal{S}_k|} \right)^{-1} \right]_{:,1}  \\
&=  \left [\vect{{D}}_{k}\vect{\widehat{H}}_{\mathcal{S}_k}\left( \vect{\widehat{H}}^{\Htran}_{\mathcal{S}_k}\vect{{D}}_{k}\vect{\widehat{H}}_{\mathcal{S}_k}+\sigmaUL\vect{{P}}_{\mathcal{S}_k}^{-1} \right)^{-1} \right]_{:,1}  \label{eq:P-RZF-combining_1}
\end{align}
where the second equality follows from the first identity of Lemma~\ref{lemma-matrix-identities} on p.~\pageref{lemma-matrix-identities} and the third equality utilizes the fact that $\vect{{D}}_{k}\vect{{D}}_{k} = \vect{{D}}_{k}$.

We call \eqref{eq:P-RZF-combining_1} \emph{\gls{P-RZF} combining}. 
This name indicates that the matrix expression in \eqref{eq:P-RZF-combining_1} has the same form as the pseudo-inverse $\vect{{D}}_{k}\vect{\widehat{H}}_{\mathcal{S}_k}( \vect{\widehat{H}}^{\Htran}_{\mathcal{S}_k}\vect{{D}}_{k}\vect{\widehat{H}}_{\mathcal{S}_k} )^{-1}$ to the partial channel matrix $\vect{\widehat{H}}_{\mathcal{S}_k}^{\Htran} \vect{{D}}_{k}$. The only difference is that the inverse has been regularized by adding the matrix $\sigmaUL \vect{{P}}_{\mathcal{S}_k}^{-1}$ containing the noise variance and transmit powers. Combining vectors based on pseudo-inverses force the interference between the UEs to zero, but might lead to large losses in the desired signal power when the UEs have similar channels. P-RZF is balancing between suppressing interference and maintaining strong desired signal powers.

The total number of complex multiplications required for channel estimation and combining vector computation with P-RZF is reported in Table~\ref{table:complexity-with-level4}. As expected, this number is independent of $K$. Hence, this scheme is also scalable. The main benefit over the P-MMSE is that an $|\mathcal{S}_k| \times |\mathcal{S}_k|$ matrix is inverted in~\eqref{eq:P-RZF-combining_1} instead of an $N|\mathcal{M}_k| \times N|\mathcal{M}_k|$ matrix as in~\eqref{eq:P-MMSE-combining}, which can substantially reduce the complexity since we typically have $N|\mathcal{M}_k| \ge |\mathcal{S}_k|$ in cell-free networks. Most likely, this benefit comes with an SE loss since, in general, the channel conditions will not be so good to every interfering UE so that the estimation errors can be ignored. The tradeoff will be evaluated later.

\subsection{Fronthaul Signaling Load for Centralized Operation}
\label{sec:level4-fronthaul}

We wrap up the analysis of the centralized operation by quantifying the required fronthaul signaling. We will do this by counting the number of complex scalars that must be sent over the fronthaul links. In practice, these scalars will have to be quantized and different types of information can be quantized using a different number of bits per scalar. However, if all the scalars are sent over the fronthaul using the same bit precision, counting the number of scalars is sufficient to compare the  signal load of different methods.

In each coherence block, AP $l$ needs to send $\tau_p N$ complex scalars representing the pilot signals $\{ \vect{y}_{t l}^{\rm pilot} : t = 1,\ldots,\tau_p \}$ and $\tau_uN$ complex scalars representing the received data signal $\vect{y}_{l}^{\rm{ul}}$. As summarized in Table~\ref{tab:signaling-level4-level3}, this requires a total of $(\tau_p + \tau_u)NL$ complex scalars, irrespective of the combining scheme used for detection at the CPU.\footnote{Note that $\tau_p N$ is an upper bound on the fronthaul signaling load for pilot signals, which is achieved in the case that AP $l$ assigns all the $\tau_p$ available pilots to the UEs that it is serving. This was assumed by the pilot assignment scheme presented in Section~\ref{sec:selecting-DCC} on p.~\pageref{sec:selecting-DCC}, but is not the case in general. In other cases, AP $l$ exchanges a lower number of complex scalars for the pilot signals through the fronthaul links.}
Since this value does not grow with $K$, we can conclude that the fronthaul signaling is scalable.

Notice that the implementations of the MMSE channel estimator and the P-MMSE combining scheme require knowledge of the spatial correlation matrices $\{\vect{R}_{kl}: k=1,\ldots,K, \, l=1,\ldots,L\}$. In a centralized network, where the APs only act as relays, this information can be acquired directly at the CPU by using one of the pilot signaling methods that already exist in Cellular Massive MIMO literature (see \cite{Sanguinetti2019a} for an overview). Hence, there is no need to exchange the spatial correlation matrices through the fronthaul links. As discussed later in Section~\ref{sec:level3-fronthaul}, the situation is different in the distributed implementation presented next, where the pilot signals are not sent to the CPU but used locally at the APs for channel estimation. Statistical information must then be acquired at the APs and gathered at the CPU using the fronthaul.

\begin{table}[t]  
	\caption{The number of complex scalars to send from the  APs  to the CPU over the fronthaul either in each coherence block or for each realization of the channel statistics (e.g., user locations). Both centralized and distributed operation are considered.}  \label{tab:signaling-level4-level3}
	\centering
	\begin{tabular}{|c|c|c|} 
		\hline
		{Implementation} & {Per coherence block}&{Statistics}    \\      \hline\hline 
		
		Centralized &   $(\tau_p+\tau_u) NL$  & $ --$     \\ \hline    
		Distributed: opt LSFD &   $\tau_u \sum\limits_{l=1}^L |\mathcal{D}_l|$  & $ \frac{3K+1}{2}\sum\limits_{l=1}^{L} |\mathcal{D}_l|$     \\ \hline   
				Distributed: n-opt LSFD &   $\tau_u \sum\limits_{l=1}^L |\mathcal{D}_l|$  & $\sum\limits_{l=1}^{L} \sum\limits_{k \in \mathcal{D}_l} \frac{3|\mathcal{S}_k|+1}{2}$  \\ \hline   
				Distributed: no LSFD &   $\tau_u \sum\limits_{l=1}^L |\mathcal{D}_l|$  & $ --$  \\ \hline  
	\end{tabular}
\end{table}

\enlargethispage{\baselineskip} 

\section{Distributed Uplink Operation}
\label{sec:level3-uplink}

Instead of entirely delegating the channel estimation and data detection to the CPU, a User-centric Cell-free Massive MIMO network can be implemented to carry out as much of the signal processing as possible in a distributed manner.
A motivating factor is that each AP can easily be equipped with a baseband processor that is (at least) as powerful as those in the UEs, thus algorithms that can make efficient use of it should be developed. The goal is to enable an ultra-dense deployment with $L \gg K$, where the CPU's capabilities are dimensioned based on the number of UEs, but largely unaffected by the number of APs. In other words, we should be able to add new APs to the network without having to upgrade the CPUs, thanks to the property that each AP comes with a local processor.
Another reason is to reduce the fronthaul signaling compared to the centralized operation and to avoid the fronthaul quantization distortion, which exists in practice but is neglected in this monograph.
Since a key property of cell-free networks is that multiple APs are involved in the service of each UE, the final data detection must be carried out at a point where the inputs from multiple APs are combined. Recall that the CPU is a logical entity so the unavoidable CPU tasks can be physically assigned to a nearby edge-cloud processor (or even one of the serving APs), so there is no need for a physical central unit.
The following two-stage procedure is called \emph{distributed operation} in this monograph and is schematically described in Figure~\ref{figure:distributed_centralized}(b).

In the first stage, each AP $l$ uses its received pilot signals $\{ \vect{y}_{t l}^{\rm pilot} : t = 1,\ldots,\tau_p \}$ to locally estimate the channels $\{\widehat{{\bf h}}_{il} : i \in {\mathcal D}_l \}$.
For each UE $k$, the AP can then use the local estimates to select a local receive combining vector ${\bf v}_{kl}$. According to~\eqref{eq:uplink-data-estimate}, the AP then computes its local estimate of $s_k$ as 
\begin{align} \label{eq:local-data-estimate-level3}
\widehat s_{kl} &= \vect{v}_{kl}^{\Htran} \vect{D}_{kl} {\bf y}_{l}^{\rm{ul}}.
\end{align}
Note that this estimate is zero for $k \not \in \mathcal{D}_l$, thus AP $l$ only needs to compute \eqref{eq:local-data-estimate-level3} for the UEs that is serves.

In the second stage, the local data estimates of all APs are gathered at the CPU where they are combined/fused into a final estimate of the UE data. The CPU computes its estimate as a linear combination of the local estimates:
\begin{align} \label{eq:global-data-estimate-level3}
\widehat{s}_k = \sum_{l=1}^{L} a_{kl}^{\star} \widehat{s}_{kl} = \sum_{l=1}^{L} a_{kl}^{\star} \vect{v}_{kl}^{\Htran} \vect{D}_{kl} {\bf y}_{l}^{\rm{ul}}
\end{align}
where $a_{kl} \in \mathbb{C}$ is the weight that the CPU assigns to the local signal estimate that AP $l$ has of the signal from UE $k$. Note that we include all the AP indices since $\widehat s_{kl}=0$ for $l\notin \mathcal{M}_k$ and the results do not change.
To limit the fronthaul signaling, the APs are only sending the local data estimates to the CPU and not the channel estimates. Hence, the CPU needs to select the weights  $\{ a_{kl} : k=1,\ldots,K, l \in \mathcal{M}_k\}$ as a deterministic function of the channel statistics. Intuitively, the CPU should assign a larger weight to an AP having a large \gls{SNR} to the UE than an AP having a small \gls{SNR}, but also the interference situation and receive combining scheme should be taken into consideration. 

The structure of the second stage resembles what is called \emph{\gls{LSFD}} and was originally used in Cellular Massive MIMO \cite{Adhikary2017a}, \cite{ashikhmin2012pilot}. Unlike the centralized operation, the two-stage procedure outlined above allows the network to make use of the distributed processing capabilities of the APs. The received signals are primarily processed at the points where they were received. Even if it the operation is not entirely distributed, it is as distributed as a cell-free network, utilizing coherent cooperation between APs, can become since only the final signal detection is done at the CPU.

The achievable SE for arbitrary local receive combining $\{\vect{v}_{kl} : k=1,\ldots,K, l\in  \mathcal{M}_k\}$ and LSFD weights will be derived next. After that, we will show how to make a judicious selection of the local combining vectors and how to compute the optimal LSFD weights. Since the preferred solutions turn out to be unscalable for networks with many UEs, we will provide scalable alternatives.

\subsection{Spectral Efficiency and Optimal LSFD Weights}
\label{sec:level3-SE}

We will now compute an SE expression for UE $k$ in the distributed operation. Plugging~\eqref{eq:DCC-uplink} into~\eqref{eq:global-data-estimate-level3} yields
\begin{align} 
\widehat{s}_k =\left(\sum_{l=1}^{L} a_{kl}^{\star} \vect{v}_{kl}^{\Htran}\vect{D}_{kl} \vect{h}_{kl}\right) s_k +\condSum{i=1}{i \neq k}{K}  \Bigg(\sum_{l=1}^{L} a_{kl}^{\star} \vect{v}_{kl}^{\Htran} \vect{D}_{kl}\vect{h}_{il}\Bigg)  s_i + n^\prime_{k}\label{eq:CPU_linearcomb-level3}
\end{align}
where $n^\prime_{k} = \sum_{l=1}^{L} a_{kl}^{\star} \vect{v}_{kl}^{\Htran}\vect{D}_{kl}\vect{n}_{l}$ represents the noise. For brevity, we define the vector $\vect{g}_{ki} \in \mathbb{C}^L$ computed as
\begin{equation}
 \vect{g}_{ki} = \begin{bmatrix}
 \vect{v}_{k1}^{\Htran}\vect{D}_{k1} \vect{h}_{i1} \\
 \vdots \\
  \vect{v}_{kL}^{\Htran} \vect{D}_{kL} \vect{h}_{iL}
  \end{bmatrix}.
  \end{equation}
 This is the vector with the receive-combined channels between UE $i$ and all APs that serve UE $k$. Using this notation, \eqref{eq:CPU_linearcomb-level3} can be expressed as 
\begin{align} 
\widehat{s}_k =\vect{a}_{k}^{\Htran}\vect{g}_{kk} s_k + \condSum{i=1}{i \neq k}{K}\vect{a}_{k}^{\Htran}\vect{g}_{ki}  s_i + n^\prime_{k} \!\!\label{eq:CPU_linearcomb_eff-level3}
\end{align}
where $\vect{a}_{k} = [ a_{k1} \, \ldots \, a_{kL} ]^{\Ttran} \in \mathbb{C}^L$ is the LSFD weight vector of UE $k$. The realizations of $\vect{g}_{ki}$ change in every coherence block, while $\vect{a}_{k}$ is supposed to be a deterministic vector that the CPU can select without knowing the channel realizations.
We notice that \eqref{eq:CPU_linearcomb_eff-level3} has the structure of a single-antenna channel where $\{\vect{a}_{k}^{\Htran}\vect{g}_{ki}: i=1,\ldots,K\}$ represent the effective channels of the different UEs. 
 Although the effective channel $\vect{a}_{k}^{\Htran}\vect{g}_{kk}$ is unknown at the CPU, we notice that its average $\mathbb{E}\{\vect{a}_{k}^{\Htran}  \vect{g}_{kk}\} = \vect{a}_{k}^{\Htran} \mathbb{E}\{ \vect{g}_{kk}\}$ is deterministic and non-zero if the receive combining is properly selected. Therefore, it can be assumed known\footnote{When dealing with ergodic capacities, all deterministic parameters can be assumed known without loss of generality, because these can be estimated using a finite number of transmission resources, while the capacity is only achieved as the amount of transmission resources goes to infinity. Hence, the estimation overhead for obtaining deterministic parameters is  negligible.}
and used to compute the following achievable SE. 

\begin{theorem} \label{theorem:uplink-capacity-level3}
An achievable SE of UE $k$ in the distributed operation is
	\begin{equation} \label{eq:uplink-rate-expression-level3}
	\begin{split}
	\mathacr{SE}_{k}^{({\rm ul},\dist)} = \frac{\tau_u}{\tau_c} \log_2  \left( 1 + \mathacr{SINR}_{k}^{({\rm ul},\dist)}  \right) \quad \textrm{bit/s/Hz}
	\end{split}
	\end{equation}
	where the effective SINR is given by
	\begin{equation} \label{eq:uplink-instant-SINR-level3}
	\mathacr{SINR}_{k}^{({\rm ul},\dist)}\! =\!  \frac{ p_{k} \left| \vect{a}_{k}^{\Htran} \mathbb{E}\{ \vect{g}_{kk}\} \right|^2  }{ 
	 \vect{a}_{k}^{\Htran} \left(
\sum\limits_{i=1}^{K} p_{i} \mathbb{E} \{  \vect{g}_{ki} \vect{g}_{ki}^{\Htran}  \} - p_{k}  \mathbb{E}\{ \vect{g}_{kk}\} \mathbb{E}\{ \vect{g}_{kk}^{\Htran}\} + 
		\vect{F}_{k}
		\right)
		\vect{a}_{k} 
	}
	\end{equation}
	and
	\begin{equation}
	\vect{F}_{k}=\sigmaUL\diag\left(\mathbb{E}\left\{ \left\|  \vect{D}_{k1}\vect{v}_{k1}  \right\|^2\right\}, \ldots, \mathbb{E}\left\{ \left\|  \vect{D}_{kL}\vect{v}_{kL}  \right\|^2\right\}\right) \in \mathbb{R}^{L \times L}.
	\end{equation}
\end{theorem}
\begin{proof}
The proof is given in Appendix~\ref{proof-of-proposition:uplink-capacity-level3} on p.~\pageref{proof-of-proposition:uplink-capacity-level3}.
\end{proof}

The pre-log factor ${\tau_u}/{\tau_c}$ in \eqref{eq:uplink-rate-expression-level3} is the fraction of each coherence block that is used for uplink data transmission. The term $\mathacr{SINR}_{k}^{({\rm ul},\dist)}$ takes the form of an effective SINR and is the ratio of the signal power term $p_{k} \left| \vect{a}_{k}^{\Htran} \mathbb{E}\{ \vect{g}_{kk}\} \right|^2 $ and a term containing noise and interference (including some self-interference from the desired signal due to the imperfect channel knowledge).
Recall that the word ``effective'' refers to that this SE is effectively the same as that of a deterministic single-antenna single-user channel where \eqref{eq:uplink-instant-SINR-level3} is the SNR and the receiver has perfect CSI. In particular, this means that the desired data signal can be encoded and the received signal can be decoded as if we would communicate over such an AWGN channel.

The SE expression in~\eqref{eq:uplink-rate-expression-level3} is very general. 
Although this monograph considers correlated Rayleigh fading channels and MMSE channel estimation, \eqref{eq:uplink-rate-expression-level3} is actually a valid SE for any channel estimator and channel model, 
in contrast to the SE expression in~\eqref{eq:uplink-rate-expression-general} which is explicitly using those assumptions. This was previously discussed in Section~\ref{subsec:alternative-bound}.

The achievable SE in~\eqref{eq:uplink-rate-expression-level3} holds for any local receive combining and LSFD vector $\vect{a}_{k}$, but we are naturally seeking a combination that makes the SE as high as possible.
We start with considering the local receive combining vectors. Since AP $l$ locally selects $\vect{v}_{kl}$ for $k \in \mathcal{D}_l$, without receiving input from other APs, there is no way for its combining scheme to be optimal from a network-wide perspective. However, the AP can do as well as it can by optimizing a local performance metric. Recall from Section~\ref{subsec:optimal-receive-combining} that the optimal receive combining in a centralized implementation minimizes the MSE in the data detection. In analogy with this result, we can let AP $l$ select the receive combining giving the best local estimate $\widehat s_{kl} = \vect{v}_{kl}^{\Htran} \vect{D}_{kl} {\bf y}_{l}^{\rm{ul}}$ in terms of minimal conditional MSE:
\begin{align}\label{eq:local-MSE}
\mathbb{E}\left \{ \left| s_{k} - \vect{v}_{kl}^{\Htran} \vect{D}_{kl}\vect{y}_{l}^{\rm ul} \right|^2  \Big| \left\{ \widehat{\vect{h}}_{il} : i=1,\ldots,K  \right\} \right\}.
\end{align}
The combining vector $\vect{v}_{kl}$ that minimizes~\eqref{eq:local-MSE} is
\begin{equation} \label{eq:L-MMSE-combining-single-AP}
\vect{v}_{kl}^{{\rm L-MMSE}} =  p_{k}  \left( \sum\limits_{i=1}^{K} p_{i} \left( \widehat{\vect{h}}_{il} \widehat{\vect{h}}_{il}^{\Htran} + \vect{C}_{il} \right) + \sigmaUL  \vect{I}_{N} \right)^{-1}   \vect{D}_{kl}  \widehat{\vect{h}}_{kl}
\end{equation}
where $\vect{C}_{il}$ is the error correlation matrix given in \eqref{eq:error-correlation}.
This can be proved by computing the conditional expectation in~\eqref{eq:local-MSE} and equating its first derivative with respect to $\vect{v}_{kl}$ to zero. One can also show that \eqref{eq:L-MMSE-combining-single-AP} is the combining vector that would maximize the SE if AP $l$ detected the data signal $s_{k}$ locally. 
We call \eqref{eq:L-MMSE-combining-single-AP} \emph{\gls{L-MMSE} combining} to distinguish it from the MMSE combining in \eqref{eq:MMSE-combining}, which is applied at the CPU in the centralized operation. L-MMSE and MMSE combining coincide (except for that the latter is padded with zeros) in the special case when only AP $l$ serves UE $k$ (i.e., $\mathcal{M}_k = \{ l \}$). Hence, the computational complexity of L-MMSE  is the same as for MMSE combining in Table~\ref{table:complexity-with-level4} but with $|\mathcal{M}_k|=1$. Since this complexity grows with $K$, we notice that L-MMSE combining is not a scalable combining scheme.

We now shift focus to the LSFD weights $\{ \vect{a}_{k} : k = 1,\ldots,K\}$. For any given local receive combining vectors, $\vect{a}_{k}$ can be optimized by the CPU to maximize the SE. Since it is a deterministic vector, it is optimized as a function of the channel statistics.
We notice that \eqref{eq:uplink-instant-SINR-level3} is a generalized Rayleigh quotient with respect to $\vect{a}_k$. Hence, we can maximize the effective SINR by using Lemma~\ref{lemma:gen-rayleigh-quotient} on p.~\pageref{lemma:gen-rayleigh-quotient}.
The optimal $\vect{a}_{k}$ is thus obtained as follows.
\begin{corollary} \label{cor:LSFD-optimal}
	The effective SINR in \eqref{eq:uplink-instant-SINR-level3} for UE $k$ is maximized by
		\begin{equation} \label{eq:LSFD-vector}
	{\vect{a}}_{k}^{\rm opt} = p_k  \left( \sum\limits_{i=1}^{K}
	p_{i} \mathbb{E} \left\{ {\vect{g}}_{ki} {\vect{g}}_{ki}^{\Htran} \right\} +  {\vect{F}}_{k} + \widetilde{\vect{D}}_k  \right)^{-1} \mathbb{E}\left\{ {\vect{g}}_{kk}\right\}
	\end{equation}
	where $\widetilde{\vect{D}}_k\in\mathbb{R}^{L\times L}$ is the diagonal matrix with the $(l,l)$th element being  one if $l\notin\mathcal{M}_k$ and zero otherwise. This leads to the maximum value
	\begin{align} \nonumber
	&\mathacr{SINR}_{k}^{({\rm ul},\dist)} =
	\\ &p_{k}  \mathbb{E}\left\{ {\vect{g}}_{kk}^{\Htran}\right\} \! \left( \sum\limits_{i=1}^{K}
	p_{i} \mathbb{E} \left\{ {\vect{g}}_{ki} {\vect{g}}_{ki}^{\Htran} \right\} -p_k \mathbb{E}\left\{ {\vect{g}}_{kk}\right\}\mathbb{E}\left\{ {\vect{g}}_{kk}^{\Htran}\right\}+ {\vect{F}}_{k}+\widetilde{\vect{D}}_k  \right)^{\!\!-1}\!\!\!\mathbb{E}\left\{ {\vect{g}}_{kk}\right\} . \label{eq:sinr-level3}
	\end{align}
\end{corollary}
\begin{proof}
We can maximize \eqref{eq:uplink-instant-SINR-level3} by using Lemma~\ref{lemma:gen-rayleigh-quotient} on p.~\pageref{lemma:gen-rayleigh-quotient} with $\vect{v} = \vect{a}_{k}$, $\vect{h} = \sqrt{p_{k}} \mathbb{E}\left\{ {\vect{g}}_{kk}\right\}$, and $\vect{B} = \sum_{i=1}^{K} p_{i} \mathbb{E} \{  \vect{g}_{ki} \vect{g}_{ki}^{\Htran}  \} - p_{k}  \mathbb{E}\{ \vect{g}_{kk}\} \mathbb{E}\{ \vect{g}_{kk}^{\Htran}\} + 
		\vect{F}_{k} + \widetilde{\vect{D}}_k$.  The term $\widetilde{\vect{D}}_k$ is not present in \eqref{eq:uplink-instant-SINR-level3} but has been added to make $\vect{B}$ positive definite and thereby invertible.\footnote{Alternatively, pseudo-inverses could have been used in \eqref{eq:LSFD-vector} and \eqref{eq:sinr-level3}, but the end result will be the same since $\widetilde{\vect{D}}_k \mathbb{E}\left\{ {\vect{g}}_{kk}\right\} = \vect{0}_L$.} This addition will not affect the result since $\widetilde{\vect{D}}_k \mathbb{E}\left\{ {\vect{g}}_{kk}\right\} = \vect{0}_L$. 
\end{proof}

This corollary provides the \gls{opt} LSFD vector that maximizes the effective SINR in~\eqref{eq:uplink-instant-SINR-level3}. The optimal vector has a structure resembling that of MMSE combining since it is also computed to maximize a generalized Rayleigh quotient.
The evaluation of ${\vect{a}}_{k}^{\rm opt}$ in \eqref{eq:LSFD-vector} requires knowledge of the $L$-dimensional statistical vector $\mathbb{E}\left\{  \vect{g}_{kk} \right\} $, the $L\times L$ Hermitian complex matrix $\mathbb{E} \left\{ {\vect{g}}_{ki} {\vect{g}}_{ki}^{\Htran}\right\}$, for $i=1,\ldots,K$, and the $L$ diagonal elements of the real-valued matrix ${\vect{F}}_{k}$. These statistical vectors and matrices can be computed at the APs, based on the received uplink signals and their choices of local combining vectors. The CPU is unable to compute them locally since it does not have access to those signals, thus the APs need to send these statistical parameters to the CPU through the fronthaul links. We will quantify the fronthaul signaling load in Section~\ref{sec:level3-fronthaul} but it is clear that the number of parameters grows with $K$, thus making the use of the opt LSFD vector unscalable according to Definition~\ref{def:scalable} on p.~\pageref{def:scalable}.

We notice that the elements of $\vect{g}_{ki}$, for all $i$, and $\vect{F}_k$ whose indices correspond to the zero rows of $\vect{D}_{kl}$ are all zero. By following the approach outlined in Remark~\ref{remark:computational-complexity-opt-MMSE}, the computation of ${\vect{a}}_{k}^{\rm opt}$ can be carried out while ignoring the $L - |\mathcal{M}_k|$ APs that are not serving UE $k$  (see also Figure~\ref{figure:block-matrices}).
The corresponding elements of $\vect{a}_k$ can be set to zero
and the computation of $\vect{a}_k^{\rm opt}$  only requires the inversion of a $ |\mathcal{M}_k| \times |\mathcal{M}_k|$ matrix and multiplication of it by an $ |\mathcal{M}_k|$-length vector, which corresponds to $|\mathcal{M}_k|^2+\frac{|\mathcal{M}_k|^3-|\mathcal{M}_k|}{3}$ complex multiplications. This is reported in Table~\ref{table:complexity-with-level3}.

\begin{remark}[Tightness of the capacity bound]  \label{remark-tightness-bounds}
The SE expression\linebreak presented in Theorem~\ref{theorem:uplink-capacity-level3} is derived based on the \gls{UatF} bound from the Cellular Massive MIMO literature \cite{massivemimobook}, \cite{Marzetta2016a} in which the channel estimates are used for receive combining but ``forgotten'' when detecting the data. While this bounding principle is artificial when applied in cellular networks, where the combining and signal detection are carried out at the same place, it makes good sense in cell-free networks where each AP carries out local combining based on its local CSI and the CPU carries out the data detection without CSI.

Signal detection without access to channel knowledge generally leads to rather poor performance since the receiver does not know the desired channel realization $\vect{a}_{k}^{\Htran} \vect{g}_{kk}$ but only its mean value $\vect{a}_{k}^{\Htran} \mathbb{E}\{ \vect{g}_{kk}\}$. However, if there is a high degree of channel hardening so that 
$\vect{a}_{k}^{\Htran} \vect{g}_{kk} \approx \vect{a}_{k}^{\Htran} \mathbb{E}\{ \vect{g}_{kk}\}$, the capacity bound in Theorem~\ref{theorem:uplink-capacity-level3} is expected to be close to the SE that we would get if $\vect{a}_{k}^{\Htran} \vect{g}_{kk}$ is perfectly known.
As shown in Section~\ref{subsec-channel-hardening} on p.~\pageref{subsec-channel-hardening}, there is no guarantee that this will be the case in cell-free networks; the degree of channel hardening can be low when  having a small number of antennas per AP \cite{Chen2018b}. In that case, the SE expression in \eqref{eq:uplink-rate-expression-level3} underestimates the practically achievable SE, but it is anyway the state-of-the-art capacity lower bound and will be used in this monograph. We will evaluate the tightness of the bound later in this section.
\end{remark}

\enlargethispage{\baselineskip} 

\subsection{Scalable Schemes for Distributed Operation}\label{sec:combining-schemes-for-distributed-operation}

While the L-MMSE combining in \eqref{eq:L-MMSE-combining-single-AP} and the opt LSFD vector in~\eqref{eq:LSFD-vector} are the preferable schemes in the distributed operation, none of these schemes is scalable. The former one has a computational complexity that increases with the number of UEs in the whole network and  the latter one requires a fronthaul signaling load that grows in the same way.
Therefore, we will now present suboptimal but scalable alternatives that are inspired by the aforementioned methods.

\subsubsection{Scalable Local Receive Combining}

We begin by revisiting the problem of selecting the combining vectors $ \vect{v}_{kl}$ for $k \in \mathcal{D}_l $ at AP $l$.  Recall that L-MMSE in \eqref{eq:L-MMSE-combining-single-AP} is not a scalable scheme since it makes use of estimates of the channels from all $K$ UEs.
To achieve scalability, the selection can instead be made as a function of the estimates $\{\widehat{\vect{h}}_{il} : i \in \mathcal{D}_l \}$ of the channels from the UEs that AP $l$ is serving. 

The simplest solution is to utilize MR combining \cite{Nayebi2016a}, \cite{Ngo2017b}, which is given by
\begin{equation}
\vect{v}_{kl}^{{\rm MR}} = \vect{D}_{kl}\widehat{\vect{h}}_{kl}.
\end{equation}
One key benefit of using this scheme is that the combining vector follows directly from the channel estimates so that no additional computations are needed. The total number of complex multiplications is  $\left(N\tau_p+N^2\right)|\mathcal{M}_k|$ and is reported in Table~\ref{table:complexity-with-level3}. Another benefit is that all the expectations in Theorem~\ref{theorem:uplink-capacity-level3} can be computed analytically, so the SE is available in closed form.

\begin{corollary} \label{cor:closed-form_ul}
	If MR combining with $\vect{v}_{kl}^{{\rm MR}} = \vect{D}_{kl} \widehat{\vect{h}}_{kl}$ is used, then the expectations in \eqref{eq:uplink-instant-SINR-level3} become
	\begin{align}  \label{eq:g_ki-expectation}
\left [ 	\mathbb{E} \left\{\vect{g}_{ki}\right\} \right ]_l=\begin{cases}  \sqrt{\eta_k\eta_i}\tau_p \tr\left( \vect{D}_{kl} \vect{R}_{il} \vect{\Psi}_{t_k l}^{-1} \vect{R}_{kl}\right) &   i\in\mathcal{P}_k \\
0 & i\notin\mathcal{P}_k  \end{cases}
	\end{align}
	with $\left[	\mathbb{E} \left\{\vect{g}_{ki} \vect{g}_{ki}^{\Htran}\right\} \right]_{lr} = \left [ 	\mathbb{E} \left\{\vect{g}_{ki}\right\} \right ]_l  \left [ 	\mathbb{E} \left\{\vect{g}_{ki}^{\star}\right\} \right ]_r $ for $ r \neq l$ and 
		\begin{align} 
\left[	\mathbb{E} \left\{\vect{g}_{ki} \vect{g}_{ki}^{\Htran}\right\} \right]_{ll} =& \eta_k\tau_p  \tr \left(  \vect{D}_{kl}\vect{R}_{il} \vect{R}_{kl} \vect{\Psi}_{t_k l}^{-1} \vect{R}_{kl}  \right) \nonumber \\ \label{eq:gki-second-moment}
&+ \begin{cases}
	\eta_k\eta_i\tau_p^{2}  \left|\tr \left(  \vect{D}_{kl} \vect{R}_{il} \vect{\Psi}_{t_k l}^{-1} \vect{R}_{kl}  \right)\right|^2   &   i\in\mathcal{P}_k
	\\
	0 & i\notin\mathcal{P}_k
	\end{cases}
	\end{align}
	while
	\begin{equation} \label{eq:F_kl}
\left[ \vect{F}_k \right]_{ll}= \sigmaUL\eta_k \tau_p  \tr\left( \vect{D}_{kl} \vect{R}_{kl} \vect{\Psi}_{t_k l}^{-1} \vect{R}_{kl}\right).
	\end{equation}
	\end{corollary}
	\begin{proof}
	The proof is available in Appendix~\ref{proof-of-cor:closed-form_ul} on p.~\pageref{proof-of-cor:closed-form_ul}.
	\end{proof}
	
The results from Corollary~\ref{cor:closed-form_ul} can be inserted into the instantaneous effective SINR in \eqref{eq:uplink-instant-SINR-level3} to obtain a closed-form SE expression. However, the final expression is lengthy so we will only present it in the case of $N=1$, where it can be substantially simplified.
	
	\begin{corollary}
	If each AP has $N=1$ antenna, then the MMSE estimate of the scalar channel $\vect{h}_{kl} \in \mathbb{C}$ has variance
	\begin{equation} \label{eq:gamma-notation}
	 \gamma_{kl} = \frac{\eta_k \tau_p \beta_{kl}^2 }{ \sum_{i \in \mathcal{P}_k } \eta_i \tau_p \beta_{il} +\sigmaUL }.
	 \end{equation}
	The instantaneous effective SINR, $\mathacr{SINR}_{k}^{({\rm ul},\dist)}$, in \eqref{eq:uplink-instant-SINR-level3} then becomes
		\begin{equation} \label{eq:uplink-instant-SINR-level3-closed}
 \frac{ p_{k} \left| 
\sum\limits_{l \in \mathcal{M}_k} a_{kl}^{\star} \gamma_{kl} \right|^2  }{ 
 \sum\limits_{i=1}^{K} p_{i}  \! \sum\limits_{l \in \mathcal{M}_k}\! \! |a_{kl}|^2 \beta_{il} \gamma_{kl} + \sum\limits_{i \in \mathcal{P}_k \setminus \{ k \}} \! p_i
\left| \sum\limits_{l \in \mathcal{M}_k} \! a_{kl}^{\star} \gamma_{kl} \sqrt{\frac{\eta_i}{\eta_k}} \frac{\beta_{il}}{\beta_{kl}}  \right|^2 \!\!
+ \sigmaUL \!\sum\limits_{l \in \mathcal{M}_k} \!\!  |a_{kl}|^2 \gamma_{kl}
	}.
	\end{equation}
	\end{corollary}
	\begin{proof}
	Using the notation in \eqref{eq:gamma-notation}, the expectations in Corollary~\ref{cor:closed-form_ul} can be rewritten as $\left[\mathbb{E} \left\{\vect{g}_{ki}\right\} \right ]_l = \sqrt{\frac{\eta_i}{\eta_k}} \frac{\beta_{il}}{\beta_{kl}} \gamma_{kl}$, for $i\in\mathcal{P}_k$ and zero for other $i$, $\left[ \vect{F}_k \right]_{ll}= \sigmaUL \gamma_{kl}$,
		\begin{align} 
\left[	\mathbb{E} \left\{\vect{g}_{ki} \vect{g}_{ki}^{\Htran}\right\} \right]_{ll} &= \beta_{il} \gamma_{kl} + \begin{cases}
\frac{\eta_i \beta_{il}^2}{\eta_k \beta_{kl}^2} \gamma_{kl}^2  &    i\in\mathcal{P}_k
	\\
	0 & i\notin\mathcal{P}_k
	\end{cases}
	\end{align}
	for $ l \in \mathcal{M}_k$ and zero for other $l$. Inserting these values into \eqref{eq:uplink-instant-SINR-level3} yields the SINR expression in \eqref{eq:uplink-instant-SINR-level3-closed}.
	\end{proof}
	
The closed-form effective SINR in \eqref{eq:uplink-instant-SINR-level3-closed} shows the key behaviors of the cell-free operation. The signal term $p_{k} | 
\sum_{l \in \mathcal{M}_k} a_{kl}^{\star} \gamma_{kl} |^2$ in the numerator contains the transmit power $p_k$ multiplied with a term containing the weighted sum of the contributions from the serving APs. Each AP contributes with a term proportional to the variance $ \gamma_{kl}$ of its channel estimate, since the combining vector is based on that estimate. The coherent combination of the contributions from the different APs is represented by the square of the sum.
 The first term in the denominator contains non-coherent interference, which means that it is a summation of the powers $p_i \beta_{il}$ of the individual interfering signals at the serving APs $l \in \mathcal{M}_k$ without any squares. Each term is multiplied with a scaling factor $|a_{kl}|^2 \gamma_{kl}$ containing the LSFD weight and the scaling made by the MR combining. The second term in the denominator is the additional coherent interference caused by pilot contamination, which has a similar shape as the signal term. The third term is the noise power.

If local MR combining is used along with the LSFD vector $\vect{a}_k = [1 \, \ldots \, 1]^{\Ttran}$, then the result of the distributed operation is equivalent to MR combining in the centralized operation. 
In other words, centralized MR combining can be implemented in a distributed fashion.
However, MR is expected to perform poorly in the presence of interference, unless a high degree of favorable propagation is achieved, which is not guaranteed in cell-free networks (see Section~\ref{subsec-favorable-propagation} on p.~\pageref{subsec-favorable-propagation}). Therefore, we will also consider combining schemes that can suppress interference while still being scalable.

\begin{sidewaystable} 
\centering
    \begin{tabular}{|c|c|c|} \hline
    Scheme & {Channel estimation} & Combining vector computation \\ \hline \hline
        opt (and n-opt) LSFD & $-$ & $|\mathcal{M}_k|^2+\frac{|\mathcal{M}_k|^3-|\mathcal{M}_k|}{3}$ \\ \hline
        L-MMSE  &  $(N\tau_p + N^2)K |\mathcal{M}_k|$ &$\frac{N^2 + N}{2} K |\mathcal{M}_k| + N^2 |\mathcal{M}_k| +\frac{N^3-N}{3} |\mathcal{M}_k| $\\   \hline
            LP-MMSE & $(N\tau_p + N^2)\sum\limits_{l \in \mathcal{M}_k}|\mathcal{D}_l|$ & $\frac{N^2 + N}{2} \!\!\!\!\sum\limits_{l \in \mathcal{M}_k}\!\!|\mathcal{D}_l| + N^2 |\mathcal{M}_k| +\frac{N^3-N}{3} |\mathcal{M}_k| $  \\ \hline
                MR & $(N\tau_p + N^2)|\mathcal{M}_k|$ & $-$ \\ \hline
    \end{tabular}
    \caption{The number of complex multiplications required per coherence block to compute the local combining vectors of UE $k$, including the computation of the channel estimates. Different combining schemes for the distributed operation are considered.} 
     \label{table:complexity-with-level3}
\end{sidewaystable}

Inspired by the scalable P-MMSE combining in \eqref{eq:P-MMSE-combining} that was designed for centralized operation, we can define the \gls{LP-MMSE} at AP $l$ based on only the channel estimates and statistics of UEs that are served by this AP (i.e., $\{\widehat{\vect{h}}_{il} : i \in \mathcal{D}_l\}$). By starting from \eqref{eq:L-MMSE-combining-single-AP} but taking a summation over only the UEs in $\mathcal{D}_l$, instead of all UEs, we obtain the local combining vector
\begin{equation} \label{eq:LP-MMSE-combining-single-AP-1}
\vect{v}_{kl}^{{\rm LP-MMSE}} =  p_{k}  \left( \sum\limits_{i \in \mathcal{D}_l} p_{i} \left( \widehat{\vect{h}}_{il} \widehat{\vect{h}}_{il}^{\Htran} + \vect{C}_{il} \right) + \sigmaUL  \vect{I}_{N} \right)^{\!-1}   \vect{D}_{kl} \widehat{\vect{h}}_{kl}.
\end{equation}
The number of complex multiplications required by LP-MMSE is reported in Table~\ref{table:complexity-with-level3}. Since this number is independent of $K$, LP-MMSE is a scalable solution as $K\to \infty$.
Compared to P-MMSE in \eqref{eq:P-MMSE-combining} that was proposed for centralized operation, LP-MMSE has a substantially lower complexity per UE since it requires to compute the inverse of an $N\times N$ matrix, rather than an $N|\mathcal{M}_k|\times N|\mathcal{M}_k|$ matrix. Note that LP-MMSE coincides with L-MMSE when AP $l$ serves all the UEs (i.e., $\mathcal{D}_l=\{1,\ldots,K\}$).
Unlike MR, the expectations in \eqref{eq:uplink-instant-SINR-level3} and \eqref{eq:sinr-level3} cannot be computed in closed form when using LP-MMSE, but can be computed numerically using Monte Carlo methods, as will be done later in this section.
	
\begin{remark}[Local P-RZF]
Since the complexity of P-MMSE is rather high in the centralized operation, we presented an alternative simplified P-RZF scheme with lower complexity. A similar simplification is not needed in the distributed operation since LP-MMSE already has a rather low complexity.
\end{remark}

\subsubsection{Scalable Large-Scale Fading Decoding}

The opt LSFD vector in Corollary~\ref{cor:LSFD-optimal} is also not scalable for implementation in large networks. The bottleneck originates from the fact that all UEs in the network affect the interference levels at all APs, thereby determining how accurate the local data estimates are.
To identify a scalable way to select ${\vect{a}}_{k}$ for UE $k$, we need to limit how many interfering UEs are considered in the computation.

The simplest solution is to give all the serving APs equal importance by setting ${\vect{a}}_{k} = [1 \, \ldots \, 1]^{\Ttran}$ \cite{Nayebi2017a}, \cite{Ngo2017b}. In this case, the CPU takes the local estimates of the data from UE $k$ and sum them up to create the final data estimate
\begin{align}\label{eq:fully-distributed-estimate}
\widehat s_{k} = \sum_{l\in\mathcal{M}_k} \widehat s_{kl}.
\end{align}
We call this approach \emph{no LSFD} since the LSFD weights are basically omitted. The reason is that the original works on Cell-free Massive MIMO considered this solution, while  LSFD weights were introduced only later. The key benefit is that no statistical parameters are needed at the CPU to compute \eqref{eq:fully-distributed-estimate}.\footnote{To be more precise, no statistical parameters are needed to compute ${\vect{a}}_{k}$ or \eqref{eq:fully-distributed-estimate}, but the data detection leading to the SE in Theorem~\ref{theorem:uplink-capacity-level3} requires that the signal strength and interference variance are known at the CPU.} The main drawback is that equal importance is given to all APs when computing \eqref{eq:fully-distributed-estimate}, even if some APs might be subject to high interference or low SNR. It can even happen that an AP reduces rather than increases the SE due to the suboptimal fusing of information at the CPU \cite{Buzzi2017b}.

The CPU should use some side-information of how accurate the respective local data estimates are, because this is what the opt LSFD vector does. Similar to the approach taken when P-MMSE combining was developed in Section~\ref{subsec:scalable-combining}, we can compute an approximately optimal solution by taking only the UEs that are served by partially the same APs as UE $k$ into consideration. If the AP selection is done properly, these should be the UEs that cause the vast majority of the interference and, thus, affect the quality of the local data estimates the most. Recall that $\mathcal{S}_k$ in \eqref{eq:S_k} denotes the set of UEs that have at least one serving AP in common with UE $k$, including itself. We then approximate the opt LSFD vector in \eqref{eq:LSFD-vector} as
\begin{align} \label{eq:partial-LSFD-vector}
&{\vect{a}}_{k}^{\rm n-opt} = p_k  \left( \sum\limits_{i\in \mathcal{S}_k}
p_{i} \mathbb{E} \left\{ {\vect{g}}_{ki} {\vect{g}}_{ki}^{\Htran} \right\} +   {\vect{F}}_{k} + \widetilde{\vect{D}}_k  \right)^{-1} \mathbb{E}\left\{ {\vect{g}}_{kk}\right\}
\end{align}
which we call the \emph{\gls{n-opt} LSFD vector} \cite{globecom_nopt}. The ``nearly optimal''  name will be demonstrated numerically later in this section.
The number of complex multiplications required for computing ${\vect{a}}_{k}^{\rm n-opt}$ is the same as those needed for computing ${\vect{a}}_{k}^{\rm opt}$ since the matrices and vectors have the same dimensions. The key difference is that the n-opt LSFD scheme requires knowledge of a lower number of statistical parameters, which is independent of $K$. This makes its fronthaul signaling load scalable, as we will elaborate on next.

\subsection{Fronthaul Signaling Load for Distributed Operation}
\label{sec:level3-fronthaul}

The fronthaul signaling required by the distributed operation can be quantified as follows. Each AP $l$ needs to send the local estimates $ \widehat{s}_{kl}$ for $k \in \mathcal{D}_l$ to the CPU, which corresponds to $\tau_u|\mathcal{D}_l|$ complex scalars per coherence block. This number equals $\tau_u \tau_p$ if we use the pilot assignment scheme from Section~\ref{sec:pilot-assignment} on p.~\pageref{sec:pilot-assignment}, where each AP serves $\tau_p$ UEs. This value does not grow with $K$ so we can conclude that this part of the distributed operation is scalable. The total fronthaul signaling is summarized in Table~\ref{tab:signaling-level4-level3}. 

Moreover, the statistical parameters for the computation of the LSFD vectors are required to be sent to the CPU and this number is different for opt LSFD and n-opt LSFD as we quantify in the following. Recall that  $\vect{g}_{ki} = [ \vect{v}_{k1}^{\Htran}\vect{D}_{k1} \vect{h}_{i1} \, \ldots \, \vect{v}_{kL}^{\Htran} \vect{D}_{kL} \vect{h}_{iL}]^{\Ttran}$. Moreover, we notice that $\mathbb{E}\left\{ \left[ \vect{g}_{ki} \right]_l \left[ \vect{g}_{ki}^{\star} \right]_r  \right\}=\mathbb{E}\left\{ \left[ \vect{g}_{ki} \right]_l\right\}\mathbb{E}\left\{ \left[ \vect{g}_{ki}^{\star} \right]_r\right\}$ for $ l\neq r$, by utilizing the independence of the channels corresponding to different APs. Hence, each AP can individually send its first- and second-order statistics to the CPU, which can then combine them to compute the required parameters.
The following statistical parameters are needed to be sent from AP $l$ to the CPU for the computation of the opt LSFD vector in \eqref{eq:LSFD-vector} for a generic UE $k\in\mathcal{D}_l$:
\begin{itemize}
	\item  $\mathbb{E}\left\{  \left[ \vect{g}_{ki} \right]_l\right\} $, for $i=1,\ldots,K$;
	\item $\mathbb{E}\left\{  \left | \left[ \vect{g}_{ki} \right]_l \right |^2\right\} $, for $i=1,\ldots,K$;
	\item $\left [ \vect{F}_k \right ]_{ll}$.
\end{itemize}
This sums up to $\left(3K+1\right)/2$ complex scalars. Each AP sends these complex scalars for each of its $|\mathcal{D}_l|$ UEs and, hence, $\left(3K+1\right)/2\sum_{l=1}^{L} |\mathcal{D}_l|$ complex scalars are needed in total. These values are summarized in Table~\ref{tab:signaling-level4-level3}. As anticipated earlier in this section, the number of statistical parameters that need to be shared when using opt LSFD increases with $K$, thus making it unscalable. 

If the n-opt LSFD vector in \eqref{eq:partial-LSFD-vector} is used instead, then AP $l$ is required to send $\left(3|\mathcal{S}_k|+1\right)/2$ complex parameters to the CPU. In that case, the required total number of complex scalars becomes $\sum_{l=1}^{L} \sum_{k \in \mathcal{D}_l} \left(3|\mathcal{S}_k|+1\right)/2$, as summarized in Table~\ref{tab:signaling-level4-level3}. Fortunately, this number does not grow with $K$ (see the discussion after \eqref{eq:Z_k-prime} for further details) and we can conclude that the combination of n-opt LSFD and any scalable local combining scheme leads to a completely scalable distributed operation in the uplink. 

Finally, in the case when no LSFD vector is used (i.e., all its elements are equal), no channel statistics need to be sent to the CPU. This case is also listed in Table~\ref{tab:signaling-level4-level3} as a worst-case baseline.

\section{Running Example}
\label{sec:running-example}

To exemplify the performance of User-centric Cell-free Massive MIMO and make comparisons with cellular networks under somewhat realistic conditions, we will now define a network setup that will be used as a running example in the remainder of this monograph.
The key parameters are given in Table~\ref{tab:system_parameters_running_example}. The total coverage area is $1$\,km $ \times \, 1$\,km and the total number of antennas is $M=LN=400$. The number of APs is either $L=400$ or $L=100$. In the former case, the number of antennas per AP will be $N=1$ whereas it is $N=4$ for the latter case. A wrap-around topology is used to mimic a large network deployment without edges, where all \glspl{AP} and \glspl{UE} are receiving interference from all directions. Unless otherwise stated, we assume that the APs are deployed uniformly at random in the coverage area. The communication takes place over a 20\,MHz bandwidth with a total receiver noise power of $-94$\,dBm (consisting of thermal noise and a noise figure of 7\,dB in the receiver hardware) at both the APs and UEs. The maximum uplink transmit power of each UE is 100\,mW
while the maximum downlink transmit power of each AP is 200\,mW. The difference models the fact that APs are connected to the electricity grid and thus less power limited.
Each coherence block consists of $\tau_c = 200$ samples. This value matches well with a 2\,ms coherence time and a 100\,kHz coherence bandwidth, which correspond to high mobility and large channel dispersion in sub-6 GHz bands, as exemplified in Section~\ref{sec:block-fading} on p.~\pageref{sec:block-fading}.
Hence, we are considering an example that supports both user mobility and outdoor propagation conditions.

\begin{table*}[t]
	\renewcommand{\arraystretch}{1.}
	\centering
	\begin{tabular}{|c|c|}
		\hline \bfseries $\!\!\!\!\!$ Parameter $\!\!\!\!\!$ & \bfseries Value\\
		\hline\hline
				Network area &  $1$\,km $ \times\,  1$\,km \\
					Network layout & $\!$Random deployment $\!$ \\
				& (with wrap-around) \\
				Number of APs &  $400$ or $100$ \\
				Number of antennas per AP  & $1$ or $4$ \\
		Number of total antennas & $M=LN=400$ \\
				Bandwidth & $B = 20$\,MHz  \\
		
		Receiver noise power & $\sigmaUL = -94$\,dBm \\
		
		Maximum uplink transmit power  & $100$\,mW\\
		Maximum downlink transmit power & $200$\,mW\\
		Samples per coherence block & $\tau_c = 200$ \\
		
		Channel gain at 1 km & $\Upsilon = -140.6$\,dB \\
		
		Pathloss exponent & $\alpha = 3.67$ \\
		Height difference between AP and UE & $10$\,m \\
		
		$\!$Standard deviation of shadow fading$\!$  & $\sigma_{\mathrm{sf}}= 4$ \\
		
		\hline
	\end{tabular}
	\caption{Key parameters of running example.}
	\label{tab:system_parameters_running_example}
\end{table*}

\enlargethispage{\baselineskip} 

The large-scale fading coefficient (channel gain) is computed in dB on the basis of the model presented in Section~\ref{subsec:large-scale-fading-model} on p.~\pageref{subsec:large-scale-fading-model}, and reported here for convenience:
\begin{equation}\label{eq:pathloss-coefficient-section5}
\beta_{kl} \,  [\textrm{dB}] = -30.5 - 36.7 \log_{10}\left( \frac{d_{kl}}{1\,\textrm{m}} \right)  + F_{kl}
\end{equation}
where $d_{kl}$ is the three-dimensional distance between \gls{AP} $l$ and \gls{UE} $k$. The APs are deployed 10\,m above the plane where the UEs are located, which acts as the minimum distance. This model matches with the 3GPP Urban Microcell model in \cite[Table~B.1.2.1-1]{LTE2017a}. The shadow fading is $F_{kl} \sim \mathcal{N}(0,4^2)$ and the terms from an AP to different UEs are correlated as \cite[Table B.1.2.2.1-4]{LTE2017a}
\begin{equation} \label{eq:shadowing-decorrelation-repeated}
\mathbb{E} \{ F_{kl} F_{ij} \} = 
\begin{cases}
4^2 2^{-\delta_{ki}/9\,\textrm{m}} & l = j \\
0 & l \neq j
\end{cases}
\end{equation}
where $\delta_{ki}$ is the distance between UE $k$ and UE $i$. The intuition behind the shadow fading correlation is that if two UEs are closely located and one of them is subject to strong shadowing caused by a large object in the propagation environment, then the other UE will likely also be subject to strong shadowing. The second row in \eqref{eq:shadowing-decorrelation-repeated} implies that each UE achieves independent shadow fading realizations from different APs, since these are deployed at fairly different locations. 
The large-scale fading model in \eqref{eq:pathloss-coefficient-section5} corresponds to a median channel gain of $-140.6$\,dB at 1\,km (the median is achieved by $F_{kl}=0$\,dB) and has a pathloss exponent $\alpha=3.67$, as reported in Table~\ref{tab:system_parameters_running_example}.

The spatial correlation matrices are generated using the local scattering model defined in Section~\ref{subsec:local-scattering-model} on p.~\pageref{subsec:local-scattering-model}. The multipath components are Gaussian distributed  around the nominal azimuth and elevation angles according to \eqref{eq:Gaussian-ASD}, where the nominal angles are computed by drawing a line between the AP and UE. We let $\sigma_{\varphi}$ and $\sigma_{\theta}$ denote the ASDs in the azimuth and elevation domains, respectively. Different values are considered throughout the simulations. We will also consider uncorrelated fading with  $\vect{R}_{kl} = \beta_{kl}  \vect{I}_{N}$ as a reference case. 

To assign the pilots and form the DCC, the joint pilot assignment and AP selection scheme from Section~\ref{sec:pilot-assignment} on p.~\pageref{sec:pilot-assignment} is used.

\subsection{Benchmark Schemes}

Most of the performance evaluations will consider the achievable SE expressions that were provided in Theorem~\ref{proposition:uplink-capacity-general} for the centralized operation and in Theorem~\ref{theorem:uplink-capacity-level3} for the distributed operation. However, we will also compare these results with some benchmark schemes. For completeness, these will be briefly presented below.

\subsubsection{Cellular Network}

To compare the performance of a cell-free network with that of a corresponding cellular network, we can treat the cellular network as a degenerate case of a cell-free network where each UE is only served by one of the APs; that is, $|\mathcal{M}_k|=1$. To compare with the best conceivable small-cell implementation, we assume the APs use the optimal receive combining. In this case, the SE can be computed as follows.

\begin{corollary} \label{theorem:uplink-capacity-general-level1}
If UE $k$ is served only by AP $l$, an achievable SE is
	\begin{equation} \label{eq:uplink-rate-expression-general-level1}
	\mathacr{SE}_{kl}^{({\rm ul},\cell)} = \frac{\tau_u}{\tau_c} \mathbb{E} \left\{ \log_2  \left( 1 + \mathacr{SINR}_{kl}^{({\rm ul},\cell)}  \right) \right\}
	\end{equation}
	where the instantaneous effective SINR is
	\begin{equation} \label{eq:uplink-instant-SINR_level1}
	\mathacr{SINR}_{kl}^{({\rm ul},\cell)}=\frac{ p_{k} \left|  \vect{v}_{kl}^{\Htran} \widehat{\vect{h}}_{kl} \right|^2  }{ 
		\condSum{i=1}{i \neq k}{K} p_{i} \left | \vect{v}_{kl}^{\Htran} \widehat{\vect{h}}_{il} \right|^2
		+ \vect{v}_{kl}^{\Htran}  \left( \sum\limits_{i=1}^{K} p_{i} \vect{C}_{il} + \sigmaUL  \vect{I}_{N} \right)  \vect{v}_{kl} 
	}
	\end{equation}
for an arbitrary receive combining $\vect{v}_{kl} \in \mathbb{C}^N$. 
	The maximum value in \eqref{eq:uplink-instant-SINR_level1} is achieved with the L-MMSE combining in \eqref{eq:L-MMSE-combining-single-AP}, leading to 
	\begin{align}
	\mathacr{SINR}_{kl}^{({\rm ul},\cell)} =  p_{k}  \widehat{\vect{h}}_{kl}^{\Htran} \left( \condSum{i=1}{i \neq k}{K} p_{i}  \widehat{\vect{h}}_{il} \widehat{\vect{h}}_{il}^{\Htran} + \sum\limits_{i=1}^{K} p_{i} \vect{C}_{il} + \sigmaUL  \vect{I}_{N} \right)^{-1} \!\!\!\!\! \!\!\widehat{\vect{h}}_{kl}. \label{eq:uplink-maximized-SINR-level1}
	\end{align}
\end{corollary}
\begin{proof}
This result follows from Theorem~\ref{proposition:uplink-capacity-general} and Corollary~\ref{cor:MMSE-combining} in the special case of $\mathcal{M}_k=\{l \}$.
\end{proof}

The SE-maximizing receive combining is L-MMSE, which is not scalable but will anyway be used as a benchmark to compare cell-free networks with the best kind of cellular network implementation.\footnote{The L-MMSE scheme is more commonly referred to as multi-cell MMSE combining in the literature on Cellular Massive MIMO \cite{BjornsonHS17}, \cite{massivemimobook}.}

The results presented above can be directly used to consider a cellular network with the same AP locations as in the cell-free counterpart. However, it can also be used to consider a Cellular Massive MIMO system, which is characterized by a substantially smaller $L$ and a substantially larger $N$ than in the cell-free network. 

In the simulations, the APs in the small-cell setup are deployed in the same locations as in the cell-free network. The system-specific parameter values for the small-cell systems are reported in Table~\ref{tab:system_parameters_running_example_small_cell}.  Hence, the number of APs will be either $L=400$ or $L=100$. In the former case, the number of antennas per AP will be $N=1$ whereas it will be $N=4$ for the latter case. 
	The same set of UEs is served by both network setups, in order to achieve a fair comparison. The small-cell system uses the same pilot assignment from Section~\ref{sec:pilot-assignment} on p.~\pageref{sec:pilot-assignment} as in the cell-free case even if only one AP serves each UE. The index of the single AP that is serving UE $k$ is determined as
	\begin{align} \label{eq:selection_small_cell}
	\ell = \underset{l\in \mathcal{M}_k}{\mathrm{arg \, max}}  \,\,\, \beta_{k l}
	\end{align}
	where $\mathcal{M}_k$ is the  DCC for UE $k$ that is found by implementing Algorithm~\ref{alg:basic-pilot-assignment} on p.~\pageref{alg:basic-pilot-assignment} for the cell-free case.

 In the Cellular Massive MIMO setup, we assume each UE is connected to the AP with the best average channel gain, i.e., $\beta_{kl}$ for UE $k$ and AP $l$, and we will specify the system parameters and how the pilot assignment is carried out later in the simulation part.

\begin{table*}[t]
	\renewcommand{\arraystretch}{1.}
	\centering
	\begin{tabular}{|c|c|}
		\hline \bfseries $\!\!\!\!\!$ Parameter $\!\!\!\!\!$ & \bfseries Value\\
		\hline\hline
		
		Number of small cells &  $400$ or $100$ \\
		Coverage area per cell &  $50$\,m $ \times\,  50$\,m or $100$\,m $ \times\,  100$\,m \\
		Number of antennas per AP  & $1$ or $4$ \\
		
		\hline
	\end{tabular}
	\caption{The parameters of the running example for the small-cell system. The APs are distributed either on a square grid or uniformly at random. The specified coverage areas are exact in the former case and computed on the average in the latter case.}
	\label{tab:system_parameters_running_example_small_cell}
\end{table*}

\subsubsection{Genie-Aided SEs With Centralized and Distributed Operations}
\label{sec:genie-aided-uplink}

The SE expressions presented in Theorem~\ref{proposition:uplink-capacity-general} for centralized operation and in Theorem~\ref{theorem:uplink-capacity-level3} for distributed operation are computed using the state-of-the-art capacity lower bounds. 
The bound for the centralized operation is reliable but Remark~\ref{remark-tightness-bounds} emphasized that the bound for the distributed operation might underestimate the actually achievable SE. This is because the bound is only tight when there is a sufficient degree of channel hardening. Unfortunately, there is no good definition of what this means. To evaluate these SE expressions, we can compare them to upper bounds. It is easy to obtain practically unachievable upper bounds by neglecting interference or changing the signal processing entirely, but this will not provide much insight into the tightness of the SE expressions presented earlier in this section.
We will instead use the following methodology: we select the receive combining using the estimated channels as before, but in the final detection step we inject perfect ``genie-aided'' channel knowledge. The SE is still computed by treating interference as noise, but the perfect CSI assumption leads to higher SE values.

We begin by considering the centralized operation and assume the channels $\{\vect{h}_k:k=1,\ldots,K\}$ are available at the CPU after the receive combining has been carried out. We can then rewrite the soft estimate of UE $k$'s uplink data in \eqref{eq:uplink-data-estimate2} as
\begin{align} \label{eq:uplink-data-estimate-level4-genie-aided}
\widehat{s}_k& =\underbrace{\vect{v}_k^{\Htran}\vect{D}_k{\vect{h}}_k s_k}_{\textrm{Desired signal}}+\underbrace{\condSum{i=1}{i \neq k}{K}\vect{v}_k^{\Htran}\vect{D}_k\vect{h}_i s_i}_{\textrm{Interference}}\;+\;\underbrace{\vect{v}_k^{\Htran}\vect{D}_k\vect{n}}_{\textrm{Noise}}
\end{align}
where the desired signal term now includes the channel $\vect{h}_k$ itself rather than its estimate. 
We then have the following result.

\begin{corollary} \label{corollary-genie-aided-level4}
	A genie-aided SE of UE $k$ in the centralized operation~is
	\begin{equation} \label{eq:genie-aided-uplink-SE-level4}
	\mathacr{SE}_{k}^{({\rm gen-ul},\cent)} = \frac{\tau_u}{\tau_c} \mathbb{E} \left\{ \log_2  \left( 1 + \mathacr{SINR}_{k}^{({\rm gen-ul},\cent)}  \right) \right\}
	\end{equation}
where
	\begin{equation} \label{eq:uplink-instant-SINR-genie-aided-level4}
	\mathacr{SINR}_{k}^{({\rm gen-ul},\cent)} = \frac{ p_{k} \left |  \vect{v}_{k}^{\Htran} \vect{D}_{k}\vect{h}_{k} \right|^2  }{ 
		\condSum{i=1}{i \neq k}{K} p_{i} \left | \vect{v}_{k}^{\Htran} \vect{D}_{k}\vect{h}_{i} \right|^2
		+ \sigmaUL\left \Vert \vect{D}_k \vect{v}_{k}\right\Vert^2  .
	}
	\end{equation}
\end{corollary}
\begin{proof}
This follows from utilizing Lemma~\ref{lemma:capacity-memoryless-random-interference} on p.~\pageref{lemma:capacity-memoryless-random-interference}.
\end{proof}

 Similarly, in the distributed operation, we can rewrite the data estimate of UE $k$ at the CPU in \eqref{eq:CPU_linearcomb_eff-level3} for final data detection as
\begin{align} 
\widehat{s}_k =\underbrace{\vect{a}_{k}^{\Htran}\vect{g}_{kk} s_k}_{\textrm{Desired signal}} + \underbrace{\condSum{i=1}{i \neq k}{K}\vect{a}_{k}^{\Htran}\vect{g}_{ki}  s_i}_{\textrm{Interference}} + \underbrace{n^\prime_{k}}_{\textrm{Noise}} \label{eq:uplink-data-estimate-level3-genie-aided}
\end{align}
where $\vect{a}_{k}\in \mathbb{C}^L$ is the LSFD vector, $n^\prime_{k} = \sum_{l=1}^{L} a_{kl}^{\star} \vect{v}_{kl}^{\Htran}\vect{D}_{kl}\vect{n}_{l}$, and $\vect{g}_{ki} = \left[ \vect{v}_{k1}^{\Htran}\vect{D}_{k1} \vect{h}_{i1} \, \ldots \, \vect{v}_{kL}^{\Htran} \vect{D}_{kL} \vect{h}_{iL}\right]^{\Ttran} \in \mathbb{C}^L$. 
If perfect CSI is available in the final data detection step, we can treat the first term in \eqref{eq:uplink-data-estimate-level3-genie-aided} as the desired signal obtained over a known channel and thus get the following result.

\begin{corollary} \label{corollary-genie-aided-level3}
A genie-aided SE of UE $k$  in the distributed operation is
	\begin{equation}\label{eq:genie-aided-uplink-SE-level3}
	\begin{split}
	\mathacr{SE}_{k}^{({\rm gen-ul},\dist)} = \frac{\tau_u}{\tau_c}\mathbb{E}\left\{ \log_2   \left( 1 + \mathacr{SINR}_{k}^{({\rm gen-ul},\dist)}  \right) \right\}
	\end{split}
	\end{equation}
where
	\begin{equation}  \label{eq:uplink-instant-SINR-genie-aided-level3}
	\mathacr{SINR}_{k}^{({\rm gen-ul},\dist)}\! =\!  \frac{ p_{k} \left| \vect{a}_{k}^{\Htran}  \vect{g}_{kk} \right|^2  }{ \!
		\!\condSum{i=1}{i \neq k}{K} p_{i}  |\vect{a}_{k}^{\Htran}\vect{g}_{ki}|^2    \!+\!  \vect{a}_{k}^{\Htran}\vect{F}_{k}^{\prime}
		\vect{a}_{k} 
	}
	\end{equation}
	with
	$$
	\vect{F}_{k}^{\prime}=\sigmaUL\diag\left( \left\|  \vect{D}_{k1}\vect{v}_{k1}  \right\|^2, \ldots,  \left\|  \vect{D}_{kL}\vect{v}_{kL}  \right\|^2\right) \in \mathbb{C}^{L \times L}.$$
\end{corollary}
\begin{proof}
This follows from utilizing Lemma~\ref{lemma:capacity-memoryless-random-interference} on p.~\pageref{lemma:capacity-memoryless-random-interference}.
\end{proof}

We reiterate that, in the simulations, the combining vectors and LSFD vectors are computed using the estimated channels, as described earlier in this section. It is only in the final data detection step that we inject the receiver with perfect CSI. In this way, we can evaluate the tightness of the capacity bounds in Theorem~\ref{proposition:uplink-capacity-general} for centralized operation and in Theorem~\ref{theorem:uplink-capacity-level3} for distributed operation, for a given receiver operation.

\section{Numerical Performance Evaluation}
\label{sec:uplink-performance-comparison}

In this section, we will quantify the uplink SE achieved by the centralized and distributed operations of User-centric Cell-free Massive MIMO, using the different combining schemes presented earlier in this section. The running example from Section~\ref{sec:running-example} will be considered in all the simulations. Comparisons will be made with cellular networks with either small cells or Massive MIMO. In addition to showing the SE, we will exemplify the fronthaul signaling load and the computational complexity of the combining schemes in the centralized and distributed operations, to demonstrate the difference between scalable and unscalable implementations.

In all the simulations in this section, we assume each UE transmits with full power both in the pilot and data transmission phases:
\begin{equation}
\eta_k=p_k=\pmax, \quad k= 1,\ldots,K. 
\end{equation}
We will consider other optimized and heuristic power control methods in Section~\ref{c-resource-allocation} on p.~\pageref{c-resource-allocation}, where it is also shown that full power transmission approximately maximizes the sum SE in the considered setup. All the APs are randomly distributed in the coverage area following an independent and uniform distribution. There are $K=40$ UEs in all the considered setups, which implies that $L \gg K$. The pilot sequence length is $\tau_p=10$ and the remaining $\tau_u=\tau_c-\tau_p=190$ samples of each coherence block are used for uplink data. The Gaussian local scattering model is used to generate the spatial correlation matrices with ASD $\sigma_{\varphi}=\sigma_{\theta}=15^{\circ}$. 

We use the following Monte Carlo simulation methodology to generate the numerical results for the cell-free and small-cell networks:
\begin{enumerate}
	\item Deploy the APs in the simulation area either in a random manner with independent uniform distribution or using a square grid. 
	\item Randomly drop the UEs one by one.
	\item Compute the distance from the considered UE to each AP by using the wrap-around topology.\footnote{This means that the north edge of the simulation area is connected to the south edge, while the west edge is connected to the east edge. Hence, there are multiple ways to draw a line between an AP and a UE, but we always consider the shortest option, which might go over an edge. The reason for having a wrap-around topology is to simulate a large-scale scenario where all the UEs effectively are in the center of the simulation area and are subject to interference from all directions.}
	\item Compute the channel gain from the considered UE to each AP by using \eqref{eq:pathloss-coefficient-section5}. The shadow fading realizations are generated using the conditional Gaussian distribution recursively from \cite[Theorem~10.2]{Kay1993a} to simulate the shadowing according to \eqref{eq:shadowing-decorrelation-repeated}.
	\item Generate spatial correlation matrices $\vect{R}_{kl}$ and estimation error correlation matrices $\vect{C}_{kl}$.
	\item Determine the pilot assignment and DCC by implementing Algorithm~\ref{alg:basic-pilot-assignment} on p.~\pageref{alg:basic-pilot-assignment}.
	\item Generate the estimated channels $\widehat{\vect{h}}_{kl}$ and use them to compute sample averages that approximate all the expectations in the SE expressions.
\end{enumerate}   

In each figure, the legend will indicate the schemes that are being used.
In some figures, we will compare the scalable operation obtained using Algorithm~\ref{alg:basic-pilot-assignment} on p.~\pageref{alg:basic-pilot-assignment} with a Cell-free Massive MIMO network, where each UE is served by all the APs. In these cases, we use ``(DCC)'' to denote the scalable operation and ``(All)'' to denote the case where all APs serve all UEs.

\subsection{Performance With Centralized Operation}\label{centralized-operation-uplink}

\begin{figure}[p!]
        \begin{subfigure}[b]{\columnwidth} \centering 
	\begin{overpic}[width=0.88\columnwidth,tics=10]{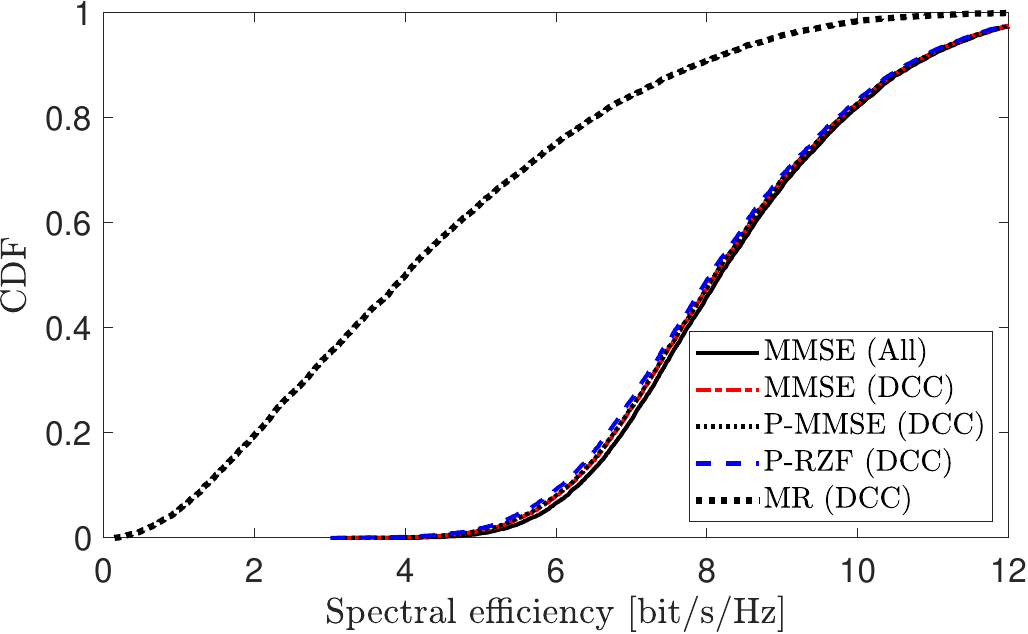}
\end{overpic} \label{figureSec5_1a} 
                \caption{$L=400$, $N=1$.} 
                \vspace{4mm}
        \end{subfigure} 
        \begin{subfigure}[b]{\columnwidth} \centering 
	\begin{overpic}[width=0.88\columnwidth,tics=10]{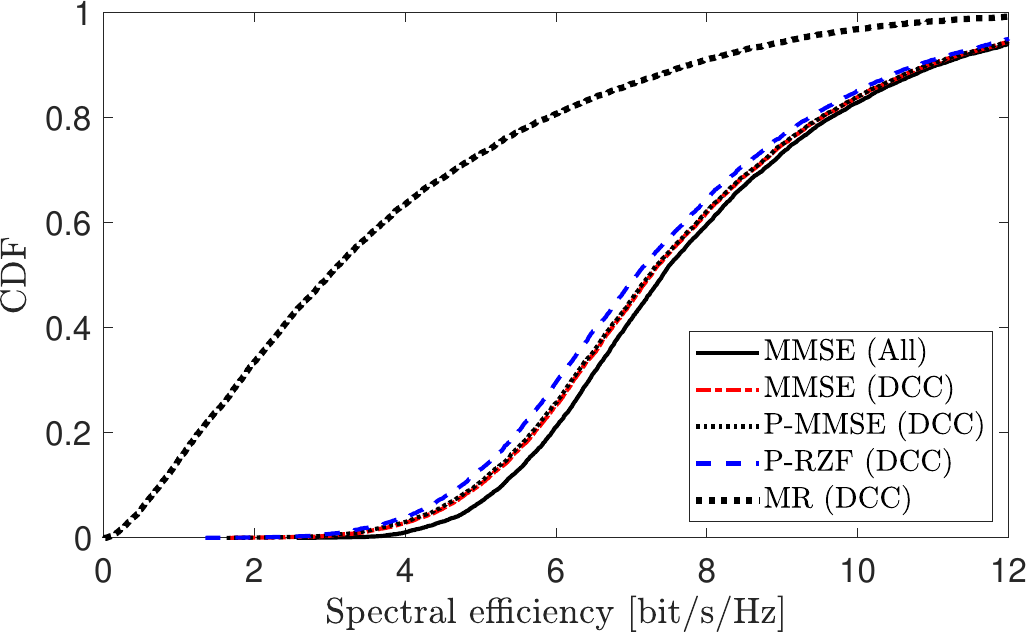}
\end{overpic} \label{figureSec5_1b} 
                \caption{$L=100$, $N=4$.} 
        \end{subfigure} 
	\caption{ CDF of the uplink SE per UE in the centralized operation. We consider $K=40$ UEs, $\tau_p=10$, and spatially correlated Rayleigh  fading with ASD $\sigma_{\varphi}=\sigma_{\theta}=15^{\circ}$. Different scalable and unscalable combining schemes are compared.}\label{figureSec5_1}
\end{figure}

In Figure~\ref{figureSec5_1}, we show the \gls{CDF} of the SE per UE with a centralized operation of User-centric Cell-free Massive MIMO. The sources of randomness that give rise to the CDF are the random AP and UE locations, as well as the shadow fading realizations. The SE is computed using \eqref{eq:uplink-rate-expression-general} for the different combining schemes described in Section~\ref{sec:level4-uplink}: MMSE, P-MMSE, P-RZF, and MR.

Figure~\ref{figureSec5_1}(a) shows the scenario with $L=400$ APs and $N=1$ antenna per AP and Figure~\ref{figureSec5_1}(b) shows the scenario with $L=100$ APs with $N=4$ antennas per AP. 
In both scenarios, the largest SE is achieved by ``MMSE (All)'', which corresponds to that all APs serve all UEs using MMSE combining. This is expected since this is the optimal scheme, however, it is not a scalable solution.
We notice that the use of ``MMSE (DCC)'' leads to a negligible performance loss, thus it is sufficient to only involve a subset of the APs in the signal detection. If we switch to using the scalable P-MMSE combining scheme, denoted as ``P-MMSE (DCC)'', we arrive at a fully scalable solution and the gap to the optimal scheme remains negligible. The P-RZF combining scheme, which is also scalable, results in a slightly lower SE than the P-MMSE scheme. Hence, by capitalizing on the large channel gain variations in cell-free networks, a properly designed scalable DCC implementation can attain almost the same performance as the optimal unscalable solution. 
We note that MR combining achieves significantly smaller SEs than the interference-suppressing schemes. Recall that MR requires a high degree of favorable propagation to achieve good results, which apparently is not available anywhere in the network.

\enlargethispage{\baselineskip} 

By comparing the performance of the two scenarios in Figures~\ref{figureSec5_1}(a) and~\ref{figureSec5_1}(b), we notice that the former case with many single-antenna APs is more desirable when having a centralized operation. The difference \linebreak is particularly large at the lower end of the CDF curves, which explains that it is beneficial to spread out the antennas as much as possible~to obtain the so-called macro-diversity gain. The reduction in the lowest SE values when switching from $L=400$, $N=1$ to $L=100$, $N=4$~are more apparent for the interference-suppressing schemes than for MR combining. Interestingly, the performance loss of using P-RZF combining is higher with $N=4$ than in the case of $N=1$. The reason is that in the first scenario with single-antenna APs, the collective estimation error correlation matrices $\vect{C}_k$ are diagonal whereas in the second case they have a block-diagonal structure with some non-zero off-diagonal elements due to the spatial correlation. When computing the P-RZF combining vector in  \eqref{eq:P-RZF-combining_0}, we neglect the estimation error correlation matrices $\vect{C}_k$ and when these are non-diagonal, this creates a more significant performance drop in comparison to the combining schemes that utilize the spatial structure of $\vect{C}_k$. On the other hand, only neglecting the matrices $\vect{C}_k$ for UEs that cause little interference, as done with P-MMSE combining, is associated with a negligible performance loss. 

\vspace{-2mm}

\subsubsection{Tightness of the SE Expressions}

\vspace{-1mm}

\begin{figure}[t!] \centering 
	\begin{overpic}[width=0.88\columnwidth,tics=10]{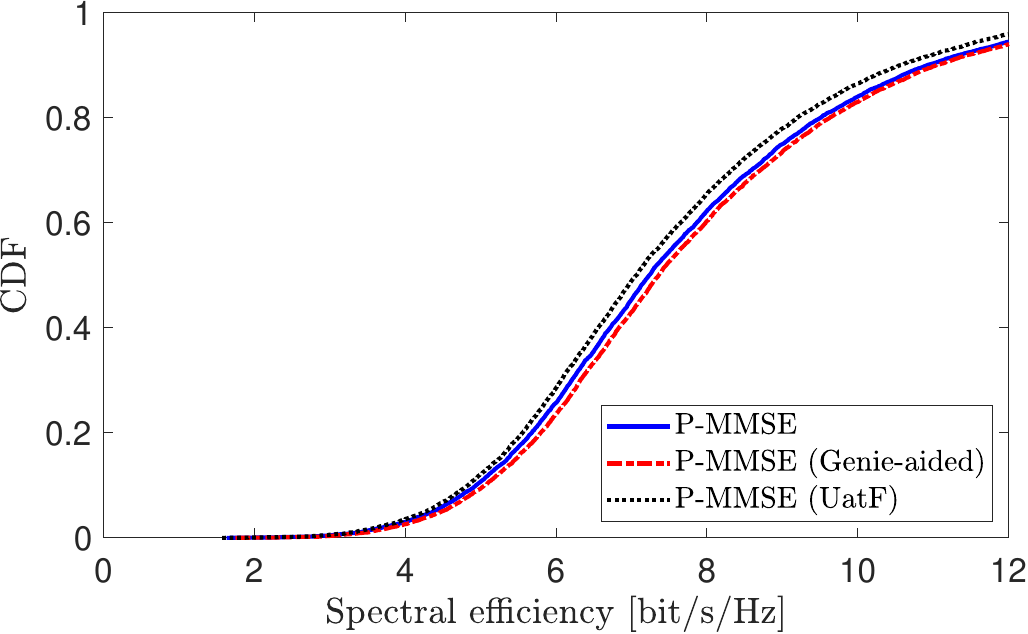}
	\end{overpic} 
	\caption{CDF of the uplink SE per UE in the centralized operation in the same scenario as in Figure~\ref{figureSec5_1}(b). We consider $L=100$, $N=4$, $K=40$, $\tau_p=10$, and spatially correlated Rayleigh  fading with ASD $\sigma_{\varphi}=\sigma_{\theta}=15^{\circ}$. The SE expression from \eqref{eq:uplink-rate-expression-general} is compared with genie-aided SE in \eqref{eq:genie-aided-uplink-SE-level4} and the UatF bound in \eqref{eq:uplink-rate-expression-general-UatF}.}    \vspace{+2mm}
	\label{figureSec5_1x} 
\end{figure}

The SE expression for centralized operation in \eqref{eq:uplink-rate-expression-general} is a lower bound on the capacity, where the signals received over the unknown parts of the channel vectors are treated as noise. To quantify the tightness of this lower bound, Figure~\ref{figureSec5_1x} considers the same simulation setup as in Figure~\ref{figureSec5_1}(b) but includes the genie-aided SE from \eqref{eq:genie-aided-uplink-SE-level4}. We focus on P-MMSE combining, which is the scalable scheme that achieves the highest SE. Since the genie-aided SE is obtained with perfect \gls{CSI} at the CPU, a higher SE is achieved compared to \eqref{eq:uplink-rate-expression-general}. However, the performance gap is very small. In the figure, we also consider the alternative SE from \eqref{eq:uplink-rate-expression-general-UatF} that is obtained by the UatF bounding technique, where no CSI is available in the final signal detection.
This simplification results in lower SE values compared to the original SE expression in \eqref{eq:uplink-rate-expression-general} but the gap is almost negligible.
We will anyway continue using the original expression in the remainder of this section, but the UatF bound will be used for optimized power control in Section~\ref{sec:optimized-power-allocation} on p.~\pageref{sec:optimized-power-allocation}.  In conclusion, the two SE expressions provided in Theorem~\ref{proposition:uplink-capacity-general} and in Theorem~\ref{proposition:uplink-capacity-general-UatF} are both representative of the practically achievable performance with centralized operation.

\subsection{Performance With Distributed Operation}

\begin{figure}[p!]
        \begin{subfigure}[b]{\columnwidth} \centering 
	\begin{overpic}[width=0.88\columnwidth,tics=10]{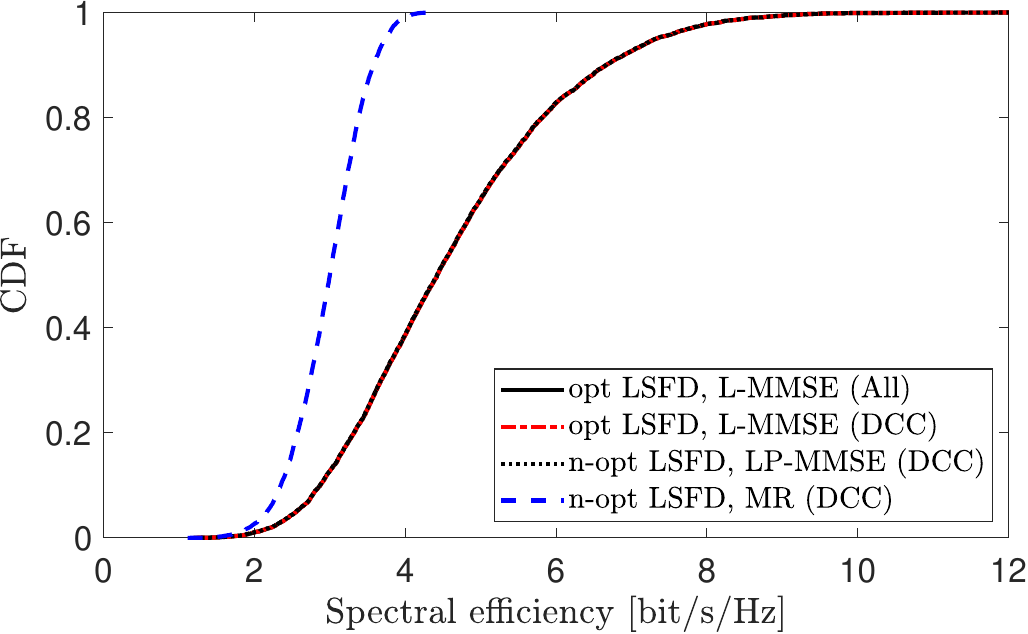}
\end{overpic} \label{figureSec5_2a}
                \caption{$L=400$, $N=1$.} 
                 \vspace{4mm}
        \end{subfigure} 
        \begin{subfigure}[b]{\columnwidth} \centering 
	\begin{overpic}[width=0.88\columnwidth,tics=10]{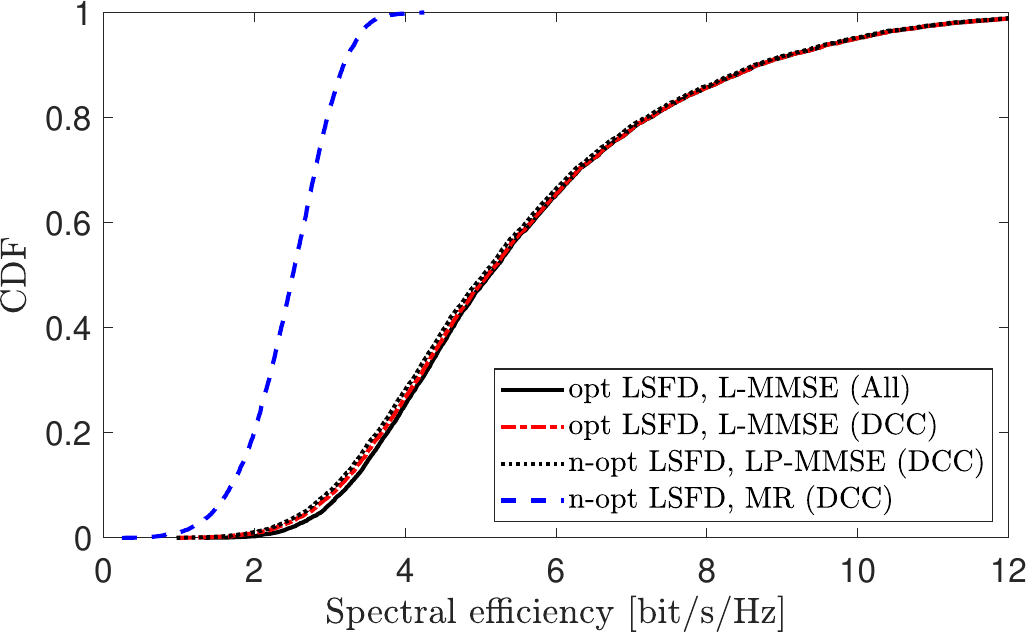}
\end{overpic} \label{figureSec5_2b} 
                \caption{$L=100$, $N=4$.} 
        \end{subfigure} 
	\caption{CDF of the uplink SE per UE in the distributed operation. We consider $K=40$ UEs, $\tau_p=10$, and spatially correlated Rayleigh fading with ASD $\sigma_{\varphi}=\sigma_{\theta}=15^{\circ}$. Different LSFD and local combining schemes are compared.}\label{figureSec5_2}
\end{figure}

We will now focus on the distributed operation. Figure~\ref{figureSec5_2} shows the \gls{CDF} of SE per UE in the same scenarios as in Figure~\ref{figureSec5_1}. Hence, 
Figure~\ref{figureSec5_2}(a) considers $L=400$ and $N=1$, while Figure~\ref{figureSec5_2}(b) considers $L=100$ and $N=4$.
The most advanced distributed implementation is to use L-MMSE combining at each AP, which is locally optimal, and then to apply the optimal LSFD weights ``opt LSFD'' at the CPU. This combination gives the highest SEs in Figure~\ref{figureSec5_2}, but it is not a scalable solution since the complexity grows unboundedly with $K$. The figure shows that we can achieve essentially the same SEs even if we make three simplifications: 1) involve only a subset of the APs in the data detection; 2) use the scalable n-opt LSFD (instead of opt LSFD); and 3) switch from L-MMSE to LP-MMSE combining.
Hence, the performance loss required to achieve a scalable implementation is negligible also in the distributed operation.
The figure also considers MR combining and we notice that there is a large SE loss compared to LP-MMSE combining. This is expected in the case of $N=4$ since LP-MMSE is using the antennas to suppress interference. However, there is also a substantial performance loss in the case of $N=1$. The reason is that MR creates larger variations in the interference power \cite{Bjornson2020a}.

\enlargethispage{\baselineskip}

By comparing the two scenarios in Figure~\ref{figureSec5_2}, we notice that the SE curves with $N=4$ are more spread out than the curves with $N=1$. The UEs with the worst channel conditions experience lower SE since there are larger gaps between the AP locations, while the UEs with the best channel conditions experience higher SE since these UEs are sensitive to interference and each AP can locally suppress interference using its array of antennas.
Note that this behavior is different from the centralized case in Figure~\ref{figureSec5_1}, where the gap between the two scenarios was larger and it was always preferable to have many APs since interference could be suppressed at the CPU using the received signals at multiple APs. One advantage of deploying $N=4$ antennas per AP rather than $N=1$ antenna (when $M=LK=400$ is fixed) is that the channel estimation is improved at each AP owing to the spatial correlation between the channel coefficients observed at the antennas in the same array. This was previously discussed in Section~\ref{sec:impact-of-spatial-correlation} on p.~\pageref{sec:impact-of-spatial-correlation}.

\subsubsection{Tightness of the SE Expressions}

\begin{figure}[t!]\centering
	\begin{overpic}[width=0.88\columnwidth,tics=10]{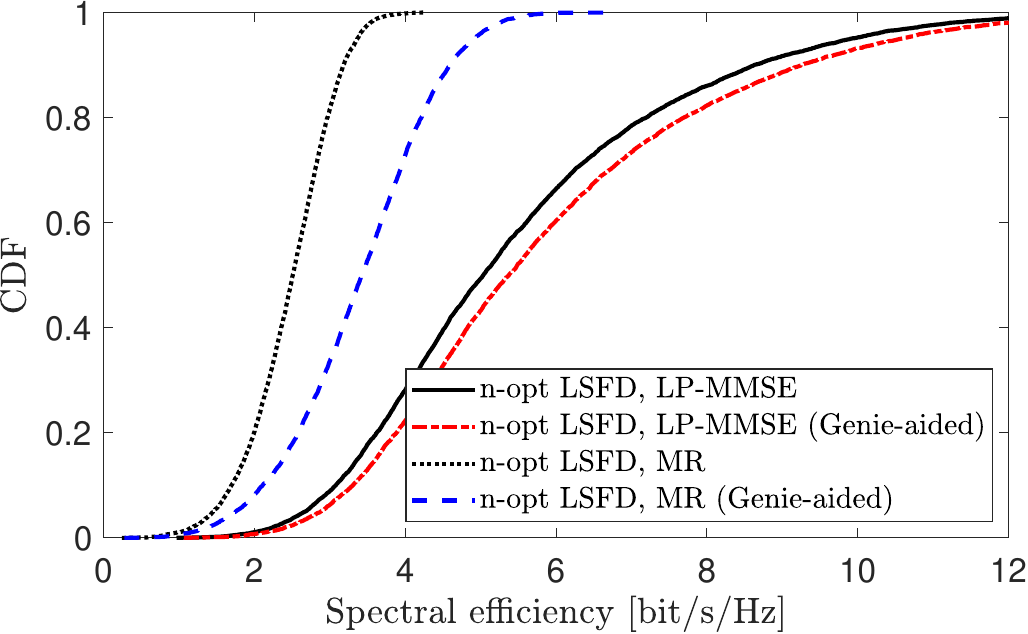}
	\end{overpic} 
	\caption{CDF of the uplink SE per UE in the distributed operation with LSFD in the same scenario as in Figure~\ref{figureSec5_2}(b). We consider $L=100$, $N=4$, $K=40$, $\tau_p=10$, and spatially correlated Rayleigh  fading with ASD $\sigma_{\varphi}=\sigma_{\theta}=15^{\circ}$. The SE expression from \eqref{eq:uplink-rate-expression-level3} is compared with genie-aided SE in \eqref{eq:genie-aided-uplink-SE-level3}.}    \vspace{+2mm}
	\label{figureSec5_2x} 
\end{figure}

As discussed in Remark~\ref{remark-tightness-bounds}, the SE expressions used in the distributed operation are developed under the assumption that the CPU does not have access to the instantaneous channel realizations but only to the channel statistics. Such capacity bounds might underestimate the achievable SE, unless there is a sufficiently high degree of channel hardening and small interference variations.
To investigate this potential issue, Figure~\ref{figureSec5_2x} shows the CDFs of the SE with the scalable schemes from Figure~\ref{figureSec5_2}(b) with $L=100$ and $N=4$. We compare these curves with the genie-aided SE from \eqref{eq:genie-aided-uplink-SE-level3}, which assumes that the same processing schemes are used but the CPU has perfect CSI when detecting the data.
The gap between the genie-aided SE and the achievable SE values is larger than in the centralized case (see Figure~\ref{figureSec5_1x}), but the difference is reasonably small when using LP-MMSE combining. However, there is a substantial gap when using MR combining. The reason for this is that MR gives rise to larger power variations in both the desired and interfering signals \cite{Bjornson2020a}, \cite{Interdonato2016a}, which the detector can deal with if they are known (as in the genie-aided case) but not when it lacks CSI.
We conclude that the SE expression that we presented is practically achievable and fairly close to what could be achieved with perfect CSI at the CPU when using LP-MMSE combining (or other interference-suppressing schemes). The gap with MR can be reduced by normalizing it as explained in \cite[Sec.~4.2.1]{massivemimobook} and \cite{Interdonato2016a}.

\subsubsection{Benefits of LSFD}

\begin{figure}[t!]\centering
	\begin{overpic}[width=0.88\columnwidth,tics=10]{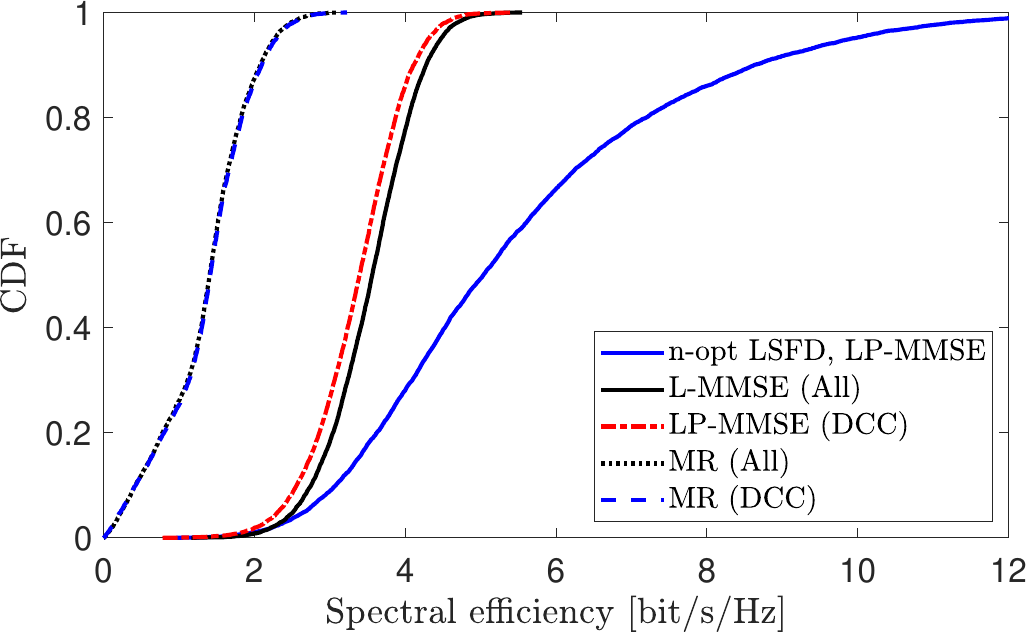}
	\end{overpic} 
	\caption{CDF of the uplink SE per UE in the distributed operation with and without LSFD. We consider the same scenario as Figure~\ref{figureSec5_2}(b) with $L=100$, $N=4$, $K=40$, $\tau_p=10$, and spatially correlated Rayleigh  fading with ASD $\sigma_{\varphi}=\sigma_{\theta}=15^{\circ}$. Different local combining schemes are compared. The scalable LSFD scheme with LP-MMSE combining from Figure~\ref{figureSec5_2}(b) is provided as a benchmark.}    \vspace{+2mm}
	\label{figureSec5_3} 
\end{figure}

In the distributed cell-free operation, it is important to give different priorities to the local data estimates obtained by different APs using LSFD at the CPU. To demonstrate this, we will now compare the performance with and without LSFD, where the latter is represented by $\vect{a}_k = [1 \, \ldots \, 1]^{\Ttran}$.
Figure~\ref{figureSec5_3} compares the CDFs of the SE per UE with MR and L-MMSE combining when all APs serve all UE with their scalable alternatives, i.e., user-centric selection of APs with MR and LP-MMSE combining. The four left-most curves are without LSFD, where all the serving APs are given equal priority, while the results for the scalable n-opt LSFD with LP-MMSE combining from Figure~\ref{figureSec5_2}(b) is included as a reference curve. The first observation is that all UEs experience a much higher SE with LSFD than without LSFD, which demonstrates that the LSFD feature is essential.
The SE is very low with MR, which can be partially addressed by adjusting the power control \cite{Ngo2017b}, but it is more efficient to use LP-MMSE combining in combination with LSFD \cite{Bjornson2019c}, \cite{Bjornson2020a}. Note that the user-centric AP selection leads to minor performance degradation when using LP-MMSE compared to L-MMSE (All), while the loss is negligible when using MR since the network is strongly interference-limited in that case.

\enlargethispage{\baselineskip}

\subsection{Comparison Between Cell-Free and Cellular Networks}

In this section, we have thus far only evaluated the performance of cell-free networks with different types of operation and different combining schemes. We will now compare cell-free networks with cellular networks, thereby extending the preliminary comparison that was made in Section~\ref{sec:benefits-cellular-networks} on p.~\pageref{sec:benefits-cellular-networks} under idealized conditions. It is inherently hard to make a fair comparison between different types of networks, but we will make an attempt by letting the total number of antennas be the same: $LN=400$. We consider both a Cellular Massive MIMO with many antennas per AP and a small-cell system with the same AP configuration as in the cell-free network.
The propagation model described in Section~\ref{sec:running-example} is used in all cases. In Table~\ref{tab:system_parameters_running_example_cellular}, we report the parameters of the Cellular Massive MIMO system where there are 4 square cells, each having an area of $500$\,m $\times$ $500$\,m. The APs are deployed at the center of the cells and they are each equipped with 100 antennas. For the cell-free network and small-cell system, either $L=400$ APs with $N=1$ antennas or $L=100$ APs with $N=4$ antennas are deployed on a symmetric square grid.

\begin{table*}[t]
	\renewcommand{\arraystretch}{1.}
	\centering
	\begin{tabular}{|c|c|}
		\hline \bfseries $\!\!\!\!\!$ Parameter $\!\!\!\!\!$ & \bfseries Value\\
		\hline\hline
		
		Number of cells &  $4$ \\
		Cell area &  $500$\,m $ \times\,  500$\,m \\
		Number of antennas per AP  & $100$ \\
		Number of UEs per cell & $10$ \\
		
		\hline
	\end{tabular}
	\caption{The parameters for the Cellular Massive MIMO system used for performance comparison. Each cell covers a square area of $500$\,m $ \times\, 500$\,m and is deployed on a grid of $2 \times 2$ cells.}
	\label{tab:system_parameters_running_example_cellular}
\end{table*}

We drop the UEs so that each AP in the Cellular Massive MIMO system serves the same number of UEs. Note that this does not mean that these 10 UEs are geometrically located in the corresponding square cell. Due to the random shadow fading, a UE can occasionally get a better channel to another \gls{AP} than the one located in its own square. To make a fair comparison for all the considered systems, we use the same UE locations and pilot assignments. In the Cellular Massive MIMO case, each UE in a cell is assigned to a unique pilot sequence from $\tau_p=10$ mutually orthogonal pilot sequences. To guarantee this condition and to be able to use the same pilot assignment, we drop the UEs in the coverage area one by one and assign the pilot to each according to Algorithm~\ref{alg:basic-pilot-assignment} on p.~\pageref{alg:basic-pilot-assignment}. Then, we check whether the serving AP in the Cellular Massive MIMO system can serve that UE with that assigned pilot (i.e., whether there is not any other UE that is served by this AP using the same pilot sequence). If this is the case, we assign this UE to the considered AP. Otherwise, we remove this randomly located UE and drop a new UE.
At the end of this UE dropping methodology, it is guaranteed that all the setups consider the same UE locations and that the Cellular Massive MIMO system is well operating: each UE is connected to the AP with the largest channel gain and the 10 UEs in each cell are using mutually orthogonal pilot sequences.

\begin{figure}[p!]
\begin{subfigure}[b]{\columnwidth} \centering 
	\begin{overpic}[width=0.88\columnwidth,tics=10]{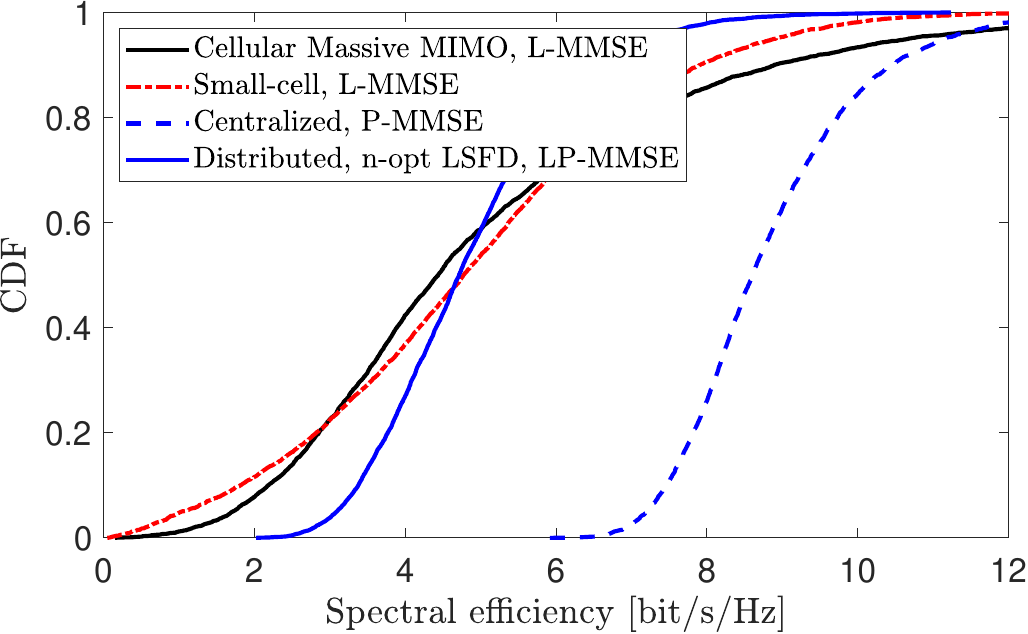}
			\put(69,16){90\% likely}
		\put(15,13.7){$- - - - - - - - - - - - - - - - - -$}
	\end{overpic} \label{figureSec5_4a} 
	\caption{$L=400$, $N=1$ for the cell-free and small-cell setups.} 
	\vspace{4mm}
\end{subfigure} 
\begin{subfigure}[b]{\columnwidth} \centering 
	\begin{overpic}[width=0.88\columnwidth,tics=10]{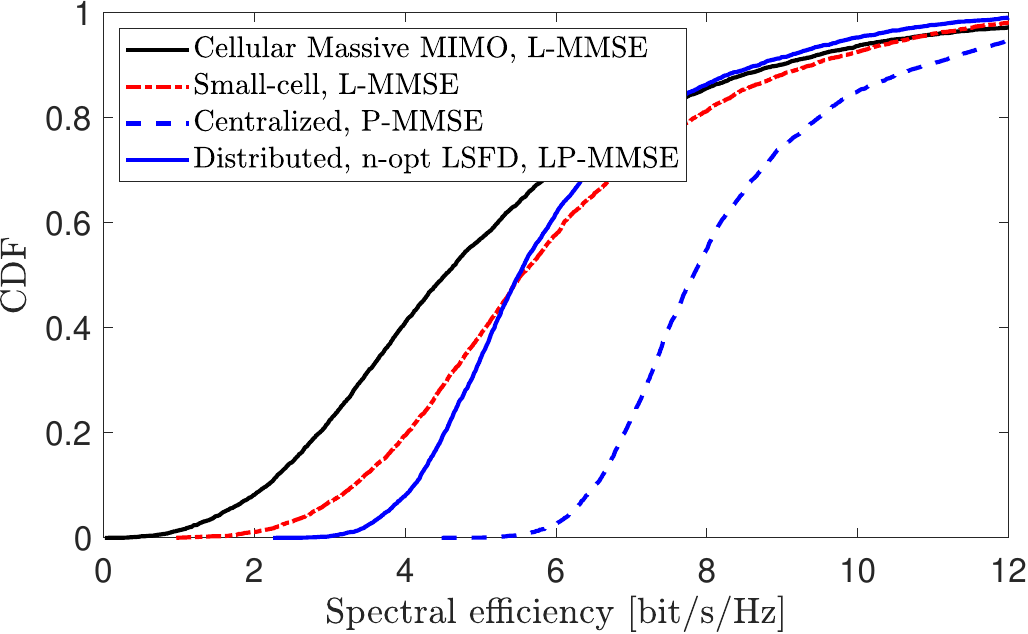}
			\put(63,16){90\% likely}
		\put(15,13.7){$- - - - - - - - - - - - - - - - - -$}
	\end{overpic} \label{figureSec5_4b} 
	\caption{$L=100$, $N=4$ for the cell-free and small-cell setups.} 
\end{subfigure} 
\caption{CDF of the uplink SE per UE for Cellular Massive MIMO, a small-cell setup, and different operations of scalable Cell-free Massive MIMO. We consider $K=40$, $\tau_p=10$, and spatially correlated Rayleigh  fading with ASD $\sigma_{\varphi}=\sigma_{\theta}=15^{\circ}$. For the Cellular Massive MIMO case, there are $L=4$ APs with $N=100$ antennas per AP.}\label{figureSec5_4}
\end{figure}

Different scalable operations of User-centric Cell-free Massive MIMO are compared with Cellular Massive MIMO and small-cell systems in Figure~\ref{figureSec5_4}. 
 The figure shows the CDF of the SE achieved by UEs at different random locations.  For the cellular systems, L-MMSE combining is used, which is the best scheme although it is unscalable. We have~also observed that the gap between the achievable and genie-aided SEs~for~the small-cell system is negligibly small and we will present only the genie-aided SE results for the small-cell systems in the remainder of this section to avoid any underestimation of the small-cell system performance.
 
Figure~\ref{figureSec5_4}(a) considers the case with $L=400$ and $N=1$.
Among all the considered network architectures and processing methods, the centralized cell-free operation with P-MMSE combining provides  the largest SEs. The CDF curve is to the right of all the other curves, which implies that all UEs benefit from using this mode of operation. The distributed cell-free operation with n-opt LSFD and LP-MMSE provides nearly the same median SE as the small-cell and Cellular Massive MIMO systems but it gives substantially higher SE for the weakest UEs (i.e., more fairness).  We can quantify the fairness gains by considering the 90\% likely SE value (where the CDF is $0.1$), which represents the SE that can be provided to 90\% of all UEs. The 90\% likely SE with the distributed cell-free operation is around 86\% and 55\% larger than in the small-cell and Cellular Massive MIMO systems, respectively. This is a natural consequence of both the architectural benefits of cell-free networks listed in Section~\ref{sec:benefits-cellular-networks} on p.~\pageref{sec:benefits-cellular-networks} and the increased channel estimation quality  observed from Figure~\ref{fig:basic-signal-gain-comparison-cellular-cellfree} on p.~\pageref{fig:basic-signal-gain-comparison-cellular-cellfree}. 
Qualitatively speaking, the results in this figure are consistent with the preliminary comparison in Figure~\ref{fig:section1_figure10} on p.~\pageref{fig:section1_figure10}. The centralized cell-free operation greatly outperforms the cellular networks, while the small-cell network is slightly preferred over Cellular Massive MIMO except in the upper and lower tails of the CDF curve. The distributed cell-free operation was not considered in Figure~\ref{fig:section1_figure10} and provides substantial gains over the cellular networks in the lower tail, but at the expense of lower performance in the upper tail.
 
\enlargethispage{\baselineskip} 
 
Figure~\ref{figureSec5_4}(b) considers the case with $L=100$ and $N=4$. The cellular curve is the same as before, the curves for the small-cell and distributed cell-free operations are shifted to the right, while the curve for the centralized cell-free operation is shifted to the left. In other words, the gap between the centralized and distributed cell-free operations reduces, thus showing that any distributed implementation should make use of multiple antennas per AP. Interestingly, the small-cell network outperforms Cellular Massive MIMO, because it combines the SNR benefits of a distributed operation with the ability to suppress interference with $N=4$ antennas per AP.
  The 90\% likely SE with the distributed cell-free operation is around 24\% and 90\% larger than in the small-cell and Cellular Massive MIMO systems, respectively. 
 In conclusion, User-centric Cell-free Massive MIMO provides a more uniform service quality among the UEs (as seen from the steeper CDF curves) and, particularly, improves the service for the UEs with the weakest channel conditions (as seen from the locations of the lower tails of the curves). The centralized operation is by far the most desirable from an SE performance perspective.

\subsection{Impact of Spatial Correlation}

\begin{figure}[t!]\centering
	\begin{overpic}[width=0.88\columnwidth,tics=10]{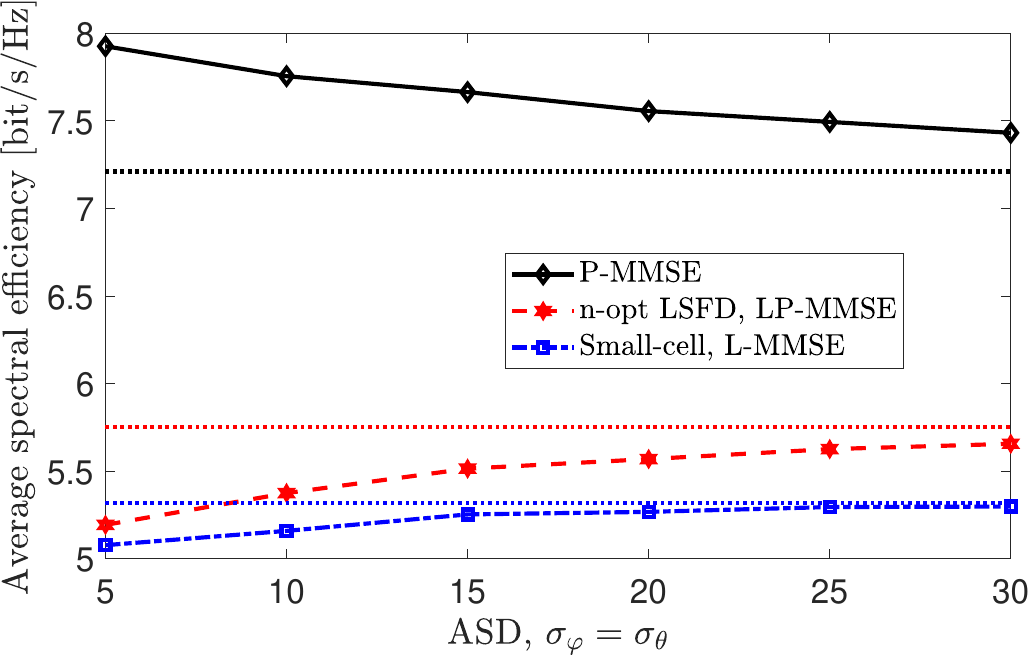}
	\end{overpic} 
	\caption{The average uplink SE per UE as a function of ASD for azimuth and elevation angles, $\sigma_{\varphi}=\sigma_{\theta}$ for different operations of Cell-free Massive MIMO and small-cell systems. We consider $L=100$, $N=4$, $K=40$, and $\tau_p=10$. The results for uncorrelated Rayleigh fading are shown as reference by the dotted lines.}    \vspace{+1mm}
	\label{figureSec5_6} 
\end{figure}

In the remaining simulations of this section, we will consider $L=100$ APs with $N=4$ antennas per AP. 
Figure~\ref{figureSec5_6} analyzes the impact of spatial channel correlation. This figure shows the average SE per UE as a function of the ASD of the azimuth and elevation angular spread. We compare a cell-free network with scalable centralized and distributed operation to a small-cell setup with L-MMSE combining. The ASD for azimuth and elevation angles are the same, $\sigma_{\varphi}=\sigma_{\theta}$, and varied on the horizontal axis. The straight dotted lines present the reference average SE in the case of uncorrelated Rayleigh fading, where the spatial correlation matrices are scaled identities. We can think of uncorrelated fading as the limit case where the ASDs tend to infinity. 
 
Note that smaller ASD means more highly correlated channels.
We notice from Figure~\ref{figureSec5_6} that spatial correlation improves the average SE in the centralized operation whereas it degrades the average SE in the distributed operation and small-cell system. 
When the ASD increases, the average SE with spatially correlated Rayleigh fading channels approaches the reference lines for the uncorrelated case; the difference is small when the ASDs are $30^\circ$. The intuition behind these results is that the centralized operation can exploit the spatial correlation to separate the UEs better (as is the case in Cellular Massive MIMO \cite[Sec.~4.1.5]{massivemimobook}) since it co-processes the signals over many antennas. However, in the distributed operation and small-cell implementation, local combining is applied at each AP and then we are mainly subject to the negative effects of spatial correlation.

Another interesting observation is that the average SE of the small-cell network and the distributed Cell-free Massive MIMO are nearly the same when the channels are highly correlated (i.e., almost \gls{LoS}), while the cell-free system benefits more from having lower spatial correlation.

\begin{figure}[t!]\centering
	\begin{overpic}[width=0.88\columnwidth,tics=10]{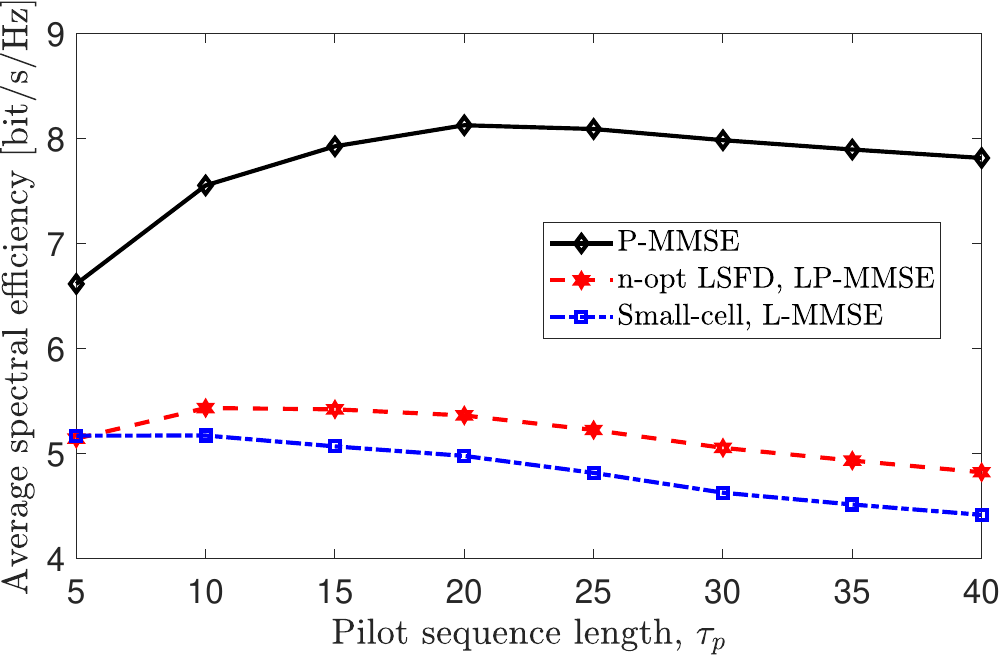}
	\end{overpic} 
	\caption{The average uplink SE per UE as a function of pilot sequence length, $\tau_p$, for different operations of Cell-free Massive MIMO and small-cell setup. We consider $L=100$, $N=4$, $K=40$, and spatially correlated Rayleigh fading with ASD $\sigma_{\varphi}=\sigma_{\theta}=15^{\circ}$. }    \vspace{+2mm}
	\label{figureSec5_7} 
\end{figure}

\subsection{Impact of the Pilot Sequence Length}

Figure~\ref{figureSec5_7} shows how the pilot sequence length, $\tau_p$, impacts the average SE per UE. We recall that a larger $\tau_p$ will result in less pilot contamination when $K$ is maintained fixed but also in a pre-log penalty since there is less room for data in every coherence block. For both the centralized and distributed cell-free operation, increasing the pilot sequence length improves the SE until a certain point, and then it decays with $\tau_p$. There are several reasons for this. First, reducing pilot contamination improves the channel estimation quality and reduces the coherent interference. Moreover, each UE is now served by more APs since each AP is allowed to serve up to $\tau_p$ UEs in the considered scalable DCC and pilot assignment scheme. However, after some point (i.e., $\tau_p=20$ in the centralized operation), the improvement saturates since the number of transmission symbols, $\tau_u$, allocated to uplink data transmission also decreases. If we continue to increase the pilot length, the SE will eventually reduce because of the pre-log penalty.
The saturation point occurs at a smaller value (i.e.,  $\tau_p=10$) in the distributed cell-free operation, while $\tau_p=5$ gives the highest SE in the small-cell system. The reason is the reduced ability to suppress interference, which does not improve much by increasing the estimation quality.

\subsection{Scalability of Fronthaul and Computations}

We will now consider the fronthaul signaling and computational complexity associated with different types of centralized and distributed schemes. We consider several random network realizations by dropping APs and UEs randomly in the simulation area and all the results in this section present the signaling and complexity per AP. Hence, they are obtained by averaging over both network realizations and all the APs. In addition, we count the operations and complex numbers once when they are common for more than one UE. 

\begin{figure}[p!]
        \begin{subfigure}[b]{\columnwidth} \centering 
	\begin{overpic}[width=0.88\columnwidth,tics=10]{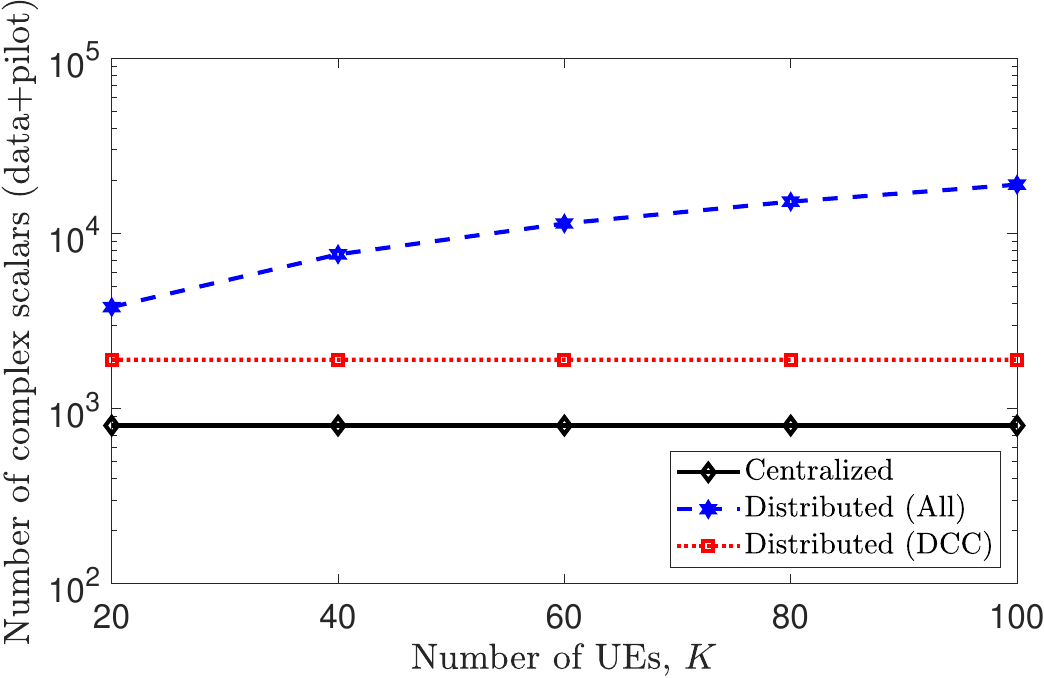}
\end{overpic} \label{figureSec5_8a} 
                \caption{Data plus pilot signals per coherence block.} 
                \vspace{4mm}
        \end{subfigure} 
        \begin{subfigure}[b]{\columnwidth} \centering 
	\begin{overpic}[width=0.88\columnwidth,tics=10]{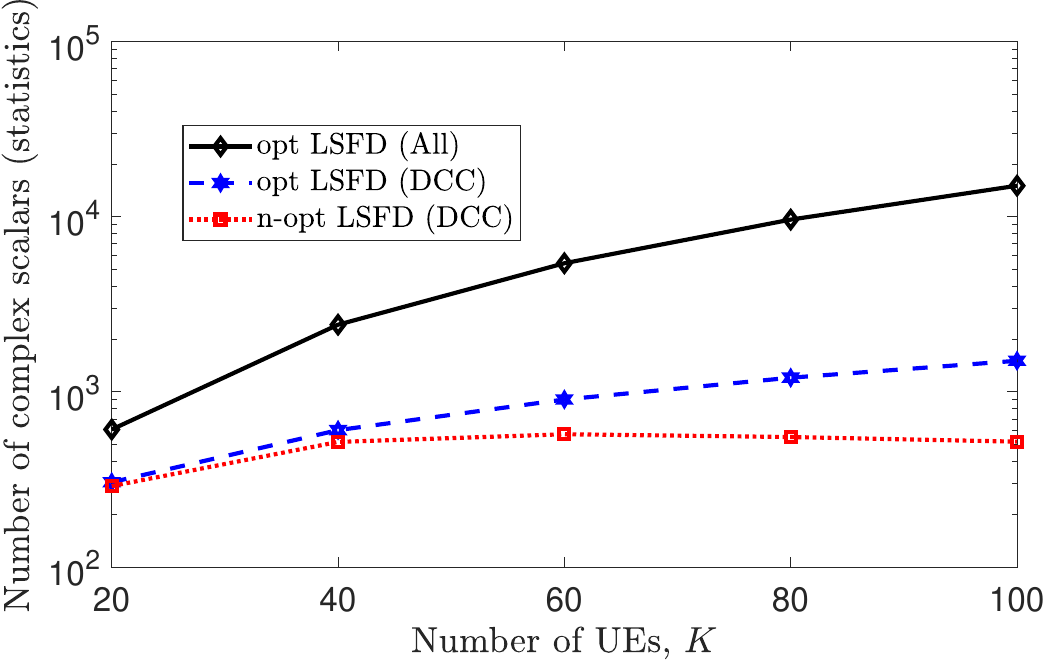}
\end{overpic} \label{figureSec5_8b}
                \caption{Statistical parameters per channel statistics realization. No statistical parameters are  exchanged with centralized operation.} 
        \end{subfigure} 
	\caption{The average number of complex scalars to send from the  APs  to the CPU over the fronthaul either in each coherence block or for each realization of the channel statistics as a function of the number $K$ of UEs. We consider $L=100$, $N=4$, and $\tau_p=10$. Both centralized and distributed operations are considered.}\label{figureSec5_8}
\end{figure}

\enlargethispage{\baselineskip} 

\subsubsection{Fronthaul Signaling}

To quantify the fronthaul signaling load of different cell-free operations, we present the average number of complex scalars to be sent from the APs to the CPU per AP in each coherence block in Figure~\ref{figureSec5_8}(a). These scalars can represent both data and pilots, depending on the operation. The number of UEs is shown on the horizontal axis and a scalable fronthaul operation is represented by having a number that does not grow with $K$.
We note that, for the centralized operation, the number of complex scalars to be sent from the APs to the CPU in each coherence block is the same irrespective of the number of UEs and DCC scheme, thus scalability is not an issue in terms of fronthaul capacity.
For the distributed operation, the number of complex scalars related to the data signals depends on the DCC. We see that it grows with $K$ if all APs serve all UEs, while it does not grow when using the proposed scalable DCC solution where each AP serves $\tau_p$ UEs.
Interestingly, the fronthaul load is larger in the distributed operations than in the centralized operation. The reason is that every AP is serving more UEs than it has antennas (i.e., $\tau_p > N$), thus the local receive combining is increasing the dimension of the signals that need to be sent to the CPU. Hence, the main purpose of a distributed operation cannot be to reduce the fronthaul capacity but to make use of distributed processing capabilities.

The number of statistical parameters that must be sent to the CPU in one realization of the network is shown in Figure~\ref{figureSec5_8}(b). We note that, for the centralized operation, no statistical parameters need to be exchanged so there are no curves for this case. 
For the distributed operation with LSFD, there is a certain number of statistical parameters to be sent from each AP to the CPU in each realization of the channel statistics. The largest number of parameters is needed when using opt LSFD and serving all UEs using all APs. The number of scalars grows rapidly with the number of UEs.
Even if we introduce DCC, the number of scalars grows with $K$ if opt LSFD is used, which indicates that the fronthaul signaling load is not scalable. However, n-opt LSFD using DCC is a scalable alternative. We notice that there is a slight decrease in the number of statistical parameters that need to be sent to the CPU in this case. The reason is that as the number of UEs increases, the average number of APs serving a generic UE $k$ decreases since each AP at most serves $\tau_p$ UEs. This in turn results in a decrease in the number of UEs that are partially served by the same APs as UE $k$, i.e., $|\mathcal{S}_k|$ in Table~\ref{tab:signaling-level4-level3}.

\begin{figure}[p!]
        \begin{subfigure}[b]{\columnwidth} \centering 
	\begin{overpic}[width=0.88\columnwidth,tics=10]{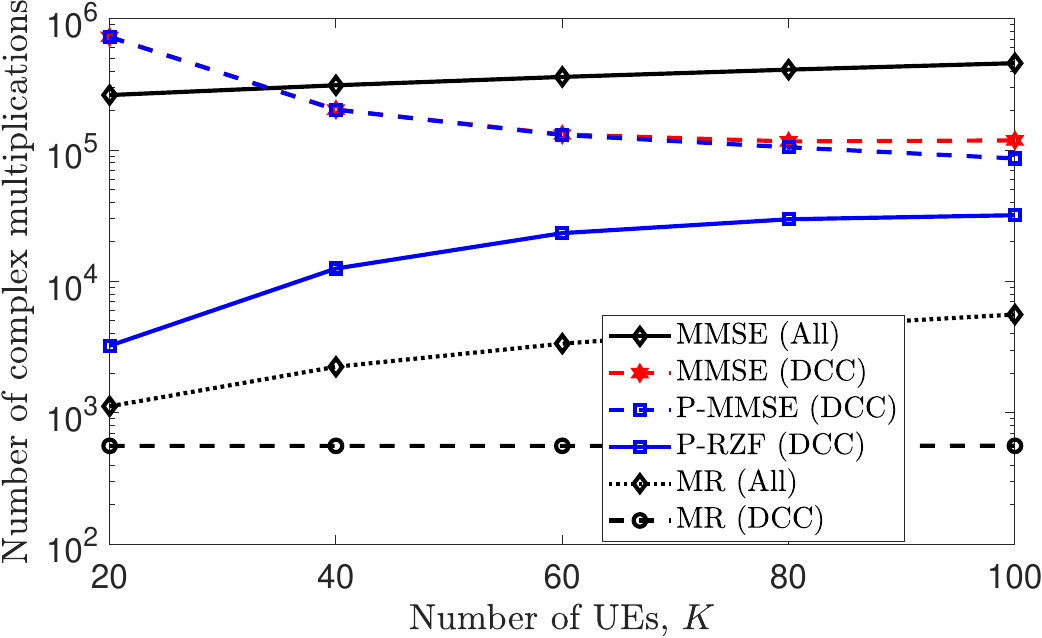}
\end{overpic} \label{figureSec5_9a}
                \caption{Centralized operation.} 
                \vspace{4mm}
        \end{subfigure} 
        \begin{subfigure}[b]{\columnwidth} \centering 
	\begin{overpic}[width=0.88\columnwidth,tics=10]{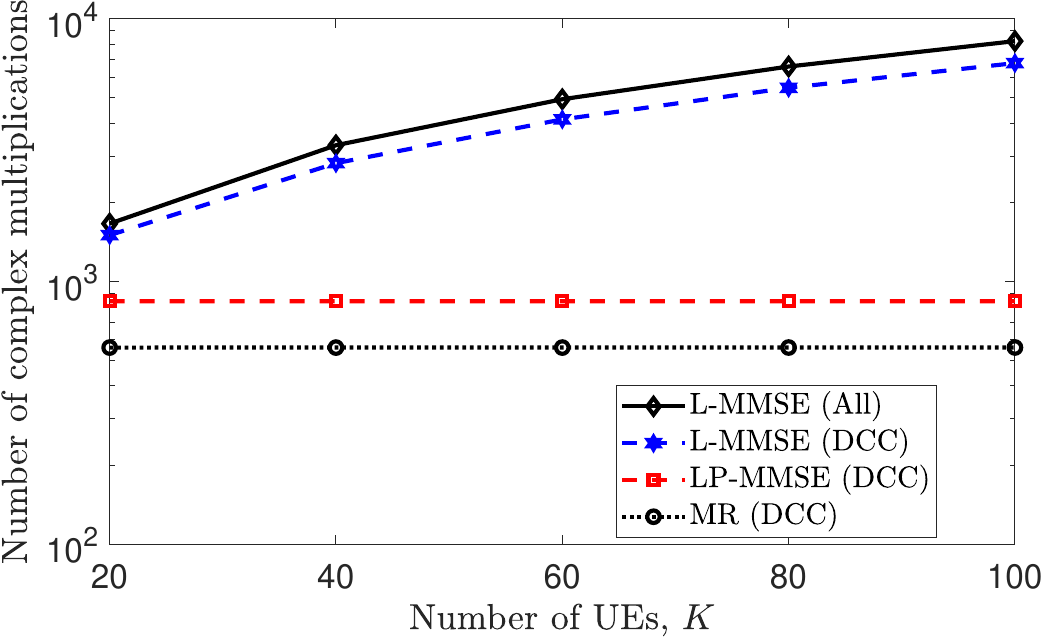}
\end{overpic} \label{figureSec5_9b}
                \caption{Distributed operation.} 
        \end{subfigure} 
	\caption{The average number of complex multiplications required per coherence block to compute the local combining vectors, including the computation of the channel estimates as a function of the number $K$ of UEs. We consider $L=100$, $N=4$, and $\tau_p=10$. Different combining schemes for the centralized and distributed operations are considered.}\label{figureSec5_9}
\end{figure}

\subsubsection{Computational Complexity}

Figure~\ref{figureSec5_9} shows the average number of complex multiplications required for the computation of different centralized and distributed combining schemes. The results for the centralized operation in Figure~\ref{figureSec5_9}(a) are based on Table~\ref{table:complexity-with-level4} whereas the results for the distributed operation in Figure~\ref{figureSec5_9}(b) are obtained from Table~\ref{table:complexity-with-level3}. Note that the matrix in  \eqref{eq:MMSE-combining} that needs to be inverted is the same when all APs serve all UEs and as all the other common operations, this is counted once for all the UEs. However, with DCC, the matrix inverse is taken separately for each UE resulting in more complex multiplications for $K=20$ compared to MMSE combining where all APs serve all UEs, as can be seen in Figure~\ref{figureSec5_9}(a). However, as the number of UEs increases, the computational complexity decreases in the DCC case while it increases when all APs serve all UEs. Hence, for a large number of UEs, the computational complexity can be reduced efficiently with a well-designed DCC. We know from previous figures that scalable methods can be utilized with a negligible SE loss and the benefit of doing that is the vast reduction in complexity observed in this figure.
Another observation is that P-RZF might be useful when the number of UEs is relatively small, however, when $K$ increases, the gap between P-MMSE and P-RZF decreases. The reason that the complexity of P-RZF increases with $K$ is that the number of UEs $|\mathcal{S}_k|$ that are served by partially the same APs as UE $k$ increases on the average when we increase $K$ from 20 to 60. Although $|\mathcal{S}_k|$ decreases after $K=60$, the overall effect of the other parameters from Table~\ref{table:complexity-with-level4} results in such a pattern in the computational complexity of P-RZF combining.

From Figure~\ref{figureSec5_9}(b), we observe that the average number of complex multiplications does not grow with $K$ for the scalable combining LP-MMSE and MR schemes. However, the total complexity grows with $K$ when using the unscalable L-MMSE schemes. Hence, scalability is important also in the distributed operation. Note that the vertical scale is substantially smaller in the distributed operation than in the centralized operation, because the distributed schemes are less computationally complex.

\newpage

\section{Summary of the Key Points in Section~\ref{c-uplink}}
\label{c-uplink-summary}

\begin{mdframed}[style=GrayFrame]

\begin{itemize}
\item The uplink of a cell-free network can be implemented with either centralized or distributed operation.

\item In the centralized operation, the APs are forwarding all the uplink signals to the CPU, which carries out the channel estimation and receive combining. This operation enables centralized combining where the signals received at all the APs that serve UE $k$ can be used to balance between obtaining a strong signal and suppressing interference.

\item The uplink SE with the centralized operation is given in Theorem~\ref{proposition:uplink-capacity-general}. It is maximized by MMSE combining, but this is not a scalable scheme. The alternative  P-MMSE, P-RZF, and MR combining schemes are scalable and represent different tradeoffs between complexity and performance.

\item In the distributed operation, each AP uses its locally received signals to compute local channel estimates and estimates of the uplink data. The local data estimates are then forwarded to the CPU, which computes a weighted average of the local estimates and then detects the data. The latter stage is called LSFD and should give higher priority to APs with good channel conditions.

\item The uplink SE with the distributed operation is given in Theorem~\ref{theorem:uplink-capacity-level3} and depends both on the local receive combining and on the LSFD weights. The optimal weights can be computed but are not scalable to implement, thus a nearly optimal but scalable alternative was derived. L-MMSE is the local receive combining that minimizes the MSE of the local data estimate, but it is not scalable. LP-MMSE and MR combining are scalable alternatives.

\end{itemize}

\end{mdframed}

\begin{mdframed}[style=GrayFrame]

\begin{itemize}

\item We defined a running example and used it to evaluate the performance of cell-free networks. We noticed that a scalable operation can be achieved with a negligible SE loss compared to the case when all APs serve all UEs, but with a substantial reduction in computational complexity.

\item The centralized operation can achieve substantially higher SE than the distributed operation, but both implementations outperform Cellular Massive MIMO and small-cell networks. The gain is particularly large for the UEs with the worst channel conditions, which demonstrates the ability of the cell-free architecture to increase the uniformity of the service. The centralized operation is associated with higher computational complexity than the distributed operation, but less fronthaul signaling.

\item If the total number of AP antennas is fixed, a centralized operation prefers to have many single-antenna APs since this reduces the average distance between a UE and its closest APs, while the interference is mitigated by centralized receive combining. In a distributed operation, there is another tradeoff: having many single-antenna APs is beneficial for the UEs with weak channel conditions, while having fewer multi-antenna APs gives a local interference suppression capability at each AP that is beneficial for the UEs with good channel conditions.

\end{itemize}

\end{mdframed}


\chapter{Downlink Operation}
\label{c-downlink}

This section describes two different downlink implementations of User-centric Cell-free Massive \gls{MIMO}, which are the counterparts of the two uplink operations described in Section~\ref{c-uplink}. Section~\ref{sec:level4-downlink} considers centralized operation in which the precoding is carried out at the CPU, 
based on the channel estimates that are obtained from the uplink pilot signals gathered from all APs. In this case, the CPU performs all the signal processing for channel estimation and precoding. The achievable SE is derived and different unscalable and scalable centralized transmit precoding schemes are presented. An uplink-downlink duality result is presented to motivate the precoding selection. 
Distributed operation is then considered in Section~\ref{sec:level2-downlink}. In this case, each AP applies precoding on the basis of locally obtained channel estimates, while the CPU is only encoding the data. The SE is analyzed and different local precoding schemes are presented. A performance comparison is provided in Section~\ref{sec:downlink-performance-comparison}, where the different schemes are analyzed in terms of SE.
The key points are summarized in Section~\ref{c-downlink-summary}.

\section{Centralized Downlink Operation}
\label{sec:level4-downlink}

As in the uplink, the most advanced downlink implementation of Cell-free Massive MIMO is a fully centralized operation, where the APs only act as remote-radio heads or relays that transmit signals that the CPU has generated and sent out over the fronthaul links. To explain the setup in further detail, we first recall the downlink system model from Section~\ref{sec:downlink-system-model-ch2} on p.~\pageref{sec:downlink-system-model-ch2}. 
The received downlink signal at UE $k$ is 
\begin{align}  \label{eq:downlink-received-UEk}
y_k^{\rm{dl}} = \sum_{l=1}^{L}  \vect{h}_{kl}^{\Htran} \vect{x}_l + n_k = \sum_{l=1}^{L}  \vect{h}_{kl}^{\Htran} \left( \sum_{i=1}^{K} \vect{D}_{il}  \vect{w}_{il} \varsigma_i \right)+ n_k
\end{align}
where $n_k \sim \CN(0,\sigmaDL)$ is the receiver noise and
\begin{align} 
\vect{x}_l = \sum_{i=1}^{K} \vect{D}_{il}  \vect{w}_{il} \varsigma_i
\label{eq:DCC-transmit-downlink2}
\end{align}
is the signal transmitted by AP $l$. This signal is created as the sum of the UEs' signals where each term intended for UE $i$ consists of the unit-power downlink data signal $\varsigma_i \in \mathbb{C}$ (with $\mathbb{E} \{|\varsigma_i|^2 \} = 1$) and the effective transmit precoding vector
\begin{equation} 
\vect{D}_{il}  \vect{w}_{il} = \begin{cases} \vect{w}_{il} & l \in \mathcal{M}_i \\
\vect{0}_{N} & l \not \in \mathcal{M}_i.
\end{cases} \label{eq:effective-precoder}
\end{equation}
Recall that $ \vect{D}_{il}=\vect{0}_{N \times N}$ implies $\vect{D}_{il}  \vect{w}_{il}=\vect{0}_N$, thus AP $l$ will only transmit to UE $i$ in the downlink if $ \vect{D}_{il} \neq \vect{0}_{N\times N}$.
By utilizing the effective precoding vectors defined in \eqref{eq:effective-precoder}, the signal in \eqref{eq:DCC-transmit-downlink2} can be equivalently expressed as a summation only over the $| \mathcal{D}_l |$ UEs served by AP $l$:
\begin{align} 
\vect{x}_l = \sum_{i=1}^{K} \vect{D}_{il}  \vect{w}_{il} \varsigma_i  = \sum_{i \in \mathcal{D}_l}  \vect{w}_{il} \varsigma_i.
\label{eq:DCC-transmit-downlink3}
\end{align}
However, we will utilize the original formulation in \eqref{eq:DCC-transmit-downlink2} since it allows us to write the received signal in \eqref{eq:downlink-received-UEk} in compact form as
\begin{align}
y_k^{\rm{dl}} =  \sum_{i=1}^{K} \begin{bmatrix} \vect{h}_{k1} \\ \vdots \\ \vect{h}_{kL} \end{bmatrix}^{\Htran} 
 \begin{bmatrix} \vect{D}_{i1}  \vect{w}_{i1} \\ \vdots \\ \vect{D}_{iL}  \vect{w}_{iL} \end{bmatrix}
 \varsigma_i + n_k =  \sum_{i=1}^{K} \vect{h}_k^{\Htran} \vect{D}_i \vect{w}_i \varsigma_i + n_k \label{eq:downlink-received-UEk-centralized}
\end{align}
where $\vect{h}_k=\left[\vect{h}_{k1}^{\Ttran} \, \ldots \, \vect{h}_{kL}^{\Ttran}\right]^{\Ttran} \in \mathbb{C}^{LN}$ is the collective channel to UE $k$ from all APs, $\vect{w}_i = \left[\vect{w}_{i1}^{\Ttran} \, \ldots \, \vect{w}_{iL}^{\Ttran}\right]^{\Ttran} \in \mathbb{C}^{LN}$ is the collective precoding vector assigned to UE $i$, and the block-diagonal matrix $\vect{D}_i = \diag(\vect{D}_{i1},\ldots,\vect{D}_{iL})$ identifies which APs are transmitting to UE $i$.

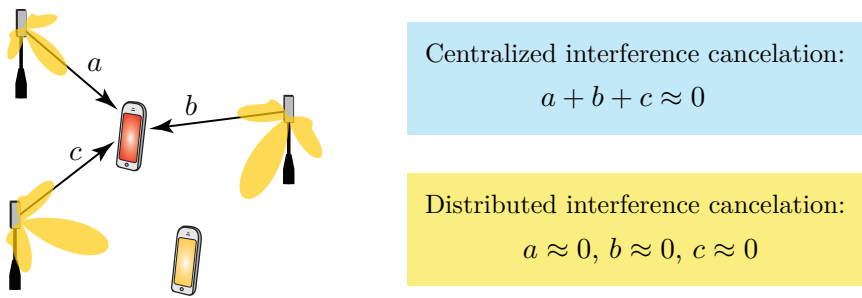
\begin{figure}[t!]
	\centering 
	\begin{overpic}[width=\columnwidth,tics=10]{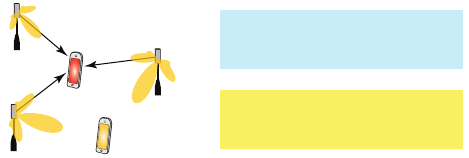}
		\put(11,26){$a$}
		\put(22,21){$b$}
		\put(9,16){$c$}
		\put(49,27){\small{Centralized interference cancelation:}}
		\put(49,10){\small{Distributed interference cancelation:}}
		\put(62,22){$a+b+c \approx  0$}
		\put(60,5){$a \approx 0$, $b \approx 0$, $c \approx 0$}
	\end{overpic}  
	\caption{When three APs transmit to a UE, they will cause interference to other neighboring UEs. In the centralized operation, the CPU can design the precoding so that the APs cancel out others' interference contributions: $a+b+c \approx 0$. In a distributed operation, where $a,b,c$ are selected without cooperation, the only way to cancel interference is to make the individual contributions small: 
		$a \approx 0$, $b \approx 0$, $c \approx 0$.}
	\label{figure:distributed-beamforming}  
\end{figure}

The system model formulation in \eqref{eq:downlink-received-UEk-centralized} provides a network-wide perspective on the downlink transmission
and is particularly useful when describing the centralized operation. 
The CPU estimates the downlink channels by exploiting the uplink-downlink channel reciprocity, which says that the uplink and downlink channels are identical within a coherence block.\footnote{Even if the physical channel is the same in the uplink and the downlink, different pieces of transceiver hardware are being used in the two directions. This can create a mismatch between the uplink and downlink channels in practice. This mismatch can be estimated and compensated for using reciprocity calibration algorithms. We will not cover that in this monograph but refer to \cite{Guillaud2005a}, \cite{Nishimori2001a}, \cite{Rogalin2014a}, \cite{Shepard2012a}, \cite{Vieira2017a}, \cite{Vieira2014b}, \cite{Zetterberg2011a} for details and algorithms.}
Hence, we can utilize the estimation results from Section~\ref{sec:MMSE-channel-estimation} on p.~\pageref{sec:MMSE-channel-estimation} to compute the MMSE estimates of the collective channels in the downlink.
The estimates are utilized by the CPU  to compute the collective precoding vectors $\{ \vect{D}_i \vect{w}_i:i=1,\ldots,K\}$ for all the $K$ UEs in the network. 
The signal $\vect{x}_l$ in \eqref{eq:DCC-transmit-downlink2} is generated for each AP using these precoding vectors and the downlink data $\{\varsigma_i:i=1,\ldots,K\}$. Each AP $l$ must only be aware of its $N$-dimensional piece $\vect{D}_{il}  \vect{w}_{il}$ of the collective precoding vector $\vect{D}_{i}  \vect{w}_{i}$ of UE $i$, but the CPU can design the collective vector so that the different pieces fit well together. In particular, the CPU can select the precoding so that the APs can cancel out each others' interference in a way that each AP is unable to figure out individually. This feature is illustrated in Figure~\ref{figure:distributed-beamforming}, where three APs are transmitting to the yellow UE but the signals are creating interference to the red UE. If the interference from the APs are represented by $a,b,c \in \mathbb{C}$, then the CPU can adjust the APs' transmit powers and phases so that $a+b+c \approx 0$. Hence, a viable solution is that each AP creates a large amount of interference but it is canceled by an equally strong amount of interference with the opposite sign from another AP: $b = -a$ and $|a|=|b| \gg 0$.
Such multi-AP interference cancelation is not possible to implement in a distributed operation where the APs are unaware of each others' channels, making $a,b,c \approx 0$ the only viable way to limit the interference.
We will return to the centralized and distributed precoding design later in this section. 

\begin{figure}[t!]
	\centering 
	\begin{overpic}[width=\columnwidth,tics=10]{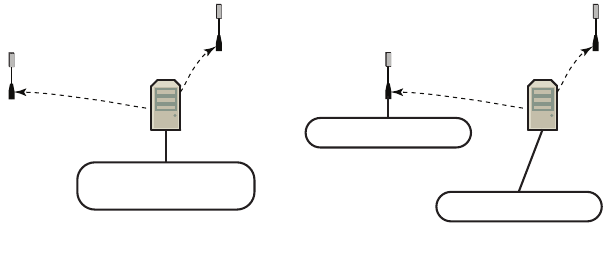}
		\put(10,28.5){$\vect{x}_l$}
		\put(0,36){\small{AP $l$}}
		\put(32,24){\small{CPU}}
		\put(70,28.5){$\varsigma_1,\ldots,\varsigma_K$}
		\put(15,13){\small{Data encoding}}
		\put(15,10){\small{Transmit precoding}}
		\put(51.5,20){\small{Transmit precoding}}
		\put(94,24){\small{CPU}}
		\put(62,36){\small{AP $l$}}
		\put(76.5,8){\small{Data encoding}}
		\put(10,0){\small{(a) Centralized operation}}
		\put(60,0){\small{(b) Distributed operation}}
	\end{overpic}  
	\caption{The downlink signal processing tasks can be divided between the APs and CPU in different ways. In the centralized operation, the data encoding and transmit precoding are done at the CPU. In the distributed operation, everything except the data encoding is done at the APs.}
	\label{figure:distributed_centralized_downlink}  
\end{figure}
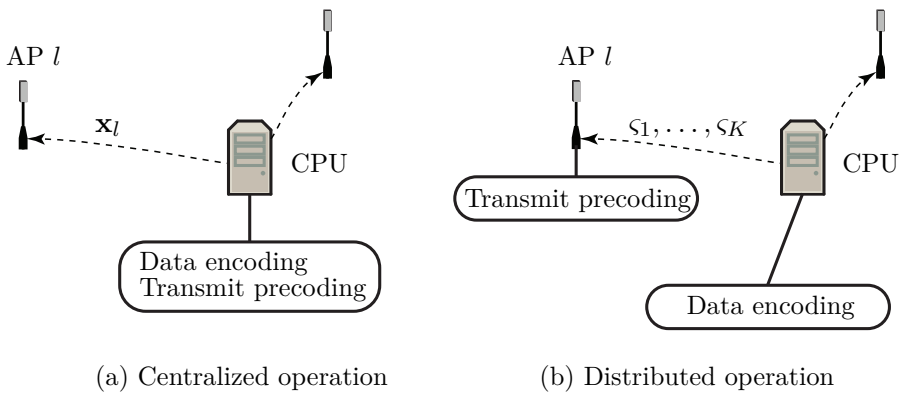

Figure~\ref{figure:distributed_centralized_downlink}(a) illustrates how the downlink signal processing is carried at the CPU in the centralized operation.
We will describe the centralized operation as if the CPU performs the baseband processing while the conversion from digital baseband to analog passband signals is carried out at each AP, but it is also possible that each AP receives a passband signal from the CPU, which is then amplified before transmission.\footnote{This can be implemented using radio-over-fiber technology.}

\enlargethispage{\baselineskip} 

The expression in \eqref{eq:downlink-received-UEk-centralized} is mathematically equivalent to the signal model of a downlink single-cell Massive MIMO system with correlated fading \cite[Sec.~2.3.2]{massivemimobook} if one treats the CPU as a transmitter equipped with $LN$ antennas.
However, some key differences exist as also previously described in the uplink:

\begin{enumerate}

\item Multiple UEs that are managed by the CPU are using the same pilot, which is normally avoided in single-cell systems.

\item The antennas are distributed at different geographical locations. Hence, the collective channel is distributed as $\vect{h}_k \sim \CN(\vect{0}_{LN}, \vect{R}_{k})$ where the spatial correlation matrix $\vect{R}_{k} = \diag(\vect{R}_{k1}, \ldots,   \vect{R}_{kL}) \in \mathbb{C}^{LN \times LN}$ has a block-diagonal structure, which is normally not the case in single-cell systems. 

\item Many elements of the effective precoding vector $\vect{D}_k\vect{w}_k$ are zero, namely the ones corresponding to APs that are not serving UE $k$.

\item The precoding vectors should be chosen to satisfy per-AP power constraints, instead of a total power constraint as in single-cell Massive MIMO systems. More precisely, we require that
\begin{equation} \label{eq:downlink-power-constraint}
\mathbb{E} \left\{ \| \vect{x}_l \|^2 \right\} \leq \rhomax, \quad l=1,\ldots,L
\end{equation}
where $\rhomax \geq 0$ denotes the maximum downlink transmit power of an AP. For notational convenience, $\rhomax$ is assumed to be the same for all APs. Note that the expectation in \eqref{eq:downlink-power-constraint} is over both the data signals and channel realizations. The motivation is that we consider a system where the available bandwidth resources are divided into many coherence blocks.  The AP's power amplifier will simultaneously transmit signals over these many blocks with independent channel realizations, thus we can exploit the ability to assign different amounts of energy to different coherence blocks by using a power constraint of the type in \eqref{eq:downlink-power-constraint}.
\end{enumerate}

These differences play an important role in the operation of cell-free networks and differentiate the SE analysis in this monograph from what can be found in the literature on Cellular Massive MIMO.

\subsection{Spectral Efficiency With Centralized Operation}
\label{sec:level4-SE-DL}

We will now derive an achievable SE expression that applies when using any precoding scheme.
The received signal in \eqref{eq:downlink-received-UEk-centralized} for UE $k$ can be divided into three terms:
\begin{align} 
y_k^{\rm{dl}}=\underbrace{\vphantom{\sum_{i=1,i\ne k}^{K}} \vect{h}_{k}^{\Htran} \vect{D}_{k}  \vect{w}_{k} \varsigma_k}_{\textrm{Desired signal}} + \underbrace{\condSum{i=1}{i \neq k}{K}  \vect{h}_{k}^{\Htran}\vect{D}_{i}  \vect{w}_{i} \varsigma_i}_{\textrm{Inter-user interference}} + \underbrace{ \vphantom{\sum_{i=1,i\ne k}^{K}} n_k.}_{\textrm{Noise}} \label{eq:DCC-downlink-three-parts}
\end{align}
The UE wants to extract the data from the first term under the presence of inter-user interference and noise, which are represented by the latter two terms. The expression $\vect{h}_{k}^{\Htran} \vect{D}_{k}  \vect{w}_{k}$ that is multiplied with the desired signal $\varsigma_k$ is called the \emph{effective downlink channel}. This terminology~refers to the fact that the precoding is effectively turning the multiple-antenna channel $\vect{h}_{k}$ into the effective single-antenna channel $\vect{h}_{k}^{\Htran} \vect{D}_{k}  \vect{w}_{k}$. The CPU has partial knowledge about the effective channel. It knows the precoding vector (since it is the one selecting it) and it also knows the partial MMSE estimate
\begin{equation} \label{eq:estimates-central-2}
\vect{D}_k \widehat{\vect{h}}_{k} 
= \begin{bmatrix} \vect{D}_{k1} \widehat{\vect{h}}_{k1} \\ \vdots \\  \vect{D}_{kL} \widehat{\vect{h}}_{kL} 
\end{bmatrix} 
\sim \CN \left( \vect{0}_{LN}, \eta_k \tau_p \vect{D}_{k} \vect{R}_{k} \vect{\Psi}_{t_k}^{-1} \vect{R}_{k} \vect{D}_{k} \right)
\end{equation}
of the channel. This partial MMSE channel estimate was previously defined in \eqref{eq:estimates-central} along with the corresponding estimation error $\vect{D}_k\widetilde{\vect{h}}_k = \vect{D}_k\vect{h}_k - \vect{D}_k\widehat{\vect{h}}_{k} \sim \CN(\vect{0}_{LN},\vect{D}_k\vect{C}_k)$  with $\vect{C}_k = \diag(\vect{C}_{k1}, \ldots, \vect{C}_{kL})$.

UE $k$ can transmit in the uplink without knowing the channel but it cannot decode the downlink data without utilizing some information about the effective channel $\vect{h}_{k}^{\Htran} \vect{D}_{k}  \vect{w}_{k}$. We have not defined any functionality for estimating this term. This is not as strange as it might seem but it is a common practice in communication theory to only send pilot signals in one direction (e.g., the uplink) and then let the entity that obtains those estimates transmit in such a way that the effective channel becomes (approximately) a positive scalar that the receiver can estimate without the need for sending pilot signals.

\begin{remark}[Pilots in only uplink direction] \label{remark:pilot-directions}
Consider a scalar channel $g \in \mathbb{C}$ which is  known at the AP but not at the UE. The AP can then precode the data signal $\varsigma$ using $g^\star/|g|$ so that the received signal becomes $g g^\star/|g| \varsigma = |g| \varsigma$. The UE still does not know the channel, but it knows that the effective channel $|g|$ is positive and it knows the power of the data signal $\varsigma$. The UE can therefore deduce the effective channel by computing the sample average power of the received signal over the received data signals in a coherence block (where the channel $|g|$ is fixed but $\varsigma$ is changing) \cite{Ngo2017a}. However, if the channel is only partially known at the AP (e.g., due to estimation errors or imperfect hardware calibration between uplink and downlink), then it might be necessary to transmit one or multiple downlink pilots as well \cite{Interdonato2019b} (e.g., dedicated demodulation reference signals or using a differential modulation scheme). For historical reasons, this is commonly done in practical networks but will not be considered in this monograph.
\end{remark}

We will compute an SE expression for the case when the UE knows the average value $\mathbb{E}\left\{\vect{h}_{k}^{\Htran}\vect{D}_{k}\vect{w}_k\right\}$ of the effective channel. This is a deterministic number and can, thus, be easily obtained in practice; deterministic numbers are always assumed known in the capacity analysis since the data transmission spans over infinitely many transmission symbols and coherence blocks.
We cannot reuse the capacity bound that was provided in Theorem~\ref{proposition:uplink-capacity-general} on p.~\pageref{proposition:uplink-capacity-general} for the centralized operation in the uplink, but instead, we will utilize the alternative SE expression in Theorem~\ref{proposition:uplink-capacity-general-UatF} on p.~\pageref{proposition:uplink-capacity-general-UatF}. We then obtain the following result.

\begin{theorem} \label{theorem:downlink-capacity-level4}
	An achievable SE of UE $k$ in the downlink with centralized operation is
	\begin{equation} \label{eq:downlink-rate-expression-level4}
	\begin{split}
	\mathacr{SE}_{k}^{({\rm dl},\cent)} = \frac{\tau_d}{\tau_c} \log_2  \left( 1 + \mathacr{SINR}_{k}^{({\rm dl},\cent)}  \right) \quad \textrm{bit/s/Hz}
	\end{split}
	\end{equation}
	where $\mathacr{SINR}_{k}^{({\rm dl},\cent)}$ is the effective SINR
	\begin{equation} \label{eq:downlink-instant-SINR-level4}
	\mathacr{SINR}_{k}^{({\rm dl},\cent)} =  \frac{ \left| \mathbb{E}\left \{ \vect{h}_{k}^{\Htran}\vect{D}_{k}{\vect{w}}_{k}\right \} \right|^2  }{  \sum\limits_{i=1}^{K} \mathbb{E} \left\{ \left|  \vect{h}_{k}^{\Htran} \vect{D}_{i} {\vect{w}}_{i} \right|^2 \right\} -  \left| \mathbb{E} \left\{  \vect{h}_{k}^{\Htran} \vect{D}_{k} {\vect{w}}_{k} \right\} \right|^2 +  \sigmaDL.
	}
	\end{equation}
\end{theorem}
\begin{proof}
The proof is given in Appendix~\ref{proof-of-theorem:downlink-capacity-level4} on p.~\pageref{proof-of-theorem:downlink-capacity-level4}.
\end{proof}

The pre-log factor ${\tau_d}/{\tau_c}$ in \eqref{eq:downlink-rate-expression-level4} is the fraction of each coherence block that is used for downlink data transmission. 
The term $\mathacr{SINR}_{k}^{({\rm dl},\cent)}$ in \eqref{eq:downlink-instant-SINR-level4} takes the form of an effective SINR, which means that the SE matches the capacity of a deterministic single-antenna single-user AWGN channel with an SNR that equals $\mathacr{SINR}_{k}^{({\rm dl},\cent)}$. It also means that the data signal can be encoded and the received signal can be decoded as if we would be communicating over such an AWGN channel. Hence, there is a known way to achieve the SE stated in Theorem~\ref{theorem:downlink-capacity-level4}.
The numerator of \eqref{eq:downlink-instant-SINR-level4} contains the square of the average effective channel. The denominator equals the total power  $\mathbb{E}\{ | y_k^{\rm{dl}}|^2\} = \sum_{i=1}^{K} \mathbb{E} \{ \left|  \vect{h}_{k}^{\Htran} \vect{D}_{i} {\vect{w}}_{i} \right|^2 \} +  \sigmaDL$ of the received signal minus the useful term from the numerator.

The SE expression in Theorem~\ref{theorem:downlink-capacity-level4} holds for any transmit precoding vector $\vect{w}_{k}$ and selection of the DCC. In fact, it also holds for any channel distribution and not only correlated Rayleigh fading, as assumed in this monograph.
The expression can be computed for any $\vect{w}_{k}$ by using Monte Carlo methods, which means that each expectation is approximated with the sample average over a large number of random realizations. More precisely, we can generate realizations of the channel estimates in a large set of coherence blocks, compute each term in \eqref{eq:downlink-instant-SINR-level4} for each realization, and then take the average over all these computations. This is the approach that we will use when evaluating the SE numerically later in this section.

\subsection{Centralized Transmit Precoding}
\label{subsec:centralized-precoding}

The SE expression in Theorem~\ref{theorem:downlink-capacity-level4} depends on the precoding of all UEs in the entire network. This makes the selection of transmit precoding vectors much more complicated than selecting receive combining vectors, which can be designed/optimized for one UE at a time. 
In particular, we noticed that the uplink SINR in \eqref{eq:uplink-instant-SINR-level4-Rayleigh} took the form of a generalized Rayleigh quotient and, thus, could be maximized with respect to the receive combining in closed form. There is no corresponding result for the downlink.
The intuition behind this difference is that the precoding determines how the signals are emitted from the APs. For whatever selected precoding, every UE will be interfered with by the transmission to the other UEs, even if the impact can hopefully be made marginal. In contrast, in the uplink, the received signals are not affected by the receive combining. However, after the reception, we take the received signals from a set of APs and process them to detect the signal from a particular UE. We can use the same received signals to detect the signals from another UE using another combining vector. 

The effective downlink SINR in \eqref{eq:downlink-instant-SINR-level4} has a very different structure than the uplink SINR in \eqref{eq:uplink-instant-SINR-level4-UatF} or \eqref{eq:uplink-instant-SINR-level4-Rayleigh},
 hence, the precoding cannot be optimized on a per-UE basis. In fact, there is not a single collection of optimal precoding vectors but it all depends on what tradeoff between the SEs achieved by the different UEs that we want to achieve.
Finding the optimal transmit precoding is computationally complicated in most cases \cite{Bjornson2013d} but there is a common heuristic that can be utilized to deduce the structure of the optimal precoding \cite{Bjornson2014d}.
It is obtained from the following uplink-downlink duality result \cite{Bjornson2020a}, which particularizes the classical duality results \cite{Boche2002a}, \cite{Tse2005a}, \cite{Viswanath2003a} for cell-free network architectures.

\begin{theorem} \label{prop:duality}
	Let $\{\vect{D}_i{\bf v}_i: i=1,\ldots,K \}$ and $\{p_i: i=1,\ldots,K \}$  denote the set of combining vectors and transmit powers, respectively, used in the uplink. 
	If the precoding vectors are selected as
	\begin{equation} \label{eq:precoding-vectors-duality-proposition}
	\vect{w}_i = \sqrt{\rho_i} \frac{{\bf v}_{i}}{\sqrt{\mathbb{E} \left\{ \| \vect{D}_{i} {\vect{v}}_{i} \|^2 \right\} }} 
	\end{equation}
	then there exists a downlink power allocation policy $\{\rho_i:  i=1,\ldots,K\}$
	with $\sum_{i=1}^{K} \rho_i/\sigmaDL=\sum_{i=1}^{K}p_i/\sigmaUL$ for which 
	\begin{align}
	\mathacr{SINR}_{k}^{({\rm dl,\cent})} = \mathacr{SINR}_{k}^{({\rm ul},\centuatf)},\quad k=1,\ldots,K
	\end{align}
	where $\mathacr{SINR}_{k}^{({\rm ul},\centuatf)}$ is the effective uplink SINR of UE $k$ from \eqref{eq:uplink-instant-SINR-level4-UatF}.
\end{theorem}
\begin{proof}
The proof is given in Appendix~\ref{proof-of-prop:duality} on p.~\pageref{proof-of-prop:duality}.
\end{proof}

This theorem shows that the effective SINRs that are achieved in the uplink (when using the SE expression from Theorem~\ref{proposition:uplink-capacity-general-UatF} on p.~\pageref{proposition:uplink-capacity-general-UatF}) are also achievable in the downlink. Consequently, we have that an achievable downlink SE for UE $k$ is
\begin{equation}
\mathacr{SE}_{k}^{({\rm dl,\cent})} = \frac{\tau_d}{\tau_c} \log_2   \left( 1 + \mathacr{SINR}_{k}^{({\rm ul},\centuatf)}   \right).
\end{equation}
To achieve this SE, the precoding must be selected according to \eqref{eq:precoding-vectors-duality-proposition}, which implies that the precoding vector for UE $i$ should be a scaled version of the combining vector that is assigned to this UE in the uplink. The scaling factors $\{\rho_i:i=1,\ldots,K\}$ represent the downlink power allocation.
The intuition behind the uplink-downlink duality in Theorem~\ref{prop:duality} is that the direction from which the signal is best received in the uplink coincides with the direction in which the signal should be transmitted in the downlink. Note that the word ``direction'' does not refer to a distinct angular direction in our three-dimensional world but the direction of a vector in the $LN$-dimensional vector space where the received and transmitted signals exist.

If the noise is the same in uplink and downlink (i.e., $\sigmaUL = \sigmaDL$), then Theorem~\ref{prop:duality} implies that the total transmit power is the same in both directions but is allocated differently between the UEs. The exact way of selecting the downlink power allocation $\{\rho_i:i=1,\ldots,K\}$ can be found in \eqref{eq:optimal-power-proof} on p.~\pageref{eq:optimal-power-proof} but we will not use this expression because there are multiple reasons why this power allocation is not used in practice.
On the one hand, each AP might be allowed to transmit with substantially higher power than each UE and we also expect that more APs than UEs exist in cell-free networks. Hence, the total available downlink power might far exceed the total uplink power and this should be utilized to achieve the maximum downlink performance. On the other hand, the power allocation obtained from the uplink-downlink duality might require that some APs transmit with very high power (beyond what is allowed by the power constraint in \eqref{eq:downlink-power-constraint}) while other APs might transmit with very low power. 
The bottom line is that we will utilize Theorem~\ref{prop:duality} as a motivation to heuristically select the downlink precoding vectors based on the uplink combining vectors according to \eqref{eq:precoding-vectors-duality-proposition}. However, \linebreak we will optimize the downlink transmit power separately instead of relying on the duality theorem. We will consider an arbitrary downlink power allocation in this section and then optimize it in Section~\ref{sec:optimized-power-allocation-downlink} on~p.~\pageref{sec:optimized-power-allocation-downlink}.

\enlargethispage{\baselineskip} 

In Section~\ref{subsec:optimal-receive-combining} on p.~\pageref{subsec:optimal-receive-combining} and Section~\ref{subsec:scalable-combining} on p.~\pageref{subsec:scalable-combining}, we presented several receive combining schemes for the centralized operation which we can utilize as the basis for the downlink precoding, motivated by the uplink-downlink duality. 
To make the transformation simple, we define the centralized precoding vector for UE $k$ as
	\begin{equation} \label{eq:precoding-centralized-expression}
	\vect{w}_k = \sqrt{\rho_k} \frac{\bar{\vect{w}}_k}{\sqrt{\mathbb{E} \left\{ \| \bar{\vect{w}}_k \|^2 \right\} }} 
	\end{equation}
where $\rho_k \geq 0 $ is the total transmit power assigned to UE $k$ from all the serving APs and $\bar{\vect{w}}_k \in \mathbb{C}^{LN}$ is an arbitrarily scaled vector that points out the direction of the precoding vector. Note that the normalization in \eqref{eq:precoding-centralized-expression} guarantees that 
\begin{equation}
\mathbb{E} \{ \| \vect{w}_k \|^2 \} = \rho_k.
\end{equation}
Any choice of $\bar{\vect{w}}_k$ will lead to a precoding scheme but scalability remains to be an important issue in the downlink operation. If we select $\bar{\vect{w}}_k$ based on a scalable uplink combining scheme, then the scalability property carries over to the downlink precoding. 

\subsubsection{MMSE Precoding}

The optimal centralized uplink operation is using MMSE combining as defined in \eqref{eq:MMSE-combining}. Based on the uplink-downlink duality, the downlink counterpart is \emph{MMSE precoding}, which is obtained from \eqref{eq:precoding-centralized-expression} using
\begin{equation} \label{eq:MMSE-precoding}
\bar{\vect{w}}_{k}^{{\rm MMSE}} =    p_{k} \left( \sum\limits_{i=1}^{K} p_{i}  \vect{D}_{k} ( \widehat{\vect{h}}_{i} \widehat{\vect{h}}_{i}^{\Htran} + \vect{C}_{i})\vect{D}_{k} + \sigmaUL \vect{I}_{LN} \right)^{\!-1} \vect{D}_{k}\widehat{\vect{h}}_{k}.
\end{equation}
While MMSE combining is provably optimal in the uplink, MMSE precoding is only optimal if we want to operate the downlink transmission in the way specified by the duality in Theorem~\ref{prop:duality}. One can then show that MMSE precoding is minimizing a sum MSE between the vector of data signals and the received signal vector of all UEs \cite{Vojcic1998a}.
In general, we want to make use of the flexibility to allocate the downlink power differently, thus we call MMSE precoding \emph{nearly optimal} rather than optimal. Intuitively, MMSE precoding will balance between transmitting a strong signal to the desired UE and limiting the interference caused to other UEs.
We will present several simplified precoding methods below and while we expect MMSE precoding to outperform them in terms of SE, there are only theoretical guarantees of this when using the power allocation specified by the duality theorem.

If MMSE combining is used in the uplink, then MMSE precoding can be used in the downlink without incurring any additional computational complexity (except for the scaling in \eqref{eq:precoding-centralized-expression} which can be absorbed into the generation of the data signals). However, we know from Table~\ref{table:complexity-with-level4} on p.~\pageref{table:complexity-with-level4} that the number of complex multiplications, regarding channel estimation and combining vector computation, grows with $K$ when using MMSE combining, thus MMSE precoding is not scalable for networks with a large number of UEs.

\subsubsection{P-MMSE and P-RZF Precoding}

Two scalable combining schemes 
 were presented in Section~\ref{subsec:scalable-combining} on p.~\pageref{subsec:scalable-combining} as an approximation of the optimal MMSE combining. These were called P-MMSE combining in \eqref{eq:P-MMSE-combining} and P-RZF combining in \eqref{eq:P-RZF-combining_1}. Motivated by the uplink-downlink duality, we can use these to develop two scalable precoding schemes.
The first one is \emph{P-MMSE precoding}, which is obtained from \eqref{eq:precoding-centralized-expression} using
\begin{equation} \label{eq:P-MMSE-precoding}
\bar{\vect{w}}_k^{{\rm P-MMSE}} =    p_k\left(\sum\limits_{i \in \mathcal {S}_k} p_{i}  \vect{D}_{k}\widehat{\vect{h}}_{i} \widehat{\vect{h}}_{i}^{\Htran}\vect{D}_{k} + {\bf{Z}}_{{\mathcal S}_k} + \sigmaUL  \vect{I}_{LN} \right)^{\!-1} \vect{{D}}_{k}\widehat{{\bf h}}_k
\end{equation}
where $\mathcal{S}_k$ from \eqref{eq:S_k} is the set of UEs served by partially the same APs as UE $k$ and ${\bf{Z}}_{{\mathcal S}_k} $ was defined in \eqref{eq:Z_k-prime} as the total estimation error correlation matrix related to those UEs. 
This precoding scheme is expected to behave similarly to MMSE precoding since it has the same structure, except that it neglects UEs that are only served by other APs, in order to reduce the computational complexity. If the DCC is properly designed, then $\mathcal{S}_k$ should include all the UEs that are within the range of influence of the APs that serve UE $k$, making the performance difference between MMSE and P-MMSE precoding small.

The inverse of an $LN \times LN$ matrix must be computed in \eqref{eq:P-MMSE-precoding} but, since only the signals transmitted from the $|\mathcal{M}_k|$ APs that serve UE $k$ matter, we can implement P-MMSE precoding by only inverting an $N |\mathcal{M}_k| \times N |\mathcal{M}_k|$ matrix (see Figure~\ref{figure:block-matrices} on p.~\pageref{figure:block-matrices} for details). If P-MMSE combining is used in the uplink, then we can use P-MMSE precoding in the downlink without  any extra computation.

As in the uplink, we can reduce the computational complexity by using \emph{P-RZF precoding} instead, which is obtained by neglecting the estimation error correlation matrix ${\bf{Z}}_{{\mathcal S}_k} $ in \eqref{eq:P-MMSE-precoding}. This enables us to rewrite the precoding expression in a more efficient form. P-RZF precoding is obtained from \eqref{eq:precoding-centralized-expression} using
\begin{align} 
\bar{\vect{w}}_{k}^{{\rm P-RZF}} =  \left [\vect{{D}}_{k}\vect{\widehat{H}}_{\mathcal{S}_k}\left( \vect{\widehat{H}}^{\Htran}_{\mathcal{S}_k}\vect{{D}}_{k}\vect{\widehat{H}}_{\mathcal{S}_k}+\sigmaUL\vect{{P}}_{\mathcal{S}_k}^{-1} \right)^{-1} \right]_{:,1}  \label{eq:P-RZF-precoding}
\end{align}
where we recall that $\vect{\widehat{H}}_{\mathcal{S}_k} \in \mathbb{C}^{LN\times |\mathcal{S}_k|}$ is obtained by stacking all the vectors $\vect{\widehat{h}}_i$ with indices $i\in \mathcal{S}_k$ with the first column being $\vect{\widehat{h}}_k$. Moreover, $\vect{{P}}_{\mathcal{S}_k} \in \mathbb{R}^{|\mathcal{S}_k| \times |\mathcal{S}_k|}$ is a diagonal matrix containing the transmit powers $p_i$ for $i\in \mathcal{S}_k$, listed in the same order as the columns of $\vect{\widehat{H}}_{\mathcal{S}_k}$.
If P-RZF combining is used in the uplink, then we can use P-RZF precoding in the downlink, without requiring any additional computational complexity. This makes it a scalable precoding scheme.

Note that the names ``P-MMSE'' and ``P-RZF'' are referring to the properties of the combining vector counterparts in the dual uplink. Intuitively, both methods will balance between transmitting a strong signal to the desired UE and limiting the interference that is caused to other UEs. The difference is that the former scheme takes the channel estimation uncertainty into account, while the latter is neglecting it to reduce the computational complexity.

\subsection{Fronthaul Signaling Load for Centralized Operation}
\label{sec:level4-fronthaul-downlink}

The centralized operation is associated with a fronthaul signaling load since all the computations are made at the CPU. The signaling related to sending the received pilot signals from the APs to the CPU is the same as for fully centralized operation in the uplink; see Section~\ref{sec:level4-fronthaul} on p.~\pageref{sec:level4-fronthaul}. Hence, we will not list them in this section to avoid counting them twice and instead focus on the signaling from the CPU to the APs that is unique to the downlink.

The only additional fronthaul signaling in the downlink is to provide each AP $l$ with the transmit signal $\vect{x}_l$ in \eqref{eq:DCC-transmit-downlink2}, which is a superposition of the precoded downlink data signals. There are $\tau_d$ such vectors to transmit per coherence block. Hence, the fronthaul signaling per AP is $\tau_dN$ complex scalars. This value is the same irrespective of the choice of precoding scheme and is summarized in Table~\ref{tab:signaling-level4-downlink}. Interestingly, the APs can be totally unaware of what precoding scheme is used and how many UEs are being served since the CPU is performing all the signal processing in the centralized operation. Note that the fronthaul signaling is independent of $K$ and, thus, it is scalable.

\begin{table}[t]  
	\caption{Number of complex scalars to be shared over the fronthaul per coherence block in the centralized downlink operation. These scalars are sent from the CPU to the APs.}  \label{tab:signaling-level4-downlink}
	\centering
	\begin{tabular}{|c|c|c|} 
		\hline
		{Scheme} & {Each coherence block}    \\      \hline\hline 
		Any precoding &   $\tau_d NL$         \\ \hline    
	\end{tabular}
\end{table}

\section{Distributed Downlink Operation}
\label{sec:level2-downlink}

A distributed operation is also possible in the downlink where almost all the processing is carried out locally at each AP. The CPU encodes the downlink data signals $\{ \varsigma_i : i =1,\ldots,K\}$ and send them to the serving APs, which carry out the remaining signal processing, as illustrated in Figure~\ref{figure:distributed_centralized_downlink}(b).
We stress that the CPU is a logical entity; the task of encoding the data signal to a given UE can be carried out anywhere in the network, thus there is no need for having a physical centralized unit.
A major benefit of the distributed operation is that we can deploy new APs without having to upgrade the computational power of the CPU since each AP contains a local processor that can carry out its associated baseband processing tasks.
In the distributed operation, each AP $l$ locally designs its transmitted signal
\begin{align} \label{eq:x_l-level2}
\vect{x}_l=\sum_{i=1}^K\vect{D}_{il}\vect{w}_{il}\varsigma_{i}
\end{align}
that was previously stated in \eqref{eq:DCC-transmit-downlink2}. To this end, AP $l$ computes its local channel estimates as discussed in Section~\ref{sec:MMSE-channel-estimation} on p.~\pageref{sec:MMSE-channel-estimation} and selects the local precoding vectors $\{\vect{w}_{il} : i\in \mathcal{D}_l \}$ based on those estimates.

\enlargethispage{\baselineskip}

Recall that, in the distributed uplink operation, we considered a two-stage operation where each AP is first processing its signals locally, followed by a second step where the CPU weighs the information obtained from the different APs together. We then considered how to optimize the weight vector. A similar operation exists in the downlink but only implicitly. The second stage is namely represented by the power allocation between the different APs, which determines which power each AP should assign to a given UE. This power allocation task is analyzed in Section~\ref{sec:optimized-power-allocation-downlink} on p.~\pageref{sec:optimized-power-allocation-downlink}, while we will consider a heuristic power allocation in this section.

\subsection{Spectral Efficiency With Distributed Operation}
\label{sec:level2-SE-downlink}

The received signal at UE $k$ is
\begin{align} 
y_k^{\rm{dl}} =& \sum_{l=1}^{L}  \vect{h}_{kl}^{\Htran}\vect{x}_l + n_k 
\nonumber \\
=&\underbrace{ \vphantom{\condSum{i=1}{i \neq k}{K}} \left( \sum_{l=1}^{L}  \vect{h}_{kl}^{\Htran}\vect{D}_{kl}  \vect{w}_{kl} \right) \varsigma_k}_{\textrm{Desired signal}}+ \underbrace{\condSum{i=1}{i \neq k}{K} \left( \sum_{l=1}^{L}  \vect{h}_{kl}^{\Htran}\vect{D}_{il}  \vect{w}_{il}   \right) \varsigma_i}_{\textrm{Inter-user interference}} + \underbrace{\vphantom{\condSum{i=1}{i \neq k}{K}} n_k}_{\textrm{Noise}} \label{eq:DCC-downlink-level2}
\end{align}
which is identical to the case of centralized downlink operation in \eqref{eq:downlink-received-UEk}. The key difference lies in the precoding selection, which now must be carried out locally at each AP using the locally available channel estimates. The information available at the UE for signal detection is the same in both cases. Hence, we can use Theorem~\ref{theorem:downlink-capacity-level4} to also compute the SE achieved by the distributed operation. We will restate the result using the distributed notation containing summations over the APs.

\begin{corollary} \label{theorem:downlink-capacity-level2}
	An achievable SE of UE $k$ in the distributed operation is
	\begin{equation} \label{eq:downlink-rate-expression-level2}
	\begin{split}
	\mathacr{SE}_{k}^{({\rm dl},\dist)} = \frac{\tau_d}{\tau_c} \log_2  \left( 1 + \mathacr{SINR}_{k}^{({\rm dl},\dist)}  \right) \quad \textrm{bit/s/Hz}
	\end{split}
	\end{equation}
	with the effective SINR given by
	\begin{equation} \label{eq:downlink-instant-SINR-level2}
	\mathacr{SINR}_{k}^{({\rm dl},\dist)} =  \frac{ \left|\sum\limits_{l=1}^L  \mathbb{E}\left \{ \vect{h}_{kl}^{\Htran}\vect{D}_{kl}{\vect{w}}_{kl}\right \} \right|^2  }{\sum\limits_{i=1}^{K} \mathbb{E} \bigg\{ \bigg| \sum\limits_{l=1}^{L}  \vect{h}_{kl}^{\Htran} \vect{D}_{il} {\vect{w}}_{il} \bigg|^2 \bigg\} -  \bigg|\sum\limits_{l=1}^L  \mathbb{E} \left\{  \vect{h}_{kl}^{\Htran} \vect{D}_{kl} {\vect{w}}_{kl} \right\} \bigg|^2 +  \sigmaDL
	}.
	\end{equation}
\end{corollary}

The SE of UE $k$ and its effective SINR can be interpreted in the same way as in the centralized case so we will not repeat the discussion here, but instead focus on the local precoding selection.

\subsection{Local Transmit Precoding}
\label{subsec:distributed-precoding}
Only a subset of the APs are transmitting a downlink data signal to a particular UE $k$. Hence, the effective precoding vectors $\vect{D}_{kl}\vect{w}_{kl}$ only need to be selected for AP $l \in \mathcal{M}_k$.
For an AP $l$ that serves UE $k$, we can express the precoding vector as 
\begin{align} \label{eq:precoding-distributed-expression}
{\vect{w}}_{kl}= \sqrt{\rho_{kl}}\frac{\bar{\vect{w}}_{kl}}{\sqrt{\mathbb{E}\left\{\left\| \bar{\vect{w}}_{kl}\right\|^2\right\}}}
\end{align} 
where $\rho_{kl} \geq 0 $ is the transmit power that AP $l$ assigns to UE $k$ and $\bar{\vect{w}}_{kl} \in \mathbb{C}^{N}$ is an arbitrarily scaled vector pointing out the direction of the precoding vector. 
Note that the normalization in \eqref{eq:precoding-distributed-expression} makes 
\begin{equation}
\mathbb{E} \left\{ \| \vect{w}_{kl} \|^2 \right\} = \rho_{kl}.
\end{equation}
Using this notation, we can effectively divide the precoding selection at AP $l$ into the following two subtasks: 

\begin{enumerate}

\item Selecting the directivity of the transmission represented by $\{ \bar{\vect{w}}_{kl} : k \in \mathcal{D}_l \} $; 

\item Selecting the power allocation represented by $\{ \rho_{kl} : k \in \mathcal{D}_l \}$. 

\end{enumerate}

The first task must be carried out in every coherence block, based on the local channel estimates, which change at this pace. In contrast, the power allocation is dominated by large-scale effects (such as pathloss) and we will, therefore, assume that the power allocation is maintained constant over many coherence blocks and computed based on the channel statistics. We will optimize the power allocation in Section~\ref{sec:optimized-power-allocation-downlink} on p.~\pageref{sec:optimized-power-allocation-downlink} and consider scalable heuristic methods in Section~\ref{sec:scalable-power-allocation} on p.~\pageref{sec:scalable-power-allocation}.

In the centralized downlink operation, we used the uplink-downlink duality in Theorem~\ref{prop:duality} to motivate that each centralized precoding vector should be selected to be parallel to the corresponding centralized combining vector. The duality result is not unique to the centralized operation: for any choice of the receive combining vectors and their resulting uplink SINRs, we can achieve the same SINR in the downlink by using the same vectors for precoding. Hence, we can utilize the local combining schemes from Section~\ref{sec:level3-uplink} on p.~\pageref{sec:level3-uplink} and normalize them to obtain local precoding vectors that should serve as reasonable heuristics. 
The  complexity of computing these vectors is the same as in the uplink. Hence, if the same vectors are used in both directions then there is no extra complexity incurred in the downlink. We will now present the downlink counterparts of the L-MMSE, LP-MMSE, and MR combining schemes that we considered in the uplink.

\subsubsection{L-MMSE Precoding}

The locally optimal operation in the uplink is L-MMSE combining, which was defined in \eqref{eq:L-MMSE-combining-single-AP}. The downlink counterpart is \emph{L-MMSE precoding} and is obtained from \eqref{eq:precoding-distributed-expression} using
\begin{equation} \label{eq:L-MMSE-precoding}
\bar{\vect{w}}_{kl}^{{\rm L-MMSE}} =  p_{k}  \left( \sum\limits_{i=1}^{K} p_{i} \left( \widehat{\vect{h}}_{il} \widehat{\vect{h}}_{il}^{\Htran} + \vect{C}_{il} \right) + \sigmaUL  \vect{I}_{N} \right)^{-1}   \vect{D}_{kl}  \widehat{\vect{h}}_{kl}.
\end{equation}
We expect this precoding method to provide the highest SE among the distributed alternatives, due to its tight connection to the L-MMSE combining method, but this cannot be formally proved. 
Roughly the same power gain from coherent precoding can be achieved by centralized MMSE precoding and L-MMSE precoding, if the same APs are utilized. However, there is a major difference when it comes to interference suppression.
As illustrated in Figure~\ref{figure:distributed-beamforming}, in the centralized operation, the serving APs can cooperate in suppressing interference; for example, by sending interfering signals with opposite phases so they cancel each other at an undesired receiver. Hence, the spatial degrees-of-freedom available for interference suppression is equal to the total number of antennas that is transmitting the signal and given by $N |\mathcal{M}_k|$ for UE $k$.
In contrast, in the distributed operation, each AP can only control the interference that itself is causing. Each AP has $N$  spatial degrees-of-freedom available for interference suppression, which is expected to be a rather small number since each AP in a Cell-free Massive MIMO system is envisioned to be equipped with few antennas. In fact, each AP might serve more UEs than it has antennas. With the distributed operation, the interference suppression capability does not increase as more APs are assigned to serving the UE.

L-MMSE precoding in \eqref{eq:L-MMSE-precoding} is not a scalable scheme since it contains a summation of all UEs in the network (see Table~\ref{table:complexity-with-level3} on p.~\pageref{table:complexity-with-level3} for the exact complexity). 
Two scalable alternatives are considered next.

\subsubsection{MR Precoding}

The first scalable option is \emph{MR precoding}, which is obtained from \eqref{eq:precoding-distributed-expression} using
\begin{equation} \label{eq:MR-precoding}
\bar{\vect{w}}_{kl}^{{\rm MR}} =  \vect{D}_{kl}  \widehat{\vect{h}}_{kl}.
\end{equation}
This scheme maximizes the fraction of the transmitted power from AP $l$ that is received at the desired UE (i.e., the numerator of the effective SINR).
However, MR precoding ignores the interference that the AP is causing, particularly among the UEs that it is serving. 
The early works on Cell-free Massive MIMO considered MR precoding with $N=1$ \cite{Nayebi2017a}, \cite{Ngo2017b}, in which case each AP has too few antennas to suppress interference. One key benefit of this scheme is that the effective SINR in Corollary~\ref{theorem:downlink-capacity-level2} can be computed in closed form.

\begin{corollary} \label{cor:closed-form-dl}
	Suppose MR precoding, as defined in \eqref{eq:MR-precoding}, is used. The expectations in \eqref{eq:downlink-instant-SINR-level2} then become
	\begin{equation}  \label{eq:signal-term-MR-closed}
	\sum\limits_{l=1}^L  \mathbb{E}\left \{ \vect{h}_{kl}^{\Htran}\vect{D}_{kl}{\vect{w}}_{kl}\right \}
 = \sum_{l=1}^L \sqrt{\rho_{kl} \eta_k \tau_p \tr( \vect{D}_{kl} \vect{R}_{kl} \vect{\Psi}_{t_k l}^{-1} \vect{R}_{kl})}
\end{equation}
\begin{align} & \nonumber
 \mathbb{E} \left\{ \left| \sum\limits_{l=1}^{L}  \vect{h}_{kl}^{\Htran} \vect{D}_{il} {\vect{w}}_{il} \right|^2 \right\}  = \sum_{l=1}^{L} \rho_{il} \frac{  \tr \left(  \vect{D}_{il} \vect{R}_{il} \vect{\Psi}_{t_i l}^{-1} \vect{R}_{il} \vect{R}_{kl}  \right) }{ \tr( \vect{R}_{il} \vect{\Psi}_{t_i l}^{-1} \vect{R}_{il})} \\ & + \begin{cases}
\left|  \sum\limits_{l=1}^{L} \sqrt{ \rho_{il} \eta_k \tau_p} \frac{  \tr \left(  \vect{D}_{il} \vect{R}_{il} \vect{\Psi}_{t_i l}^{-1} \vect{R}_{kl}  \right) }{ \sqrt{ \tr( \vect{R}_{il} \vect{\Psi}_{t_i l}^{-1} \vect{R}_{il}) } } \right|^2 &    i \in \mathcal{P}_k
\\
0 &  i \notin \mathcal{P}_k. 
\end{cases} \label{eq:interference-term-MR-closed}
\end{align} 
\end{corollary}
\begin{proof}
By inserting the MR precoding expression into \eqref{eq:downlink-instant-SINR-level2}, we can observe that the expectations that appear are the same as in Corollary~\ref{cor:closed-form_ul} on p.~\pageref{cor:closed-form_ul} (except that some indices must be interchanged).
\end{proof}

The expressions in Corollary~\ref{cor:closed-form-dl} can be inserted into the effective SINR in \eqref{eq:downlink-instant-SINR-level2} to obtain a closed-form SE expression. However, the final expression is lengthy. Hence, we will focus on discussing its properties instead of presenting it in a single equation.

The signal term in the numerator of the effective SINR is the square of \eqref{eq:signal-term-MR-closed}. We can recognize $\eta_k \tau_p \vect{R}_{kl} \vect{\Psi}_{t_k l}^{-1} \vect{R}_{kl}$ from Corollary~\ref{cor:MMSE-estimation} on p.~\pageref{cor:MMSE-estimation} as the correlation matrix of the MMSE channel estimate $\widehat{\vect{h}}_{kl}$. Hence, the signal term can be equivalently expressed as
\begin{equation}
\left(\sum_{l=1}^{L}  \sqrt{\rho_{kl} \mathbb{E}\{ \| \vect{D}_{kl} \widehat{\vect{h}}_{kl} \|^2 \} }\right)^2
\end{equation}
and grows when the channel estimation quality improves. To make the signal term large, it is important to maximize the estimation quality, which can be achieved by limiting the pilot contamination effect in the ways discussed in Section~\ref{sec:impact-on-estimation-accuracy} on p.~\pageref{sec:impact-on-estimation-accuracy}. The contributions from the serving APs are coherently combined, which is represented by the fact that the square roots of their contributions are added up followed by taking the square of the sum. This operation leads to a power gain as compared to the case when the contributions are added up directly in the power domain. As a simple example of this, suppose $a$ and $b$ are the power of the contributions from two different APs. We can then notice that
\begin{equation}
(\sqrt{a}+\sqrt{b})^2 = a+b+2\sqrt{ab} > a+b
\end{equation}
for any strictly positive values of $a$ and $b$.

The interference term caused by the signal transmission to UE $i$ is given in \eqref{eq:interference-term-MR-closed}. This expression is more complicated to interpret than the signal term but we can recognize terms of the kind $ \vect{R}_{il} \vect{\Psi}_{t_i l}^{-1} \vect{R}_{il}$ which are proportional to the correlation matrix of the channel estimate $\widehat{\vect{h}}_{il}$. This is natural since this channel estimate is used for MR precoding to UE $i$.
The first interference term in \eqref{eq:interference-term-MR-closed} contains a product $\vect{R}_{il} \vect{R}_{kl} $ between the correlation matrices of the interfering UE and the desired UE. If these UEs have very different spatial correlation properties (e.g., if the dominant eigenvectors are orthogonal), then the interference will be smaller than if they have matching correlation matrices. There is also an additional interference term, which is only included if UE $i$ uses the same pilot as UE $k$. This term is called coherent interference since it also contains a summation of square roots followed by a squaring. This pilot-contaminated term becomes equal to the signal term if $i=k$.

The effective SINR expression with MR precoding can be substantially simplified in the case of $N=1$ antenna per AP.

	\begin{corollary}
	If each AP has $N=1$ antenna, then the MMSE estimate of the scalar channel $\vect{h}_{kl}\in \mathbb{C}$ has variance
	\begin{equation} \label{eq:gamma-notation-repeated}
	 \gamma_{kl} = \frac{\eta_k \tau_p \beta_{kl}^2 }{ \sum_{i \in \mathcal{P}_k } \eta_i \tau_p \beta_{il} +\sigmaUL }.
	 \end{equation}
	The effective SINR in \eqref{eq:downlink-instant-SINR-level2} with MR precoding then becomes
		\begin{equation} \label{eq:downlink-instant-SINR-level2-closed}
\mathacr{SINR}_{k}^{({\rm dl},\dist)} =  \frac{  \left( 
\sum\limits_{l \in \mathcal{M}_k} \sqrt{\rho_{kl} \gamma_{kl}} \right)^2  }{ \sum\limits_{i=1}^{K} \sum\limits_{l \in \mathcal{M}_i} \rho_{il} \beta_{kl} +
\sum\limits_{i \in \mathcal{P}_k \setminus \{ k \}} \left(  \sum\limits_{l \in \mathcal{M}_i} \sqrt{ \rho_{il} \gamma_{kl}}  \right)^2
+  \sigmaDL
	}.
	\end{equation}
	\end{corollary}
	\begin{proof}
	Using the notation in \eqref{eq:gamma-notation-repeated}, the expectations in Corollary~\ref{cor:closed-form-dl} can be rewritten as 
$\sum_{l=1}^L  \mathbb{E}\left \{ \vect{h}_{kl}^{\Htran}\vect{D}_{kl}{\vect{w}}_{kl}\right \}
 = \sum_{l \in \mathcal{M}_k} \sqrt{\rho_{kl} \gamma_{kl}}$ and 
\begin{align} & \nonumber
 \mathbb{E} \left\{ \left| \sum\limits_{l=1}^{L}  \vect{h}_{kl}^{\Htran} \vect{D}_{il} {\vect{w}}_{il} \right|^2 \right\}  = \sum_{l \in \mathcal{M}_i} \rho_{il} \beta_{kl} + \begin{cases}
\left|  \sum\limits_{l \in \mathcal{M}_i} \sqrt{ \rho_{il} \gamma_{kl}}  \right|^2 &    i \in \mathcal{P}_k
\\
0 & i \notin \mathcal{P}_k.
\end{cases}
\end{align} 
Inserting these values into \eqref{eq:downlink-instant-SINR-level2} yields the SINR expression in \eqref{eq:downlink-instant-SINR-level2-closed}, where absolute values of positive terms have been replaced by parenthesis. 
	\end{proof}

The closed-form effective SINR in \eqref{eq:downlink-instant-SINR-level2-closed} shows the key behaviors of the cell-free operation even clearer than for $N>1$. The signal term $( 
\sum_{l \in \mathcal{M}_k} \sqrt{\rho_{kl} \gamma_{kl}} )^2$ in the numerator contains a coherent combination of the signal contributions from all the serving APs, which leads to a power gain.
 The first term in the denominator contains non-coherent interference, which means that it is a summation of the powers of the individual signals transmitted to all the UEs from their respective serving APs. The second term is the additional coherent interference caused by pilot contamination. The third term is the noise power.

MR precoding is expected to work well if there is a high degree of favorable propagation, which is not guaranteed to be the case in cell-free networks (see Section~\ref{subsec-favorable-propagation} on p.~\pageref{subsec-favorable-propagation}). If each AP is equipped with multiple antennas, then we can use a local precoding scheme that is also capable of suppressing interference. The aforementioned unscalable L-MMSE precoding is an example of this but there are scalable alternatives that are suitable for cell-free networks.

\subsubsection{LP-MMSE Precoding}

We can achieve a scalable approximation of L-MMSE precoding by only considering the UEs that the AP serves, as was done for the uplink in \eqref{eq:LP-MMSE-combining-single-AP-1}. We call this \emph{LP-MMSE precoding} and obtain it from \eqref{eq:precoding-distributed-expression} using
\begin{equation} \label{eq:LP-MMSE-precoding-single-AP-1}
\bar{\vect{w}}_{kl}^{{\rm LP-MMSE}} =  p_{k}  \left( \sum\limits_{i \in \mathcal{D}_l} p_{i} \left( \widehat{\vect{h}}_{il} \widehat{\vect{h}}_{il}^{\Htran} + \vect{C}_{il} \right) + \sigmaUL  \vect{I}_{N} \right)^{\!-1}   \vect{D}_{kl} \widehat{\vect{h}}_{kl}.
\end{equation}
If AP $l$ is serving all the UEs in its area of influence, LP-MMSE precoding will be approximately equal to L-MMSE precoding. However, the major benefit is that LP-MMSE was previously shown to be a scalable scheme. If it is used in both uplink and downlink, then no additional  computations are required for creating the downlink precoding vectors.

\begin{remark}[Other precoding schemes]
The literature contains other precoding schemes than those considered in this monograph, in particular, different variations on MMSE and L-MMSE precoding where the regularization terms are selected differently or are removed, as in the case of ZF precoding \cite{Interdonato2020a}, \cite{Nguyen2017a}. Moreover, different APs can use different schemes, to reduce the overall computational complexity or protect certain ``sensitive'' UEs from interference. Interestingly, in the special case when each AP has more antennas than there are pilots (i.e., $N> \tau_p$) and the channels are subject to uncorrelated Rayleigh fading, there are several local ZF schemes for which the SE can be computed in closed form \cite{Interdonato2020a}.
\end{remark}

\subsection{Fronthaul Signaling Load for Distributed Operation}
\label{sec:level2-fronthaul-downlink}

The only downlink processing that is carried out at the CPU in the distributed operation is the encoding of the data signals. The CPU needs to send a portion of the data signals $\{ \varsigma_k : k = 1, \ldots, K\}$ to each AP, corresponding to the UEs that the AP is serving. AP $l$ needs to receive $\tau_d|\mathcal{D}_l|$ complex scalars per coherence block. This number is independent of the choice of precoding scheme. A key difference from the centralized operation is that the fronthaul signaling requirement of an AP is proportional to the number of UEs that it serves rather than the number of antennas.
The total number of complex scalars is summarized in Table~\ref{tab:signaling-level2-downlink}.

\begin{table}[t]  
	\caption{Number of complex scalars to be shared over the fronthaul per coherence block in the distributed downlink operation. These scalars are sent from the CPU to the APs.}  \label{tab:signaling-level2-downlink}
	\centering
	\begin{tabular}{|c|c|c|} 
		\hline
		{Scheme} & {Each coherence block}    \\      \hline\hline 
		Any precoding &  $\tau_d \sum\limits_{l=1}^L |\mathcal{D}_l|$         \\ \hline    
	\end{tabular}
\end{table}

\section{Numerical Performance Evaluation}
\label{sec:downlink-performance-comparison}

We will now compare the downlink SE achieved by the centralized and distributed precoding schemes described earlier in this section. We will continue the running example defined in Section~\ref{sec:running-example} on p.~\pageref{sec:running-example}.
There are many power allocation parameters that need to be selected to run simulations. We will consider power allocation optimization in detail in Section~\ref{sec:optimized-power-allocation-downlink} on p.~\pageref{sec:optimized-power-allocation-downlink}, while only heuristic power allocation schemes are utilized in this section. We will elaborate more on heuristic power allocation in Section~\ref{sec:scalable-power-allocation} on p.~\pageref{sec:scalable-power-allocation}.

In the centralized operation, we use a specific version of the scalable centralized power allocation in Section~\ref{sec:scalable-power-allocation} on p.~\pageref{sec:scalable-power-allocation} with the transmit power allocated to each UE as
\begin{align}\label{eq:fractional_power_centralized_DL}
\rho_k=\rhomax \frac{\left(\sqrt{\sum\limits_{l\in\mathcal{M}_k}\beta_{kl}}\right)^{-1}\!\!\!\left(\sqrt{\omega_k}\right)^{-1}}{\underset{\ell\in \mathcal{M}_k}{\max} \  \sum\limits_{i\in \mathcal{D}_{\ell}}   \left(\sqrt{\sum\limits_{l\in\mathcal{M}_i}\beta_{il}}\right)^{-1}\!\!\!\sqrt{\omega_i}}
\end{align}
with
\begin{align}
\omega_k = \underset{\ell\in\mathcal{M}_k}{\max} \ \mathbb{E} \left\{ \left\| \bar{\vect{w}}_{k\ell}^{\prime} \right\|^2 \right\}
\end{align} 
where $\bar{\vect{w}}_{k\ell}^{\prime} \in \mathbb{C}^{N}$ is the portion of the normalized centralized precoding vector $\frac{\bar{\vect{w}}_k}{\sqrt{\mathbb{E} \left\{ \| \bar{\vect{w}}_k \|^2 \right\} }}$ from \eqref{eq:precoding-centralized-expression} that corresponds to AP $\ell$. The precoding scheme will determine how this power is distributed between the different APs, but it is typically the closest APs that contribute with the majority of the power. The normalization factor in the denominator is selected to make sure that none of the APs will transmit with more power than the maximum value $\rhomax$.  With this scheme, the transmit powers of the UEs are selected inversely proportional to the square-root of their total channel gains from their serving APs. The details of the general version of this scalable power allocation scheme and how each per-AP power constraint is satisfied are shown in  Section~\ref{sec:scalable-power-allocation} on p.~\pageref{sec:scalable-power-allocation}. 

In the distributed operation, a specific version of the distributed power allocation scheme that was proposed in  \cite{Interdonato2019a} will be used:
\begin{equation} \label{eq:fractional_power_DL-section6}
\rho_{kl} = \begin{cases} \rhomax \frac{ \sqrt{\beta_{kl}}}{\sum_{i \in \mathcal{D}_l} \sqrt{\beta_{il}}} & k \in \mathcal{D}_l\\
0 &  k \notin \mathcal{D}_l
\end{cases}
\end{equation}
where the per-AP transmit power constraints are satisfied by construction. With this scheme, each AP will assign more power to the UEs that it has good channels to, than to UEs with worse channels. The general version of this heuristic scheme is discussed in more detail in Section~\ref{sec:scalable-power-allocation} on p.~\pageref{sec:scalable-power-allocation}.

Unless otherwise stated, the main simulation parameters are as follows. There are $K=40$ UEs 
 and the pilot sequence length is $\tau_p=10$. The remaining $\tau_d=\tau_c-\tau_p=190$ transmission symbols of each coherence block are used for downlink data only. The Gaussian local scattering model is used to generate the spatial correlation matrices with ASD $\sigma_{\varphi}=\sigma_{\theta}=15^{\circ}$. 
We use the same Monte Carlo simulation methodology as described in Section~\ref{sec:uplink-performance-comparison} on p.~\pageref{sec:uplink-performance-comparison} to generate the numerical results. In each figure, the legend will indicate the schemes that are being used.  We use ``(DCC)'' to denote the DCC implementation with Algorithm~\ref{alg:basic-pilot-assignment} on p.~\pageref{alg:basic-pilot-assignment}  and ``(All)'' to denote the case where all APs serve all UEs.

\subsection{Benchmark Schemes}
\label{sec:genie-aided-downlink}

The SE expressions provided in Theorem~\ref{theorem:downlink-capacity-level4} and Corollary~\ref{theorem:downlink-capacity-level2} are computed under the assumption that the receiving UE is applying a simple receiver structure that treats the average effective channel $\mathbb{E}\left \{ \vect{h}_{k}^{\Htran}\vect{D}_{k}{\vect{w}}_{k}\right \}$ as the true channel. To evaluate the tightness of this capacity bound, we will compare it with a benchmark where the UE has perfect channel knowledge, which has been obtained in some genie-aided manner. If the gap between the expressions is small, then the previously provided SEs are good performance metrics. We note that the effective downlink SINRs in the centralized and distributed operations, given in \eqref{eq:downlink-instant-SINR-level4} and \eqref{eq:downlink-instant-SINR-level2}, respectively, have the same structure. The difference is in the selection of the transmit precoding vectors $\{{\vect{w}}_{il}:i\in \mathcal{D}_l,l=1,\ldots,L\}$. Hence, we will obtain a common genie-aided SE expression, which can be used in both cases.

If the channels $\{\vect{h}_{kl}:k\in \mathcal{D}_l,l=1,\ldots,L\}$ are known at UE $k$, we can rewrite the received signal at UE $k$ in \eqref{eq:downlink-received-UEk} as
\begin{align} 
y_k^{\rm{dl}}& =\underbrace{ \vphantom{\condSum{i=1}{i \neq k}{K}} \sum_{l=1}^{L}  \vect{h}_{kl}^{\Htran} \vect{D}_{kl}  {\vect{w}}_{kl} \varsigma_k}_{\textrm{Desired signal}} + \underbrace{\condSum{i=1}{i \neq k}{K}  \sum_{l=1}^{L} \vect{h}_{kl}^{\Htran}\vect{D}_{il}  {\vect{w}}_{il} \varsigma_i}_{\textrm{Interference}} +\underbrace{ \vphantom{\condSum{i=1}{i \neq k}{K}} n_k}_{\textrm{Noise}} .\label{eq:DCC-downlink-2-genie-aided}
\end{align}
The first term contains the desired signal, while we treat the remaining terms as noise in line with the capacity bounds considered previously in this section. We then have the following genie-aided SE expression.

\begin{corollary} \label{corollary-genie-aided-downlink}
	A genie-aided SE of UE $k$ is
	\begin{equation} \label{eq:genie-aided-downlink-SE}
	\mathacr{SE}_{k}^{({\rm gen-dl})} = \frac{\tau_d}{\tau_c} \mathbb{E} \left\{ \log_2  \left( 1 + \mathacr{SINR}_{k}^{({\rm gen-dl})}  \right) \right\}
	\end{equation}
	where
	\begin{equation} \label{eq:downlink-instant-SINR-genie-aided}
	\mathacr{SINR}_{k}^{({\rm gen-dl})} = \frac{ \left|\sum\limits_{l=1}^L   \vect{h}_{kl}^{\Htran}\vect{D}_{kl} {\vect{w}}_{kl} \right|^2  }{  \condSum{i=1}{i \neq k}{K} \bigg| \sum\limits_{l=1}^{L} \vect{h}_{kl}^{\Htran} \vect{D}_{il} {\vect{w}}_{il} \bigg|^2   +  \sigmaDL}.
	\end{equation}
\end{corollary}
\begin{proof}
	This follows from utilizing Lemma~\ref{lemma:capacity-memoryless-random-interference} on p.~\pageref{lemma:capacity-memoryless-random-interference}.
\end{proof}

\enlargethispage{1.2\baselineskip} 

\subsection{Performance With Centralized Operation}

\begin{figure}[p!]
	\begin{subfigure}[b]{\columnwidth} \centering 
		\begin{overpic}[width=0.88\columnwidth,tics=10]{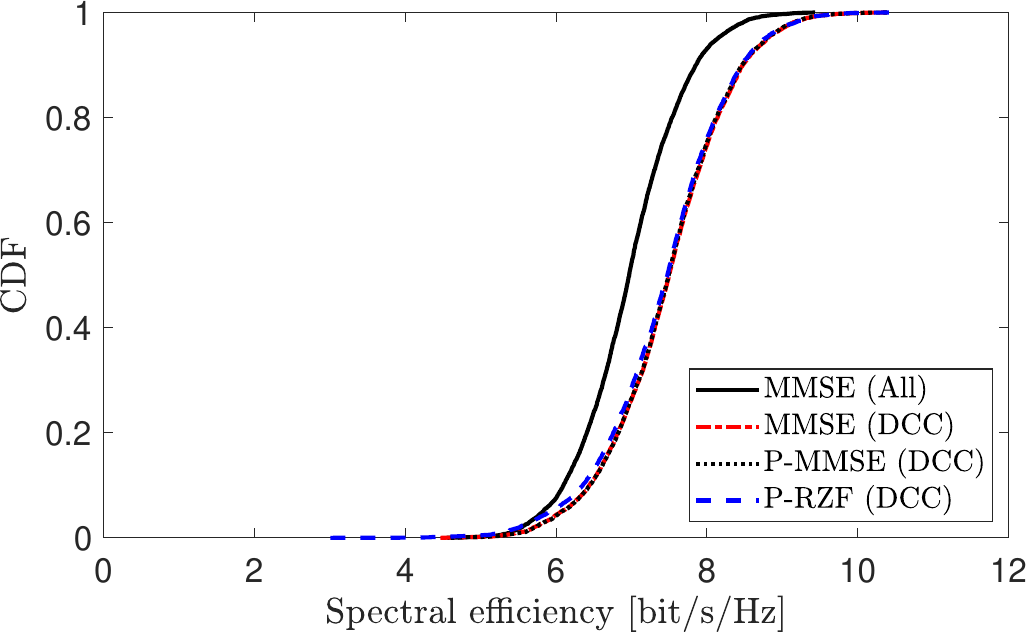}
		\end{overpic} \label{figureSec6_1a}
		\caption{$L=400$, $N=1$.} 
		\vspace{4mm}
	\end{subfigure} 
	\begin{subfigure}[b]{\columnwidth} \centering 
		\begin{overpic}[width=0.88\columnwidth,tics=10]{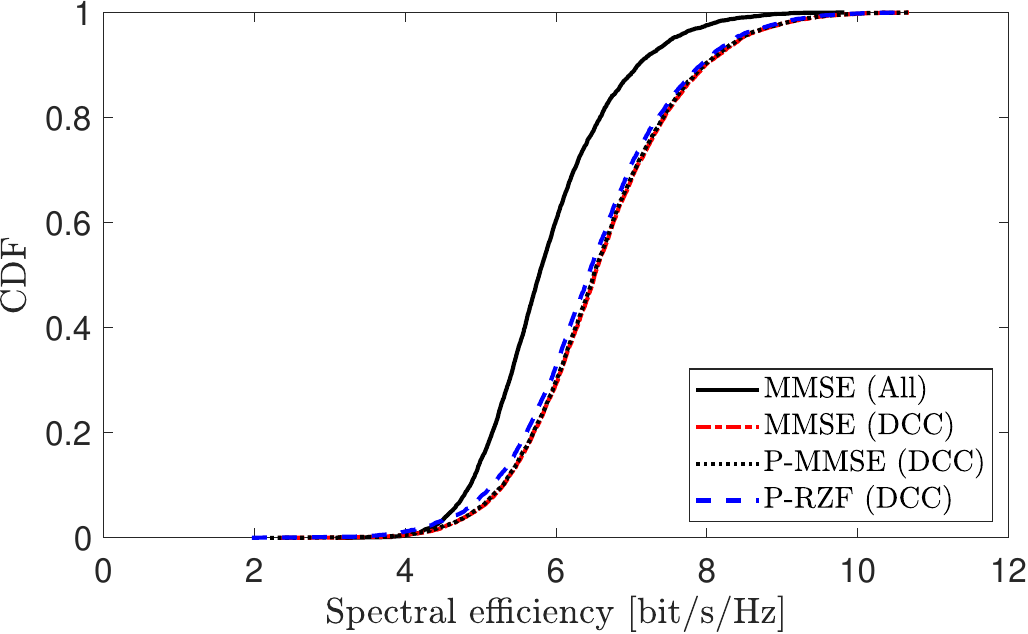}
		\end{overpic} \label{figureSec6_1b}
		\caption{$L=100$, $N=4$.} 
	\end{subfigure} 
	\caption{ CDF of the downlink SE per UE in the centralized operation. We consider $K=40$, $\tau_p=10$, and spatially correlated Rayleigh  fading with ASD $\sigma_{\varphi}=\sigma_{\theta}=15^{\circ}$. Different precoding schemes are compared.}\label{figureSec6_1}
\end{figure}

In Figure~\ref{figureSec6_1}, we show the \gls{CDF} of the downlink SE per UE in the centralized operation. The randomness  that gives rise to the CDF is due to the AP and UE locations, as well as to the shadow fading realizations. The SE is computed using \eqref{eq:downlink-rate-expression-level4} in Theorem~\ref{theorem:downlink-capacity-level4} for the different centralized precoding schemes described in Section~\ref{sec:level4-downlink}: MMSE, P-MMSE, and P-RZF. We stress that these are three heuristic precoding schemes that are motivated, via the uplink-downlink duality in Theorem~\ref{prop:duality}, by the good performance of their uplink counterparts. Although the strict duality is only valid for specific transmit power coefficients, these precoding schemes perform satisfactorily also for other power allocations. In this part, we use the heuristic power allocation scheme in \eqref{eq:fractional_power_centralized_DL}.

Figure~\ref{figureSec6_1}(a) considers the scenario with $L=400$ APs and $N=1$ antenna per AP and Figure~\ref{figureSec6_1}(b) considers the scenario with $L=100$ APs with $N=4$ antennas per AP. For both scenarios, the key observation is that MMSE precoding using all the APs provides smaller SE than the schemes  where the DCC framework is used to restrict how many UEs that each AP is serving. Transmitting from all the APs is not an efficient way when it comes to the downlink operation. Instead, selecting the best APs is a more efficient way of achieving a power allocation that suppresses interference. 
The difference between MMSE, P-MMSE, and P-RZF is almost negligible.
Recall that MMSE precoding is not a scalable scheme but, fortunately, the fully scalable P-MMSE and P-RZF precoding schemes provide almost the same SE. 

When we switch to the second scenario in Figure~\ref{figureSec6_1}(b) with less densely deployed APs, we see a SE drop for all UEs, except in the upper tail. Hence, we conclude that the former scenario with many single- antenna APs is more desirable for the centralized downlink operation. This is in line with our previous observations in Section~\ref{centralized-operation-uplink} on p.~\pageref{centralized-operation-uplink} for the centralized uplink operation.

\begin{figure}[t!]\centering
	\begin{overpic}[width=0.88\columnwidth,tics=10]{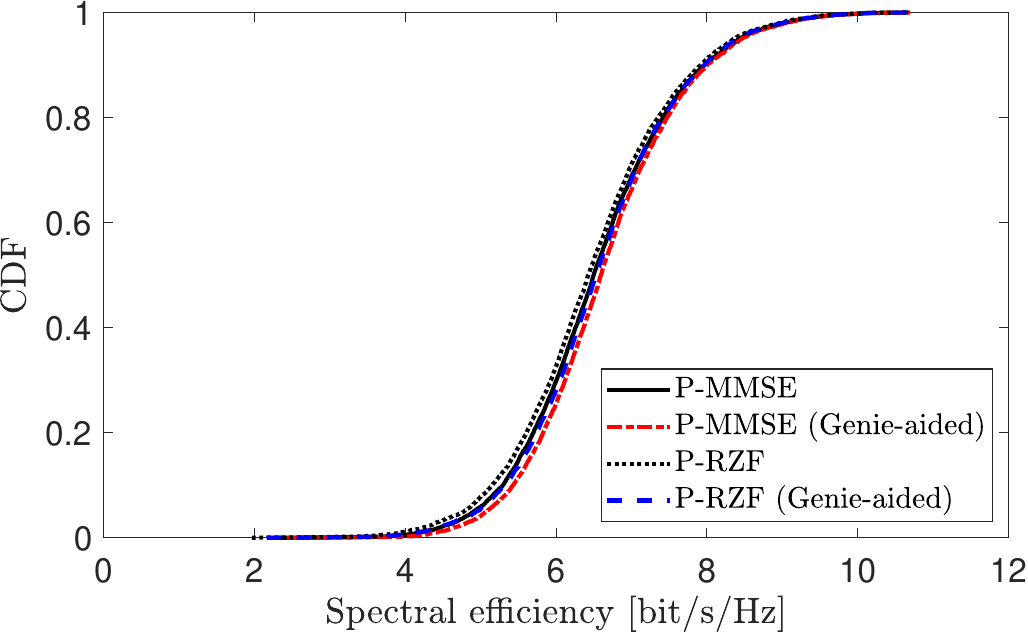}
	\end{overpic} 
	\caption{CDF of the downlink SE per UE in the centralized operation of the same scenario as in Figure~\ref{figureSec6_1}(b). We consider $L=100$, $N=4$, $K=40$, $\tau_p=10$, and spatially correlated Rayleigh  fading with ASD $\sigma_{\varphi}=\sigma_{\theta}=15^{\circ}$. The SE expression from \eqref{eq:downlink-rate-expression-level4} is compared with genie-aided SE in \eqref{eq:genie-aided-downlink-SE}.}    \vspace{+2mm}
	\label{figureSec6_1x} 
\end{figure}

\subsubsection{Tightness of the SE Expression}

The  SE expression for centralized operation in \eqref{eq:downlink-rate-expression-level4} is derived under the assumption that the receiver treats the average effective channel as the true channel. This results in a valid lower bound on the capacity but there is a risk that an improved receiver will perform better, particularly, if there is a low degree of channel hardening.
To quantify the tightness of the SE expression in \eqref{eq:downlink-rate-expression-level4}, Figure~\ref{figureSec6_1x} considers the same simulation setup as in Figure~\ref{figureSec6_1}(b) but also includes the genie-aided SEs from \eqref{eq:genie-aided-downlink-SE}, which are obtained when the UEs have perfect CSI. We consider the scalable P-MMSE and P-RZF precoding schemes. In both cases, the performance gap to the genie-aided curve is virtually non-existing. Hence, similar to the uplink, we conclude that a centralized operation leads to a high degree of channel hardening in the running example, so that it is sufficient for the receiver to only know the average effective channel when detecting data.
However, one can create simulation setups where the gap is larger (cf.~\cite{Chen2018b}) and then the downlink channel estimation methods discussed in Remark~\ref{remark:pilot-directions} are needed to close the gap.

\subsection{Performance With Distributed Operation}

\begin{figure}[p!]
	\begin{subfigure}[b]{\columnwidth} \centering 
		\begin{overpic}[width=0.88\columnwidth,tics=10]{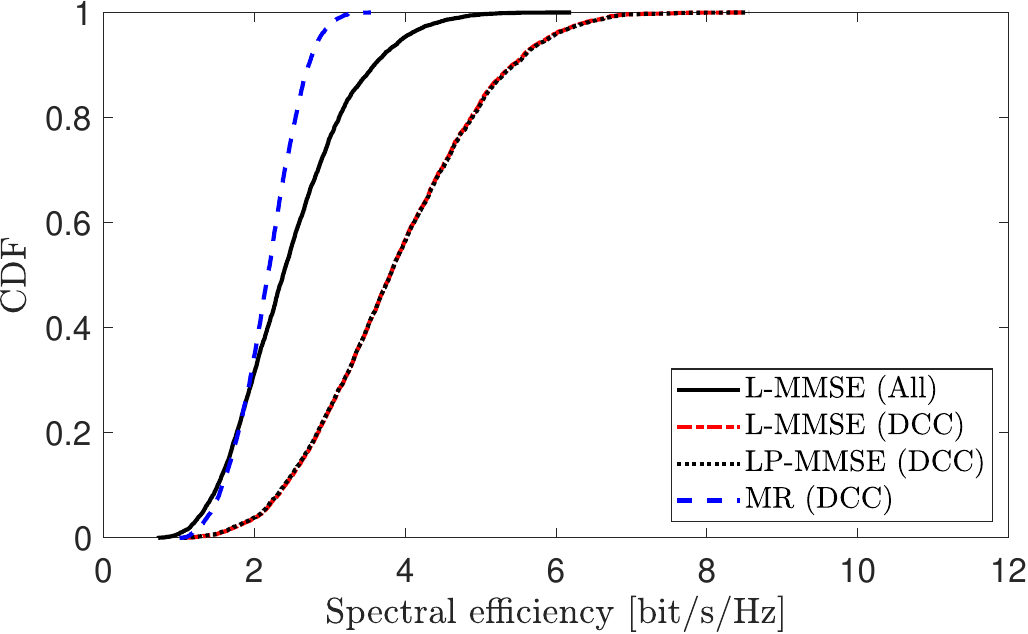}
		\end{overpic} \label{figureSec6_3a}
		\caption{$L=400$, $N=1$.} 
		\vspace{4mm}
	\end{subfigure} 
	\begin{subfigure}[b]{\columnwidth} \centering 
		\begin{overpic}[width=0.88\columnwidth,tics=10]{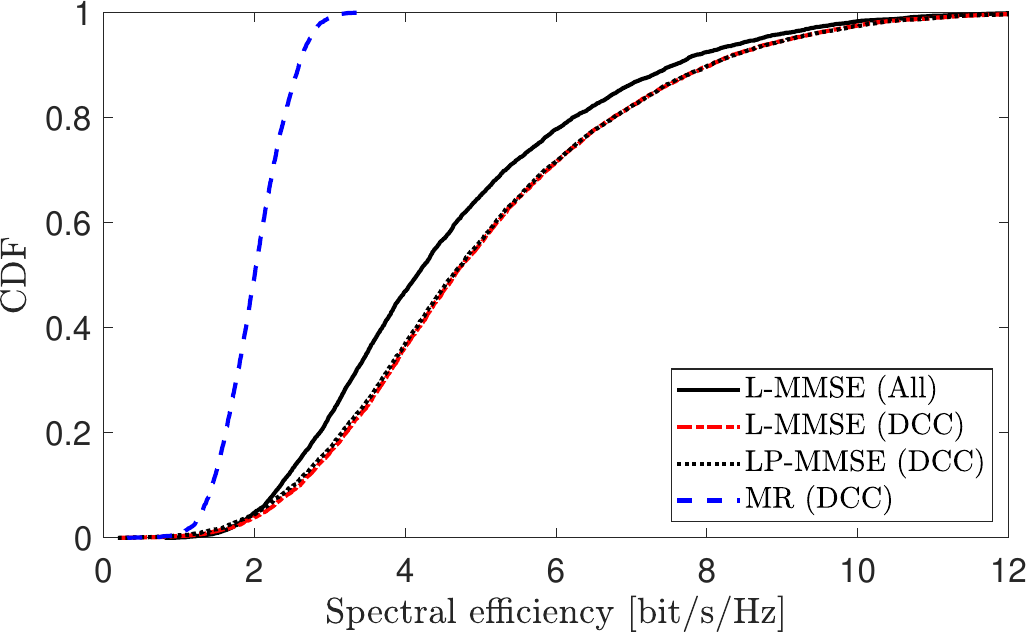}
		\end{overpic} \label{figureSec6_3b}
		\caption{$L=100$, $N=4$.} 
	\end{subfigure} 
	\caption{CDF of the downlink SE per UE in the distributed operation. We consider $K=40$, $\tau_p=10$, and spatially correlated Rayleigh  fading with ASD $\sigma_{\varphi}=\sigma_{\theta}=15^{\circ}$. Different precoding schemes are compared.}\label{figureSec6_3}
\end{figure}

We will now consider the distributed operation. Figure~\ref{figureSec6_3} shows the CDF of the SE per UE in the same scenarios as in Figure~\ref{figureSec6_1} with the heuristic power allocation method stated in \eqref{eq:fractional_power_DL-section6}.
Figure~\ref{figureSec6_3}(a) assumes $L=400$ and $N=1$, while Figure~\ref{figureSec6_3}(b) considers $L=100$ and $N=4$. In both scenarios, the L-MMSE precoding scheme that uses all the APs results in significantly lower SE compared to the DCC implementation with L-MMSE and its scalable version LP-MMSE. This unexpected result is caused by the assumed power allocation scheme and can be explained as follows. In general, most of the UEs have negligibly small channel gains to a particular AP and, hence, transmitting to all the UEs is like screaming and anyway only being heard by the closest UEs. The AP is essentially taking power that could have been used to serve the nearby UEs and assign it to faraway UEs, which mainly results in extra interference to the nearby UEs.
In this case, the DCC essentially provides an improved heuristic power allocation scheme where the power is allocated only to the UEs that are assigned to an AP with good channel conditions \cite{Buzzi2017b}.

Similar to the uplink, both figures demonstrate that we can achieve roughly the same SE by using the scalable LP-MMSE precoder instead of the unscalable L-MMSE precoder. When comparing the two scenarios, we notice that the performance is greatly improved when considering fewer APs with multiple antennas.
This is an expected result since both L-MMSE and LP-MMSE are taking the interference and estimation errors into account and they can suppress interference better with $N=4$ antennas per AP. 
The improvements are largest in the upper tail of the CDF curves, where interference is the limiting factor, and this also means that the SE curves are more spread out with $N=4$.
These conclusions are, qualitatively speaking, in line with those made for the uplink in Section~\ref{sec:uplink-performance-comparison} on p.~\pageref{sec:uplink-performance-comparison}.
Finally, we note that the aforementioned properties do not apply to MR, which performs almost equally bad in both scenarios.

\begin{figure}[t!]\centering
	\begin{overpic}[width=0.88\columnwidth,tics=10]{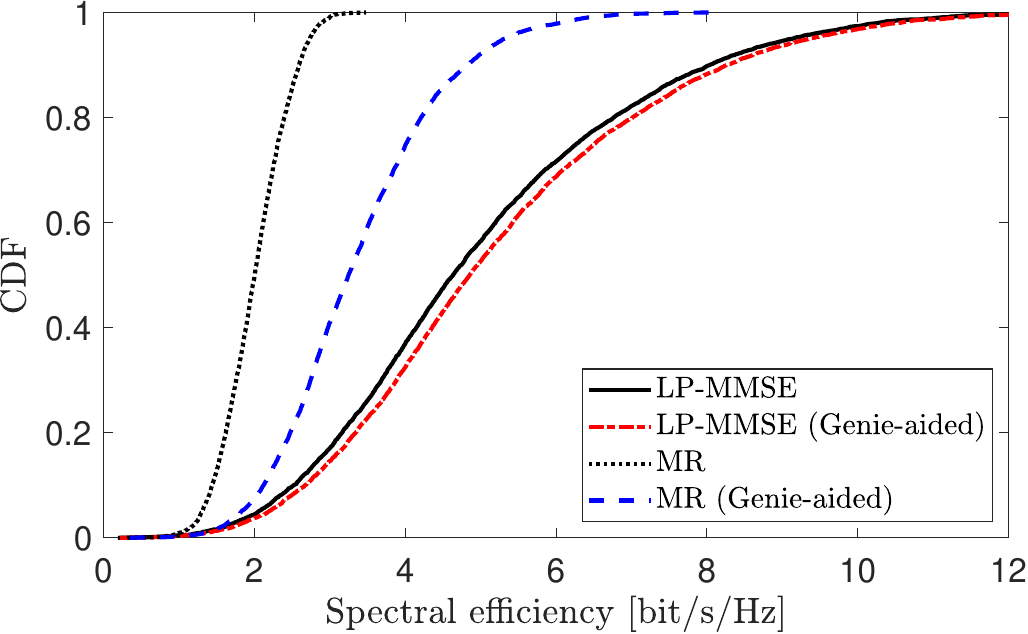}
	\end{overpic} 
	\caption{CDF of the downlink SE per UE in the distributed operation in the same scenario of Figure~\ref{figureSec6_3}(b). We consider $L=100$, $N=4$, $K=40$, $\tau_p=10$, and spatially correlated Rayleigh  fading with ASD $\sigma_{\varphi}=\sigma_{\theta}=15^{\circ}$. The SE expression from \eqref{eq:downlink-rate-expression-level2} is compared with genie-aided SE in \eqref{eq:genie-aided-downlink-SE}.}    \vspace{+2mm}
	\label{figureSec6_3x} 
\end{figure}

\subsubsection{Tightness of the SE Expression}

As in the centralized case, the SE expression used in the distributed operation assumes that the UEs have only access to the mean of the effective channels. To quantify the tightness of the provided SE results, compared to what could be achieved with the refined downlink channel estimation schemes described in Remark~\ref{remark:pilot-directions}, we will compare with the genie-aided SE in \eqref{eq:genie-aided-downlink-SE}, which assumes the same precoding schemes are used at the APs but the UEs are having access to perfect CSI when detecting the downlink signals.

Figure~\ref{figureSec6_3x} shows the CDF of the SE with the scalable precoding schemes from Figure~\ref{figureSec6_3}(b) with $L=100$ and $N=4$.
The genie-aided results match rather well with the achievable SE values provided by \eqref{eq:downlink-rate-expression-level2} when using LP-MMSE precoding. 
The gap is slightly larger than in the centralized case (see Figure~\ref{figureSec6_1x}) but nevertheless small.
Hence, for LP-MMSE precoding, we can conclude that the provided SE expression is both practically achievable and fairly close to what could be achieved with perfect CSI at the UEs.

\begin{figure}[p!]
	\begin{subfigure}[b]{\columnwidth} \centering 
		\begin{overpic}[width=\columnwidth,tics=10]{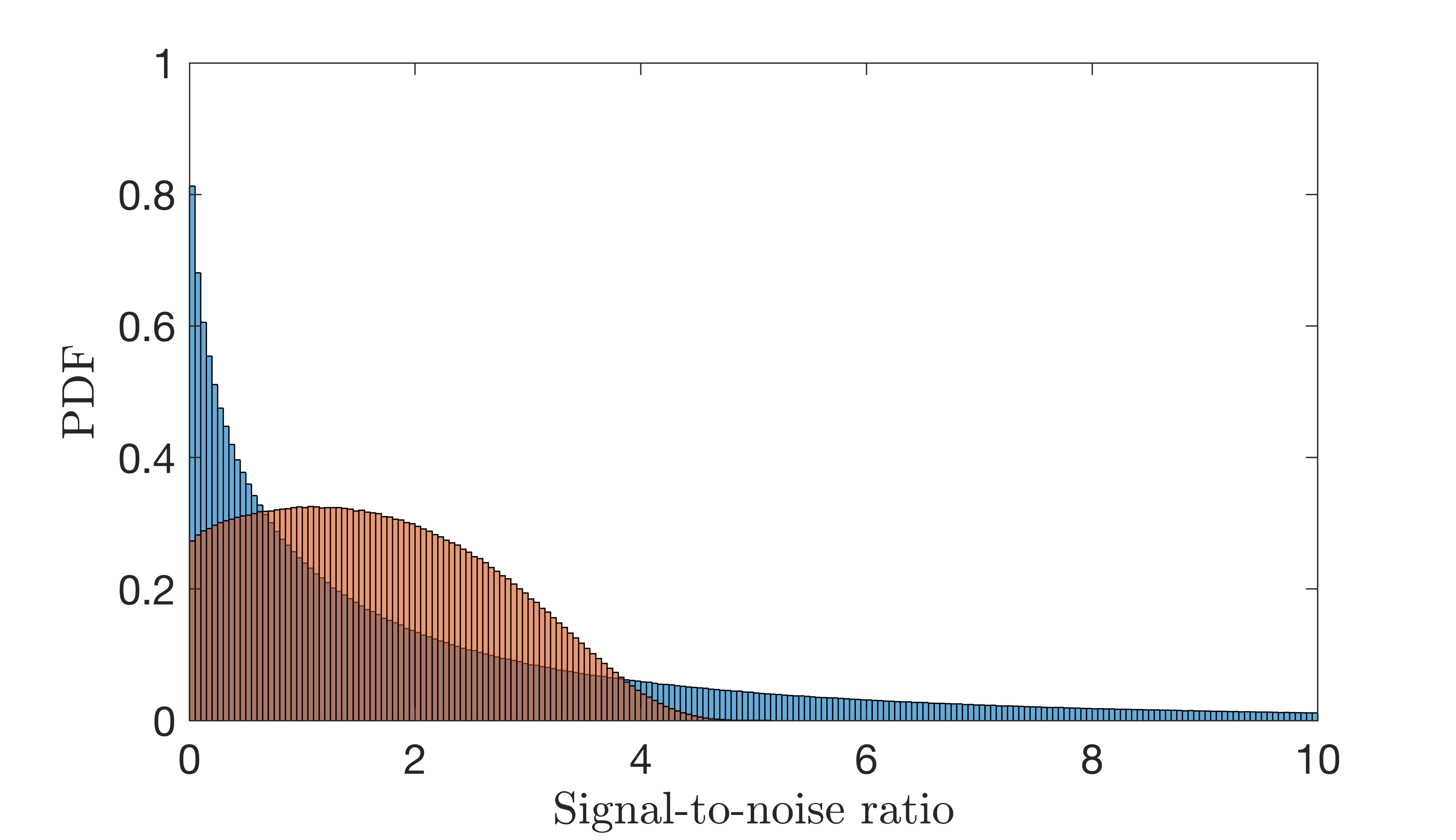}
		\put(19,34){MR}
		\put(34,28){LP-MMSE}
		\put(21,33){\vector(-1,-1){5}}
		\put(35,27){\vector(-1,-1){5}}
		\end{overpic} \label{figureSec6_signalgain}
		\caption{Variations in the received signal power.} 
		\vspace{4mm}
	\end{subfigure} 
	\begin{subfigure}[b]{\columnwidth} \centering 
		\begin{overpic}[width=\columnwidth,tics=10]{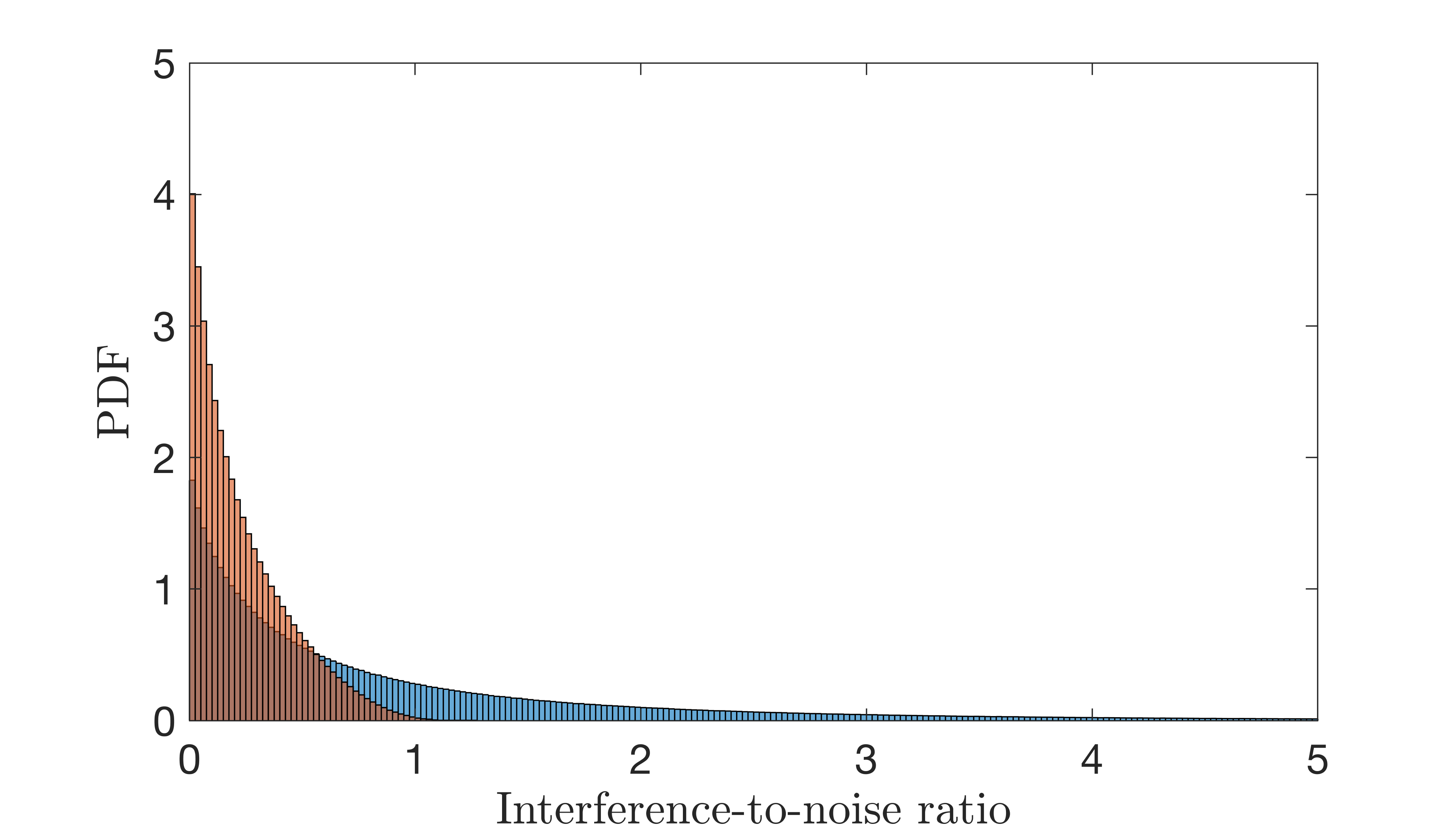}
		\put(20,35){LP-MMSE}
		\put(34,17){MR}
		\put(21,34){\vector(-1,-1){5}}
		\put(35,16){\vector(-1,-1){5}}
		\end{overpic} \label{figureSec6_interfgain}
		\caption{Variations in the received interference power.} 
	\end{subfigure} 
	\caption{Distribution of the signal and interference powers when using MR or LP-MMSE precoding in a simple setup with $L=1$ AP, $N=2$ antennas, $K=2$ UEs, perfect CSI, and uncorrelated Rayleigh fading channels. The distributions are widely different: LP-MMSE gives rise to bounded support while MR provides large variations, which makes it harder to find tight capacity bounds.}\label{figureSec6_signalgain_interfgain}
\end{figure}
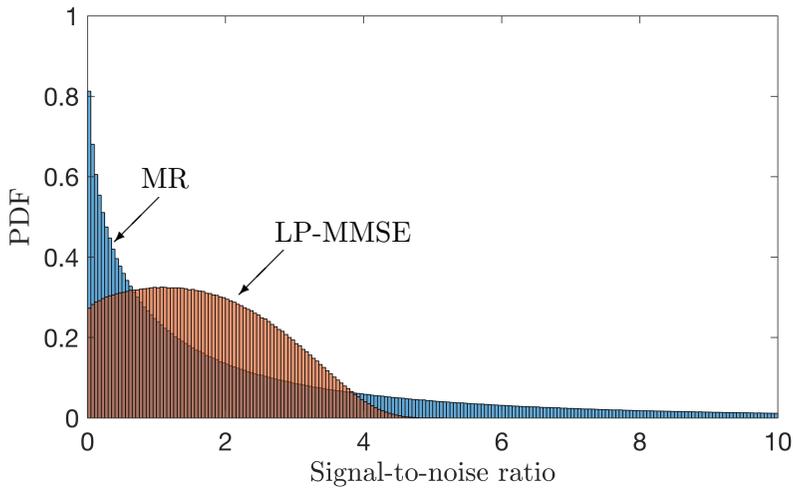
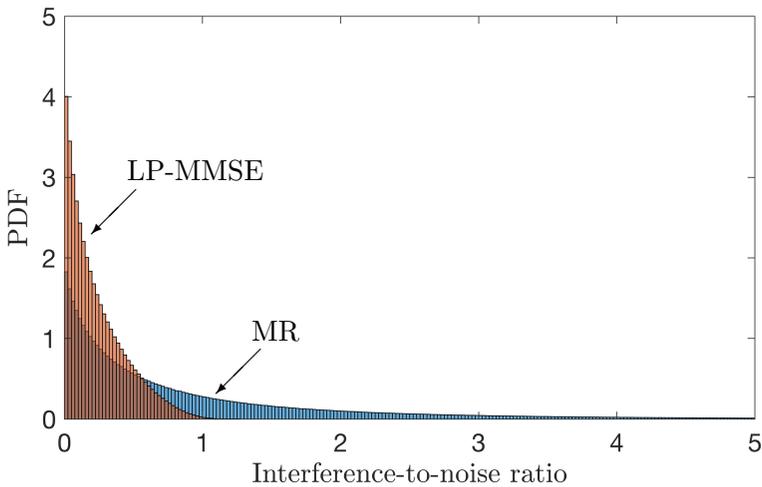

There is a substantial gap between the achievable and genie-aided SE expressions when using MR precoding. To explain the intuition behind this result, Figure~\ref{figureSec6_signalgain_interfgain} shows the variations in signal and interference power in a toy example with $L=1$ AP, $N=2$ antennas, $K=2$ UEs, perfect CSI, and uncorrelated Rayleigh fading channels. Figure~\ref{figureSec6_signalgain_interfgain}(a) shows the PDF of the signal power (normalized by the noise power) when considering different channel realizations and either MR or LP-MMSE precoding. MR provides almost twice the average signal power as compared with LP-MMSE, but even when accounting for that, it is clear that LP-MMSE has bounded support while MR has a distribution with a long tail. Similar behaviors can be observed in Figure~\ref{figureSec6_signalgain_interfgain}(b), where the interference power caused to the other UE is considered. LP-MMSE gives substantially smaller values, due to its interference suppression, and also a distribution with bounded support, while MR gives rise to large variations. 
It is the large variations in the signal and interference power that cause the gap between the achievable and genie-aided SE expressions, which implies that more refined downlink estimation schemes or capacity bounds are needed when using MR; see \cite{Caire2017a}, \cite{Chen2018b}, \cite{Interdonato2019b}. Note that these results are in line with the uplink counterpart from Section~\ref{sec:uplink-performance-comparison} on p.~\pageref{sec:uplink-performance-comparison}.

\subsection{Impact of the Number of UEs}

\begin{figure}[p!]
	\begin{subfigure}[b]{\columnwidth} \centering 
		\begin{overpic}[width=0.88\columnwidth,tics=10]{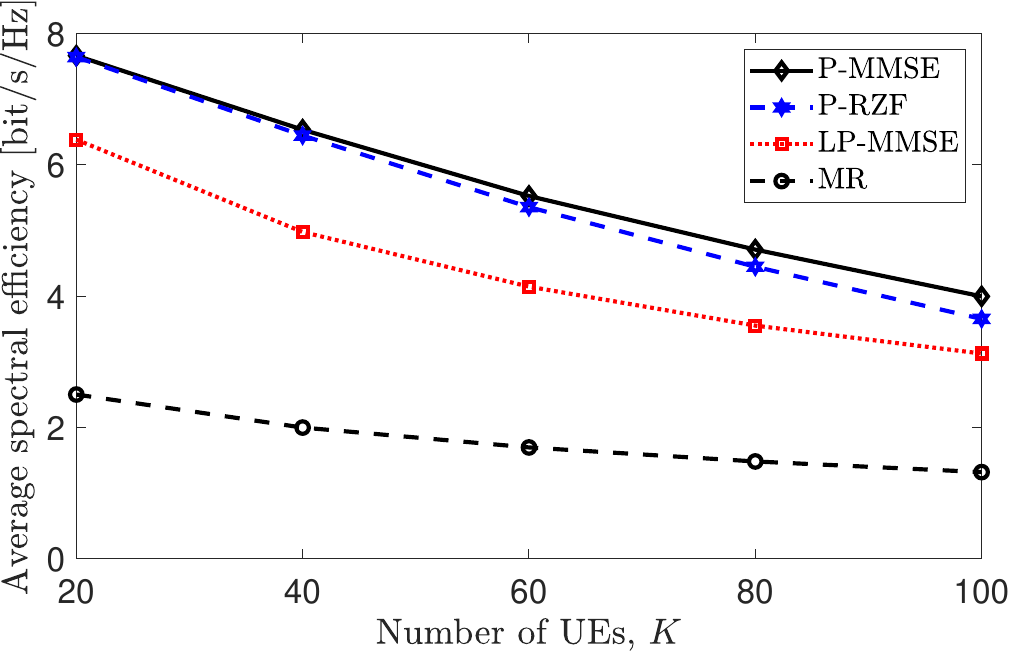}
		\end{overpic} \label{figureSec6_4a}
		\caption{The average SE per UE.} 
		\vspace{4mm}
	\end{subfigure} 
	\begin{subfigure}[b]{\columnwidth} \centering 
		\begin{overpic}[width=0.88\columnwidth,tics=10]{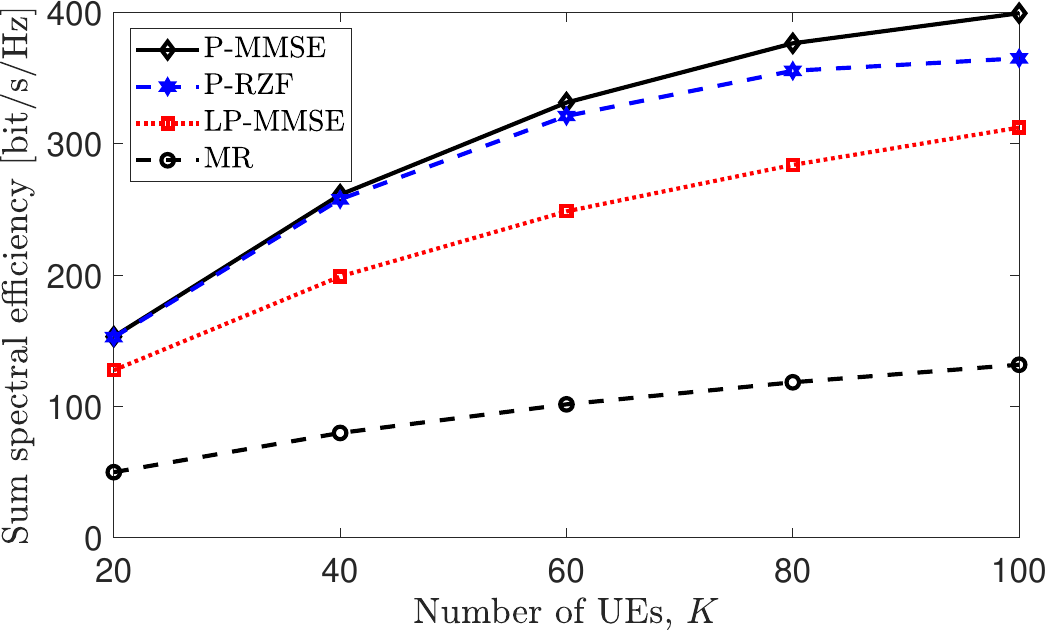}
		\end{overpic} \label{figureSec6_4b}
		\caption{The average sum SE.} 
	\end{subfigure} 
	\caption{The average downlink SE per UE and the sum SE as a function of the number of UEs $K$ for different operations of Cell-free Massive MIMO. We consider $L=100$, $N=4$, $\tau_p=10$, and spatially correlated Rayleigh fading with ASD $\sigma_{\varphi}=\sigma_{\theta}=15^{\circ}$.}\label{figureSec6_4}
\end{figure}

We will now analyze the impact of the number of UEs on network performance. The setup with $L=100$ APs with $N=4$ antennas per AP is considered.
We plot the average SE per UE in Figure~\ref{figureSec6_4}(a) whereas the average sum SE is reported in Figure~\ref{figureSec6_4}(b). Centralized operation with P-MMSE or P-RZF precoding is compared with distributed operation with LP-MMSE or MR precoding. 
The centralized schemes outperform the distributed ones, in accordance with the previous results. P-MMSE provides a slightly larger SE than P-RZF, as also observed in Figure~\ref{figureSec6_1}. In addition, LP-MMSE provides substantially higher SE than MR, since it can suppress interference. Note that the pilot length is $\tau_p=10$ irrespective of the number of UEs. Hence, the pilot contamination increases with $K$, but despite the extra interference, we observe no more than a two-fold reduction in SE per UE when we increase the number of UEs from $K=20$ and $K=100$ in Figure~\ref{figureSec6_4}(a). 
The benefit of multiplexing more UEs is that the sum SE increases almost linearly with $K$ in Figure~\ref{figureSec6_4}(b). This shows how a cell-free network is capable of serving a huge number of UEs.

\subsection{Impact of Spatial Correlation}

\begin{figure}[t!]\centering
	\begin{overpic}[width=0.88\columnwidth,tics=10]{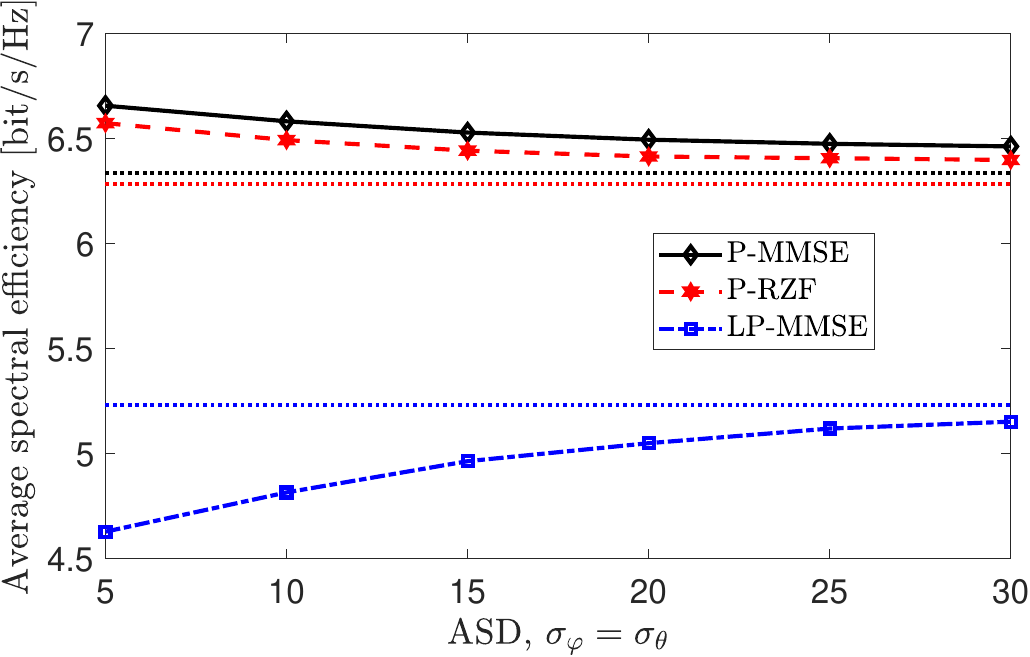}
	\end{overpic} 
	\caption{The average downlink SE per UE as a function of ASD for azimuth and elevation angles, $\sigma_{\varphi}=\sigma_{\theta}$ for different operations of Cell-free Massive MIMO. We consider $L=100$, $N=4$, $K=40$, and $\tau_p=10$. The results for uncorrelated Rayleigh fading are shown as reference by the dotted lines.}  
	\label{figureSec6_5} 
\end{figure}

We conclude this section by analyzing the impact that spatial channel correlation has on the setup with $L=100$, $N=4$, and $K=40$. The ASD is the same in the azimuth and elevation domains: $\sigma_{\varphi}=\sigma_{\theta}$.
Figure~\ref{figureSec6_5} shows the average SE per UE as a function of the ASD, for different precoding schemes.
For each scheme, the corresponding straight dotted line shows the performance achieved when having uncorrelated Rayleigh fading (i.e., no spatial correlation). The lines corresponding to the uncorrelated Rayleigh fading follow the same order as the other lines for each scheme. We notice that spatial correlation degrades the average SE in the distributed operation while it is advantageous in the centralized case, since a smaller ASD corresponds to a more highly correlated channel. This is consistent with previous observations in the uplink; see Figure~\ref{figureSec5_6} on p.~\pageref{figureSec5_6}.
When the ASD increases, the average SE approaches the reference case with no spatial correlation.

\enlargethispage{\baselineskip}

Recall that there are several tradeoffs in connection with spatial correlation. In Section~\ref{sec:hardening-favorable} on p.~\pageref{sec:hardening-favorable}, we showed that the degree of channel hardening decreases with an increasing spatial correlation, which is a negative impact of correlation.
There are also some benefits from correlation.
The level of favorable propagation between two UEs is maximized when the UEs' channels are highly spatially correlated but have very different dominant eigenspaces. Furthermore, the channel estimation quality is improved with increased spatial correlation, as analyzed in Section~\ref{sec:impact-of-spatial-correlation} on p.~\pageref{sec:impact-of-spatial-correlation}. 
When putting these properties together, it is clear that the negative effects of spatial correlation dominate over the positive effects when using a distributed operation where each AP selects a low-dimensional precoding vector that cannot exploit the spatial correlation to a sufficient extent.

Finally, we note that the gap between P-MMSE and P-RZF is slightly larger under spatial correlation than with uncorrelated fading. Recall that the difference between these schemes is that P-MMSE is utilizing the estimation error correlation matrices $\vect{C}_k$, which affects the precoding direction in the correlated fading case. However, under uncorrelated fading, they are scaled identity matrices that can be lumped together with the noise term, thereby reducing their impact.

We compared the fronthaul signaling load with the centralized and distributed uplink operations in Fig.~\ref{figureSec5_8}(a) on p.~\pageref{figureSec5_8a}. We notice that the fronthaul signaling loads for data and pilots are the same in the downlink as in the uplink 
when $\tau_d=\tau_u$. This is seen by comparing Tables~\ref{tab:signaling-level4-downlink} and \ref{tab:signaling-level2-downlink} with Table~\ref{tab:signaling-level4-level3} on p.~\pageref{tab:signaling-level4-level3}. Hence, we will not provide any additional numerical results related to this.

\newpage

\section{Summary of the Key Points in Section~\ref{c-downlink}}
\label{c-downlink-summary}

\begin{mdframed}[style=GrayFrame]

\begin{itemize}
\item The downlink of a cell-free network can be implemented with either centralized or distributed operation.

\item In the centralized operation, the CPU is selecting the precoding vectors and computes the signals to be transmitted, while the APs are only taking care of the physical transmission. This operation enables centralized precoding where the signals transmitted from multiple APs can be coherently received at desired UEs while also suppressing each other at the undesired UEs.

\item The downlink SE with the centralized operation is given in Theorem~\ref{theorem:downlink-capacity-level4}. The optimal precoding is computationally intractable to obtain, but the uplink-downlink duality in Theorem~\ref{prop:duality} establishes that a scaled version of the uplink combining vector is a good heuristic for the corresponding downlink precoding vector. The MMSE, P-MMSE, and P-RZF precoding schemes are defined in this way, whereof the latter two admit scalable implementations.

\item In the distributed operation, each AP is receiving the downlink data from the CPU and designs the locally transmitted signal using local precoding based on the locally available channel estimates. 
The signals transmitted from the serving APs are coherently received at the desired UE, but the interference suppression capability is reduced compared to the centralized operation since each AP can only suppress the interference that itself is generating. This is a limiting factor since each AP is envisioned to have few antennas.

\end{itemize}

\end{mdframed}

\newpage

\begin{mdframed}[style=GrayFrame]

\begin{itemize}

\item The downlink SE with the distributed operation is given in Corollary~\ref{theorem:downlink-capacity-level2}. The uplink-downlink duality motivates using the local combining vectors from the uplink as precoding vectors in the downlink. L-MMSE, LP-MMSE, and MR precoding are defined in this way, whereof the latter two admit scalable implementations.

\item We compared the performance of different centralized and distributed implementations of cell-free networks using the running example. In line with the uplink, scalable precoding schemes in both centralized and distributed operations can achieve almost the same SE as their unscalable counterparts, but with much lower complexity.

\item The precoding schemes are motivated by the uplink-downlink duality, thus there is no guarantee that the optimal combining scheme (from the virtual uplink system) provides the highest performance when used for precoding in the downlink.

\item In the centralized operation, it is preferable to have many single-antenna APs when there is a fixed total number of  antennas, just as in the uplink.
In the distributed operation, UEs with good channel conditions benefit from having a smaller number of multi-antenna APs, which can use local precoding schemes such as LP-MMSE to suppress interference.

\item Increasing the number of UEs in the network improves the sum SE in both the centralized and distributed operations, although the average SE per UE decreases due to the extra interference. Scalable cell-free networks have the capability of serving a large number of UEs with a fixed complexity.

\end{itemize}

\end{mdframed}


\chapter{Spatial Resource Allocation}
\label{c-resource-allocation} 
\vspace{-1cm}
This section describes some important considerations for the optimization, operation, and implementation of User-centric Cell-free Massive MIMO in practical networks. While previous sections have presented the foundations, this section is more related to the latest trends and developments. New algorithms and insights are likely to arise in the future, beyond what is presented here.
The general theme is spatial resource allocation, which refers to the allocation of transmit powers and pilot signals to the UEs, the design of cooperation clusters, and the provisioning of fronthaul resources among the spatially distributed UEs and APs. Section~\ref{sec:optimized-power-allocation} describes power optimization algorithms for maximizing the sum SE and max-min fairness utility functions in both uplink and downlink. Since these algorithms solve network-wide optimization problems, they are not scalable but serve as theoretical benchmarks.
 Next, Section~\ref{sec:scalable-power-allocation} presents heuristic power allocation alternatives, which are designed to be scalable and efficient. The optimized and scalable power allocation methods are compared in Section~\ref{sec:comparison-power-optimzation}.
  In Section~\ref{sec:pilot-assignment-advanced} and Section~\ref{sec:selecting-DCC-advanced}, we provide brief overviews of algorithms for pilot assignment and DCC selection, respectively. Section~\ref{sec:implementation-constraints} discusses important implementation constraints that appear in practical deployments, including having limited fronthaul capacity. Finally, a summary of the key points is provided in Section~\ref{c-resource-allocation-summary}.

\section{Transmit Power Optimization}
\label{sec:optimized-power-allocation}

The transmit power coefficients have so far been treated as heuristically selected constants, but these can be optimized to maximize a network-wide utility function. Section~\ref{sec:optimization-algorithms} on p.~\pageref{sec:optimization-algorithms} provided the theoretical foundations for maximizing the max-min fairness and sum SE utilities. 
These represent two structured ways of selecting an operating point at the outer boundary of the SE region, as was illustrated in Figure~\ref{figure:basic_rate_region} on p.~\pageref{figure:basic_rate_region}.
In this section, we will consider these utilities in the cell-free context and maximize them with respect to the uplink and downlink power coefficients under their respective power constraints. We will mainly use fixed-point and weighted MMSE-based algorithms that take the channel statistics and choice of combining/precoding scheme as inputs. These algorithms can be implemented at the CPU and the same solution can be applied as long as the channel statistics remain the same. Hence, there is no need to adapt the transmit power on a  coherence block basis.
We will consider the uplink and downlink separately.

\subsection{Uplink Power Optimization}
\label{sec:optimized-power-allocation-uplink}

We begin by considering the uplink power optimization.
UE $k$ transmits its uplink data with a power $p_k$, which can be selected as any number between $0$ and the maximum  power $\pmax$. The procedure of selecting an appropriate uplink power is often called \emph{power control} since it can be viewed as controlling how much different UEs cut down on their powers, compared to $\pmax$, to maximize a particular utility function.

\enlargethispage{0.75\baselineskip} 

Several different SE expressions for the centralized and distributed operation were derived in Section~\ref{c-uplink} on p.~\pageref{c-uplink}. In this section, we consider the SE for the centralized case in Theorem~\ref{proposition:uplink-capacity-general-UatF} on p.~\pageref{proposition:uplink-capacity-general-UatF} and the SE for the distributed case in Theorem~\ref{theorem:uplink-capacity-level3} on p.~\pageref{theorem:uplink-capacity-level3}. Both results were derived using the \gls{UatF} bounding technique~\cite[Theorem~4.4]{massivemimobook} and, therefore, share a common structure that enables us to optimize the transmit powers using the same algorithms.\footnote{We have previously used the SE expression in Theorem~\ref{proposition:uplink-capacity-general} on p.~\pageref{proposition:uplink-capacity-general} for analyzing the uplink performance in the centralized operation. This expression  gives slightly larger SE values that better represent the practically achievable communication performance, but since the transmit powers only appear inside of expectations, the expression is not amendable for power optimization. The power allocation algorithms that we develop in this section are optimally designed for the more conservative capacity bounds in Theorem~\ref{proposition:uplink-capacity-general-UatF} on p.~\pageref{proposition:uplink-capacity-general-UatF} and in Theorem~\ref{theorem:uplink-capacity-level3} on p.~\pageref{theorem:uplink-capacity-level3}, but can also be used along with the bound in Theorem~\ref{proposition:uplink-capacity-general}.}

We can gather all the uplink powers in a vector $\vect{p}=[p_1 \, \ldots \, p_K]^{\Ttran}$ and notice that they affect all the UEs. The uplink SE of UE $k$ depends on $\vect{p}$ via its effective SINR. The numerator of the SINR depends on the power $p_k$ of its desired signal and the interference term in the denominator depends on all the power coefficients in $\vect{p}$. The effective SINR for UE $k$ in the centralized and distributed uplink operation can be jointly expressed in the generic form
\enlargethispage{\baselineskip} 
\begin{align}\label{eq:SINR-uplink}
\mathrm{SINR}_k\left(\vect{p}\right)=\frac{b_kp_k}{\vect{c}_k^{\Ttran}\vect{p}+\sigma_k^2}
\end{align}
as a function of the vector $\vect{p}$ with all transmit power coefficients.
The difference between the two types of operation lies in the values of the parameters $b_k \geq 0$, $\vect{c}_k=[c_{k1} \, \ldots \, c_{kK}]^{\Ttran}\in \mathbb{R}_{\geq 0}^K$, and $\sigma_k^2 \geq 0$. They represent the average channel gain of the desired signal, the vector with the average channel gains for the respective interfering signals, and the effective noise variance, which are given as
\begin{align}
&b_k=\begin{cases} \left| \mathbb{E} \left\{ \vect{v}_k^{\Htran}\vect{D}_k\vect{h}_k\right\}\right|^2 & \text{ for centralized operation} \\
\left| \vect{a}_k^{\Htran}\mathbb{E} \left\{ \vect{g}_{kk}\right\}\right|^2 & \text{ for distributed operation}, \end{cases} \quad \forall k \\
&c_{kk}=\begin{cases} \mathbb{E} \left\{\left|  \vect{v}_k^{\Htran}\vect{D}_k\vect{h}_k\right|^2\right\}-b_k & \text{ for centralized operation} \\
\mathbb{E} \left\{ \left| \vect{a}_k^{\Htran}\vect{g}_{kk}\right|^2\right\}-b_k & \text{ for distributed operation}, \end{cases} \quad \forall k\\
&c_{ki}=\begin{cases} \mathbb{E} \left\{\left|  \vect{v}_k^{\Htran}\vect{D}_k\vect{h}_i\right|^2\right\} & \text{ for centralized operation} \\
\mathbb{E} \left\{ \left| \vect{a}_k^{\Htran}\vect{g}_{ki}\right|^2 \right\} & \text{ for distributed operation}, \end{cases} \quad \forall k, \forall i\neq k \\
&\sigma_k^2=\begin{cases} \sigmaUL\mathbb{E} \left\{ \left\Vert \vect{D}_k\vect{v}_k\right\Vert^2\right\} & \text{ for centralized operation} \\
 \vect{a}_k^{\Htran}\vect{F}_k\vect{a}_{k} & \text{ for distributed operation}, \end{cases} \quad \forall k.
\end{align}
The generic SINR expression in \eqref{eq:SINR-uplink} has a structure that we recognize from Lemma~\ref{lemma:generic-SINR-expression} on p.~\pageref{lemma:generic-SINR-expression} with $\left|\mathrm{DS}_k(\vect{p})\right|^2 = b_kp_k$ being the signal term in the numerator, $\sum_{i=1}^K\mathbb{E}\left\{\left|\mathrm{I}_{ki}\left(\vect{p}\right)\right|\right\}^2 = \vect{c}_k^{\Ttran}\vect{p}$ being the total interference power, and $\sigma_k^2$ is the noise power. 
Hence, we can use the optimization algorithms presented in Section~\ref{sec:optimization-algorithms} on p.~\pageref{sec:optimization-algorithms}, which apply for arbitrary instances of the SINR expression from Lemma~\ref{lemma:generic-SINR-expression}, 
to solve two power optimization problems: max-min SE fairness and sum SE maximization. We stress that the algorithms presented below are applying to centralized operation with arbitrary receive combining and distributed operation with arbitrary local receive combining and LSFD weights. This does not mean that the same transmit powers are optimal in all situations. Since different network operations lead to different values of $b_k$, $\vect{c}_k$, and $\sigma_k^2$, the optimal transmit powers will naturally be different even if they are obtained using the same algorithms.

\subsubsection{Uplink Max-Min SE Fairness}

We first consider the max-min SE fairness problem, which was formulated for the general case in \eqref{eq:maxmin-fairness-problem}. In the considered uplink scenario, there are $K$ individual transmit power constraints and the max-min fairness optimization problem can be particularized as
\begin{align} \label{eq:maxmin-fairness-problem-uplink}
&\underset{\vect{p} \geq \vect{0}_K}{\textrm{maximize}} \quad \min\limits_{k\in \{1,\ldots,K\}} \quad \frac{b_kp_k}{\vect{c}_k^{\Ttran}\vect{p}+\sigma_k^2} \\
&\textrm{subject to} \quad p_k\leq \pmax, \quad k=1,\ldots,K.  \nonumber
\end{align}
This problem has the same structure as the generic problem in \eqref{eq:maxmin-fairness-problem} with $R=K$ and $\vect{a}_r$ 
having a one in the $r$th entry and zeros elsewhere for $r=1,\ldots,K$. Under the following conditions, the solution to \eqref{eq:maxmin-fairness-problem-uplink} can be computed by using the fixed-point algorithm in Algorithm~\ref{alg:fixed-point} on p.~\pageref{alg:fixed-point}.

\begin{lemma}
The three conditions in Lemma~\ref{lemma:fixed-point} on p.~\pageref{lemma:fixed-point} are satisfied for $\mathrm{SINR}_k(\vect{p})$ in \eqref{eq:SINR-uplink} if the coefficients $b_k$, $c_{ki}$ for all $i \neq k$, and $\sigma_k^2$ are strictly positive. Hence, we can use Algorithm~\ref{alg:fixed-point} to solve \eqref{eq:maxmin-fairness-problem-uplink}.
\end{lemma}

The requirements stated in the lemma are mild and basically say that the processed signal for every UE includes a non-zero fraction of the transmitted power from all other UEs and that the noise power is non-zero.
By particularizing Algorithm~\ref{alg:fixed-point}  for the problem at hand, we obtain Algorithm~\ref{alg:fixed-point-uplink}, which converges to the optimal solution of the max-min fairness problem in \eqref{eq:maxmin-fairness-problem-uplink}. 
All UEs will have equal SE at the optimum and at least one UE will use the maximum power $\pmax$.
Algorithm~\ref{alg:fixed-point-uplink} converges quickly and has relatively low computational complexity since it only involves iterative closed-form updates of the variables. However, its complexity grows with $K$, thus it is not scalable.

\begin{algorithm}
	\caption{Fixed-point algorithm for solving the uplink max-min fairness problem in \eqref{eq:maxmin-fairness-problem-uplink}.} \label{alg:fixed-point-uplink}
	\begin{algorithmic}[1]
		\State {\bf Initialization:} Set arbitrary initial power $\vect{p}>\vect{0}_K$ and the solution accuracy $\epsilon>0$
		\While{$\underset{k\in\{1,\ldots,K\}}{\max} \textrm{SINR}_k\left(\vect{p}\right)-\underset{k\in\{1,\ldots,K\}}{\min} \textrm{SINR}_k\left(\vect{p}\right) > \epsilon$} 
		\State $p_k \gets \frac{p_k}{\textrm{SINR}_k\left(\vect{p}\right)}$, $k=1,\ldots,K$
		\State $\vect{p} \gets \frac{p_{\rm max}}{\underset{k\in\{1,\ldots,K\}}{\max}p_k}\vect{p}$
		\EndWhile
		\State {\bf Output:} Optimal transmit powers $\vect{p}$
		\State Max-min SE $\underset{k\in\{1,\ldots,K\}}{\min} \frac{\tau_u}{\tau_c}\log_2\left(1+ \textrm{SINR}_k\left(\vect{p}\right)\right)$
	\end{algorithmic}
\end{algorithm}

\subsubsection{Uplink Sum SE Maximization}

We will now consider the sum SE maximization problem, which is formulated in the uplink as
\begin{align} \label{eq:sum-SE-maximization-uplink}
&\underset{\vect{p} \geq \vect{0}_K}{\textrm{maximize}} \quad \sum\limits_{k=1}^K  \log_2\left(1+\frac{b_kp_k}{\vect{c}_k^{\Ttran}\vect{p}+\sigma_k^2}\right) \\
&\textrm{subject to} \quad p_k\leq \pmax, \quad k=1,\ldots,K . \nonumber
\end{align}
This problem has the same structure as the generic one in \eqref{eq:sum-SE-maximization}. Hence, we can conclude that \eqref{eq:sum-SE-maximization-uplink} is not convex, however, a local optimum can be attained by the block coordinate descent algorithm given in Algorithm~\ref{alg:sum-SE} on p.~\pageref{alg:sum-SE}.
That algorithm was developed based on a weighted MMSE reformulation of the sum SE maximization problem, where the MSE in the data detection was stated in \eqref{eq:ek-MSE}. For the particular SINR expression considered in this section, the corresponding MSE can be particularized as
\begin{align} 
e_k\left(\vect{p}, u_k\right)=u_{k}^2\left(b_kp_k+\vect{c}_k^{\Ttran}\vect{p}+\sigma_k^2\right) -2u_{k}\sqrt{b_kp_k}+1  \label{eq:ek-MSE-uplink}
\end{align}
where we have used the fact that $b_k$ is real-valued and positive and $u_k$ is also real-valued since all the coefficients in the SINR expression are real-valued.
The solution to the optimization problem in Step 6 of Algorithm \ref{alg:sum-SE} on p.~\pageref{alg:sum-SE} can be obtained in closed-form as
\begin{align} \label{eq:pk-update-uplink}
p_k=\min\left(\pmax, \frac{b_kd_k^2u_k^2}{\left(d_ku_k^2b_k+\sum\limits_{i=1}^Kd_iu_i^2c_{ik}\right)^2}  \right)
\end{align}
by treating $\sqrt{p_1},\ldots,\sqrt{p_K}$ as the optimization variables and decomposing the problem into $K$ independent subproblems each of which is a quadratic minimization under a bound constraint. The steps of the block coordinate descent algorithm for the uplink sum SE maximization can then be simplified to Algorithm~\ref{alg:sum-SE-uplink}, which is an iterative algorithm where all the updates are based on closed-form expressions. 
The algorithm converges quickly and has relatively low computational complexity, but the complexity grows with $K$ so the algorithm is not scalable. Note that Algorithm~\ref{alg:sum-SE-uplink} only guarantees to find a local optimum to \eqref{eq:sum-SE-maximization-uplink}, thus different initializations may lead to different solutions.

A typical property of sum SE maximization problems is that the optimal solution might assign zero power to some UEs, which naturally leads to zero SE. This can occur for UEs that have weak channels to all APs, however, we will show in Section~\ref{sec:comparison-power-optimzation}  that this is not an issue that appears when applying the algorithm to the running example. We can be sure that at least one UE will use the maximum power $\pmax$ at the optimum solution.

\begin{algorithm}
	\caption{Block coordinate descent algorithm for solving the sum SE maximization problem in \eqref{eq:sum-SE-maximization-uplink}.} \label{alg:sum-SE-uplink}
	\begin{algorithmic}[1]
		\State {\bf Initialization:} Set the solution accuracy $\epsilon>0$
		\State Set an arbitrary feasible power vector $\vect{p}$
		\While{  $\sum_{k=1}^K\Big(d_{k}e_{k}\left(\vect{p},u_k\right)-\ln\left( d_{k}\right)\Big)$  is either improved more than $\epsilon$ or not improved at all}
		\State $u_{k} \gets \frac{\sqrt{b_kp_k}}{b_kp_k+\vect{c}_k^{\Ttran}\vect{p}+\sigma_k^2}, \quad k=1,\ldots,K$
		\State $d_k \gets \frac{1}{u_{k}^2\left(b_kp_k+\vect{c}_k^{\Ttran}\vect{p}+\sigma_k^2\right) -2u_{k}\sqrt{b_kp_k}+1}, \quad k=1,\ldots,K$
		\State $p_k \gets \min\left(\pmax, \frac{b_kd_k^2u_k^2}{\left(d_ku_k^2b_k+\sum\limits_{i=1}^Kd_iu_i^2c_{ik}\right)^2}  \right), \quad k=1,\ldots,K$
		\EndWhile
		\State {\bf Output:} Optimal transmit powers $\vect{p}$
		\State Sum SE $\frac{\tau_u}{\tau_c}\sum_{k=1}^K\log_2\left(d_k\right)$
	\end{algorithmic}
\end{algorithm}

\begin{remark}[Pilot power optimization]
The optimization problems in \eqref{eq:maxmin-fairness-problem-uplink} and \eqref{eq:sum-SE-maximization-uplink} optimize the transmit powers $p_1,\ldots,p_K$ that are used for uplink data transmission. The uplink transmission also involves the pilot powers $\eta_1,\ldots,\eta_K$, which were implicitly assumed to be constant (they appear at the inside of $b_k$, $\vect{c}_k$, and $\sigma_k^2$). If the pilot assignment is carried out properly, then pilot transmission with maximum power $\eta_1=\ldots=\eta_K= \pmax$ is expected to be a nearly optimal solution. There are a few papers that also consider pilot power optimization. For example, \cite{Mai2018} optimizes the pilot powers by minimizing the maximum NMSE in the channel estimation, but without considering the performance in the data transmission phase. 
The joint pilot and data power optimization is considered in \cite{Masoumi2018} for single-antenna APs, but only an approximation of the max-min fairness problem is solved. Further research on the joint pilot and data power optimization problem is required.
\end{remark}

\subsection{Downlink Power Optimization}
\label{sec:optimized-power-allocation-downlink}

We will now consider the downlink power optimization. AP $l$ transmits the signal $\vect{x}_l =  \sum_{i=1}^{K} \vect{D}_{il}  \vect{w}_{il} \varsigma_i $ defined in \eqref{eq:DCC-transmit-downlink2} and is assumed to have a maximum transmit power of $\rhomax$. This power can be arbitrarily divided between the UEs and this procedure is often called \emph{power allocation}. An AP can decide to not use all of its power to not cause unnecessary interference, similar to the situation in the uplink. 
The centralized and distributed operation will be handled differently in this section, even if the power constraints are the same in both cases. The reason is that the precoding vectors are coupled between the APs in the centralized case, while they are not in the distributed case.

In the centralized operation, the centralized precoding vectors
\begin{equation}
\vect{w}_k = \sqrt{\rho_k} \frac{\bar{\vect{w}}_k}{\sqrt{\mathbb{E} \left\{ \left\| \bar{\vect{w}}_k \right\|^2 \right\} }}
\end{equation}
from \eqref{eq:precoding-centralized-expression} are used for $k=1,\ldots,K$. There are $K$ power allocation coefficients $\{\rho_k:k=1,\ldots,K\}$ to optimize, where $\rho_k \geq 0$ represents the total downlink power allocated to UE $k$ from all the serving APs.

In the distributed operation, the local precoding vectors for UE $k$ are defined in \eqref{eq:precoding-distributed-expression}  as
\begin{equation}
{\vect{w}}_{kl}= \sqrt{\rho_{kl}}\frac{\bar{\vect{w}}_{kl}}{\sqrt{\mathbb{E}\left\{\left\| \bar{\vect{w}}_{kl}\right\|^2\right\}}}
\end{equation}
for $l \in \mathcal{M}_k$. There are $\sum_{k=1}^K |\mathcal{M}_k|$ power allocation coefficients $\{\rho_{kl} : l\in \mathcal{M}_k, k=1,\ldots,K\}$ to optimize, where $\rho_{kl} \geq 0$ denotes the downlink  power that AP $l$ is assigning to UE $k$. 
Since there are more coefficients~in the distributed case, we can expect the optimization to be more complex.

\enlargethispage{\baselineskip}

\subsubsection{Optimized Centralized Downlink Power Allocation}
\label{sec:optimized-power-allocation-downlink-scheme1}

We begin with the centralized operation for which the downlink SE is given by Theorem~\ref{theorem:downlink-capacity-level4} on p.~\pageref{theorem:downlink-capacity-level4}. We notice that the effective SINR, $\mathacr{SINR}_{k}^{({\rm dl},\cent)}$, achieved by UE $k$ can be expressed as a function of the downlink power coefficients $\boldsymbol{\rho}=[\rho_1 \, \ldots \, \rho_K]^{\Ttran}$ as
\begin{align}\label{eq:SINR-downlink}
\mathacr{SINR}_{k}^{({\rm dl},\cent)}\left(\boldsymbol{\rho}\right)=\frac{\tilde{b}_k\rho_k}{\tilde{\vect{c}}_k^{\Ttran}\boldsymbol{\rho}+\sigmaDL}
\end{align}
where $\tilde{b}_k$ is the average channel gain of the desired signal and $\tilde{\vect{c}}_k=[\tilde{c}_{k1} \, \ldots \, \tilde{c}_{kK}]^{\Ttran}\in \mathbb{R}_{\geq 0}^K$ is a vector containing the average channel gains for the respective interfering signals.
The specific values of these parameters in the centralized downlink operation are
\begin{align}
&\tilde{b}_k=  \frac{\left| \mathbb{E} \left\{ \vect{h}_{k}^{\Htran}\vect{D}_{k}\bar{\vect{w}}_{k}\right\} \right|^2 }{ \mathbb{E} \left\{ \| \bar{\vect{w}}_k \|^2 \right\}}, \quad \forall k \\
&\tilde{c}_{kk}= \frac{ \mathbb{E} \left\{\left|  \vect{h}_{k}^{\Htran}\vect{D}_{k}\bar{\vect{w}}_{k}\right|^2\right\} }{ \mathbb{E} \left\{ \| \bar{\vect{w}}_k \|^2 \right\}} -\tilde{b}_k, \quad \forall k\\
&\tilde{c}_{ki}= \frac{ \mathbb{E} \left\{\left|  \vect{h}_{k}^{\Htran}\vect{D}_{i}\bar{\vect{w}}_{i}\right|^2\right\} }{ \mathbb{E} \left\{ \| \bar{\vect{w}}_i \|^2 \right\} },  \quad \forall k, \forall i\neq k .
\end{align}
Recall that $\{\bar{\vect{w}}_{k}: k=1,\ldots,K\}$ are vectors that point out the directions of the centralized precoding vectors. MMSE, P-MMSE, and P-RZF precoding were considered in Section~\ref{subsec:centralized-precoding} on p.~\pageref{subsec:centralized-precoding} and represent different ways of selecting $\bar{\vect{w}}_{k}$. The choice of precoding scheme will affect the values of $\tilde{b}_k$ and $\tilde{\vect{c}}_k$, but the same optimization algorithms can be utilized for any precoding since the SINRs are always given by \eqref{eq:SINR-downlink}.

We note that the effective downlink SINR in \eqref{eq:SINR-downlink} has the same structure as the effective uplink SINR in \eqref{eq:SINR-uplink}, except that the $K$ transmit power coefficients are now contained in a vector that we denote $\boldsymbol{\rho}$. Hence, we can apply almost the same optimization algorithms to solve the max-min fairness and sum SE maximization problems, with the only structural difference that we have one power constraint per AP. 
Let $\bar{\vect{w}}_{kl}^{\prime} \in \mathbb{C}^N$ denote the portion of the normalized centralized precoding vector $\bar{\vect{w}}_k /\sqrt{\mathbb{E} \{ \left\| \bar{\vect{w}}_k \right\|^2 \} }$ corresponding to AP $l$, so we have
\begin{align} \label{eq:normalized_centralized_precoding}
& \frac{\bar{\vect{w}}_k}{\sqrt{\mathbb{E} \left\{ \left\| \bar{\vect{w}}_k \right\|^2 \right\} }} = \begin{bmatrix} \bar{\vect{w}}_{k1}^{\prime} \\ \vdots \\  \bar{\vect{w}}_{kL}^{\prime} \end{bmatrix}.
\end{align}
Note that $\sum_{l=1}^{L} \mathbb{E} \{ \| \bar{\vect{w}}_{kl}^{\prime} \|^2 \} = 1$ by construction, while the individual vectors $\{\bar{\vect{w}}_{kl}^{\prime}: l\in \mathcal{M}_k\}$ can have arbitrary average squared norms between 0 and 1. The value $\mathbb{E} \{ \| \bar{\vect{w}}_{kl}^{\prime} \|^2 \} \in [0,1]$ represents the fraction of the total transmit power assigned to UE $k$ that will be sent from AP $l$. Hence, the power constraint for AP $l$ can be formulated as
\begin{equation} \label{eq:power-constraint-centralized}
\sum_{k\in \mathcal{D}_l}\rho_k\mathbb{E} \left\{ \left\| \bar{\vect{w}}_{kl}^{\prime} \right\|^2 \right\} \leq \rhomax.
\end{equation}

We first consider the max-min SE fairness problem that is formulated for the general case in \eqref{eq:maxmin-fairness-problem}. In the special case at hand, there are $L$ per-AP transmit power constraints of the kind in \eqref{eq:power-constraint-centralized}. 
The max-min fairness optimization problem can then be particularized as
\begin{align} \label{eq:maxmin-fairness-problem-downlink}
&\underset{\boldsymbol{\rho} \geq \vect{0}_K}{\textrm{maximize}} \quad \min\limits_{k\in \{1,\ldots,K\}} \quad \frac{\tilde{b}_k\rho_k}{\tilde{\vect{c}}_k^{\Ttran}\boldsymbol{\rho}+\sigmaDL} \\
&\textrm{subject to} \quad \sum_{k\in \mathcal{D}_l}\rho_k \mathbb{E} \left\{ \left\| \bar{\vect{w}}_{kl}^{\prime} \right\|^2\right\}\leq \rhomax, \quad l=1,\ldots,L.  \nonumber
\end{align}
The above problem has the same structure as the generic problem in \eqref{eq:maxmin-fairness-problem} with $R=L$ and $\vect{a}_r$ is the vector whose elements are
 $\mathbb{E} \{ \| \bar{\vect{w}}_{kr}^{\prime} \|^2\}$ for UEs that are served by AP $r$ and zero elsewhere. Note that the three conditions in Lemma~\ref{lemma:fixed-point} on p.~\pageref{lemma:fixed-point} are satisfied, just as in uplink, and we can thus design a similar fixed-point algorithm that converges to the optimal solution of \eqref{eq:maxmin-fairness-problem-downlink}. The specific steps are provided in Algorithm~\ref{alg:fixed-point-downlink}. This algorithm will converge quickly but the complexity grows with $K$. This means that it is not scalable to optimize the powers in this way in a large cell-free network.
As usual for max-min fairness problems, all UEs will obtain the same SE at the optimum. 

\begin{algorithm}
	\caption{Fixed-point algorithm for solving the centralized downlink max-min fairness problem in \eqref{eq:maxmin-fairness-problem-downlink}.} \label{alg:fixed-point-downlink}
	\begin{algorithmic}[1]
		\State {\bf Initialization:} Set arbitrary initial power $\boldsymbol{\rho}>\vect{0}_K$ and the solution accuracy $\epsilon>0$
		\While{$\underset{k\in\{1,\ldots,K\}}{\max} \textrm{SINR}_k\left(\boldsymbol{\rho}\right)-\underset{k\in\{1,\ldots,K\}}{\min} \textrm{SINR}_k\left(\boldsymbol{\rho}\right) > \epsilon$} 
		\State $\rho_k \gets \frac{\rho_k}{\textrm{SINR}_k\left(\boldsymbol{\rho}\right)}$, $k=1,\ldots,K$
		\State $\boldsymbol{\rho} \gets \frac{\rho_{\rm max}}{\underset{l\in\{1,\ldots,L\}}{\max}\sum\limits_{k\in \mathcal{D}_l}\rho_k \mathbb{E} \{ \| \bar{\vect{w}}_{kl}^{\prime} \|^2\}}\boldsymbol{\rho}$
		\EndWhile
		\State {\bf Output:} Optimal transmit powers $\boldsymbol{\rho}$
			\State Max-min SE $\underset{k\in\{1,\ldots,K\}}{\min} \frac{\tau_d}{\tau_c}\log_2\left(1+ \textrm{SINR}_k\left(\boldsymbol{\rho}\right)\right)$
	\end{algorithmic}
\end{algorithm}

We now consider the downlink sum SE maximization problem in the centralized operation, which  is formulated as
\begin{align}  \label{eq:sum-SE-maximization-downlink}
&\underset{\boldsymbol{\rho} \geq \vect{0}_K}{\textrm{maximize}} \quad \sum\limits_{k=1}^K  \log_2\left(1+\frac{\tilde{b}_k\rho_k}{\tilde{\vect{c}}_k^{\Ttran}\boldsymbol{\rho}+\sigmaDL}\right) \\
&\textrm{subject to} \quad \sum_{k\in \mathcal{D}_l}\rho_k \mathbb{E} \left\{ \left\| \bar{\vect{w}}_{kl}^{\prime} \right\|^2\right\}\leq \rhomax, \quad l=1,\ldots,L. \nonumber
\end{align}
This problem also resembles the uplink counterpart with the main structural difference being the power constraints that are coupled between the optimization variables. We notice that \eqref{eq:sum-SE-maximization-downlink} is a problem of the kind in \eqref{eq:sum-SE-maximization} with  $R=L$ and $\vect{a}_r$ defined in the same way as for the max-min fairness problem above.
This implies that the sum SE maximization problem is not convex but a local optimal solution can be obtained by particularizing the block coordinate descent algorithm in Algorithm~\ref{alg:sum-SE} on p.~\pageref{alg:sum-SE} for the problem at hand. This algorithm utilizes a weighted MMSE reformulation of the sum SE maximization problem and the corresponding MSE in \eqref{eq:ek-MSE} becomes
\begin{align} 
&e_k\left(\boldsymbol{\rho}, u_k\right)=u_{k}^2\left(\tilde{b}_k\rho_k+\tilde{\vect{c}}_k^{\Ttran}\boldsymbol{\rho}+\sigmaDL\right) -2u_{k}\sqrt{\tilde{b}_k\rho_k}+1  \label{eq:ek-MSE-downlink}
\end{align}
in the considered downlink problem. Algorithm~\ref{alg:sum-SE-downlink} provides the resulting algorithm and \eqref{eq:sum-SE-maximization-weighted-MMSE-subproblem-DL} in Step 6 contains a subproblem that needs to be solved in every iteration.
If we treat $\{\sqrt{\rho_k}:k=1,\ldots,K\}$  as the optimization variables, \eqref{eq:sum-SE-maximization-weighted-MMSE-subproblem-DL} becomes a convex quadratically-constrained quadratic programming problem. Although a closed-form solution does not exist in general due to the multiple power constraints involving the same variables, any convex solver can be utilized to solve this subproblem \cite{Boyd2004a}.

\begin{remark}[General-purpose convex solvers] \label{remark:convex-solvers}
A convex optimization problem that lacks a closed-form solution can be solved either by a dedicated solver, which is developed to exploit the special structure of the problem at hand, or by a general-purpose solver that has been optimized for a wide class of problems. The former approach is preferred from a runtime efficiency perspective, while the latter approach is preferred for faster code development since the finer implementation details are abstracted away.
We took the second approach when comparing power allocation schemes in Section~\ref{sec:comparison-power-optimzation}.
More precisely, we made use of the solver SDPT3 \cite{SDPT3} and wrote the code using CVX \cite{cvx2014}, an interface for  specifying convex problems and connect them to a solver. 
An example of the first approach is \cite{Chakraborty2020a}.
\end{remark}

\begin{algorithm}
	\caption{Block coordinate descent algorithm for solving the sum SE maximization problem in \eqref{eq:sum-SE-maximization-downlink}.} \label{alg:sum-SE-downlink}
	\begin{algorithmic}[1]
		\State {\bf Initialization:} Set the solution accuracy $\epsilon>0$
		\State Set an arbitrary feasible initial power vector $\boldsymbol{\rho}$
		\While{  $\sum_{k=1}^K\Big(d_{k}e_{k}\left(\boldsymbol{\rho},u_k\right)-\ln\left( d_{k}\right)\Big)$  is either improved more than $\epsilon$ or not improved at all}
		\State $u_{k} \gets \frac{\sqrt{\tilde{b}_k\rho_k}}{\tilde{b}_k\rho_k+\tilde{\vect{c}}_k^{\Ttran}\boldsymbol{\rho}+\sigmaDL}, \quad k=1,\ldots,K$
		\State $d_k \gets 1\Big/ \!\left(u_{k}^2\left(\tilde{b}_k\rho_k+\tilde{\vect{c}}_k^{\Ttran}\boldsymbol{\rho}+\sigmaDL\right) -2u_{k}\sqrt{\tilde{b}_k\rho_k}+1\right), \quad k=1,\ldots,K$
			\State Solve the following convex problem for the current values of  $u_k$ and $d_k$:
		\begin{align}
		&\underset{\boldsymbol{\rho} \geq \vect{0}_K}{\textrm{minimize}} \quad \sum_{k=1}^Kd_{k}e_{k}\left(\boldsymbol{\rho},u_k\right) \label{eq:sum-SE-maximization-weighted-MMSE-subproblem-DL} \\
		&\hspace{0cm}\textrm{subject to} \quad \sum_{k\in \mathcal{D}_l}\rho_k\mathbb{E} \left\{ \left\| \bar{\vect{w}}_{kl}^{\prime} \right\|^2\right\}\leq \rhomax, \quad l=1,\ldots,L  \nonumber
		\end{align}
		\State Update $\boldsymbol{\rho}$ by the obtained solution to \eqref{eq:sum-SE-maximization-weighted-MMSE-subproblem-DL}.
		\EndWhile
		\State {\bf Output:} Optimal transmit powers $\boldsymbol{\rho}$
			\State Sum SE $\frac{\tau_d}{\tau_c}\sum_{k=1}^K\log_2\left(d_k\right)$
	\end{algorithmic}
\end{algorithm}

\enlargethispage{\baselineskip}

\subsubsection{Optimized Distributed Downlink Power Allocation}
\label{sec:optimized-power-allocation-downlink-scheme2}

In the distributed downlink operation, there are $\sum_{k=1}^K |\mathcal{M}_k|$ power allocation coefficients to optimize: $\{\rho_{kl} :  l\in \mathcal{M}_k, k=1,\ldots,K\}$. 
This number is larger than $K$ (as in the centralized operation), except in the extreme case when each UE is only served by one AP. For example, if each AP serves $\tau_p$ UEs (one per pilot), then there will be $L \tau_p$ power coefficients to optimize in the distributed operation.

The same SE expression is used as in the centralized operation, but the effective SINRs have a different dependence on the optimization variables than in the centralized case.
To obtain tractable formulations, we first introduce a new set of optimization variables:
\begin{equation} \label{eq:square-roots-powers}
\tilde{\rho}_{kl}=\sqrt{\rho_{kl}}\geq0
\end{equation}
for $k\in \mathcal{D}_l$ and $l=1,\ldots,L$.
These are the square roots of the transmit power coefficients. There is a one-to-one correspondence in \eqref{eq:square-roots-powers} between the power variables, thus optimization with respect to the new variables $\{\tilde{\rho}_{kl} :  l\in \mathcal{M}_k, k=1,\ldots,K\}$ is equivalent to the optimization with respect to the original ones. The benefit of the new variables is that 
the effective downlink SINR for UE $k$ in \eqref{eq:downlink-instant-SINR-level2} can be expressed as
\begin{align}\label{eq:SINR-downlink-2}
\mathacr{SINR}_{k}^{({\rm dl},\dist)}\left(\left\{\tilde{\boldsymbol{\rho}}_i\right\}\right)=\frac{\left|\tilde{\vect{b}}_k^{\Ttran}\tilde{\boldsymbol{\rho}}_k\right|^2}{\sum\limits_{i=1}^K\tilde{\boldsymbol{\rho}}_i^{\Ttran}\tilde{\vect{C}}_{ki}\tilde{\boldsymbol{\rho}}_i-\left|\tilde{\vect{b}}_k^{\Ttran}\tilde{\boldsymbol{\rho}}_k\right|^2+\sigmaDL}
\end{align}
where
\begin{align}
 &\tilde{\boldsymbol{\rho}}_k=\left [ \tilde{\rho}_{k1} \, \ldots \, \tilde{\rho}_{kL} \right ]^{\Ttran} \in \mathbb{R}_{\geq 0}^L, \quad \forall k \\
  &\tilde{\vect{b}}_k \in \mathbb{R}_{\geq 0}^L, \nonumber \\
   & \left[ \tilde{\vect{b}}_{k} \right]_{l}=\begin{cases}  \frac{\mathbb{E} \left\{ \vect{h}_{kl}^{\Htran}\bar{\vect{w}}_{kl}\right\}}{\sqrt{\mathbb{E}\left\{\left\| \bar{\vect{w}}_{kl}\right\|^2\right\}}} &  l\in \mathcal{M}_k \\
  0 & l \notin \mathcal{M}_k, \end{cases} \quad \forall k \\
  & \tilde{\vect{C}}_{ki}\in \mathbb{C}^{L \times L}, \nonumber \\
  &\left[ \tilde{\vect{C}}_{ki} \right]_{lr}=\begin{cases}\frac{\mathbb{E} \left\{ \vect{h}_{kl}^{\Htran}\bar{\vect{w}}_{il}\bar{\vect{w}}_{ir}^{\Htran}\vect{h}_{kr}\right\}}{\sqrt{\mathbb{E}\left\{\left\| \bar{\vect{w}}_{il}\right\|^2\right\}}\sqrt{\mathbb{E}\left\{\left\| \bar{\vect{w}}_{ir}\right\|^2\right\}}} &  l\in \mathcal{M}_i, r\in \mathcal{M}_i \\ 0 & l\notin \mathcal{M}_i \textrm{ or } r\notin \mathcal{M}_i, \end{cases} \quad \forall k,i .
\end{align}
The elements of $\tilde{\vect{b}}_k$ are assumed to be real and non-negative, which is satisfied when using L-MMSE, LP-MMSE, and MR precoding. This condition can also be satisfied without loss of generality for any other type of precoding by rotating the phase of the precoding vectors \cite{Bengtsson2001a}. 

\pagebreak
The max-min fairness problem can now be formulated for the distributed downlink operation as
\begin{align} \label{eq:maxmin-fairness-problem-downlink-t}
&\underset{\tilde{\boldsymbol{\rho}}_k\geq \vect{0}_L, \forall k, t \geq 0}{\textrm{maximize}} \quad t \\
&\textrm{subject to} \quad  \frac{\left|\tilde{\vect{b}}_k^{\Ttran}\tilde{\boldsymbol{\rho}}_k\right|^2}{\sum\limits_{i=1}^K\tilde{\boldsymbol{\rho}}_i^{\Ttran}\tilde{\vect{C}}_{ki}\tilde{\boldsymbol{\rho}}_i-\left|\tilde{\vect{b}}_k^{\Ttran}\tilde{\boldsymbol{\rho}}_k\right|^2+\sigmaDL}\geq t, \quad k=1,\ldots,K \nonumber \\
&\hspace{21mm} \sum_{k\in \mathcal{D}_l}\tilde{\rho}_{kl}^2\leq \rhomax, \quad l=1,\ldots,L\nonumber
\end{align}
by utilizing the auxiliary variable $t$ that represents the minimum SINR among all UEs. Note that the power constraint 
$\sum_{k\in \mathcal{D}_l}\tilde{\rho}_{kl}^2\leq \rhomax$ of AP $l$ sums up all the squared coefficients related to this AP.
We notice that \eqref{eq:maxmin-fairness-problem-downlink-t} is an instance of the generic problem formulation in \eqref{eq:maxmin-fairness-problem-t}, thus an optimal solution to \eqref{eq:maxmin-fairness-problem-downlink-t} can be obtained using Algorithm~\ref{alg:bisection} on p.~\pageref{alg:bisection}. 
This algorithm contains a bisection search over $t$ where a solution to \eqref{eq:maxmin-fairness-feasibility2} is computed for each candidate value. By including a minimization of the total transmit power in the objective as in \eqref{eq:maxmin-fairness-feasibility2} to improve convergence rate, the subproblem for the problem at hand can be expressed in the second-order cone programming form:
\begin{align} \label{eq:maxmin-fairness-feasibility-downlink}
&\underset{\tilde{\boldsymbol{\rho}}_k\geq \vect{0}_L, \forall k}{\textrm{minimize}} \quad \sum_{k=1}^{K} \| \tilde{\boldsymbol{\rho}}_k\|^2  \\
&\textrm{subject to} \quad   \left \Vert \begin{bmatrix}   \tilde{\vect{C}}_{k1}^{\frac12}\tilde{\boldsymbol{\rho}}_1 \\ \vdots \\ \tilde{\vect{C}}_{kK}^{\frac12}\tilde{\boldsymbol{\rho}}_K \\ \sqrt{\sigmaDL} \end{bmatrix} \right \Vert \leq \sqrt{\frac{1+ t^{\rm candidate}}{ t^{\rm candidate}}}\tilde{\vect{b}}_k^{\Ttran}\tilde{\boldsymbol{\rho}}_k, \quad k=1,\ldots,K \nonumber \\
&\hspace{21mm} \left\| \begin{bmatrix}  \tilde{\rho}_{1l} & \ldots &  \tilde{\rho}_{Kl} \end{bmatrix}  \right\| \leq \sqrt{\rhomax}, \quad l=1,\ldots,L \nonumber \\
&\hspace{21mm} \tilde{\boldsymbol{\rho}}_k \geq \vect{0}_L, \quad k=1,\ldots,K \nonumber
\end{align}
where we implicitly set $\tilde{\rho}_{kl}$ to zero for $k\notin\mathcal{D}_l$ in the second constraint.

\pagebreak 
This reformulation is achieved by noticing that the SINR constraint $\mathacr{SINR}_{k}^{({\rm dl},\dist)}\left(\left\{\tilde{\boldsymbol{\rho}}_i\right\}\right) \geq t^{\rm candidate}$ can be equivalently expressed as
\begin{equation}
\frac{ ( \tilde{\vect{b}}_k^{\Ttran}\tilde{\boldsymbol{\rho}}_k )^2 }{
\left \Vert \begin{bmatrix}   \tilde{\vect{C}}_{k1}^{\frac12}\tilde{\boldsymbol{\rho}}_1 & \ldots & \tilde{\vect{C}}_{kK}^{\frac12}\tilde{\boldsymbol{\rho}}_K & \sqrt{\sigmaDL} \end{bmatrix} \right \Vert^2 - 
(\tilde{\vect{b}}_k^{\Ttran}\tilde{\boldsymbol{\rho}}_k)^2} \geq t^{\rm candidate}
\end{equation}
and then rearranged as in the first constraint of \eqref{eq:maxmin-fairness-feasibility-downlink}. This is a trick that first appeared in \cite{Bengtsson2001a} and then was used for Cell-free Massive MIMO in \cite{Ngo2017b}. By utilizing the subproblem in \eqref{eq:maxmin-fairness-feasibility-downlink}, we obtain Algorithm~\ref{alg:bisection-downlink-case}.

\begin{algorithm}
	\caption{Bisection search algorithm for solving the max-min fairness problem in \eqref{eq:maxmin-fairness-problem-downlink-t}.} \label{alg:bisection-downlink-case}
	\begin{algorithmic}[1]
		\State {\bf Initialization:} Set the solution accuracy $\epsilon>0$
		\State Set the initial lower and upper bounds for the max-min SINR:
		\State $t^{\rm lower}\gets0$
		\State $t^{\rm upper}\gets \underset{k\in \{1,\ldots,K\}}{\min}\quad |\mathcal{M}_k|\rhomax\frac{\tilde{\vect{b}}_k^{\Ttran}\tilde{\vect{b}}_k}{\sigmaDL}$
		\State Initialize solution variables: $\tilde{\boldsymbol{\rho}}_k^{\rm opt}=\vect{0}_L$ for $k=1,\ldots,K$, $t^{\rm opt}=0$
		\While{$t^{\rm upper}-t^{\rm lower}>\epsilon$}
		\State $t^{\rm candidate}\gets \frac{t^{\rm lower}+t^{\rm upper}}{2}$
		\State Solve the following convex problem:
\begin{align} \label{eq:maxmin-fairness-feasibility-downlink2}
&\underset{\tilde{\boldsymbol{\rho}}_k\geq \vect{0}_L, \forall k}{\textrm{minimize}} \quad \sum_{k=1}^{K} \| \tilde{\boldsymbol{\rho}}_k\|^2  \\
&\textrm{subject to} \quad   \left \Vert \begin{bmatrix}   \tilde{\vect{C}}_{k1}^{\frac12}\tilde{\boldsymbol{\rho}}_1 \\ \vdots \\ \tilde{\vect{C}}_{kK}^{\frac12}\tilde{\boldsymbol{\rho}}_K \\ \sqrt{\sigmaDL} \end{bmatrix} \right \Vert \leq \sqrt{\frac{1+ t^{\rm candidate}}{ t^{\rm candidate}}}\tilde{\vect{b}}_k^{\Ttran}\tilde{\boldsymbol{\rho}}_k, \quad k=1,\ldots,K \nonumber \\
&\hspace{21mm} \left\| \begin{bmatrix}  \tilde{\rho}_{1l} & \ldots &  \tilde{\rho}_{Kl} \end{bmatrix}  \right\| \leq \sqrt{\rhomax}, \quad l=1,\ldots,L \nonumber \\
&\hspace{21mm} \tilde{\boldsymbol{\rho}}_k \geq \vect{0}_L, \quad k=1,\ldots,K \nonumber
\end{align}
		\If{\eqref{eq:maxmin-fairness-feasibility-downlink2} is feasible}
		\State $t^{\rm lower}\gets t^{\rm candidate}$
		\State $\tilde{\boldsymbol{\rho}}_k^{\rm opt} \gets \tilde{\boldsymbol{\rho}}_k$, which is the solution to \eqref{eq:maxmin-fairness-feasibility-downlink2} for $k=1,\ldots,K$
		\Else
		\State  $t^{\rm upper}\gets t^{\rm candidate}$
		\EndIf
		\EndWhile
		\State {\bf Output:} Optimal square-roots of the transmit powers $\tilde{\boldsymbol{\rho}}_1^{\rm opt},\ldots,\tilde{\boldsymbol{\rho}}_K^{\rm opt}$, $t^{\rm opt}=\underset{k\in \{1,\ldots,K\}}{\min} \mathacr{SINR}_{k}\left(\left\{\tilde{\boldsymbol{\rho}}_i^{\rm opt}\right\}\right)$
		\State Max-min SE $\frac{\tau_d}{\tau_c}\log_2\left(1+ t^{\rm opt}\right)$
	\end{algorithmic}
\end{algorithm}

It remains to solve the downlink sum SE maximization problem for the distributed operation, which  can be formulated as
\begin{align}\label{eq:sum-SE-maximization-downlink-2}
&\underset{\tilde{\boldsymbol{\rho}}_k \geq \vect{0}_L,\forall k}{\textrm{maximize}} \quad \sum\limits_{k=1}^K  \log_2\left(1+\frac{\left|\tilde{\vect{b}}_k^{\Ttran}\tilde{\boldsymbol{\rho}}_k\right|^2}{\sum\limits_{i=1}^K\tilde{\boldsymbol{\rho}}_i^{\Ttran}\tilde{\vect{C}}_{ki}\tilde{\boldsymbol{\rho}}_i-\left|\tilde{\vect{b}}_k^{\Ttran}\tilde{\boldsymbol{\rho}}_k\right|^2+\sigmaDL}\right) \\
&\textrm{subject to} \quad \sum_{k\in \mathcal{D}_l}\tilde{\rho}_{kl}^2\leq \rhomax, \quad l=1,\ldots,L.   \nonumber
\end{align}
We recognize \eqref{eq:sum-SE-maximization-downlink-2} as an instance of the generic sum SE maximization problem in \eqref{eq:sum-SE-maximization}. Due to Lemma~\ref{theorem:weighted-MMSE-convergence} on p.~\pageref{theorem:weighted-MMSE-convergence}, a local optimal solution to \eqref{eq:sum-SE-maximization-downlink-2} can be obtained by the block coordinate descent algorithm given in Algorithm~\ref{alg:sum-SE-downlink-2}. This algorithm is designed based on a weighted MMSE reformulation of the sum SE maximization problem, where the MSE in \eqref{eq:ek-MSE} becomes
\begin{align} 
&e_k\left(\{\tilde{\boldsymbol{\rho}}_i\}, u_k\right)=u_{k}^2\left(\sum\limits_{i=1}^K\tilde{\boldsymbol{\rho}}_i^{\Ttran}\tilde{\vect{C}}_{ki}\tilde{\boldsymbol{\rho}}_i+\sigmaDL\right) -2u_{k}\tilde{\vect{b}}_k^{\Ttran}\tilde{\boldsymbol{\rho}}_k+1  \label{eq:ek-MSE-downlink-2}
\end{align}
for the considered downlink problem. Step 6 in Algorithm~\ref{alg:sum-SE-downlink-2} contains a subproblem that needs to be solved at each iteration. It is a convex quadratically-constrained quadratic programming problem, which can be solved using any numerical solver for convex problems. See Remark~\ref{remark:convex-solvers} for some suggested tools.

\begin{algorithm}
	\caption{Block coordinate descent algorithm for solving the sum SE maximization problem in \eqref{eq:sum-SE-maximization-downlink-2}.} \label{alg:sum-SE-downlink-2}
	\begin{algorithmic}[1]
		\State {\bf Initialization:} Set the solution accuracy $\epsilon>0$
		\State Set arbitrary feasible initial powers $\{\tilde{\boldsymbol{\rho}}_k\}$
		\While{  $\sum_{k=1}^K\Big(d_{k}e_{k}\left(\{\tilde{\boldsymbol{\rho}}_i\},u_k\right)-\ln\left( d_{k}\right)\Big)$  is either improved more than $\epsilon$ or not improved at all}
		\State $u_{k} \gets \frac{\tilde{\vect{b}}_k^{\Ttran}\tilde{\boldsymbol{\rho}}_k}{\sum\limits_{i=1}^K\tilde{\boldsymbol{\rho}}_i^{\Ttran}\tilde{\vect{C}}_{ki}\tilde{\boldsymbol{\rho}}_i+\sigmaDL}, \quad k=1,\ldots,K$
		\State $d_k \gets \frac{1}{u_{k}^2\left(\sum\limits_{i=1}^K\tilde{\boldsymbol{\rho}}_i^{\Ttran}\tilde{\vect{C}}_{ki}\tilde{\boldsymbol{\rho}}_i+\sigmaDL\right) -2u_{k}\tilde{\vect{b}}_k^{\Ttran}\tilde{\boldsymbol{\rho}}_k+1}, \quad k=1,\ldots,K$
		\State Solve the following problem for the current values of  $u_k$ and $d_k$:
		\begin{align}
		&\underset{\tilde{\boldsymbol{\rho}}_k \geq \vect{0}_L, \forall k}{\textrm{minimize}} \quad \sum_{k=1}^Kd_{k}\left(u_{k}^2\left(\sum\limits_{i=1}^K\tilde{\boldsymbol{\rho}}_i^{\Ttran}\tilde{\vect{C}}_{ki}\tilde{\boldsymbol{\rho}}_i+\sigmaDL\right) -2u_{k}\tilde{\vect{b}}_k^{\Ttran}\tilde{\boldsymbol{\rho}}_k\right) \label{eq:sum-SE-maximization-weighted-MMSE-subproblem-DL-2} \\
		&\hspace{0cm}\textrm{subject to} \quad \sum_{k\in \mathcal{D}_l}\tilde{\rho}_{kl}^2\leq \rhomax, \quad l=1,\ldots,L    \nonumber
		\end{align}
		\State Update $\{\tilde{\boldsymbol{\rho}}_k\}$ by the obtained solution to \eqref{eq:sum-SE-maximization-weighted-MMSE-subproblem-DL-2}.
		\EndWhile
		\State {\bf Output:} Optimal square-roots of the transmit powers $\tilde{\boldsymbol{\rho}}_1,\ldots,\tilde{\boldsymbol{\rho}}_K$
		\State Sum SE $\frac{\tau_d}{\tau_c}\sum_{k=1}^K\log_2\left(d_k\right)$
	\end{algorithmic}
\end{algorithm}

\begin{remark}[Other utility functions and scheduling]
There are other utility functions than max-min fairness and sum SE that can be maximized. For example, one can introduce user-specific weights in the utility functions to get the weighted max-min fairness and weighted sum SE utilities. The algorithms presented in this monograph can be extended to handle these cases as well.
Moreover, there are utility functions that have a totally different structure, such as the geometric mean of the SEs (also known as proportional fairness) and the harmonic mean of the SEs \cite{Bjornson2013d}, \cite{Farooq2020a}, \cite{Luo2008a}.
However, apart from limiting the length of this monograph, there are two strong reasons for why we only covered the max-min fairness and sum SE utilities.
The first reason is that the early works  \cite{Nayebi2017a}, \cite{Ngo2017b} on Cell-free Massive MIMO emphasized the importance of the max-min fairness utility, and the resulting ability of cell-free networks to deliver uniformly good service over large coverage areas. This is why we considered that metric.
The second reason is that the weighted sum SE maximization problem is of main importance in practical networks, where the UEs have data queues of limited lengths and packets are arriving at random to these queues.
One can then formulate dynamic resource allocation problems that take the queue dynamics into account \cite{Georgiadis2006a}, \cite{Shirani2010a} and irrespective of what the long-term utility function is, the problems that need to be solved regularly are weighted sum SE maximization problem where the weights are computed based on the lengths of the data queues and the long-term utility. The first step into dynamic resource allocation for cell-free networks was taken in \cite{Chen2020b}, which only considers the uplink. There appear to be many open research questions related to the interplay between scheduling and power allocation in Cell-free Massive MIMO.
\end{remark}

\section{Scalable Distributed Power Optimization}
\label{sec:scalable-power-allocation}

The algorithms presented in Section~\ref{sec:optimized-power-allocation} jointly optimize the transmit powers for all UEs to maximize a network-wide utility function.
Since there are $K$ SINRs to optimize and at least $K$ optimization variables, it is unavoidable that the computational complexity becomes unscalable as $K \to \infty$.
In fact, the complexity of any network-wide power allocation optimization grows unboundedly with $K$, which makes them unscalable according to Definition~\ref{def:scalable} on p.~\pageref{def:scalable}. Hence, to obtain a scalable power allocation scheme that is practically implementable in large cell-free networks, we need to devise distributed and heuristic schemes, where each AP or UE makes a local decision with limited involvement of the other APs/UEs. For example, a device can make decisions based on the CSI that it can acquire locally.
 
It is easy to design a scalable scheme; for example, we can let every UE transmit with full power in the uplink and let every AP allocate its power equally among the UEs it serves in the downlink.
The challenge is to identify heuristic schemes that also provide reasonably good SEs, according to network-wide utility functions. This basically boils down to numerically evaluating different heuristic schemes against the optimized baselines of the kind presented in Section~\ref{sec:optimized-power-allocation}. In this section, we will present some selected recent approaches to distributed power allocation. We refer to \cite{Bjornson2011a}, \cite{Bjornson2010c}, \cite{Interdonato2019a}, \cite{Nayebi2017a}, \cite[Sec.~3.4.4]{Bjornson2013d}, for further examples and stress that further research is needed in this direction.

\enlargethispage{\baselineskip} 

\vspace{-2mm}
\subsection{Scalable Uplink Power Control}

Fractional power control is a classical heuristic in the uplink of multi-user systems, particularly in cellular networks \cite{Simonsson2008a}, \cite{Whitehead1993a}. It builds on the principle of using power control to compensate for a fraction of the pathloss differences within each cell. Power control is a balancing act between utilizing the pathloss differences to provide the cell-center UEs with high SEs (at the expense of causing interference to other UEs and cells) and compensating for the pathloss differences to improve the SE for cell-edge UEs (at the expense of forcing cell-center UEs to reduce their power and thereby their SE).
The characteristic feature of fractional power control is that each UE computes its transmit power as a function of only its channel gain to the serving AP, but since the same function is utilized by all UEs, the interference statistics can also be implicitly tuned by an appropriate function selection \cite{Whitehead1993a}.

The power control situation is different in cell-free networks compared to cellular networks since each UE has multiple serving APs. A UE can have a strong channel to one serving AP but a weak channel to another serving AP, while another UE experiences the opposite situation. \linebreak \emph{Should any of the UEs reduce its power in this situation to limit~the~mutual interference and, if yes, how much?} There are no simple~answers~but there are algorithms that work fairly well and these are essentially adding up the channel gains from the serving APs as if there was only one AP.

A fractional power control algorithm for cell-free networks was proposed and motivated in \cite{Nikbakht2019a}, \cite{Nikbakht2020a} for a setup where all APs serve all UEs using MR combining, but it can be easily adapted to the case when each AP serves a subset of the UEs using an arbitrary combining scheme \cite{Chen2020a}. UE $k$ selects its uplink transmit power as
\enlargethispage{\baselineskip} 
\begin{align}\label{eq:fractional_power_UL}
p_k= \pmax\frac{\left(\sum\limits_{l\in\mathcal{M}_k}\beta_{kl}\right)^{\upsilon}}{\underset{i\in\{1,\ldots,K\}}{\max}\left(\sum\limits_{l\in\mathcal{M}_i}\beta_{il}\right)^{\upsilon}}
\end{align}
where the exponent $\upsilon$ dictates the power control behavior. The denominator in \eqref{eq:fractional_power_UL} makes sure that $p_k \in [0,\pmax]$.
Note that $\sum_{l\in\mathcal{M}_i}\beta_{il}$ is the total channel gain from UE $i$ to the APs that serve it.
If $\upsilon=0$, all UEs transmit with maximum power as assumed in the simulations~in Section~\ref{sec:uplink-performance-comparison} on p.~\pageref{sec:uplink-performance-comparison}. If $\upsilon=-1$, each UE is fully~compensating~for~the variations in the total channel gain among the UEs so that $p_i\sum_{l\in\mathcal{M}_i}\beta_{il}$ becomes the same for  $i=1,\ldots,K$. This can be viewed as an approximation of max-min fairness power control. Fractional power control traditionally consists of finding a tradeoff between these extremes by selecting $\upsilon \in [-1,0]$ so this is the range considered in \cite{Nikbakht2019a}, \cite{Nikbakht2020a}.~To~identify~an approximation of sum SE maximization, we also need to consider $\upsilon > 0$ so more power is used by the UEs with good channel conditions. Another possible generalization is to replace $\beta_{kl}$ with $\beta_{kl}-\tr(\vect{C}_{kl})/N$ to take the performance loss due to channel estimation errors into account.

\enlargethispage{\baselineskip}

One problem with the fractional power control scheme described above is that it is not scalable since we need to compute the maximum of $K$ terms in the denominator of \eqref{eq:fractional_power_UL}. 
However, we can easily modify \eqref{eq:fractional_power_UL} such that the number of the terms in the denominator does not grow with $K$ and thereby achieve scalability.
In cellular networks, this is normally done by replacing the denominator with a constant representing the worst-case cell-edge conditions (e.g., what is the lowest channel gain that a connected UE can have). This makes good sense in symmetric cellular deployments, where this value is roughly the same in every cell, while it can be an overly pessimistic approximation in asymmetric cell-free networks.
 Inspired by the scalable combining schemes from Section~\ref{subsec:scalable-combining} on p.~\pageref{subsec:scalable-combining}, one option is to only compute the maximum with respect to the indices of the UEs that are partially served by the same APs. This makes intuitive sense in large networks where each UE is mostly affected by the closest UEs.
Using the set $\mathcal{S}_k$ defined in \eqref{eq:S_k}, for which $|\mathcal{S}_k|$ does not grow unboundedly when $K$ goes to infinity, a scalable version of the fractional power control in \eqref{eq:fractional_power_UL} is obtained by selecting the transmit power of UE $k$ as
\begin{align}\label{eq:fractional_power_UL-scalable}
p_k= \pmax\frac{\left(\sum\limits_{l\in\mathcal{M}_k}\beta_{kl}\right)^{\upsilon}}{\underset{i\in\mathcal{S}_k}{\max}\left(\sum\limits_{l\in\mathcal{M}_i}\beta_{il}\right)^{\upsilon}}.
\end{align}
Depending on the value of $\upsilon$, we can heuristically aim at optimizing different utility functions. To find the desired value, the network designer can simulate the CDF of the per-UE SE for different values of $\upsilon$ and then determine which value gives the most desirable tradeoff between high fairness and high sum SE.

\begin{remark}[Minimizing the signal-to-interference ratio]
Fractional power control is a heuristic scheme in the sense that it is not directly connected to maximizing a utility function of the individual SEs. However, under certain conditions, such power control can be shown to minimize the variations in the \gls{SIR} experienced by UEs at random locations. For example, \cite{Whitehead1993a} proved that if the channel gains are Gaussian distributed in the decibel scale and have equal variance, then fractional power control with $\upsilon=-1/2$  minimizes the variance of the SIR in a two-UE setup. A generalization of this result to cell-free networks is provided in \cite{Nikbakht2020a} and serves as a motivation for why fractional power control makes practical sense to shape the CDF curve of the SIR experienced at random locations, even if the connections to the CDF of the per-UE SE or to utility functions such as the sum SE and max-min fairness are heuristic.
\end{remark}

\subsection{Scalable Centralized Downlink Power Allocation}

In the centralized downlink operation, the transmit powers that different APs are transmitting with to a given UE are coupled through the centralized precoding vectors
\begin{equation} \label{eq:normalized-precoding-vector}
\frac{\bar{\vect{w}}_k}{\sqrt{\mathbb{E} \left\{ \left\| \bar{\vect{w}}_k \right\|^2 \right\} }} = \begin{bmatrix} \bar{\vect{w}}_{k1}^{\prime} \\ \vdots \\  \bar{\vect{w}}_{kL}^{\prime} \end{bmatrix}, 
\quad k=1,\ldots,K.
\end{equation}
If one AP increases its power to UE $k$, then all the other APs that serve this UE must do the same to keep the same direction of the precoding vector. Otherwise, the ability for the APs to cancel each others' interference (as illustrated in Figure~\ref{figure:distributed-beamforming} on p.~\pageref{figure:distributed-beamforming}) is lost. This ability is one of the key benefits of the centralization operation and should not be lost if the performance gain over the distributed operation should be retained.
Recall from \eqref{eq:power-constraint-centralized} that the power constraint at AP $l$ is
\begin{equation} \label{eq:power-constraint-centralized2}
\sum_{k\in \mathcal{D}_l}\rho_k\mathbb{E} \left\{ \left\| \bar{\vect{w}}_{kl}^{\prime} \right\|^2 \right\} \leq \rhomax.
\end{equation}
Since $\sum_{l\in \mathcal{M}_k} \mathbb{E} \{ \| \bar{\vect{w}}_{kl}^{\prime} \|^2 \} = 1$, the fraction $\mathbb{E} \{ \| \bar{\vect{w}}_{kl}^{\prime} \|^2 \}  \in [0,1]$ that an arbitrary AP $l \in \mathcal{M}_k$ allocates to UE $k$ is generally much smaller than $\rho_k$. 
This must be accounted for so that all the serving APs can select a common power value that satisfies all of their power constraints.
A simple heuristic solution is to assign the same power
\begin{equation}\label{eq:equal_power_allocation_DL}
\rho_k = \frac{\rhomax}{\tau_p}
\end{equation}
to all UEs, as proposed in \cite{Bjornson2020a}.
 In this case, the normalized precoding vector in \eqref{eq:normalized-precoding-vector}
 determines how this power is distributed between the APs, and all APs are guaranteed to satisfy their power constraints since they serve at most $\tau_p$ UEs and will at most allocate $\rhomax/\tau_p$ to each of them. Since the computation of \eqref{eq:equal_power_allocation_DL} is independent of $K$, this network-wide equal power allocation scheme is scalable.

A more general heuristic can be obtained by taking inspiration from the fractional uplink power control in  \eqref{eq:fractional_power_UL-scalable} by selecting the downlink power allocation coefficients proportionally to the total channel gain from the serving APs: 
\begin{align}\label{eq:fractional_power_centralized_DL-scalable}
\rho_k\propto \left(\sum\limits_{l\in\mathcal{M}_k}\beta_{kl}\right)^{\upsilon}.
\end{align}
The exponent $\upsilon \in [-1,1]$ in \eqref{eq:fractional_power_centralized_DL-scalable} determines the power allocation behavior. The proportionality constant must be selected so that all the power constraints are satisfied, which can be done as follows:
\begin{align}\label{eq:fractional_power_centralized_DL-scalable1b}
\rho_k=\rhomax \frac{\left(\sum\limits_{l\in\mathcal{M}_k}\beta_{kl}\right)^{\upsilon}}{\underset{\ell\in \mathcal{M}_k}{\max} \  \sum\limits_{i\in \mathcal{D}_{\ell}}   \left(\sum\limits_{l\in\mathcal{M}_i}\beta_{il}\right)^{\upsilon}}.
\end{align}
If $\upsilon=0$ and $|\mathcal{D}_{\ell}| = \tau_p$, then \eqref{eq:fractional_power_centralized_DL-scalable1b} turns into the network-wide equal power allocation scheme in \eqref{eq:equal_power_allocation_DL}.
 More power is allocated to the UEs with higher total channel gains if $\upsilon>0$, while it is the other way around if $\upsilon<0$. 
 The former is resembling the main characteristics of power allocation for maximum sum SE, while the latter is resembling max-min fairness.
 
The denominator in \eqref{eq:fractional_power_centralized_DL-scalable1b} makes sure that all the power constraints are satisfied since
$\sum_{i \in \mathcal{D}_l} \rho_i \leq \rhomax$ for $l=1,\ldots,L$.
However, this is a conservative design for the worst-case situation that each UE is only served by only one AP, which is generally not the case in cell-free networks. The consequence is that all APs will operate far below their maximum power.
Suppose we know the largest fraction of $\rho_k$ that any of the serving APs can be assigned to transmit:
\begin{align}
\omega_k = \underset{\ell\in\mathcal{M}_k}{\max} \ \mathbb{E} \left\{ \left\| \bar{\vect{w}}_{k\ell}^{\prime} \right\|^2 \right\}.
\end{align} 
We can then use it as an additional tuning parameter and change \eqref{eq:fractional_power_centralized_DL-scalable} into
\begin{align}\label{eq:fractional_power_centralized_DL-scalable2}
\rho_k\propto \frac{\left(\sum\limits_{l\in\mathcal{M}_k}\beta_{kl}\right)^{\upsilon}}{\omega_k^{\kappa}}
\end{align}
where the exponent $0\leq\kappa\leq1$ is a parameter that reshapes the ratio of power allocation between different UEs. The motivation for such a scaling factor is as follows. When the power allocation in \eqref{eq:fractional_power_centralized_DL-scalable1b} is used and $\omega_k \approx 1/| \mathcal{M}_k |$ (i.e., the smallest value it can take), then all the serving APs will approximately transmit with power $\rho_k /| \mathcal{M}_k |$ to UE $k$ but manage their power constraints as if they transmitted with power $\rho_k$.
To prevent this situation to some extent, we can scale the original power coefficient of each UE inversely proportional to $\omega_k^{\kappa}$ as in \eqref{eq:fractional_power_centralized_DL-scalable2}. The intuition is that each AP should instead 
manage its power constraint as if it transmits with power $\rho_k \omega_k^\kappa$ (note that $\rho_k \omega_k^\kappa \geq \rho_k \omega_k$).
To satisfy the power constraint at each AP, we select the proportionality constant in \eqref{eq:fractional_power_centralized_DL-scalable2} to obtain
\begin{align}\label{eq:fractional_power_centralized_DL-scalable3}
\rho_k=\rhomax \frac{\left(\sum\limits_{l\in\mathcal{M}_k}\beta_{kl}\right)^{\upsilon}\omega_k^{-\kappa}}{\underset{\ell\in \mathcal{M}_k}{\max} \  \sum\limits_{i\in \mathcal{D}_{\ell}}   \left(\sum\limits_{l\in\mathcal{M}_i}\beta_{il}\right)^{\upsilon}\omega_i^{1-\kappa}}.
\end{align}
The following chain of inequalities demonstrates that the power constraint is satisfied at each AP $l^{\prime}$:
\begin{align}\label{eq:fractional_power_centralized_DL-scalable3_power}
\sum_{k\in\mathcal{D}_{l^{\prime}}}\rho_k\mathbb{E} \left\{ \left\| \bar{\vect{w}}_{kl^{\prime}}^{\prime} \right\|^2 \right\}&\leq \sum_{k\in\mathcal{D}_{l^{\prime}}}\rho_k\omega_k =\sum\limits_{k\in\mathcal{D}_{l^{\prime}}} \frac{\rhomax\left(\sum\limits_{l\in\mathcal{M}_k}\beta_{kl}\right)^{\upsilon}\omega_k^{1-\kappa}}{\underset{\ell\in \mathcal{M}_k}{\max} \  \sum\limits_{i\in \mathcal{D}_{\ell}}   \left(\sum\limits_{l\in\mathcal{M}_i}\beta_{il}\right)^{\upsilon}\omega_i^{1-\kappa}} \nonumber\\
& \leq  \rhomax\frac{\sum\limits_{k\in\mathcal{D}_{l^{\prime}}}\left(\sum\limits_{l\in\mathcal{M}_k}\beta_{kl}\right)^{\upsilon}\omega_k^{1-\kappa}}{\underset{\ell\in \bigcap\limits_{k\in \mathcal{D}_{l^{\prime}}}\!\!\!\mathcal{M}_k}{\max} \  \sum\limits_{i\in \mathcal{D}_{\ell}}   \left(\sum\limits_{l\in\mathcal{M}_i}\beta_{il}\right)^{\upsilon}\omega_i^{1-\kappa}} \nonumber\\
&\leq  \rhomax
\end{align}
where the last inequality follows from the fact the set $\bigcap_{k\in \mathcal{D}_{l^{\prime}}}\!\!\mathcal{M}_k$ includes $l^{\prime}$, thus the numerator is smaller or equal to the denominator.

We stress that \eqref{eq:fractional_power_centralized_DL-scalable3} is a scalable power allocation scheme since the computational complexities associated with the terms in the numerator and denominator do not grow with $K$. When evaluating the downlink performance in the running example in Section~\ref{sec:downlink-performance-comparison} on p.~\pageref{sec:downlink-performance-comparison}, we used \eqref{eq:fractional_power_centralized_DL-scalable3} with $\upsilon = -0.5$ and $\kappa=0.5$. In Section~\ref{sec:comparison-power-optimzation}, we will investigate numerically how to select the parameters $\upsilon \in [-1,1]$ and $\kappa \in [0,1]$.

\subsection{Scalable Distributed Downlink Power Allocation}

In the distributed downlink operation, the power allocation contains multiple power coefficients per UE. On the one hand, this increases the complexity compared to uplink power control, but on the other hand, it becomes easier to find suitable tradeoffs. If a UE is served by one AP with a strong channel and by one AP with a weak channel, then it can be assigned widely different powers from these APs in the downlink, while it is less obvious if it should transmit with high or low power in the uplink.
From a UE's perspective, it is preferable to be allocated more downlink power from APs with strong channels than APs with weak channels, since this leads to a higher SNR than the opposite allocation.
From an AP's perspective, it is natural to prioritize transmitting to UEs that it has good channels to, compared to UEs with weak channels, because the opposite strategy would cause large interference to the UEs with good channels. 
These basic principles can be utilized to determine a distributed heuristic power allocation scheme where each AP determines its own transmit power allocation independently of the other APs.
Since the number of UEs that are served by a particular AP does not grow with $K$ for a scalable cell-free network (e.g., it serves at most $\tau_p$ UEs when we use Algorithm~\ref{alg:basic-pilot-assignment} on p.~\pageref{alg:basic-pilot-assignment}), this kind of distributed power allocation is scalable. 

A distributed power allocation scheme was proposed in  \cite{Interdonato2019a} for a system with $N=1$. We will present a generalization for arbitrary $N$ using the notation of this monograph. With this power allocation policy, AP $l$ selects its downlink powers $\rho_{1l}, \ldots, \rho_{Kl}$ as
\begin{equation} \label{eq:fractional_power_DL}
\rho_{kl} = \begin{cases} \rhomax \frac{f\left(\mathcal{G}_{kl}\right)}{\sum_{i \in \mathcal{D}_l} f\left(\mathcal{G}_{il}\right)} &  k \in \mathcal{D}_l\\
0 &  k \notin \mathcal{D}_l
\end{cases}
\end{equation}
where $f(\cdot)$ is a pre-determined function and the input is given by the channel statistics which, for the channel between UE $i$ and AP $l$, is
\begin{equation}
\mathcal{G}_{il} = \{ \vect{R}_{il}, \vect{C}_{il} \}.
\end{equation}
The intuition is that $f\left(\mathcal{G}_{il}\right)$ should somehow determine the relative importance of transmitting to UE $i$, while the term in the denominator of \eqref{eq:fractional_power_DL} normalizes the transmit powers so that the AP is always using its maximum power: $\sum_{k \in \mathcal{D}_l} \rho_{kl} = \rhomax$.

\enlargethispage{-\baselineskip}

Many different power allocation policies can be formulated according to \eqref{eq:fractional_power_DL}. If we select $f\left(\mathcal{G}_{il}\right)=1$, every UE is equally important and we obtain per-AP equal power allocation with $\rho_{kl}=\frac{\rhomax}{|\mathcal{D}_l|}$.
Another option is $f\left(\mathcal{G}_{il}\right) = (\beta_{il})^{\upsilon}$ \cite{Interdonato2019a}, which turns \eqref{eq:fractional_power_DL} into
\begin{equation} \label{eq:fractional_power_DL-2}
\rho_{kl} = \begin{cases} \rhomax \frac{ (\beta_{kl})^{\upsilon}}{\sum_{i \in \mathcal{D}_l} (\beta_{il})^{\upsilon}} &  k \in \mathcal{D}_l\\
0 &  k \notin \mathcal{D}_l
\end{cases}
\end{equation}
where the exponent $\upsilon$ dictates the power allocation behavior. If $\upsilon=0$, we obtain the per-AP equal power allocation. If $\upsilon=1$, we give~higher emphasis to the UEs according to their respective channel gains. This leads to allocating more power to the UEs with better channel qualities. If $\upsilon=-1$, the power allocation is inversely proportional~to~the~channel gain, so that each of the served UEs will obtain the same received~power.
This might seem like a good feature from a fairness perspective, but~it is generally not since the UEs with good channel conditions will then~be subject to high interference.
A more opportunistic power allocation~with $\upsilon \in [0,1]$ seems to be preferred to make efficient use of the fact that~each UE typically has a good channel from some APs and a worse channel from other APs \cite{Interdonato2019a}. 
Hence, even if  \eqref{eq:fractional_power_DL-2} has an expression that~resembles the uplink fractional power control expression in \eqref{eq:fractional_power_UL}, the~intended operation is very different: we want to emphasize SNR differences instead of mitigating them.
When evaluating the downlink performance~in~the running example in Section~\ref{sec:downlink-performance-comparison} on p.~\pageref{sec:downlink-performance-comparison}, we used \eqref{eq:fractional_power_DL-2} with $\upsilon = 0.5$.

Power allocation policies of the kind in \eqref{eq:fractional_power_DL} are scalable by design since only the UEs in $\mathcal{D}_l$ are considered. Another good feature is that every UE will be allocated a non-zero power from all its serving APs, leading to a non-zero SE. Estimation errors can be taken into account by selecting $f\left(\mathcal{G}_{il}\right) = (\beta_{il}-\tr(\vect{C}_{il})/N)^{\upsilon}$ instead of  $f\left(\mathcal{G}_{il}\right) = (\beta_{il})^{\upsilon}$, with the consequence of moving power from UEs with lower channel quality to UEs with better channel quality.
For any choice of the function $f(\cdot)$, we can identify the desired value of $\upsilon$ in a given network by simulating CDF curves of the per-UE SE for UEs at different locations. By plotting different curves for different values of $\upsilon$, we can determine which value gives the most desirable tradeoff between high fairness and high~sum SE.

\section{Comparison of Power Optimization Schemes}
\label{sec:comparison-power-optimzation}

We will now compare the scalable power control/allocation schemes from Section~\ref{sec:scalable-power-allocation} with the optimized benchmarks from Section~\ref{sec:optimized-power-allocation}.
To this end, we continue the running example that was defined in Section~\ref{sec:running-example} on p.~\pageref{sec:running-example}. The specific system parameters are: $L=100$ APs, $N=4$ antennas per AP, and $K=40$ UEs. We consider the user-centric implementation based on the DCC formation algorithm presented in Section~\ref{sec:pilot-assignment} on p.~\pageref{sec:pilot-assignment} with different scalable processing schemes. The channels follow spatially correlated Rayleigh fading with ASDs $\sigma_{\varphi}=\sigma_{\theta}=15^{\circ}$.

When we compare the performance of different schemes, we will show CDF curves of the SEs that the UEs achieve at different random locations. When we did the same thing in previous chapters, the curves achieved by the different schemes that we compared were often non-intersecting so that we could easily conclude that the rightmost curve was the preferable one. 
To achieve an efficient scalable implementation, we would pick the rightmost scheme among the scalable alternatives. The situation will be different when we compare different power optimization schemes in this section because we will get curves that intersect, which demonstrates that different schemes are preferable for different types of UEs. Note that the lower tail of a CDF curve represents the performance achieved by the UEs with the worst channel conditions, while the upper tail represents the performance achieved by the UEs with the best channel conditions. One way to measure fairness is by looking at the steepness of the CDF curve; if the curve is steep, then the SE difference between UEs with the worst and best channel conditions is small. However, one should also bear in mind that a network that gives zero SE to everyone features great fairness (everyone gets the same performance) but it is not practically desirable. Hence, the network designer will eventually have to select a scheme that provides the right tradeoff between fairness and sum SE.

\subsection{Comparison of Uplink Power Control Schemes}

We first consider the uplink and compare both optimized and scalable schemes.
As optimized power control schemes, we consider the max-min fairness and sum SE maximization algorithms from Section~\ref{sec:optimized-power-allocation-uplink}, namely Algorithm~\ref{alg:fixed-point-uplink} and Algorithm~\ref{alg:sum-SE-uplink}, respectively. The corresponding results are respectively denoted by ``\gls{MMF}'' (max-min fairness) and ``SumSE''. In addition, we consider three heuristic but scalable schemes. The first one is that all UEs transmit with full power, which is denoted by ``Full''. Moreover, we consider the fractional power control scheme in \eqref{eq:fractional_power_UL-scalable} with exponents $\upsilon=0.5$ or $\upsilon=-0.5$ and denote it  by ``\gls{FPC}''. Note that the case with $\upsilon=0.5$ allows UEs with better channel conditions to transmit with higher power whereas the case with $\upsilon=-0.5$ does the opposite.

\label{sec:performance-comparison-power-allocation}
\begin{figure}[p!]
	\begin{subfigure}[b]{\columnwidth} \centering 
		\begin{overpic}[width=0.88\columnwidth,tics=10]{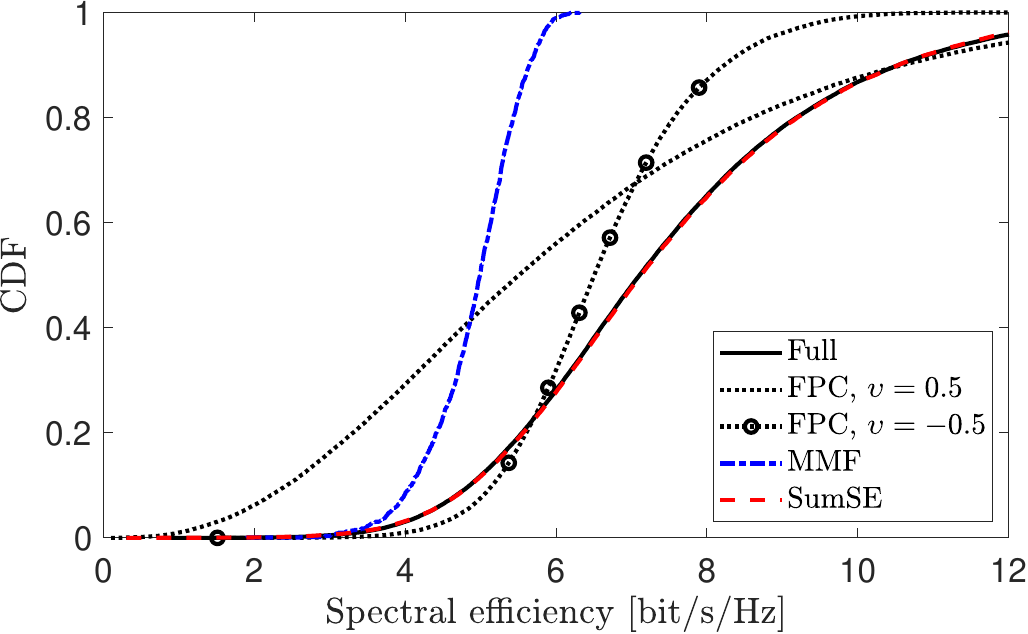}
		\end{overpic} \label{figureSec7_1a}
		\caption{Centralized uplink operation with P-MMSE combining.} 
		\vspace{4mm}
	\end{subfigure} 
	\begin{subfigure}[b]{\columnwidth} \centering 
		\begin{overpic}[width=0.88\columnwidth,tics=10]{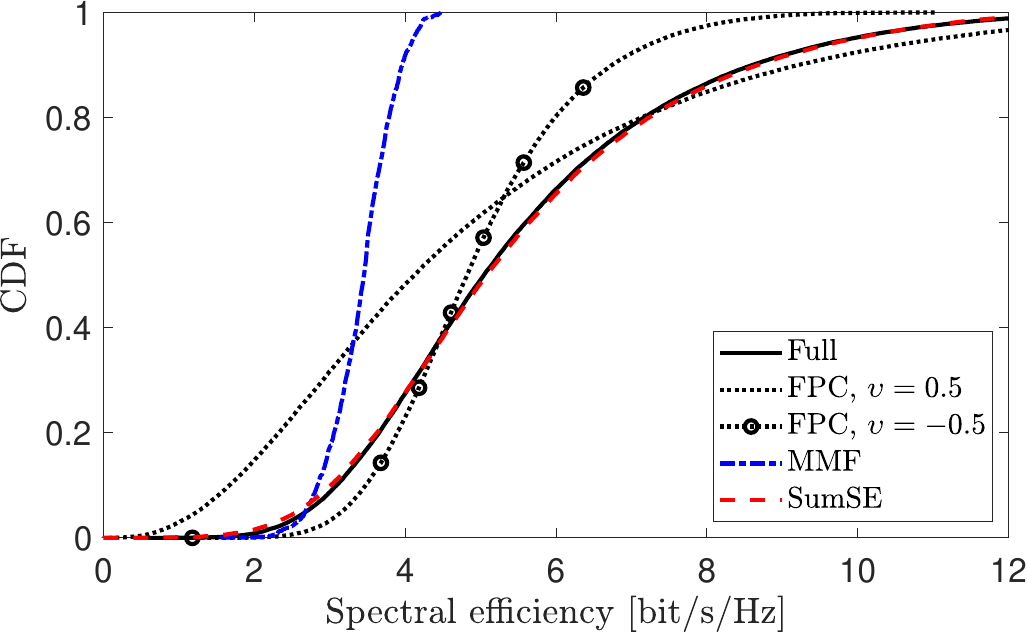}
		\end{overpic} \label{figureSec7_1b}
		\caption{Distributed uplink operation with n-opt LSFD and LP-MMSE combining.} 
	\end{subfigure} 
	\caption{CDF of the uplink SE per UE in the centralized and distributed operation with different optimized and heuristic power control methods. We consider $L=100$, $N=4$, $K=40$, $\tau_p=10$, and spatially correlated Rayleigh  fading with ASD $\sigma_{\varphi}=\sigma_{\theta}=15^{\circ}$.}\label{figureSec7_1}
\end{figure}

Figure~\ref{figureSec7_1} compares the uplink SE of the power control methods described above. We consider a scalable centralized operation with P-MMSE combining in Figure~\ref{figureSec7_1}(a) and a scalable distributed operation with n-opt LSFD and LP-MMSE combining in Figure~\ref{figureSec7_1}(b). In both cases, the sum SE maximizing power control provides the same performance as with full-power transmission. In fact, when the initial point to Algorithm~\ref{alg:sum-SE-uplink} is that everyone transmits with maximum power, then the objective function is not improved in the next iteration. Although Algorithm~\ref{alg:sum-SE-uplink} attains only a local optimum, we tried several different random initializations and observed that the algorithm always converged to the solution where everyone transmits with full power. 
This does not mean that the full-power transmission will maximize the sum SE in any Cell-free Massive MIMO system, but it demonstrates that the considered network setup is very capable of suppressing interference so that everyone can transmit at maximum power. If we, hypothetically speaking, would add a UE with extremely bad channel conditions to the network, then the sum SE maximization might allocate zero power to it since this utility function does not provide any performance guarantees.

When it comes to max-min fairness power control, we notice~that~the lower tail of the corresponding CDF curve begins at the highest number among all the studied schemes. This is expected since the performance of the most unfortunate UE in the entire network is maximized. However, by zooming in at the lower tail of the CDF curves, we can see that~max-min fairness only results in a marginal improvement compared to the sum SE curve, and the price to pay is substantially smaller~SEs~for~the vast majority of UEs.
In fact, the fractional power control~method~with $\upsilon=-0.5$ achieves nearly the same SE for the most unfortunate UEs, while not sacrificing the SE for the other UEs to the same extent.~When we switch to fractional power control with $\upsilon=0.5$, we notice that~its~behavior is similar to sum SE maximization for the UEs with good channel conditions, but not as good for the UEs with weaker channel conditions.

In summary, maximizing the sum SE will result in a CDF curve that is almost entirely to the right of the competing schemes. Hence, it is generally the preferred choice and this power optimization can be implemented in a scalable manner by letting all UEs transmit with full power. This is why we utilized this scheme for performance comparisons in Section~\ref{sec:uplink-performance-comparison} on p.~\pageref{sec:uplink-performance-comparison}.
If we want a higher level of fairness, in terms of increasing the SE for the most unfortunate UEs, then a fractional power control method with $\upsilon=-0.5$ is a good scalable option. It must be thus clear that the max-min fairness problem focuses on an extreme type of fairness that only helps one UE in the network, at the expense of everyone else. This effect is particularly evident in a large cell-free network where most UEs are barely affecting the most unfortunate UE but anyway are forced to cut down on their transmit powers.

\subsection{Comparison of Downlink Power Allocation Schemes}

We continue with the comparison of optimized and scalable power allocation schemes by considering the downlink. We begin by studying the centralized operation where we use Algorithm~\ref{alg:fixed-point-downlink} for max-min fairness power allocation and Algorithm~\ref{alg:sum-SE-downlink} for sum SE maximization. We also consider the network-wide equal power allocation scheme in \eqref{eq:equal_power_allocation_DL} and the fractional power allocation method in \eqref{eq:fractional_power_centralized_DL-scalable3} as scalable benchmarks. Recall that the latter method was considered in the simulations in Section~\ref{sec:downlink-performance-comparison} on p.~\pageref{sec:downlink-performance-comparison}. These methods are denoted by ``MMF'', ``SumSE'', ``Equal'', and ``\gls{FPA}'' respectively.

\begin{figure}[p!]
	\begin{subfigure}[b]{\columnwidth} \centering 
		\begin{overpic}[width=0.88\columnwidth,tics=10]{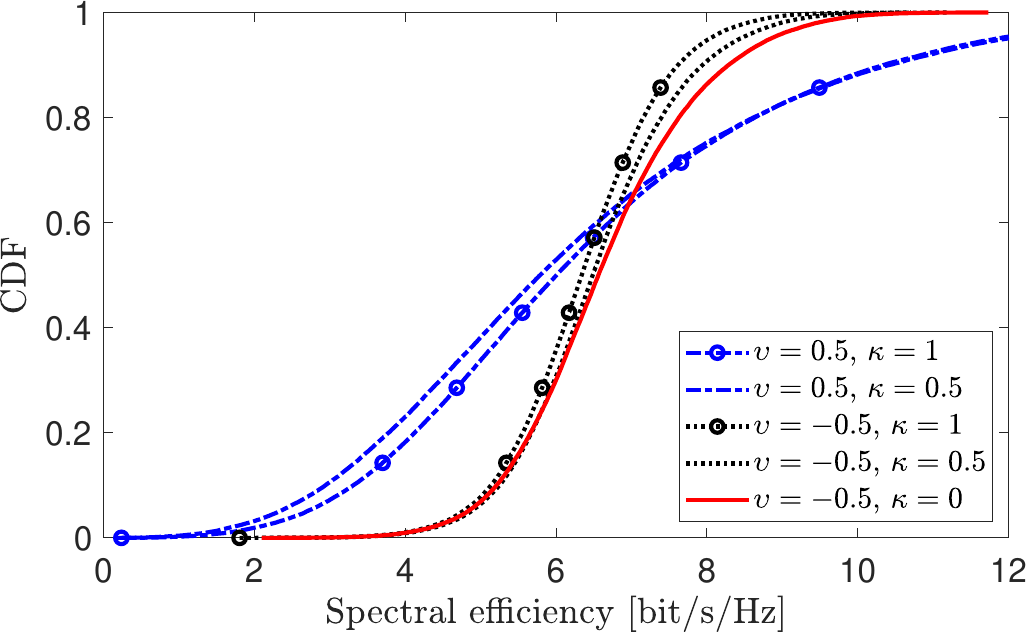}
		\end{overpic} \label{figureSec7_2a}
		\caption{Fractional power allocation with different parameter values.} 
		\vspace{4mm}
	\end{subfigure} 
	\begin{subfigure}[b]{\columnwidth} \centering 
		\begin{overpic}[width=0.88\columnwidth,tics=10]{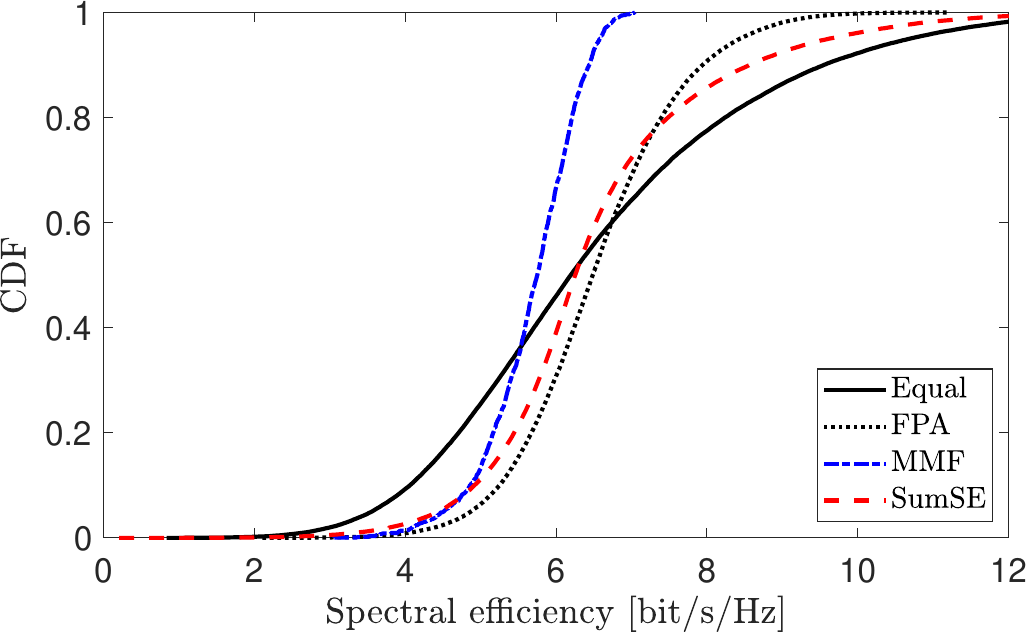}
			\put(24,15.7){90\% likely}
			\put(13,13.7){$- - - - - - - - - - - - $}
		\end{overpic} \label{figureSec7_2b}
		\caption{Centralized downlink operation with several power allocation schemes.} 
	\end{subfigure} 
	\caption{CDF of the downlink SE per UE in the centralized operation with different optimized and heuristic power allocation methods. We consider P-MMSE precoding, $L=100$, $N=4$, $K=40$, $\tau_p=10$, and spatially correlated Rayleigh  fading with ASD $\sigma_{\varphi}=\sigma_{\theta}=15^{\circ}$. }\label{figureSec7_2}
\end{figure}

Figure~\ref{figureSec7_2} shows the CDFs of the SE per UE for the P-MMSE~precoding scheme, which is scalable. In Figure~\ref{figureSec7_2}(a), we consider the effect of tuning parameters in the FPA scheme, i.e., the exponents $\upsilon$ and $\kappa$ in \eqref{eq:fractional_power_centralized_DL-scalable3}. When $\upsilon$ is positive, more emphasis is put on the UEs with higher total channel gains whereas it is the other way around for a negative $\upsilon$. Hence, $\upsilon<0$ mimics the max-min fairness optimization. There is an additional tuning parameter $\kappa$ that reshapes the power allocation to account for the unequal contributions that the serving APs provide to the centralized precoding vector. From the figure, we notice that the effect of $\upsilon$ on the SE spread is more substantial compared to $\kappa$. When $\upsilon=0.5$, the more fortunate UEs attain higher SE compared to the case of $\upsilon=-0.5$. On the other hand, $\upsilon=-0.5$ is more preferable to provide more uniform service quality. Another observation is that the selection of $\kappa$ affects the SE differently for the different values of $\upsilon$. When $\upsilon=0.5$, it is better to use a higher $\kappa$ while the reverse is true when $\upsilon=-0.5$.

\enlargethispage{\baselineskip} 
 
In Figure~\ref{figureSec7_2}(b), we compare the CDFs of the SE per UE for the optimized and scalable power allocation methods described above. We use the fractional power allocation from Figure~\ref{figureSec7_2}(a) with $\upsilon=-0.5$ and $\kappa=0.5$, as we did in the results of Section~\ref{sec:downlink-performance-comparison} on p.~\pageref{sec:downlink-performance-comparison}. This is an option that gives the best performance in the lower tail.
We notice that the max-min fairness-based power allocation begins at the largest value and has the smallest difference between its lower and upper tails, which are two properties that are expected from this type of power allocation. 
However, as in the uplink case studied in Figure~\ref{figureSec7_1}, this type of fairness results in a significant performance drop for all UEs except the most unfortunate ones.
In fact, with sum SE maximization or fractional power allocation, it is possible to serve nearly all of the UEs with higher SE and only a few with reduced SE in comparison with the max-min fairness solution. Hence, we conclude that the sum SE metric is preferable over the max-min fairness metric both when it comes to average performance and fairness. Fractional power allocation provides a good heuristic tradeoff.
A key observation is that most of the UEs attain higher SE with the scalable fractional power allocation than with sum SE maximization and the 90\% likely SE (where the CDF curve is 0.1) is better than with max-min fairness. Hence, this scheme strikes a good tradeoff between fairness and sum SE.

\enlargethispage{\baselineskip}

Different from the uplink case, sum SE maximization generally performs better than the network-wide equal power allocation. The CDF for the network-wide equal power allocation is the rightmost one in the upper tails, but there is a substantial gap to the other curves for all other UEs. For example, the 90\% likely SE is around 1.5\,bit/s/Hz higher when considering fractional power allocation.

\begin{figure}[t!]\centering
	\begin{overpic}[width=0.88\columnwidth,tics=10]{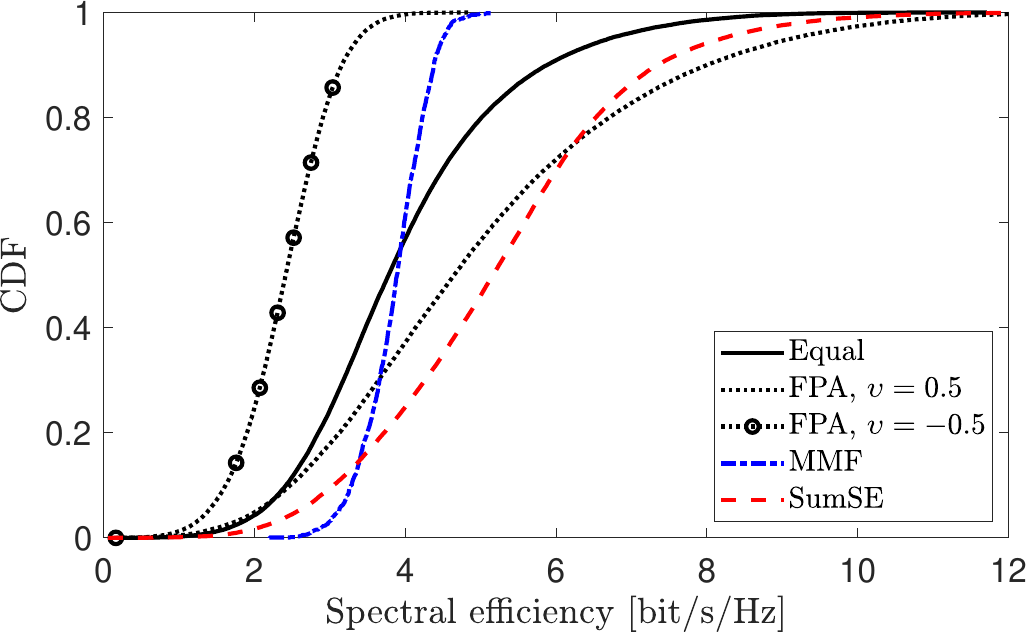}
			\put(40,17.7){85\% likely}
		\put(13,15.8){$- - - - - - - - - - - - $}
	\end{overpic} 
	\caption{CDF of the downlink SE per UE in the distributed operation with different optimized and heuristic power allocation methods. Scalable LP-MMSE precoding is used. We consider $L=100$, $N=4$, $K=40$, $\tau_p=10$, and spatially correlated Rayleigh  fading with ASD $\sigma_{\varphi}=\sigma_{\theta}=15^{\circ}$.  }   
	\label{figureSec7_3} 
\end{figure}

In Figure~\ref{figureSec7_3}, we consider the distributed operation with the scalable LP-MMSE precoding. When it comes to the optimized power allocation schemes, we consider the max-min fairness and sum SE maximization algorithms in Algorithm~\ref{alg:bisection-downlink-case} and Algorithm~\ref{alg:sum-SE-downlink-2}, respectively. As scalable alternatives, we consider per-AP equal power allocation with $\rho_{kl}=\frac{\rhomax}{|\mathcal{D}_l|}$ and the heuristic scheme in \eqref{eq:fractional_power_DL-2}. We denote the latter one by ``FPA'' in analogy with the conceptually similar fractional uplink power control approach in \eqref{eq:fractional_power_UL-scalable} and we consider two different exponents: $\upsilon=0.5$ and $\upsilon=-0.5$. Note the case with $\upsilon=0.5$ leads to that each AP allocates more power to the UEs with better channel conditions whereas $\upsilon=-0.5$ leads to the opposite effect.

Most of the UEs benefits from sum SE maximization in Figure~\ref{figureSec7_3}, however, max-min fairness can significantly improve the SE achieved by the most unfortunate UEs in the network. The UEs with the 15\% worst channel conditions (below 85\% likely SE) achieve higher SE with max-min fairness than with sum SE maximization, which is a much higher percentage than we have observed for uplink power control and centralized downlink power allocation.
We further notice that using the exponent $\upsilon=-0.5$ in the heuristic scheme in \eqref{eq:fractional_power_DL-2} is not an efficient approach, since the CDF curve is to the left of all the other curves. On the other hand, the case with $\upsilon=0.5$, which we utilized for performance comparison in Section~\ref{sec:downlink-performance-comparison} on p.~\pageref{sec:downlink-performance-comparison}, is much better than the per-AP equal power allocation. The reason is that less interference is created when each UE obtains most of its power from the AP that it has the best channel to, compared to when all the serving APs are transmitting with equal power. The gap between the case of $\upsilon=0.5$ and the sum SE maximization scheme is relatively small. 

We conclude that we can choose between sum SE maximization and max-min fairness in the distributed downlink operation depending on the tradeoff between average SE and fairness that we want to obtain. The  heuristic scheme in \eqref{eq:fractional_power_DL-2} is a reasonable approximation of the sum SE maximization case but further research on scalable power allocation schemes is needed.

\begin{remark}[Machine learning for efficient power optimization]
Most of the signal processing problems that have been considered in previous sections of this monograph have known optimal solutions, as well as computationally scalable approximations that seem to perform very well. The presented algorithms for channel estimation, precoding, and combining in centralized and distributed operations are examples of successful man-made algorithms.
In contrast, the downlink power allocation and uplink power control problems are still not solved from a practical perspective. 
The optimization algorithms presented earlier in this section have computational complexities that grow polynomially with the number of UEs, which is an algorithmic deficiency \cite{Bjornson2020b}. The man-made heuristic schemes presented earlier are not bad but there is a substantial gap to the optimal benchmark algorithms, partially because the heuristic schemes have access to less information and partially because it is inherently hard to design heuristic schemes that work well everywhere in the network.
Machine learning might be the tool that is needed to overcome these deficiencies and some first steps are taken in \cite{Bashar2020c}, \cite{Chakraborty2019a}, \cite{Nikbakht2020b}, \cite{Yan2020a}, \cite{Zhao2020a}. A deep learning approach can potentially find shortcuts in existing centralized optimization algorithms, thereby lowering the computational complexity while only suffering from small performance penalties. Other utility functions than those considered in this monograph can also be considered.
When it comes to distributed power allocation schemes (or other schemes that have access to less information than the benchmark algorithms), there is no need to have the same policy at every AP but the specific characteristics of the local propagation environment around each AP can be learned to enhance the performance. The benefit from this can hopefully outweigh the drawback of only having access to local information at the AP.
This research direction is only in its infancy.
\end{remark}

\enlargethispage{\baselineskip}

\vspace{-2mm}
\section{Pilot Assignment}
\label{sec:pilot-assignment-advanced}

When multiple UEs are simultaneously using the same pilot signal, there will be pilot contamination that reduces the estimation quality of the pilot-sharing UEs' channels and increases the mutual interference in the data transmission phase. This issue is not unique to cell-free networks but appears also in cellular networks. Pilot contamination has received particular attention in the literature of Cellular Massive MIMO \cite{Adhikary2017a}, \cite{ashikhmin2012pilot}, \cite{BjornsonHS17}, \cite{Jose2011b}, \cite{Marzetta2010a}, \cite{Mueller2014b}, \cite{Rusek2013a}, \cite{Sanguinetti2019a}, \cite{Yin2016a}, mainly because the resulting additional interference grows with the number of AP antennas. 
The easiest approach to limit the pilot contamination issue is by selecting the pilot-sharing UEs in a judicious manner. To this end, the standard approach in cellular networks is  to associate each cell with a predetermined subset of the pilots. This subset can, for instance, be selected so that neighboring cells are using different subsets. 
Each AP can then assign the pilots arbitrarily to the UEs that reside in its cell. Hence, the cellular structure is exploited to make the pilot assignment relatively straightforward to implement in a distributed fashion. Such methods cannot be used in a cell-free network and, therefore, there is a need for developing new algorithms for pilot assignment.

The general behaviors of pilot contamination in cell-free networks were exemplified in Section~\ref{sec:impact-on-estimation-accuracy} on p.~\pageref{sec:impact-on-estimation-accuracy}. A key insight was that we want to avoid that two UEs that are close to the same set of APs are assigned to the same pilot. A naive approach to pilot assignment is random allocation \cite{Interdonato2018}, where we generate a random integer from 1 to $\taupu$ for every UE and assign this UE to the pilot with the matching index. A benefit of this approach is that it requires no coordination between different UEs or APs, while the main drawback is that the probability that a UE will use the same pilot as its geographically closest neighbor is $1/\taupu$. This is the worst-case situation that should be avoided by using a more structured pilot assignment algorithm.

\subsection{Utility-Based Pilot Assignment}

The key to structured pilot assignment is to define a utility function $f(t_1,\ldots,t_K)$ that represents the goal of the pilot assignment and takes the indices $t_1,\ldots,t_K \in \{ 1,\ldots,\tau_p\}$ of the pilots assigned to the different UEs as input.
The utility can, for example, be defined so that it is maximized when the extra interference caused by pilot contamination is minimized \cite{Ngo2017b} or when the sum SE is maximized \cite{Liu2020a}. In any case, we have the following general problem formulation:
\begin{align} \label{eq:pilot-assignment-problem}
\underset{t_1,\ldots,t_K}{\textrm{maximize}} & \quad f(t_1,\ldots,t_K) \\
\textrm{subject to} &\quad t_k \in \{ 1, \ldots, \tau_p \}, \quad k=1,\ldots,K. \nonumber
\end{align}
This is a combinatorial problem and we can, thus, find the optimal pilot assignment for a given utility function by an exhaustive search over all $\taupu^K$ possible pilot assignment.\footnote{This number can be slightly reduced by exploiting that it only matters which UEs use the same pilot and not what the index of their pilot is.}
The complexity grows exponentially with $K$, which makes it challenging to implement an exhaustive search in a practical network with many UEs, even if it would only be used for offline benchmarking. We need to be able to solve the pilot assignment problem regularly since the utility will depend on which UEs are currently active in the network.

Instead of finding the global optimum to \eqref{eq:pilot-assignment-problem}, we can design an algorithm that finds a decent suboptimal solution. One approach is to design a greedy algorithm that optimizes the utility with respect to one UE at a time. The algorithm can either consider each UE once or iterate until convergence. In small networks where most UEs can be assigned to unique pilots, the algorithm in \cite{Sabbagh2018a} can be used to find suitable UE pairs that can reuse pilots.
The greedy algorithm in \cite{Ngo2017b} is first assigning pilots randomly to the UEs, followed by an iterative procedure where each UE determines if the extra interference caused by pilot contamination can be reduced by switching to another pilot. 
A variation of that greedy algorithm is proposed in \cite{Zhang2018b} by also making use of the geographical locations of the UEs.
A UE clustering algorithm is proposed in \cite{Attarifar2018a} to dynamically divide the network into geographical clusters in which each pilot is only used once. This principle resembles the cellular approach to the pilot assignment problem but makes use of the actual UE locations.
A refined clustering algorithm that utilizes the physical distances between APs and UEs is proposed in \cite{Chen2020a}.
Yet another clustering algorithm is proposed in \cite{Femenias2019a} but it is using the inner products between the vectors $[\beta_{k1} \, \ldots \, \beta_{kL}]^{\Ttran}$ of different UEs as a similarity metric instead of location-based parameters.

Several of the aforementioned algorithms make direct use of the observation that two closely located UEs should not be assigned to the same pilot. However, it is important to bear in mind that it is the channel gains that matter when determining the pilot contamination and even if these are strongly correlated with the UE locations, there can be large variations, which are modeled by shadow fading. 

The mathematical literature contains many combinatorial algorithms that can potentially be applied to pilot assignment. Tabu-search is an algorithm that was used in \cite{Liu2020a} for SE-maximizing pilot assignment. The main principle is to iteratively explore small variations on the current assignment and select a preferable one. A list of previously selected solutions is kept to avoid going back to them. The Hungarian algorithm was utilized in \cite{Buzzi2020a} and tuned to optimize different utilities.

\enlargethispage{1.2\baselineskip}

It is generally hard to make a fair comparison of pilot assignment algorithms because they might  optimize different utilities, they might get stuck in different local optima in different setups, and their computational complexity can be widely different. However, one can conclude that network-wide algorithms are preferably implemented at the CPU and the complexity will grow at least linearly with $K$, which makes their implementation unsuitable for large networks. Our experimental experience is that the most important aspect of a pilot assignment in cell-free networks is to avoid the worst assignments where closely located UEs use the same pilot. This is fairly easy to achieve in a network with $L \gg K$ since each pilot will be reused quite sparsely in the network, as seen from the APs' perspective. The coherent interference that made pilot contamination a major concern in the Cellular Massive MIMO literature is likely not a major issue in cell-free networks, where there also are many antennas but each UE is only served by a small subset.

\enlargethispage{\baselineskip}

\subsubsection{Scalable Pilot Assignment}

If a pilot assignment algorithm should be scalable, it probably must be implemented by a local interaction between a UE and its neighboring APs. Any algorithm that attempts to maximize a network-wide utility and exploits network-wide information will require an immense complexity to evaluate the utility. The pilot assignment algorithm presented in Algorithm~\ref{alg:basic-pilot-assignment} on p.~\pageref{alg:basic-pilot-assignment} is an attempt to design such a scalable algorithm. It originates from \cite{Bjornson2020a} where the main idea was to connect the pilot assignment for a given UE to how it accesses the network (as it is usually done in cellular networks). When the UE is becoming active, it selects a neighboring AP and this AP locally determines which pilot is the most appropriate for the UE to use. More precisely, it computes/measures the amount of pilot interference at each of the pilots and assigns the UE to the pilot with the least interference. This likely corresponds to the pilot where the pilot-sharing UEs are furthest away from the AP, which makes intuitive sense: we want each pilot to be reused as sparsely as possible in space. The algorithm is by no means optimal but has been used throughout this monograph and has provided  good results.

Further research on scalable pilot assignment is certainly needed and since it is a type of clustering problem, machine learning might be a suitable tool for developing efficient algorithms.

\section{Selection of Dynamic Cooperation Clusters}
\label{sec:selecting-DCC-advanced}

The DCC framework is restricting which APs are allowed to serve a given UE in the uplink and downlink. It is unavoidable that this leads to lower SE than in a network where any AP can serve any UE, since we have fewer degrees-of-freedom when optimizing the system. However, there are  strong reasons for introducing these restrictions. As stressed earlier in this monograph, scalability in terms of computational complexity and fronthaul signaling is one reason.
Another reason is to reduce the energy consumption \cite{Ngo2018a} or to limit the delay spread of the downlink signals, which increases with the distance between the UE and the serving AP that is furthest away \cite{Bjornson2013d}, \cite{Zhang2008a}. Intuitively, the SE loss can be kept small if each UE is served by all the APs within its area of influence, but the question is what this means in practice.

Some general guidelines for DCC selection  were provided in \cite[Sec.~4.7]{Bjornson2013d}. Firstly, each UE should have a ``master AP'' that it is anchored to. This AP is required to serve the UE, to guarantee a non-zero SE, while service from other APs is provisioned based on availability (e.g., if these APs are not overloaded with serving other UEs). Secondly, the UEs should be selected from a user-centric perspective, as emphasized throughout this monograph.
Thirdly, different UEs can be assigned to different numbers of APs depending on the local propagation conditions. For example, a UE that has a very good channel to one AP might only need to be served by that AP, or a few more, while a UE that is in between many APs or is subject to much interference will require multiple APs to boost the SNR and/or suppress interference.
Finally, \cite{Bjornson2013d} suggests that one should use the channel quality rather than the geographical locations when measuring proximity to different APs and determining which APs should serve the UE.

There are many possible user-centric clustering algorithms. While it is possible to design network-wide algorithms where all UEs are jointly considered, this will be an unscalable solution where the complexity grows with $K$. Hence, we will concentrate on approaches that consider one UE at a time.

Recall that $\mathcal{M}_k \subset \{ 1, \ldots, L \}$ is the subset of APs that serves UE $k$. 
One approach for selecting the subset is
\begin{equation} \label{eq:threshold-DCC}
\mathcal{M}_k = \{ l = 1,\ldots,L : \beta_{kl} \geq \Delta \}
\end{equation}
where $\Delta>0$ is a constant parameter \cite{Bjornson2011a}. This means that the UE is served by all APs that have a large-scale fading coefficient $\beta_{kl} $ that is above a threshold specified by $\Delta$.
A variation of this approach is to predetermine that each UE $k$ should be served by a certain number of APs. These APs are then selected as the ones with the largest values on $\beta_{kl}$ \cite{Buzzi2017a}. This corresponds to tuning the threshold $\Delta$ in \eqref{eq:threshold-DCC} for each UE to get the predetermined number of serving APs.
Alternatively, the number of serving APs can be selected on a per-UE basis to make sure that
\begin{equation}
\frac{\sum_{l \in \mathcal{M}_k } \beta_{kl}}{\sum_{l=1 }^{L} \beta_{kl}} \geq \delta
\end{equation}
where $\delta>0$ is a threshold representing the fraction of the total received power that  UE $k$ would receive in the downlink \cite{Ngo2018a}. For example, if $\delta= 0.95$, then the UE will be served by the subset of APs that has the strongest channels and which contributes to more than 95\% of the total received power at the UE.

The aforementioned approaches are user-centric and make intuitive sense, but are not taking into account how many UEs a given AP can practically serve. There are two important types of limitations. The first limitation is scalability. Each AP has a limited processing power so it can only manage a limited number of UEs in a distributed operation and only transmit and receive data related to a limited number of UEs over its fronthaul link. The second limitation is pilot contamination. It is only reasonable for an AP to serve one UE per pilot \cite{Bjornson2020a}, \cite{Sabbagh2018a}, because otherwise the weaker of the pilot-sharing UEs will be subject to strong interference. An exception to this rule is if the pilot-sharing UEs have very different spatial correlation matrices so that the AP can separate their channels in the estimation phase based on that information.

If we need to select the clusters $\mathcal{M}_1,\ldots,\mathcal{M}_K$ to comply with the per-AP constraint of the type $|\mathcal{D}_l| \leq \tau_p$ (where $\mathcal{D}_l$ is the set of UEs served by AP $l$), we cannot apply any of the approaches mentioned above. More precisely, we cannot make our selection from a purely user-centric perspective but must take the limitations caused by the network architecture into account.
One scalable solution is to let the pilot assignment algorithm determine the clusters \cite{Bjornson2020a}. For example, each AP can serve (up to) one UE per pilot, namely the UE that has the largest large-scale fading coefficient $\beta_{kl}$ among those that use the pilot. This approach is scalable but the performance will rely on the fact that the pilot assignment problem has been solved appropriately. This was the approach taken in Algorithm~\ref{alg:basic-pilot-assignment} on p.~\pageref{alg:basic-pilot-assignment} and followed throughout this monograph.
We observed that the SE loss compared to serving all UEs by all APs is small, at least in the running example with $L \gg K$. 
However, further research on scalable cooperation cluster formation is needed, particularly for challenging cases with many UEs in certain areas of the network.

\section{Implementation Constraints}
\label{sec:implementation-constraints}

Scalability has been the main focus when we developed the foundations of User-centric Cell-free Massive MIMO in this monograph. The basic definition of scalability, according to Definition~\ref{def:scalable} on p.~\pageref{def:scalable}, is that the signal processing complexity and fronthaul signaling per AP and UE remain finite as $K \to \infty$. This implies that once we have deployed an AP and connected it to its CPU, we will not have to update the local processors and fronthaul infrastructure as we are deploying additional APs, increase the coverage area, or serve a larger number of UEs. However, there are further implementation constraints to consider when building practical networks. We will briefly mention a few of these constraints in this section to stress that further research is needed in these directions.

\subsection{Low-Cost Components}
To enable a ubiquitous deployment of APs, it is important to use compact low-cost components, which might be based on UE-grade chipsets rather than conventional hardware for cellular infrastructure. Practical transceiver components are subject to hardware impairments of different kinds \cite{Schenk2008a}, including non-linearities in power amplifiers, mismatches in mixers, finite-resolution analog-to-digital and digital-to-analog conversion, and phase noise in local oscillators. These effects lead to signal distortion which in some cases can be modeled as a signal power loss plus an uncorrelated additive distortion term, using the Bussgang decomposition \cite{Bussgang1952a}, \cite{Demir2020a}.
The impact of such distortion on cell-free networks has been analyzed in \cite{Masoumi2020a}, \cite{Zhang2018a}, \cite{Zheng2020a} for general hardware impairments, while the special case of quantization distortion in each \gls{ADC} was considered in \cite{Hu2019a}. The obtained SE expressions can be utilized to get a better sense of the practically achievable SE and to optimize the transmit power based on these expressions.
The analysis in the aforementioned papers follows the methodology that was developed in the Cellular Massive MIMO literature
\cite{massivemimobook} or approximations thereof. These models are relatively simple and future research should consider more detailed models.

\subsection{Quantization of Fronthaul Signaling}
We have focused on limiting the number of scalar numbers that need to be transmitted over the fronthaul links per coherence block. However, in practice, these signals must also be quantized before being transmitted over the fronthaul. While the quantization distortion caused by hardware impairments (such as finite-resolution ADCs) is applied on a sample-by-sample basis and largely uncontrollable, the fronthaul compression can be optimized using appropriately designed compression formats that are applied to a block of signal samples. The signal distortion can potentially be limited by making use of rate-distortion theory. Some papers in this direction are \cite{Bashar2020c}, \cite{Bashar2019b}, \cite{Boroujerdi2019a}, \cite{Femenias2019a}, \cite{Maryopi2019a}, \cite{Masoumi2020a}, \cite{Parida2018a}. An interesting consideration that appears when having fronthaul compression is that it matters where certain computations are carried out. If a processing task, such as channel estimation or signal detection, is carried out locally at the AP, the result will be more accurate than if the received signals are first compressed and sent to the CPU and then processed in the same way. 
For example, in a centralized operation, we can choose between estimate-and-quantify and quantify-and-estimate protocols \cite{Bashar2019b}, \cite{Maryopi2019a}. Similar to the case of hardware impairments, SE expressions that are developed by taking the fronthaul compression into account can be used for optimizing the transmit power or other resource allocation tasks.

\enlargethispage{\baselineskip} 

\subsection{Radio Stripes \& Other Constrained Fronthaul Architectures}

The structure of the fronthaul might also be constrained in the sense that each AP might not have a dedicated fronthaul link but a shared connection with a subset of other APs. For example, in the radio stripes architecture \cite{Interdonato2018}, the APs are integrated into fronthaul cables that are connected to a CPU at one end, which also provides a power supply. A fronthaul signal from the outmost AP needs to travel via all the other APs before reaching the CPU. The benefit of this architecture is that amount of cabling can be greatly reduced. In a deployment with a particular fronthaul architecture, the structure can be utilized to optimize the fronthaul signaling and signal processing. For example, the APs can send signals to their neighboring APs which enables a co-processing and interference mitigation \cite{Shaik2020a}. MMSE combining can be implemented in a sequential fashion \cite{Shaik2021a}.

\subsection{Synchronization of APs}

Proper synchronization of the distributed APs is necessary for coherent uplink and downlink transmission. The APs must not be phase-synchronized since this is momentarily achieved in every coherence block through the channel estimation, which estimates the combined effect of the propagation channels and the phase-shifts induced by the hardware. However, the cooperating APs must be  synchronized in time and frequency; see \cite{Jeong2020a} for a recent review of how to achieve that in Cellular Massive MIMO systems and \cite{Etzlinger2018a} for a more general review. Perfect synchronization has been assumed in this monograph, but the actual implementation can be challenging since the APs are distributed. Some initial algorithms for cell-free networks are described in \cite{Cheng2013a}, \cite{Interdonato2018}, but further work is needed on this topic. The user-centric cooperation clusters might relax the synchronization requirements since only the geographically closest APs must be well synchronized, while more distant APs that are serving non-overlapping sets of UEs do not require the same level of synchronization.

\newpage
\enlargethispage{\baselineskip} 
\section{Summary of the Key Points in Section~\ref{c-resource-allocation}}
\label{c-resource-allocation-summary}

\begin{mdframed}[style=GrayFrame]

\begin{itemize}
\item UEs in cell-free networks have conflicting performance goals due to the interference and shared power budgets. Power allocation can be used to find a tradeoff between them. 

\item Any algorithm that jointly optimizes the transmit powers in uplink or downlink to maximize a network-wide utility function becomes unscalable as $K\to \infty$ since there are $K$ SINRs to optimize and at least $K$ optimization variables. 

\item To obtain a practically implementable power allocation that is scalable, some kind of distributed heuristic scheme is needed, where each AP or UE makes a local decision based on the channel knowledge that it can acquire locally. Fractional power control is a classical heuristic that can be also used in cell-free networks.

\item Sum SE maximization is often the preferable optimization criterion since it attains high SEs for the UEs with good channel conditions and satisfactorily SEs for UEs with bad channel conditions. An alternative metric is the max-min fairness, which provides (slightly) higher SE for the UE with the worst conditions but a substantial SE loss for most other UEs.

\item In the uplink, the sum SE  is maximized when all UEs transmit with full power. Hence, there is no need to run a network-wide optimization problem in the centralized and distributed uplink operation. To emphasize fairness, fractional power control can be utilized with a negative exponent.

\item In the centralized downlink operation, fractional power allocation with a negative exponent provides higher SE to the most of the UEs than the sum SE maximization and it also achieves a good balance between fairness and sum SE. 

\end{itemize}

\end{mdframed}

\begin{mdframed}[style=GrayFrame]

\begin{itemize}

\item In the distributed downlink simulations, the fractional power allocation with a positive exponent performs much better than per-AP equal power allocation. There is a gap to the sum SE maximization-based scheme but it is relatively small.

\item The pilot contamination effect can be reduced by assigning pilots to UEs to limit the interference. The optimal pilot assignment is a combinatorial problem whose complexity grows exponentially with $K$. However, greedy algorithms that operate at one UE at a time can be developed. They are scalable and guarantee reasonably good performance.

\item An unscalable network where all APs serve all UEs provides the highest SEs, but the dynamic cooperation clusters can be selected to achieve a scalable operation with a small performance loss. Any user-centric clustering algorithm where all UEs are jointly considered is unscalable as $K\to \infty$. If properly designed, approaches that consider one UE at a time can provide good performance while being scalable.

\item Scalability is not the only requirement for building practical networks. The need for enabling a ubiquitous deployment of APs makes it preferable to use compact low-cost components, whose hardware imperfections must be taken into account in the analysis and design. The fronthaul signal compression is another limiting factor. The structure of the fronthaul might also impose limitations since some APs might share a connection with a subset of other APs. Finally, proper synchronization of the distributed APs is necessary for coherent uplink and downlink transmission. These are topics (among many others) that require further investigation.

\end{itemize}

\end{mdframed}

\begin{acknowledgements}

We would first like to thank our students and collaborators in the areas related to this monograph. Without the results, encouragements, and insights obtained through our joint research during the last decade, it wouldn't have been possible to write this monograph.
We are grateful for the constructive feedback from the reviewers, which helped us to focus our final editing efforts at the right places. In particular, we would like to thank Angel Lozano, Jiayi Zhang, Mahmoud Zaher, and Yasaman Khorsandmanesh for giving detailed comments.

\"{O}zlem Tu\u{g}fe Demir and Emil Bj\"ornson have been supported by the Wallenberg AI, Autonomous Systems and Software Program (WASP) funded by the Knut and Alice Wallenberg Foundation. Emil Bj\"ornson has also been supported by the Excellence Center at Link\"oping -- Lund in Information Technology (ELLIIT), the Center for Industrial Information Technology (CENIIT), the Swedish Research Council, and the Swedish Foundation for Strategic Research. Luca Sanguinetti has been partially supported by the University of Pisa under the PRA Research Project CONCEPT, and by the Italian Ministry of Education and Research (MIUR) in the framework of the CrossLab project (Departments of Excellence).

\end{acknowledgements}

\appendix

\chapter{Notation and Abbreviations}

\section*{Mathematical Notation}

Upper-case boldface letters are used to denote matrices (e.g., $\vect{X},\vect{Y}$), while column vectors are denoted with lower-case boldface letters (e.g., $\vect{x},\vect{y}$). Scalars are denoted by lower/upper-case italic letters (e.g., $x,y,X,Y$) and sets by calligraphic letters (e.g., $\mathcal{X},\mathcal{Y}$). 

The following mathematical notations are used:

\begin{longtable}{ll}
   $\mathbb{C}^{N \times M}$ & The set of complex-valued $N \times M$ matrices\\
   $\mathbb{R}^{N \times M}$ & The set of real-valued $N \times M$ matrices\\
   $\mathbb{C}^{N},\mathbb{R}^{N}$ & Short forms of $\mathbb{C}^{N \times 1}$ and $\mathbb{R}^{N \times 1}$ for vectors \\
  $ \mathbb{R}_{\geq 0}^{N}$ & The set of non-negative members of $\mathbb{R}^{N}$\\
   $x \in \mathcal{S}$ & $x$ is a member of the set $\mathcal{S}$\\
   $x \not \in \mathcal{S}$ & $x$ is not a member of the set $\mathcal{S}$\\
   $\{ x \in \mathcal{S} : P \}$ & The subset of $\mathcal{S}$ containing all members \\ & that satisfy a property $P$\\
   $[\vect{x}]_i$ & The $i$th element of a vector $\vect{x}$ \\
   $[\vect{X}]_{ij}$ & The $(i,j)$th element of a matrix $\vect{X}$ \\
   $[\vect{X}]_{:,1}$ & The first column of a matrix $\vect{X}$ \\
   $\diag(\cdot)$ & $\diag(x_1,\ldots,x_N)$ is a diagonal matrix with \\ & the scalars $x_1,\ldots,x_N$ on the diagonal, \\
   &   $\diag(\vect{X}_1,\ldots,\vect{X}_N)$ is a block diagonal matrix with  \\
             &  the matrices $\vect{X}_1,\ldots,\vect{X}_N$ on the diagonal\\
   $\vect{X}^{\star}$ &  The complex conjugate of $\vect{X}$\\
   $\vect{X}^{\Ttran}$ &  The transpose of $\vect{X}$\\
   $\vect{X}^{\Htran}$ &  The conjugate transpose of $\vect{X}$\\
   $\vect{X}^{-1}$ & The inverse of a square matrix $\vect{X}$\\
   $\vect{X}^{\frac12}$ & The square-root of a square matrix $\vect{X}$\\
   $\Re(x)$ & Real part of  $x$ \\
   $\imagunit$ & The imaginary unit\\
   $|x|$   &  Absolute value (or magnitude) of a scalar variable $x$\\
   $e$ & Euler's number ($e \approx 2.718281828$) \\
   $\log_a(x)$ & Logarithm of $x$ using the base $a >0$\\
   $\sin(x)$ & The sine function of $x$ \\   
   $\cos(x)$ & The cosine function of $x$ \\   
   $\tr( \vect{X})$ & Trace of a square matrix $\vect{X}$ \\
   $\mathcal{N}(\vect{x},\vect{R})$ & The real Gaussian distribution \\ & with mean $\vect{x}$ and covariance matrix $\vect{R}$ \\
   $\CN(\vect{0},\vect{R})$ & The circularly symmetric complex Gaussian   \\ &  distribution with zero mean and correlation matrix $\vect{R}$,  \\ & where circular symmetry means that if $\vect{y} \sim \CN(\vect{0},\vect{R})$ \\ & then $e^{\imagunit \phi}\vect{y} \sim \CN(\vect{0},\vect{R})$ for any given $\phi$ \\
   $\mathbb{E}\{ {x} \}$ & The expectation of a random variable $x$ \\
   $\mathbb{V}\{ {x} \}$ & The variance of a random variable $x$ \\
   $\|\vect{x}\|$ & The $L_2$-norm $\|\vect{x}\| = \sqrt{ \sum_i | [\vect{x}]_i |^2 }$ of  a vector $\vect{x}$ \\
   $\vect{I}_M$ & The $M \times M$ identity matrix \\
   $\vect{1}_N$ & The $N \times 1$ matrix (i.e., vector) with only ones \\
   $\vect{1}_{N\times M}$ & The $N \times M$ matrix with only ones \\
   $\vect{0}_{M}$ & The $M \times 1$ matrix (i.e., vector) with only zeros \\
       $\vect{0}_{N \times M}$ & The $N \times M$ matrix with only zeros \\
\end{longtable}

\section*{Abbreviations}

The following acronyms and abbreviations are used in this monograph:
\newpage
\enlargethispage{\baselineskip}
\printnoidxglossaries

\chapter{Useful Lemmas}\label{appendix1}
\vspace*{-1.2in}

This appendix contains a few classical results related to matrices, which are utilized to prove the results in other parts of the monograph.

\begin{lemma}[Matrix inversion lemma] \label{lemma:matrix-inversion}
	Consider the matrices $\vect{A} \in \mathbb{C}^{N_1 \times N_1}$, $\vect{B} \in \mathbb{C}^{N_1 \times N_2}$, $\vect{C} \in \mathbb{C}^{N_2 \times N_2}$, and $\vect{D} \in \mathbb{C}^{N_2 \times N_1}$. The following identity holds if all the involved inverses exist:
	\begin{equation}
	(\vect{A} + \vect{B} \vect{C} \vect{D} )^{-1} = \vect{A}^{-1} - \vect{A}^{-1} \vect{B} (\vect{D} \vect{A}^{-1}  \vect{B} + \vect{C}^{-1}  )^{-1} \vect{D} \vect{A}^{-1} .
	\end{equation}
\end{lemma}

\begin{lemma} \label{lemma-matrix-identities}
For matrices $\vect{A} \in \mathbb{C}^{N_1 \times N_2}$ and $\vect{B} \in \mathbb{C}^{N_2 \times N_1}$, it holds that
\begin{align} 
\left( \vect{I}_{N_1} + \vect{A} \vect{B} \right)^{-1} \vect{A} &= \vect{A} \left( \vect{I}_{N_2} +  \vect{B} \vect{A} \right)^{-1} \label{eq:matrix-identity-commutation-inverse} \\
\tr \left( \vect{A} \vect{B}  \right) &= \tr \left(\vect{B}  \vect{A}  \right).   \label{eq:matrix-identity-trace-commutation} 
\end{align}
\end{lemma}

\begin{lemma} \label{lemma-nonorthogonal-positive-semidefinite}
	For the non-zero positive semi-definite matrix $\vect{A} \in \mathbb{C}^{N \times N}$ and positive definite matrix $\vect{B} \in \mathbb{C}^{N \times N}$, their inner product is strictly positive:
	\begin{equation}
	\tr\left(\vect{A}\vect{B}\right)>0.
	\end{equation}
\end{lemma}
\begin{proof}
	Consider the eigendecomposition of $\vect{A}=\vect{U}\vect{\Lambda}\vect{U}^{\Htran}$ where $\vect{\Lambda}=\diag\left(\lambda_1,\ldots,\lambda_N\right)$ with $\lambda_n\geq0$, for $n=1,\ldots,N$. Let $\vect{u}_n$ denote the $n$th column of the eigenvalue matrix $\vect{U}$. Then, we have
	\begin{equation}
	\tr\left(\vect{A}\vect{B}\right)=\sum_{n=1}^N\lambda_n\vect{u}_n^{\Htran}\vect{B}\vect{u}_n
	\end{equation}
	that is strictly greater than zero since $\vect{B}$ is a positive definite matrix and at least one eigenvalue of $\vect{A}$ is positive.
\end{proof}

\begin{lemma}\label{lemma:scaling-mse}
	For the positive semi-definite matrices $\vect{A} \in \mathbb{C}^{N \times N}$, $\vect{C} \in \mathbb{C}^{N \times N}$, and positive definite matrix $\vect{B} \in \mathbb{C}^{N \times N}$, the following inequality holds:
	\begin{equation} \label{eq:scaling-mse-lemma}
	\tr\left(\vect{A}\left(\vect{B}+\vect{C}\right)^{-1}\right)\leq \tr\left(\vect{A}\vect{B}^{-1}\right)
	\end{equation}
	where the equality holds only when $\vect{C}\vect{B}^{-1}\vect{A}=\vect{0}_{N \times N}$.
\end{lemma}
\begin{proof}
	Consider the eigendecomposition of $\vect{C}=\vect{U}\vect{\Lambda}\vect{U}^{\Htran}=\vect{U}_1\vect{\Lambda}_1\vect{U}_1^{\Htran}$, where $\vect{U}\in \mathbb{C}^{N \times N}$ is the unitary matrix of eigenvectors and $\vect{\Lambda}=\diag\left(\lambda_1,\ldots,\lambda_N\right)$ with the eigenvalues $\lambda_n\geq0$, for $n=1,\ldots,N$. $\vect{U}_1\in \mathbb{C}^{N \times r}$ and $\vect{\Lambda}_1\in \mathbb{C}^{r\times r}$ are the partitions of $\vect{U}$ and $\vect{\Lambda}$, respectively corresponding to the positive eigenvalues. Applying the matrix inversion lemma (Lemma~\ref{lemma:matrix-inversion}) to the inverse of $\vect{B}+\vect{U}_1\vect{\Lambda}_1\vect{U}_1^{\Htran}$, we can express the left side of the inequality in  \eqref{eq:scaling-mse-lemma} as 
	\begin{align}
	&\tr\left(\vect{A}\left(\vect{B}+\vect{C}\right)^{-1}\right)\nonumber\\
	&=\tr\left(\vect{A}\vect{B}^{-1}\right)-\tr\left(\vect{A}\vect{B}^{-1}\vect{U}_1\left(\vect{U}_1^{\Htran}\vect{B}^{-1}\vect{U}_1+\vect{\Lambda}_1^{-1}\right)^{-1}\vect{U}_1^{\Htran}\vect{B}^{-1}\right)
	\end{align}
	that is strictly less than $\tr\left(\vect{A}\vect{B}^{-1}\right)$ if $\vect{U}_1^{\Htran}\vect{B}^{-1}\vect{A}\vect{B}^{-1}\vect{U}_1$ is non-zero by Lemma~\ref{lemma-nonorthogonal-positive-semidefinite} noting that the matrix $\left(\vect{U}_1^{\Htran}\vect{B}^{-1}\vect{U}_1+\vect{\Lambda}_1^{-1}\right)^{-1}$ is positive definite. If $\vect{U}_1^{\Htran}\vect{B}^{-1}\vect{A}\vect{B}^{-1}\vect{U}_1=\vect{0}_{r \times r}$ that is equivalent to $\vect{C}\vect{B}^{-1}\vect{A}=\vect{0}_{N \times N}$, then both sides of 	\eqref{eq:scaling-mse-lemma} are equal.
\end{proof}

\begin{lemma} \label{lemma:fourth-order-moment-Gaussian-vector}
Consider the vector $\vect{a} \sim \CN(\vect{0}_{N}, \vect{A} )$, with covariance matrix $\vect{A} \in \mathbb{C}^{N \times N}$, and any deterministic matrix $\vect{B} \in \mathbb{C}^{N \times N}$. It holds that
\begin{equation} \label{eq:lemma:fourth-order-moment-Gaussian-vector}
\mathbb{E}\{ | \vect{a}^{\Htran}  \vect{B} \vect{a} |^2  \} =  |\tr( \vect{B} \vect{A} ) |^2 + \tr( \vect{B} \vect{A} \vect{B}^{\Htran} \vect{A} ) .
\end{equation}
\end{lemma}
\begin{proof}
Note that $\vect{a} = \vect{A}^{\frac12} \vect{w}$ for $\vect{w} = [w_1 \, \ldots \, w_N]^{\Ttran} \sim \CN(\vect{0}_{N}, \vect{I}_N )$, thus we can write 
\begin{equation} \label{eq:lemma:fourth-order-moment-Gaussian-vector1}
\mathbb{E}\{ | \vect{a}^{\Htran}  \vect{B} \vect{a} |^2  \} = \mathbb{E}\{ | \vect{w}^{\Htran} (\vect{A}^{\Htran})^{\frac12}  \vect{B} \vect{A}^{\frac12}  \vect{w} |^2  \} = \mathbb{E}\{ | \vect{w}^{\Htran} \vect{C} \vect{w} |^2  \}
\end{equation}
where we defined  $\vect{C} = (\vect{A}^{\Htran})^{\frac12}  \vect{B} \vect{A}^{\frac12}$. Let $c_{n_1,n_2}$ denote the element in $\vect{C}$ on row $n_1$ and column $n_2$.
Using this notation, we can expand \eqref{eq:lemma:fourth-order-moment-Gaussian-vector1} as
\begin{align} \
\notag &\mathbb{E}\{ | \vect{w}^{\Htran} \vect{C} \vect{w} |^2  \} = \sum_{n_1=1}^{N} \sum_{n_2=1}^{N} \sum_{m_1=1}^{N} \sum_{m_2=1}^{N} \mathbb{E}\{
w_{n_1}^{\star} c_{n_1,n_2} w_{n_2}w_{m_1} c_{m_1,m_2}^{\star} w_{m_2}^{\star} \} 
\end{align}
\begin{align}
&\overset{(a)}{=}  \sum_{n=1}^{N} \mathbb{E}\{
|w_{n}|^4  \} |c_{n,n}|^2
+
\sum_{n=1}^{N}
\condSum{m=1}{m \neq n}{N}
  \mathbb{E}\{
|w_{n}|^2 \} \mathbb{E}\{ |w_{m}|^2  \} c_{n,n} c_{m,m}^{\star} 
\notag \\ &+
\sum_{n_1=1}^{N}
\condSum{n_2=1}{n_2 \neq n_1}{N} \mathbb{E}\{
|w_{n_1}|^2 \} \mathbb{E}\{ |w_{n_2} |^2 \} |c_{n_1,n_2}|^2
\notag \\ & \overset{(b)}{=}  \sum_{n=1}^{N} 2 |c_{n,n}|^2
+
\sum_{n=1}^{N}
\condSum{m=1}{m \neq n}{N}
 c_{n,n} c_{m,m}^{\star} +
\sum_{n_1=1}^{N}
\condSum{n_2=1}{n_2 \neq n_1}{N}  |c_{n_1,n_2}|^2
\notag \\ & =
\sum_{n=1}^{N}
\sum_{m=1}^{N}
 c_{n,n} c_{m,m}^{\star} +
\sum_{n_1=1}^{N}
\sum_{n_2=1}^{N}  |c_{n_1,n_2}|^2
\notag \\ & \overset{(c)}{=} |\tr(\vect{C})|^2 + \tr(\vect{C} \vect{C}^{\Htran})
\end{align}
where $(a)$ utilizes that circular symmetry implies that $\mathbb{E}\{
w_{n_1}^{\star} w_{n_2}w_{m_1} w_{m_2}^{\star} \}$ is only non-zero when the terms with conjugates have matching indices to the terms without conjugates. The first expression is given by $n_1 = n_2 = m_1 = m_2$, the second term is given by $n_1 = n_2$ and $m_1 = m_2$ with $n_1 \neq m_1$, and the third term is given by $n_1 = m_1$ and $n_2 = m_2$ with $n_1 \neq n_2$.
In $(b)$, we utilize that $\mathbb{E}\{ |w_{n}|^2  \} = 1$ and $\mathbb{E}\{ |w_{n}|^4  \} = 2$.
In $(c)$, we write the sums of elements in $\vect{C}$ using the trace. The resulting expression is equivalent to  \eqref{eq:lemma:fourth-order-moment-Gaussian-vector}, which is shown by replacing $\vect{C}$ with $\vect{A}$ and $\vect{B}$ and utilizing the fact that $\tr( \vect{C}_1 \vect{C}_2 ) = \tr(  \vect{C}_2 \vect{C}_1 )$ for any matrices $\vect{C}_1, \vect{C}_2$ such that  $\vect{C}_1$ and  $\vect{C}_2^{\Ttran}$ have the same dimensions.
\end{proof}

\chapter{Collection of Proofs}\label{appendix2}
\vspace*{-1.8in}

This appendix contains proofs of lemmas, theorems, and corollaries that were deemed to long to be placed in the main body of the monograph.

\vspace{-3mm}

\section{Proofs from Section~\ref{c-channel-estimation}}
We report below the proofs from Section~\ref{c-channel-estimation}. 
\subsection{Proof of Lemma~\ref{lemma:nmse-concave}}\label{proof-of-lemma:nmse-concave}
To prove this result, it is enough to show that the Hessian matrix $D^2\mathacr{NMSE}\left(\boldsymbol{\lambda}\right)$ is negative definite for $\lambda_n\geq 0$. $D^2\mathacr{NMSE}\left(\boldsymbol{\lambda}\right)$ is given as
	\begin{align}
	\!\!\!\!D^2\mathacr{NMSE}\left(\boldsymbol{\lambda}\right)=\frac{1}{N\beta}\diag \left( \frac{-2\eta\tau_p\sigmaUL}{\left(\eta\tau_p\lambda_1+\sigmaUL\right)^3}, \ldots, \frac{-2\eta\tau_p\sigmaUL}{\left(\eta\tau_p\lambda_N+\sigmaUL\right)^3} \right)
	\end{align}
	and is negative definite since it is diagonal with strictly negative entries.
	\vspace{-3mm}
	
\subsection{Proof of Lemma~\ref{lemma:nmse-eigenvalue-distribution}}\label{proof-of-lemma:nmse-eigenvalue-distribution}
Note that all the elements of $\boldsymbol{\lambda}$ and $\boldsymbol{\lambda^\prime}$ are identical except the $(r-1)$th and the $r$th ones. Hence, the difference between $\mathacr{NMSE}\left(\boldsymbol{\lambda}\right)$ and $\mathacr{NMSE}\left(\boldsymbol{\lambda^\prime}\right)$ results from the summation terms in \eqref{eq:nmse-without-pilot-contamination} for $n \in \{ r-1, r\}$. Using this property, $\mathacr{NMSE}\left(\boldsymbol{\lambda}\right)- \mathacr{NMSE}\left(\boldsymbol{\lambda^\prime}\right)$ is
\begingroup
\allowdisplaybreaks
	\begin{align}
	&\frac{\eta\tau_p}{N\beta}\left( \frac{(\lambda_{r-1}+\lambda_r)^2}{\eta\tau_p(\lambda_{r-1}+\lambda_r)+\sigmaUL}-\frac{\lambda_{r-1}^2}{\eta\tau_p\lambda_{r-1}+\sigmaUL}-\frac{\lambda_r^2}{\eta\tau_p\lambda_r+\sigmaUL}\right) \nonumber\\
	&\overset{(a)}{=}\frac{1}{N\beta}\left( \frac{(x+y)^2}{x+y+c}-\frac{x^2}{x+c}-\frac{y^2}{y+c}\right) \nonumber \\
	& = \frac{\left(x+y\right)^2\left(xy+c\left(x+y+c\right)\right)-\left(x^2\left(y+c\right)+y^2\left(x+c\right)\right)\left(x+y+c\right)}{N\beta \left(x+y+c\right)\left(x+c\right)\left(y+c\right)} \nonumber \\
	&\overset{(b)}{=} \frac{\left(x+y\right)^2xy+2cxy\left(x+y+c\right)-xy\left(x+y\right)\left(x+y+c\right)}{N\beta \left(x+y+c\right)\left(x+c\right)\left(y+c\right)}  \nonumber \\
	& \overset{(c)}{=} \frac{2cxy\left(x+y+c\right)-cxy\left(x+y\right)}{N\beta \left(x+y+c\right)\left(x+c\right)\left(y+c\right)}  \overset{(d)}{>}0 
 	\end{align}
 	\endgroup
	where we have introduced $x=\lambda_{r-1}$, $y=\lambda_r$, and the constant $c=\sigmaUL/(\eta\tau_p)$ in $(a)$ for simplicity. In $(b)$ and $(c)$, we have canceled the common terms $(x^2+y^2)c(x+y+c)$ and $\left(x+y\right)^2xy$, respectively, in the numerator. Finally, the result in $(d)$ is obtained from the fact that $x>0$, $y>0$, and $c>0$.

	\section{Proofs from Section~\ref{c-uplink}}
	We report below the proofs from Section~\ref{c-uplink}. 
\subsection{Proof of Theorem~\ref{proposition:uplink-capacity-general}}\label{proof-of-proposition:uplink-capacity-general}
	
	The processed signal in \eqref{eq:uplink-data-estimate2} can be treated as the discrete memoryless interference channel in Lemma~\ref{lemma:capacity-memoryless-random-interference} on p.~\pageref{lemma:capacity-memoryless-random-interference} with a random channel response $h=\vect{v}_k^{\Htran}\vect{D}_k\widehat{\vect{h}}_k $, the input $x=s_k$, the output $y=\vect{v}_k^{\Htran}\vect{D}_k\vect{y}^{\rm{ul}}$, and the realization  $u=\{\vect{D}_k\widehat{\vect{h}}_i:i=1,\ldots,K\}$ that affects the conditional variance of the interference. The effective noise $\vect{v}_k^{\Htran}\vect{D}_k\vect{n}$ may not be Gaussian and all the interference and noise are included in $\upsilon$ with $n=0$ in Lemma~\ref{lemma:capacity-memoryless-random-interference} on p.~\pageref{lemma:capacity-memoryless-random-interference}. The input power is $p=\mathbb{E}\{\left|s_k\right|^2\}=p_k$. The term $\upsilon$ is given by
	\begin{align}
		\upsilon=\condSum{i=1}{i \neq k}{K}\vect{v}_k^{\Htran}\vect{D}_k\widehat{\vect{h}}_i s_i+\sum_{i=1}^K\vect{v}_k^{\Htran}\vect{D}_k\widetilde{\vect{h}}_i s_i+\vect{v}_k^{\Htran}\vect{D}_k\vect{n}.
		\end{align}
	To prove the theorem, we need to show that the requirements in Lemma~\ref{lemma:capacity-memoryless-random-interference} on p.~\pageref{lemma:capacity-memoryless-random-interference} are satisfied and then compute the conditional variance $p_{\upsilon}(h,u)=\mathbb{E}\{\left|\upsilon\right|^2|h,u\}$. First, we notice that the realizations of $h$ and $u$ are known at the CPU and $\upsilon$ has conditional zero-mean given $(h,u)$, i.e., $\mathbb{E}\{\upsilon|h,u\}=0$ since the  symbols $\{s_i:i=1,\ldots,K\}$ and the noise vector $\vect{n}$ are independent of the channel estimates and zero-mean estimation errors. The conditional variance given $(h,u)$ is
	\begingroup
	\allowdisplaybreaks
	\begin{align}
	p_{\upsilon}(h,u)&=\mathbb{E}\left\{\left|\upsilon\right|^2|h,u\right\}=\mathbb{E}\left\{\left|\upsilon\right|^2|\left\{\vect{D}_k\widehat{\vect{h}}_i\right\}\right\}\nonumber\\
	&=\condSum{i=1}{i \neq k}{K}p_i\left|\vect{v}_k^{\Htran}\vect{D}_k\widehat{\vect{h}}_i\right|^2+\sum_{i=1}^Kp_i\vect{v}_k^{\Htran}\vect{D}_k\mathbb{E}\left\{\widetilde{\vect{h}}_i\widetilde{\vect{h}}_i^{\Htran}\right\}\vect{D}_k\vect{v}_k\nonumber\\&\quad+\vect{v}_k^{\Htran}\vect{D}_k\mathbb{E}\{\vect{n}\vect{n}^{\Htran}\}\vect{D}_k\vect{v}_k
	 \nonumber \\
	 &=\condSum{i=1}{i \neq k}{K} p_i\left|\vect{v}_k^{\Htran}\vect{D}_k\widehat{\vect{h}}_i\right|^2+\sum_{i=1}^Kp_i\vect{v}_k^{\Htran}\vect{D}_k\vect{C}_i\vect{D}_k\vect{v}_k+\sigmaUL\vect{v}_k^{\Htran}\vect{D}_k\vect{D}_k\vect{v}_k
		\end{align}
		\endgroup
which is equal to  the denominator of \eqref{eq:uplink-instant-SINR-level4}. In the derivation, we have used the fact that the individual terms of $\upsilon$ are mutually uncorrelated and the combining vector $\vect{v}_k$ depends only on the channel estimates, which are independent of the estimation errors. As a final requirement for using Lemma~\ref{lemma:capacity-memoryless-random-interference}, we note that the input signal $x=s_k$ is conditionally uncorrelated with $\upsilon$ given $(h,u)$, i.e., $\mathbb{E}\{s_k^{\star}\upsilon|\{\vect{D}_k\widehat{\vect{h}}_i\}\}=0$ due to the independence of the different symbols and the zero-mean channel estimation errors. 

 As a last step, we note that only the fraction $\tau_u/\tau_c$ of the samples is used for uplink data transmission, which results in the lower bound on the capacity that is stated in the theorem and measured in bit/s/Hz.
 
 \subsection{Proof of Theorem~\ref{proposition:uplink-capacity-general-UatF}}\label{proof-of-proposition:uplink-capacity-general-UatF}

By adding and subtracting the term $\mathbb{E} \{ \vect{v}_k^{\Htran}\vect{D}_k\vect{h}_k \} s_k$, 
the received signal in \eqref{eq:uplink-data-estimate2} can alternatively be expressed as
	\begin{equation}
	\label{eq:uplink-data-estimate2-UatF}
\widehat{s}_k = \mathbb{E} \{ \vect{v}_k^{\Htran}\vect{D}_k\vect{h}_k \} s_k + \upsilon_k
\end{equation}
where
\begin{align} \nonumber
\upsilon_k =& \left( \vect{v}_k^{\Htran}\vect{D}_k\vect{h}_k - \mathbb{E} \{ \vect{v}_k^{\Htran}\vect{D}_k\vect{h}_k \} \right) s_k +\condSum{i=1}{i \neq k}{K}\vect{v}_k^{\Htran}\vect{D}_k\vect{h}_i s_i + \vect{v}_k^{\Htran}\vect{D}_k\vect{n}.
	\end{align}
This is a deterministic channel (since $\mathbb{E} \{ \vect{v}_k^{\Htran}\vect{D}_k\vect{h}_k \}$ is constant) with the additive interference plus noise term $\upsilon_k$, which has zero mean since $\{s_i:i=1,\ldots,K\}$ and $\vect{n}$ have zero mean. Although $\upsilon_k$ contains the desired signal $s_k$, it is uncorrelated with it since
\begin{equation}
\mathbb{E}\left\{ s_k^{\star} \upsilon_k \right\} = \underbrace{\mathbb{E}\left\{\left( \vect{v}_k^{\Htran}\vect{D}_k\vect{h}_k - \mathbb{E} \{ \vect{v}_k^{\Htran}\vect{D}_k\vect{h}_k \} \right)  \right\} }_{=0}  \mathbb{E}\{ | s_k|^2 \} = 0.
\end{equation}
Therefore, we can apply Lemma~\ref{lemma:capacity-memoryless-deterministic-interference} on p.~\pageref{lemma:capacity-memoryless-deterministic-interference} with $h =  \mathbb{E} \{ \vect{v}_k^{\Htran}\vect{D}_k\vect{h}_k \}$, $x= s_k$, $p=p_k$, $\upsilon = \upsilon_k$, and $\sigma^2=0$.
By noting that the signals of different UEs are independent and that the noise contributions at different APs are independent, we have that
\begin{equation}
\mathbb{E} \{ |\upsilon_k |^2 \} = \sum\limits_{i=1}^K p_{i}  \mathbb{E} \left\{| \vect{v}_{k}^{\Htran} \vect{D}_{k}{\vect{h}}_{i} |^2\right\}
		- p_{k} \left |\mathbb{E} \left\{  \vect{v}_{k}^{\Htran} \vect{D}_{k}{\vect{h}}_{k} \right\}\right|^2  + \sigmaUL \mathbb{E} \left\{\| \vect{D}_k \vect{v}_{k} \|^2\right\}.
\end{equation}
The SE expression presented in the theorem then follows from Lemma~\ref{lemma:capacity-memoryless-deterministic-interference}.  
 As a last step, we note that only the fraction $\tau_u/\tau_c$ of the samples is used for uplink data transmission, which results in the lower bound on the capacity that is stated in the theorem and measured in bit/s/Hz.

\subsection{Proof of Theorem~\ref{theorem:uplink-capacity-level3}}\label{proof-of-proposition:uplink-capacity-level3}

Since the CPU does not have knowledge of the channel estimates, it needs to treat the average channel gain $\vect{a}_{k}^{\Htran}\mathbb{E}\{\vect{g}_{kk}\}$ as the true deterministic channel. Hence, the signal model is
\begin{equation} \label{eq:CPU-decoding-level3}
\widehat{s}_k = \vect{a}_{k}^{\Htran}\mathbb{E}\{\vect{g}_{kk}\}  s_k + \upsilon_k
\end{equation}
which is a deterministic channel with the additive interference-plus-noise term
\begin{align}  \upsilon_k 
=  \big(\vect{a}_{k}^{\Htran}\vect{g}_{kk} -  \vect{a}_{k}^{\Htran} \mathbb{E}\{ \vect{g}_{kk}\}\big) s_k+ 
\condSum{i=1}{i \neq k}{K} \vect{a}_{k}^{\Htran}  \vect{g}_{ki} s_i +  n^\prime_{k}.
\end{align}
The term $\upsilon_k$ has zero mean and is uncorrelated with the signal term in \eqref{eq:CPU-decoding-level3} since
\begin{align}
\underbrace{\mathbb{E}\left\{ \vect{a}_{k}^{\Htran}\vect{g}_{kk} -  \vect{a}_{k}^{\Htran} \mathbb{E}\left\{ \vect{g}_{kk}\right\}\right\}}_{=0}\mathbb{E}\left\{\left|s_k\right|^2\right\} = 0.
\end{align}
Therefore, we can apply Lemma~\ref{lemma:capacity-memoryless-deterministic-interference} on p.~\pageref{lemma:capacity-memoryless-deterministic-interference} with $h =  \vect{a}_{k}^{\Htran}\mathbb{E}\{\vect{g}_{kk}\}$, $x= s_k$, $p=p_k$, $\upsilon = \upsilon_k$, and $\sigma^2=0$.
By noting that the signals of different UEs are independent and that the received noise at different APs are independent, we have that
\begin{equation}
\mathbb{E} \{ |\upsilon_k |^2 \} = \sum\limits_{i=1}^{K} p_{i} \mathbb{E} \{  |\vect{a}_{k}^{\Htran}\vect{g}_{ki}|^2  \} - p_{k} \left| \vect{a}_{k}^{\Htran} \mathbb{E}\{ \vect{g}_{kk}\} \right|^2 \!+\!  \vect{a}_{k}^{\Htran}\vect{F}_{k}
		\vect{a}_{k} .
\end{equation}
The SE expression presented in the theorem then follows from Lemma~\ref{lemma:capacity-memoryless-deterministic-interference}.  As a last step, we note that only the fraction $\tau_u/\tau_c$ of the samples is used for uplink data transmission, which results in the lower bound on the capacity that is stated in the theorem and measured in bit/s/Hz.

\subsection{Proof of Corollary~\ref{cor:closed-form_ul}}\label{proof-of-cor:closed-form_ul}
	
	The proof consists of a direct computation of the three types of expectations that appear in \eqref{eq:uplink-instant-SINR-level3}. We begin with
	\begin{align} \nonumber
	\left [ \mathbb{E} \left\{\vect{g}_{ki}\right\} \right ]_l &=	\mathbb{E} \left\{  \vect{v}_{kl}^{\Htran} \vect{D}_{kl} \vect{h}_{il} \right\} = \tr \left( \vect{D}_{kl} \mathbb{E} \left\{    \widehat{\vect{h}}_{il} \widehat{\vect{h}}_{kl}^{\Htran} \right\} \right) \\ &= \begin{cases}
	\sqrt{\eta_k\eta_i}\tau_p  \tr \left(  \vect{D}_{kl}  \vect{R}_{il}\vect{\Psi}_{t_kl}^{-1}\vect{R}_{kl} \right) &   i\in\mathcal{P}_k \\
	0 & i\notin\mathcal{P}_k
	\end{cases}
	\end{align}
	where the second equality follows from the second identity of Lemma~\ref{lemma-matrix-identities} on p.~\pageref{lemma-matrix-identities} and the fact that $\widetilde{\vect{h}}_{il}$ and $\widehat{\vect{h}}_{kl}$ are independent. The third equality follows from \eqref{eq:cross-correlation-channel-estimates} and gives the expression in \eqref{eq:g_ki-expectation}.
Similarly, 
\begin{align}
\left[ \vect{F}_k \right]_{ll}= \sigmaUL\mathbb{E} \left\{ \left\|  \vect{D}_{kl} \vect{v}_{kl} \right\|^2\right\} = \sigmaUL \tr \left( \vect{D}_{kl} \mathbb{E} \left\{    \widehat{\vect{h}}_{kl} \widehat{\vect{h}}_{kl}^{\Htran} \right\} \right) =\sigmaUL \left [ \mathbb{E} \left\{\vect{g}_{kk}\right\} \right ]_l
\end{align}
	where we used the second identity from Lemma~\ref{lemma-matrix-identities} and then identify the expression of $\left [ \mathbb{E} \left\{\vect{g}_{kk}\right\} \right ]_l$.
This gives us the expression in \eqref{eq:F_kl}.

It remains to compute the elements of $\mathbb{E}\left\{ \vect{g}_{ki} \vect{g}_{ki}^{\Htran}  \right\}$. We observe that $\mathbb{E}\left\{ \left[ \vect{g}_{ki} \right]_l \left[ \vect{g}_{ki}^{\star} \right]_r  \right\}=\mathbb{E}\left\{ \left[ \vect{g}_{ki} \right]_l\right\}\mathbb{E}\left\{ \left[ \vect{g}_{ki}^{\star} \right]_r\right\}$ for $ r\neq l$ due to the independence of the channels of different APs. Hence, it only remains to compute
\begin{equation}
\left[	\mathbb{E} \left\{\vect{g}_{ki} \vect{g}_{ki}^{\Htran}\right\} \right]_{ll}  =
\mathbb{E} \left\{\widehat{\vect{h}}_{kl}^{\Htran}\vect{D}_{kl} \vect{h}_{il}\vect{h}_{il}^{\Htran} \vect{D}_{kl}  \widehat{\vect{h}}_{kl} \right\} = \tr\left(\vect{D}_{kl}  \mathbb{E} \left\{ \vect{h}_{il}\vect{h}_{il}^{\Htran}  \vect{D}_{kl}\widehat{\vect{h}}_{kl} \widehat{\vect{h}}_{kl}^{\Htran} \right\} \right)
\end{equation}
where we utilized the second identity from Lemma~\ref{lemma-matrix-identities}.

If $i \not \in \mathcal{P}_k$, we can utilize the independence of $\widehat{\vect{h}}_{kl}$ and $ \vect{h}_{il}$ to
obtain
\begin{align} \nonumber
 \tr\left(\vect{D}_{kl}  \mathbb{E} \left\{ \vect{h}_{il}\vect{h}_{il}^{\Htran} \vect{D}_{kl} \widehat{\vect{h}}_{kl} \widehat{\vect{h}}_{kl}^{\Htran} \right\} \right) &= \tr\left(\vect{D}_{kl}  \mathbb{E} \left\{  \vect{h}_{il}\vect{h}_{il}^{\Htran} \right\}\vect{D}_{kl}\mathbb{E} \left\{  \widehat{\vect{h}}_{kl} \widehat{\vect{h}}_{kl}^{\Htran} \right\} \right)  \\ 
 &= \eta_k \tau_p \tr \left( \vect{D}_{kl} \vect{R}_{il}   \vect{R}_{kl} \vect{\Psi}_{t_k l}^{-1} \vect{R}_{kl} \right)
\end{align}
where we (for brevity) omitted one $\vect{D}_{kl}$ term that does not affect the result.

If $i \in \mathcal{P}_k$, we notice that
	\begingroup
\allowdisplaybreaks
\begin{align} \nonumber
& \tr\left(\vect{D}_{kl}  \mathbb{E} \left\{ \vect{h}_{il}\vect{h}_{il}^{\Htran} \vect{D}_{kl} \widehat{\vect{h}}_{kl} \widehat{\vect{h}}_{kl}^{\Htran} \right\} \right) \\ 
 &=  \tr\left(\vect{D}_{kl}  \mathbb{E} \left\{ \widehat{\vect{h}}_{il} \widehat{\vect{h}}_{il}^{\Htran} \vect{D}_{kl} \widehat{\vect{h}}_{kl} \widehat{\vect{h}}_{kl}^{\Htran} \right\} \right) 
 + \tr\left(\vect{D}_{kl}  \mathbb{E} \left\{ \widetilde{\vect{h}}_{il} \widetilde{\vect{h}}_{il}^{\Htran} \vect{D}_{kl} \widehat{\vect{h}}_{kl} \widehat{\vect{h}}_{kl}^{\Htran} \right\} \right)
\end{align}
 \endgroup
where the equality follows from separating $\vect{h}_{il}$ into its estimate and estimation error. The second term becomes $ \eta_k \tau_p  \tr ( \vect{D}_{kl} \vect{C}_{il} \vect{R}_{kl} \vect{\Psi}_{t_k l}^{-1} \vect{R}_{kl} )$ by utilizing the independence of estimates and estimation error and omitting one $\vect{D}_{kl}$ term. The first term is computed by utilizing the result from \eqref{eq:relation-i-k-estimates} to rewrite the estimate as  $\widehat{\vect{h}}_{il}   =\sqrt{ \frac{\eta_i}{\eta_k} }  \vect{R}_{il}\vect{R}_{kl}^{-1}   \widehat{\vect{h}}_{kl}$: 
\begin{align} \nonumber
&\tr\left(\vect{D}_{kl}  \mathbb{E} \left\{ \widehat{\vect{h}}_{il} \widehat{\vect{h}}_{il}^{\Htran} \vect{D}_{kl} \widehat{\vect{h}}_{kl} \widehat{\vect{h}}_{kl}^{\Htran} \right\} \right)   \\ \nonumber
&= \frac{\eta_i}{\eta_k} \tr\left(\vect{D}_{kl}  \mathbb{E} \left\{ \vect{R}_{il}(\vect{R}_{kl})^{-1} \widehat{\vect{h}}_{kl} \widehat{\vect{h}}_{kl}^{\Htran} (\vect{R}_{kl})^{-1} \vect{R}_{il} \vect{D}_{kl} \widehat{\vect{h}}_{kl} \widehat{\vect{h}}_{kl}^{\Htran} \right\} \right) \\ \nonumber
&= \frac{\eta_i}{\eta_k}  \mathbb{E} \{ |\widehat{\vect{h}}_{kl}^{\Htran} \vect{D}_{kl}   \vect{R}_{il}(\vect{R}_{kl})^{-1} \widehat{\vect{h}}_{kl}|^2 \} \\
&= \eta_k\eta_i\tau_p^{2}  \left|\tr \left(  \vect{D}_{kl} \vect{R}_{il} \vect{\Psi}_{t_k l}^{-1} \vect{R}_{kl}  \right)\right|^2 + \eta_k \tau_p  \tr \left( \vect{D}_{kl} (\vect{R}_{il} - \vect{C}_{il})\vect{R}_{kl} \vect{\Psi}_{t_k l}^{-1} \vect{R}_{kl}\right)
\end{align}
where the last step follows from Lemma~\ref{lemma:fourth-order-moment-Gaussian-vector} on p.~\pageref{lemma:fourth-order-moment-Gaussian-vector} and some algebra. By adding these two terms together, we finally obtain \eqref{eq:gki-second-moment}.
Note that the proof holds even if $\vect{R}_{kl}$ is non-invertible because $\vect{R}_{kl}^{-1}  \widehat{\vect{h}}_{kl} = \sqrt{\eta_k \tau_p} \vect{R}_{kl}^{-1} \vect{R}_{kl} \vect{\Psi}_{t_kl}^{-1} \vect{y}_{t_kl}^{{\rm pilot}} = \sqrt{\eta_k \tau_p} \vect{\Psi}_{t_kl}^{-1} \vect{y}_{t_kl}^{{\rm pilot}}$, where there is no inversion. We are only using the inversion to shorten the notation.

	\section{Proofs from Section~\ref{c-downlink}}
	We report below the proofs from Section~\ref{c-downlink}.
\subsection{Proof of Theorem~\ref{theorem:downlink-capacity-level4}}\label{proof-of-theorem:downlink-capacity-level4}

Since the UE $k$ only has knowledge of the average of the effective channel, $\mathbb{E}\left\{\vect{h}_{k}^{\Htran}\vect{D}_{k}\vect{w}_k\right\}$, the received signal in \eqref{eq:DCC-downlink-three-parts} at UE $k$ can be expressed as
	\begin{align} \label{eq:UE-decoding-level4}
	y_k^{\rm{dl}} &= \mathbb{E}\left\{\vect{h}_{k}^{\Htran}\vect{D}_{k}\vect{w}_k\right\} \varsigma_k + \upsilon_k +  n_k 
	\end{align}
	which is a deterministic channel with the additive noise $n_k$ and the additive interference term
	\begin{align}  \upsilon_k 
	= & \left(\vect{h}_{k}^{\Htran}\vect{D}_{k}\vect{w}_k-\mathbb{E}\left\{\vect{h}_{k}^{\Htran}\vect{D}_{k}\vect{w}_k\right\} \right) \varsigma_k + 
	\condSum{i=1}{i \neq k}{K} \vect{h}_{k}^{\Htran}\vect{D}_{i}\vect{w}_i \varsigma_i.
	\end{align}
	The $\upsilon_k$ term has zero mean (since $\varsigma_i$ has zero mean) and although it contains the desired signal $\varsigma_k$, it is uncorrelated with it since
\begin{equation}
\mathbb{E}\left\{ \varsigma_k^{\star} \upsilon_k \right\} = 
\underbrace{\mathbb{E} \left\{ \left(\vect{h}_{k}^{\Htran}\vect{D}_{k}\vect{w}_k-\mathbb{E}\left\{\vect{h}_{k}^{\Htran}\vect{D}_{k}\vect{w}_k\right\} \right)  \right\}}_{=0} \mathbb{E}\{ | \varsigma_k|^2\} = 0.
\end{equation}
Therefore, we can apply Lemma~\ref{lemma:capacity-memoryless-deterministic-interference} on p.~\pageref{lemma:capacity-memoryless-deterministic-interference} with $h =\mathbb{E}\left\{\vect{h}_{k}^{\Htran}\vect{D}_{k}\vect{w}_k\right\} $, $x= \varsigma_k$, $p=1$, $\upsilon = \upsilon_k$, and $\sigma^2=\sigmaDL$.
By noting that the signals of different UEs are independent, we have that
\begin{equation}
\mathbb{E} \{ |\upsilon_k |^2 \} = \sum\limits_{i=1}^{K} \mathbb{E} \left\{ \left|  \vect{h}_{k}^{\Htran} \vect{D}_{i} \vect{w}_{i} \right|^2 \right\} -  \left| \mathbb{E} \left\{  \vect{h}_{k}^{\Htran} \vect{D}_{k} \vect{w}_{k} \right\} \right|^2.
\end{equation}
The SE expression presented in the theorem then follows from Lemma~\ref{lemma:capacity-memoryless-deterministic-interference} on p.~\pageref{lemma:capacity-memoryless-deterministic-interference}.  
 As a last step, we note that only the fraction $\tau_d/\tau_c$ of the samples is used for downlink data transmission, which results in the lower bound on the capacity that is stated in the theorem and measured in bit/s/Hz.

\subsection{Proof of Theorem~\ref{prop:duality}}\label{proof-of-prop:duality}

	Let $\gamma_{k} = \mathacr{SINR}_{k}^{({\rm ul},\centuatf)}$ denote the value of the effective SINR in \eqref{eq:uplink-instant-SINR-level4-UatF} for the uplink powers $\{p_i: i=1,\ldots,K\}$ and combining vectors $\{ \vect{D}_i {\bf v}_{i}: i=1,\ldots,K\}$. We want to show that $\gamma_{k} = \mathacr{SINR}_{k}^{({\rm dl,\cent})}$ is achievable in the downlink when \eqref{eq:precoding-vectors-duality-proposition} is satisfied for all $i$. Plugging \eqref{eq:precoding-vectors-duality-proposition} into \eqref{eq:downlink-instant-SINR-level4} yields the following SINR constraints:
	\begin{align} 
	\gamma_k  =\frac{{\rho_{k}}\bigg|    \mathbb{E} \left\{ \vect{h}_{k}^{\Htran}  \frac{\vect{D}_{k}{\bf v}_{k}}{\sqrt{\mathbb{E} \{ \| \vect{D}_k{\vect{v}}_{k} \|^2 \}}}  \right\} \bigg|^2 }{ \sum\limits_{i=1}^{K} {\rho_{i}}\mathbb{E} \left\{ \left|  \vect{h}_{k}^{\Htran}  \frac{ \vect{D}_{i} {\bf v}_{i}}{\sqrt{\mathbb{E} \left\{ \left\| \vect{D}_i{\vect{v}}_{i} \right\|^2 \right\}}}   \right|^2 \right\} - {\rho_{k}} \left|   \mathbb{E} \left\{ \vect{h}_{k}^{\Htran}  \frac{ \vect{D}_{k} {\bf v}_{k}}{\sqrt{\mathbb{E} \left\{ \left\| \vect{D}_k{\vect{v}}_{k} \right\|^2\right\}}}  \right\} \right|^2 + \sigma_{\rm dl}^2 }\label{eq:downlink-SINR-level4-proof}
	\end{align}
	for $k=1,\ldots,K$. We define the diagonal matrix $\vect{\Gamma} \in \mathbb{R}^{K \times K}$ with the $k$th diagonal element being
	\begin{align}
	[\vect{\Gamma}]_{kk} =\frac{1}{\gamma_k}\left|    \mathbb{E} \left\{ \vect{h}_{k}^{\Htran}  \frac{\vect{D}_{k} {\bf v}_{k}}{\sqrt{\mathbb{E} \{ {\vect{v}}_{k}^{\Htran}\vect{D}_{k} {\vect{v}}_{k} \}}}  \right\} \right|^2
	\end{align}
	and let $\vect{\Sigma} \in \mathbb{R}^{K \times K}$ be the matrix whose $(k,i)$th element is
	\begin{align}
	[\vect{\Sigma}]_{ki} = \mathbb{E} \left\{ \left|  \vect{h}_{k}^{\Htran}  \frac{\vect{D}_{i}{\bf v}_{i}}{\sqrt{\mathbb{E} \{ {\vect{v}}_{i}^{\Htran}\vect{D}_{i} {\vect{v}}_{i} \}}}   \right|^2 \right\} - \begin{cases}
	0 & i \ne k \\
	\gamma_k[\vect{\Gamma}]_{kk}& i=k.
	\end{cases}
	\end{align}
	Using these matrices, the SINR constraint in \eqref{eq:downlink-SINR-level4-proof} can be expressed as
	\begin{align}
[\vect{\Gamma}]_{kk} = \frac{ \sum\limits_{i=1}^{K}\rho_i [\vect{\Sigma}]_{ki} + \sigma_{\rm dl}^2 }{\rho_k}.
	\end{align}
	By rearranging this equation, we obtain
$\sigma_{\rm dl}^2 = \rho_k [\boldsymbol{\Gamma}]_{kk} - \sum\nolimits_{i=1}^{K}\rho_i [\boldsymbol{\Sigma}]_{ki}$.
	The $K$ constraints can be written in matrix form as $
	{\bf 1}_K\sigma_{\rm dl}^2 = \left(\boldsymbol{\Gamma} - \boldsymbol{\Sigma}\right)\boldsymbol{\rho}$ 
	with $\boldsymbol{\rho} =[\rho_1 \, \ldots \, \rho_K]^{\Ttran}$ being the downlink transmit power vector. The SINR constraints in \eqref{eq:downlink-SINR-level4-proof} are thus satisfied if 
	\begin{align}\label{eq:optimal-power-proof}
	\boldsymbol{\rho} = \left(\boldsymbol{\Gamma} - \boldsymbol{\Sigma}\right)^{-1}{\bf 1}_K\sigma_{\rm dl}^2.
	\end{align}
	This is a feasible power if $\boldsymbol{\Gamma} - \boldsymbol{\Sigma}$ is invertible, which always holds when ${\bf p}=[p_1 \, \ldots \, p_K]^{\Ttran}$ is feasible. To show this, we notice that the $K$ uplink SINR conditions can be expressed in a similar form where $\boldsymbol{\Sigma}$ is replaced by $\boldsymbol{\Sigma}^{\Ttran}$
	such that ${\bf p} = \left(\boldsymbol{\Gamma} - \boldsymbol{\Sigma}^{\Ttran}\right)^{-1}{\bf 1}_K\sigma_{\rm ul}^2$.
	Since the eigenvalues of $\boldsymbol{\Gamma} - \boldsymbol{\Sigma}$ and $\boldsymbol{\Gamma} - \boldsymbol{\Sigma}^{\Ttran}$ are the same and the uplink SINR conditions are satisfied by assumption, we can always select the downlink powers according to \eqref{eq:optimal-power-proof}. Substituting ${\bf 1}_K = \frac{1}{\sigma_{\rm ul}^2}\left(\vect{\Gamma} - \vect{\Sigma}^{\Ttran}\right){\bf p}$ into \eqref{eq:optimal-power-proof} yields 
	\begin{align}
	\boldsymbol{\rho} = \frac{\sigma_{\rm dl}^2}{\sigma_{\rm ul}^2}\left(\boldsymbol{\Gamma} - \boldsymbol{\Sigma}\right)^{-1}\left(\boldsymbol{\Gamma} - \boldsymbol{\Sigma}^{\Ttran}\right){\bf p}. \label{eq:downlink-duality-power}
	\end{align}
	The total transmit power condition now follows from direct computation by noting that ${\bf 1}_K^{\Ttran}\left(\vect{\Gamma} - \vect{\Sigma}\right)^{-1}{\bf 1}_K={\bf 1}_K^{\Ttran}\left(\vect{\Gamma} - \vect{\Sigma}^{\Ttran}\right)^{-1}{\bf 1}_K$.
	
To complete the proof, we need to show the power allocation coefficients obtained by \eqref{eq:downlink-duality-power} are positive. Please see \cite[Proof of Theorem~4.8]{massivemimobook} for the final technical details.

\backmatter  

\printbibliography

\end{document}